\documentstyle[psfig,equations]{aipproc3}

\catcode`@=11\def\tableofcontents{\section*{\contentsname
\@mkboth{\uppercase{\contentsname}}{\uppercase{\contentsname}}}%
\@starttoc{toc}}
\catcode`@=12

\input epsf

\begin{document}
\large
\begin{flushright}
FERMILAB-Conf-00/279-T  \\
SCIPP 00/37     \\
hep--ph/0010338 \\
October 31, 2000 \\
\end{flushright}
\normalsize
\title{Report of the Tevatron Higgs Working Group}

\author{


{\bf Marcela Carena}\\ {\it Fermi National Accelerator Laboratory, P.O.
Box 500, Batavia, IL 60510} \\[2.mm]
{\bf John S. Conway}\\ {\it Department of Physics and Astronomy,
Rutgers University, Piscataway,
NJ 08854}\\[2.mm]
{\bf Howard E. Haber}\\ {\it Santa Cruz Institute for Particle
Physics, University of California, Santa Cruz, CA 95064}\\[2.mm]
{\bf John D. Hobbs}\\ {\it SUNY at Stony Brook, Department of Physics,
Stony Brook, NY 11794} \\[6.mm]

\underline{Working Group Members} \\ \vspace{0.1in}

\begin{tabbing}
\hspace*{3.6in} \= \hspace*{3.6in} \kill
     Michael~Albrow (Fermilab)     \> 
     Howard~Baer (Florida State)   \\
     Emanuela~Barberis (LBNL)  \>
     Armando~A.~Barrientos~Bendez\a'{u} (Hamburg) \\
     Pushpalatha~Bhat (Fermilab)  \>
     Alexander~Belyaev (Moscow State) \\ 
     Csaba~Bal\a'{a}zs (Hawaii) \>
     Wasiq~Bokhari (Pennsylvania)  \\
     Francesca~Borzumati (KEK)  \>
     Dhiman~Chakraborty (Stony Brook)  \\       
     J.~Antonio~Coarasa (Aut\a`{o}noma Barcelona) \>
     Ray~Culbertson (Fermilab)  \\
     Regina~Demina (Kansas State)  \>          
     J.~Lorenzo~D\a'{\i}az-Cruz (Puebla) \\ 
     Duane~Dicus (Texas, Austin)  \>
     Bogdan~Dobrescu (Yale) \\ 
     Tommaso~Dorigo (INFN Padova)   \>
     Herbi~Dreiner (Bonn) \\
     Keith~Ellis (Fermilab)  \>
     Henry~Frisch (Chicago)  \\          
     David~Garcia (CERN)    \>
     Russell~Gilmartin (Florida State) \\ 
     Mar\a'{\i}a~Concepci\a'{o}n~Gonzalez-Garc\a'{\i}a (Val\a`{e}ncia) \>
     Jaume~Guasch (Karlsruhe)  \\
     Anna~Goussiou (Stony Brook)  \>
     Tao~Han (Wisconsin)            \\   
     Brian~W.~Harris (Argonne)    \>
     Hong-Jian~He (Texas, Austin)   \\
     David~Hedin (Northern Illinois)  \>
     Sven~Heinemeyer (Brookhaven)   \\
     Ulrich~Heintz (Boston)   \>
     Wolfgang~Hollik (Karlsruhe)  \\
     Richard~Jesik (Indiana)    \>
     Ben~Kilminster (Rochester)  \\
     Bernd~A.~Kniehl (Hamburg)    \>
     Jean-Lo\"{\i}c~Kneur (Montpellier)  \\
     Mark~Kruse (Rochester)     \>
     Stephen~Kuhlmann (Argonne)   \\
     Stefano~Lami (INFN Pisa)  \>
     Greg~Landsberg (Brown)   \\
     Sergio~M.~Lietti  (S\~{a}o Paulo) \>
     Dmitri~Litvintsev (Fermilab)  \\
     Charles~Loomis (California, Santa Cruz)   \>
     Arnaud~Lucotte (ISN Grenoble)   \\
     Konstantin~Matchev (CERN)   \>
     Stephen~Mrenna (California, Davis)  \\          
     Pasha Murat (Fermilab)    \>
     Sergio~F.~Novaes (S\~{a}o Paulo) \\ 
     Nir~Polonsky (MIT)   \>
     Harrison~Prosper (Florida State)  \\
     Alexander~Pukhov (Moscow State)   \>
     Alberto~Ribon (INFN Padova)   \\
     Maria~Roco (Fermilab)         \>
     Andrey~Rostovtsev (DESY)  \\
     Michael~Schmitt (Northwestern)  \>        
     Vladimir~Sirotenko (Fermilab)    \\
     Robert~Snihur (University College London)   \>
     Joan~Sol\a`{a} (Aut\a`{o}noma Barcelona)  \\ 
     Michael~Spira (Paul Scherrer Institute)  \>
     Tim~Stelzer (Illinois, Urbana)   \\
     Zack~Sullivan (Argonne)    \>
     Tim~M.P.~Tait (Argonne)    \\
     Xerxes~Tata (Hawaii)     \>        
     Andr\a'{e}~S.~Turcot (Brookhaven)   \\
     Juan~Valls (Rutgers)       \>  
     Sini\v{s}a~Veseli (Fermilab)    \\
     Roc\a'{\i}o~Vilar (Cantabria)  \>      
     Gordon~Watts (Washington, Seattle)  \\ 
     Carlos E.M.~Wagner (Argonne and Chicago)  \>
     Georg~Weiglein (CERN)     \\
     Scott~Willenbrock (Illinois, Urbana)  \> 
     John~Womersley (Fermilab)   \\
     Weiming~Yao (LBNL)           \>
     Chien-Peng~Yuan (Michigan State)   \\
     Dieter~Zeppenfeld (Wisconsin)  \>       
     Ren-Jie~Zhang (Wisconsin)       \\
\end{tabbing}

}



\newcommand{\met}{\,/\!\!\!\!E_{T}}
\newcommand{\et}{E_T}
\newcommand{\pt}{p_T}

\newcommand{\pp}{p\bar{p}}
\newcommand{\qq}{q\bar{q}}
\newcommand{\bb}{b\bar{b}}
\newcommand{\cc}{c\bar{c}}

\newcommand{\ttbar}{t\bar{t}}
\newcommand{\lpm}{\ell^+\ell^-}
\newcommand{\nn}{\nu\bar{\nu}}

\newcommand{\trilep}{\ell^\pm\ell^{\prime\pm}\ell^\mp}
\newcommand{\lsdilep}{\ell^\pm\ell^\pm jj}

\newcommand{\lsim}{\stackrel{<}{\sim}}
\newcommand{\gsim}{\stackrel{>}{\sim}}

\newcommand{\etal}{{\em et al.}}


\maketitle

\begin{abstract}
Despite the success of the Standard Model (SM), which provides a
superb description of a wide range of experimental particle physics
data, the dynamics responsible for electroweak symmetry breaking is
still unknown.  Its elucidation remains one of the primary goals of
future high energy physics experimentation.  Present day global fits to
precision electroweak data based on the Standard Model favor the
existence of a weakly-interacting scalar Higgs boson, which is a
remnant of elementary scalar dynamics that drives electroweak symmetry
breaking.  The only known viable theoretical framework
incorporating light elementary scalar fields employs ``low-energy''
supersymmetry, where the scale of supersymmetry breaking is
${\cal O}$(1~TeV).  The Higgs sector of the Minimal
Supersymmetric extension of the Standard Model (MSSM) is of particular
interest because it predicts the existence of a light CP-even neutral
Higgs boson with a mass below about 130~GeV.  Moreover, over a
significant portion of the MSSM parameter space, the properties of
this scalar are indistinguishable from those of the SM Higgs boson.

In Run 2 at the Tevatron, the upgraded CDF and D\O\ experiments will
enjoy greatly enhanced sensitivity in the search for the SM Higgs
boson and the Higgs bosons of the MSSM.  This report presents the
theoretical analysis relevant for Higgs physics at the Tevatron
collider and documents the Higgs Working Group simulations to estimate
the discovery reach of an upgraded Tevatron for the SM and MSSM Higgs
bosons.  Based on a simple detector simulation, we have determined the
integrated luminosity necessary to discover the SM Higgs in the mass
range 100--190~GeV.  The first phase of the Run~2 Higgs search, with a
total integrated luminosity of 2 fb$^{-1}$ per detector, will provide
a 95\% CL exclusion sensitivity comparable to that expected at the
end of the LEP2 run.  With 10 fb$^{-1}$ per detector, this exclusion
will extend up to Higgs masses of 180~GeV, and a tantalizing $3\sigma$
effect will be visible if the Higgs mass lies below 125~GeV.  With 25
fb$^{-1}$ of integrated luminosity per detector, evidence for SM Higgs
production at the 3$\sigma$ level is possible for Higgs masses up to
180~GeV.  However, the discovery reach is much less impressive for
achieving a 5$\sigma$ Higgs boson signal.  Even with 30~fb$^{-1}$ per
detector, only Higgs bosons with masses up to about 130~GeV can be
detected with 5$\sigma$ significance.  These results can also be
re-interpreted in the MSSM framework and yield the required
luminosities to discover at least one Higgs boson of the MSSM Higgs
sector.  With 5--10~fb$^{-1}$ of data per detector, it will be
possible to exclude at 95\% CL nearly the entire MSSM Higgs parameter
space, whereas 20--30~fb$^{-1}$ is required to obtain a 5$\sigma$
Higgs discovery over a significant portion of the parameter space.
Moreover, in one interesting region of the MSSM parameter space (at
large $\tan\beta$), the associated production of a Higgs boson and a
$b\bar b$ pair is significantly enhanced and provides potential for
discovering a non-SM-like Higgs boson in Run~2.  Further studies
related to charged Higgs boson searches and exploiting other search
modes of the neutral Higgs bosons are underway and may enhance the
above discovery potential.
\end{abstract}

\clearpage
\pagestyle{plain}
\pagenumbering{roman}
\parskip 0.15ex
\tableofcontents
\parskip 0pt plus 1pt

\clearpage
\pagenumbering{arabic}

\section{Theoretical Aspects of Higgs Physics at the Tevatron}
\makeatletter
\def\@cite#1#2{{[{#1}]\if@tempswa\typeout
{IJCGA warning: optional citation argument
ignored: `#2'} \fi}}


\newcount\@tempcntc
\def\@citex[#1]#2{\if@filesw\immediate\write\@auxout{\string\citation{#2}}\fi
  \@tempcnta\z@\@tempcntb\m@ne\def\@citea{}\@cite{\@for\@citeb:=#2\do
    {\@ifundefined
       {b@\@citeb}{\@citeo\@tempcntb\m@ne\@citea\def\@citea{,}{\bf ?}\@warning
       {Citation `\@citeb' on page \thepage \space undefined}}%
    {\setbox\z@\hbox{\global\@tempcntc0\csname b@\@citeb\endcsname\relax}%
     \ifnum\@tempcntc=\z@ \@citeo\@tempcntb\m@ne
       \@citea\def\@citea{,}\hbox{\csname b@\@citeb\endcsname}%
     \else
      \advance\@tempcntb\@ne
      \ifnum\@tempcntb=\@tempcntc
      \else\advance\@tempcntb\m@ne\@citeo
      \@tempcnta\@tempcntc\@tempcntb\@tempcntc\fi\fi}}\@citeo}{#1}}
\def\@citeo{\ifnum\@tempcnta>\@tempcntb\else\@citea\def\@citea{,}%
  \ifnum\@tempcnta=\@tempcntb\the\@tempcnta\else
   {\advance\@tempcnta\@ne\ifnum\@tempcnta=\@tempcntb \else \def\@citea{--}\fi
    \advance\@tempcnta\m@ne\the\@tempcnta\@citea\the\@tempcntb}\fi\fi}
\makeatother

\def\mpl{M_{\rm PL}}
\def\beq{\begin{equation}}
\def\eeq{\end{equation}}
\def\crr{\crcr\noalign{\vskip .1in}}
\newcommand{\mathbold}[1]{\mbox{\boldmath $\bf#1$}}
\def\beqa{\begin{eqalignno}}
\def\eeqa{\end{eqalignno}}
\def\beqno{\begin{eqalignno}}
\def\eeqno{\end{eqalignno}}
\def\ifmath#1{\relax\ifmmode #1\else $#1$\fi}
\def\half{\ifmath{{\textstyle{1 \over 2}}}}
\def\lsim{\mathrel{\raise.3ex\hbox{$<$\kern-.75em\lower1ex\hbox{$\sim$}}}}
\def\gsim{\mathrel{\raise.3ex\hbox{$>$\kern-.75em\lower1ex\hbox{$\sim$}}}}
\def\eq#1{eq.~(\ref{#1})}
\def\fig#1{fig.~\ref{#1}}
\def\Fig#1{Fig.~\ref{#1}}
\def\figs#1#2{figs.~\ref{#1} and \ref{#2}}
\def\Figs#1#2{Figs.~\ref{#1} and \ref{#2}}
\def\Ref#1{ref.~\cite{#1}}
\def\Refs#1#2{refs.~\cite{#1} and \cite{#2}}
\def\Rref#1{Ref.~\cite{#1}}
\def\Rrefs#1#2{Refs.~\cite{#1} and \cite{#2}}
\def\eqs#1#2{eqs.~(\ref{#1})--(\ref{#2})}
\def\Eq#1{Eq.~(\ref{#1})}
\def\Eqs#1#2{Eqs.~(\ref{#1})--(\ref{#2})}
\def\eqns#1#2{eqs.~(\ref{#1}) and (\ref{#2})}
\def\tanb{\tan\beta}
\def\sinb{\sin\beta}
\def\cosb{\cos\beta}
\def\sina{\sin\alpha}
\def\cosa{\cos\alpha}
\def\sinbma{\sin(\beta-\alpha)}
\def\cosbma{\cos(\beta-\alpha)}
\def\sinbmaii{\sin^2(\beta-\alpha)}
\def\cosbmaii{\cos^2(\beta-\alpha)}
\def\hsm{h_{\rm SM}}
\def\mhsm{m_{h_{\rm SM}}}
\def\hl{h}
\def\ha{A}
\def\hh{H}
\def\hpm{H^\pm}
\def\mha{m_{\ha}}
\def\mhl{m_{\hl}}
\def\mhh{m_{\hh}}
\def\mhpm{m_{\hpm}}
\def\mhmax{m_h^{\rm max}}
\def\mz{m_Z}
\def\mw{m_W}
\def\mww{m_W^2}
\def\mzz{m_Z^2}
\def\mt{m_t}
\def\mb{m_b}
\def\mstopa{M_{\widetilde t_1}}
\def\mstopb{M_{\widetilde t_2}}
\def\msusy{M_{\rm S}}
\def\msusyy{M_{\rm S}^2}
\def\MSUSY{M_{\rm SUSY}}
\def\MSUSYY{M^2_{\rm SUSY}}
\def\mgut{M_{\rm X}}
\def\ie{{\it i.e.}}
\def\eg{{\it e.g.}}
\def\etc{{\it etc.}}
\def\vs{{\it vs.}}
\def\opcit{{\it op.~cit.}}
\def\SM{Standard Model}
\def\phm{\phantom{-}}
\def\ls#1{\ifmath{_{\lower1.5pt\hbox{$\scriptstyle #1$}}}}
\def\app#1#2#3{{\sl Act. Phys. Pol. }{\bf B#1} (#2) #3}
\def\apa#1#2#3{{\sl Act. Phys. Austr.}{\bf #1} (#2) #3}
\def\ppnp#1#2#3{{\sl Prog. Part. Nucl. Phys. }{\bf #1} (#2) #3}
\def\npb#1#2#3{{\sl Nucl. Phys. }{\bf B#1} (#2) #3}
\def\jpa#1#2#3{{\sl J. Phys. }{\bf A#1} (#2) #3}
\def\plb#1#2#3{{\sl Phys. Lett. }{\bf B#1} (#2) #3}
\def\prd#1#2#3{{\sl Phys. Rev. }{\bf D#1} (#2) #3}
\def\pR#1#2#3{{\sl Phys. Rev. }{\bf #1} (#2) #3}
\def\prl#1#2#3{{\sl Phys. Rev. Lett. }{\bf #1} (#2) #3}
\def\prc#1#2#3{{\sl Phys. Reports }{\bf #1} (#2) #3}
\def\cpc#1#2#3{{\sl Comp. Phys. Commun. }{\bf #1} (#2) #3}
\def\nim#1#2#3{{\sl Nucl. Inst. Meth. }{\bf #1} (#2) #3}
\def\pr#1#2#3{{\sl Phys. Reports }{\bf #1} (#2) #3}
\def\sovnp#1#2#3{{\sl Sov. J. Nucl. Phys. }{\bf #1} (#2) #3}
\def\jl#1#2#3{{\sl JETP Lett. }{\bf #1} (#2) #3}
\def\jet#1#2#3{{\sl JETP Lett. }{\bf #1} (#2) #3}
\def\zpc#1#2#3{{\sl Z. Phys. }{\bf C#1} (#2) #3}
\def\ptp#1#2#3{{\sl Prog.~Theor.~Phys.~}{\bf #1} (#2) #3}
\def\nca#1#2#3{{\sl Nouvo~Cim.~}{\bf#1A} (#2) #3}
\def\hpa#1#2#3{{\sl Helv.~Phys.~Acta~}{\bf #1} (#2) #3}
\def\aop#1#2#3{{\sl Ann.~of~Phys.~}{\bf #1} (#2) #3}
\def\fP#1#2#3{{\sl Fortschr.~Phys.~}{\bf #1} (#2) #3}
\def\9{\phantom 0}     


\subsection{Introduction: the Quest for the Origin of Electroweak
Symmetry Breaking}

With the discovery of the top quark at the Tevatron \cite{Tevatron},
the Standard Model of particle physics appears close to final
experimental verification.  Ten years of precision measurements of
electroweak observables at LEP, SLC and the Tevatron have failed to
find any definitive departures from Standard Model predictions
\cite{ELW,Arno,strom}.  In some cases, theoretical predictions have been
checked with an accuracy of one part in a thousand or better.  The
consistency of these calculations is evidenced by the excellent
agreement between the value of the top quark mass measured directly at
the Tevatron, and the corresponding value deduced from precisely
measured electroweak observables at LEP and SLC that are sensitive to
top-quark loop radiative corrections.

Although the global analysis of electroweak
observables provides a superb fit to the Standard Model predictions,
there is still no {\it direct}
experimental evidence for the underlying dynamics responsible for
electroweak symmetry breaking.  The observed masses of the $W^\pm$
and $Z$ bosons can be understood as a consequence
of three Goldstone bosons ($G^\pm$ and $G^0$) that
end up as the longitudinal components of the gauge bosons.
However, the origin of the
Goldstone bosons still demands an explanation.  The electroweak symmetry
breaking dynamics that is employed by the Standard Model posits a
self-interacting complex doublet of scalar fields, which consists of four
real degrees of freedom.  Renormalizable interactions are
arranged in such a way that the neutral component of the scalar doublet
acquires a vacuum expectation value, $v=246$~GeV, which sets the scale
of electroweak symmetry breaking.
Consequently, three massless Goldstone bosons are generated, while
the fourth scalar degree of freedom that remains in the physical spectrum
is the CP-even neutral Higgs boson of the Standard Model.
It is further assumed in the SM that the scalar doublet also couples to fermions through
Yukawa interactions.  After electroweak symmetry breaking, these interactions
are responsible for the generation of quark and charged
lepton masses.

The self-interacting scalar field is only one  model of
electroweak symmetry breaking; other approaches,
based on very different dynamics, are also possible.
For example, one can introduce new fermions and new dynamics
({\it i.e.}, new forces), in
which the Goldstone bosons are a consequence of the strong
binding of the new fermion fields \cite{techni}.  Present
experimental data are not sufficient to identify with certainty the
nature of the dynamics responsible for electroweak symmetry breaking.
The quest to understand electroweak symmetry breaking requires continued
experimentation at present and future
colliders: the upgraded Tevatron, the
LHC and proposed lepton colliders under development.

As described above, the Standard Model is clearly a very good
approximation to the physics of elementary particles and their
interactions at an energy scale of ${\cal O}(100)$~GeV and below.  However,
theoretical considerations teach us that the Standard Model is not the
ultimate theory of the fundamental particles and their interactions.
At an energy scale above the Planck scale, $\mpl\simeq 10^{19}$~GeV, quantum
gravitational effects become significant
and the Standard Model must
be replaced by a more fundamental theory that incorporates
gravity.\footnote{Similar conclusions
also apply to recently proposed extra-dimensional theories in which
quantum gravitational effects can become
significant at energies scales as low as ${\cal{O}}$(1 TeV) \cite{ED}.}
It is also possible that the Standard Model breaks down at some energy scale
(called $\Lambda$) below the Planck scale.
In this case, the Standard Model degrees of freedom are no longer
adequate for describing the theory above $\Lambda$ and new physics
must become relevant.  One possible signal of this occurrence lies in
the behavior of the Standard Model couplings.
The Standard Model is not an
asymptotically free theory since some of the couplings ({\it e.g.}, the
U(1) gauge coupling, the Higgs--top-quark Yukawa coupling, and the Higgs
self-coupling) eventually blow up at some high energy scale.   Among
these couplings, only the Higgs self-coupling may blow up at an energy
scale below $\mpl$.  Of course, there may be other
experimental or theoretical hints that
new degrees of freedom exist at some high energy scale below $\mpl$.
For example, the recent experimental evidence for neutrino
masses of order $10^{-2}$~eV or below cannot be strictly explained in
the Standard Model.  Yet, one can easily write down a dimension-5
operator that is suppressed by $v/\Lambda$, which is responsible for
neutrino masses.  If $m_\nu=10^{-2}$~eV, then one obtains as a rough estimate
$\Lambda\sim 10^{15}$~GeV.


It is clear from the above discussion that
the Standard Model is not a {\it fundamental} theory;
at best, it is an {\it effective field theory}~\cite{EFT}.
At an energy scale below $\Lambda$, the Standard Model (with
higher-dimension operators to parameterize the physics generated at
$\Lambda$) provides an extremely good description of all observable
phenomena.  Therefore, an essential question that future experiments
must address is: what is the minimum scale $\Lambda$ at which new
physics beyond the Standard Model must enter?
The search for the origin of electroweak symmetry breaking and the quest
to identify $\Lambda$ are intimately tied together.  We can consider two
scenarios.  In the first scenario, electroweak symmetry breaking
dynamics results in the existence of a single Higgs boson as posited by
the Standard Model.  In this case, one would ask whether new
phenomena beyond the Standard Model must enter at an energy scale $\Lambda$
that is accessible to experiment.  In the second scenario, electroweak
symmetry breaking dynamics does not result in a weakly-coupled Higgs
boson as assumed in the Standard Model.  In this case, the effective
theory that describes current data is a theory that contains the
Standard Model fields {\it excluding} the
Higgs boson.  In such an approach, the latter effective field theory
{\it must} break down at $\Lambda\simeq{\cal O}(1~{\rm TeV})$
in order to restore the unitarity of the theory, and new
physics associated with the electroweak symmetry breaking dynamics must
enter.

Although current data provides no {\it direct} evidence to distinguish
between the two scenarios just described, there is
indirect evidence that could be interpreted as favoring the first
approach.  Namely, the global Standard Model fit to electroweak data
takes the Higgs mass as a variable to be fitted.  The results of the LEP
Electroweak Working Group analysis yields \cite{Arno}:
\beq
\mhsm=67^{+60}_{-33}~{\rm GeV}\,.
\eeq
In fact, direct searches at LEP show no evidence for the Higgs boson,
and imply that $\mhsm> 107.9$~GeV \cite{LEPHiggs}
at the 95\% CL.\footnote{Preliminary
results from the 2000 LEP run show no clear evidence of the Higgs
boson, with a corresponding Higgs mass lower limit of 113.2~GeV (at
95\% CL) \cite{LEPHiggsprelim,tomjunk}.}
Thus, it probably is more useful
to quote the 95\% CL upper limit that is obtained in the global Standard
Model fit \cite{Arno}:~\footnote{The 95\% CL upper limit can change
by as much as $\sim 20$~GeV depending on the analysis.  
More recent upper Higgs mass limits, ranging between 170~GeV and 210~GeV, were
presented at the ICHEP 2000 meeting in Osaka, Japan and reviewed
in \Ref{strom}.}
\beq \label{smhiggslim}
\mhsm<188~{\rm GeV}\quad {\rm at}~95\%~{\rm CL}\,.
\eeq
These results reflect the logarithmic sensitivity to the Higgs mass via
the virtual Higgs loop contributions to the various electroweak
observables.  The Higgs mass range above is consistent
with a weakly-coupled Higgs scalar that is expected to emerge from the
Standard Model scalar dynamics (although the Standard Model does not
predict the mass of the Higgs boson; rather it relates it to the
strength of the scalar self-coupling).

Henceforth, we shall take the above result as an indication that
the Standard Model (with a weakly-coupled Higgs boson as suggested
above) is the appropriate effective field theory at the 100~GeV scale.
If this is the case, then the eventual discovery of the Higgs boson will
have a profound effect on the determination of $\Lambda$, the scale at
which the Standard Model must break down.  The key parameter for
constraining $\Lambda$
is the Higgs mass, $\mhsm$.  If $\mhsm$ is too large, then the
Higgs self-coupling blows up at some scale $\Lambda$ below the
Planck scale \cite{hambye}.  If $\mhsm$ is too small, then the Higgs
potential develops
a second (global) minimum at a large value of the scalar field of order
$\Lambda$ \cite{quiros}.  Thus new physics must enter at a scale
$\Lambda$
or below in order that the true minimum of the theory correspond to the
observed SU(2)$\times$U(1) broken vacuum with $v=246$~GeV.
Thus, given a value of
$\Lambda$, one can compute the minimum and maximum Higgs mass allowed.
The results of this computation (with shaded bands indicating the
theoretical uncertainty of the result) are illustrated in \fig{trivial}.

\begin{figure}
  \begin{center}
\centerline{\psfig{file=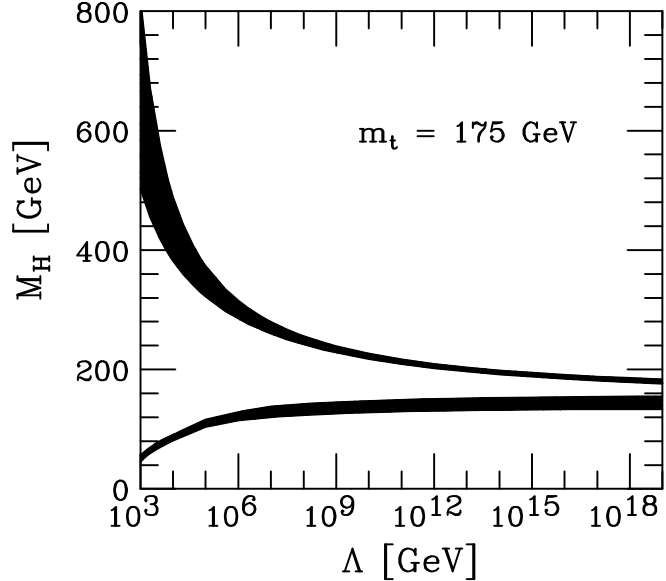,width=9cm}}
  \end{center}
  \caption[0]{\label{trivial} The lower \protect\cite{quiros} and the
upper \protect\cite{hambye} Higgs mass bounds as a function of the
energy scale $\Lambda$ at which the Standard Model breaks down,
assuming $M_t=175$~GeV and $\alpha_s(m_Z)=0.118$.  The shaded
areas above reflect the theoretical uncertainties in the
calculations of the Higgs mass bounds.  This figure is taken from
\protect\Ref{Riesselmann}.}
\end{figure}

Three Higgs boson mass ranges are of particular
interest:
\begin{enumerate}
\item 110~GeV~$\lsim \mhsm\lsim$~130~GeV
\item 130~GeV~$\lsim \mhsm\lsim$~180~GeV
\item 180~GeV~$\lsim \mhsm\lsim$~190~GeV
\end{enumerate}
In mass range 1, the Higgs boson mass lies above the present
direct LEP search limit.
Moreover, if the Higgs boson were discovered in this range, then
\fig{trivial} would imply that $\Lambda<\mpl$.  Finally, as we shall explain
in Section~I.C, this is the mass range expected in the minimal
supersymmetric model. Mass range 2 corresponds to a range of Higgs
masses which would be consistent with $\Lambda=\mpl$.
In such a scenario, the Standard Model could in principle remain viable
all the way up to the Planck scale.
Finally, in mass range 3, we are
still consistent with the 95\% CL Higgs mass limit quoted
in \eq{smhiggslim}.
Again, a Higgs boson discovered in this mass range would imply that
$\Lambda<\mpl$.

\subsection{The Standard Model Higgs Boson}

In the Standard Model, the Higgs mass is given by: $\mhsm^2=\lambda
v^2$, where $\lambda$ is the Higgs self-coupling parameter.  Since
$\lambda$ is unknown at present, the value of the Standard Model Higgs
mass is not predicted.  However, other theoretical considerations,
discussed in Section~I.A, place constraints on the Higgs mass
as exhibited in \fig{trivial}.  In contrast, the Higgs couplings to
fermions and gauge bosons are predicted by the theory.  In particular,
the Higgs couplings are proportional to the corresponding particle
masses, as shown in \fig{smcouplings}.  The vertices of
\fig{smcouplings}
govern the most important features of Higgs phenomenology at colliders.
In Higgs production and decay processes, the dominant mechanisms involve
the coupling of the Higgs boson to the $W^\pm$, $Z$ and/or
the third generation quarks and leptons.
It should be noted that a $\hsm gg$ ($g$=gluon)
coupling is induced by virtue of a one-loop graph
in which the Higgs boson couples to a virtual $t\bar t$ pair.
Likewise, a $\hsm\gamma\gamma$ coupling is generated, although in this
case the one-loop graph in which the Higgs boson couples to
a virtual $W^+W^-$ pair is the dominant contribution.   Further details
of Standard Higgs boson properties are given in \Ref{hhg}.
\vskip 0.2in
\begin{figure}[htb]
\centering
\centerline{\psfig{file=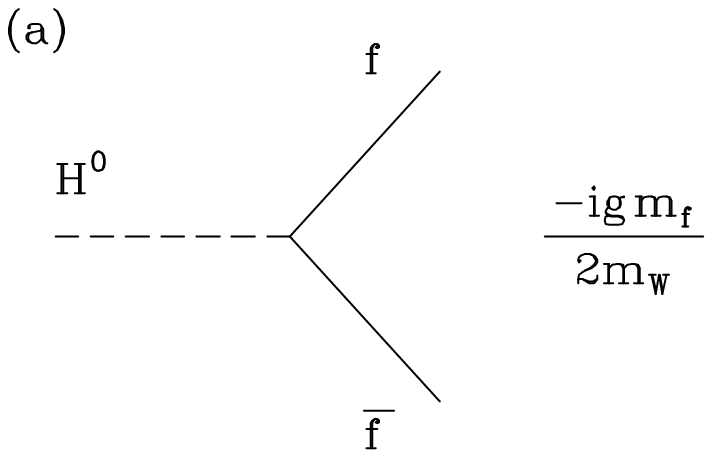,width=5cm,angle=90}
\hskip1pc \psfig{file=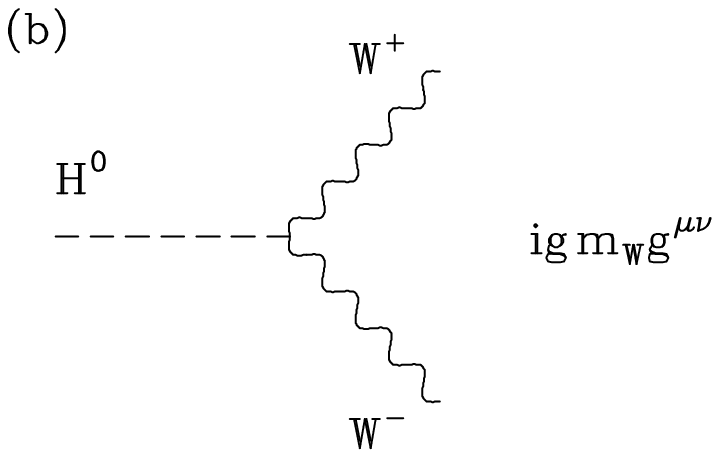,width=5cm,angle=90}
\hskip1pc
\psfig{file=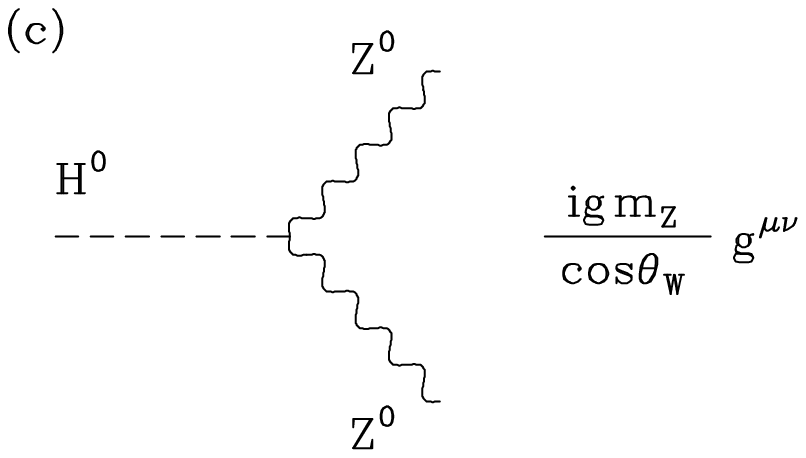,width=5cm,angle=90}}
\vskip1pc
\caption[0]{\label{smcouplings}Standard Model Higgs boson
interactions at tree-level.}
\end{figure}

\subsubsection{Present Status of the Standard Model Higgs Boson Search}

Before turning to the relevant \SM\ Higgs production processes and
decay modes at the Tevatron,
we briefly comment on the expected status of the Higgs search
at the end of the final year of the LEP2 collider run in 2000.
In 2000, the LEP2 collider operated at a variety of
center-of-mass energies between 203 and 208 GeV (with most data taken at
204.9 GeV and 206.8 GeV).  The total integrated luminosity 
exceeds
100~pb$^{-1}$ per experiment.  If no Higgs signal is seen, then the
95\%~CL lower limit on the \SM\ Higgs mass is projected to be
$\mhsm\gsim 115$~GeV.  In order to discover the Higgs boson at the
$5\sigma$ level, one must have $\mhsm\lsim 112$~GeV.
These results are based on \fig{SM95}~$(a)$ and $(b)$, taken from
\Ref{egross}.

\begin{figure}[!ht]
\centering
\centerline{\psfig{file=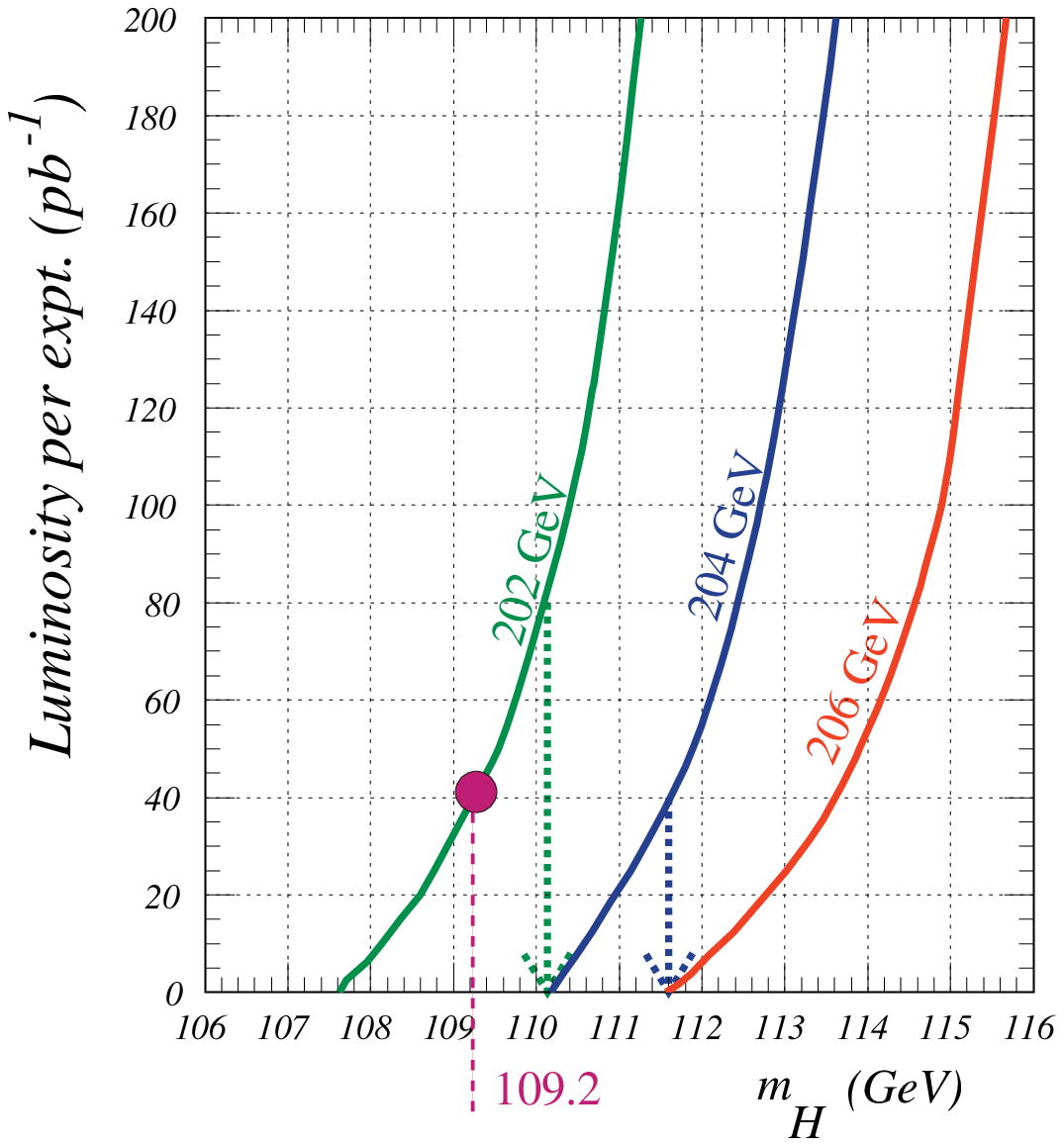,width=8cm}
\hfill
\psfig{file=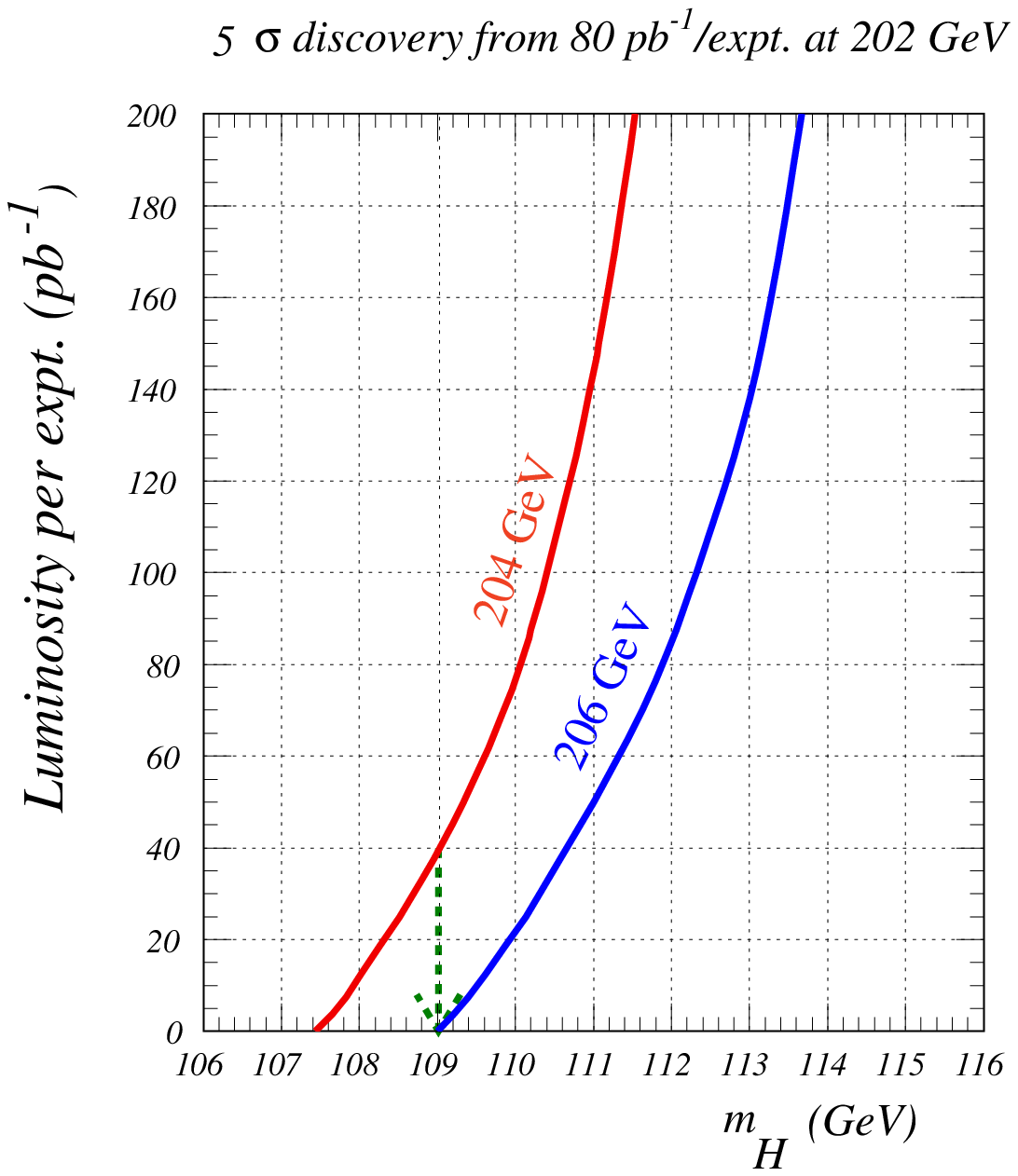,width=8cm}}
\vskip1pc
\caption[0]{ \label{SM95} Projected LEP2 limits on the Standard Model Higgs boson
mass (in GeV) as a function of integrated luminosity per experiment,
for different assumptions concerning the operation of
the machine: (a)  95\% CL exclusion and
             (b)  $5\sigma$ discovery.
This figure is taken from \Ref{egross}.}
\end{figure}
One of the goals of the Higgs Working Group is to
examine the potential for the upgraded Tevatron to extend the LEP2
Higgs search.
We now briefly survey the dominant Higgs decay and production processes
most relevant to the Higgs search at the upgraded Tevatron.

\subsubsection{Standard Model Higgs Boson Decay Modes}

The branching ratios for the dominant decay modes of a Standard Model
Higgs boson are shown as a function of Higgs boson mass in \fig{fg:1}
and table~\ref{tableBR} \cite{9}.

\vskip6pt\noindent
\underline{$\hsm\to f\bar f$} \\[0.2cm]
For Higgs boson masses below about 130 GeV, the
decay $\hsm\to b\bar b$
dominates, while the decay $\hsm\to \tau^+\tau^-$ can also be
phenomenologically relevant.
\begin{figure}
\begin{center}
\centerline{\psfig{file=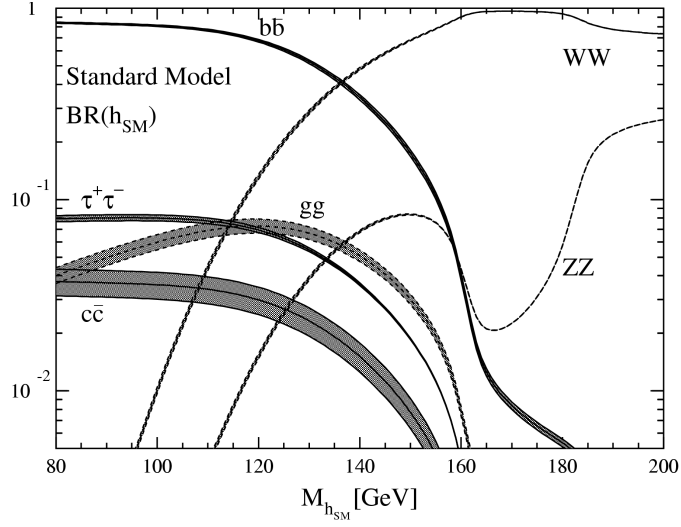,width=9cm}}
\end{center}
\caption[0]{\label{fg:1}  Branching ratios of the dominant decay
modes of the \SM\ Higgs boson \cite{9}.  These results
have been obtained with the program HDECAY \cite{hdecay}, and include
QCD corrections beyond the leading order. The shaded bands represent the
variations due to the uncertainties in the input parameters:
$\alpha_s(M_Z^2) = 0.120 \pm 0.003$, $\overline{m}_b(M_b) = 4.22
\pm 0.05$~GeV, $\overline{m}_c(M_c) = 1.22 \pm 0.06$~GeV, $M_t =
174 \pm 5$~GeV.}
\end{figure}

\begin{table}[htbp]
\caption [0]
{\label{tableBR} Table of branching ratios of the dominant decay
modes of the \SM\ Higgs boson \cite{9}. The branching ratios listed
above correspond to the central values of $\alpha_s$ and the quark
mass parameters given in the caption of \fig{fg:1}.}
\begin{center}
\begin{tabular}{c c c c c c c c}
$\mhsm$ [GeV] & $b \bar{b}$ & $\tau^+ \tau^-$ & $ gg$ & $c \bar{c}$ &
$WW^{(*)}$ & $ZZ^{(*)}$ & $\gamma \gamma$ \\[1ex]
\tableline
\980 &  0.841 &  $7.94\times 10^{-2}$ & $4.00\times 10^{-2}$ & $3.73
\times 10^{-2}$  & $5.71 \times 10^{-4}$ & $1.58 \times 10^{-4}$ & $9.01
\times 10^{-4}$  \rule{0in}{3ex} \\
\990 &  0.829 &  $8.02\times 10^{-2}$ & $4.96\times 10^{-2}$ & $3.67
\times 10^{-2}$  & $1.88 \times 10^{-3}$ & $3.95 \times 10^{-4}$ & $1.20
\times 10^{-3}$   \\
100 &  0.811 &  $8.00\times 10^{-2}$ & $5.99\times 10^{-2}$ & $3.58
\times 10^{-2}$  & $1.01 \times 10^{-2}$ & $1.06 \times 10^{-3}$ & $1.55
\times 10^{-3}$  \\
110 &  0.769 & $7.73\times 10^{-2}$ & $6.89\times 10^{-2}$  &
$3.39\times 10^{-2}$ & $4.41 \times 10^{-2}$ &$ 4.12 \times 10^{-3}$ &
$1.92 \times 10^{-3}$\\
120 &  0.677 &  $6.93 \times 10^{-2}$ & $7.27\times 10^{-2}$ & $2.99\times
10^{-2}$ &  0.132 & $1.52\times 10^{-2}$ & $2.21 \times 10^{-3}$ \\
130 &  0.526 & $5.46 \times 10^{-2}$  & $6.67 \times 10^{-2}$  &
$2.32\times 10^{-2}$ & 0.287 &$3.88\times 10^{-2}$ & $2.25\times 10^{-3}$ \\
140 &  0.342 & $3.61 \times 10^{-2}$  &$5.08 \times 10^{-2}$ &
$1.51\times 10^{-2}$ & 0.483 &$6.81 \times 10^{-2}$ &$ 1.96 \times 10^{-3}$ \\
150 &  0.175 & $1.87\times 10^{-2}$  & $3.01 \times 10^{-2}$& $7.71\times
10^{-3}$ & 0.681 & $8.35\times 10^{-2}$ & $1.40\times 10^{-3}$ \\
160 &   $3.97 \times 10^{-2}$& $4.29\times 10^{-3}$ &$7.85 \times
10^{-3}$  & $1.75\times 10^{-3}$ &0.901 &$4.35 \times 10^{-2}$
&$5.54\times 10^{-4}$ \\
170 &   $8.35 \times 10^{-3}$ &$9.12\times 10^{-4}$  &$1.89\times
10^{-3}$   &$3.68 \times 10^{-4}$ &0.965 &$2.25\times 10^{-2}$ & $1.50
\times 10^{-4}$\\
180 &  $5.35 \times 10^{-3}$  &$5.90\times 10^{-4}$  &$1.37\times
10^{-3}$   & $2.35 \times 10^{-4}$ & 0.935 &$5.75\times 10^{-2}$ &
$1.02\times 10^{-4}$ \\
190 &  $3.38 \times 10^{-3}$  &$3.77  \times 10^{-4}$ &$9.82\times
10^{-4}$   & $1.49\times 10^{-4}$ &0.776 &0.219 &$6.71 \times 10^{-5}$\\
200 &  $2.57 \times 10^{-3}$ &$2.89 \times 10^{-4}$  &$8.40 \times
10^{-4}$   &$1.13 \times 10^{-4}$ &0.735 &0.261 &$5.25 \times 10^{-5}$
\\
\end{tabular}
\vspace{.1in}
\end{center}
\end{table}

These results
have been obtained with the program HDECAY \cite{hdecay}, and include
QCD corrections beyond the leading
order \cite{DSZ}.\footnote{The leading electroweak
corrections are small in the Higgs mass range of interest and may be
safely neglected.}
The QCD corrections are significant for the $\hsm\to b\bar b, 
c\bar c$ decay widths due
to large logarithmic contributions.  The dominant part of these
corrections can be absorbed by evaluating
the running quark mass at a scale equal to the Higgs mass. In
order to gain a
consistent prediction of the partial decay widths one has to use
$\overline{\rm MS}$ masses, $\overline{m}_Q(M_Q)$ [where $M_Q$ is the
corresponding quark pole mass], obtained by fits to
experimental data. The evolution of $\overline{m}_Q(M_Q)$ to
$\overline{m}_Q(\mhsm)$ is controlled by the renormalization
group equations for the running $\overline{\rm MS}$ masses.
A recent analysis of this type can be found in \Ref{quarkmasses}.
For example, for $\alpha_s(\mz)=0.118\pm 0.003$, 
the following $\overline{\rm MS}$ quark masses are obtained:
$\overline{m}_b(\mz)=3.00\pm 0.11$~GeV and $\overline{m}_c(\mz)=
0.677^{+.056}_{-.061}$~GeV.  Using
$m_\tau=1.777$~GeV, it follows that the following hierarchy of Higgs
branching ratios is expected for Higgs masses of order 100~GeV:
${\rm BR}(\tau^+\tau^-)/{\rm BR}(b\bar b)\simeq 0.1$ and
${\rm BR}(c\bar c)/{\rm BR}(b\bar b)\simeq 0.04$.  Note that the
large decrease in the charm quark mass due to QCD running is responsible
for suppressing ${\rm BR}(c\bar c)$ relative to ${\rm BR}(\tau^+\tau^-)$,
in spite of the color enhancement of the former, thereby reversing the
naively expected hierarchy.
In \fig{fg:1}, the shaded bands indicate the theoretical uncertainty in
the predicted branching ratios.  These arise primarily from the
uncertainty in $\alpha_s$, and to a lesser extent the uncertainty in the
quark masses.


\vskip6pt\noindent
\underline{$\hsm\to gg$} \\[0.2cm]
Though one--loop suppressed, the decay $\hsm\to gg$ is competitive
with other decays in the relevant Higgs mass region because of the
large top Yukawa
coupling and the color factor.  The partial width for this decay is
primarily of interest because it determines the $gg\to\hsm$ production
cross-section.

\vskip6pt\noindent
\underline{$\hsm\to WW,ZZ$} \\[0.2cm]
For Higgs boson masses above about 110 GeV, the decay mode
$\hsm\to WW$, where (at least) one of the $W$ bosons is
off-shell (denoted henceforth by $WW^*$) becomes relevant.
Above 135 GeV, this is the dominant decay mode
\cite{13,16}.
The corresponding Higgs branching ratio to $ZZ^{(*)}$ is less useful
for the Tevatron Higgs search, while constituting the gold-plated mode
for the Higgs search at the LHC \cite{lhchiggs} when both $Z$ bosons
decay to electrons or muons.

\subsubsection{Standard Model Higgs Boson Production at the Tevatron}

This section describes the most important Higgs production processes at
the Tevatron.  The relevant cross-sections are depicted in
\fig{fg:4} (based on computer programs available from M.~Spira
\cite{hxsec}), and the corresponding numerical results are given in 
table~\ref{tableSIG} \cite{9,DSSW,private}.
 
Combining these Higgs production mechanisms with the decays discussed
in the previous section, one obtains the most promising signatures.

\begin{figure}[htb]
\begin{center}
\centerline{\psfig{file=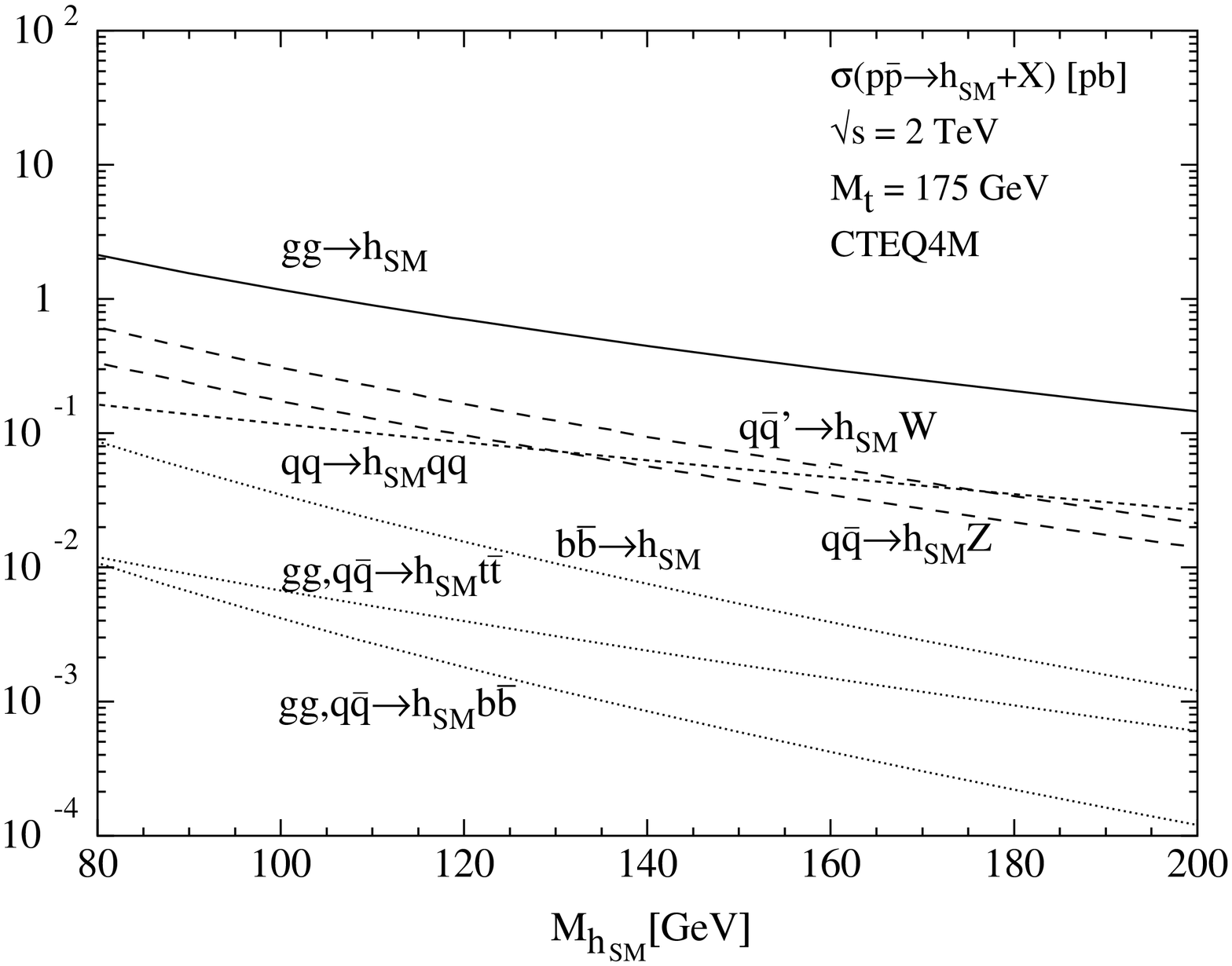,width=9cm,angle=-90}}
\end{center}
\vskip-1.5pc
\caption[0]{\label{fg:4} Higgs production cross-sections 
(in units of pb) at the Tevatron
[$\sqrt{s}=2$ TeV] for the various production mechanisms as a function of the
Higgs mass, taken from \Refs{9}{private}.
The full NLO QCD-corrected results are shown for the gluon 
fusion $gg \to \hsm$, vector boson fusion $qq\to V^*V^*qq \to \hsm qq$
[here, $qq$ refers to both $ud$ and $q\bar q$ scattering, although the
latter dominates at the Tevatron],
Higgs-strahlung processes $q\bar q \to V^* \to V\hsm$ (where
$V=W^\pm$, $Z$) and $b\bar b\to\hsm$.  
Tree-level cross-sections are exhibited for
Higgs production processes in association with heavy quark pairs:
$gg,q\bar q \to \hsm t\bar t, \hsm b\bar b$.  In the latter case,
the cross-section has been computed with the running 
Higgs--bottom-quark Yukawa
coupling evaluated at the corresponding Higgs mass.  See
text for further discussion.}
\end{figure}

\begin{table}[htbp]
\vspace*{1pc}
\caption [0]
{\label{tableSIG} Table of Higgs production cross-sections
(in units of pb) at the Tevatron
[$\sqrt{s}=2$ TeV] for the various production mechanisms as a function of the
Higgs mass.  For further details, see the caption of \fig{fg:4}.
}
\begin{center}
\setlength{\tabcolsep}{4pt}
\begin{tabular}{lc c c c c c c }
\multicolumn{1}{c}{$\mhsm$ [GeV]} & $gg\to\hsm$ & $q\bar q'\to\hsm W$ &  $q\bar q\to\hsm Z$
& $qq\to \hsm qq$ & $gg$, $q\bar q\to\hsm t\bar t$
&  $gg$, $q\bar q\to\hsm b\bar b$ & $b\bar b\to\hsm$
\\[1ex] \tableline
\noalign{\vskip1ex}
\980 &  2.132 &  0.616 & 0.334 & 0.163 & $1.11 \times 10^{-2}$ & $1.08
\times 10^{-2}$ & $8.67\times 10^{-2}$  \\
\990 &  1.557 &  0.431 & 0.238 & 0.138 & $8.31 \times 10^{-3}$ & $6.60
\times 10^{-3}$ & $5.40\times 10^{-2}$  \\
100 &  1.170 &  0.308 & 0.173 & 0.117 & $6.29 \times 10^{-3}$ & $4.17
\times 10^{-3}$ & $3.47\times 10^{-2}$   \\
110 &  0.900 & 0.224 & 0.128  & 0.100 & $4.81 \times 10^{-3}$
&$ 2.71 \times 10^{-3}$ & $2.30\times 10^{-2}$ \\
120 &  0.704 &  0.165 & $9.65\times 10^{-2}$ & $8.54\times
10^{-2}$ &  $3.71\times 10^{-3}$ & $1.80 \times 10^{-3}$ 
& $1.55\times 10^{-2}$ \\
130 &  0.558 & 0.124  & $7.35 \times 10^{-2}$  & $7.32\times 10^{-2}$
& $2.88\times 10^{-3}$ & $1.22\times 10^{-3}$ & $1.07\times 10^{-2}$ \\
140 &  0.448 & $9.37 \times 10^{-2}$  &$5.65 \times 10^{-2}$ &
$6.29\times 10^{-2}$ & $2.24 \times 10^{-3}$ &$ 8.45 \times 10^{-4}$ 
& $7.53 \times 10^{-3}$ \\
150 &  0.364 & $7.18\times 10^{-2}$  & $4.40 \times 10^{-2}$&
$5.42\times 10^{-2}$ & $1.76\times 10^{-3}$ & $5.92\times 10^{-4}$ 
& $5.38\times 10^{-3}$ \\
160 &  0.298 & $5.54\times 10^{-2}$ &$3.45 \times
10^{-2}$  & $4.69\times 10^{-2}$ &$1.39 \times 10^{-3}$
&$4.21\times 10^{-4}$  & $3.89\times 10^{-3}$ \\
170 & 0.247 & $4.32\times 10^{-2}$  &$2.73\times
10^{-2}$   & $4.06\times 10^{-2}$ &$1.10\times 10^{-3}$ &
$3.03\times 10^{-4}$ & $2.86\times 10^{-3}$ \\
180 & 0.205 &$3.39\times 10^{-2}$  &$2.17\times
10^{-2}$   & $3.52 \times 10^{-2}$ & $8.77\times 10^{-4}$ &
$2.20\times 10^{-4}$ & $2.12\times 10^{-3}$ \\
190 &  0.172  &$2.68  \times 10^{-2}$ &$1.74\times
10^{-2}$   & $3.06\times 10^{-2}$  & $7.00\times 10^{-4}$ &
$1.62 \times 10^{-4}$ & $1.59\times 10^{-3}$ \\
200 &  0.145  &$2.13 \times 10^{-2}$  &$1.40 \times 10^{-2}$ &
$2.66 \times 10^{-2}$ & $5.62\times 10^{-4}$ &$1.20 \times 10^{-4}$
& $1.20\times 10^{-3}$ \\[1ex] 
\end{tabular}
\vspace{.1in}
\end{center}
\end{table}

\break
\vskip6pt\noindent
\underline{$q\bar q\to V^*\to V\hsm$ [$V=W^\pm$ or $Z$]} \\[0.2cm]
Given sufficient luminosity, the most promising
\SM\ Higgs discovery mechanism at the Tevatron for $\mhsm\lsim 130$~GeV
consists of $q\bar q$ annihilation into a virtual $V^*$ ($V=W$ or $Z$),
where the virtual $V^*\to V\hsm$ followed by $\hsm\to b\bar b$ and
the leptonic decay of the $V$ \cite{Stange:1994ya}.
The cross-section for $q\bar q\to W^\pm\hsm$ (summed over both $W$
charge states) reaches values of $0.3$---$0.02$~pb for
$100~{\rm GeV}\lsim\mhsm\lsim 200$~GeV as shown in \fig{fg:4},
taken from \Ref{9}.
The corresponding $q\bar q\to Z\hsm$ cross-section is roughly a
factor of two lower over the same Higgs mass range.  The QCD corrections
to $\sigma(q\bar q\to V\hsm)$ coincide with those
of the Drell-Yan process and increase the cross-sections by about 30\%
\cite{23,23a0,23a}. The
theoretical uncertainty is estimated to be about $15\%$ from the
remaining scale dependence. The dependence on different sets of parton
densities is rather weak and also leads to a variation of the
production cross-sections by about 15\%.

In order to discover a Higgs signal in $q\bar q\to V\hsm\to V
b\bar b$ at the Tevatron, one must be able to separate the signal from an
irreducible Standard Model $Vb\bar b$ background.  The kinematic
properties of the signal and background are not identical, so by
applying appropriate cuts, a statistically significant signal can be
extracted given sufficient luminosity.  Although the Standard Model
$Vb\bar b$ signal can be studied experimentally, a reliable theoretical
computation of the predicted $Vb\bar b$ differential cross-section is
an essential ingredient of the Tevatron Higgs search.


At lowest order in QCD perturbation theory, the rate for $Vb\bar b$
production is of ${\cal O}(\alpha_s^2)$.  The appropriate energy scale
of the strong coupling constant is not fixed at lowest order, leading
to a significant ambiguity in the theoretical predictions.  Recently,
the QCD next-to-leading order (NLO) corrections to the $Wb\bar b$
differential cross-section were calculated \cite{kellis}.  The
higher order corrections greatly reduce the scale dependence of the
predicted rate. Therefore, we now have a better absolute prediction
from theory, and one might be tempted to simply rescale the
tree-level cross-section used above to match the NLO result for the
overall rate.  However, the NLO calculation may also change the shape of
kinematic distributions from the lowest order results.  In particular,
the NLO calculation determines more accurately the kinematics of the
$b\bar b$ pair, and thus permits an extrapolation of the $M_{b\bar b}$
distribution to higher values than have been measured.  Unfortunately,
the NLO cross-sections are generally not reliable for describing {\it
details} of the kinematic distributions, since these are sensitive to
the details of the hadronization and fragmentation of the final state.
In order to overcome this deficiency, a consistent treatment
that combines the NLO calculations with the Monte Carlo simulations is
required.   The analyses presented in Section II of this report have
not yet taken advantage of the new information provided by the NLO
result.

The signatures of Higgs
production in the $V\hsm$ channel are governed by the corresponding
decays of the Higgs and vector boson.  The dominant decay mode of the
Higgs boson in the mass range of $\mhsm\lsim 135$~GeV is $\hsm\to b\bar
b$; in this case, the leptonic decays of
the final state $W$ and $Z$ (these include the missing energy signature
associated with $Z\to \nu\bar\nu$) serve as a trigger for the $V\hsm$
events and significantly reduce QCD backgrounds.
The detection of the Higgs signal via
the more copious four-jet final states
resulting from hadronic decays of the $W$ and $Z$ is severely
hampered by huge irreducible backgrounds.

For $\mhsm\gsim 135$~GeV, the
Higgs decay mode $\hsm\to W^+W^-$ (where one $W$ is off-shell if
$\mhsm<2\mw$) becomes dominant.  In this case, the final state consists
of three gauge bosons, $VW^+W^-$ ($V=W^\pm$ or $Z$), and the like-sign
di-lepton signature becomes the primary signature for Higgs discovery.

\vskip6pt\noindent
\underline{$gg\to \hsm$} \\[0.2cm]
The gluon fusion processes proceeds primarily through
a top quark triangle loop \cite{29a,sally,19}, and
is the dominant neutral Higgs
production mechanism at the Tevatron, with cross-sections 
of roughly $1.0$---$0.1$~pb for
$100~{\rm GeV}\lsim\mhsm\lsim 200$~GeV,
as shown in \fig{fg:4}.
The two-loop QCD corrections enhance the gluon fusion cross-section by
about 60---100\% \cite{19}. These are dominated by
soft and collinear gluon radiation in the \SM\
\cite{29}.\footnote{Multiple soft-gluon emission also has a large effect
on the Higgs boson transverse momentum distribution \cite{balazs}.}
The remaining scale dependence results in a theoretical
uncertainty of about $20\%$. The dependence of the gluon fusion 
cross-section on different parton densities yields roughly an additional
15\% uncertainty in the theoretical prediction.
The analytical QCD corrections to Higgs boson plus one jet production
have recently
been evaluated in the limit of heavy top quarks, but there is no
numerical analysis so far \cite{30}.

The signature $gg\to\hsm\to b\bar b$ is not promising
at the Tevatron due to the
overwhelming QCD background of $b\bar b$ production.
The $gg\to\hsm\to\tau^+\tau^-$ signature, although not thoroughly
studied, probably requires a missing $E_T$ resolution beyond the
capabilities of the upgraded CDF and D\O\ detectors.
For $\mhsm\gsim 135$~GeV, the $\hsm\to WW^*$ decay channel
becomes dominant and provides a potential Higgs discovery mode for the
Tevatron.  The strong
angular correlations of the final state leptons resulting from
$WW^*$ is one of the crucial ingredients for this
discovery channel \cite{5,dreiner}.

\vskip6pt\noindent
\underline{$V^*V^*\to \hsm$ [$V=W^\pm$ or $Z$]} \\[0.2cm]
Vector boson fusion is a shorthand notation for the full
$q\bar q\to
q\bar q\hsm$ process, where the quark and anti-quark both radiate 
virtual vector bosons ($V^*$)
which then annihilate to produce the Higgs boson.  Vector
boson fusion via $ud\to du\hsm$ (and its charge-conjugate process)
is also possible, although the
relative contribution is small at the Tevatron.  In \fig{fg:4},
all contributing
processes are included, labeled $qq\to qq\hsm$ for simplicity.
The resulting \SM\ cross-sections are in the range $0.1$---$0.03$~pb for
$100~{\rm GeV}\lsim\mhsm\lsim 200$~GeV.
The QCD corrections enhance the cross-section by about 10\% \cite{23a,32}.

The modest $V^*V^*$ fusion cross-section precludes observation of any of
the rare SM Higgs decay modes in $qq\to qq\hsm$ events
at the Tevatron. For example, for
$\mhsm = 120$~GeV and 30~fb$^{-1}$ of data, only six
$\hsm\to \gamma\gamma$ events are expected
from the production process $qq\to qq\hsm$.
Similarly, under the same assumptions, only eleven di-lepton events
resulting from $\hsm\to\tau^+\tau^-$ are expected from the same data
sample prior to any acceptance cuts that
are required to reduce the
large $Z\to\tau^+\tau^-$, $W^+W^-$ or $t\bar t$ backgrounds. Typically
these cuts reduce the Higgs signal by another order of magnitude or
more~\cite{wbfLHC}.

For $\mhsm\lsim 135$~GeV, the dominant Higgs decay channel is $\hsm\to
b\bar b$.  In $qq\to qq\hsm$ events, the Higgs boson would appear
as a $\bar bb$ invariant mass peak in 4-jet events in which two of the jets
are identified as $b$-quarks.  Note that the $qqb\bar b$ final state
can also arise from $W\hsm$ and $Z\hsm$ events.  These channels are
very difficult to detect due to the very large QCD backgrounds.  In
the case of $V^*V^*$ fusion, because
there are two forward jets in the full
process $qq \to V^*V^*qq \to \hsm + qq$, one may hope to
be able to somewhat suppress the QCD backgrounds by appropriate cuts.
A preliminary analysis is presented in Section~II.B.4c.
The initial conclusion is that
this channel does not appear to be promising at
the upgraded Tevatron.

\vskip6pt\noindent
\underline{$q\bar q, gg\to \hsm Q\overline Q$, $Q=t,b$} \\[0.2cm]
The theoretical predictions for the cross-section of associated
production of a Higgs boson and a heavy quark pair, 
$Q\bar Q$, ($Q=t,b$)\footnote{At Tevatron energies,
the $q\bar q$ annihilation contribution
dominates over $gg$-fusion for $t\bar t\hsm$ production, whereas
the reverse is true for $b\bar b\hsm$ production.  
Both mechanisms are included in the results exhibited in \fig{fg:4}\
and in table~\ref{tableSIG}.} are shown in \fig{fg:4}\
and in table~\ref{tableSIG}.

The tree-level cross-section for $gg$, $q\bar q\to t\bar t\hsm$
is displayed in \fig{fg:4} and in 
table~\ref{tableSIG}.  Although the full NLO
QCD result is not presently known, the NLO QCD corrections are known
in the limit of $\mhsm\ll m_t$ \cite{24}. In this limit the cross-section 
factorizes into the production of a $t\bar t$ pair, which is
convolved with a splitting function for Higgs radiation $t\to t\hsm$,
resulting in an increase of the cross-section by about 20--60\%.
However, since this equivalent Higgs approximation is only valid to
within a factor of two, this result may not be sufficiently reliable.

The case of associated production of a Higgs boson and a bottom
quark pair ($b\bar b$) is more subtle and requires a separate discussion.
For values
of $\mhsm\gg m_b$ (which is satisfied in practice), the total
inclusive cross-section is known at next-to-leading order
\cite{DSSW}.  The leading-order process is $b\bar b\to \hsm$,
where the initial $b$ quarks reside in the proton sea.  Since the
$b$ quark sea arises from the splitting of gluons into $b\bar b$
pairs, the final state is actually $\hsm b\bar b$, but the final
state $b$ quarks tend to reside at low transverse momentum. The
next-to-leading-order correction to this process is modest,
indicating that the perturbative expansion is under control.
Large logarithms, $\ln(\mhsm^2/m^2_b)$, which arise in the
calculation are absorbed into the $b$ distribution functions.  The
calculation also makes evident that the Yukawa coupling of the
Higgs to the bottom quarks should be evaluated from the running
mass evaluated at the Higgs mass scale, $\overline m_b(\mhsm)$
[as in the case of the QCD-corrected decay rate for $\hsm\to
b\bar b$], rather than the running mass evaluated at the $b$
quark mass or the physical (``pole'') mass.  

The relation between the $b\bar b\to\hsm$ cross-section and
the $gg\to b\bar b\hsm$ cross-section requires some
clarification.  For example, it is not correct to simply
add the results of the $b\bar b\to\hsm$ cross-section and the
cross-section for $b\bar b\hsm$ associated production, since
logarithmic terms [proportional to $\ln(\mhsm^2/m^2_b)$] in the latter
have been incorporated into the $b$-quark distribution functions that
appear in the former.  However, one can subtract out these logarithmic
terms.  The resulting {\it subtracted} $b\bar b\hsm$ cross-section
(which can be negative) can be added to the previously obtained $b\bar
b\to\hsm$ cross-section without double counting \cite{soper,25}.
Likewise, the processes $gb\to b\hsm$ and $g\bar b\to\bar b\hsm$ also
contribute and the relevant logarithms must be subtracted before
adding the corresponding cross-sections \cite{25}.  Nevertheless, we
have checked that for $\mhsm\gg m_b$ the correction due to the
subtracted $b\bar b\hsm$ cross-section is small \cite{DSSW,private}.  Thus,
the NLO QCD-corrected $b\bar b\to\hsm$ cross-section should provide a
fairly reliable estimate to the total inclusive $\hsm b\bar b$
cross-section.\footnote{For a completely consistent ${\cal
O}(\alpha_s^2)$ computation, one would need to include two-loop (NNLO)
QCD corrections to $b\bar b\to \hsm$ and the one-loop (NLO)
QCD corrections to $gb\to b\hsm$ (and its charge-conjugate process). 
[Note that for $\mhsm\gg m_b$, the $b$-quark parton distribution
function is of order $\alpha_s\ln(\mhsm^2/m^2_b)\sim {\cal O}(1)$.]
The modest size of the NLO corrections
to $b\bar b\to\hsm$ suggest that the ${\cal O}(\alpha_s^2)$ 
corrections are probably small.}


In practice, the {\it total} inclusive cross-section for Higgs
production in association with a bottom quark pair is not 
measurable at the Tevatron, since one must observe one or both of the
final-state $b$ quarks in order to isolate the signal above other
Standard Model
backgrounds.  Hence one or both of the final state $b$ quarks must
be produced at large transverse momentum.  This is a fraction of
the total inclusive cross-section, depending on the minimum
transverse momentum required on one or both $b$ quarks.  The
leading-order process for Higgs production in association with a
pair of high transverse momentum $b$ quarks is $q\bar q, gg \to
\hsm b\bar b$.  Since this process is known only at leading order,
the theoretical uncertainty in the cross-section is large.
Clearly, it is very important to compute the NLO QCD corrections for the
differential cross-section (as a function
of the final state $b$ quark transverse momenta) for $b\bar b\hsm$
production.  This computation is not yet available in the literature.


The tree-level $gg$, $q\bar q\to b\bar b\hsm$ 
cross-section (as a function of $\mhsm$) shown in \fig{fg:4}
has been computed by fixing
the scales of the initial (CTEQ4M \cite{CTEQ}) parton densities, 
the running coupling $\alpha_s$ and the running Higgs--bottom-quark
Yukawa coupling (or equivalently, the running $b$-quark mass) at the
value of the corresponding $\hsm$ mass.  In particular, by employing
the running $b$ quark mass evaluated at $\mhsm$, we are implicitly
resumming large logarithms associated with the QCD-corrected Yukawa
coupling.  Thus, the tree-level $b\bar b\hsm$ cross-section
displayed in \fig{fg:4} implicitly includes a part of the QCD
corrections to the full inclusive cross-section.\footnote{The effect
of using the running $b$-quark mass [$\overline m_b(\mhsm)$]  
as opposed to the $b$-quark pole mass
[$M_b\simeq 5$~GeV] is to {\it reduce} the cross-section by roughly a 
factor of two.}  Nevertheless, as emphasized above,
the most significant effect of the QCD
corrections to $b\bar b\hsm$ production arises from the kinematical
region where the  $b$ quarks are emitted near the forward
direction.  In fact, large logarithms arising in
this region spoils the convergence of the 
QCD perturbation series since $\alpha_s\ln(\mhsm^2/m_b^2)\sim {\cal
O}(1)$.  These large logarithms (already present at lowest order)
must be resummed to all orders, and this resummation
is accomplished by the generation of the $b$-quark distribution
function as described above.  Thus, the QCD-corrected {\it fully} inclusive 
$b\bar b\hsm$ cross-section is well approximated by $b\bar b\to\hsm$ and its
QCD corrections.  The latter is also exhibited in \fig{fg:4} 
and is seen to be roughly an order of magnitude larger than
the tree-level $b\bar b\hsm$ cross-section.    
Of course, this result is not very relevant for the Tevatron
experimental searches in which transverse momentum cuts on the
$b$-jets are employed.  Ultimately, one needs the QCD-corrected
{\it differential} cross-section for $b\bar b\hsm$ (as a function of the final
state $b$-quark transverse momenta) in order to do realistic simulations of
the Higgs signal in this channel.

Based on our best estimates for the $\hsm Q\bar Q$ production
cross-sections shown in \fig{fg:4}, we conclude that in
the \SM\ both processes of Higgs radiation off top and bottom quarks
have very  small event rates.  However, in some
extensions of the \SM, the coupling of the Higgs boson to $b\bar b$
can be significantly enhanced.  This leads to enhanced $\hsm b\bar b$
production which may be observable at an upgraded Tevatron with
sufficient luminosity.  A concrete example (the minimal supersymmetric
model at large $\tan\beta$) will be discussed further in Section~I.C.6a.

To detect enhanced $b\bar b\hsm$ production, we assume that the dominant
decay of the Higgs boson is into $b\bar b$ pairs.  Thus, the signal must
be extracted from events with four $b$-quarks (suggesting the need for
tagging at least three of the $b$-jets).
Theoretical estimates of the $b\bar b b\bar b$ Standard Model
background rates are less reliable than the corresponding $W b\bar b$
rates discussed above.  Heavy quark production rates are typically
difficult to estimate, and the situation is
complicated by demanding 3 or 4 $b$--tagged jets.
Improved theoretical modeling and detailed comparisons with Run 2
Tevatron data will be essential to improve the present background
estimates.

\subsection{Higgs Bosons in Low-Energy Supersymmetry}

Although the Higgs mass range 130~GeV~$\lsim\mhsm\lsim 180$~GeV appears
to permit an effective Standard Model
that survives all the way to the Planck scale, most theorists consider
such a possibility unlikely.  This conclusion is based on
the ``naturalness'' \cite{natural} argument as follows.  In
an effective field theory, all parameters of the low-energy theory
({\it i.e.} masses and couplings) are calculable in terms of parameters
of a more fundamental, renormalizable theory that
describes physics at the energy scale $\Lambda$.  All
low-energy couplings and fermion masses are logarithmically sensitive to
$\Lambda$.  In contrast, scalar squared-masses are {\it quadratically}
sensitive to $\Lambda$.  Thus, in this framework, the observed Higgs
mass (at one-loop) has the following form:
\beq \label{natural}
m_h^2= (m_h^2)_0+cg^2\Lambda^2\,,
\eeq
where $(m_h)_0$ is a parameter of the fundamental theory and $c$ is a
constant, presumably of ${\cal O}(1)$, which is calculable within the
low-energy effective theory.  The ``natural'' value
for the scalar squared-mass is $g^2\Lambda^2$.  Thus, the expectation
for $\Lambda$ is
\beq \label{tevscale}
\Lambda\simeq {m_h\over g}\sim {\cal O}(1~{\rm TeV})\,.
\eeq
If $\Lambda$ is significantly larger than 1~TeV (often called the
hierarchy problem in the literature), then the only way
to find a Higgs mass associated with the scale of electroweak symmetry
breaking is to have an ``unnatural'' cancellation between the two terms
of \eq{natural}.
This seems highly unlikely given that the two terms of
\eq{natural} have completely different origins.  As a cautionary note,
what appears to be unnatural today might be shown in the future to be
a natural consequence of dynamics or symmetry at the Planck scale.
Only further experimentation that tests our present
understanding can determine if the quest for new physics at
scales $\Lambda\ll\mpl$ is ultimately fruitful.

A viable theoretical framework that incorporates weakly-coupled Higgs
bosons and satisfies the constraint of \eq{tevscale} is that of
``low-energy'' or ``weak-scale'' supersymmetry \cite{Nilles85,Haber85,smartin}.
In this framework, supersymmetry is
used to relate fermion and boson masses (and interactions).  
Since fermion masses are only
logarithmically sensitive to $\Lambda$, boson masses will exhibit the
same logarithmic sensitivity if supersymmetry is exact.  Since no
supersymmetric partners of Standard Model particles have been
found, we know that supersymmetry cannot be an exact symmetry of nature.
Thus, we identify $\Lambda$ with the supersymmetry-breaking scale.  The
naturalness constraint of \eq{tevscale} is still relevant, so in the
framework of low-energy supersymmetry, the scale of supersymmetry
breaking should not be much larger than about 1~TeV in order that the
naturalness of scalar masses be preserved.  The supersymmetric extension
of the Standard Model would then replace the Standard Model as the
effective field theory of the TeV scale.  One could then ask: at what
scale does this model break down?  The advantage of the supersymmetric
approach is that the effective low-energy supersymmetric theory {\it
can} be valid all the way up to the Planck scale, while still being
natural!

In order to begin our study of Higgs bosons in low-energy supersymmetry,
we need a specific model framework.  The simplest realistic model of
low-energy supersymmetry is a minimal supersymmetric extension of the
Standard Model (MSSM), which employs the minimal supersymmetric particle
spectrum \cite{habermssm}.  Higgs phenomenology in the MSSM and the
corresponding
discovery reach at the Tevatron was a primary focus of the Higgs
Working Group.  Non-minimal supersymmetric approaches are also of interest and
have been addressed in part by the Beyond the MSSM Working Group of this
workshop \cite{btmssm}.

\subsubsection{The Tree-Level Higgs Sector of the MSSM}

Both hypercharge $Y=-1$ and $Y=+1$ Higgs doublets are
required in any Higgs sector of
an anomaly-free supersymmetric extension of the Standard Model.
The supersymmetric structure of the theory also requires (at least) two
Higgs doublets to generate mass for both ``up''-type and ``down''-type
quarks (and charged leptons) \cite{Inoue82,Gunion86}.
Thus, the MSSM
contains the particle spectrum of a two-Higgs-doublet extension of the
Standard Model and the corresponding supersymmetric partners
\cite{Haber85,smartin}.

The two-doublet Higgs sector \cite{hhgchap4} contains eight scalar
degrees of freedom:
one complex $Y=-1$ doublet, {\boldmath $\Phi_d$}$=(\Phi_d^0,\Phi_d^-)$
and one complex $Y=+1$ doublet, {\boldmath
$\Phi_u$}$=(\Phi_u^+,\Phi_u^0)$.  The notation reflects the
form of the MSSM Higgs sector coupling to fermions: $\Phi_d^0$
[$\Phi_u^0$] couples exclusively to down-type [up-type] fermion pairs.
When the Higgs potential is minimized, the neutral components of the
Higgs fields acquire vacuum expectation values:\footnote{The
phases of the Higgs fields can be chosen such that the vacuum
expectation values are real and positive.  That is, the tree-level MSSM
Higgs sector conserves CP, which implies that the neutral Higgs mass
eigenstates possess definite CP quantum numbers.}
\beq
\langle {\mathbold{\Phi_d}} \rangle={1\over\sqrt{2}} \left(
\begin{array}{c} v_d\\ 0\end{array}\right), \qquad \langle
{\mathbold{\Phi_u}}\rangle=
{1\over\sqrt{2}}\left(\begin{array}{c}0\\ v_u
\end{array}\right)\,,\label{potmin}
\eeq
where the normalization has been chosen such that
$v^2\equiv v_d^2+v_u^2={4\mw^2/ g^2}=(246~{\rm GeV})^2$.
Spontaneous electroweak symmetry breaking results in
three Goldstone bosons, which
are absorbed and become the longitudinal components of
the $W^\pm$ and $Z$.  The remaining five physical Higgs particles
consist of a charged Higgs pair
\vspace*{-1pc}
\beq \label{hpmstate}
\hpm=\Phi_d^\pm\sinb+ \Phi_u^\pm\cosb\,,
\eeq
one CP-odd scalar
\vspace*{-1pc}
\beq \label{hastate}
\ha= \sqrt{2}\left({\rm Im\,}\Phi_d^0\sinb+{\rm Im\,}\Phi_u^0\cosb
\right)\,,
\eeq
and two CP-even scalars:
\vspace*{-1pc}
\beqno
\hl &=-(\sqrt{2}\,{\rm Re\,}\Phi_d^0-v_d)\sin\alpha+
(\sqrt{2}\,{\rm Re\,}\Phi_u^0-v_u)\cos\alpha\,,\nonumber\\
\hh &=(\sqrt{2}\,{\rm Re\,}\Phi_d^0-v_d)\cos\alpha+
(\sqrt{2}\,{\rm Re\,}\Phi_u^0-v_u)\sin\alpha\,,
\label{scalareigenstates}
\eeqno
(with $\mhl\leq \mhh$).
The angle $\alpha$ arises when the CP-even Higgs
squared-mass matrix (in the $\Phi_d^0$---$\Phi_u^0$ basis) is
diagonalized to obtain the physical CP-even Higgs states (explicit
formulae will be given below).

The supersymmetric structure of the theory imposes constraints on the
Higgs sector of the model.  For example, the Higgs
self-interactions are not independent parameters; they can be expressed
in terms of the electroweak gauge coupling constants.  As a result,
all Higgs sector parameters at tree-level
are determined by two free parameters: the ratio of the two
neutral Higgs field vacuum expectation values,
\beq \label{tanbdef}
\tanb\equiv {v_u\over v_d}\,,
\eeq
and one Higgs mass, conveniently chosen to be $\mha$. In particular,
\beq
\mhpm^2 =\mha^2+\mw^2\,,
\label{susymhpm}
\eeq
and the CP-even Higgs bosons $\hl$ and $\hh$ are eigenstates of the
following squared-mass matrix
\beq
{\cal M}_0^2 =    \left(
\matrix{\mha^2 \sin^2\beta + m^2_Z \cos^2\beta&
           -(\mha^2+m^2_Z)\sin\beta\cos\beta \cr
  -(\mha^2+m^2_Z)\sin\beta\cos\beta&
  \mha^2\cos^2\beta+ m^2_Z \sin^2\beta }
   \right)\,.\label{kv}
\eeq
The eigenvalues of ${\cal M}_0^2$ are
the squared-masses of the two CP-even Higgs scalars
\beq
  m^2_{H,h} = \half \left( \mha^2 + m^2_Z \pm
                  \sqrt{(\mha^2+m^2_Z)^2 - 4m^2_Z \mha^2 \cos^2 2\beta}
                  \; \right)\,,\label{kviii}
\eeq
and $\alpha$ is the angle that diagonalizes the CP-even Higgs
squared-mass matrix.
From the above results, one obtains:
\beq
\cos^2(\beta-\alpha)={\mhl^2(\mz^2-\mhl^2)\over
\mha^2(\mhh^2-\mhl^2)}\,.
\label{cbmasq}
\eeq
In the convention where $\tanb$ is positive ({\it i.e.},
$0\leq\beta\leq\pi/2$), the angle $\alpha$ lies in the range
$-\pi/2\leq\alpha\leq 0$.

An important consequence of \eq{kviii} is that there is an upper bound
to the mass of the light CP-even Higgs boson, $\hl$.  One finds that:
\beq \label{kx}
\mhl^2\leq\mz^2\cos 2\beta\leq\mz^2\,.
\eeq
This is in marked contrast to the Standard Model, in which the theory
does not constrain the value of $\mhsm$ at tree-level.  The origin of
this difference is easy to ascertain.  In the \SM, $\mhsm^2=\lambda v^2$
is proportional to the Higgs self-coupling $\lambda$, which is a free
parameter.  On the other hand, all Higgs self-coupling
parameters of the MSSM are related to the squares of the electroweak
gauge couplings.

Note that the Higgs mass inequality [\eq{kx}] is saturated in the limit
of large $\mha$.  In the limit of $\mha\gg\mz$, the expressions for the
Higgs masses and mixing angle simplify and one finds
\beqno
\mhl^2 &\simeq\ \mz^2\cos^2 2\beta\,,\nonumber \\[3pt]
\mhh^2 &\simeq\ \mha^2+\mz^2\sin^2 2\beta\,,\nonumber \\[3pt]
\mhpm^2& = \ \mha^2+\mw^2\,,\nonumber \\[3pt]
\cos^2(\beta-\alpha)&\simeq\ {\mz^4\sin^2 4\beta\over 4\mha^4}\,.
\label{largema}
\eeqno
Two consequences are immediately apparent.
First, $\mha\simeq\mhh
\simeq\mhpm$, up to corrections of ${\cal O}(\mz^2/\mha)$.  Second,
$\cos(\beta-\alpha)=0$ up to corrections of ${\cal O}(\mz^2/\mha^2)$.
This limit is known as the {\it decoupling} limit \cite{decoupling}
because when $\mha$ is
large, one can focus on an effective low-energy theory below the scale
of $\mha$ in which the effective Higgs sector consists only of one
CP-even Higgs boson, $\hl$.  As we shall demonstrate below, the
tree-level couplings of $\hl$ are precisely those of the Standard
Model Higgs boson when $\cos(\beta-\alpha)=0$.
From \eq{largema}, one can also derive:
\begin{equation}  \label{cotalf}
\cot\alpha = -\tan\beta - \frac{2\mz^2}{\mha^2} \tan\beta \cos
2\beta + {\cal O}\left(\frac{\mz^4}{\mha^4}\right)\,.
\end{equation}
This result will prove useful in evaluating the CP-even Higgs boson
couplings to fermion pairs in the decoupling limit.

The phenomenology of the Higgs sector depends in detail on
the various couplings of the Higgs bosons to gauge bosons, Higgs
bosons and fermions.  The couplings of the two CP-even Higgs bosons
to $W$ and $Z$ pairs
are given in terms of the angles $\alpha$ and $\beta$ by
\beqno
g\ls{\hl VV}&=g\ls{V} m\ls{V}\sinbma \nonumber \\[3pt]
           g\ls{\hh VV}&=g\ls{V} m\ls{V}\cosbma\,,\label{vvcoup}
\eeqno
where
\beq
g\ls V\equiv\begin{cases}
g,& $V=W\,$,\\ g/\cos\theta_W,& $V=Z\,$. \end{cases}
\label{hix}
\eeq
There are no tree-level couplings of $\ha$ or $\hpm$ to $VV$.
Next, consider the couplings of one gauge boson to two
neutral Higgs bosons:
\beqno g\ls{\hl\ha Z}&={g\cosbma\over 2\cos\theta_W}\,,\nonumber \\[3pt]
           g\ls{\hh\ha Z}&={-g\sinbma\over 2\cos\theta_W}\,.
           \label{hvcoup}
\eeqno
From the expressions above, we see
that the following sum rules must hold separately for $V=W$ and $Z$:
\beqno
g_{\hh V V}^2 + g_{\hl V V}^2 &=
                      g\ls{V}^2m\ls{V}^2\,,\nonumber \\[3pt]
g_{\hl\ha Z}^2+g_{\hh\ha Z}^2&=
             {g^2\over 4\cos^2\theta_W}\,, \nonumber \\[3pt]
 g^2_{\phi ZZ} + 4m^2_Z g^2_{\phi \ha Z}& = {g^2m^2_Z\over
        \cos^2\theta_W}\,,\qquad \phi=\hl, \hh \,.\label{hxi}
\label{sumruletwo}
\eeqno
Similar considerations also hold for
the coupling of $\hl$ and $\hh$ to $W^\pm H^\mp$.  We can summarize
the above results by noting that the coupling of $\hl$ and $\hh$ to
vector boson pairs or vector--scalar boson final states is proportional
to either $\sin(\beta-\alpha)$ or $\cos(\beta-\alpha)$ as indicated
below \cite{hhg}:
\beq
\renewcommand{\arraycolsep}{2cm}
\let\us=\underline
\begin{array}{ll}
\us{\cos(\beta-\alpha)}&  \us{\sin(\beta-\alpha)}\\ [3pt]
\noalign{\vskip3pt}
       \hh W^+W^-&        \hl W^+W^- \\
       \hh ZZ&            \hl ZZ \\
       Z\ha\hl&          Z\ha\hh \\
       W^\pm H^\mp\hl&  W^\pm H^\mp\hh \\
       ZW^\pm H^\mp\hl&  ZW^\pm H^\mp\hh \\
       \gamma W^\pm H^\mp\hl&  \gamma W^\pm H^\mp\hh
\end{array}
\label{littletable}
\eeq
Note in particular that {\it all} vertices
in the theory that contain at least
one vector boson and {\it exactly one} non-minimal Higgs boson state
($\hh$, $\ha$ or $\hpm$) are proportional to $\cos(\beta-\alpha)$.
This can be understood as a consequence of unitarity sum rules which
must be satisfied by the tree-level amplitudes of the
theory \cite{cornwall,wudka}.

In the MSSM, the Higgs tree-level couplings to fermions obey the
following
property: $\Phi_d^0$ couples exclusively to down-type fermion pairs and
$\Phi_u^0$ couples exclusively to up-type fermion pairs. This pattern
of Higgs-fermion couplings defines the
Type-II two-Higgs-doublet model \cite{wise,hhg}.
The gauge-invariant Type-II Yukawa interactions (using 3rd family
notation) are given by:
\begin{equation} \label{typetwo}
-{\cal L}_{\rm Yukawa}= h_t\left[\bar t P_L t \Phi^0_u-\bar t P_L b
\Phi^+_u\right] + h_b\left[\bar b P_L b \Phi^0_d-\bar b P_L t
\Phi^-_d\right] + {\rm h.c.}\,,
\end{equation}
where $P_L\equiv\half(1-\gamma_5)$ is the left-handed projection
operator.  [Note that ($\overline\Psi_1 P_L \Psi_2)^\dagger=
\overline\Psi_2 P_R \Psi_1$, where $P_R\equiv\half(1+\gamma_5)$.]
Fermion masses are generated when the neutral Higgs components acquire
vacuum expectation values.  Inserting \eq{potmin} into \eq{typetwo}
yields a relation between the quark masses and the Yukawa couplings:
\beqa
h_b =& {\sqrt{2}\,m_b\over v_d}={\sqrt{2}\, m_b\over
v\cos\beta}\,,\label{hbdef} \\
h_t =& {\sqrt{2}\,m_t\over v_u}={\sqrt{2}\, m_t\over v\sin\beta}\,.
\label{htdef}
\eeqa
Similarly, one can define the Yukawa coupling of the Higgs boson to
$\tau$-leptons (the latter is a down-type fermion).  The couplings
of the physical Higgs bosons to the third generation fermions is
obtained from \eq{typetwo} by using \eqs{hpmstate}{scalareigenstates}.
In particular, the couplings of the neutral Higgs bosons to $f\bar f$
relative to the Standard Model
value, $gm_f/2\mw$, are given by
\begin{eqaligntwo}
 \label{qqcouplings}
\hl b\bar b \;\;\; ({\rm or}~ \hl \tau^+ \tau^-):&~~~ -
{\sin\alpha\over\cos\beta}=\sin(\beta-\alpha)
-\tan\beta\cos(\beta-\alpha)\,,\nonumber\\[3pt]
\hl t\bar t:&~~~ \phm{\cos\alpha\over\sin\beta}=\sin(\beta-\alpha)
+\cot\beta\cos(\beta-\alpha)\,,\nonumber\\[3pt]
\hh b\bar b \;\;\; ({\rm or}~ \hh \tau^+ \tau^-):&~~~
\phm{\cos\alpha\over\cos\beta}=
\cos(\beta-\alpha)
+\tan\beta\sin(\beta-\alpha)\,,\nonumber\\[3pt]
\hh t\bar t:&~~~ \phm{\sin\alpha\over\sin\beta}=\cos(\beta-\alpha)
-\cot\beta\sin(\beta-\alpha)\,,\nonumber\\[3pt]
\ha b \bar b \;\;\; ({\rm or}~ \ha \tau^+
\tau^-):&~~~\phm\gamma_5\,{\tan\beta}\,,
\nonumber\\[3pt]
\ha t \bar t:&~~~\phm\gamma_5\,{\cot\beta}\,,
\end{eqaligntwo}
(the $\gamma_5$ indicates a pseudoscalar coupling), and the
charged Higgs boson couplings to fermion pairs
(with all particles pointing into the vertex) are given by
\begin{eqaligntwo}
\label{hpmqq}
g_{H^- t\bar b}= &~~~ {g\over{\sqrt{2}\mw}}\
\Bigl[m_t\cot\beta\,P_R+m_b\tan\beta\,P_L\Bigr]\,,
\nonumber\\[3pt]
g_{H^- \tau^+ \nu}= &~~~ {g\over{\sqrt{2}\mw}}\
\Bigl[m_{\tau}\tan\beta\,P_L\Bigr]\,.
\end{eqaligntwo}

We next consider the behavior of the Higgs couplings at large
$\tan\beta$.  This limit is of particular interest since
at large $\tan\beta$, some of the Higgs couplings to down-type
fermions can be significantly enhanced.\footnote{In models of low-energy
supersymmetry, there is some theoretical prejudice that suggests that
$1 < \tanb \lsim m_t/m_b$, with the fermion running masses evaluated
at the electroweak scale.
For example, $\tanb\lsim 1$
[$\tanb > \mt/\mb$] is disfavored
since in this case, the Higgs--top-quark [Higgs--bottom-quark]
Yukawa coupling blows up at an
energy scale significantly below the Planck scale.}
Let us examine two
particular large $\tan\beta$ regions of interest.  (i) If $\mha\gg\mz$,
then the decoupling limit is reached, in which $|\cos(\beta-\alpha)|\ll
1$ and $\mhh\simeq\mha$.
From \eqs{largema}{qqcouplings}, it follows that the $b\bar b\hh$
and $b\bar b\ha$ couplings have equal strength and are significantly
enhanced (by a factor of
$\tanb$) relative to the $b\bar b\hsm$ coupling, whereas the $VV\hh$
coupling is negligibly
small. In contrast, the values of the $VV\hl$ and $b\bar b\hl$ couplings
are equal to the corresponding couplings of the \SM\ Higgs boson.
To show that the value of the $b\bar b\hl$ coupling
[see \eq{qqcouplings}]
reduces to that of $b\bar b\hsm$ in the decoupling limit,
note that \eq{largema} implies that $|\tan\beta\cos(\beta-\alpha)|\ll 1$
when $\mha\gg \mz$ even when $\tan\beta\gg 1$.
(ii) If $\mha\lsim \mz$ and $\tanb\gg 1$, then $|\sin(\beta-\alpha)|\ll
1$ [see \fig{cosgraph}] and $\mhl\simeq\mha$.
In this case, the $b\bar b\hl$
and $b\bar b\ha$ couplings have equal strength and are significantly
enhanced (by a factor of
$\tanb$) relative to the $b\bar b\hsm$ coupling, while the $VV\hl$
coupling is negligibly small.  Using \eq{sumruletwo} it follows that
the $VV\hh$ coupling is equal in strength to the $VV\hsm$ coupling.
However, the value of the $b\bar b\hh$ coupling can differ
from the corresponding $b\bar b\hsm$ coupling when $\tanb\gg 1$
[since in case (ii), where $|\sin(\beta-\alpha)|\ll 1$,
the product $\tan\beta\sin(\beta-\alpha)$ need not
be particularly small]. Note that in both cases above,
only two of the three neutral Higgs bosons have enhanced
couplings to $b\bar b$.
\begin{figure}[!htb]
\centering
\centerline{\psfig{file=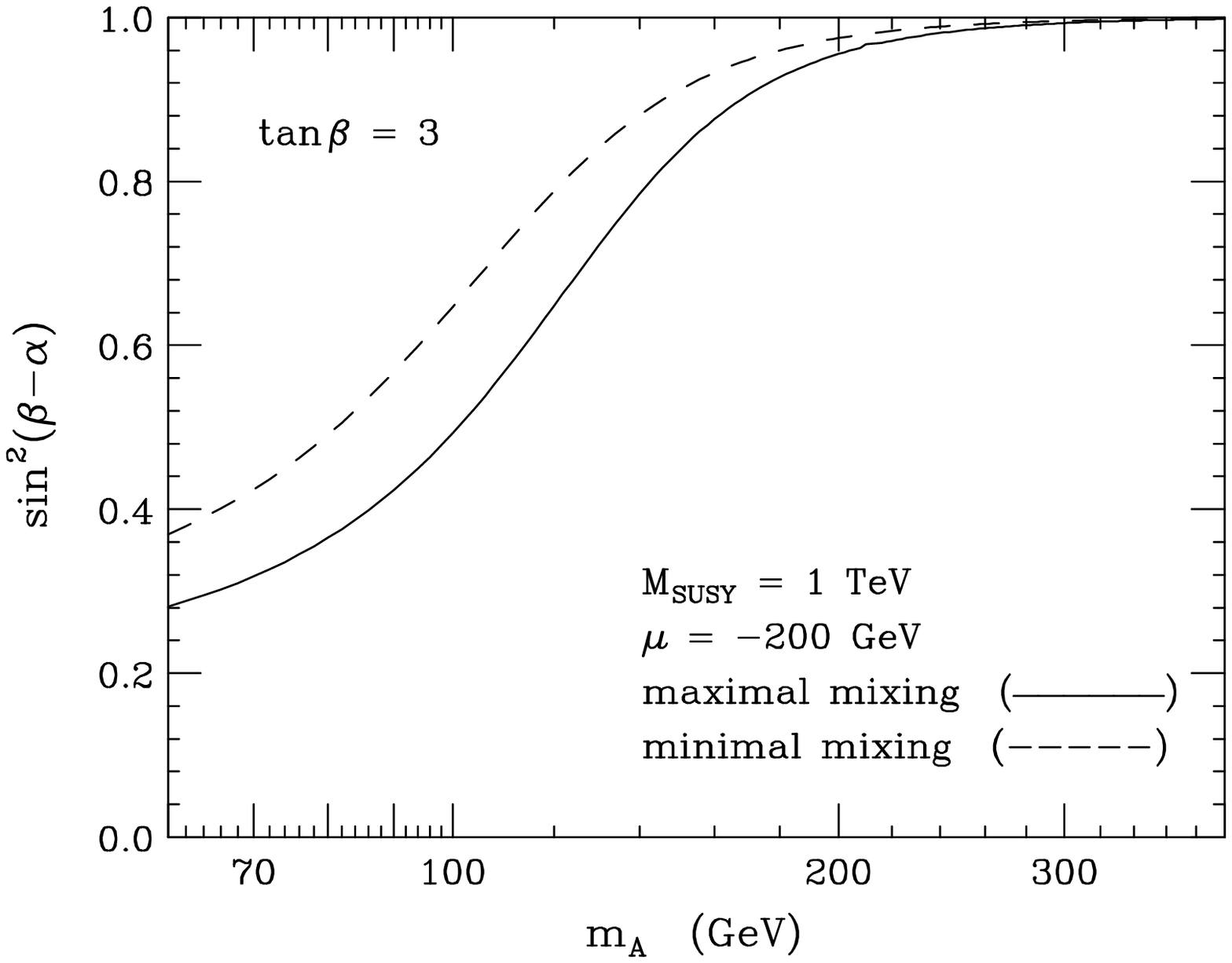,width=8cm}
\hfill
\psfig{file=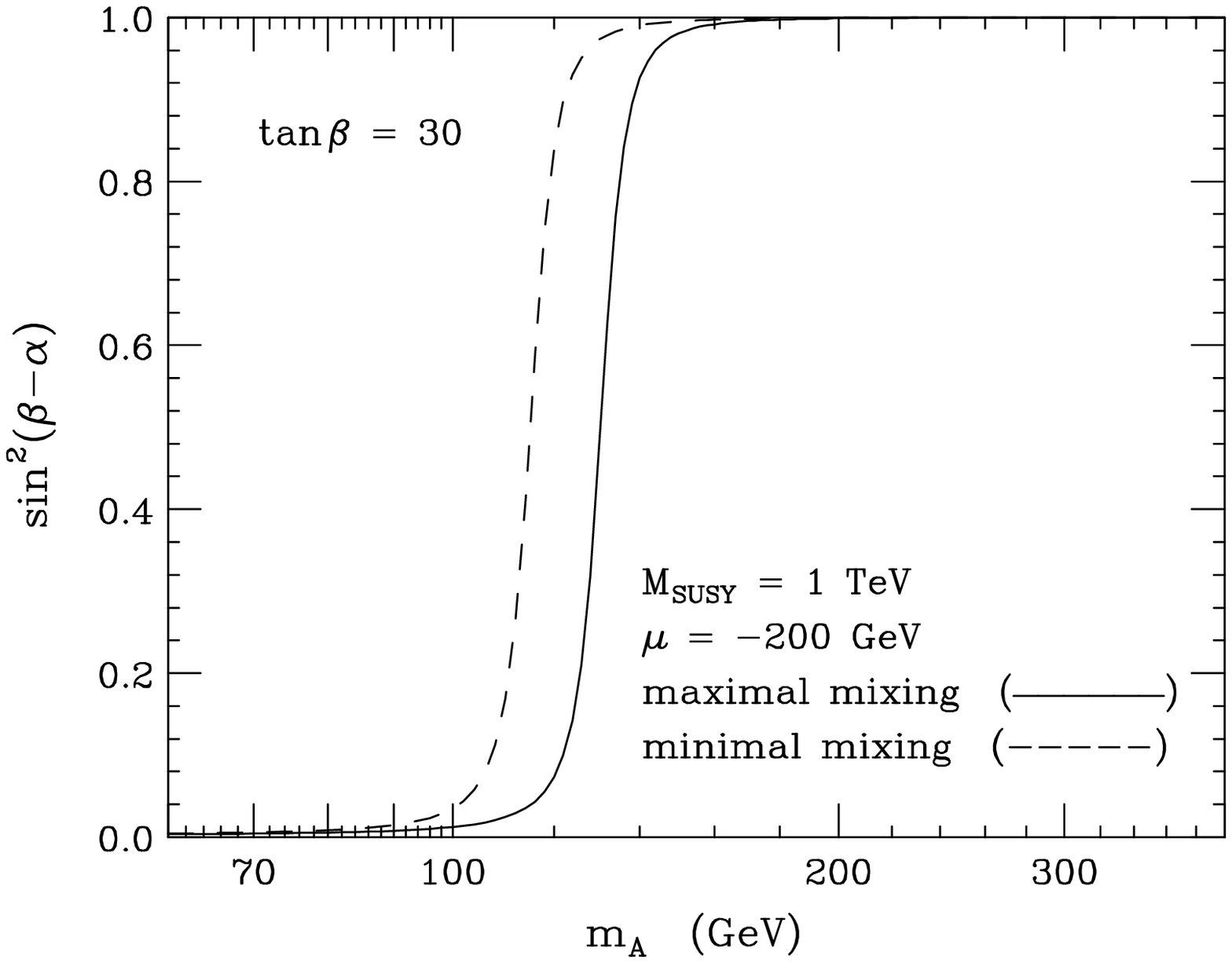,width=8cm}}
\vskip1pc
\caption[0]{\label{cosgraph} The value of $\sin^2(\beta-\alpha)$
is shown as a function of
$\mha$ for two choices of $\tan\beta = 3$ and $\tan\beta = 30$.
When radiative-corrections are included, one can define an approximate
loop-corrected angle $\alpha$ as a function of $\mha$, $\tan\beta$ and
the MSSM parameters.  In the figures above, we
have incorporated radiative corrections, assuming that
$\MSUSY\equiv M_Q=M_U=M_D=1$~TeV.  In addition,
two extreme cases for the squark mixing parameters
are shown (see Section~I.C.2 for further discussion of the
radiative corrections and their dependence on the supersymmetric
parameters). The decoupling effect expected from
\eq{largema}, in which $\sin^2(\beta-\alpha)\simeq 1$ for $\mha\gg m_Z$,
continues to hold even when radiative corrections are included.
}
\end{figure}

The decoupling limit of $\mha\gg\mz$ is effective for all values of
$\tan\beta$.  It is easy to check that the pattern of all Higgs
couplings displayed in \eqs{vvcoup}{qqcouplings}
respects the decoupling limit.  That is, in the limit where $\mha\gg\mz$,
$\cos(\beta-\alpha)={\cal O}(\mz^2/\mha^2)$, which means that
the $\hl$ couplings to
Standard Model particles approach values corresponding precisely
to the couplings of the Standard Model Higgs boson.
The region of MSSM Higgs sector parameter space in which the decoupling
limit applies is large, because
$\sin(\beta-\alpha)$ approaches one quite rapidly once
$\mha$ is larger than about 200~GeV, as shown in \fig{cosgraph}.
As a result, over a significant region of the MSSM parameter space, the
search for the lightest CP-even Higgs boson
of the MSSM is equivalent to the search for the \SM\
Higgs boson.  This result is more general; in
many theories of non-minimal Higgs sectors, there is a significant
portion of the parameter space that approximates the decoupling limit.
Consequently, simulations of the \SM\ Higgs signal are also relevant for
exploring the more general Higgs sector.

\subsubsection{The Radiatively-Corrected MSSM Higgs Sector: (a) Higgs
masses}
\label{RCMSSM}

So far, the discussion has been based on a tree-level analysis of the
Higgs sector.  However, radiative corrections can have a significant
impact on the predicted values of Higgs masses and couplings.
The radiative corrections involve both loops of Standard Model particles
and loops of supersymmetric partners.  
The dominant effects arise from
loops involving the third generation quarks and squarks
and are proportional to the corresponding Yukawa couplings.  
Thus, we first review the parameters that control the masses
and mixing of the third-generation squarks.  (We shall neglect
intergenerational mixing effects, which have little impact on the
discussion that follows.)

For each left-handed and right-handed quark of fixed flavor, $q$,
there is a corresponding supersymmetric partner
$\widetilde q_L$ and $\widetilde q_R$, respectively.  These are the
so-called interaction eigenstates, which mix according to the squark
squared-mass matrix. The mixing angle that diagonalizes the squark mass
matrix will be denoted by $\theta_{\widetilde q}$.   The
squark mass eigenstates, denoted by $\widetilde q_1$ and $\widetilde q_2$,
are obtained by diagonalizing the following $2\times 2$ matrix
\begin{equation}
 \left(  
\matrix{M_{Q}^2+m_f^2+D_L & m_f X_f \crr
   m_f X_f & M_{R}^2+m_f^2+D_R}
\right)  \,,
\label{stopmatrix}
\end{equation}
where $D_L\equiv (T_{3f}-e_f\sin^2\theta_W)\mz^2\cos2\beta$ and
$D_R\equiv e_f\sin^2\theta_W\mz^2\cos2\beta$.  In addition,
$f=t$, $M_R\equiv M_U$, $e_t=2/3$ and
$T_{3f}=1/2$ for the top-squark (or {\it stop}) mass matrix, while
$f=b$, $M_R\equiv
M_D$, $e_b=-1/3$ and $T_{3f}=-1/2$ for the bottom-squark
(or {\it sbottom}) mass matrix.
The squark mixing parameters are given by
\beqa
 X_t & \equiv A_t-\mu\cot\beta\,,\nonumber \\
 X_b & \equiv A_b-\mu\tan\beta\,.
\eeqa
Thus, the top-squark and bottom-squark masses and mixing angles
depend on the supersymmetric Higgs mass parameter $\mu$ and
the soft-supersymmetry-breaking parameters: $M_Q$, $M_U$, $M_D$,
$A_t$ and $A_b$.\footnote{For simplicity, we shall take $A_t$,
$A_b$ and $\mu$ to be real parameters.  That is, we are neglecting
possible CP-violating effects that can enter the MSSM Higgs sector
via radiative corrections. For a discussion of the implications of
such effects, see \Refs{cpcarlos}{cpcarlos2}.}

The radiative corrections to the Higgs squared-masses have been computed
by a number of techniques, and using a variety of approximations such as
the effective potential at one-loop
\cite{early-veff,veff,berz,erz} and
two-loops \cite{zhang,espizhang,EZ2} [only the
${\cal O}(m_t^2 h_t^2\alpha_s)$ and ${\cal O}(m_t^2 h_t^4)$
two-loop results are known], and diagrammatic methods
\cite{turski,hhprl,brig,madiaz,1-loop,completeoneloop,hempfhoang,weiglein}.
Complete one-loop diagrammatic computations of the MSSM Higgs masses
have been presented by a number of groups \cite{completeoneloop};
the resulting expressions are quite complex,
and depend on all the parameters of the MSSM.
Partial two-loop diagrammatic
results are also known~\cite{hempfhoang,weiglein}.
These include the ${\cal O}(m_t^2 h_t^2 \alpha_s)$ contributions to
the neutral CP-even Higgs boson squared-masses in the on-shell
scheme~\cite{weiglein}.

One of the most striking effects of the radiative corrections to the
MSSM Higgs sector is the modification of the upper bound of the
light CP-even Higgs mass, as first noted in \Refs{early-veff}{hhprl}.  
Consider the region of parameter space
where $\tanb$ is large and $\mha\gg\mz$.  In this limit, the
{\it tree-level} prediction for $\mhl$ corresponds to its theoretical
upper bound, $\mhl=\mz$.  Including radiative corrections, the theoretical
upper bound is increased.  The dominant effect arises from an incomplete
cancellation of the top-quark and top-squark loops (these
effects actually cancel in the exact supersymmetric limit).\footnote{In
certain regions of parameter space (corresponding to large $\tanb$ and
large values of $\mu$), the
incomplete cancellation of the bottom-quark and bottom-squark loops can
be as important as the corresponding top sector contributions.  For
simplicity, we ignore this contribution in \eq{deltamh} below.}
The qualitative behavior of the radiative corrections can be most easily
seen in the large top squark mass limit, where in addition, the
splitting of the two diagonal entries and the off-diagonal entry
of the top-squark squared-mass matrix are both small in comparison to
the average of the two stop squared-masses:
\beq \label{mstwo}
\msusyy\equiv\half(\mstopa^2+\mstopb^2)\,.
\eeq
In this case, the upper bound on the lightest CP-even Higgs
mass is approximately given by
\beq \label{deltamh}
\mhl^2\lsim \mz^2+{3g^2\mt^4\over
8\pi^2\mw^2}\left[\ln\left({M_S^2\over\mt^2}\right)+{x_t^2}
\left(1-{x_t^2\over 12}\right)\right]\,,
\eeq
where $x_t\equiv X_t/M_S$.

The more complete treatments of the radiative corrections
cited above show that
\eq{deltamh} somewhat overestimates the true upper bound of $\mhl$.
Nevertheless, \eq{deltamh} correctly reflects some noteworthy features
of the more precise result.  First, the increase of the
light CP-even Higgs mass bound beyond $\mz$ can be significant.  This is
a consequence of the $m_t^4$ enhancement of the one-loop radiative
correction.
Second, the dependence of the light Higgs mass on the stop mixing
parameter $X_t$ implies that (for a given value of
$\msusy$) the upper bound of the light Higgs mass
initially increases with $X_t$ and reaches its {\it maximal} value
$X_t=\sqrt{6}\msusy$.  This point is referred to as the {\it maximal
mixing} case (whereas $X_t=0$ corresponds to the {\it
minimal mixing} case).  In a more complete computation that includes
both two-loop logarithmic and non-logarithmic corrections, the $X_t$
values corresponding to maximal and minimal mixing are 
shifted and exhibit an asymmetry under $X_t\to -X_t$ as shown in
\fig{mhxt}.
In the numerical
analysis presented in this and subsequent figures in this
section, we assume for simplicity that the third generation
diagonal soft-supersymmetry-breaking 
squark squared-masses are degenerate: $\MSUSY\equiv
M_Q=M_U=M_D$, which defines the parameter $\MSUSY$.\footnote{We also
assume that $\MSUSY\gg\mt$, in which case it follows that
$M_S^2\simeq\MSUSY^2$ up to corrections of ${\cal O}(\mt^2/\MSUSY^2)$.}

\begin{figure}
  \begin{center}
\centerline{\psfig{file=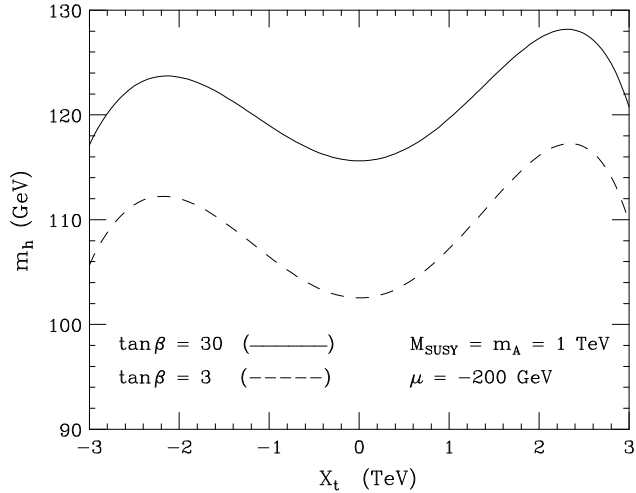,height=6.5cm}}
  \end{center}
  \caption[0]{\label{mhxt} The radiatively corrected
light CP-even Higgs mass is plotted
as a function of $X_t$, where $X_t\equiv A_t-\mu\cot\beta$,
for $\mha=1$~TeV and two choices of $\tanb=3$ and 30.
%
Here, we have taken $M_t=174.3$~GeV, and we have assumed that
the diagonal soft squark squared-masses are degenerate:
$\MSUSY\equiv M_Q=M_U=M_D=1$~TeV.
}
\end{figure}

Third, note the logarithmic sensitivity to the top-squark
masses.  Naturalness arguments that underlie
low-energy supersymmetry imply that the supersymmetric particles masses
should not be larger than a few TeV.  Still, the precise upper bound on
the light Higgs mass depends on the specific choice for the upper limit
of the stop masses.  The dependence of the light Higgs mass obtained by
the more complete computation (to be discussed further below) as a
function of $\MSUSY$ is shown in \fig{mhmsusy}.
\begin{figure}[!ht]
\centering
\centerline{\psfig{file=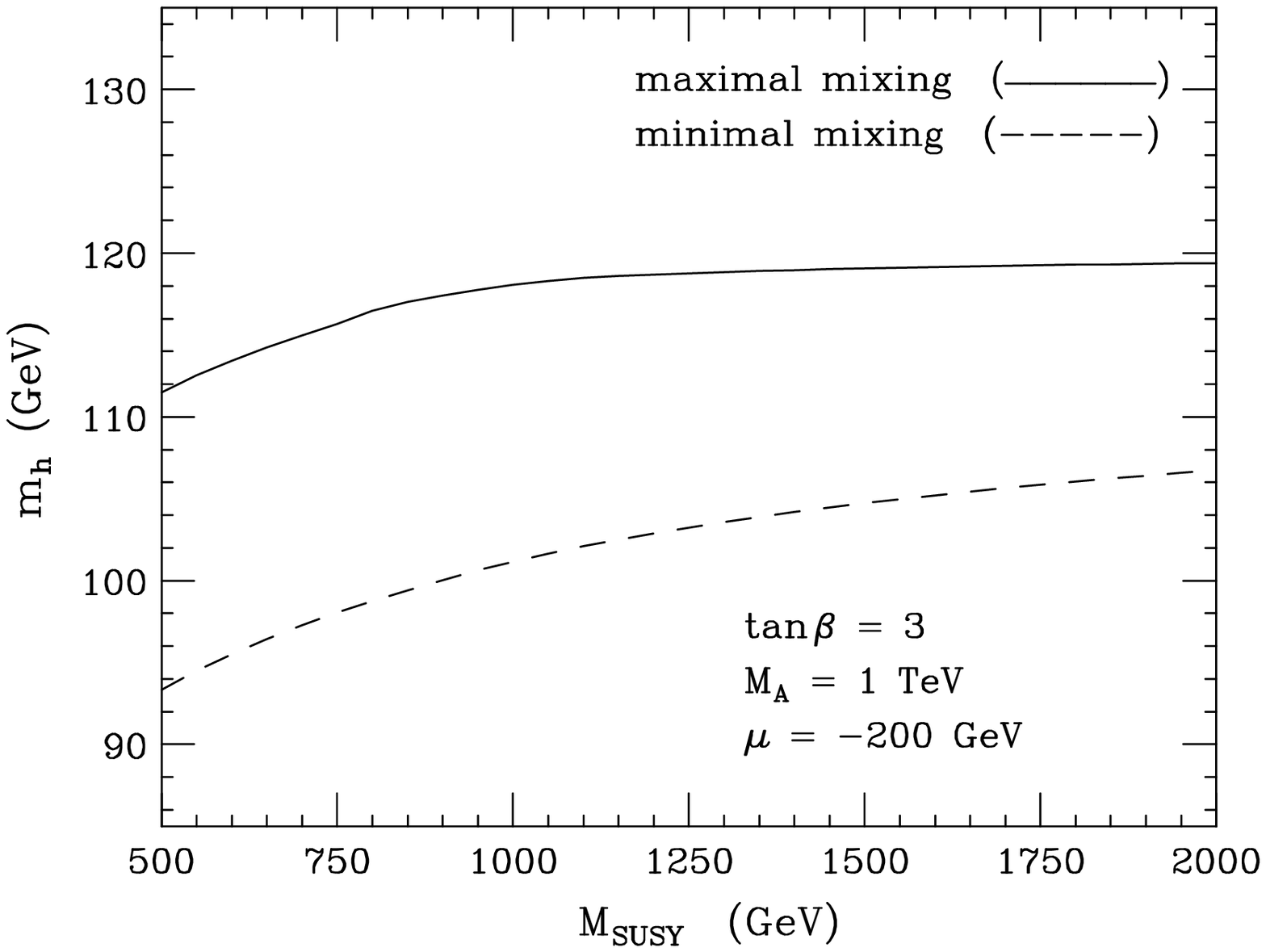,width=8cm}
\hfill
\psfig{file=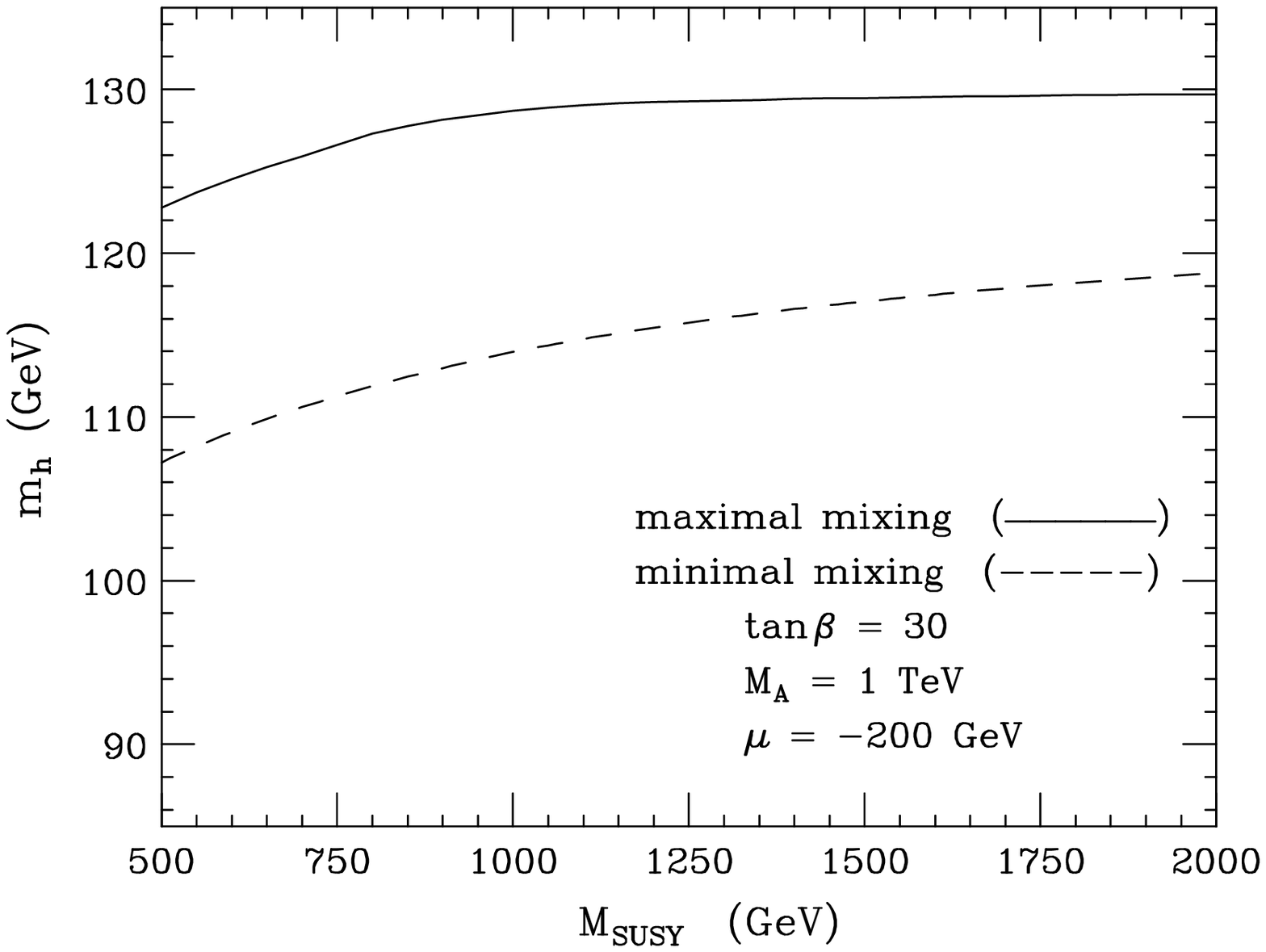,width=8cm}}
\vskip1pc
  \caption[0]{\label{mhmsusy} The radiatively corrected light CP-even
Higgs mass is plotted as a function of $\MSUSY\equiv M_Q=M_U=M_D$,
for $\mha= 1$~TeV and two choices of $\tanb=3$ and
$\tanb=30$.  Maximal mixing and minimal mixing are defined according
to the value of $X_t$ that yields the maximal and minimal Higgs mass
as shown in \fig{mhxt}.  Here, we have chosen $M_t=174.3$~GeV,
while the supersymmetric parameters in the maximal and minimal mixing
cases are chosen according to the first two benchmark scenarios of
\Ref{benchmark} (displayed in table~\ref{benchmarks} of Section~III.B.2).
}
\end{figure}

As noted above, the largest contribution to the one-loop radiative
corrections is enhanced by a factor of $m_t^4$ and grows
logarithmically with the top squark mass.  Thus, higher order
radiative corrections can be non-negligible for large top
squark masses, in which case the large logarithms must be resummed.
The renormalization group (RG) techniques for resumming the leading
logarithms have been developed by a number of
authors \cite{rge,2loopquiros,llog,carena}.
The computation of the RG-improved
one-loop corrections requires numerical integration of a coupled set of
RG equations~\cite{llog}.
Although this procedure has been
carried out in the literature, the analysis is unwieldy
and not easily amenable to large-scale Monte-Carlo studies.
It turns out that over most of the parameter range, it is sufficient to
include the leading and sub-leading logarithms at two-loop order.
(Some additional non-logarithmic terms, which cannot be ascertained by
the renormalization group method, must also be included \cite{chhhww}.)
Compact analytic expressions have been obtained for the dominant one and
two-loop contributions to the matrix elements of the
radiatively-corrected CP-even Higgs squared-mass matrix:
\beq \label{calmatrix}
{\cal M}^2\equiv \left(
\matrix{{\cal M}_{11}^2 &  {\cal M}_{12}^2
\crr {\cal M}_{12}^2 &  {\cal M}_{22}^2 } \right)
={\cal M}_0^2+\delta {\cal M}^2\,,
\eeq
where the tree-level contribution ${\cal M}_0^2$ was given in \eq{kv}
and $\delta {\cal M}^2$ is the contribution from the radiative
corrections.
Diagonalizing this matrix yields radiatively-corrected values for
$\mhl^2$, $\mhh^2$ and the CP-even Higgs mixing angle $\alpha$.
Explicit expressions for the ${\cal M}_{ij}^2$, given in
\Refs{carena}{hhh}, include the dominant leading and sub-leading
logarithms at two-loop order (the latter are generated by an
iterative solution to the RG-equations).  Also included are the
leading effects at one loop of the supersymmetric thresholds and the
corresponding two-loop logarithmically enhanced terms, which can again
be determined by iteration of the RG-equations.
The most important
effects of this type are squark mixing effects in the third
generation.   The procedures described above produce a
prediction for the Higgs mass in terms of running parameters in the
$\overline{\rm MS}$ scheme.  It is
a simple matter to relate these parameters to the corresponding
on-shell parameters used in the diagrammatic
calculations~\cite{espizhang,chhhww}.

Additional non-logarithmic two-loop
contributions, which can generate a non-negligible shift in
the Higgs mass (of a few GeV), must also be included.\footnote{An 
improved procedure for computing the 
radiatively-corrected neutral Higgs mass matrix and the charged Higgs 
mass in a self-consistent way (including possible CP-violating
effects), which incorporates
one-loop supersymmetric threshold corrections to the 
Higgs--top-quark and Higgs--bottom-quark Yukawa couplings,
can be found in \Ref{cpcarlos2}.} 
A compact analytical expression that incorporates these effects
at ${\cal O}(m_t^2 h_t^2 \alpha_s)$ was given in \Ref{compact}.
An important source of such contributions are the
one-loop supersymmetric threshold
corrections to the relation between the Higgs--top-quark and 
Higgs--bottom-quark Yukawa couplings and the corresponding quark masses
[see, {\it e.g.}, \eq{yukbmass}].  These
generate a non-logarithmic two-loop shift of the radiatively corrected
Higgs mass proportional to the corresponding squark mixing parameters.
One consequence of these contributions \cite{chhhww} is the asymmetry
in the predicted value of $\mhl$ under $X_t\to -X_t$ as noted 
in \fig{mhxt}.
Recently, the computation of $\mhl$ has been
further refined by the inclusion of genuine two-loop corrections of
${\cal O}(m_t^2 h_t^4)$ \cite{EZ2}.
These non-logarithmic corrections, which depend on the stop
mixing parameters, can slightly increase the Higgs mass. This
improvement is not yet implemented in the figures shown in this section.

The numerical results displayed in figs.~\ref{cosgraph}--\ref{massvsma}
are based on the calculations of \Refs{carena}{hhh},
with improvements as described in \Refs{weiglein}{chhhww}.
The supersymmetric parameters in the maximal and minimal mixing
cases have been chosen according to the first two benchmark scenarios of
\Ref{benchmark} (displayed in table~\ref{benchmarks} of Section~III.B.2).
Of particular interest is the
upper bound for the lightest CP-even Higgs mass ($\mhl$).  At fixed
$\tan\beta$, the maximal value of
$\mhl$ is reached for $\mha\gg\mz$ (see \fig{massvsma}).
Taking $\mha$ large, \fig{mhtanb} illustrates that the
maximal value of the lightest CP-even Higgs mass
bound is realized at large $\tanb$ in the case of maximal mixing.
For each value of $\tan\beta$,
we denote the maximum value of $\mhl$
by $\mhmax(\tan\beta)$ [this value also depends on the
third-generation squark mixing parameters].
Allowing for the
uncertainty in the measured value of $\mt$ and the
uncertainty inherent in the theoretical analysis,
one finds for $\MSUSY\lsim 2$~TeV that $\mhl\lsim
\mhmax=\mhmax(\tan\beta\gg 1)$,
where
\beqa \label{mhmaxvalue}
\mhmax&\simeq  122~{\rm GeV}, \quad
\mbox{if top-squark mixing is minimal,} \nonumber \\
\mhmax&\simeq  135~{\rm GeV}, \quad
\mbox{if top-squark mixing is maximal.}
\eeqa
In practice, parameters leading to maximal
mixing are not expected in typical models of
supersymmetry breaking. Thus, in general, the  upper bound
on the lightest Higgs boson mass is expected to be
somewhere between the two extreme limits quoted above.
Cross-checks among various programs \cite{cwprog,hehprog,feynhiggs}
and rough estimates of higher order corrections not yet computed
suggest that the results for Higgs masses should be accurate to within
about 2 to 3 GeV over the parameter ranges displayed
in figs.~\ref{mhxt}---\ref{massvsma}.

\begin{figure}
  \begin{center}
\centerline{\psfig{file=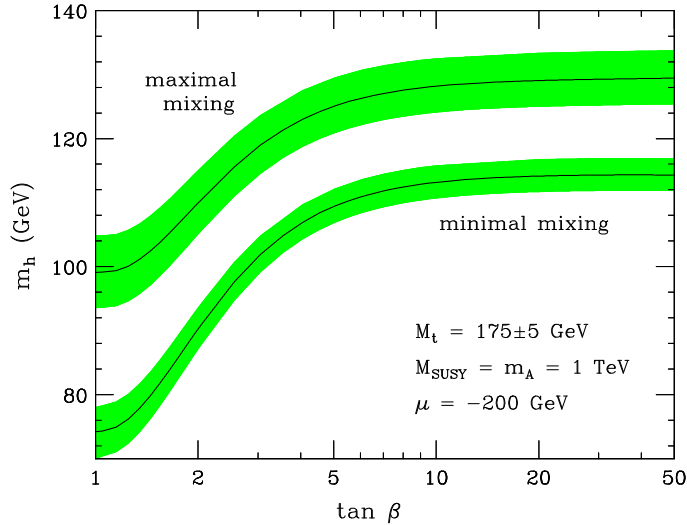,width=9cm}}
  \end{center}
  \caption[0]{\label{mhtanb} The radiatively corrected
light CP-even Higgs mass is plotted
as a function of $\tanb$, for $\MSUSY\equiv M_Q=M_U=M_D=1$~TeV and
$\mha=1$~TeV,
for the maximal mixing [upper band] and minimal mixing [lower band]
benchmark cases [see caption to \fig{mhmsusy}].  The impact of the top
quark
mass is exhibited by the shaded bands; the central value corresponds
to $M_t=175$~GeV, while the upper [lower] edge of the bands
correspond to increasing [decreasing] $M_t$ by 5~GeV. }
\end{figure}


In \fig{massvsma}, we exhibit the masses of
the CP-even neutral and the charged Higgs masses as a function of
$\mha$.  The squared-masses of the lighter and heavier neutral CP-even
Higgs are related by
\beq \label{mhmaxtb}
\mhh^2\cosbmaii+\mhl^2\sinbmaii=[\mhmax(\tan\beta)]^2\,.
\eeq
Note that $\mhh\geq\mhmax$ for all values of
$\mha$ and $\tan\beta$ [where $\mhmax$ is to be evaluated depending on
the top-squark mixing, as indicated in \eq{mhmaxvalue}].
It is interesting to consider the behavior of the
CP-even Higgs masses in the large $\tan \beta$ regime.
For large values of $\tanb$ and for $\mha/\tan\beta \ll
\mhmax(\tan\beta)$, the off-diagonal elements of
the Higgs squared-mass matrix ${\cal M}^2$
become small compared to the diagonal elements
$|{\cal M}_{12}^2| \ll    {\cal M}_{11}^2 + {\cal M}_{22}^2 $;
${\cal M}_{12}^4 \ll    {\cal M}_{11}^2 {\cal M}_{22}^2 $.
Hence the two CP-even Higgs squared-masses
are approximately given by the diagonal elements of ${\cal M}^2$.
As before, we employ the notation where
$\mhmax$ refers to the asymptotic
value of $m_h$ at large $\tanb$ and $\mha$ (the actual numerical
value of $\mhmax$ depends primarily on the assumed values of
the third generation squark mass and mixing parameters).
If $\mha>\mhmax$, then $\mhl\simeq\mhmax$ and
$\mhh\simeq\mha$, whereas if $\mha<\mhmax$, then
$\mhl\simeq\mha$ and $\mhh\simeq\mhmax$.
This behavior can be seen in \fig{massvsma}.

\begin{figure}
  \begin{center}
\centerline{\psfig{file=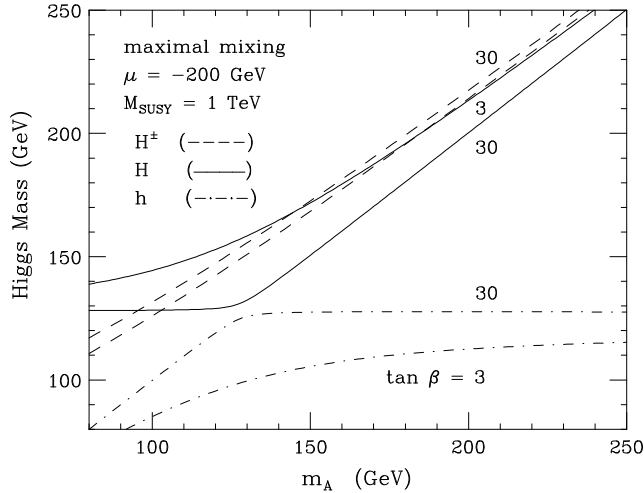,height=6.5cm}}
  \end{center}
  \caption[0]{\label{massvsma} Lightest CP-even Higgs mass ($\mhl$),
heaviest CP-even Higgs mass ($\mhh$) and charged Higgs mass ($\mhpm$) as
a function of $\mha$ for two choices of $\tan\beta=3$ and
$\tan\beta=30$.  Here, we have taken $M_t=174.3$~GeV, and we have assumed
that the diagonal soft squark squared-masses are degenerate:
$\MSUSY\equiv M_Q=M_U=M_D=1$~TeV.  In addition, we choose the
other supersymmetric parameters
corresponding to the maximal mixing benchmark scenario [see caption to
\fig{mhmsusy}].  The slight increase in the charged Higgs mass as
$\tan\beta$ is increased from 3 to 30 is a consequence of the
radiative corrections.
}
\end{figure}

\subsubsection{The Radiatively-Corrected MSSM Higgs Sector: (b) Higgs
couplings}

Radiative corrections also significantly modify the tree-level values
of the Higgs boson couplings to fermion pairs and to vector boson pairs.
As discussed above, the tree-level Higgs couplings depend crucially on
the value of $\sinbma$.  In first approximation,
when radiative corrections of the Higgs
squared-mass matrix are computed, the diagonalizing angle $\alpha$ is
renormalized.  Thus, one may compute a radiatively-corrected value for
$\sinbma$.  This provides one important source of the radiative
corrections of the Higgs couplings.   In \fig{cosgraph}, we show the
effect of radiative corrections on the
value of $\sinbma$ as a function of $\mha$  for different values of the
squark mixing parameters and $\tanb$.  One can then
simply insert the radiatively corrected value of $\alpha$ into
eqs.~(\ref{vvcoup}), (\ref{hvcoup}) and
(\ref{qqcouplings}) to obtain radiatively-improved couplings of
Higgs bosons to vector bosons and to fermions.

To better understand the behavior of the Higgs boson couplings
to fermion pairs, consider the radiatively corrected CP-even Higgs
squared-mass matrix ${\cal M}^2$ [\eq{calmatrix}].  The complete
expressions for the individual matrix elements (even after a
series of approximations in which many sub-leading terms are
dropped) are quite involved, and we will not display them here.
However, it is useful to examine the leading contributions to the
off-diagonal element, ${\cal{M}}^2_{12}$. After including the
dominant one--loop corrections induced by the top-squark
sector\footnote{If $\tanb$ is large, bottom-squark sector effects
may become important.  For simplicity, these effects will be
neglected in the formulae exhibited in this section.}, together
with the two--loop, leading--logarithm effects\footnote{This
result is only valid if the splitting between the two top-squark
squared-masses is small compared to $\msusyy$. In addition, the
conditions $2m_t|A_t|$, $2m_t|\mu|< \msusyy$ must also be
fulfilled.} \cite{CMW1,CMW2},
\beqa
{\cal M}^2_{12} & \simeq  -\left[\mha^2 + \mz^2 \right]
\sin\beta \cos\beta
\nonumber\\
& -  \left[
\frac{3h_t^4 v^2}{48\pi^2 M_S} {\mu x_t}\left(6-{x_t
a_t}\right)\sin^2\beta - \frac{3 h_t^2 \mzz}{32 \pi^2}
{\mu x_t} \right]
\left[ 1 +{9 h_t^2 - 32 g_s^2\over
32\pi^2}\ln\left({M_S^2\over\mt^2}\right)\right],
\label{matel}
\eeqa
where $a_t\equiv A_t/M_S$, $x_t\equiv X_t/M_S$, $g_s$ is the QCD
coupling constant and  $h_t$ is the Higgs--top-quark Yukawa coupling
[\eq{htdef}].  
%
The mixing angle $\alpha$ can be determined by diagonalizing the CP-even
Higgs squared-mass matrix [\eq{calmatrix}]:
\beq
\sin\alpha\cos\alpha = {{\cal M}_{12}^2 \over
\sqrt{({\rm Tr}{\cal M}^2)^2-4\det{\cal M}^2}}\,,
\qquad\qquad
\cos^2\alpha-\sin^2\alpha = {{\cal M}_{11}^2-{\cal M}_{22}^2 \over
\sqrt{({\rm Tr}{\cal M}^2)^2-4\det{\cal M}^2}}\,.
\label{sin2alpha}
\eeq
It follows that in the limit where ${\cal M}_{12}^2\to 0$,
either $\sin\alpha\to 0$ (if ${\cal M}_{11}^2>{\cal M}_{22}^2$) or
$\cos\alpha\to 0$ (if ${\cal M}_{11}^2<{\cal M}_{22}^2$).
As a result, some of the Higgs boson couplings
to quark and lepton pairs [\eq{qqcouplings}] can be strongly
suppressed, if radiative corrections suppress the value of
${\cal M}_{12}^2$.\footnote{Although ${\cal M}_{12}^2$ is {\it negative}
at tree level (implying that $-\pi/2\leq\alpha\leq 0$), it is possible
that radiative corrections flip the sign of ${\cal M}_{12}^2$ [see
\eq{matel}].  Thus, the range of the radiatively corrected angle
$\alpha$ can be taken to be $-\pi/2\leq\alpha\leq\pi/2$.}

If ${\cal M}_{12}^2\simeq 0$ and $\tanb$ is large (values of
$\tan\beta \gsim 5$ are sufficient), the resulting pattern of Higgs
couplings is easy to understand.  In this limit,
${\cal M}^2_{11}\simeq\mha^2$ and ${\cal M}^2_{22}\simeq \mhmax$,
as noted at the end of section~I.C.2.  Two cases must be treated
separately depending on the value of $\mha$.
First, if $\mha < \mhmax$, then
$\sin\alpha \simeq - 1$ (assuming ${\cal M}_{12}^2 <0$ as in tree-level),
$\cos\alpha \simeq 0$ and $\sinb\simeq \cos(\beta - \alpha) \simeq 1$.
In this case, the lighter CP-even Higgs boson
$\hl$ is roughly aligned along the $\Phi_d^0$ direction and
the heavier CP-even Higgs boson
$\hh$ is roughly aligned along the $\Phi_u^0$ direction [see
\eq{scalareigenstates}].  In particular, the coupling of $\hh$ to
$b\bar b$ and $\tau^+\tau^-$ is significantly diminished (since
down-type fermions couple to $\Phi_d^0$), while the $\hh VV$ couplings
[\eq{vvcoup}] are approximately equal to those of the Standard Model
[since $\cosbmaii \simeq 1$].
Consequently, the branching ratios of $\hh$ into
$gg$, $\gamma\gamma$, $c\bar c$, and $W^{+}W^{-}$ can be
greatly enhanced over \SM\ expectations~\cite{CMW1,CMW2,Wells,bdhty}.
Second, if $\mha > \mhmax$ then $\sin\alpha \simeq 0$ and
$\sinb\simeq\cosa\simeq \sin(\beta - \alpha) \simeq 1$ and
the previous considerations for $\hh$ apply now to $\hl$.

For moderate or large values of
$\tan\beta$, the vanishing of ${\cal M}_{12}^2$ leads
to the approximate numerical relation \cite{CMW1}:
\begin{equation}
\left[{\mha^2\over \mz^2}
+ 1 \right] \simeq
{\mu x_t \tanb \over 100M_S} \left(
{2 a_t x_t} -
11 \right)
\left[ 1 -\frac{15}{16 \pi^2}  \log\left({M_S^2\over\mt^2}\right) \right],
\label{suppression}
\end{equation}
where we have replaced $h_t$, $g_s$
and the weak gauge couplings by their approximate numerical values at the
weak scale.
For low values of $\mha$, or large values of the
mixing parameters, a cancellation can easily
take place for large values
of $\tan\beta$.  For instance, if $M_S \simeq
1$ TeV,
$\mu=- X_t=M_S$,
and $m_A  \simeq m_h \simeq 80$ GeV, a cancellation can
take place for $\tan\beta \simeq 28$, with
$m_H \simeq 117$ GeV. The heaviest
CP--even Higgs boson has Standard Model--like couplings
to the gauge bosons
[$\cos^2(\beta-\alpha)\simeq 1$], but the branching ratios for
decays into $W^{\pm}$ bosons, gluons and charm quarks are
enhanced with respect to the SM case:
${\rm BR}(W^+W^-)=0.34$, ${\rm BR}(gg)=0.27$ and ${\rm BR}(c\bar c)=0.11$.

Although it is difficult to have exact cancellation of the off-diagonal element
${\cal M}_{12}^2$, in many regions of the supersymmetric parameter
space,  important suppressions may be present.
Generically, the $m_t^4$--dependent radiative corrections
to ${\cal M}_{12}^2$ depend strongly on the sign of the product
$\mu X_t$ ($A_t \simeq X_t$ for large $\tan\beta$
and moderate $\mu$) and on the value of $|A_t|$.
For the same value of $X_t$, a change in the
sign of $\mu$ can lead to observable variations in the branching
ratio for the Higgs boson decay into bottom quarks. If $|a_t|\lsim
\sqrt{11/2}$, the absolute value of the off--diagonal matrix
element, and hence, the coupling of bottom
quarks to the Standard Model--like Higgs boson tends to be suppressed
(enhanced) for values of
$\mu A_t < 0$  ($\mu A_t > 0$).
For larger values of $|a_t|$, the suppression (enhancement)
occurs for the opposite sign of $\mu A_t$.

In addition to the radiative corrections to couplings that
enter via the renormalization of the CP-even Higgs mixing angle
$\alpha$, there is another source of radiative corrections to Higgs
couplings that is potentially important at large $\tanb$.  Such
corrections depend on the details of the MSSM spectrum (which enter via
loop-effects).  The corrections we wish to explore now are those that
arise in the relation between $m_b$ and $\tanb$.  At tree-level, the
Higgs couplings to $b\bar b$ are proportional to the
Higgs--bottom-quark Yukawa coupling $h_b$ [\eq{hbdef}].
Deviations from the tree-level relation, \eq{hbdef},
due to radiative corrections are calculable
and finite \cite{hffsusyqcd,deltamb,deltamb1,deltamb2}.  One of the
fascinating properties of such corrections is that in certain cases
the corrections do {\it not} vanish in the limit of large supersymmetric
mass parameters.  These corrections grow with $\tanb$ and therefore can
be significant in the large $\tanb$ limit.
%
In the supersymmetric limit, bottom quarks only couple to $\Phi_d^0$.
However, supersymmetry is broken and
the bottom quark will receive a small coupling
to $\Phi_u^0$ from radiative corrections,
\begin{equation}
-{\cal L}_{\rm Yukawa} \simeq h_b \Phi_d^0 b \bar{b} + (\Delta h_b)
\Phi_u^0 b\bar{b}\,.
\label{couplings}
\end{equation}
Because the Higgs doublet acquires a vacuum
expectation value, the bottom quark mass receives
an extra contribution equal to
$(\Delta h_b) v_u$.  Although $\Delta h_b$ is
one--loop suppressed with respect to $h_b$,
for sufficiently large values of $\tan\beta$ ($v_u \gg v_d$)
the contribution to the bottom quark mass of both terms in
\eq{couplings} may be comparable in size. This induces a
large modification in the tree level relation [\eq{hbdef}],
\beq
m_b = {h_b v_d\over\sqrt{2}} (1+\Delta_b)\,, \qquad
\label{yukbmass}
\eeq
where $\Delta_b \equiv (\Delta h_b)\tan\beta/h_b$.
The function $\Delta_b$ contains two main
contributions, one from a bottom squark--gluino loop
(depending on the two bottom squark masses $M_{\tilde b_1}$
and $M_{\tilde b_2}$ and the gluino mass $M_{\tilde g}$) and another one
from a
top squark--higgsino loop (depending on the two top squark masses
$M_{\tilde t_1}$ and $M_{\tilde t_2}$ and the higgsino mass parameter
$\mu$).  The explicit form of $\Delta_b$ at one--loop in the limit of
$M_S \gg m_b$ is given by \cite{deltamb,deltamb1,deltamb2}:
\beq
\Delta_b \simeq {2\alpha_s \over 3\pi} 
M_{\tilde g}\mu\tan\beta~I(M_{\tilde b_1},
M_{\tilde b_2},M_{\tilde g})
 + {Y_t \over 4\pi} A_t\mu\tan\beta~I(M_{\tilde t_1},M_{\tilde t_2},\mu),
\label{deltamb}
\eeq
where $\alpha_s=g_s^2/4\pi$,
$Y_t\equiv h_t^2/4\pi$, and contributions proportional to the
electroweak gauge couplings have been neglected.  In addition,
the function $I$ is defined by
\beq
I(a,b,c) = {a^2b^2\ln(a^2/b^2)+b^2c^2\ln(b^2/c^2)+c^2a^2\ln(c^2/a^2) \over
(a^2-b^2)(b^2-c^2)(a^2-c^2)},
\eeq
and is manifestly positive.
Note that the Higgs coupling proportional to $\Delta h_b$ is a
manifestation of the broken supersymmetry in the low energy theory;
hence, $\Delta_b$ does not decouple
in the limit of large values of the supersymmetry breaking masses. Indeed,
if all supersymmetry breaking mass parameters (and $\mu$)
are scaled by a common factor, the correction
$\Delta_b$ remains constant.

Similarly to the case of the bottom quark, the relation between $m_\tau$ and
the Higgs--tau-lepton Yukawa coupling $h_\tau$ is modified:
\beq
m_\tau = {h_\tau v_d\over\sqrt{2}} (1+\Delta_\tau).
\eeq
The correction $\Delta_\tau$ contains a contribution from a
tau slepton--neutralino loop (depending on the two stau masses
$M_{\tilde \tau_1}$ and $M_{\tilde \tau_2}$ and the
mass parameter of the $\widetilde B$ (``bino'') component
of the neutralino, $M_{1}$) and a
tau sneutrino--chargino loop (depending on the tau sneutrino mass
$M_{\tilde \nu_\tau}$, the mass parameter of the $\widetilde W^\pm$
component of the chargino, $M_{2}$, and $\mu$).
It is given by \cite{deltamb1,deltamb2}:
\beq
\Delta_\tau = {\alpha_1 \over 4\pi} M_1\mu\tan\beta\,
I(M_{\tilde\tau_1},
M_{\tilde\tau_2},M_1) + {\alpha_2 \over 4\pi} M_2\mu\tan\beta \,
I(M_{\tilde\nu_\tau},M_2,\mu),
\eeq
where $\alpha_2\equiv g^2/4\pi$ and $\alpha_1\equiv g^{\prime 2}/4\pi$
are the electroweak gauge couplings.
Since corrections to $h_\tau$ are proportional to $\alpha_1$ and
$\alpha_2$, they are  expected to be smaller than
the corrections to $h_b$.  For example,
for $\tan\beta \lsim 50$, one finds
$\Delta_\tau < 0.15$.

From \eq{couplings} we can obtain the couplings of the physical neutral Higgs
bosons to $b\bar b$.  First, consider the CP-odd Higgs boson
[\eq{hastate}].  From \eq{couplings}, we obtain for the $\ha b\bar b$
coupling:
\beq \label{lintabb}
{\cal L}_{\rm int} = - i g_{\ha b\bar b} A \bar{b} \gamma_5 b
\eeq
with
\beq
g_{\ha b\bar b}
= h_b \sin\beta + \Delta h_b \cos\beta \simeq h_b
\sin\beta = {\sqrt{2}m_b\over (1 + \Delta_b) v}\tan\beta\,,
\label{bhCP}
\eeq
where we have used the result of \eq{yukbmass} for $h_b$, and we have
discarded a term of ${\cal O}(\Delta_b/\tan^2\beta)$.
Similarly, for the CP-even Higgs bosons [\eq{scalareigenstates}],
we obtain for the $\hl b\bar b$ and $\hh b\bar b$ couplings:
\begin{equation}
{\cal L}_{\rm int} =  g_{\hl b\bar b}\hl b \bar{b} +
 g_{\hh b\bar b}\hh b \bar{b}
\end{equation}
with
\begin{equation}
 g_{\hl b\bar b}  \simeq -\frac{\sqrt{2}m_b \sin\alpha}{v \cos\beta}\
{1\over 1+\Delta_b}
\left[ 1 - \frac{\Delta_b}{\tan\alpha \tan\beta} \right]\,,
\label{barhb}
\end{equation}
\begin{equation}
 g_{\hh b\bar b} \simeq \frac{\sqrt{2}m_b \cos\alpha}{v \cos\beta}
{1\over 1+\Delta_b}
\left[ 1 + \frac{\Delta_b\tan\alpha}{\tan\beta} \right]\,.
\label{tildehb}
\end{equation}
~\\

The sign of  $\Delta_b$
is governed by the sign of $M_{\tilde{g}}  \mu$,
since the  bottom-squark gluino loop  gives the dominant contribution
to Eq. (\ref{deltamb}). Henceforth, we define  $M_{\tilde{g}}$ to be
positive.  Then for $\mu >0$ ($\mu<0$), the radiatively
corrected coupling  $g_{\ha b\bar b}$ in \eq{bhCP}
is suppressed (enhanced) with respect to
its  tree level value.  In contrast, the radiative corrections to
$g_{\hl b\bar b}$ and $g_{\hh b\bar b}$ [\eqns{barhb}{tildehb}]
have a more complicated
dependence on the supersymmetric parameters due to the dependence on
the CP-even mixing angle $\alpha$.
Since $\alpha$ and  $\Delta_b$  are governed by different
combinations of the supersymmetry breaking parameters, it is difficult
to exhibit in a simple way the behavior
of the radiatively corrected couplings of the CP-even Higgs bosons
to the bottom quarks as a function of the MSSM parameters.

It is interesting to study different limits of the above couplings.
For $m_A \gg \mhmax$,
the lightest CP--even Higgs boson should behave like
the \SM\ Higgs boson.  To verify this assertion, one can use the result
for the CP-even mixing angle $\alpha$ in the decoupling limit
[\eq{cotalf}].  Plugging this result into \eq{barhb}, one indeed finds
that in the limit of large $\mha$,
$g_{\hl b\bar b}  = \sqrt{2}m_b/v$, which is precisely the coupling of
the \SM\ Higgs boson to $b\bar b$ \cite{loganetal}.
Moreover, in the same limit of large $\mha$,
$g_{\hh b\bar b} \simeq h_b \sin \beta [1 + {\cal O}
(\Delta_b/\tan^2\beta)]$.  When $\mha$ approaches $\mhmax$,
due to the large $1/\cos\beta$ factor appearing in the definition
of the Yukawa coupling [\eq{barhb}],
a small departure from $\cos \beta \simeq - \sin \alpha$ can induce large
departures of the coupling $g_{\hl b\bar b}$
from the \SM\ value.
In contrast, for $\mha \ll\mhmax$, $\sin\beta \simeq -\sin\alpha \simeq 1$.
In this case, $g_{\hl b\bar b}\simeq
h_b \sin \beta [1 + {\cal O}(\Delta_b/\tan^2\beta)]$, while
the $\hh$ coupling to $b\bar b$ may deviate substantially from
the corresponding coupling of the \SM\ Higgs boson.

As discussed above, in the large $\tan\beta$
regime, the off--diagonal elements of the Higgs squared-mass
matrix can receive large radiative corrections
with respect to the tree--level value.
If the bottom-quark and tau-lepton mass corrections
are large, then the bottom-quark and tau-lepton couplings to either
$\hl$ or $\hh$  do not vanish when $\sin 2\alpha=0$,
but are given by $\Delta h_b$ and
$\Delta h_{\tau}$, respectively.  This result is exhibited explicitly in
\eqs{barhb}{tildehb}.  For example, in the limit of
$\sin\alpha = 0$, \eq{barhb} yields
\begin{equation}
g_{\hl b\bar b}={\sqrt{2} m_b\Delta_b \over
v\sinb(1+\Delta_b)}=\Delta h_b\,. \label{barhb2}
\end{equation}
The last step above is a consequence of \eq{yukbmass} and the
definition of $\Delta_b$ in terms of $\Delta h_b$.  Likewise, in
the limit of $\cosa=0$, \eq{tildehb} yields $g_{\hh b\bar
b}=\Delta h_b$.  In both cases, the Higgs couplings to $b\bar b$
are much smaller than the corresponding \SM\ coupling only if
$|\Delta_b| \ll 1$.\footnote{In a full one-loop computation, one must
take into account the momentum dependence of the two-point functions,
which leads to an effective momentum-dependent mixing angle.
Nevertheless, suppressions of Higgs branching ratios persist, depending
on the choice of the supersymmetric parameters, as shown in
\Ref{hffsusyprop}.}
Similar results apply to the Higgs couplings to $\tau^+\tau^-$.

If $\Delta_b\simeq {\cal O}(1)$ (which is possible if $\tanb\gg
1$), then the Higgs couplings to $b\bar b$ are not particularly
small in the limiting cases considered above.  However, a strong
suppression of the couplings can still take place if the value of
the CP-even Higgs mixing angle $\alpha$ is slightly shifted away
from the limiting values considered above.  For example,
\eq{barhb} implies that $g_{\hl b\bar b}\simeq 0$ if
$\tan\alpha=\Delta_b/\tanb$. Inserting this result into the
corresponding expression for the $\hl\tau^+\tau^-$ coupling, it
follows that
\begin{equation}
g_{\hl\tau^+\tau^-} = \frac{\sqrt{2}m_{\tau}\cosa}{v\sin\beta} \left(
\frac{ \Delta_{\tau} - \Delta_b}{ 1 + \Delta_{\tau}} \right)\,.
\end{equation}
Similarly, \eq{tildehb} implies
that $g_{\hh b\bar b}\simeq 0$ if $\tan\alpha=-\tanb/\Delta_b$.
Inserting this result into the corresponding expression for the
$\hh\tau^+\tau^-$ coupling, it follows that
\begin{equation}
g_{\hh\tau^+\tau^-} = \frac{\sqrt{2}m_{\tau}\sina}{v\sin\beta}\left(
\frac{ \Delta_{\tau} - \Delta_b}{ 1 + \Delta_{\tau}} \right)\,.
\end{equation}
In both cases, we see that although the Higgs coupling to $b\bar b$ can
be strongly suppressed for certain parameter choices, the corresponding
Higgs coupling to $\tau^+\tau^-$ may be  typically unsuppressed if
$\tan\beta$ is very large and $\Delta_b\simeq {\cal O}(1)$.  In such
cases, the $\tau^+\tau^-$ decay mode can be the dominant Higgs decay
channel for the CP-even Higgs boson with SM-like couplings to gauge
bosons.


To recapitulate, a cancellation in the off--diagonal element of
the CP-even Higgs squared-mass matrix (as a consequence of the
radiative corrections) can lead to a strong suppression of the
Higgs boson coupling to $b\bar b$ and $\tau^+\tau^-$. In general,
this implies a significant increase in the corresponding Higgs
branching ratios into gauge bosons and charm quark pairs. However,
for very large values of $\tan\beta$ and values of the
bottom-quark mass corrections $\Delta_b$ of order one, the Higgs
branching ratio into $\tau^+\tau^-$ may increase in the regions in
which the $b\bar b$ decays are strongly suppressed.

A similar analysis can be used to derive radiatively corrected couplings
of the charged Higgs boson to fermion pairs \cite{chhiggstotop2,eberl}.  
The tree-level couplings
of the charged Higgs boson to fermion pairs
[\eq{hpmqq}] are modified accordingly by replacing
$m_b \rightarrow m_b/(1 + \Delta_b)$ and
$m_{\tau} \rightarrow m_{\tau}/(1 + \Delta_{\tau})$, 
respectively.

\subsubsection{Present Status of the MSSM Higgs Boson Searches}

Before turning to the relevant MSSM Higgs production processes
and decay modes at the
Tevatron, we shall survey the expected status of the MSSM Higgs
search at the end of the final LEP2 collider run.  At LEP2,
the MSSM Higgs boson production processes are $e^+e^- \rightarrow Z^*
\to Z\hl$ and $\hl \ha$.  In very tiny regions of MSSM parameter
space, the production of a $\hh$ instead of a $\hl$ is kinematically
possible and is also considered.
Since the $ZZ\hl$ coupling is proportional to $\sin(\beta-\alpha)$,
while the $Z\hl\ha$ coupling is proportional to $\cos(\beta-\alpha)$,
these two processes are complementary throughout the
$\mha$ {\it vs.} $\tan\beta$ plane.
The large $\tanb$ ($\tanb
>3$), low $\mha$ region [where $\cosbma$ is large, close to one] is
covered via $\hl \ha$ searches while the large $\mha$ region is
covered via $\hl Z$ searches.
The large $\mha$ region corresponds to the decoupling limit [where
$\sinbma\simeq 1$], and the $\hl Z$ production cross~section
approaches its SM value.  The discovery and
exclusion limits in the latter case can be determined from the
corresponding results of the Standard Model Higgs boson shown in
\fig{SM95}. In the low $\mha$
and low $\tanb$ region both the $Z\hl$ and $\hl\ha$ production channels
contribute.

At present, the published LEP bounds for the MSSM CP-even Higgs bosons are:
$\mhl> 88.3$~GeV and $\mha> 88.4$~GeV at 95\% CL \cite{LEPHiggs}.
These results correspond to the large $\tan\beta$ region in which $Z\hl$
production is suppressed.  The Higgs mass limits then arise from the
non-observation of $\hl\ha$ production.  These limits are expected to
improve slightly when the LEP2 data from 2000 are fully analyzed.  
Projected coverage of the MSSM parameter space via
Higgs searches after the final run of LEP2 are shown in
\fig{mh200} \cite{pjanot}.
The projected 95\% CL exclusion contours 
in the $(a) \tanb$ {\it vs.} $\mhl$ and $(b) \tanb$ {\it vs.} $\mha$
planes are exhibited.
To make such plots, the value of $\MSUSY$ and
the squark mixing, which affect the radiatively corrected Higgs
masses and couplings, must be specified.
In general, the most conservative limits
in the $\tan\beta$ {\it vs.} $\mhl$ [or $\mha$] plane are
obtained in the case of maximal mixing chosen for these
plots.\footnote{The contours shown in \fig{mh200} do not employ the most
recent set of two-loop Higgs mass radiative corrections (discussed in
Section~I.C.2).  When these corrections are included, the theoretical
upper bound on $\mhl$ is slightly increased.  This implies that the
region of MSSM parameter space {\it not} ruled out by the projected 
LEP Higgs searches is slightly larger than the one shown in \fig{mh200}.} 
However, there are some
regions of MSSM parameter space, in which the Higgs couplings to
vector boson pairs or fermion pairs may be suppressed and small holes
may appear for other choices of the supersymmetric parameters 
(see {\it e.g.} refs.~\cite{benchmark} and \cite{tomjunk}) in areas
that are otherwise covered in the maximal mixing case shown in
\fig{mh200}.  


\begin{figure}[!ht]
\centering
\centerline{\psfig{file=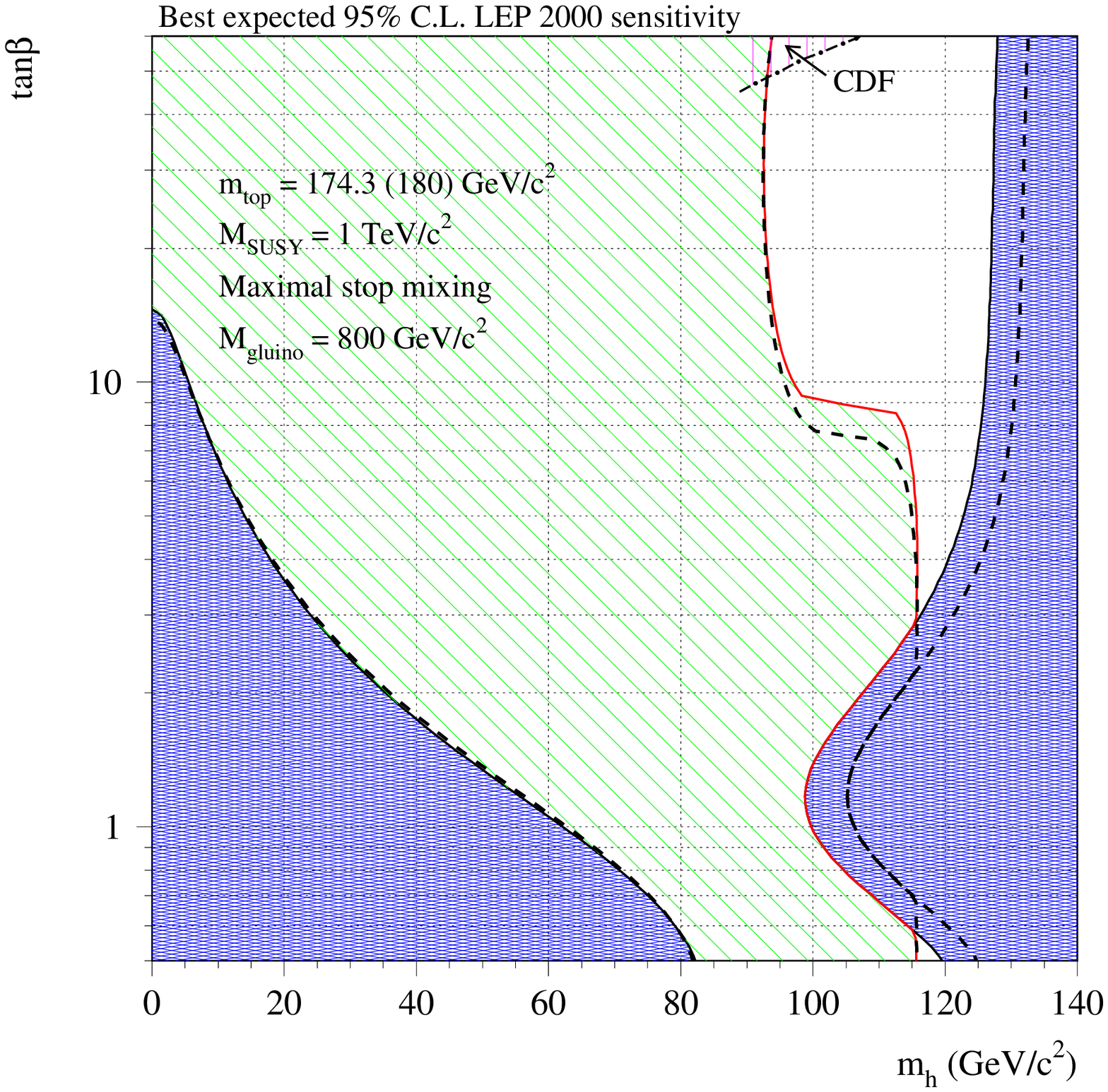,width=8cm}
\hfill
\psfig{file=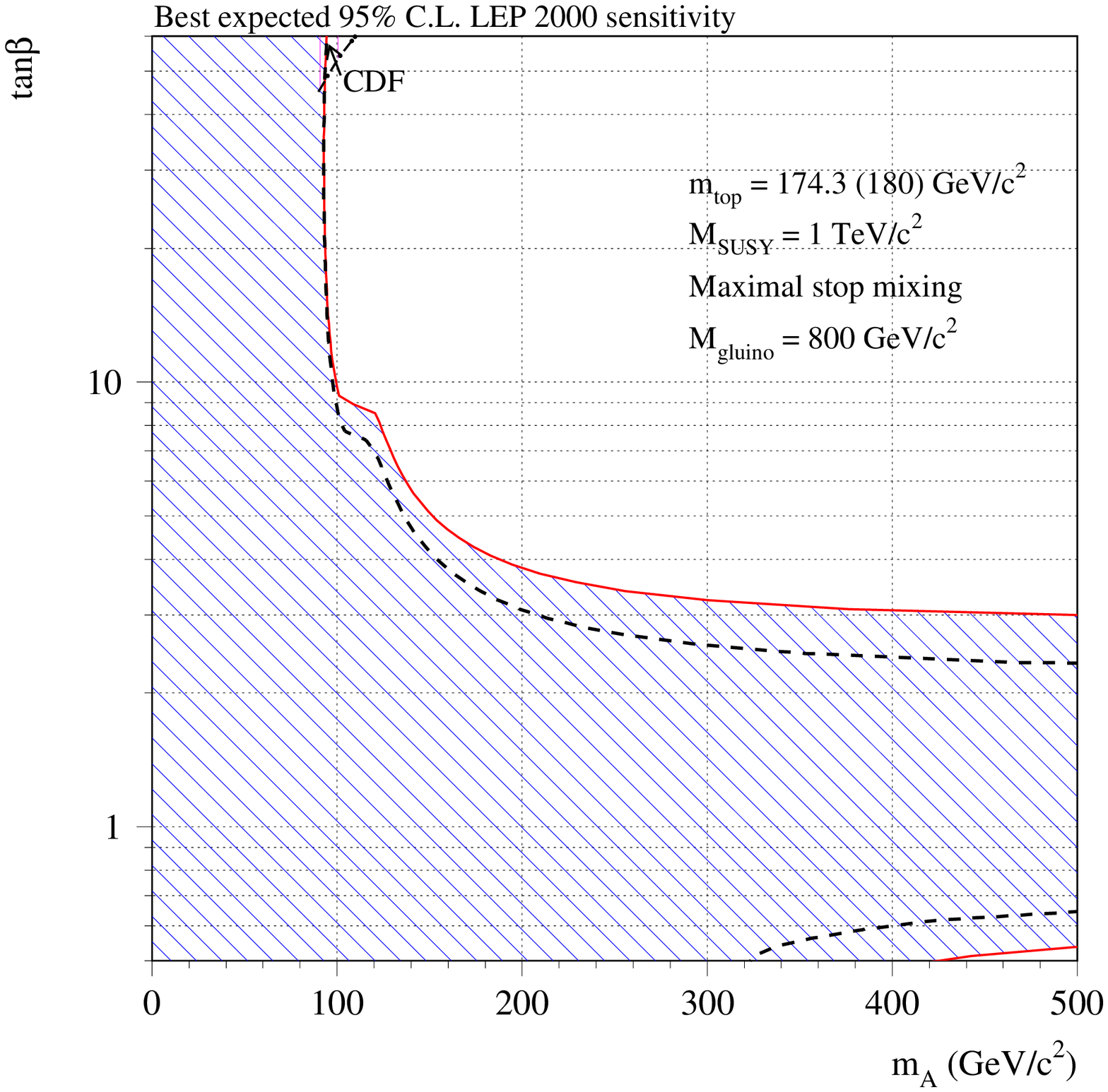,width=8cm}}
\vskip1pc
\caption[0]{\label{mh200} Projected LEP2 contours of the 95\% C.L. exclusion
limits for MSSM Higgs sector parameters as a function of $\tan\beta$ and
(a) $\mhl$ and (b) $\mha$ (in GeV), taken from \Ref{pjanot}.
In (a), the dark shaded region is
theoretically forbidden.  The light shaded regions in (a) and (b) are
expected to be ruled out after the final year of LEP2 running.  The
solid [dashed] line boundary between the light shaded and unshaded
region corresponds to an assumed value of $M_t=174.3$ [180]~GeV, which
governs the size of the radiative corrections to the Higgs masses.
The contours shown have been obtained assuming an average top-squark
squared-mass of $(1~{\rm TeV})^2$ and maximal top-squark mixing.}
\end{figure}

From \fig{mh200} $(a)$ and $(b)$,
it follows that, independent of the value
of $\tanb$, LEP2 will be sensitive to a light Higgs boson
with a mass up to about 93 GeV \cite{pjanot},
depending slightly on the final luminosity collected.
For low values of $\tanb$, the coverage is better and the search
will cover up to the theoretically largest allowed value of the
lightest Higgs mass, which is obtained for large values of the CP-odd
Higgs mass. In this decoupling limit,
the lightest Higgs has SM-like properties.  The maximal LEP2
reach for a light Higgs with SM-like couplings to the $Z$ boson is the same
as the SM one, namely up to a mass of about 112 GeV for a $5\sigma$ discovery.

The anticipated final LEP lower bounds on the SM-like Higgs mass
significantly constrain the MSSM parameter space.
In particular, it is possible to derive a lower bound on $\tanb$ as a
function of the stop masses and mixing angles.
As an example, \fig{ccpw}
shows the bounds on $\tanb$ that will be obtained in the case that no Higgs
boson is found at LEP with a mass below 115~GeV in the $Z \hl$ channel.
This figure illustrates that the lower bound on $\tan\beta$ will range
between about 2.3 and 5 depending on the value of the stop masses and
mixing.  It is interesting to note that even for very large values
of the stop masses and mixing angles, the bound on $\tanb$ resulting
from the direct bound on the Higgs mass becomes stronger than the
bound on $\tanb$ that one can get by the requirement of perturbative
consistency of the theory up to scales of order $M_{\rm PL}$ (associated
with the infrared fixed point solution of the top Yukawa coupling)
\cite{CCPW}.
The constraints on the
MSSM parameters that will be available from Higgs searches
after the final run of LEP2 will be useful for guiding supersymmetric
and Higgs particle searches at the Tevatron and the LHC.

\begin{figure}[!ht]
\begin{center}
\centerline{\psfig{file=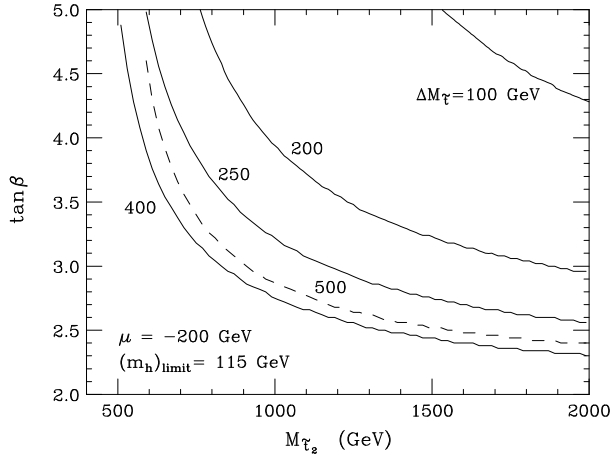,width=8cm}}
\end{center}
\vspace*{-1pc}
\caption[0]{\label{ccpw} Bounds on $\tanb$ obtained for
$M_t$ = 175 GeV and for a lower bound on the Higgs mass
of 115 GeV, as a function of the heavier stop mass $M_{\tilde{t}_2}$,
for different values of the stop mass splitting $\Delta M_{\tilde t}\equiv
M_{\tilde{t}_2}-M_{\tilde{t}_1}$ ranging from 0 to 500 GeV
(subject to the constraint that the lighter top squark mass is greater
than 65~GeV).
As $\Delta M_{\tilde t}$  increases, the contours move to the left
until the 400~GeV contour is reached.  (The contours between 300
GeV and 450 GeV are all nearly on top of the 400~GeV contour.)  The
dashed line [the 500 GeV contour] shows that the contours begin to move
back to the right as $\Delta M_{\tilde t}$ increases further.
This plot is an updated version of one that was first shown in
\Ref{CCPW}, and was produced using \Ref{feynhiggs}.
  }
\end{figure}

\medskip

One of the goals of the Tevatron Higgs Working Group is to
examine the potential for the upgraded Tevatron to extend the LEP2
MSSM Higgs search.  We begin by examining the most promising channels
for MSSM Higgs discovery at the upgraded Tevatron.

\subsubsection{MSSM Higgs Boson Decay Modes}

In the MSSM, we must consider the decay properties of three neutral
Higgs bosons ($\hl$, $\hh$ and $\ha$) and one charged Higgs pair
($\hpm$).  In the region of parameter space where $\mha\gg m_Z$
and the masses of supersymmetric particles are large, the
decoupling limit applies, and we find that the properties of $\hl$
are indistinguishable from the Standard Model Higgs
boson.\footnote{If supersymmetric particles are light, then
the decoupling limit does not strictly apply even in the limit of
$\mha\gg m_Z$.  In particular, the $\hl$ branching ratios are
modified, if the decays of $\hl$ into supersymmetric particles are
kinematically allowed.  In addition, 
if light superpartners exist that can couple
to photons and/or gluons, then the one-loop $gg$ and $\gamma\gamma$
decay rates would also deviate from the corresponding Standard Model
Higgs decay rates due to the extra contribution of the
light superpartners appearing in the loops.}
In this
case, the discussion of Section~I.A.2 applies, and the decay
properties of $\hl$ are precisely those of the Standard Model Higgs
boson.  In this case, the heavier Higgs states, $\hh$, $\ha$ and
$\hpm$, are roughly mass degenerate and cannot be observed at the
upgraded Tevatron.

For values of $\mha\sim {\cal O}(\mz)$, all Higgs boson states lie
below 200~GeV in mass, and could in principle be accessible at an
upgraded Tevatron (given sufficient luminosity).  In this parameter
regime, there is a significant area of the parameter space in which
none of the neutral Higgs boson decay properties approximates that of
the Standard Model Higgs boson.  For example, when
$\tan\beta$ is large, supersymmetry-breaking effects can significantly
modify the $b\bar b$ and/or the $\tau^+\tau^-$ decay rates with
respect to those of the Standard Model Higgs boson.
Additionally, the Higgs bosons can decay into new channels, either
containing lighter Higgs bosons or supersymmetric particles.
In the following,
the decays of the neutral Higgs bosons $\hl$, $\hh$ and
$\ha$ (sometimes denoted collectively by $\phi$) and 
the decays of charged Higgs
bosons are discussed with particular emphasis on differences
from Standard Model expectations.

As discussed earlier (see \fig{mh200}), if the
Higgs boson has not yet been discovered by the time the upgraded
Tevatron begins its run, then the anticipated LEP limits will rule out
values of $\tan\beta$ below about 2.5.\footnote{LEP will not be able to
rule out small values of $\tan\beta\simeq 0.5$ in a narrow range of
MSSM parameter space.  However, theoretical arguments favor values of
$\tan\beta\gsim 1$, so we do not consider further the possibility of
$\tan\beta\lsim 1$.}
Thus, in the following discussion, two cases are considered to illustrate
the difference between ``low'' and ``high'' $\tan\beta$:
$\tan\beta=6$ and 30.

\vskip6pt\noindent
\centerline{a.~Neutral Higgs Boson Decays}
\vskip6pt

\vskip6pt\noindent
\underline{$\phi\to f\bar f$} \\[0.2cm]
In the MSSM, the decay modes $\hl,\hh,\ha \to b\bar b$, $\tau^+\tau^-$
dominate
the neutral Higgs decay modes for large $\tanb$, while for small $\tanb$
they are important for neutral Higgs masses $\lsim 150$ GeV as can be
seen from \fig{fg:2}a--c.
As in the \SM\ case, the QCD corrections~\cite{hffsmqcd}
reduce the partial
decay widths into $b,c$ quarks by about 50--75\% as a result of the
running quark masses, while they are moderate for decays into top
quarks \cite{DSZ}.
The dominant Higgs propagator corrections of ${\cal O}(G_F m_t^4/m_W^2)$
and ${\cal O}(G_F \alpha_s m_t^4/m_W^2)$ can to a good approximation be
absorbed into the effective mixing angle $\alpha$~\cite{hffsusyprop}. As
explained above, as a consequence of this universal correction the
coupling of $\hl$ to $b\bar b$ and $\tau^+\tau^-$
can be strongly suppressed for
small $\mha$ and large $\tan\beta$. For the decays into $b\bar b$,
the supersymmetric-QCD
corrections~\cite{hffsusy1l,hffsusyqcd,bfpt,CMW1,eberl,hffsusyprop,loganetal}
can be very significant for large values of $\mu$ and $\tan\beta$.
Their dominant effect manifests
itself in the relation between $m_b$ and the Yukawa couplings
[resulting in \eqns{barhb}{tildehb}].
The dominant electroweak corrections from
higgsinos to $\hl \to f \bar f$ enter in the same way.
The remaining
process-specific electroweak one-loop corrections are typically at the
level of only a few percent~\cite{hffsusy1l}.

\begin{figure}
\centering
\centerline{\psfig{file=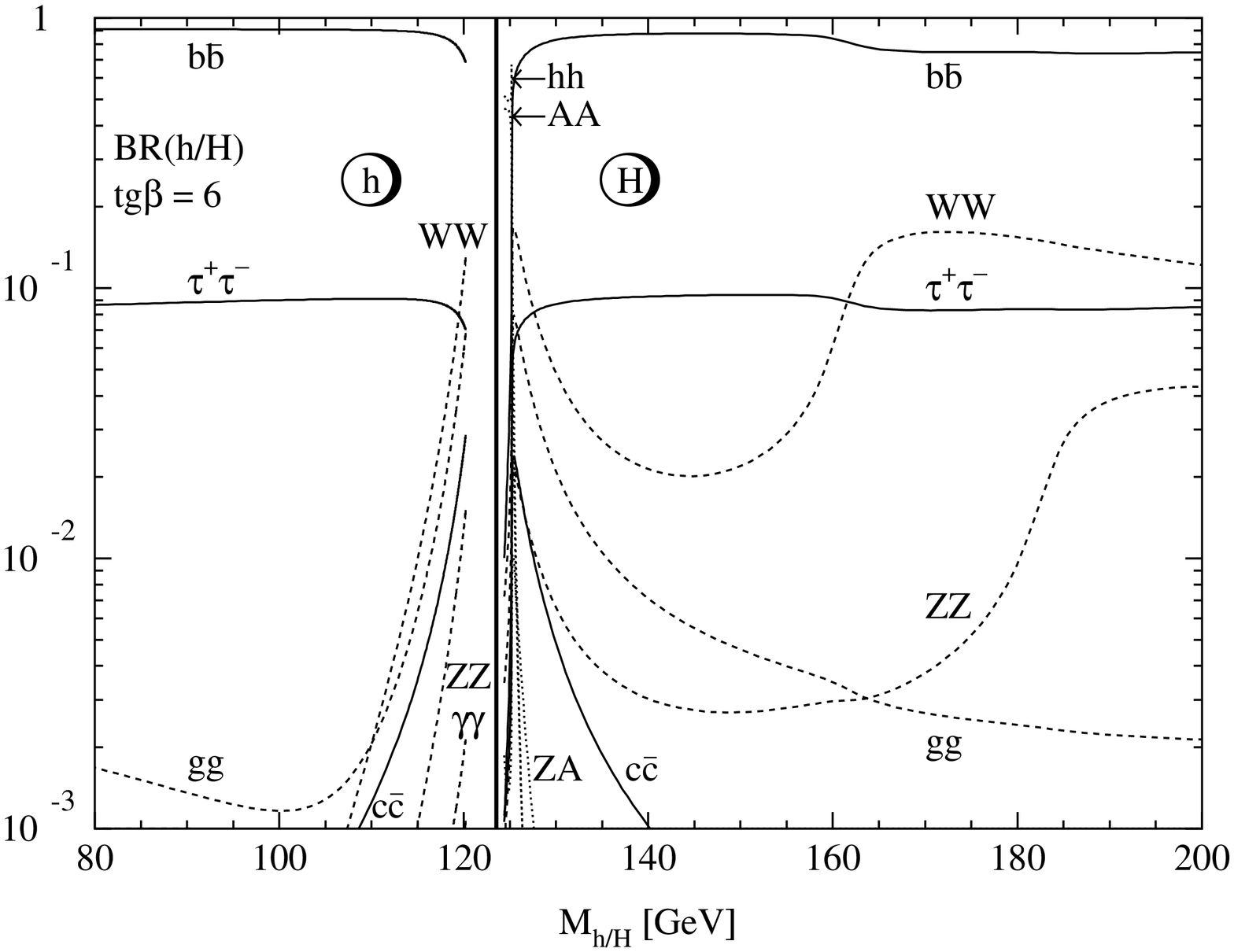,width=8cm,angle=-90}
\hfill
\psfig{file=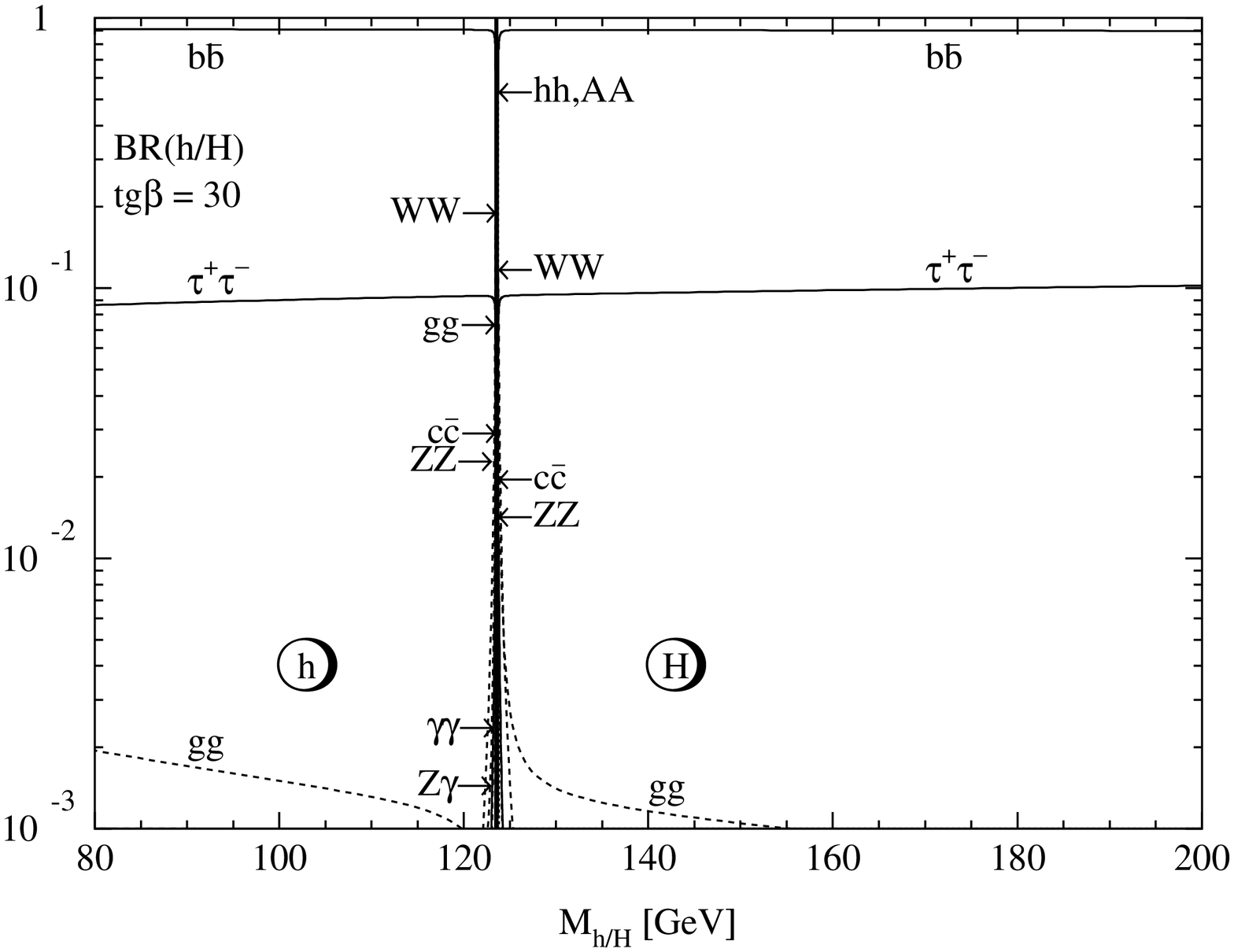,width=8cm,angle=-90}}
\vskip2pc
\centerline{\psfig{file=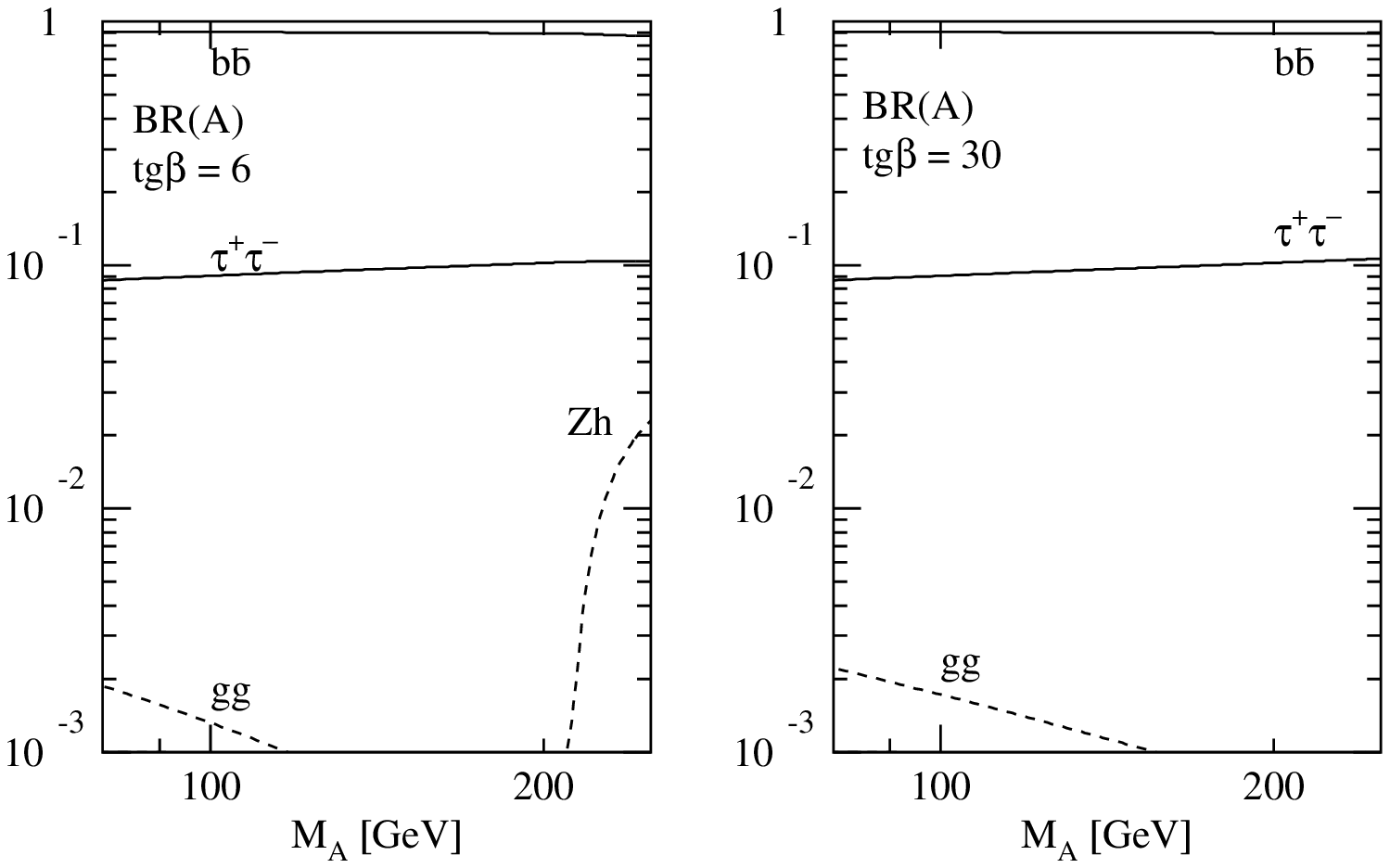,width=8cm,angle=-90}
\hfill
\psfig{file=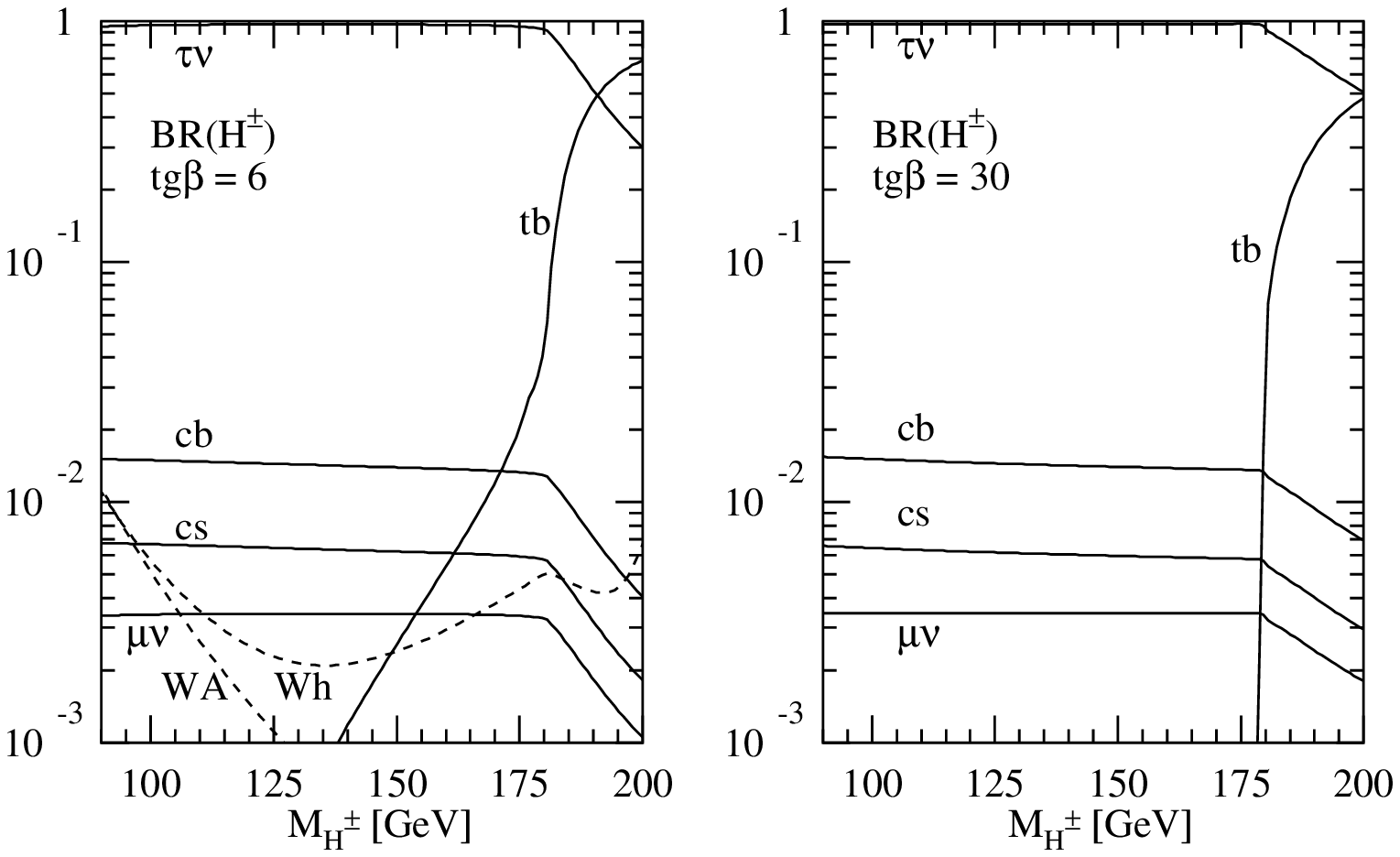,width=8cm,angle=-90}}
\vskip1pc
\caption[0]{\label{fg:2} Branching ratios of the MSSM Higgs bosons
(a) $\hl$ and $\hh$, with $\tan\beta=6$ (b)  $\hl$ and $\hh$, with
$\tan\beta=30$ (c) $\ha$, with $\tan\beta=6$ and $30$ and
(d) $\hpm$ with $\tan\beta=6$ and $30$.  The Higgs masses and branching ratios
are sensitive to the third generation squark spectrum through
radiative corrections.  The above plots were made under the assumption
that the average top and bottom squark masses are 1~TeV and top-squark
mixing is maximal.  In addition, other supersymmetric parameters are
chosen such that there are no supersymmetric particle decay modes in
the Higgs mass ranges shown above.}
\end{figure}

\vskip6pt\noindent
\underline{$\phi\to WW$, $ZZ$} \\[0.2cm]
In the MSSM, the decays $\hl, \hh\to WW,ZZ$ are suppressed by
kinematics and by angle factors [\eq{vvcoup}].  Hence, these decays 
are less important than in the
SM. Their branching ratios turn out to be sizeable only for small and
moderate $\tanb$
or in the decoupling regime, where the light scalar Higgs particle $\hl$
reaches the upper bound of its mass.

\break
\vskip6pt\noindent
\underline{$\phi\to gg$, $\gamma\gamma$} \\[0.2cm]
The gluonic (photonic) decays of the Higgs bosons $\hl,\hh,\ha\to gg
(\gamma\gamma)$ reach branching ratios of $\sim 10\%$ ($\sim 10^{-2}$) in the
MSSM and are expected to be unimportant at the Tevatron.
A light top squark can enhance the $\gamma\gamma$ branching ratio
when its coupling has the opposite sign to the top quark contribution,
and it will reduce the $gg$ branching ratio in this case.

In addition to the above decay modes, which also arise in the Standard
Model, there exist new Higgs decay channels that involve scalars of
the extended Higgs sector and supersymmetric final states.
The unambiguous observation of these modes would clearly constitute
direct evidence of new physics beyond the Standard Model.

\vskip6pt\noindent
\underline{$\hh\to \hl\hl, \ha\ha$; $\hl\to \ha\ha$} \\[0.2cm]
The decay $\hl\to\ha\ha$ is kinematically 
possible only for small values of $\mha$,
which would seem to have been ruled out by the present LEP 
Higgs mass bounds.  However,
there may be small regions in the MSSM parameter space, in which this
decay is still kinematically allowed, consistent with the LEP Higgs
search. The one-loop radiatively corrected rate for $\hl\to\ha\ha$ has
been computed in \Ref{htoaa}.

For $\mhh\lsim 200$~GeV, the decay modes $\hh\to \hl\hl$, $\ha\ha$ are
significant in the MSSM only for a very small range of $\hh$ masses
(as shown in \fig{fg:2}).  However, the process $\hh\to\hl\hl$ is
not kinematically possible given the present LEP bounds on $\mhl$.
The present LEP limits on $\mhl$ and $\mha$ also suggest that
$\hh\to\ha\ha$ is ruled out, although there may be 
small regions of the MSSM parameter space (as noted above) 
in which this latter decay mode is still allowed.
For completeness, we note that for
values of $\tanb\lsim 5$, the branching ratio of $\hh\to \hl\hl$
is dominant for a Higgs mass range of $200~{\rm GeV}\lsim\mhh\lsim 2m_t$.
The dominant radiative corrections to this decay arise from
the corrections to the self-interaction $\lambda_{\hh\hl\hl}$ in the
MSSM and are large \cite{7}.

\vskip6pt\noindent
\underline{$\hh\to\ha Z$, $\hpm W^\mp$; $\ha\to\hl Z$} \\[0.2cm]
The decay modes $\hh\to \ha Z$, $\hpm W^\mp$ and
$\ha\to \hl Z$ are important for
small $\tanb$ below the $t\bar t$ threshold.  Below the
corresponding thresholds, decays into off-shell Higgs and gauge bosons turn out
to be important especially for small $\tanb$ \cite{13}.  However, the
MSSM parameter space where these effects are important for the
Tevatron Higgs search are close to being ruled out by the anticipated
LEP limits.  In particular, present LEP limits already imply that
the two-body decays $\hh\to \ha Z$, $\hpm W^\mp$ are kinematically
forbidden.  Similarly, the decay $\ha\to\hl Z$ can be relevant at the
Tevatron only for $\mha\gsim 200$~GeV and $\tan\beta$ near its
experimental lower limit.


\vskip6pt\noindent
\underline{\it $\phi\to$~supersymmetric particles} \\[0.2cm]
Higgs decays into
charginos, neutralinos and third-generation squarks and sleptons can become
important, once they are kinematically allowed \cite{13a}.
For Higgs masses below 130~GeV, the range of supersymmetric parameter space
in which supersymmetric decays are dominant is rather narrow
(when the current bounds on supersymmetric particle masses is taken
into account).  One interesting possibility is a significant branching
ratio of $\hl\to\widetilde\chi^0 \widetilde\chi^0$, which could arise
for values of $\mhl$ near its upper theoretical limit (assuming the
decay is kinematically possible).  Such an invisible decay mode would
clearly complicate the Higgs search at the Tevatron.
Note that if supersymmetric particles are directly produced and observed
at the Tevatron, then new sources of Higgs boson production (either in
association with a supersymmetric particle or in the decay chains of
supersymmetric particle decays) may exist \cite{scottkon}.


\vskip6pt
\centerline{b.~Charged Higgs Boson Decays}
\vskip6pt

The anticipated constraints on the MSSM Higgs parameter space from LEP
(see \fig{mh200}) restrict the possible values of $\mhpm$ and
$\tan\beta$.  Using the tree-level relation of \eq{susymhpm}, it
follows that $\mhpm\gsim 120$~GeV.\footnote{Using the
radiatively-corrected expression for the charged Higgs mass does not
significantly change this estimate.}  Moreover, if no Higgs signal is
found at LEP, $\tan\beta$ will be constrained to be
larger than roughly 2.5.   
As will be seen in the next section, the prospects for detecting
charged Higgs bosons at the upgraded Tevatron are poor unless they
are light enough to appear among the decay products of the top quark.
In this case, $t\bar t$ production followed by $t\to bH^+$ (or
$\bar t\to \bar b H^-$) provides the most promising mechanism for
discovering a charged Higgs boson at the Tevatron.  These decays are
kinematically allowed when $\mhpm<m_t-m_b$, implying that
$\mha\lsim 150$~GeV.  Thus, there is a relevant range of MSSM
parameter space ($95~{\rm GeV}\lsim \mha\lsim 150$~GeV and
$\tan\beta\gsim 5$) in which the charged Higgs boson of the MSSM
is accessible to the Tevatron Higgs search.  

However, one can take a broader view.  In non-supersymmetric two-Higgs
doublet models, the phenomenology of the Higgs sector depends on the
pattern of Higgs couplings to fermion-pairs.  Type-II two-Higgs
doublet models possess the same Higgs-fermion couplings as in the
MSSM.  Models with a different Higgs-fermion coupling pattern can also
be constructed.
In non-supersymmetric two-Higgs-doublet models, all the
Higgs masses and the angles $\alpha$ and
$\beta$ are independent parameters.  Thus, the
restrictions on the Higgs parameters derived above 
in the case of the MSSM Higgs
sector do not apply in the more general non-supersymmetric model.
and it is quite simple to extend the charged Higgs phenomenology of
the MSSM to the more general case.  Additional complications arise
only when the neutral Higgs bosons of the model affect the production
and/or decay modes of the charged Higgs boson.
Furthermore, using present experimental and theoretical
knowledge, the indirect lower bound on the charged
Higgs mass coming from measurement of the inclusive $b \rightarrow s
\gamma$ decay is about 165 GeV \cite{bsga}
in the non-supersymmetric Type-II two-Higgs-doublets model. 
In the MSSM, no such limit exists for $\mhpm$, since 
supersymmetric particles can be exchanged in loops
contributing to $b \rightarrow s \gamma$ decay; these effects can
approximately cancel out the corresponding charged-Higgs 
exchange contribution in some regions of MSSM parameter space.

\vskip6pt\noindent
\underline{\it $\hpm\to ff'$} \\[0.2cm]
%
%
Because charged Higgs couplings are
proportional to fermion masses [see \eq{hpmqq}], the decays to third
generation quarks and leptons are dominant.  In particular,
for $\mhpm<m_t+m_b$ (so that the channel $H^+\to t\bar b$ is closed),
$H^+\to\tau^+\nu_\tau$ is favored if $\tan\beta\gsim 1$, while
$H^+\to c\bar s$ is favored only if $\tan\beta$ is small.
Indeed, ${\rm BR}(H^+\to\tau^+\nu_\tau) \simeq 1$ if $\tan\beta\gsim 5$.
The  branching fractions of the two-body decay modes of
the charged Higgs boson are shown in \fig{fg:2}d as a function of  $\mhpm$,
and   also in \fig{fg:br_h}
as a function of $\tan\beta$ for several values of $\mhpm$.
These results apply generally to Type-II two-Higgs doublet models.
For $\mhpm\gsim 180$---$200$~GeV, the decay
$H^+\to t\bar b\to W^+b \bar b$ is the dominant decay mode.

The results of \fig{fg:2}d and
\fig{fg:br_h} include QCD quantum effects in all decays when relevant,
besides in the three body decay mode $H^{\pm}\to W^{\pm} b\bar b$.
The QCD radiative corrections have been computed
in \Ref{chhiggsdecay}.
Their effect on the charged Higgs branching
ratios may be significant only in the region of $\tan\beta$ where the
$c\bar s$ and $\tau^+\nu_\tau$ decay modes are competitive or for
sufficiently large values of $\tan \beta$ for the decay mode $H^+\to t\bar b$.
In principle, depending on the supersymmetric parameter space,
virtual supersymmetric (SUSY) particle exchange can also influence the
branching ratios.  The computation of radiative 
SUSY-QCD and SUSY-electroweak loop effects to the width of $H^+\to t\bar b$
can be found in \Refs{chhiggstotop}{chhiggstotop2}. In particular,
the latter includes the
resummation of the large QCD quantum effects and the dominant
contributions for 
SUSY quantum effects, which are those associated to
the redefinition of the Yukawa couplings due to bottom mass
corrections (as discussed in section~I.B.3).

\vskip6pt\noindent
\underline{\it $H^\pm\to W^\pm b\bar b$} \\[0.2cm]
The decay $H^+\to t^\ast b$ (where $t^\ast$ is an off-shell
$t$-quark) can become competitive (and even dominant) with respect to
the two body fermion decays.  
Only the tree-level prediction has been displayed
in \fig{fg:br_h}.  QCD radiative corrections have been computed
in \Ref{chhiggsdecay3}.  It is found that the pure QCD corrections
raise the decay width of $H^+\to W^+b\bar b$ by about $12\%$, and the
SUSY-QCD  corrections (due to gluino and third generation
squark exchanges) can be comparable or even larger than the pure QCD
corrections in some regions of the MSSM parameter space.

\vskip6pt\noindent
\underline{$H^\pm\to W^\pm\hl$} \\[0.2cm]
The decay $H^+\to W^+\hl$ is kinematically allowed only for charged
Higgs bosons that are heavier than the top quark.
Hence, this mode is
not relevant for the mass region which is most likely to be accessible
at the Tevatron.  The related
decays $H^+\to W^*\ha$, $W^*\hl$, where the $W$  is off-shell, can be
sizeable for small $\tanb$ and $M_{H^+}< m_t+m_b$ \cite{bordjo}.  However, as
discussed above, this region of the MSSM parameter space is expected
to be ruled out by the end of the LEP run.  In more general
non-supersymmetric two Higgs doublet models, these modes could be
present, although their predicted rates are more model dependent
(depending on unconstrained parameters from the neutral Higgs sector).

\begin{figure}[!ht]
\centerline{\psfig{figure=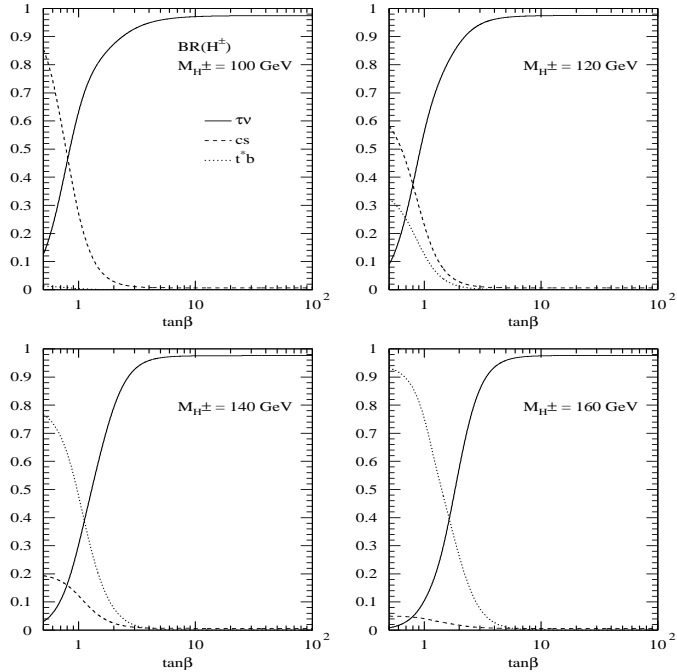,height=9.0cm,width=9.0cm}}
\caption{Branching fractions of 
charged Higgs bosons as functions of $\tan\beta$:
BR$(H^+\to\tau^+\nu_\tau)$ (solid), BR$(H^+\to c\bar s)$ (dashed),
and BR$(H^+\to W^+b\bar b)$ (dotted), for four
values of $\mhpm$:
(a) 100 GeV, (b) 120 GeV, (c) 140 GeV, and (d) 160 GeV.}
\label{fg:br_h}
\end{figure}

\vskip6pt\noindent
\underline{$H^\pm\to$~supersymmetric particles} \\[0.2cm]
Charged Higgs decay into charginos/neutralinos, and third generation
squarks and sleptons can be important once they are kinematically
allowed.  However, for $\mhpm\lsim m_t-m_b$ (most relevant for
Tevatron phenomenology), the area of MSSM parameter space in which the
two-body supersymmetric decay modes are kinematically allowed is not
large \cite{baeretal,bordjo}.
Still, this scenario cannot be ruled out at present (the most
likely supersymmetric decay mode is $\hpm\to\widetilde\chi^\pm
\widetilde\chi^0$).

\subsubsection{MSSM Higgs Boson Production at the Tevatron}

\vskip6pt\noindent
\centerline{a.~Neutral Higgs Boson Production Processes}

\vskip6pt\noindent
\underline{$q\bar q\to V^*\to V\phi$ $[V=W,Z]$, [$\phi=\hl,\hh$]}
\\[0.2cm]
Over most of the MSSM Higgs
parameter space, one of the two CP-even Higgs bosons has very
suppressed couplings to $VV$, while the other one couples to $VV$ with
Standard Model strength.  For the latter scalar state,
the process $q\bar q\to V^*\to V + \phi$ [$V=W^\pm$ or $Z$]
can be important.  Note that the CP-even scalar
$\phi$ has SM like-couplings to the vector bosons in two cases:
(i)~in the decoupling regime for the lightest Higgs boson, where
$\phi=\hl$ and (ii)~for large $\tanb$ and low $\mha$, where 
$\phi=\hh$.  In either case, the corresponding scalar 
$\phi$ has a mass less than about 130~GeV.
This behavior is evident from the total cross-section curves shown in
\fig{fg:5}a,b.  These cross-sections include first-order QCD
corrections; the corresponding $K$ factors are approximately 30\%
\cite{23}.  The SUSY-QCD corrections due to the
exchange of virtual squarks and gluinos are not included; these are
known to be small \cite{djospi}.

\begin{figure}
\centering
\centerline{\psfig{file=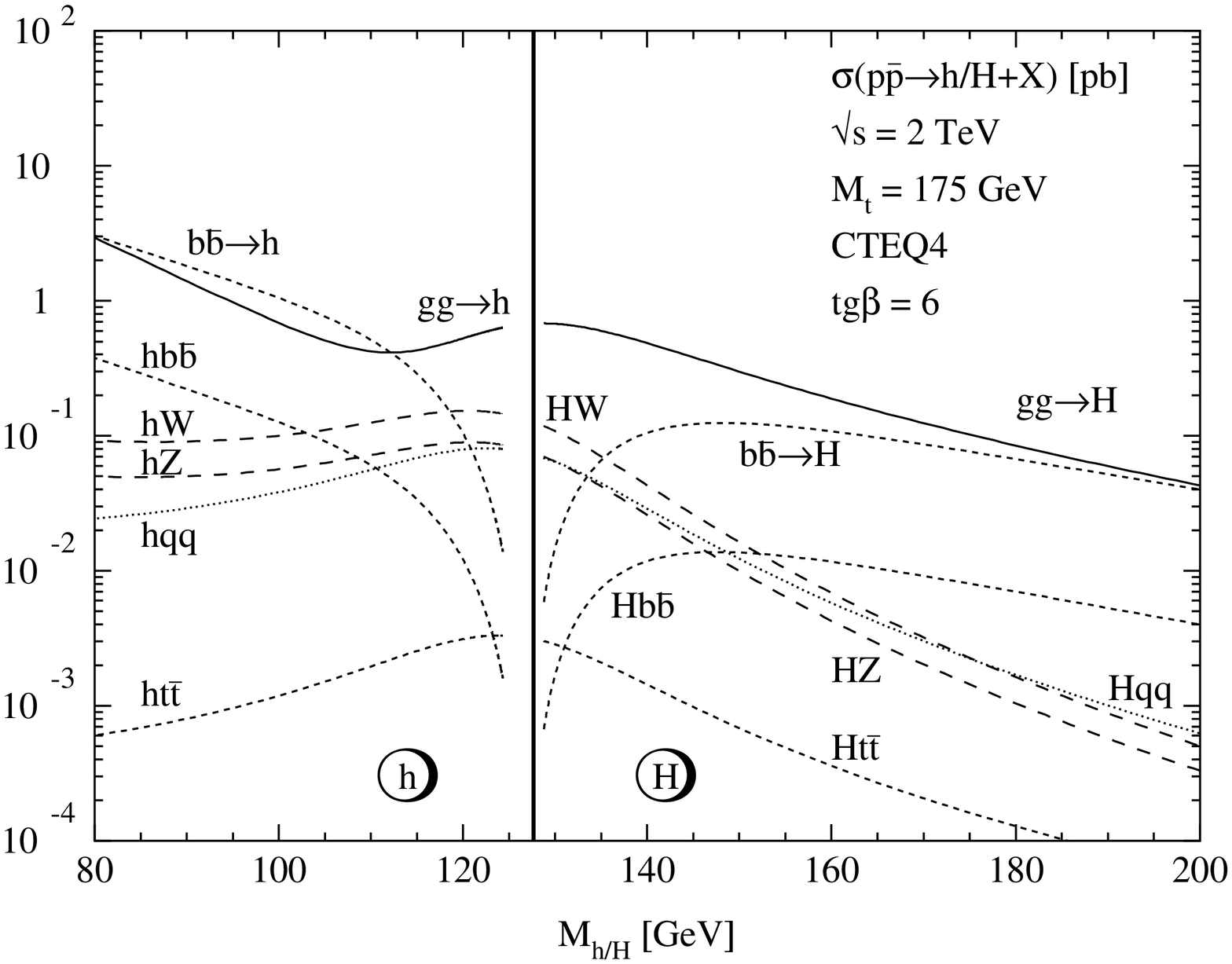,width=8cm,angle=-90}
\hfill
\psfig{file=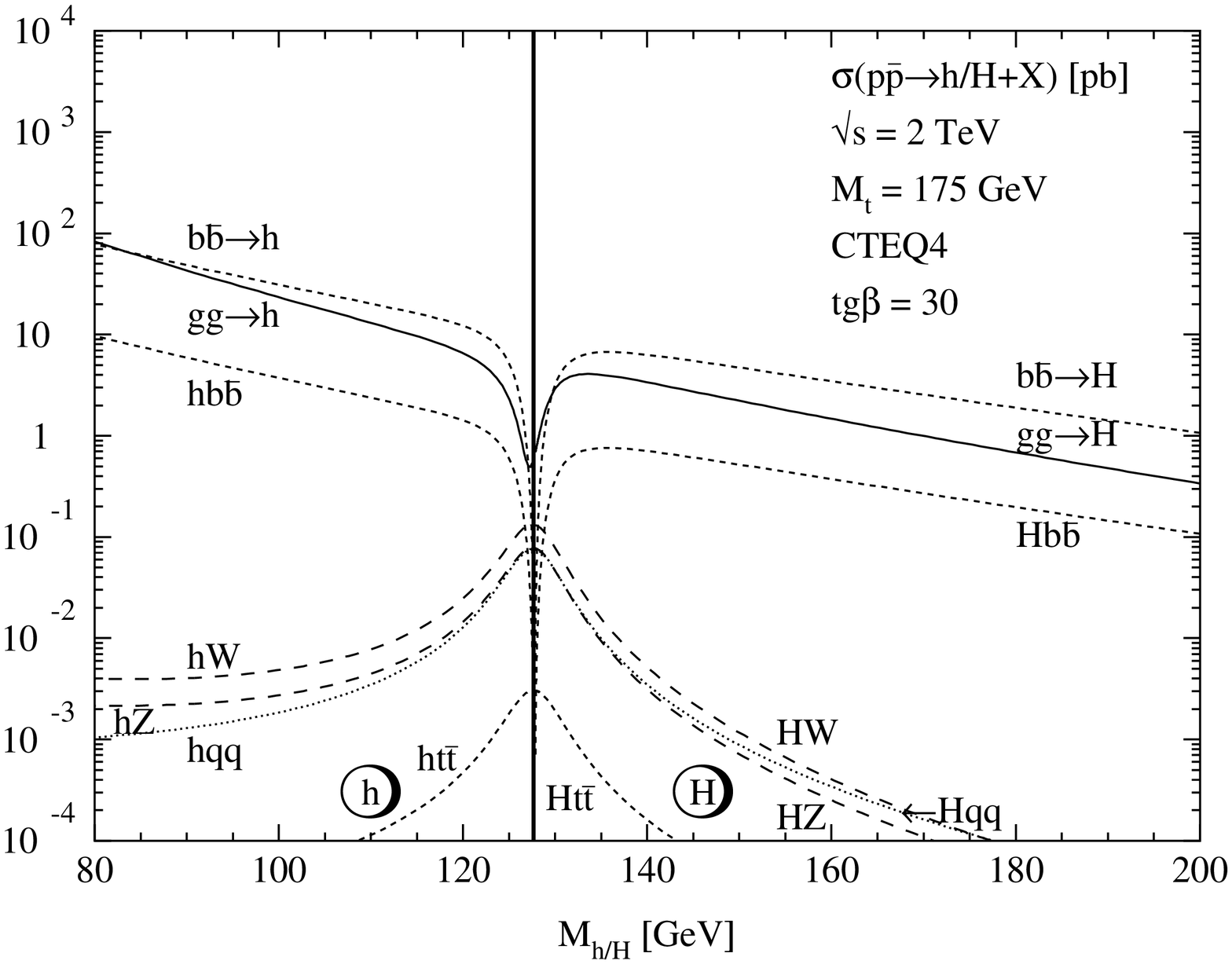,width=8cm,angle=-90}}
\vskip2pc
\centerline{\psfig{file=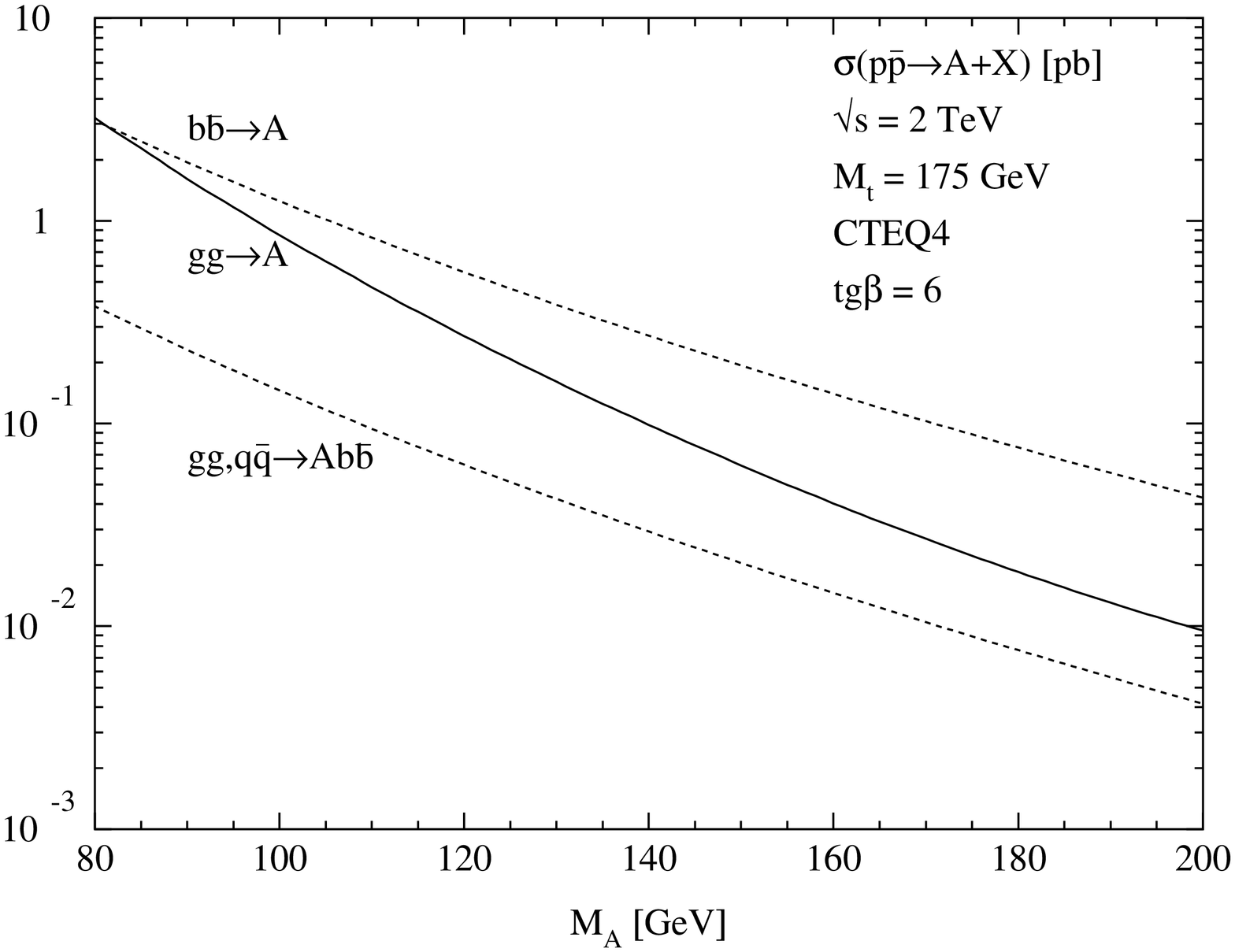,width=8cm,angle=-90}
\hfill
\psfig{file=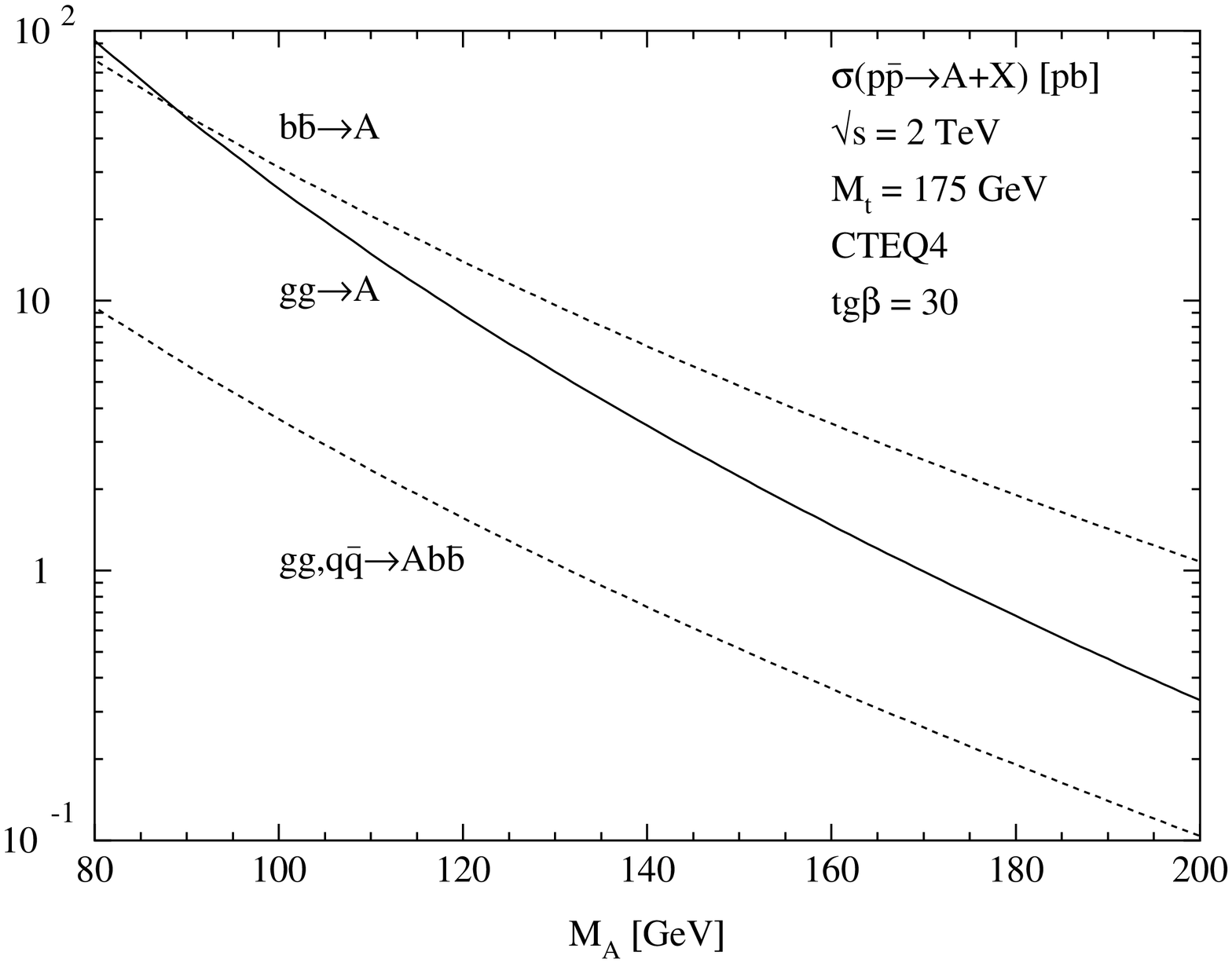,width=8cm,angle=-90}}
\vskip1pc
\caption[0]{\label{fg:5} Neutral MSSM Higgs production cross-sections 
at the Tevatron [$\sqrt{s}=2$ TeV] for gluon fusion $gg\to \phi$,
vector-boson fusion $qq\to qqV^*V^* \to qqh$,
$qqH$, vector-boson bremsstrahlung $q\bar q\to V^* \to \hl V/\hh V$ and
the associated
production $gg,q\bar q \to \phi b\bar b/ \phi t\bar t$ including all known
QCD corrections, where $\phi=\hl$, $\hh$ or $\ha$.   
As in \fig{fg:4}, in the vector boson fusion process,
$qq$ refers to both $ud$ and $q\bar q$ scattering.
The four panes exhibited above show 
(a) $h,H$ production for $\tanb=6$, (b) $\hl$, $\hh$
production for
$\tanb=30$, (c) $\ha$ production for $\tanb=6$, (d) $A$ production for
$\tanb=30$.}
\end{figure}

\vskip6pt\noindent
\underline{$gg\to \phi$ [$\phi=\hl$, $\hh$, $\ha$]} \\[0.2cm]
The gluon fusion processes are
mediated by heavy top and bottom quark triangle loops and the
corresponding supersymmetric partners \cite{29a,spiramssm,21}. Gluon fusion is
the dominant neutral Higgs
production mechanism at the Tevatron, even for large $\tanb$, with
cross-sections ranging from $0.03$--$30$~pb
(for $100~{\rm GeV}\lsim m_\phi\lsim 200$~GeV), 
as shown in \fig{fg:5}. However,
as in the case of the \SM\ Higgs boson, the
dominant $b\bar b$ final states are
overwhelmed by the huge QCD background of $b\bar b$ production.
Hence, it is not possible to detect $gg\to\phi\to b\bar b$
at the Tevatron.  Only the
$\tau^+\tau^-$ decay modes may be promising for large $\tanb$,
especially if the Higgs bosons are produced in association with a jet.

As in the case of the $gg\to\hsm$, the two-loop QCD corrections are
dominated by soft and collinear gluon radiation for small $\tanb$ in the MSSM
\cite{29}. The $K$ factor
remains nearly the same after including squark loops, since the dominant
soft and collinear gluon effects are universal, thus suppressing the (s)quark
mass dependence \cite{21}.

\break
\vskip6pt\noindent
\underline{$V^*V^*\to \phi$ [$V=W^\pm$ or $Z$; $\phi=\hl$, $\hh$]} \\[0.2cm]
As noted previously, one typically finds that one of
the two CP-even Higgs bosons has very
suppressed couplings to $VV$, while the other one couples to $VV$ with
Standard Model strength.
The process $qq\to qqV^*V^* \to qq\phi$
(CP-even Higgs production via vector boson fusion, $V=W^\pm$ or $Z$)
can be significant only for the SM-like scalar state.
In either case the corresponding scalar $\phi$ has a mass less than
about 130~GeV.  In this mass range, observation of the Higgs boson
via this channel at the Tevatron will be extremely difficult.
The corresponding cross-sections are shown in
\fig{fg:5}a,b.  These cross-sections include first-order QCD
corrections; the corresponding $K$ factors are small, enhancing the
cross-section by roughly 10\%
\cite{32}.  The corresponding SUSY-QCD corrections are
known to be small \cite{djospi}, and are not included.

\vskip6pt\noindent
\underline{$q\bar q, gg\to \phi t\bar t,
\phi b\bar b$ [$\phi=\hl$, $\hh$, $\ha$]} \\[0.2cm]
In the MSSM, Higgs boson
radiation off bottom quarks becomes important for large $\tanb$ with
cross-sections of order 1 pb for $\hl$ and $\ha$ with masses of
order
100 GeV and 0.1 pb for  $\hh$ and $\ha$ with masses of
order  200 GeV,
see \fig{fg:5}. Thus, the
theoretical predictions, including full NLO computations, are
crucial to do realistic simulations of the MSSM Higgs signals in these
channels.\footnote{As mentioned in the corresponding discussion for
$q\bar q$, $gg\to b\bar b\hsm$ [see Section~I.B.3], one also needs to
evaluate $gb \to b\phi$ and $b\bar b\to \phi$ with suitable subtraction
of the logarithms due to quasi-on-shell quark exchange (to avoid double
counting) in order to obtain the total inclusive cross-section for
$\phi$ production.  For example, the $\tan\beta$ enhancement
can lead to copious $s$-channel production of Higgs bosons via $b$-quark
fusion \cite{bhy}.  A complete calculation of this type, fully
consistent up to ${\cal O}(\alpha_s^2)$ does not yet exist in the
literature.}
Moreover, as discussed in Section~I.B.3, SUSY-QCD and SUSY-electroweak
corrections to the $b \bar b \phi$ coupling as those shown
in eqs.~(\ref{bhCP}), (\ref{barhb}) and (\ref{tildehb})
play a very important role in enhancing or suppressing (depending on
the exact supersymmetric spectrum) these production
cross-sections at large $\tanb$ \cite{CMW1,CMW2,bdhty}.

\vskip6pt\noindent
\underline{\it Higgs Boson Pair Production} \\[0.2cm]
Light scalar Higgs pair production $gg\to \hl \hl$ yields a
cross-section at the Tevatron with $\sigma \gsim 10$ fb \cite{33,34}.
The cross-section for $q\bar q, gg\to \hl\ha$ is of similar size in
some
regions of the MSSM parameter space [see \fig{fg:hpair}]. Since
the process $gg\to \hh\to \hl\hl$ is sensitive to the trilinear
coupling $\lambda_{\hh\hl\hl}$ it is important for a partial
reconstruction
of the Higgs potential. One may hope that the dominant $b\bar bb\bar b,
b\bar b\tau^+\tau^-$ final states can be extracted from the QCD
backgrounds due to the different event topologies. The two-loop QCD
corrections have recently been calculated [for $gg$ initial states in
the limit of heavy top quarks, thus leading to a reliable result for
small $\tanb$]. They enhance the $gg\to hh, hA$ production cross-sections 
by about 70--90\% and the Drell--Yan-like $q\bar q \to
\hl\ha$ cross-section by about 30\% \cite{34}.  SUSY-QCD corrections,
considered in \Ref{djospi}, are small and not included here.

\begin{figure}

  \begin{center}
\centerline{\psfig{file=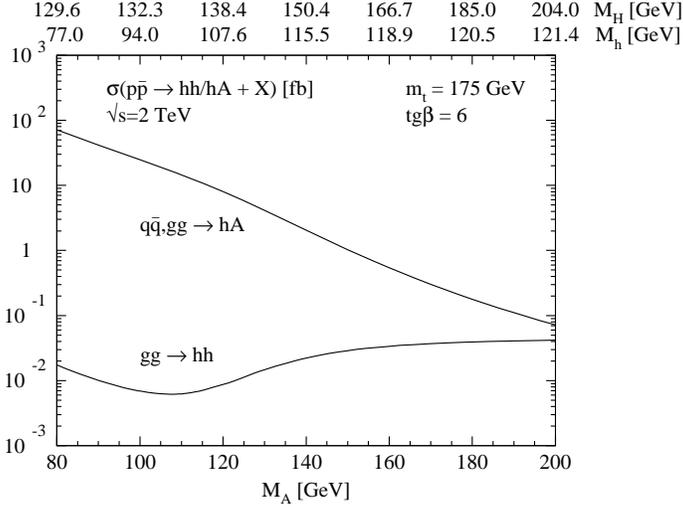,width=9cm,angle=-90}}
  \end{center}
  \caption[0]{\label{fg:hpair}  QCD corrected production cross-sections of
$hh, hA$ pairs at the Tevatron [$\sqrt{s}=2$ TeV] as a function of the
pseudoscalar Higgs mass for $\tanb = 6$. The labeled axis 
at the top of the figure exhibits the
corresponding values of the light and heavy scalar Higgs masses $\mhl$,
$\mhh$.}
\end{figure}

\vskip6pt\noindent
\centerline{b.~Charged Higgs Boson Production Processes}

\vskip6pt\noindent
\underline{\it $t\to bH^+$} \\[0.2cm]
If $\mhpm<m_t-m_b$, then the charged Higgs boson $H^\pm$ can be
produced in the decay of the top quark via $t\to bH^+$ (and $\bar t\to
\bar b H^-$) \cite{Roy1,Roy2,TDECAY,bordjo}.  The $t \to bH^+$ decay
mode can be competitive with the dominant \SM\ decay mode, $t\to
bW^+$, depending on the value of $\tan\beta$.  Here, the case of a
general Type-II two-Higgs doublet model [based on the Higgs-fermion
couplings of \eq{qqcouplings}] is considered.  Since the Higgs sector
of the MSSM is a Type-II two-Higgs-doublet model, the following
considerations also apply to the MSSM under the assumption that there
are no open charged Higgs decay channels into supersymmetric final
states, {\it i.e.}  ${\rm BR}(t\to bH^+)+{\rm BR}(t\to bW^+) = 1$.
\Fig{fg:br_t} shows
the $t\to bH^+$ branching fraction as a function of
$\tan\beta$ for various choices of $\mhpm$.
Note that ${\rm BR}(t\to b H^+)$ is significant for very small or
very large values of $\tan\beta$, while it is suppressed for
intermediate values of $\tan\beta$.

\begin{figure}
\centering
\centerline{\psfig{file=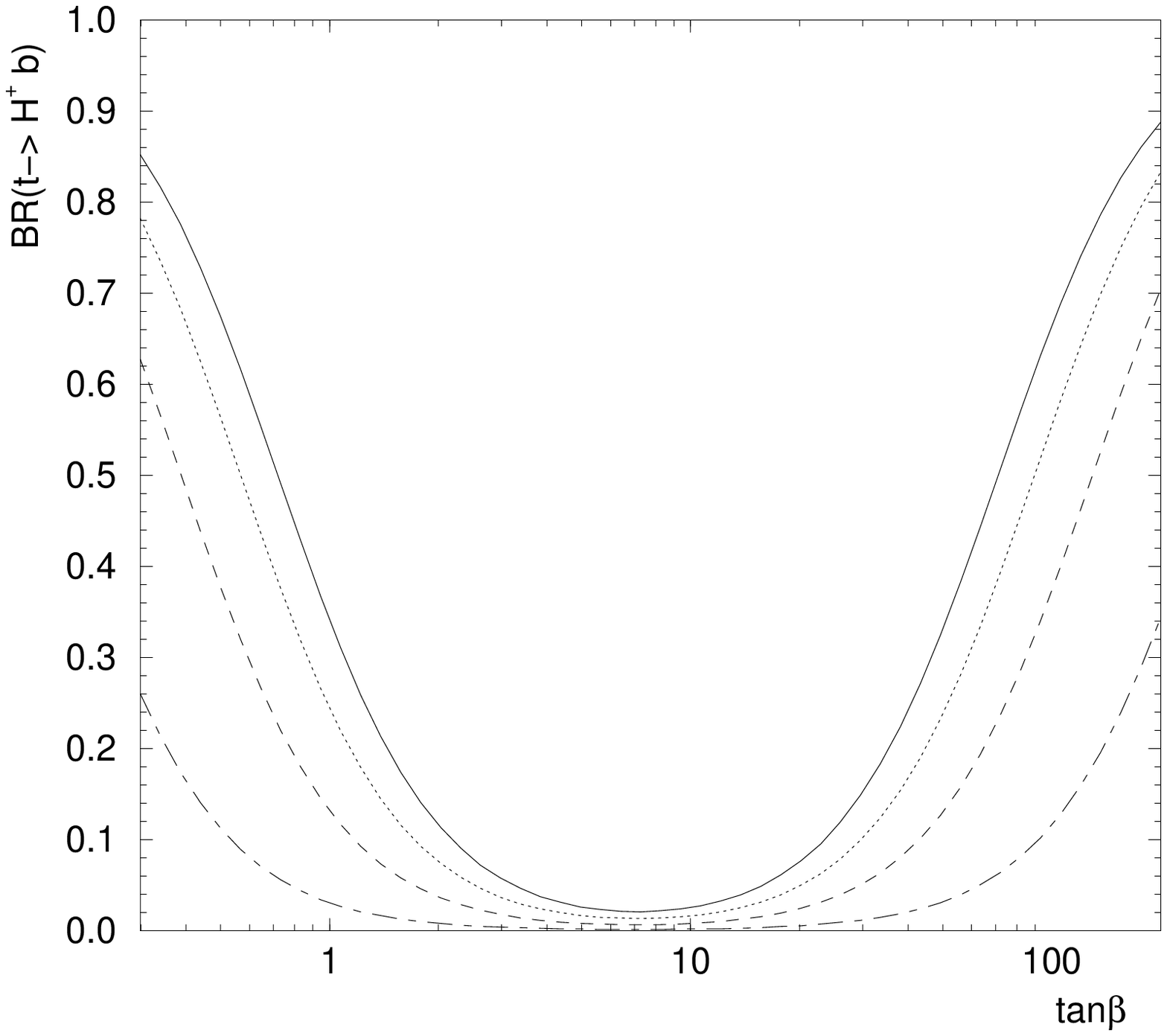,width=8cm,angle=0}
\hfill
\psfig{file=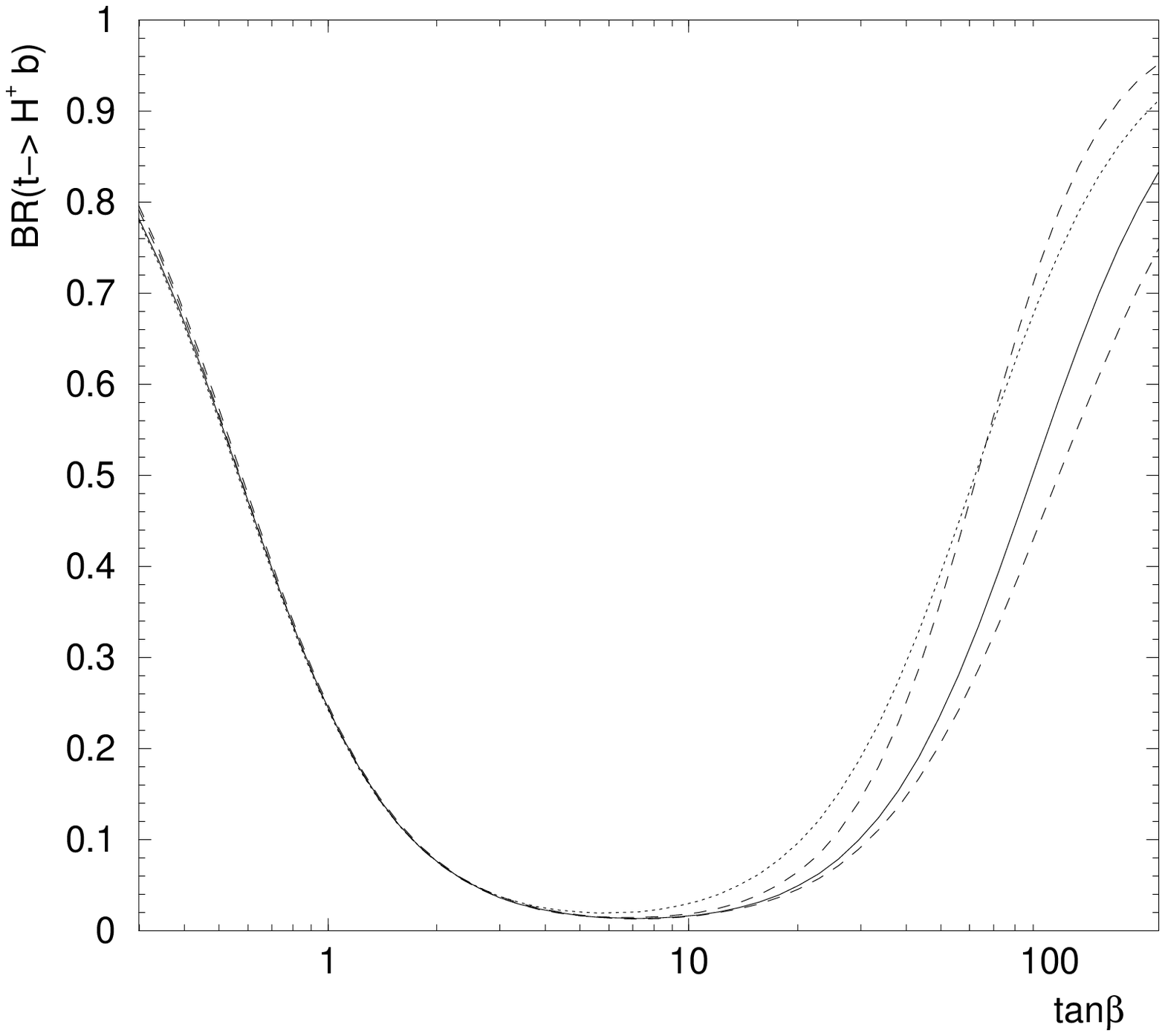,width=8cm,angle=0}}
\vskip1pc
\caption[0]{\label{fg:br_t} (a)
BR($t\to H^+ b$) as a function of $\tan\beta$ for
four values of $\mhpm$: 100 GeV (solid), 120 GeV (dotted),
140 GeV (dashed), and 160 GeV (dot--dashed), with $M_t = 175$~GeV and
$m_b=5$~GeV in the general two-Higgs-doublet model (without supersymmetry).
One-loop QCD corrections are included.
(b) BR($t\to H^+ b$) in
the MSSM for $\mhpm=120$~GeV.  The tree-level prediction is
depicted by the dotted curve.  The one-loop QCD corrected result
[from (a)] is shown by the solid curve, and the two dashed curves
incorporate additional one-loop MSSM radiative corrections for two
different choices of MSSM parameters as described in the text.}
\end{figure}

The results of \fig{fg:br_t}(a) incorporate
one-loop QCD corrections (first computed in \Ref{toptochhiggsdecay})
and are thus applicable to the general Type-II two-Higgs doublet model.
The QCD corrections to the $t\to bH^+$ branching
ratio are sizeable for values of $\tan \beta \geq 20$.
In the MSSM, one-loop virtual supersymmetric particle
exchange can also significantly influence the top quark branching
ratios, depending on the choice of supersymmetric parameters.
As an example, in \fig{fg:br_t}(b), the effects of one-loop MSSM
radiative corrections is exhibited by the dashed lines for the case of
$\mhpm=120$~GeV and the following supersymmetric parameters:
$m_{\tilde g}=M_2=m_{\tilde
{b}_{1}}= m_{\tilde{t}_{1}}=|A_{t}|=|A_{b}|=1$~TeV and
$|\mu|=300$~GeV.  The
upper [lower] dashed curve corresponds to a negative [positive] value
of $\mu$ and positive [negative] values of $A_t$ and $A_b$.
A full one-loop calculation of $\Gamma (t\rightarrow
H^{+}\,b)$  in the MSSM including
all sources of large Yukawa couplings was presented in
Ref \cite{susytoptochhiggsdecay}.
As explained in section~I.B.3, the dominant
SUSY-QCD corrections (due to gluino exchange)
and SUSY-electroweak corrections are those
associated with the redefinition of the Yukawa couplings due to
$b$-quark mass renormalization effects.
A treatment including
resummation of the leading QCD quantum effects and the dominant
contributions from loop effects arising from supersymmetric particle
exchange can be found in ref.~\cite{chhiggstotop2}.

The total cross-section for charged Higgs production in the
mass region under consideration is then given by:

\begin{equation} \label{sighplus}
\sigma(p\bar p\to \hpm+X)=\left(1-[{\rm BR}(t\to bW^+)]^2\right)
\sigma(p\bar p\to t\bar t+X)\,.
\end{equation}
With $\sigma(p \bar p \rightarrow t \bar t) \simeq 7$ pb at $\sqrt{s} = 2.0$
TeV \cite{ttbarxsec},
roughly 1400 $t\bar t$ events per detector will be produced in
Run 2 of the Tevatron.  Folding in the top quark branching ratio given
in \fig{fg:br_t}, it is a simple matter to compute the inclusive charged
Higgs cross-section for $\mhpm<m_t-m_b$.  For larger values of $\mhpm$,
the top quark that produces the charged Higgs boson must be off-shell,
and one must compute the full $2\to 3$ processes $p\bar p\to
H^+\bar t b+X$ and $p\bar p \to H^-t \bar b+X$.  This computation will be
described in the next section.  Note that if one evaluates
$\sigma(p\bar p\to H^+\bar t b+X)$ in the region of $\mhpm<m_t-m_b$,
one obtains the {\it single} charged Higgs
inclusive cross-section,
$\sigma(p\bar p\to H^+ +X)={\rm BR}(t\to bH^+)\sigma(p\bar p
\to t\bar t+X)$ (which is plotted in \fig{TeVsum}\footnote{In 
this computation, the renormalization and
factorization scales have been chosen to be $\mu=2m_t$.}), 
rather than full charged Higgs inclusive cross-section
of \eq{sighplus}.\footnote{The latter is not quite a factor of two
larger than the former since $X$ can contain a charged Higgs boson; one
must subtract off $[{\rm BR}(t\to bH^+)]^2\sigma(p\bar p\to t\bar t+X)$
to avoid double-counting.}  
As expected, in the
region of $\mhpm<m_t-m_b$, the single inclusive charged Higgs
cross-section is completely dominated by on-shell $t\bar t$ production
followed by the decay of one of the top quarks into a charged Higgs
boson.

\begin{figure}
\centering
\centerline{\psfig{file=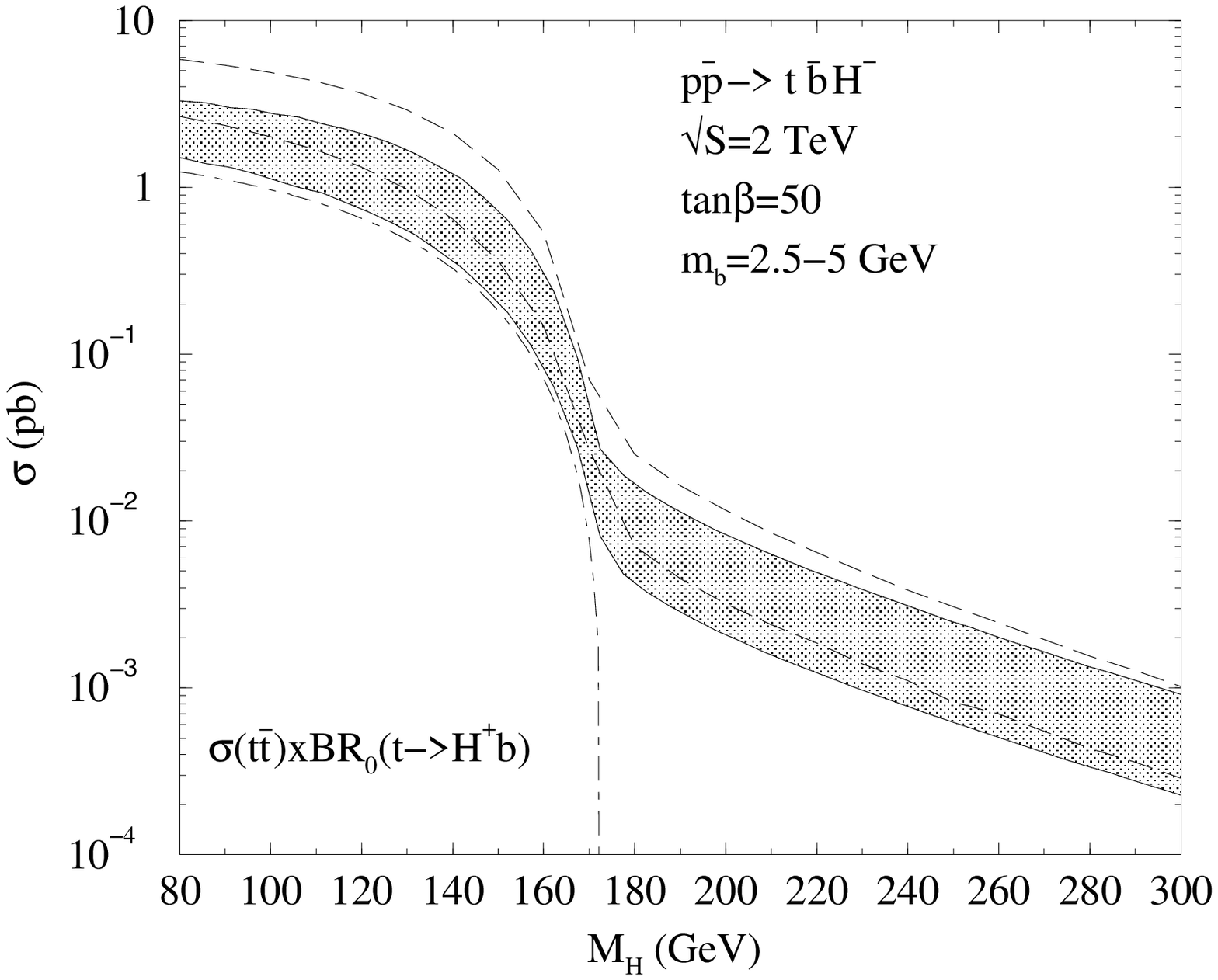,width=8cm,angle=0}
\hfill
\psfig{file=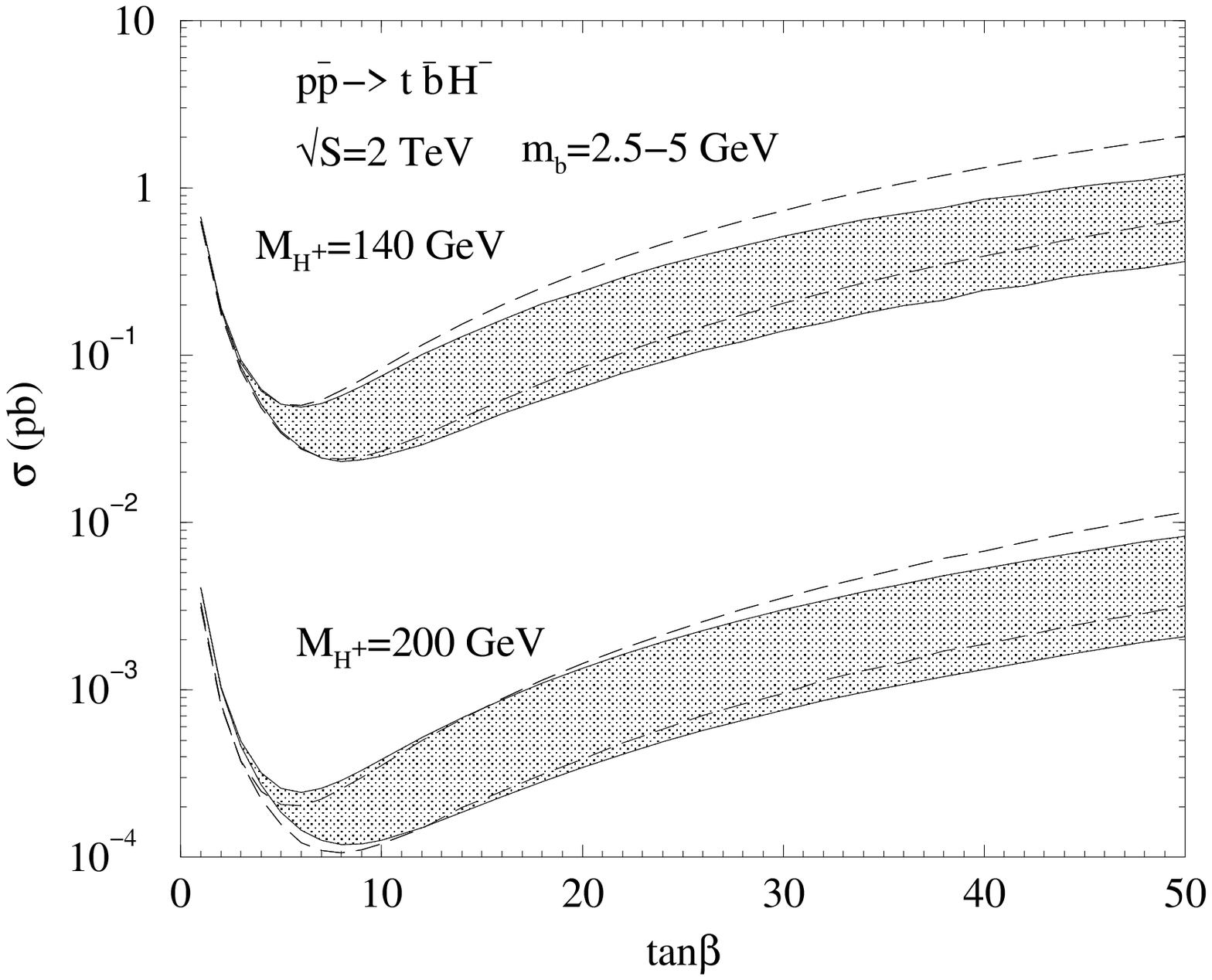,width=8cm,angle=0}}
\vskip1pc
\caption[0]{\label{TeVsum} (a) The leading-order production cross-section
$\sigma(p\bar{p}\to \bar b t H^- +X)$, for $\sqrt{s} = 2$~TeV and
$M_t=175$~GeV, as a function of the charged Higgs mass, is shown for
$\tan\beta = 50$.
The cross-section is obtained by adding the
contribution of the $2 \to 2$ processes, $g b \to t H^-$, to those of
the $2 \to 3$ processes, $ g g \to t \bar{b} H^-$ and
$ q \bar{q} \to t \bar{b} H^-$ (suitably subtracted to avoid double
counting as explained in the text).  By definition, no final state
$\bar b$ is detected in the $2 \to 2$ process (in this case, the initial
$b$ was made via $g\to b\bar b$ which implies that the attendant $\bar
b$ is produced in the forward direction), while the final state
$\bar b$ is detected in the $2 \to 3$ process.
The solid lines correspond to
the prediction of a general two-Higgs doublet with no supersymmetry.
The shaded band corresponds to
a variation of the Higgs--fermion Yukawa coupling corresponding to
a $b$-quark mass which is varied in the interval 2.5~GeV~$\leq m_b\leq
5$~GeV, which corresponds to the difference of the running $b$-quark
mass evaluated at the Higgs mass and the $b$-quark pole mass.   This
uncertainty is meant to represent the uncertainty of the unknown QCD
corrections.  The leading-order parton distribution functions
CTEQ4M \cite{CTEQ} are used.  The dot-dashed line corresponds to
the tree-level value of $\sigma(p\bar p\to t\bar t+X)\times {\rm
BR}(t\to bH^+)$.  Finally, the dashed lines exhibit the effect of
the leading MSSM radiative corrections, for a set of MSSM parameters
as described in the text.
(b) The production cross-section for
$\sigma(p\bar{p}\to \bar b t H^- +X)$ is shown as a function of $\tan\beta$
for two values of the charged Higgs mass.  For the case of
$\mhpm=140$~GeV, the top quark that decays into the observed charged
Higgs boson is on-shell, while for $\mhpm=200$~GeV, the corresponding
top quark is off-shell.  The interpretation of the solid and dashed
lines and the shaded area are the same as in (a).
These results are based on the calculations of
\Refs{chhiggstotop}{francesca}.}
\end{figure}

\vskip6pt\noindent
\underline{\it $gb\to tH^-$; $gg$, $q\bar q\to t\bar b H^-$} \\[0.2cm]
If $\mhpm>m_t-m_b$, then charged Higgs boson production occurs
mainly through radiation off a third generation quark.
Single charged Higgs associated
production proceeds via the $2\to 3$ partonic processes
$gg,\, q\bar q\to t\bar b H^-$ (and the charge conjugate final state).
As in the case of $b\bar b\hsm$ production,
large logarithms $\ln(\mhpm^2/m^2_b)$ arise
for $\mhpm\gg m_b$ due to quasi-on-shell
$t$-channel quark exchanges, which can be resummed by absorbing them
into the $b$-quark parton densities.  Thus, the proper procedure for
computing the charged Higgs production cross-section is to add the
cross-sections for $gb\to tH^-$ and $gg\to t\bar b H^-$
and subtract out the large logarithms accordingly
from the calculation of the $2\to 3$ process \cite{soper,OT}.
This procedure avoids double-counting of the large logarithms at
${\cal O}(\alpha_s)$, and correctly resums the leading logs to all
orders.  In particular,
the contribution to the total cross-section coming from the kinematical
region of the gluon-initiated $2\to 3$ process in which one of the
two gluons splits into a pair of $b$-quarks (one of which is
collinear with the initial proton or antiproton), is incorporated into
the $b$-quark parton density.  A cruder calculation would
omit the contribution of the $2\to 2$ process and simply include
the results of the unsubtracted $2\to 3$ process.  The latter procedure
would miss the resummed leading logs that are incorporated into the
$b$-quark density.  However, the numerical difference between the two
procedures is significant only for $\mhpm\gg m_t$.

The single inclusive charged Higgs cross-section at the Tevatron is
shown in \fig{TeVsum}(a) as a function of the charged Higgs mass, for
$\tan\beta=50$.
These results are based on the calculations of
\Refs{chhiggstotop}{francesca} and include the contributions of 
the $2\to 2$ process
and suitably subtracted $2\to 3$ process as described above.
\Fig{TeVsum}(b) depicts the single inclusive charged Higgs cross-section 
as a function of $\tan\beta$, for
$\mhpm=140$ and $200$~GeV.
In the computations that produced these two figures, the solid lines
correspond to the prediction of a general two-Higgs doublet model
(without supersymmetry).  Only tree-level diagrams are evaluated, with
the Higgs--fermion Yukawa coupling determined by a $b$-quark mass
which is varied in the interval 2.5~GeV~$\leq m_b\leq 5$~GeV
(corresponding to the shaded band).  The renormalization
and factorization scales were fixed at the
threshold value $m_t + m_{H^\pm}$.  The variation of these scales
would change the values of the cross-section presented here.
For example, a variation in the interval between $(m_t +\mhpm)/2$ and
$2(m_t + \mhpm)$, can produce deviations up to $\pm 30\%$ with respect
to the values shown in the figures.  In low-energy supersymmetric models,
the effects of the leading SUSY-electroweak corrections can
significantly modify the tree-level $H^+ t\bar b$ vertex depending on
the values of the MSSM parameters.  This is illustrated by the dashed
lines in \fig{TeVsum} which are displaced from the corresponding solid
lines due to the virtual supersymmetric effects.\footnote{The leading
MSSM effects shown in \fig{TeVsum} have been implemented only in the
computation of the $2\to 3$ process, $gg$, $q\bar q\to t\bar b H^-$.}
The effects
displayed correspond to the following supersymmetric parameters:
$m_{\tilde{g}}=300$~GeV, $m_{\tilde
{b}_{1}}=250$~GeV, $m_{\tilde{t}_{1}}=200$~GeV,
$A_{t}=A_{b}=300$~GeV, and $\left|\mu\right|=150$~GeV.  The upper
[lower] dashed curve corresponds to a positive [negative] value of
$\mu$.

\vskip6pt\noindent
\underline{\it $q\bar q$, $gg\to H^+ H^-$} \\[0.2cm]
Pair production of charged Higgs bosons occurs via Drell-Yan $q\bar q$
annihilation.  The dominant contribution, which  arises from the
annihilation of $u$ and $d$ quarks into a virtual photon or $Z$, is
independent of $\tan\beta$.  Some $\tan\beta$ dependence enters through
$b\bar b$ annihilation via $t$-channel top-quark exchange, although this
effect is more than one order of magnitude suppressed relative to the
dominant contribution.  The tree-level result for $\sigma(p\bar p\to H^+
H^- +X)$ for $\sqrt{s}=2$~TeV, taken from \Ref{kniehl1}, is shown in
\fig{kniehl}.  These results are obtained with
the Higgs--fermion Yukawa coupling based on
a fixed $b$-quark pole mass of $M_b=4.7$~GeV.   Note that the inclusive
$H^+H^-$ cross-section lies below the cross-section for single charged
Higgs associated production
(see \fig{TeVsum}) over the entire Higgs mass range shown.

\vskip6pt\noindent
\underline{\it $b\bar b$, $gg\to \hpm W^\mp+X$} \\[0.2cm]
Associated production of a charged Higgs boson and a $W^\pm$ can occur
via $b\bar b$ annihilation and $gg$-fusion \cite{kniehl2}.  The
contribution of
$b\bar b$ annihilation to $\sigma(p\bar p\to \hpm W^\mp+X)$ (both charge
state pairs are included)
is shown as function of the charged Higgs mass for $\tan\beta =
6$ and 30 in \fig{kniehl}.
The $b\bar b$ annihilation always
dominates over $gg$-fusion
except in the parameter regime where $\tan\beta$ is close to one
and the charged Higgs mass is greater than about 200 GeV.
The $gg$ fusion contribution is greatly suppressed
if $\tan\beta\gsim 6$, independent of the value of the charged Higgs
mass.

Given the small number of events predicted (before cuts, efficiencies
and backgrounds are taken into account), it seems unlikely that this
process would provide a charged Higgs discovery signature at the
Tevatron.

\begin{figure}

  \begin{center}
\centerline{\psfig{file=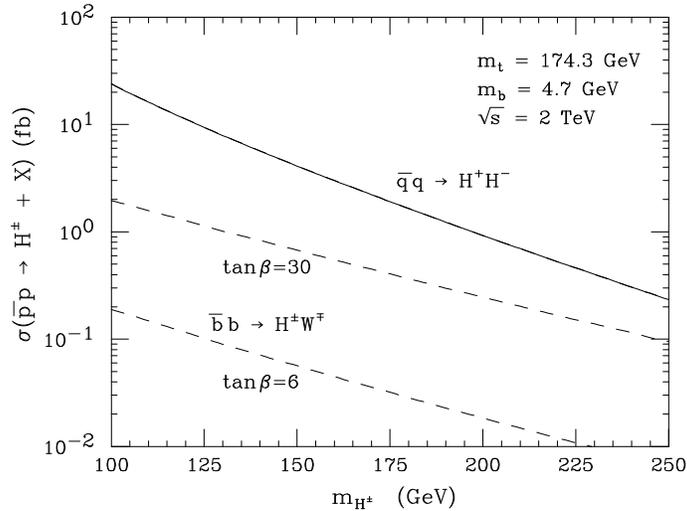,width=9cm}}
  \end{center}
  \caption[0]{\label{kniehl}
Total cross-section (in fb) for (i) $p\bar p\to H^+ H^-+X$
\cite{kniehl1} (solid line) and
(ii) $p\bar p\to \hpm W^\mp +X$ \cite{kniehl2} (dashed lines)
as a function of $\mhpm$ for $\tan\beta=6$ and 30.  The
dependence of process (i) on $\tan\beta$ is negligible.
The contribution of $b\bar b$ annihilation to process (ii) dominates
over the $gg$ fusion scattering mechanism.
$M_t=174.3$~GeV and a fixed $b$-quark pole mass
of $M_b=4.7$~GeV are used to fix the Higgs--fermion Yukawa coupling.
The leading-order parton
distribution functions CTEQ5L \cite{CTEQ} are used.}
\end{figure}

\newpage

\subsection{New Higgs Physics Beyond the Standard Model/MSSM}

\vspace{0.1in}


In the previous section, the particular case of the Minimal
Supersymmetric Standard Model was discussed in great detail.
A more agnostic approach to the study of physics beyond the
Standard Model is to construct an effective Lagrangian.
The effective Lagrangian is a 
model--independent approach that
describes the effects of new physics associated with an
energy scale $\Lambda$ which lies beyond the scale where
experiments are performed.  The effective Lagrangian can be
constructed out of higher dimensional (hence non--renormalizable)
operators and depends only
on the particle content of the low energy theory.
The discussion of this section is limited to the
case of having a light Higgs boson
contained in these operators. Hence, a linearly realized
\cite{linear,hisz} $SU_L(2) \times U_Y(1)$ invariant effective
Lagrangian is assumed to describe the bosonic sector of the SM,
keeping the fermionic sector unchanged.

There are eleven dimension--6 operators involving the gauge
bosons and the Higgs scalar field which respect local $SU_L(2)
\times U_Y(1)$, $C$ and $P$ symmetries \cite{linear}.  Six of
these operators either affect only the Higgs self--interactions
or contribute to the gauge boson two--point functions at tree
level and are severely constrained from low energy physics below
the present sensitivity of high energy experiments \cite{hisz}.
From the remaining five ``blind'' operators, four affect the
Higgs couplings and can be written as
\cite{linear,hisz},
\begin{equation}
{\cal L}_{\mbox{eff}} 
= \frac{1}{\Lambda^2} \Bigl[
 f_W (D_{\mu} \Phi)^{\dagger} \hat{W}^{\mu \nu} (D_{\nu} \Phi)
 + f_B (D_{\mu} \Phi)^{\dagger} \hat{B}^{\mu \nu} (D_{\nu} \Phi)
+ f_{WW} \Phi^{\dagger} \hat{W}_{\mu \nu} \hat{W}^{\mu \nu} \Phi
+ f_{BB} \Phi^{\dagger} \hat{B}_{\mu \nu} \hat{B}^{\mu \nu} \Phi
  \Bigr]\,,
\label{lagrangian}
\end{equation}
where $\Phi$ is the Higgs field doublet,   $\hat{B}_{\mu\nu} = i
(g'/2) B_{\mu \nu}$, and $\hat{W}_{\mu \nu} = i (g/2) \sigma^a
W^a_{\mu \nu}$ with $B_{\mu \nu}$ and $ W^a_{\mu \nu}$ being the
field strength tensors of the $U(1)$ and $SU(2)$ gauge fields
respectively.
Anomalous $H\gamma\gamma$, $HZ\gamma$, $HZZ$ and $HWW$
couplings are generated by Eq.~(\ref{lagrangian}), which modify the
Higgs boson production and decay \cite{hagiwara2}. In the unitary
gauge they are given by
\beqa
{\cal L}_{\mbox{eff}}^{\mbox{H}} &=
g_{H \gamma \gamma} H A_{\mu \nu} A^{\mu \nu} +
g^{(1)}_{H Z \gamma} A_{\mu \nu} Z^{\mu} \partial^{\nu} H
+ g^{(2)}_{H Z \gamma} H A_{\mu \nu} Z^{\mu \nu}
+ g^{(1)}_{H Z Z} Z_{\mu \nu} Z^{\mu} \partial^{\nu} H \nonumber \\
&+ g^{(2)}_{H Z Z} H Z_{\mu \nu} Z^{\mu \nu} +
g^{(2)}_{H W W} H W^+_{\mu \nu} W^{- \, \mu \nu}
+ g^{(1)}_{H W W} \left (W^+_{\mu \nu} W^{- \, \mu} \partial^{\nu} H
+ \mbox{h.c.} \right)\,,
\label{H}
\eeqa
where $A(Z)_{\mu \nu} = \partial_\mu A(Z)_\nu - \partial_\nu
A(Z)_\mu$. The effective couplings $g_{H \gamma \gamma}$,
$g^{(1,2)}_{H Z \gamma}$, and $g^{(1,2)}_{H Z Z}$  and
$g^{(1,2)}_{H WW}$ are related to the coefficients of the
operators appearing in Eq.~(\ref{lagrangian}) and can be found  elsewhere
\cite{hagiwara2}. Of special interest is the
Higgs couplings to two photons, which is given by
\begin{equation}
g_{H \gamma \gamma} =
- \left( \frac{g \sin^2\theta_W M_W}{2 \Lambda^2} \right)
                      (f_{BB} + f_{WW})  \,.
\label{g}
\end{equation}

Eq.~(\ref{lagrangian}) also generates  new contributions to
the triple gauge boson vertex \cite{linear,hisz}. The operators
proportional to ${f}_{W}$  ${f}_{B}$ give rise to both anomalous
Higgs--gauge boson couplings and to new triple and quartic
gauge boson self--couplings. On the other hand,
the operators proportional to
${f}_{WW}$ and ${f}_{BB}$ only affect $HVV$ couplings
and cannot be constrained by the study of anomalous trilinear
gauge boson couplings.

A summary of the present bounds on the couplings introduced in
\eqs{lagrangian}{H} from searches at the Tevatron and LEP,
and an analysis of the possible bounds attainable at the upgraded Tevatron
will be discussed in Section III.C.

New physics can also modify the couplings of Higgs bosons to fermions.
For example, in the topcolor model \cite{chill} the coupling of a
composite charged Higgs boson to charm and bottom quarks can be so
large that the signal of the charged Higgs boson via the $s$-channel
process is spectacular \cite{bhy}.  A similar conclusion applies to
the generic type-III two-Higgs-doublet model, in which both Higgs
doublets couple to up-type and down-type
fermions.\footnote{Constraints on Type-III two-Higgs doublet models
have been recently addressed in \Ref{typethree}.}

One can develop a more model independent approach 
for the effects of new physics on 
fermion--Higgs boson couplings by following the
chiral Lagrangian methods as in the case of the vector boson--Higgs boson
couplings discussed above.  Such a generic analysis has yet to be
applied to the Tevatron search for Higgs bosons.

\section{Experimental Studies}

  \subsection{The Tevatron in Run 2}		
In the next run of the Tevatron collider, with the new Main Injector
and Antiproton Recycler, the machine will deliver more than an order
of magnitude more instantaneous luminosity, and ultimately more than
two orders of magnitude integrated luminosity to the CDF and D\O\
experiments.  This prospect greatly enhances the discovery potential
for new particles, since more integrated luminosity translates into
more sensitivity at higher masses.  As will be seen below, this is
particularly important for the search for the Higgs boson, since the
discovery reach is limited by statistics.

In Run 1 the Tevatron operated with six proton and six antiproton
bunches, colliding every 3.5 $\mu$sec.  Instantaneous luminosities at
the beginning of each store typically reached
1.6$\times$10$^{31}$/cm$^2$/sec, or about 2~pb$^{-1}$ per week
integrated luminosity delivered to the experiments.  The main
limitations were in the proton and antiproton beam currents, and the
antiproton stacking rate.

In Run 2 the new Main Injector will provide emittance much better
matched with that of the Tevatron, greatly improving the antiproton
transfer efficiency to the Tevatron, and also will give a threefold
increase in the antiproton production rate.  At the end of each
Tevatron store, beams will be decelerated and the antiprotons
extracted into the new Antiproton Recycler, and re-used in the
subsequent store.  This will effectively double the antiproton
stacking rate.

In addition the Tevatron will switch to 36-bunch operations, which
will greatly alleviate the problem of multiple interactions, keeping
it at roughly the same level as in Run 1.  The machine energy will
increase from 1.8 TeV to 2.0 TeV in the center of mass; this typically
increases physics cross sections by about 30-40\%.

The initial goal of Run 2 (called Run 2a) is to deliver $>$2 fb$^{-1}$ by then end of
2002, with instantaneous luminosities of up to
2$\times$10$^{32}$/cm$^2$/sec.  With 36$\times$36 bunches the crossing
time will be 396 nsec; however both CDF and D\O\ have reconfigured
their front-end electronics to accommodate 132-nsec crossing times,
since with subsequent improvements the Tevatron will operate with more
bunches, with this collision crossing time.

Further improvements to the luminosity will require successful
research and development in the areas of improved antiproton
availability and controlling the effect of the beam-beam interaction.
If these lead to luminosities in the range of
5$\times$10$^{32}$/cm$^2$/sec, then in Run 2b the Tevatron could
deliver 15~fb$^{-1}$ by then end of 2007 \cite{holmes}.

This report presents results of feasibility studies for Higgs searches
at the Tevatron during the upcoming data taking.  During Run 1,
initial Higgs searches were made by both CDF \cite{juano,cdfrun1higgs} and
D\O \cite{hobbsfit}.  Figure~\ref{f-cdf-runI} shows the Run 1 Standard
Model Higgs cross section limits from CDF in each final state studied
and also the limit obtained by combining all channels.  In addition to
the Run 1 experimental results, initial Run 2 feasibility studies were
performed in refs.~\cite{tev2k,snowmass96,tev33}.  These studies
considered production of the Standard Model Higgs boson with $M_H <
130$~GeV, and concluded that $25\ \mathrm{fb}^{-1}$ was needed to
discover a 120~GeV Higgs boson at the $5\sigma$ level.  The present study
improves upon the existing work in a number of significant ways. 
In this report, a improved background analysis and a
more realistic detector simulation has been employed.  In addition, 
results from all the leading channels were combined, and new channels
relevant for the search for Higgs bosons of higher mass were
examined.  We first discuss in detail the simulation and analysis methods
used in this report.  We then focus on the search for the Standard Model
Higgs boson in a variety of channels.  Finally, we consider  
new search techniques relevant for 
Higgs bosons that arise in theories beyond the Standard Model. 

\begin{figure}
  \begin{center}
    \parbox{5.0in}{\epsfxsize=\hsize\epsffile[70 130  520 670]{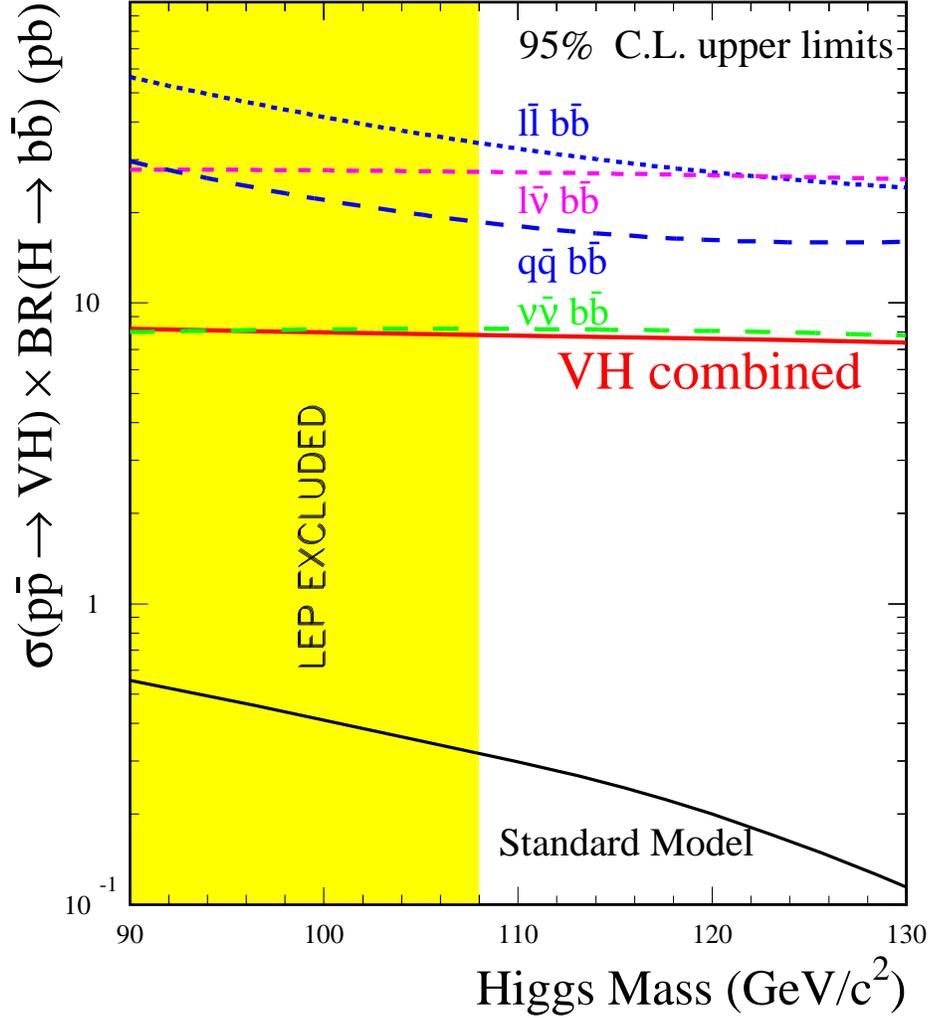}}
  \end{center}
  \caption[0]{CDF Run 1 95\% CL upper limits on the cross section times branching 
           ratio for production of a Standard Model Higgs boson in association 
           with a $W$ or $Z$, with Higgs decaying to $\bb$ \cite{cdfrun1higgs}.  The four dashed 
           curves show the limits from the individual search channels, and the
           solid curve labeled ``$VH$ combined'' shows the limit from combining
           all four channels.  The expectation from Standard Model Higgs production 
           is shown by the solid curve labeled ``Standard Model.''}
  \label{f-cdf-runI}
\end{figure}

  \subsection{Simulation and Analysis Methods}		
The goal of the Higgs Group experimental studies is to estimate the
discovery reach for the Standard Model and MSSM Higgs bosons in Run 2
and beyond at the Tevatron.  This is ultimately expressed in terms of
the integrated luminosity required to either exclude the Higgs with
95\% confidence if it does not exist, or discover it with some statistical
significance, $3\sigma$ or $5\sigma$ for example, if it does exist at
some mass.

Both CDF and D\O\ will be significantly upgraded in Run 2, and
among the most significant enhancements relevant for the Higgs search
are 
\begin{itemize}
  \item three-dimensional charged particle tracking in both 
        experiments, 
  \item precise silicon vertex detectors for heavy quark tagging, 
  \item charged track triggering using the silicon detectors,
  \item efficient electron and muon identification over the 
        central and forward regions, and
  \item good jet energy resolution with multiple detector
        elements.
\end{itemize}

Estimating the integrated luminosity thresholds requires knowledge of
the signal production cross section and acceptance, identification
efficiencies, and backgrounds.  At the time of the Workshop, neither
CDF nor D\O\ has had full simulation programs of the detector
available.  The Workshop participants agreed to estimate the Higgs
reach based on an average of the expected CDF and D\O\ detector
performance, as represented in a simple simulation, called SHW,
described below.

Also, neither the final $b$-tagging efficiency nor the $\bb$ mass
resolution for Run 2 is known, and part of the motivation for the 
Workshop was to determine the potential gain from future work on these
very important factors in the Higgs reach.  Studies of the tagging
efficiency and mass resolution are presented below, and serve as the
basis for the assumptions made in the individual signal channel studies
and in the SHW simulation program.

We emphasize, however, that obtaining optimal $b$-tagging efficiency,
Higgs mass resolution, and experimental backgrounds will ultimately
rely, in each experiment, upon detailed studies of the data to be
collected in Run 2.  Our aim is to show how the final results depend
on these crucial factors, with reasonably optimistic projections for
what we might ultimately attain with a great deal of hard work in
the coming years.

    \subsubsection{SHW Simulation}			\vspace{0.1in}
\small
\begin{center}
{\it John Conway, 
     Ray Culbertson, 
     Regina Demina,
     Ben Kilminster,
     Mark Kruse, 
     Cal Loomis,
     Konstantin Matchev,
     Maria Roco} \\
\end{center}
\normalsize\nopagebreak

The SHW simulation provides a simple simulation of the ``average''
detector response of the upgraded CDF and D\O\  detectors.  The program
provides an interface to the ISAJET~\cite{ISAJET} and PYTHIA~\cite{PYTHIA}
Monte Carlo programs,
and stores events (internally and externally) using the STDHEP
package.  SHW thus provides a simple interface to these Monte Carlo
programs, allowing people to perform event generation, simple detector
simulation, and event analysis in the same program.  

A parameterized simulation such as this, by its nature, will not fully
simulate all the effects of the real detectors.  But this program
provides a reasonably close approximation for signal and background
acceptance studies, and allows the most promising channels and
analyses to be compared.  Where possible, the results presented in
subsequent sections are compared with actual data and extrapolations
of the full Run 1 detector simulation.  The good agreement found in
these comparisons gives confidence that the SHW simulation is accurate
enough for the studies in this Workshop, considering the uncertainties
in detector performance and reconstruction techniques.

SHW begins by generating events (or reading events from a file), and
then determining what charged tracks and calorimeter energy deposits
the detector would record.  From that information the program then
makes lists of trigger and reconstructed ``objects'' including high-ET
photons, electrons, muons, hadronically decaying taus, jets (including
$b$- and $c$-jet tagging), and heavy stable charged particles.  The
user of the program can put her/his event analysis code in a routine
called from within this framework.

The simulation implemented in SHW operates by taking the list of generated 
final state particles, and for each one simulating the response of the 
detector in terms of charged particle tracking and calorimeter energy 
deposits.  This information, in turn, is used to perform particle 
identification much as in the actual experimental data analyses.
 
\vspace{0.2in}
{\bf Tracking} \\ \nopagebreak

A final state charged particle will leave a track which can be
reconstructed with high efficiency if it traverses the central region
of the detector.  The SHW package does not simulate magnetic
deflection, which is very small for high-$\pt$ tracks (such as
energetic leptons) and does not contribute significantly to the
uncertainty in the reconstruction of jet energies.  Any final state
charged particle with $|\eta|<2$ and $\pt > 300$ MeV/$c$ is assumed to
have a 97\% probability of producing a track, which is recorded in the
track list.  Track momentum resolution is simulated by including
Gaussian smearing in sagitta simulated as if the track were in a 1.4
Tesla magnetic field.  The gaussian width comes from assuming a
1-meter tracker with a resolution of 0.08\%/GeV $\sigma_{\pt}/\pt^2$ at
high $\pt$.

\vspace{0.2in}
{\bf Calorimetry} \\ \nopagebreak

The calorimeter represented by SHW is segmented in a number of cells 
in eta and phi, controlled by parameters in the include file.  Both the size 
of the cells and the range of coverage in pseudorapidity are set by these 
parameters.  The default was 28 cells in azimuth, and 80 cells in the 
pseudorapidity range $-4 < \eta < +4$, reflecting an ``average'' of the 
CDF and D\O\ calorimeter segmentations.

Each calorimeter tower has an electromagnetic section and a hadronic section, 
in which some fraction of a particle's energy is deposited.  The response of 
the calorimeter in the Run 1 simulation programs provides the basis for 
determining how much energy each particle deposits in a calorimeter tower.  
For example, using a Monte Carlo simulation of single pions, it is determined 
that on average 25\% of the energy is deposited (about half of the time; the 
rest of the time no energy is deposited) in the electromagnetic section, and 
the rest in the hadronic section.  The single pion simulation shows also that 
in about 5\% of cases the particle passes through the non-instrumented region 
between calorimeter cells (``cracks``).  

The energy resolution of the electromagnetic calorimeter is assumed to
be 20\%/$\sqrt{E}$, with $E$ in GeV, for photons and electrons, and
80\%/$\sqrt{E}$ for hadrons.  Muons deposit minimum-ionizing energy of
0.5 GeV in the electromagnetic calorimeter, and 2.5 GeV in the
hadronic calorimeter.  The program assumes that a given particle
deposits all its energy in a single cell.
   
Figure~\ref{shw_cal} shows the electromagnetic and hadronic
calorimeter response as a function of incident particle energy for a
CDF Monte Carlo simulation of single electrons, muons, and pions.
These distributions motivate the choice of parameters in the SHW
calorimeter simulation.

\begin{figure}
  \begin{center}
    \parbox{2.1in}{\epsfxsize=\hsize\epsffile[0 0 500 700]{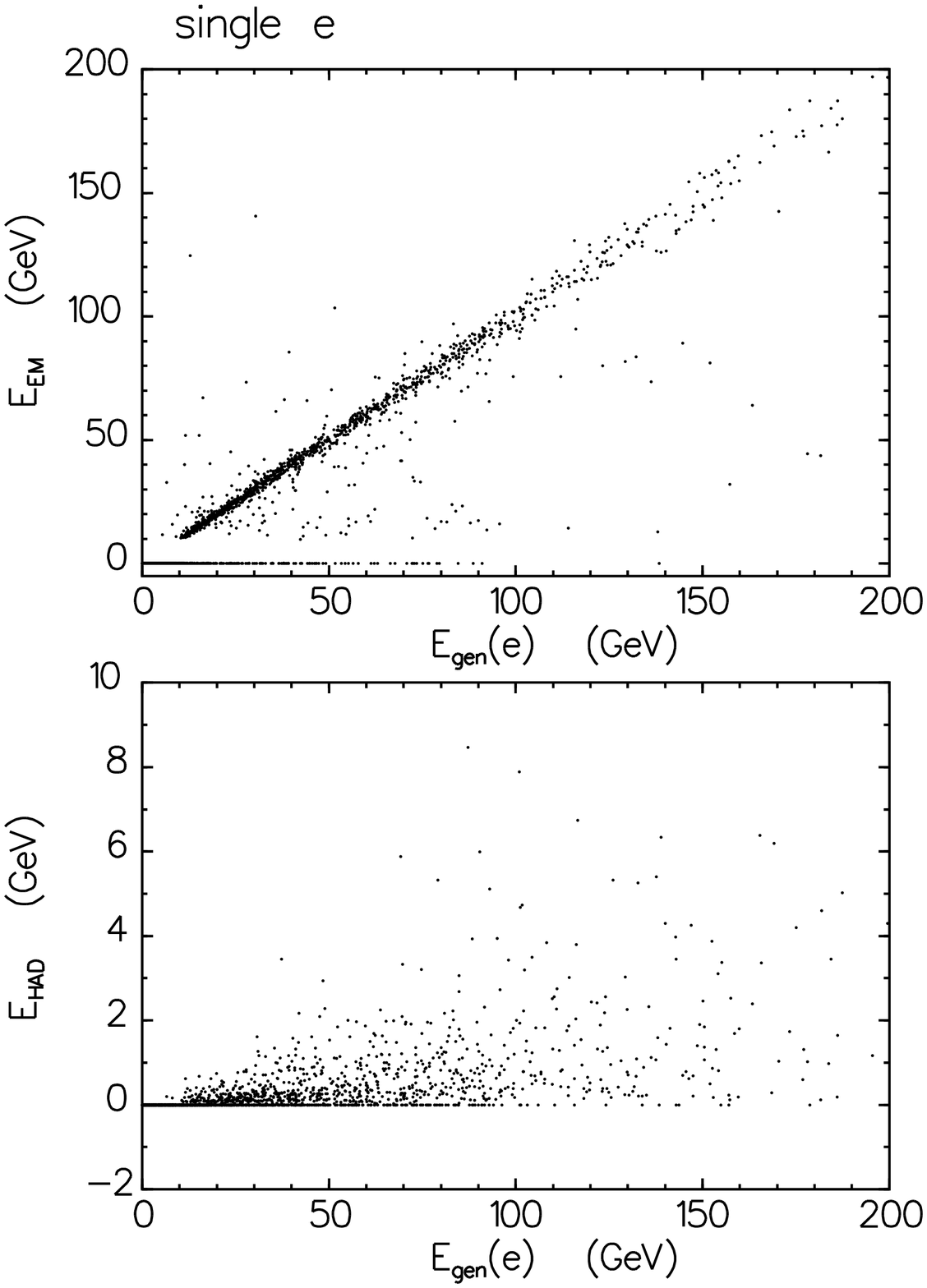}}
    \parbox{2.1in}{\epsfxsize=\hsize\epsffile[0 0 500 700]{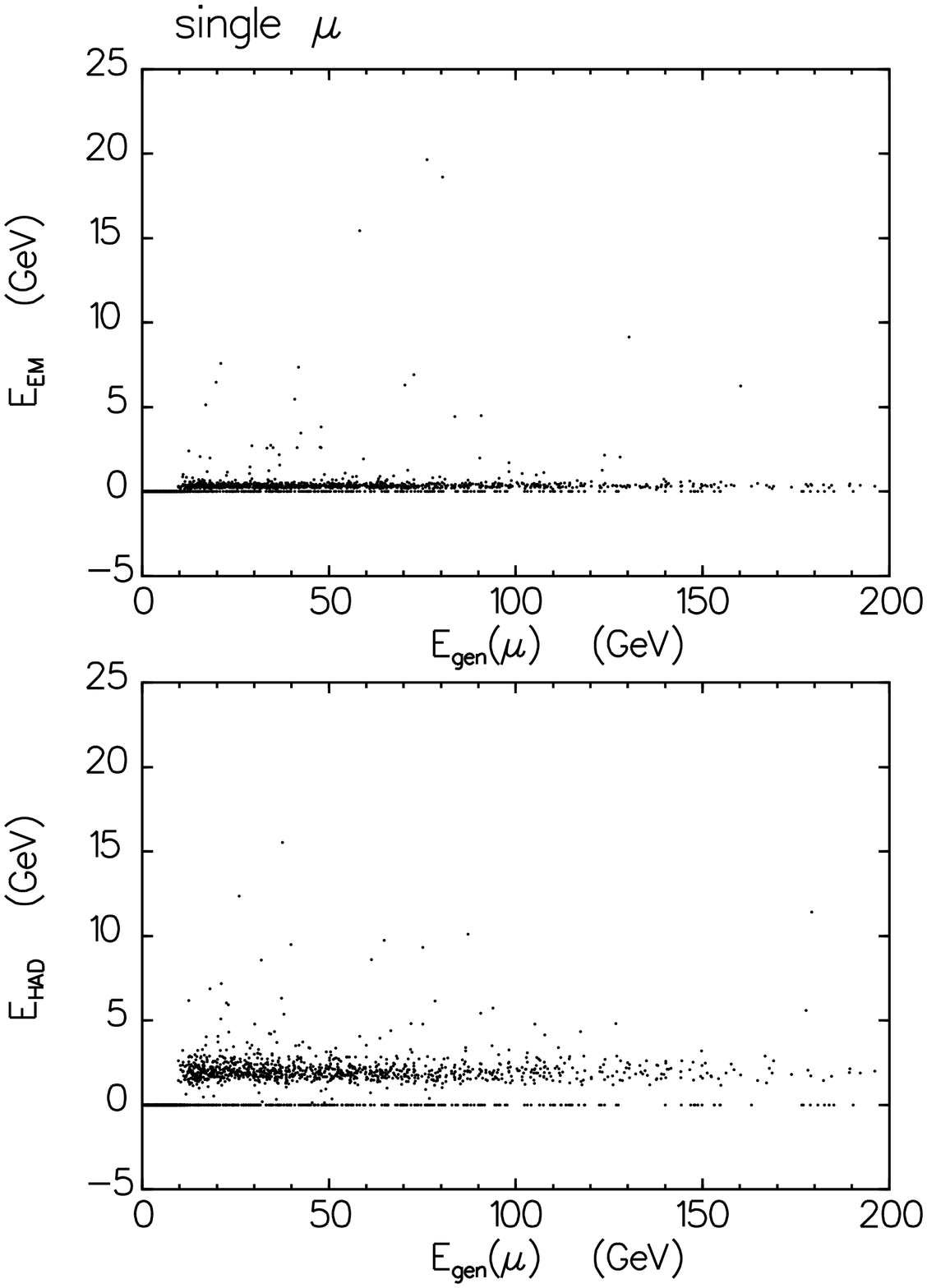}}
    \parbox{2.1in}{\epsfxsize=\hsize\epsffile[0 0 500 700]{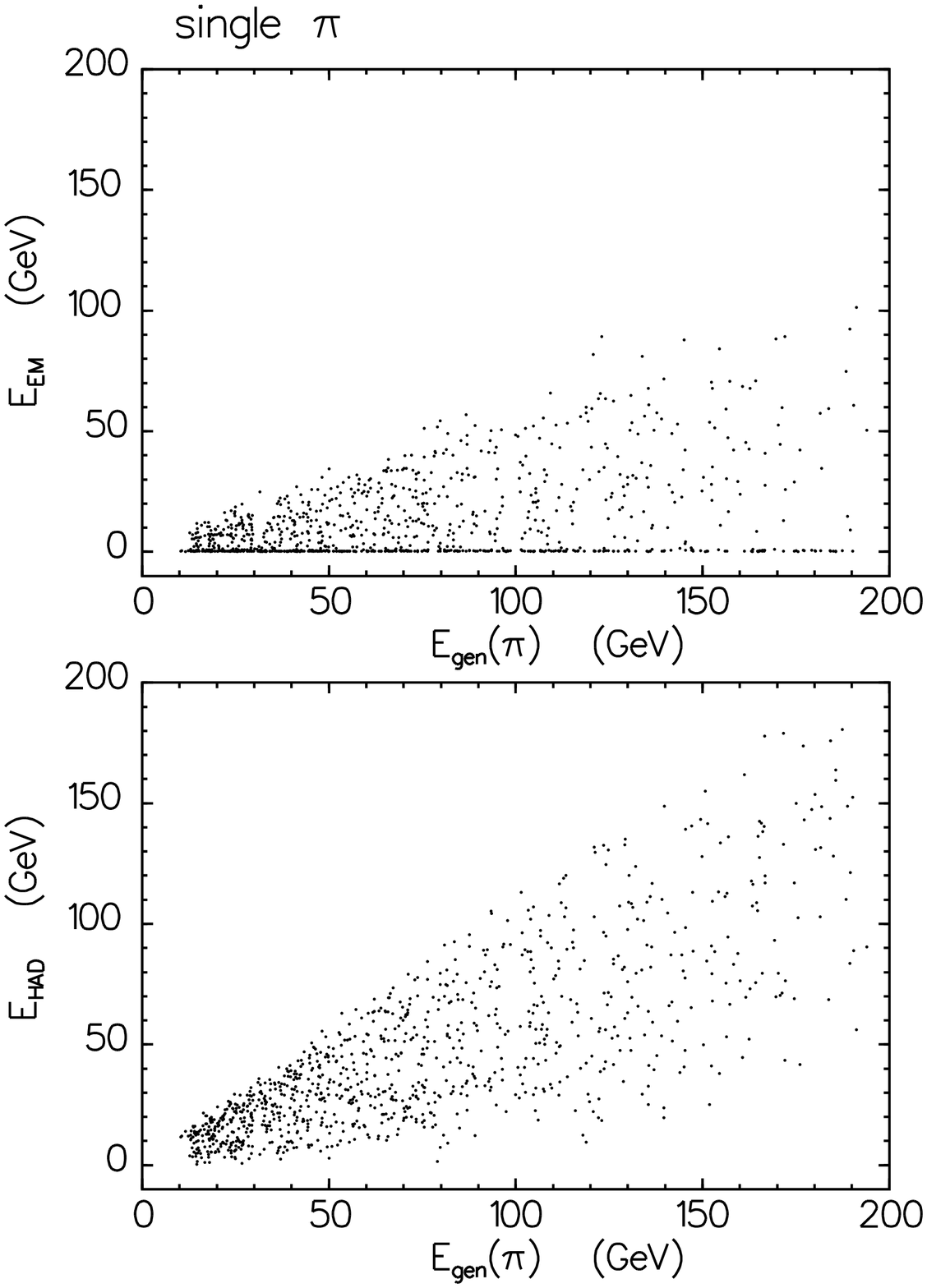}}
  \end{center}
  \caption{Calorimeter response to electrons, muons, and pions from the
           CDF Run 1 detector simulation.  The vertical axis is the 
           the energy deposited in the calorimeter, and the horizontal
           axis is the energy of the particle.}
  \label{shw_cal}
\end{figure}

Once calorimeter tower energies are determined, the program performs a
simple clustering algorithm, assigning towers with energy deposits to
one of a list of calorimeter clusters formed from scanning the
calorimeter tower energy array iteratively for the tower with maximum
$\et$, and adding to the cluster thus formed any tower with energy
above threshold and within $\Delta R$=0.4, where $\Delta
R^2=\Delta\eta^2+\Delta\phi^2$.
  
The program also calculates the total missing $\et$ from the individual 
calorimeter towers, performing a vector sum of the individual tower 
energies in the transverse plane.  

\break
\vspace{0.2in}
{\bf Trigger} \\ \nopagebreak

The CDF and D\O\  triggers rely on using fast information from the
tracking and calorimeter to signal the presence of photons, leptons,
jets, missing $\et$, etc.  The SHW simulation uses the list of charged
tracks and calorimeter tower energy deposits to arrive at a list of
trigger ``objects'' including electromagnetic deposits, jets, missing
ET, etc.  These can be required in combination to simulate the actual
online triggers foreseen in Run 2.

\vspace{0.2in}
{\bf Physics objects} \\ \nopagebreak

Typical analyses at hadron colliders begin by identifying photons, electrons, 
muons, hadronic taus, jets, and so forth and then demanding the presence of 
certain combinations of these objects depending on the final state to be 
selected.  The SHW package performs this by making identification requirements
on the calorimeter clusters, tracks, and raw calorimeter tower energies.  

The simulation identifies photons and electrons by finding calorimeter
towers with $\et > 10$ GeV in the range $|\eta| < 2$, having a ratio
of hadronic to electromagnetic energy less than 0.125.  The transverse
energy in a cone of 0.4 around this tower must be less than 10\% of
that in the central tower.  In the case of photons, there must be no
track within $\Delta R < 0.15$, or one track if it has $\pt < 1$ GeV/$c$.
For electrons, there must be a track within $\Delta R < 0.15$, and the
highest $\pt$ such track must have $0.5 < \et/\pt < 1.5$.  To simulate 
gaps between calorimeter cells, electrons and photons are dropped if they
are near the calorimeter cell boundaries in $\phi$.  

Muons are identified by finding tracks coming from generated muons, and
applying efficiency factors for fiducial coverage and typical reconstruction
efficiency.  The overall efficiency for muons is roughly 85\%.  No
fake muons are generated from random tracks.  

Hadronically decaying taus are identified by seeking calorimeter
deposits in the region $|\eta|<1.5$ associated with one or three
charged tracks within 10$^\circ$ of the calorimeter cluster centroid.
Hadronic jets are rejected by demanding no other tracks in an annular
region between 10$^\circ$ and 30$^\circ$ centered on the calorimeter
cluster centroid.  There must be at least one track with $\pt > 5$
GeV/$c$.  Electrons are rejected by demanding that
$E_{cal}/E_{trk}(1-E_{cal}^{em}/E_{cal}^{had})>0.25$.  The efficiency
for hadronically decaying taus varies with the event jet environment
from about 40\% to 60\%.

Jet clusters are formed from the list calorimeter tower energies.  The
algorithm seeks the maximum-$\et$ tower not assigned to a cluster, then
adds to the new cluster any adjacent towers within a specified cone
in $\Delta R$ (default 0.4) around the seed tower.  Jets can be tagged
as $b$- or $c$-jets by first determining the true jet type by comparing
the reconstructed jet with generated quark and gluon directions and
energies.

There are two default algorithms available for heavy quark tagging,
based on the CDF jet probability algorithm, and an algorithm which
counts high-impact-parameter tracks in jets.  The efficiencies for
these algorithms are parametrized as a function of jet $\et$.
Figure~\ref{shw_tag} shows the tagging rates per taggable jet for
these parametrizations, for jets within the fiducial region.

Note, however, that some of the analyses presented in the following
sections make use of parametrizations of the $b$-tagging efficiency
which differ from the default parametrizations in SHW.  In this scheme
there are ``tight'' and ``loose'' tagging efficiencies, as depicted in
Figure~\ref{shw_tl_tag}.  This algorithm is based on projections
from the CDF Run~1 tagging rates, incorporating the improved vertexing
possible with the new silicon detectors, and using both the secondary
vertex (SECVTX) algorithm and soft lepton (SLT) tagging.

\begin{figure}
  \begin{center}
    \parbox{3.0in}{\epsfxsize=\hsize\epsffile[0 470 500 700]{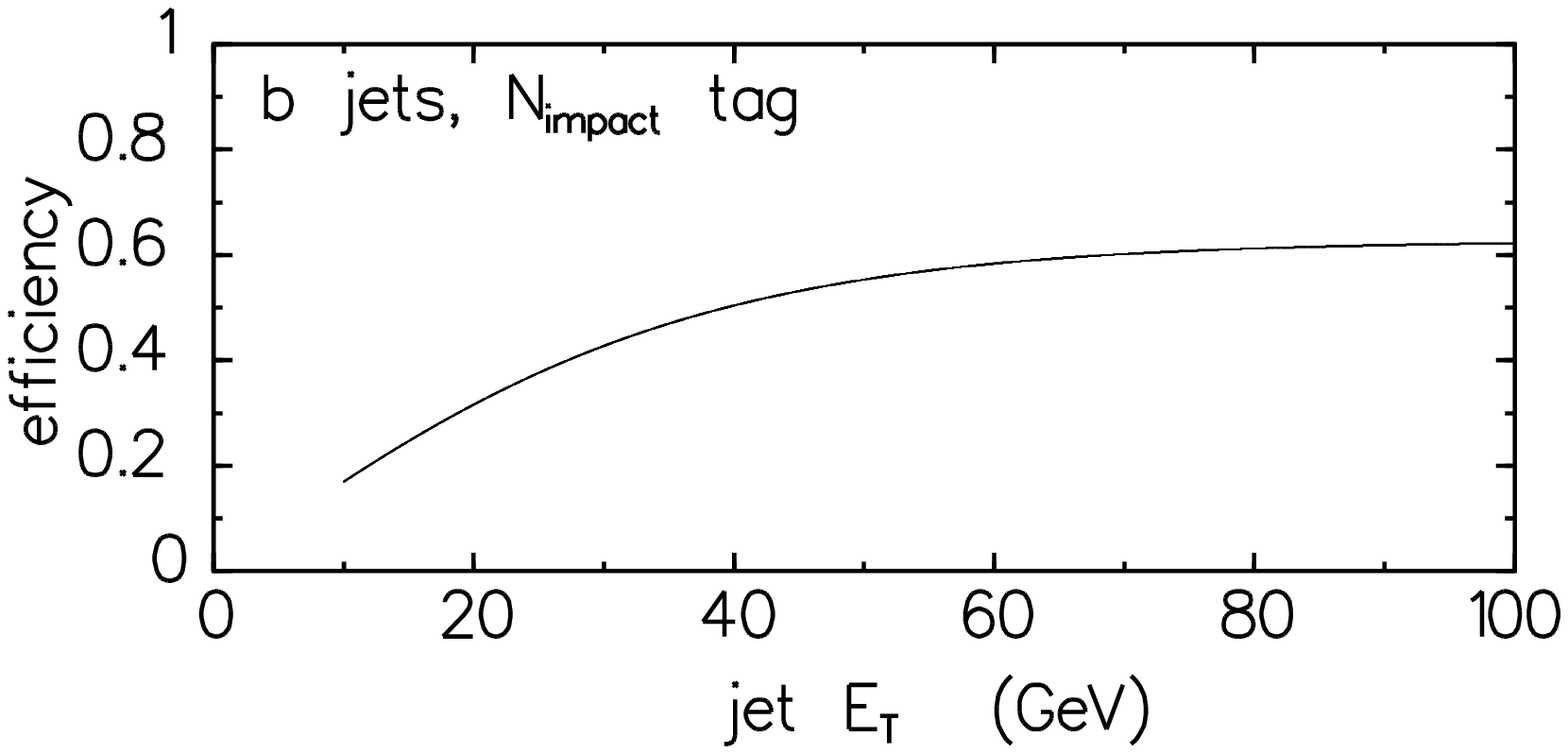}}\\
    \parbox{3.0in}{\epsfxsize=\hsize\epsffile[0 470 500 700]{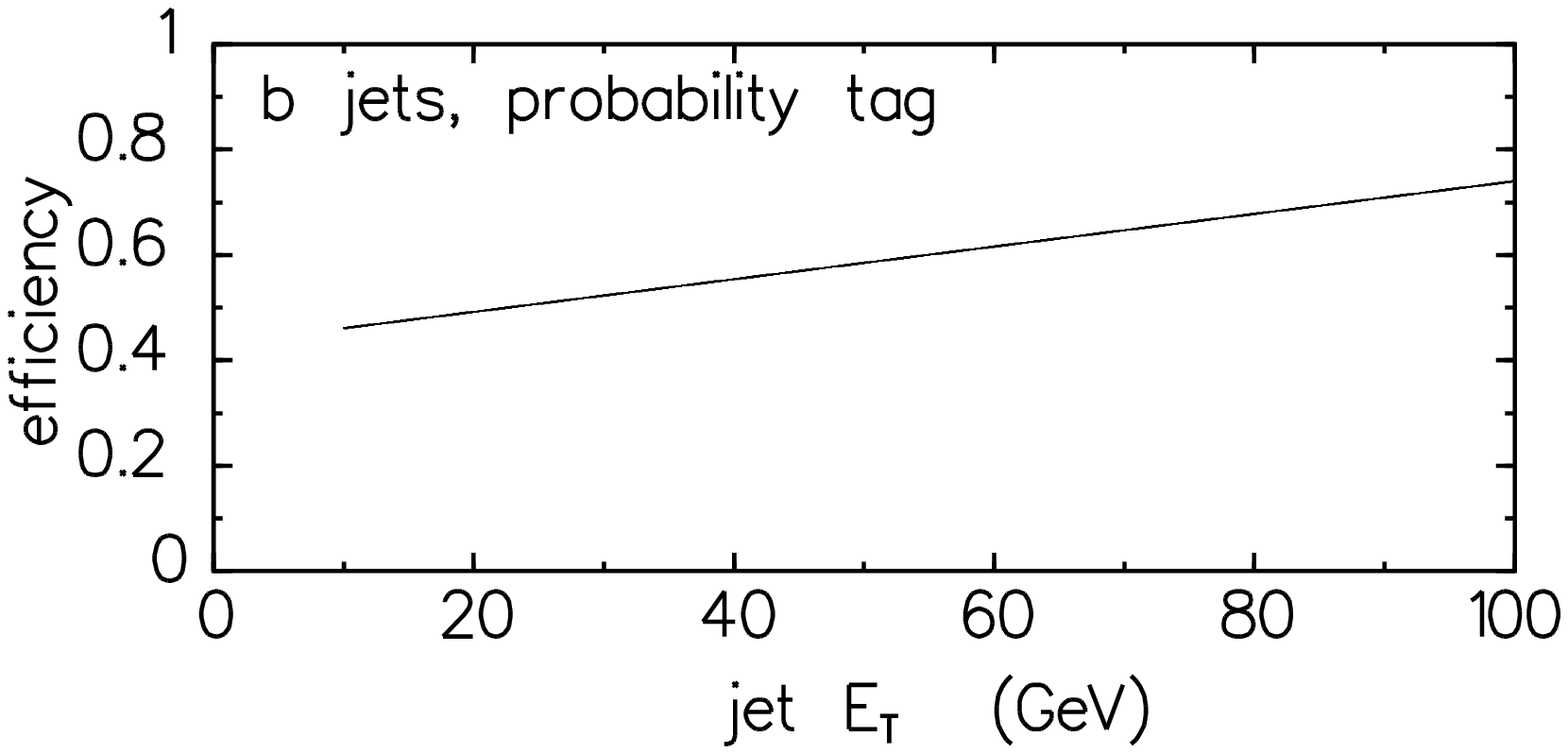}}\\
    \parbox{3.0in}{\epsfxsize=\hsize\epsffile[0 470 500 700]{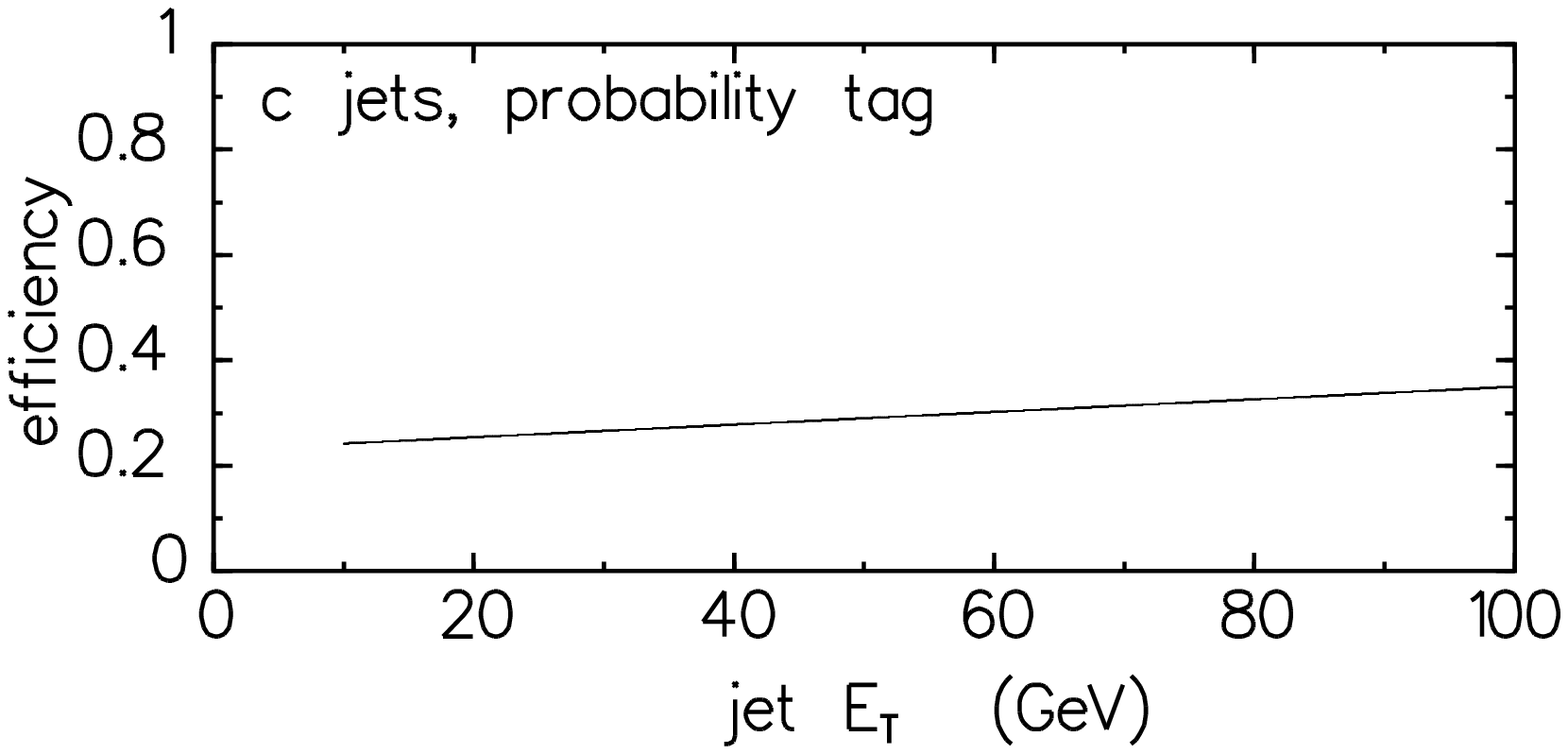}}
  \end{center}
  \caption{Parametrized default SHW tagging efficiencies per taggable jet 
           for $b$ and $c$ jets.}
  \label{shw_tag}
\end{figure}

\begin{figure}
  \begin{center}
    \parbox{3.0in}{\epsfxsize=\hsize\epsffile{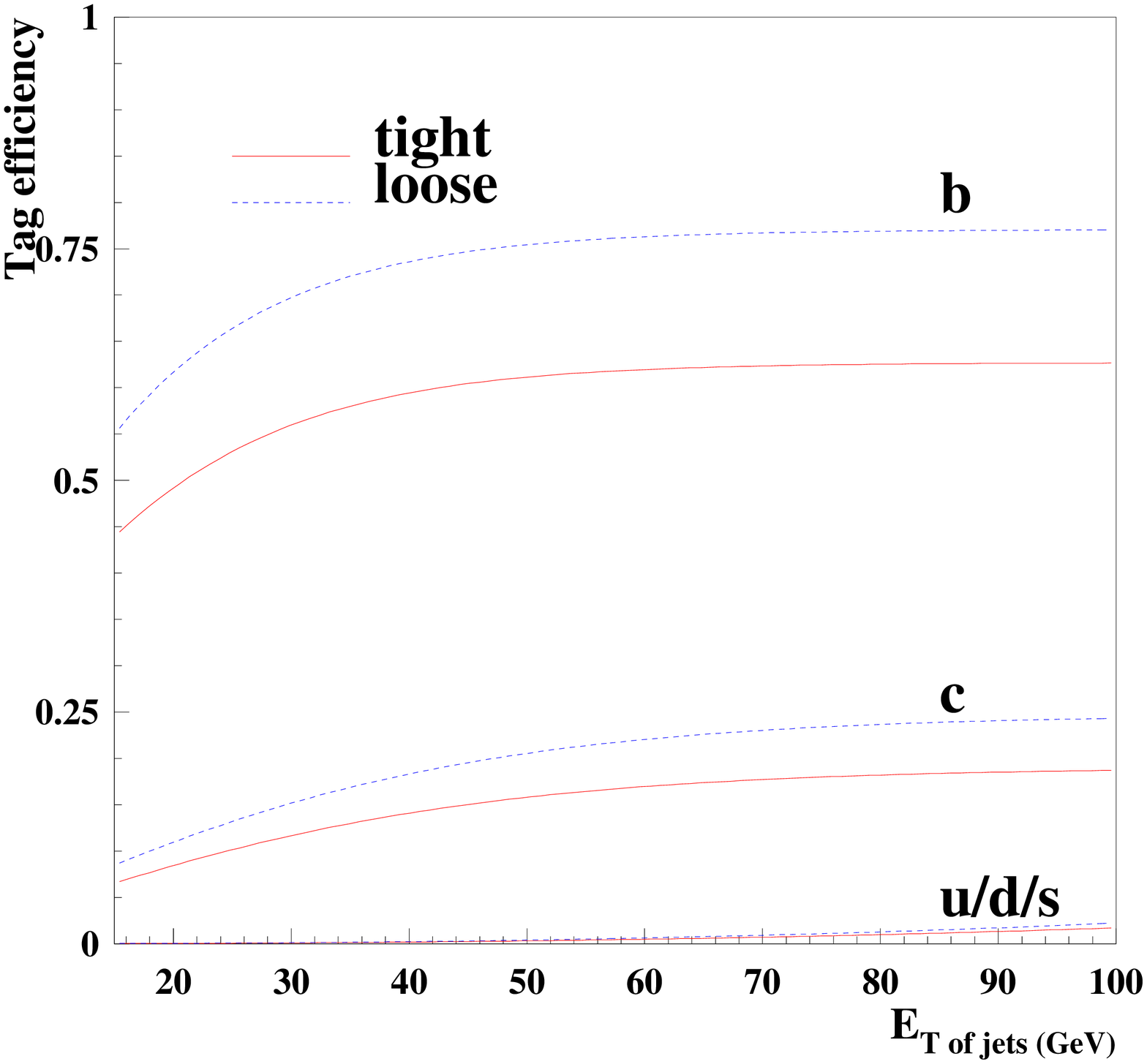}}
    \parbox{3.0in}{\epsfxsize=\hsize\epsffile{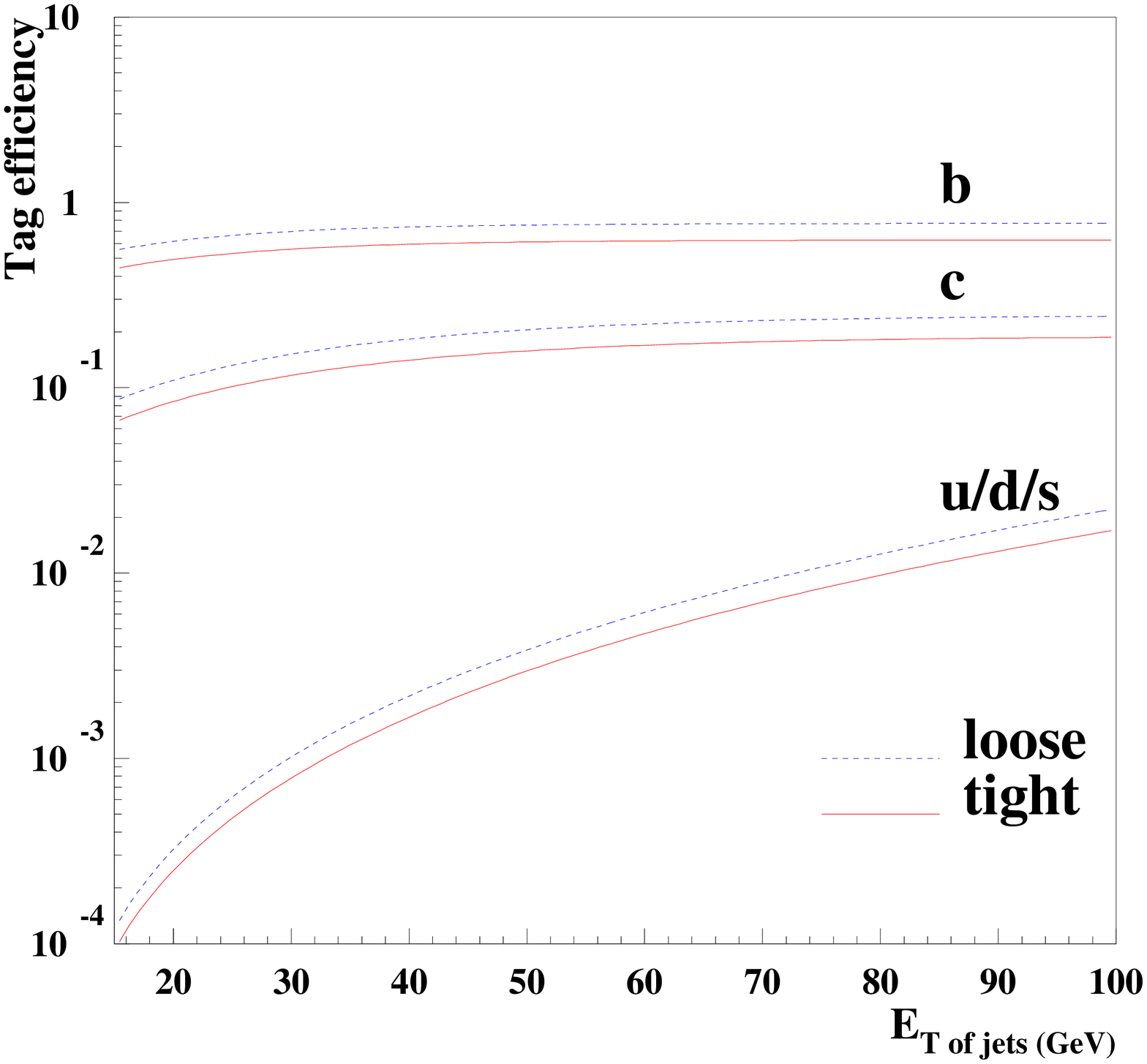}}
  \end{center}
  \caption{Parametrization of tight/loose tagging efficiency (per
           taggable jet) for $b$, $c$, and
           light quark jets used in some analyses. The left plot shows
           the efficiencies on a linear scale, and the right plot on a
           log scale.  These curves are based on extrapolations of the 
           CDF Run~1 efficiencies.}
  \label{shw_tl_tag}
\end{figure}

    \subsubsection{Study of $\bb$ Mass Resolution}	\small
\begin{center}
{\it T. Dorigo,
     S. Kuhlmann,
     R. Snihur,
     G. Watts,
     J. Womersley} \\
\end{center}
\normalsize \nopagebreak

For Higgs masses less than about 135~GeV the most promising channels
in Run~2 at the Tevatron involve the decay of the Higgs to $b\overline
b$, such as the process $\pp\to WH\to \ell\nu\bb$.  The signal is an
excess of events having $\bb$ invariant mass peaked at the Higgs mass.
Obtaining the best possible invariant $\bb$ mass resolution is
therefore crucial in establishing a signal.

This section presents the results of two studies of the dijet mass resolution
in the upgraded Run 2 detectors.  

\newpage
\vspace{0.2in} \large
{\bf CDF Dijet Mass Resolution Studies} \\ \normalsize \nopagebreak

The $b\bar{b}$ decay of the Higgs boson dominates if $M_H <
135~GeV/c^2$. Our ability to extract this particle from the large QCD
background in Run 2 (for instance in the $W b \bar{b}$ final state,
when associated $WH$ production is sought) will therefore depend
critically on the resolution we can attain on the Higgs boson mass,
reconstructed from the measured $b$-quark jet energies: both the
possibility to see a bump in a dijet mass spectrum and the option to
apply a mass window cut as a selection tool will strictly depend on
the attainable $b\bar{b}$ mass resolution.

A study of jet--energy resolution in QCD dijet events with one
electromagnetic jet was carried out by CDF.  This study improved upon
the calorimeter--only jet energy measurement typically used in Run 1
CDF (and D\O) analyses by including information from charged--particle
momenta measurements and shower maximum detectors.
Figure~\ref{f-kuhlfig} shows the jet energy resolution for both the
calorimeter--only jet energy algorithm and the improved algorithm.
This shows roughly a 30\% improvement in energy resolution.

\begin{figure}
  \begin{center}\hbox{
  \parbox{3.25in}{\epsfxsize=\hsize\epsffile{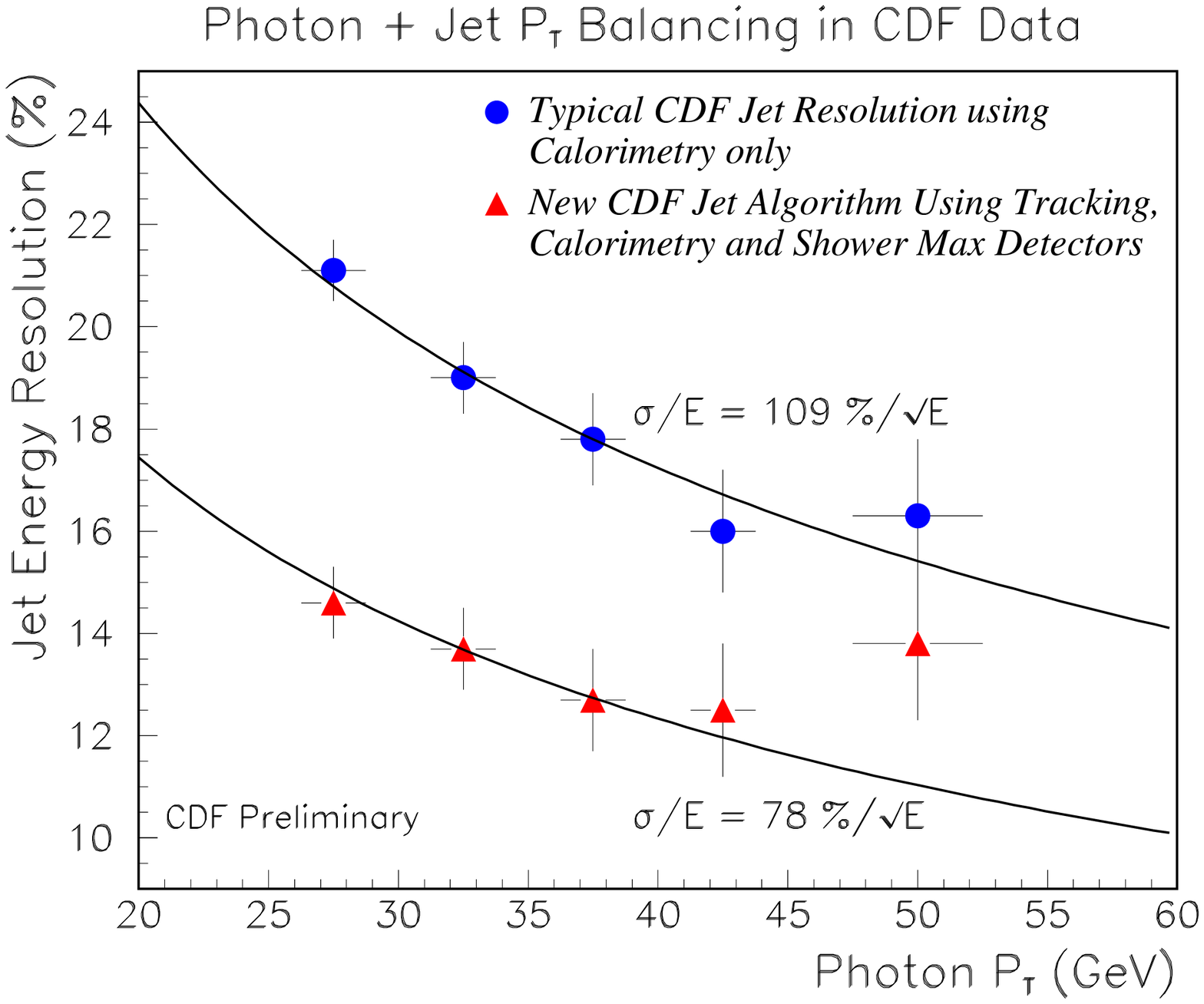}}
  \parbox{3.25in}{\epsfxsize=\hsize\epsffile{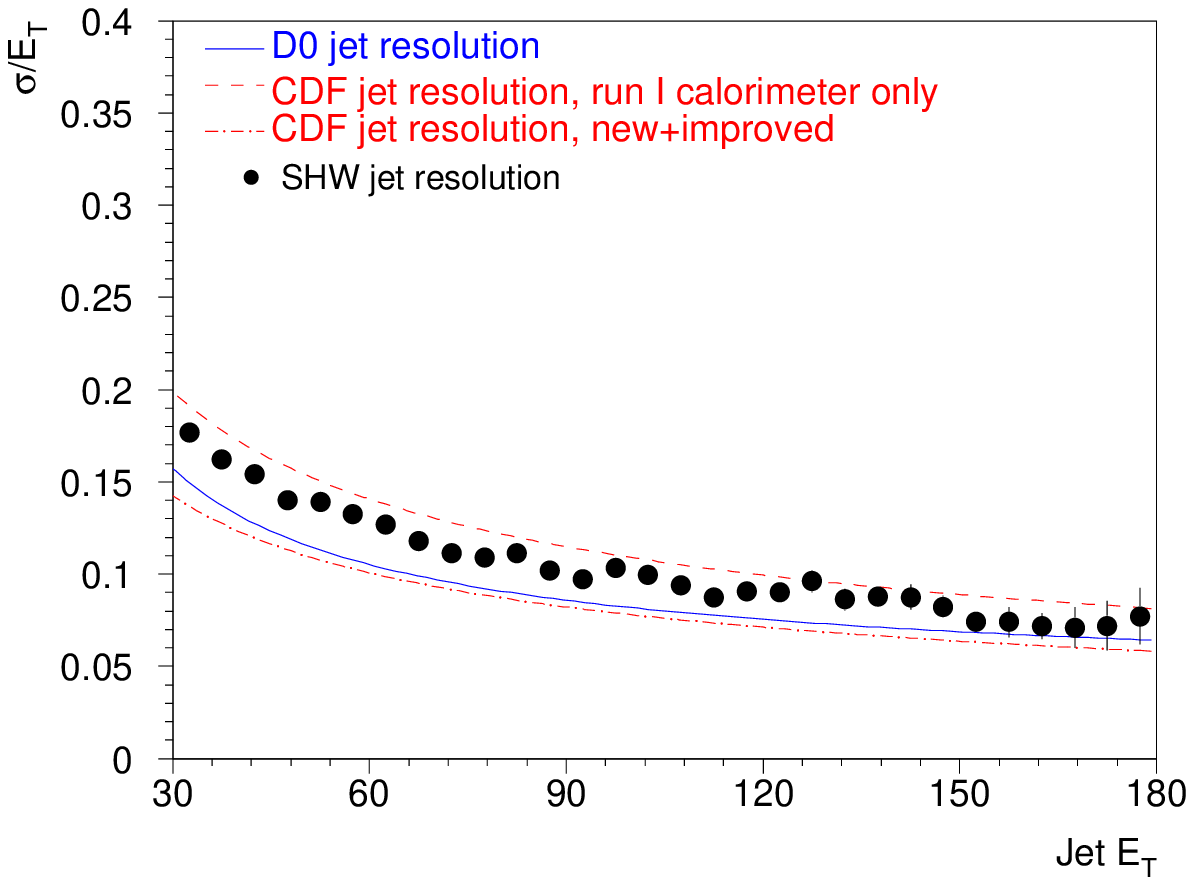}}}
  \end{center} 
  \caption{The left-hand panel shows CDF jet energy resolution 
           computed for a calorimeter--only 
           measurement similar to Run 1 results and for an
           improved algorithm which incorporates energy measurements 
           from additional detector systems.  The right-hand panel shows
           the resolution obtained from SHW compared with the two CDF
           resolutions and the standard D\O\ jet energy resolution.  It
           is clear that both experiments have achieved energy resolutions 
           significantly better than that used in this study.}
  \label{f-kuhlfig}
\end{figure}

A study of the dijet mass resolution in $Z \to b \bar{b}$ events was
performed by CDF in the context of a search for that process in events
collected by an inclusive muon trigger.  Two million $Z \to b \bar{b}$
decays were generated with the PYTHIA 5.7 Monte Carlo, and filtered as
the data.
 
After the application of the standard CDF jet correction routine, many
observable quantities were studied as a function of the difference
between measured jet momenta and originating parton momenta, in the
hope of using the measured values of these observables to help
reducing the energy mismeasurement.  The quantities that were found
most useful for this purpose were the following: the muon momentum,
the missing transverse energy projection along each jet, and the jet
charged fraction.

The muon momentum is needed in correcting the energy of jets
originating from the semileptonic decays of $b$-quarks, because the
minimum ionizing muons do not contribute linearly to the energy
measured in the calorimeter. The missing $E_T$, projected along the
direction of the jets in the transverse plane, provides useful
information on the momentum taken away by the neutrino and on a
possible fluctuation of the energy measurement. The charged fraction
of the jets (defined as the ratio between the total momentum of
reconstructed charged tracks inside the jet cone and the jet momentum
measured in the calorimeter) also helps reduce the uncertainty in the
energy measurement.

By properly accounting for the value of these observables, it was
possible to reduce the relative width of the dijet mass distribution,
$\sigma_M / M_{jj}$, by nearly $50\%$ (see Figure~\ref{mass_evol}, left).
  
\begin{figure}
  \begin{center}
    \parbox{3.0in}{\epsfxsize=\hsize\epsffile{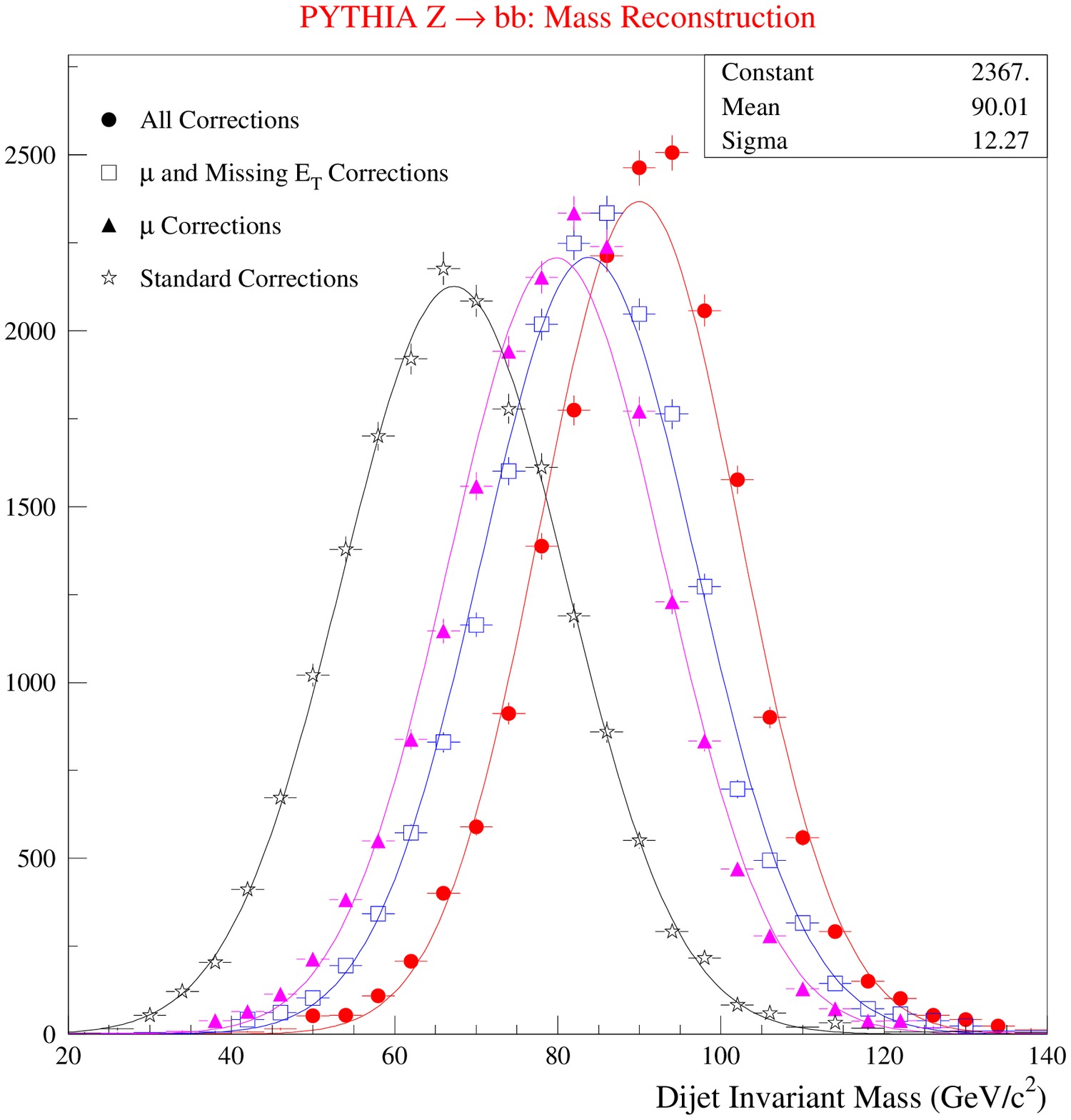}}
    \parbox{3.0in}{\epsfxsize=\hsize\epsffile{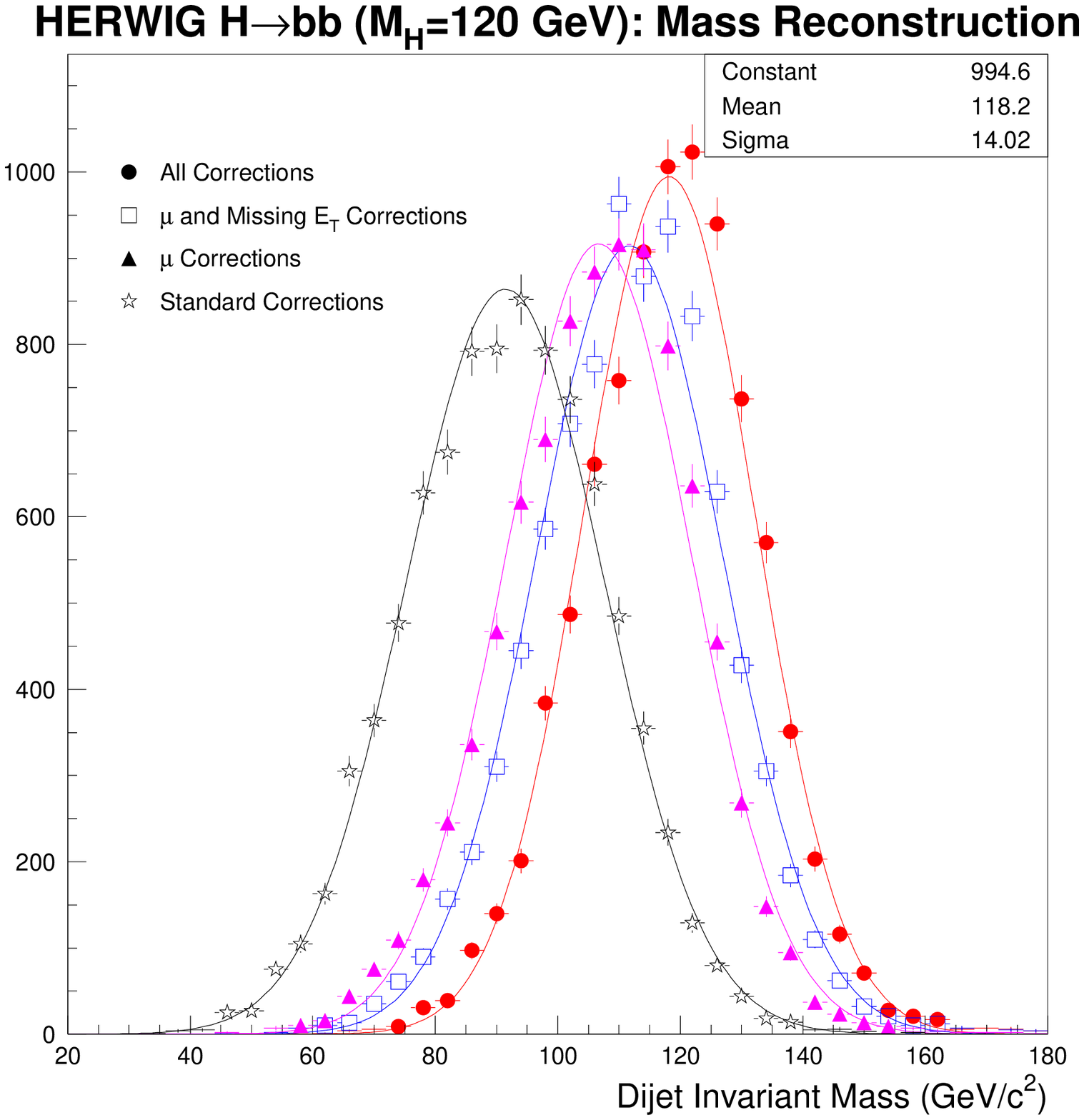}}
  \end{center}
  \caption{ The four gaussian fits show the improvement of the mass 
            reconstruction
	    for simulated $Z \to b\bar{b}$ decays (PYTHIA 5.7, left) 
            and simulated $H \to b\bar{b}$ decays (HERWIG 5.6, right) 
            when the observable characteristics of the $b$-quark 
            decays are properly taken into account in the mass 
            reconstruction.  The CDF Run~1 simulation was used in this
            study.}
  \label{mass_evol}
\end{figure}

The validity of these corrections was tested with HERWIG \cite{herwig}
to check the
effect of a different fragmentation model, and no differences were
found.  The corrections were also checked on a simulated sample of $gg
\to H \to b\bar{b}$ decays, with $M_H=120~GeV$, and were found to be
equally useful for a different resonance mass (Figure~\ref{mass_evol},
right).

Finally, it was possible to show that a real sample of $Z \to b\bar{b}$
decays extracted from Run 1 data behaved as expected from the simulation:
the reconstructed excess of events over background predictions indeed
showed an average value and width in agreement with Monte Carlo predictions
both before and after the application of the jet corrections 
(Figure~\ref{excess_evol}).

\begin{figure}
  \begin{center}
    \parbox{3.0in}{\epsfxsize=\hsize\epsffile{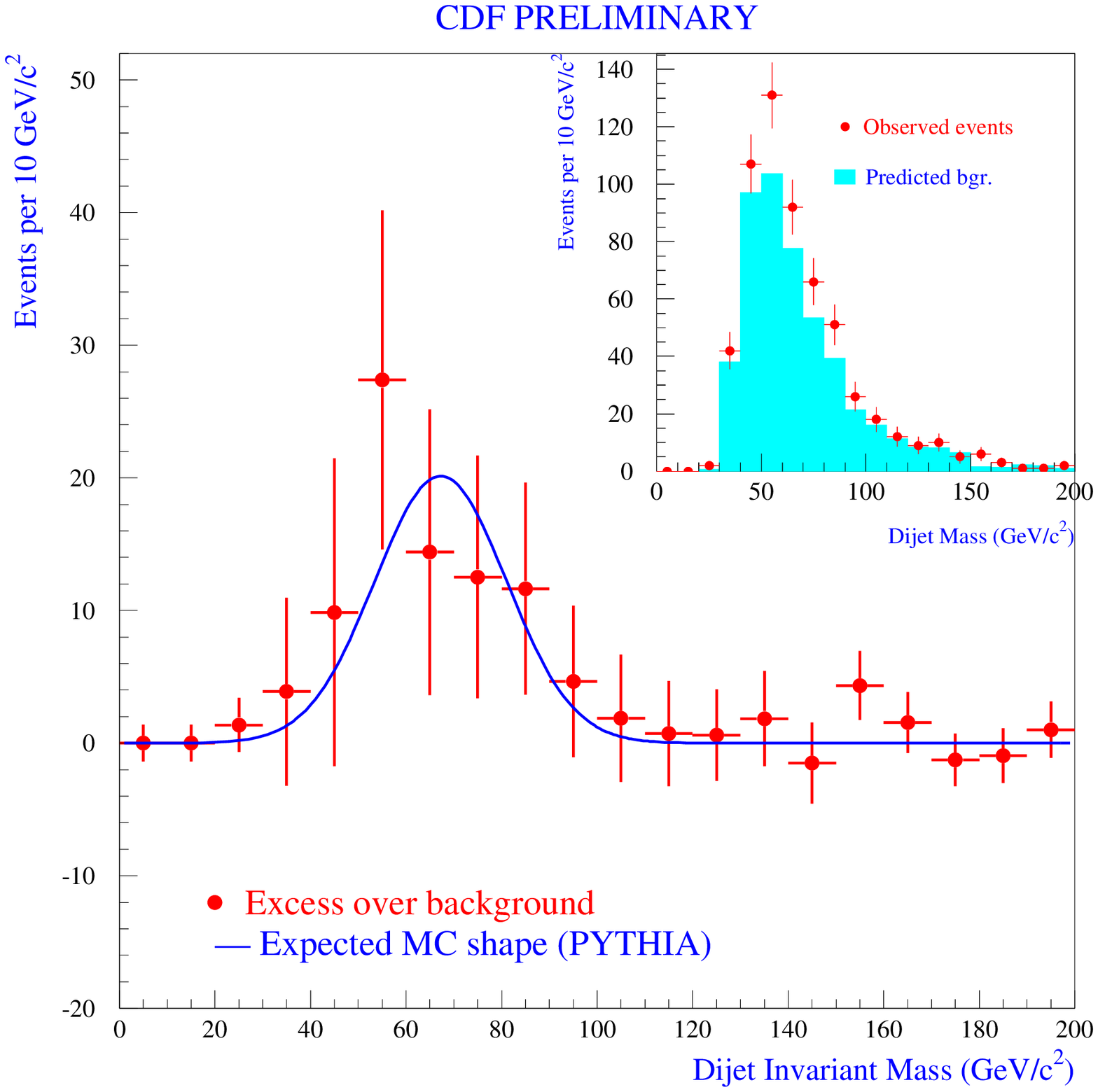}}
    \parbox{3.0in}{\epsfxsize=\hsize\epsffile{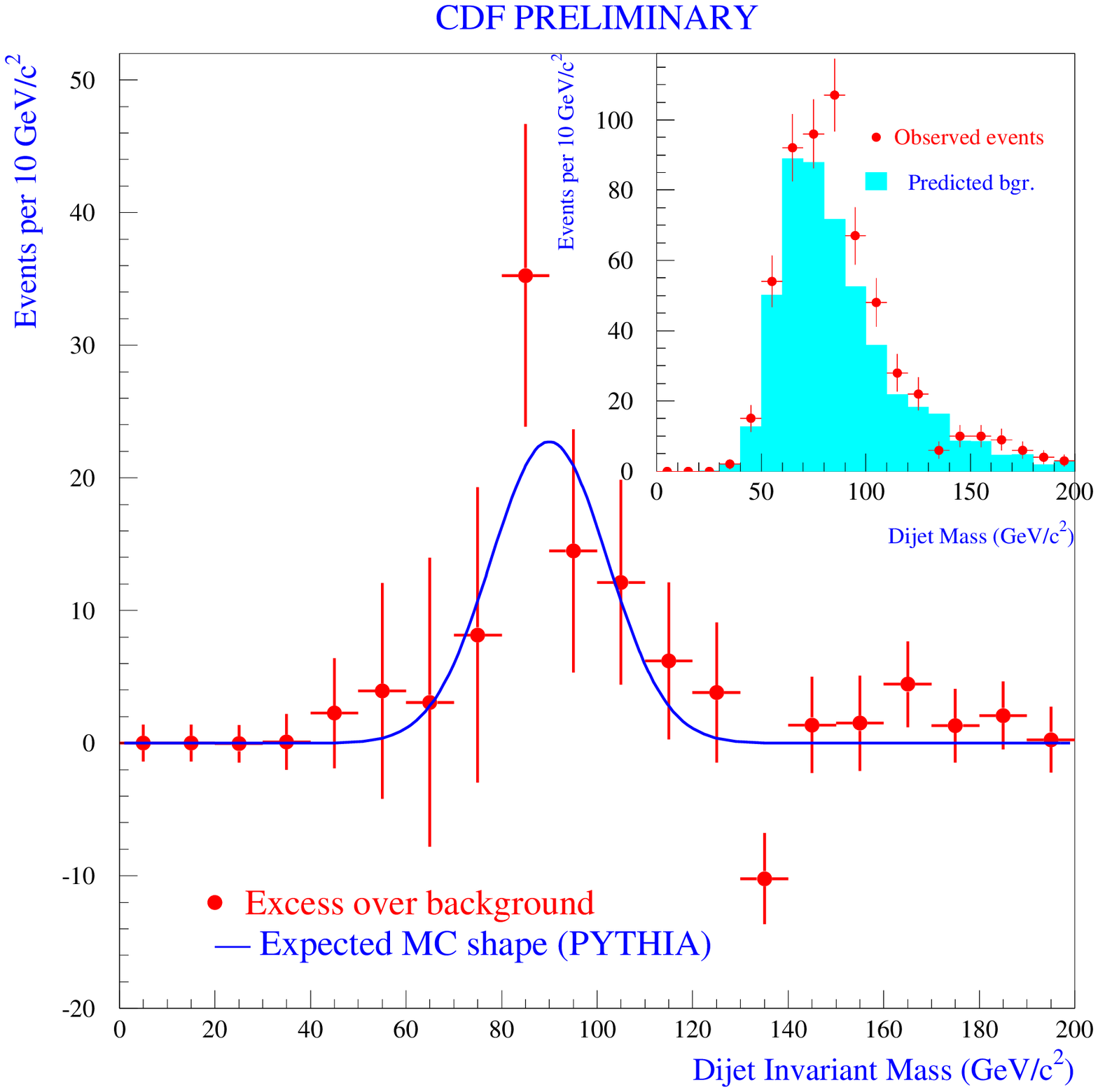}}
  \end{center}
  \caption{ The comparisons between observed data and background predictions
            (insets) allow the extraction of gaussian excesses due to the 
            $Z \to b \bar{b}$ decay both before (left) and after (right)
            the use of the $b$-quark specific jet corrections. 
            The excess behaves as predicted by Monte 
            Carlo expectations in the two cases. }  
  \label{excess_evol}
\end{figure}

The corrections described above were specifically intended for a pair 
of back-to-back $b$-quarks, when one of the two decays to a muon. 
Their success in improving the significance of the $Z \to b\bar{b}$ signal,
however, demonstrates the usefulness of the chosen approach, 
and spurs new studies aimed at different final state
topologies for the search for new resonances.


\vspace{0.2in} \large
{\bf D\O\ Dijet Mass Study} \\ \normalsize \nopagebreak

This study is based on events generated using PYTHIA with a Higgs mass
of 90~GeV simulated in the D\O\ detector.  The $W$ decay was forced to
$\mu \nu$ in order to give a relatively ``clean'' dijet Higgs signal
in the calorimeter.  Events were generated with both initial and final
state radiation (ISR and FSR) as default, but we also generated
samples with neither ISR nor FSR, with only ISR, and with only FSR,
for comparison purposes.  Events were then passed through the D0GEANT
detector simulation which provides a full simulation of the
calorimeter response and resolution.  (This is the Run~1,
fortran-based detector simulation).  The digitized events were then
passed through the Run~1 calorimeter reconstruction and jet finding
packages.

In the analyses presented here, we required the following standard jet
selection cuts:
\[E_T > 15\ {\rm GeV\ and}\ |\eta|< 2.0.\]

The reconstructed mass distributions were characterized using a fit
function formed of a gaussian convoluted with two
exponentials~\cite{hobbsfit}.  The fits typically yield a $\chi^2$ per
degree of freedom between 1 and 2.  The bin containing the peak and 
the bin separation corresponding to the FWHM were found by hand.

Two jet clustering algorithms were investigated.  The first is the
standard D\O\ cone algorithm, with a cone size of $\Delta R =0.7$.
The second is the $k_T$ algorithm, with $D$ parameter settings of 0.5
and 1.0 ($D$ is a close approximation to the $\Delta R$ in the cone
algorithm.)  In the $k_T$ algorithm each hit calorimeter cell forms a
vector.  Vectors are combined with each other to build a jet by
joining the two vectors with the smallest relative $p_T$ to form a new
vector.  Combination stops when all vectors are separated by $\Delta R
> D$ \cite{ktalg}.

Our attempt throughout this study is to demonstrate the factors
governing mass resolution, rather than to present final results on the
resolution that can be attained.  Much work will have to be done in
deriving energy scale corrections before final resolution numbers are
finalized.

\vspace{0.2in}
{\bf Parton, Particle and Calorimeter Level Resolutions} \\ \nopagebreak

To demonstrate the effects of hadronization, radiation, and of the
calorimeter energy measurement, we have applied the same $k_T$
algorithm at the parton, particle and calorimeter cell levels.  We
used the D\O\ Run 1 $k_T$ jet finder and excluded the muon and
neutrino from the $W$ decay from the parton and particle level jets.
Two values of the jet size parameter $D$ were used, 0.5 and 1.0.
Figure~\ref{skip_id_1_page1} shows the mass resolution obtained at the
parton level. Without any gluon radiation, the Higgs mass is
reconstructed perfectly (left hand pane); gluon radiation degrades
this somewhat, as shown in the right hand pane which has both initial
state radiation (ISR) and final state radiation (FSR) included.

\begin{figure}
  \begin{center}
    \parbox{3.0in}{\epsfxsize=\hsize\epsffile{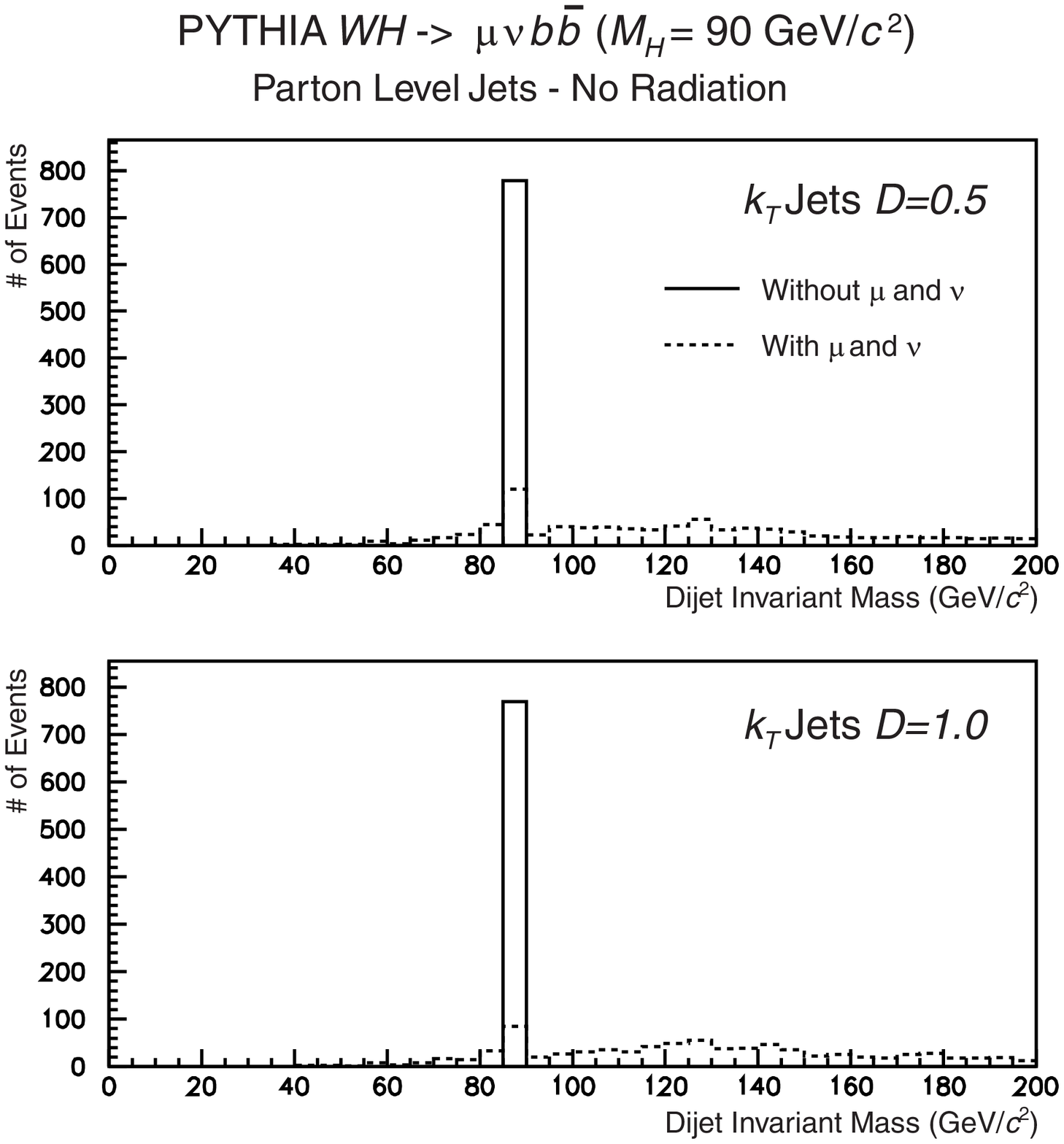}}
    \parbox{3.0in}{\epsfxsize=\hsize\epsffile{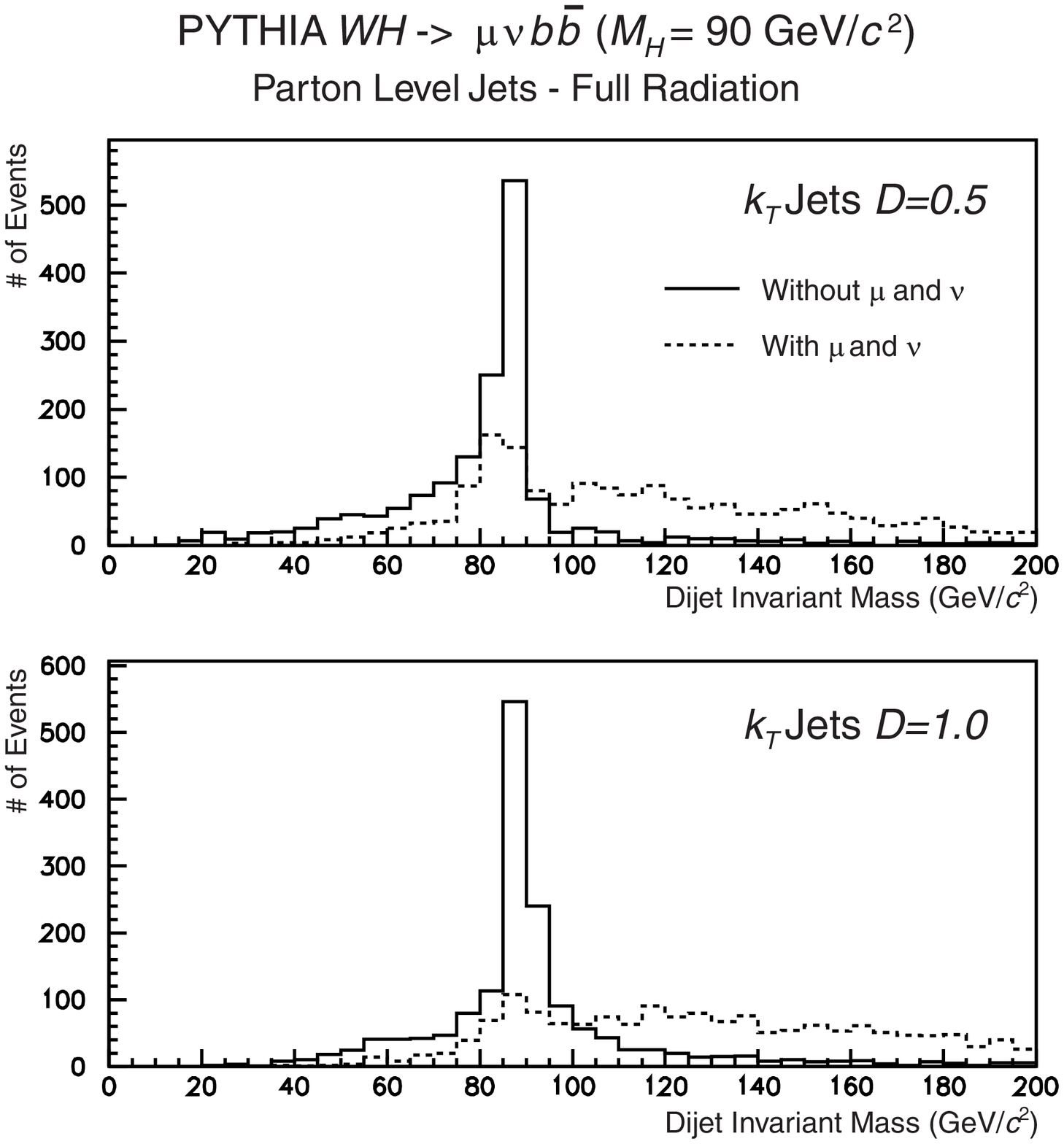}}
  \end{center}
  \caption{The dijet invariant mass using reconstructed parton level $k_T$ 
     jets. The upper two plots are $D=0.5$ and the lower two are $D=1.0$; 
     the left hand column has no initial or final-state radiation and the 
     right hand column is a simulation will both on.  The solid line does 
     not include muons and neutrinos in the jet; the dashed line does.}
  \label{skip_id_1_page1}
\end{figure}

Hadronization introduces a low-side tail to the mass
distributions, as may be seen in Figure~\ref{skip_id_2_page1}; 
it is somewhat washed out when ISR and FSR 
are included.  At this point
the larger $D$ value starts to give a better resolution since it
clusters more of the final state particles into the reconstructed jets.

\begin{figure}
  \begin{center}
    \parbox{3.0in}{\epsfxsize=\hsize\epsffile{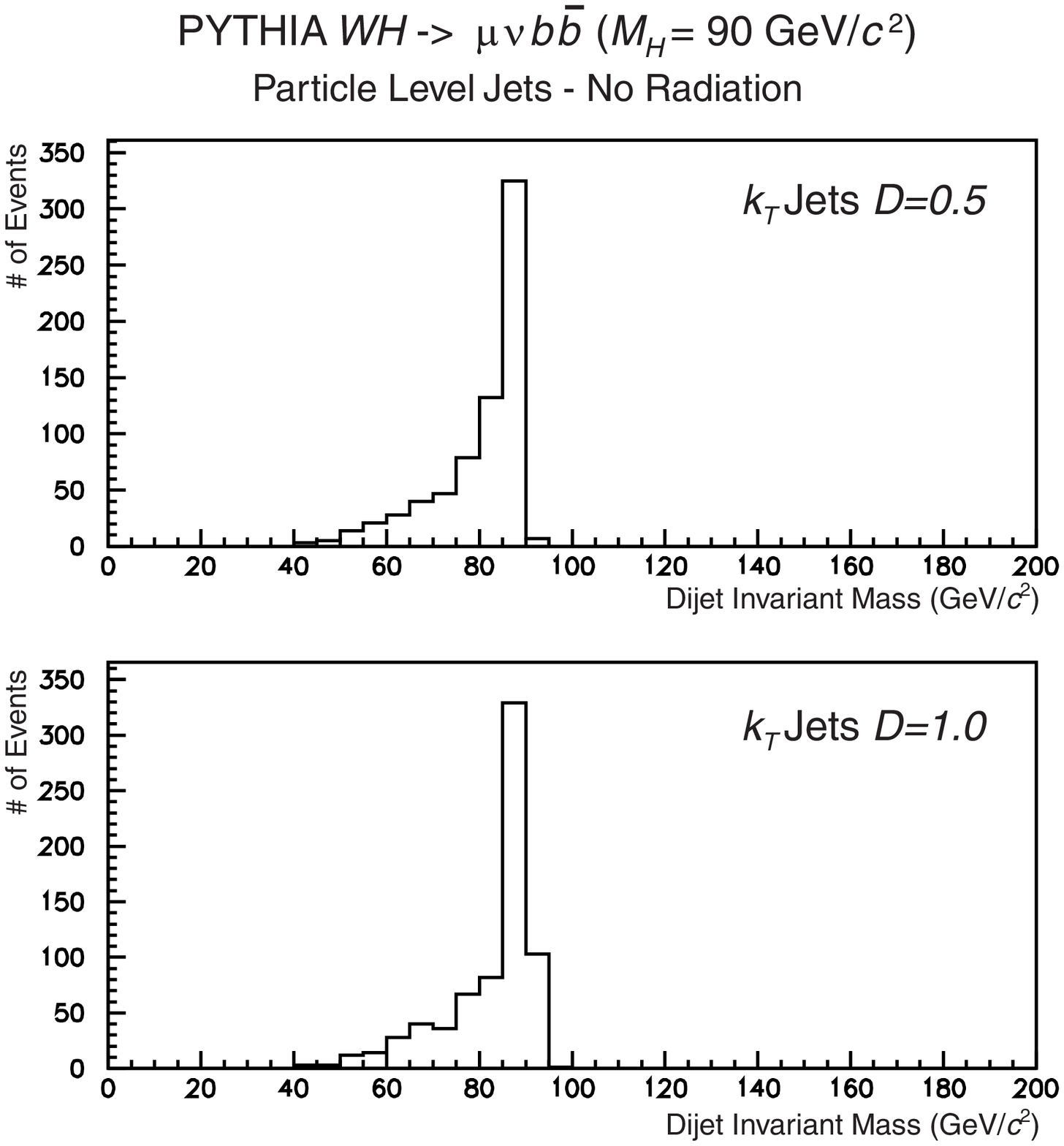}}
    \parbox{3.0in}{\epsfxsize=\hsize\epsffile{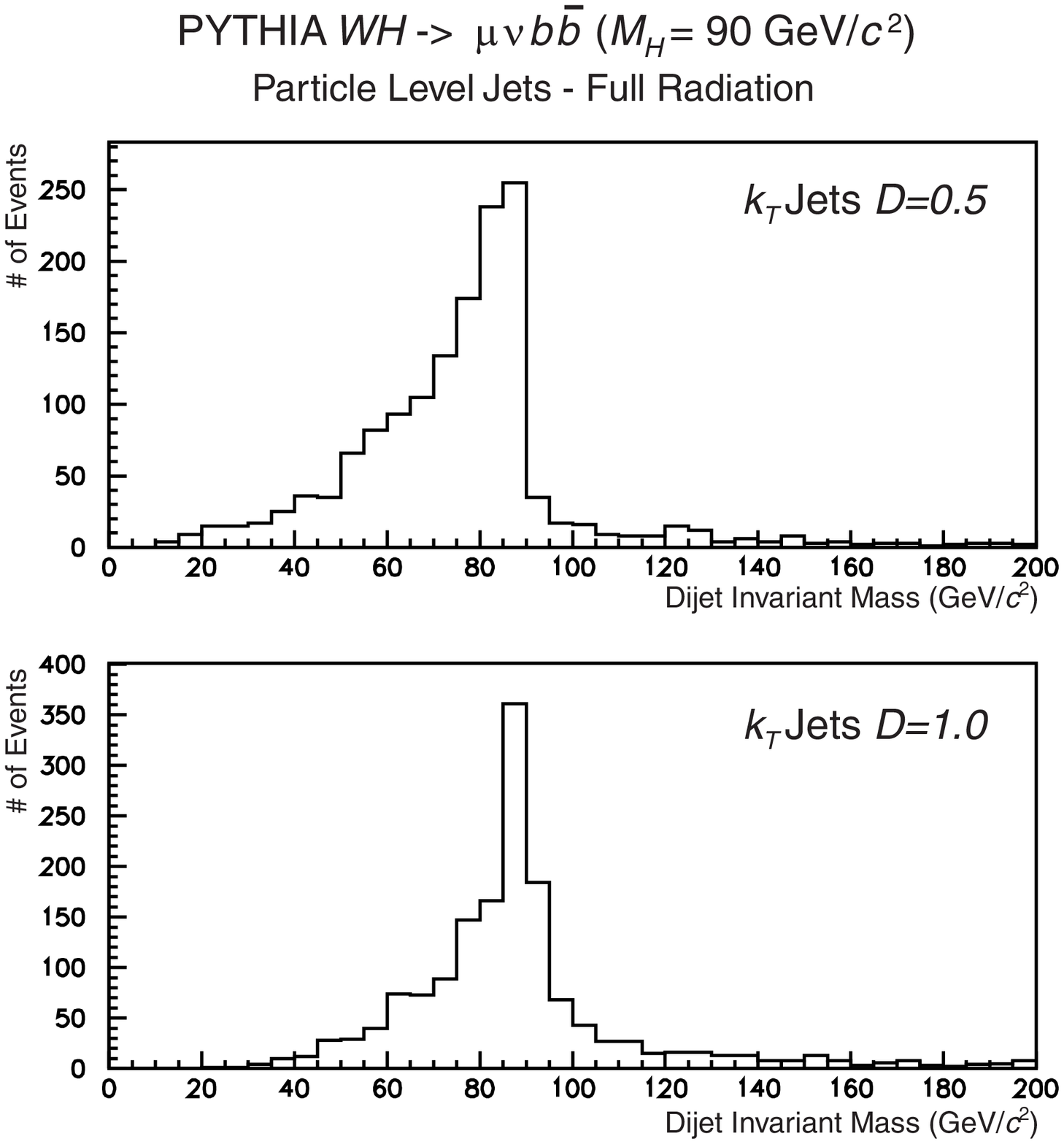}}
  \end{center}
  \caption{The dijet invariant mass calculated using reconstructed particle 
      level $k_T$ jets. The upper two plots are $D=0.5$ and the lower two 
      are $D=1.0$; the left hand column has no initial or final-state 
      radiation and the right hand column is a simulation with both on. }
  \label{skip_id_2_page1}
\end{figure}

Finally at the calorimeter level we obtain the distributions
in Figure~\ref{skip_id_6_page4} which correspond to what would
be measured in the detector.  We will now investigate the
calorimeter level distributions in more detail.

\begin{figure}
  \begin{center}
    \parbox{3.0in}{\epsfxsize=\hsize\epsffile{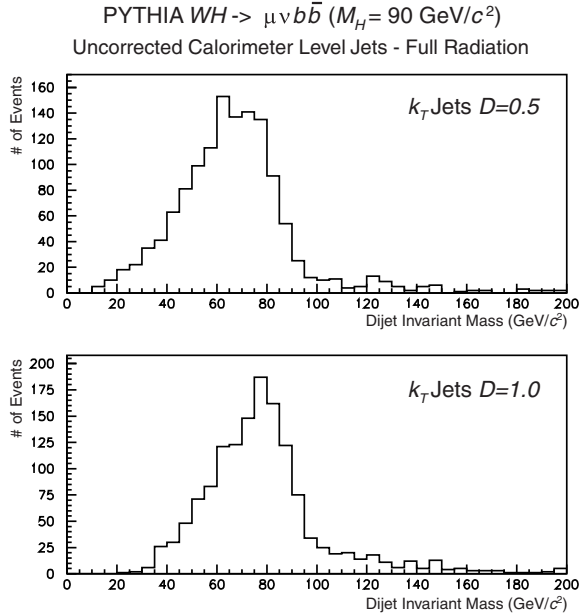}}
  \end{center}
  \caption{The dijet invariant mass calculated using uncorrected 
     calorimeter $k_T$ jets. The upper plot is $D=0.5$ and the lower is 
     for $D=1.0$. Both initial-state and final-state radiation are simulated.}
  \label{skip_id_6_page4}
\end{figure}

\vspace{0.2in}
{\bf Dijet Mass Definition} \\ \nopagebreak

We compared two different dijet mass estimators:
\[M_{4v}^2 = (E_{1} + E_{2})^{2} - (p_{x1} + p_{x2})^{2} + (p_{y1} + p_{y2})^{2} + (p_{z1} + p_{z2})^{2}\]
and
\[M_{jj}^2 =  2 E_{T1} E_{T2} ( \cosh(\eta_{1} - \eta_{2} )
	- \cos(\phi_{1} - \phi_{2}) ). \]

For the particular case of $R=0.7$ cone jets, and plotting the mass
of the leading two jets in the event,
the two quantities are shown in figure \ref{new_t6_15_c_un_1812}.
They yield very similar fractional widths though they are
not bin-by-bin identical.  In most of what follows we shall use $M_{4v}$.

\begin{figure}
  \begin{center}
    \parbox{3.0in}{\epsfxsize=\hsize\epsffile{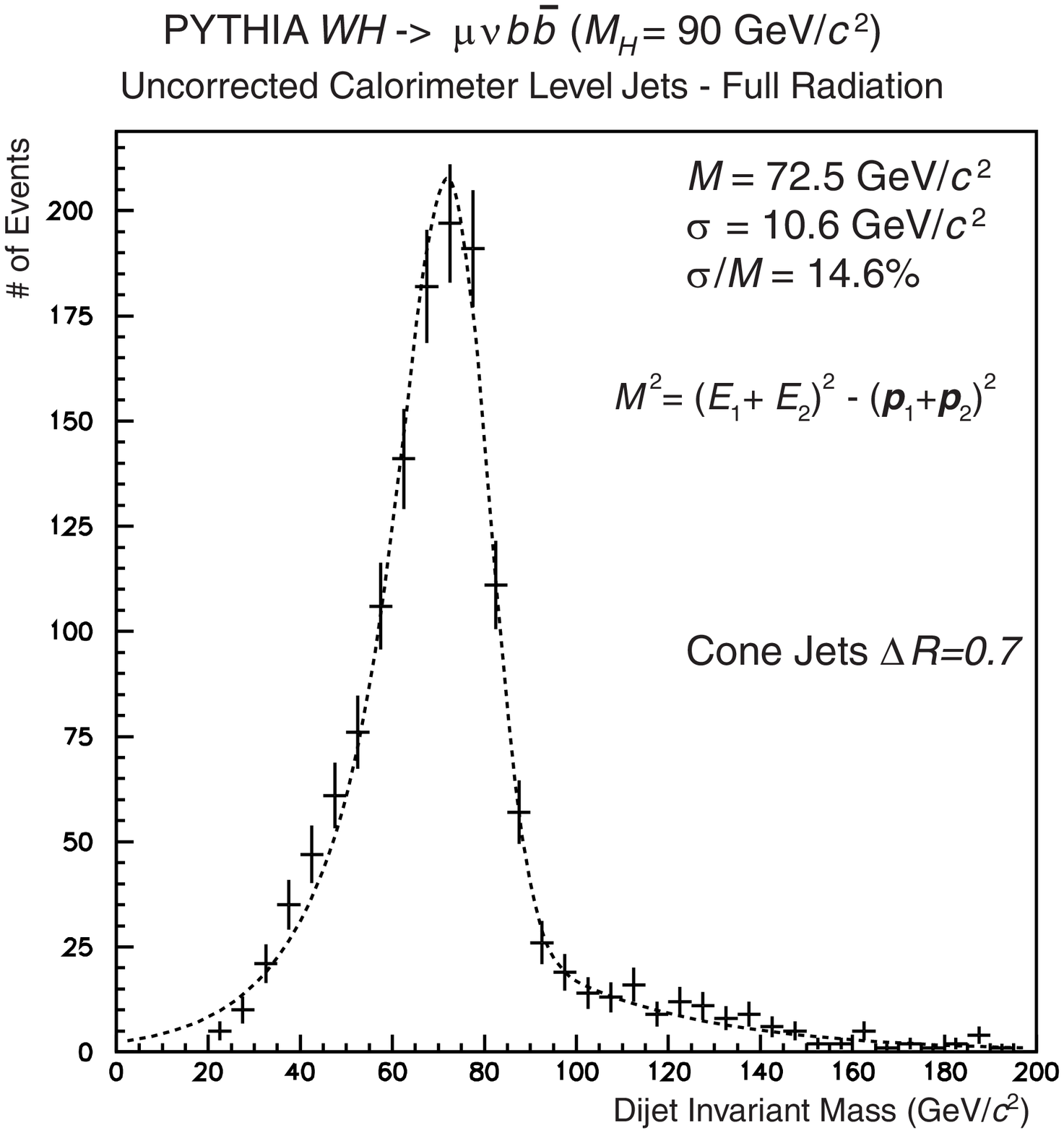}}
    \parbox{3.0in}{\epsfxsize=\hsize\epsffile{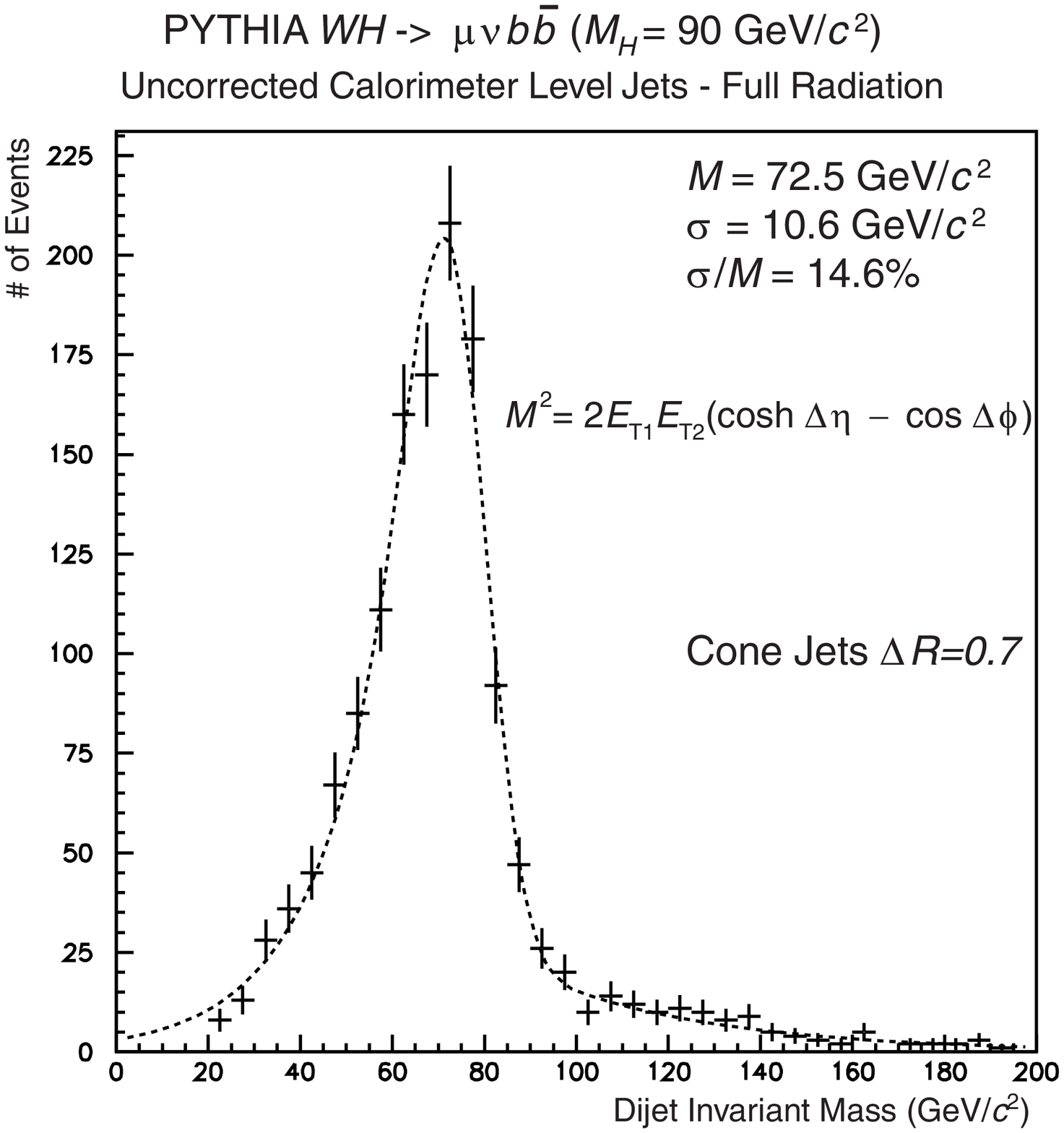}}
  \end{center}
  \caption{
     A comparison of two methods of calculating the dijet mass of the Higgs
     decay. The $M_{4v}$ calculation is on the left, and the $M_{jj}$ is on 
     the right.}
  \label{new_t6_15_c_un_1812}
\end{figure}

\vspace{0.2in}
{\bf Energy Scale Corrections} \\ \nopagebreak

We applied the D\O\ Run~1 jet energy scale corrections to the
jets before forming the invariant mass.  To our 
disappointment, the mass resolution was degraded about 10\% by
these ``corrections''. 
This is perhaps not too surprising since the corrections were
derived for jets at much higher $E_T$ and were also
not optimized for $b$-jets
in any way.  The D\O\ top mass analysis used a different set of jet 
corrections which would probably be more appropriate here, but these
were not available in our analysis package.  In what follows we shall
use uncorrected jet energies.

\vspace{0.2in}
{\bf Cone and $k_T$ Jets Compared} \\ \nopagebreak

One of the hopes of using the $k_T$ algorithm was that the mass resolutions
would be better.  The $k_T$ algorithm should do a good job of recombining
final state radiation with the parent parton jet because it recombines
particles with a small invariant mass. The $k_T$ jet mass resolution shown in
this paper is worse than the resolutions found by the cone jet algorithm
except in the high luminosity case with $D=0.5$.  The mass resolutions from $k_T$ 
are larger than for the cone, as can be seen in Figure~\ref{new} for
$D=1.0$ and 0.5.
The larger $D$ value gives a better resolution and a 
peak mass closer to the true Higgs mass.  
A complete analysis of the 
$k_T$ jet mass resolution, including jet energy corrections, in progress
at D\O.

\begin{figure}
  \begin{center}
    \parbox{3.0in}{\epsfxsize=\hsize\epsffile{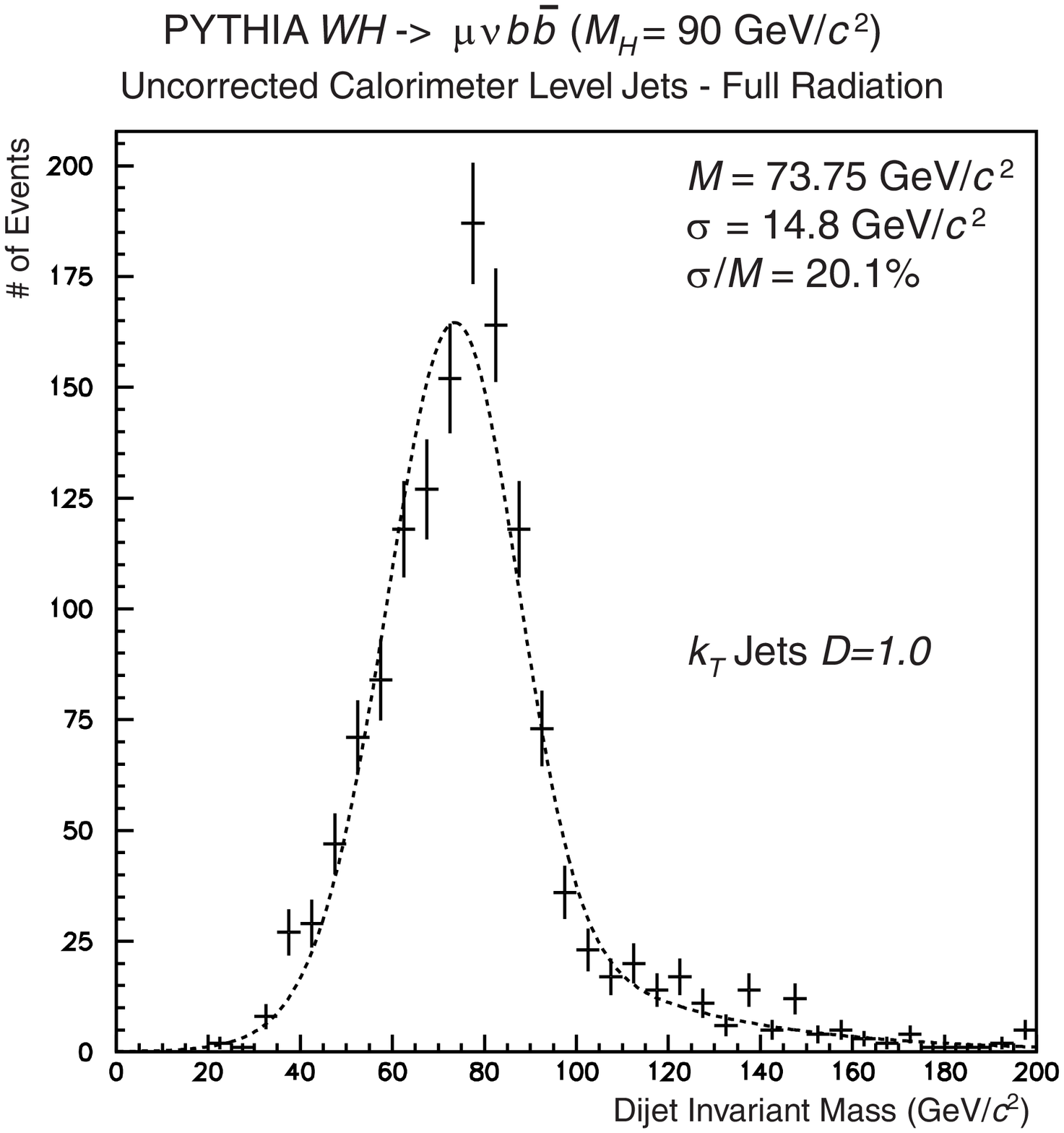}}
    \parbox{3.0in}{\epsfxsize=\hsize\epsffile{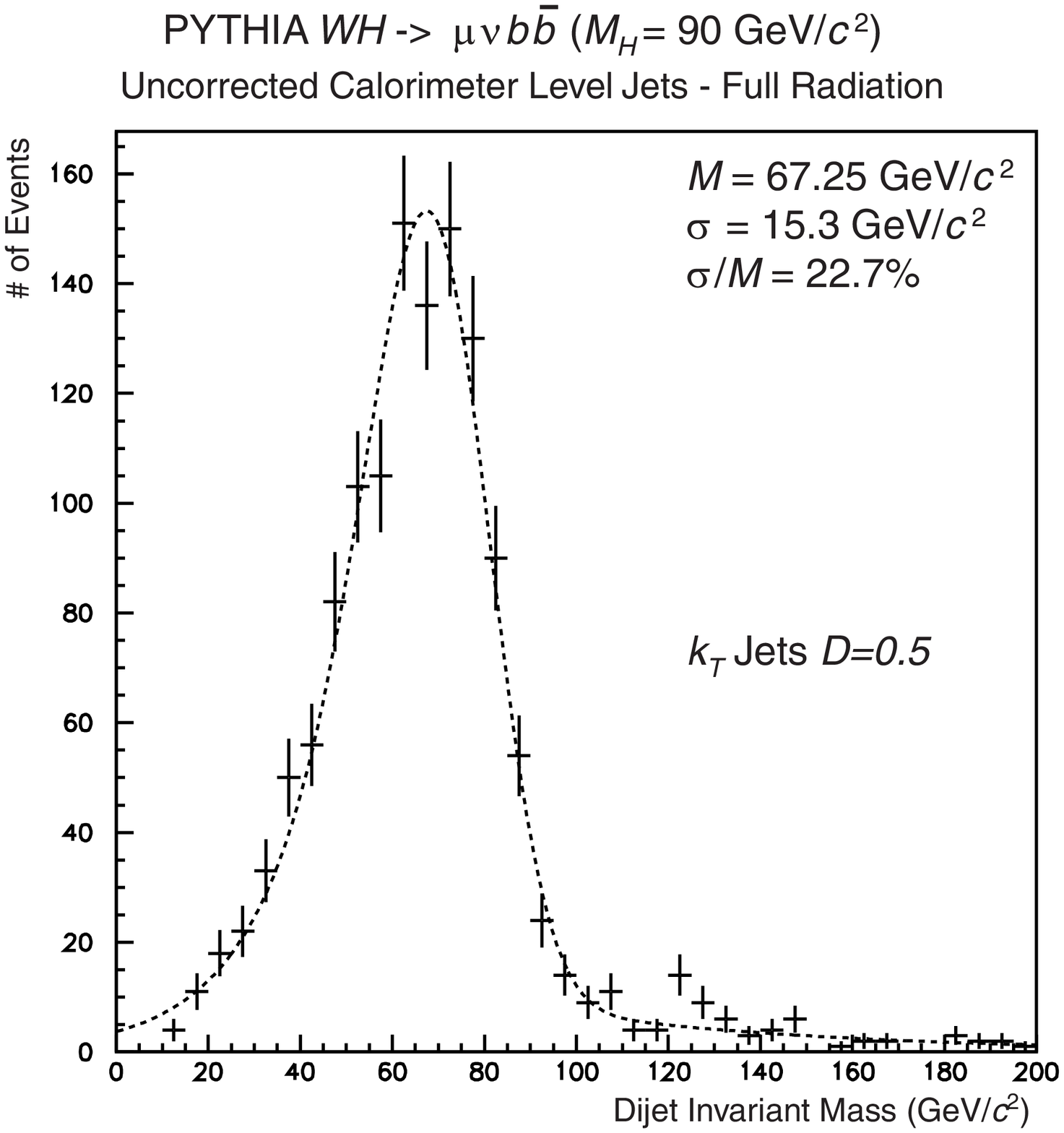}}
  \end{center}
  \caption{The dijet invariant mass calculated using uncorrected calorimeter 
     $k_T$ jets with $D=1.0$ (left hand plot) and $D=0.5$ (right hand plot). 
     Both initial and final state radiation were simulated.}
  \label{new}
\end{figure}

\vspace{0.2in}
{\bf Initial and Final State Radiation} \\ \nopagebreak

\begin{figure}
  \begin{center}
    \parbox{3.0in}{\epsfxsize=\hsize\epsffile{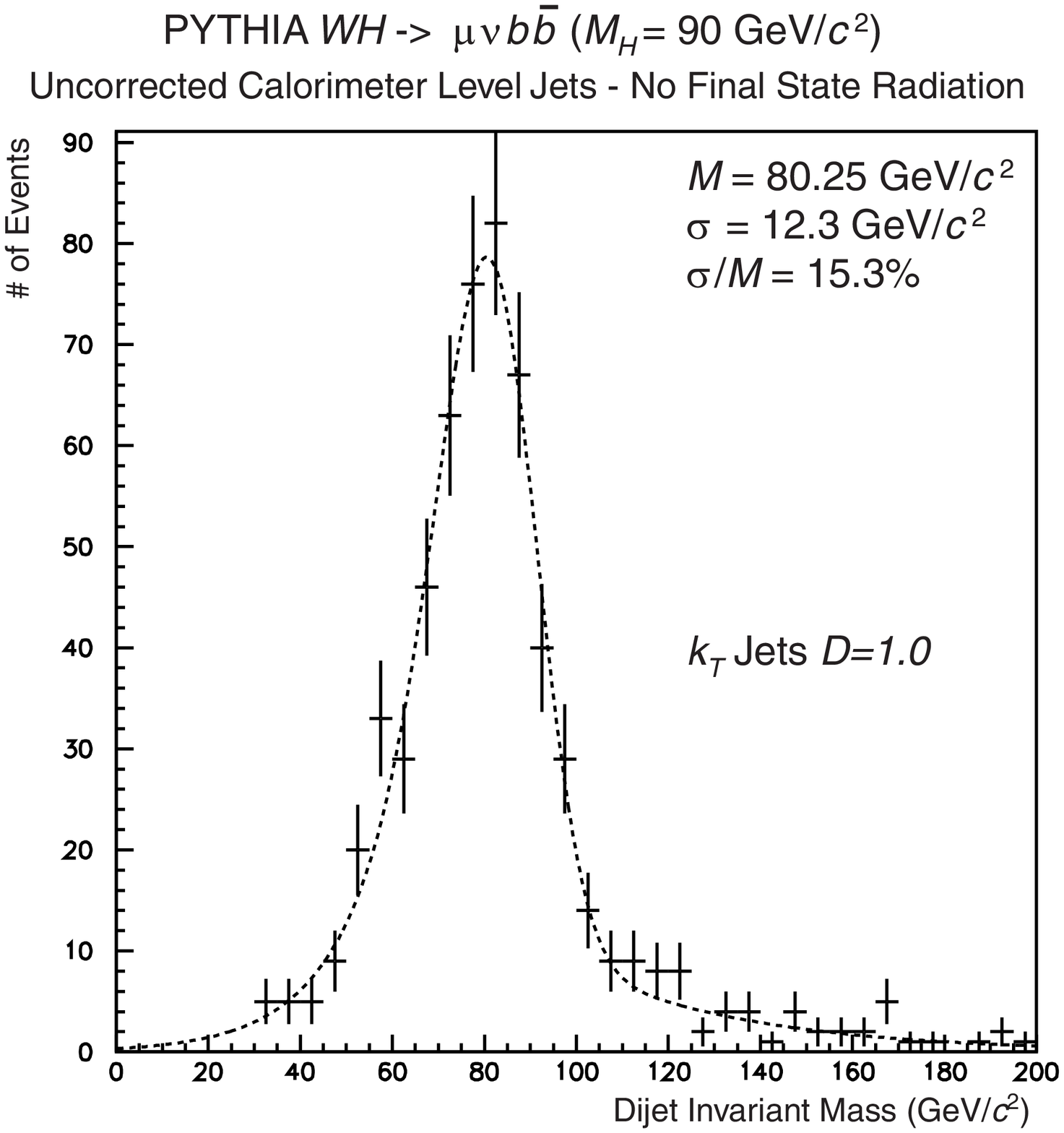}}
    \parbox{3.0in}{\epsfxsize=\hsize\epsffile{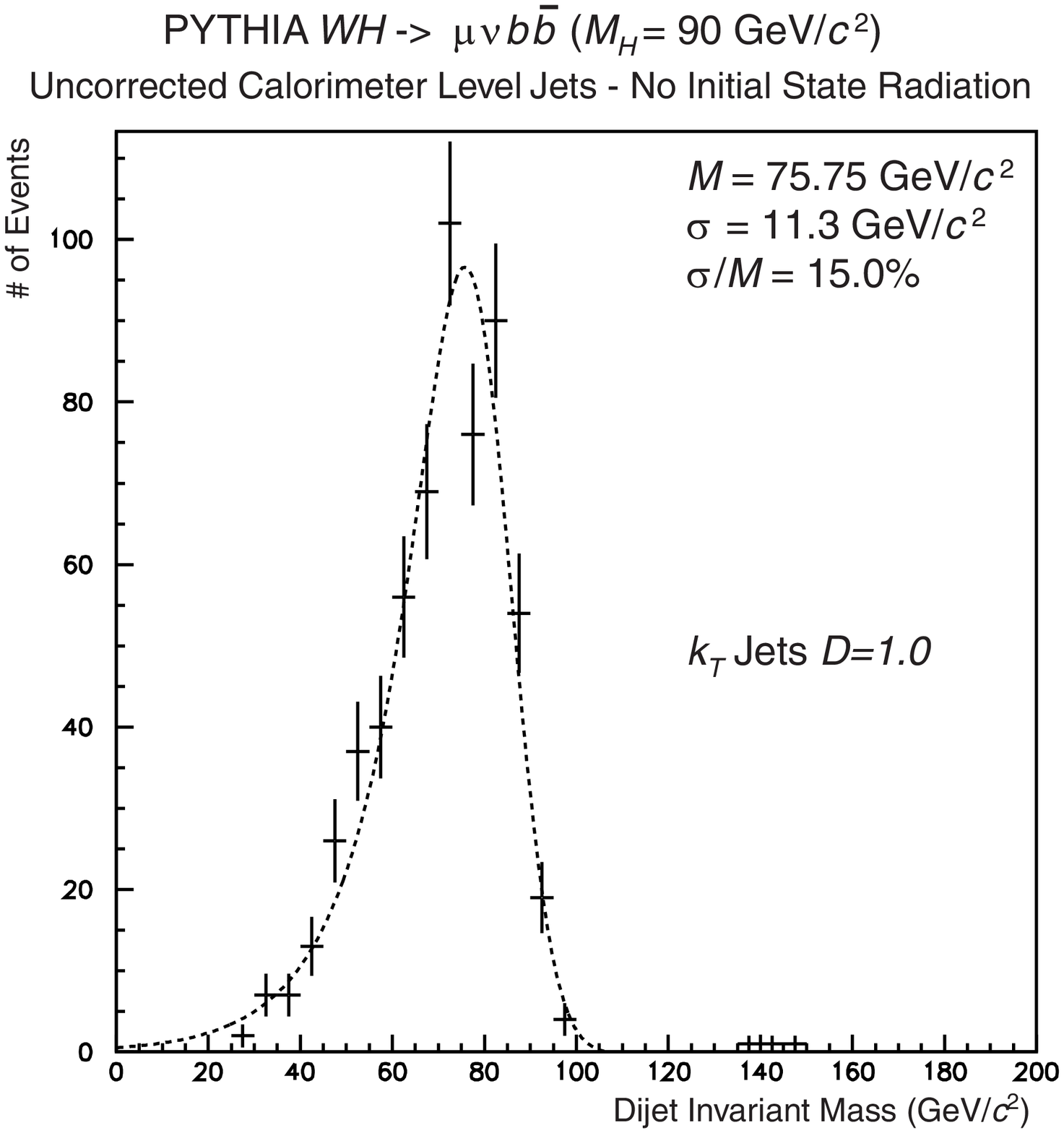}}
  \end{center}
  \caption{The dijet invariant mass calculated using uncorrected calorimeter 
     $k_T$ jets with $D=1.0$. The left hand plot has no final state radiation 
     and the right hand plot has no initial state radiation. These should be 
     compared to the left hand plot of Figure~\ref{new} which has 
     both final and initial state radiation simulated.}
  \label{nofsr}
\end{figure}

Comparison of Figure~\ref{new} with Figure~\ref{nofsr}(left), which was
generated without final state radiation (FSR), shows  
that FSR contributes a low-end tail, 
because the jet algorithms fail to cluster in all the radiation.
Initial State Radiation (ISR) on the other hand, 
contributes a high-end tail,
because the jet algorithms cluster in extra (unwanted) radiation,
as shown by Figure~\ref{nofsr}(right) which was generated without ISR.
Comparison of the figures will also show that ISR and FSR
bias the peak position by a few GeV as well as contributing to
the resolution.

For $D=1.0$, the ISR component of the mass resolution is roughly equivalent
to the FSR piece; for $D=0.5$, FSR contributes relatively 
more (giving a worse resolution overall).
In principle, the $D$ parameter could be optimized to
accept as much FSR as possible while rejecting ISR.

\vspace{0.2in}
{\bf $b$-quark matching} \\ \nopagebreak

To simulate the effect of requiring a $b$-tag within each of the jets
used for the mass reconstruction, we required that the leading jets
be the closest jets to the $b$ and $\overline b$ parton directions.
This significantly improves the resolution, because it cuts out events
where one of the leading jets comes from FSR (and is hence likely to
match a gluon rather than the  $b$ or $\overline b$).  The improvement
is seen both for cone jets and $k_T$ jets in
Figure~\ref{new_t6_15_c_un_11812}.

\begin{figure}
  \begin{center}
    \parbox{3.0in}{\epsfxsize=\hsize\epsffile{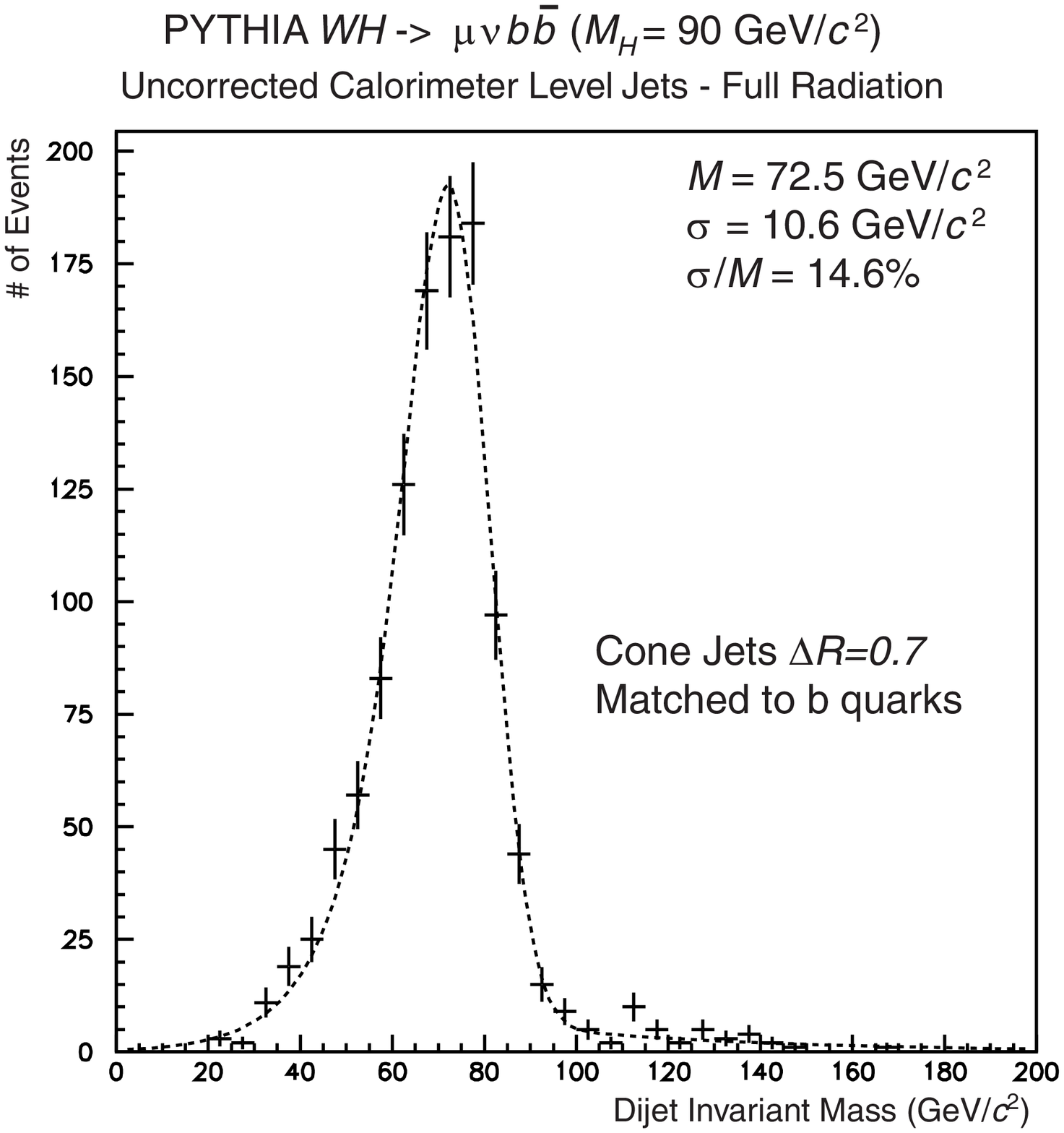}}
    \parbox{3.0in}{\epsfxsize=\hsize\epsffile{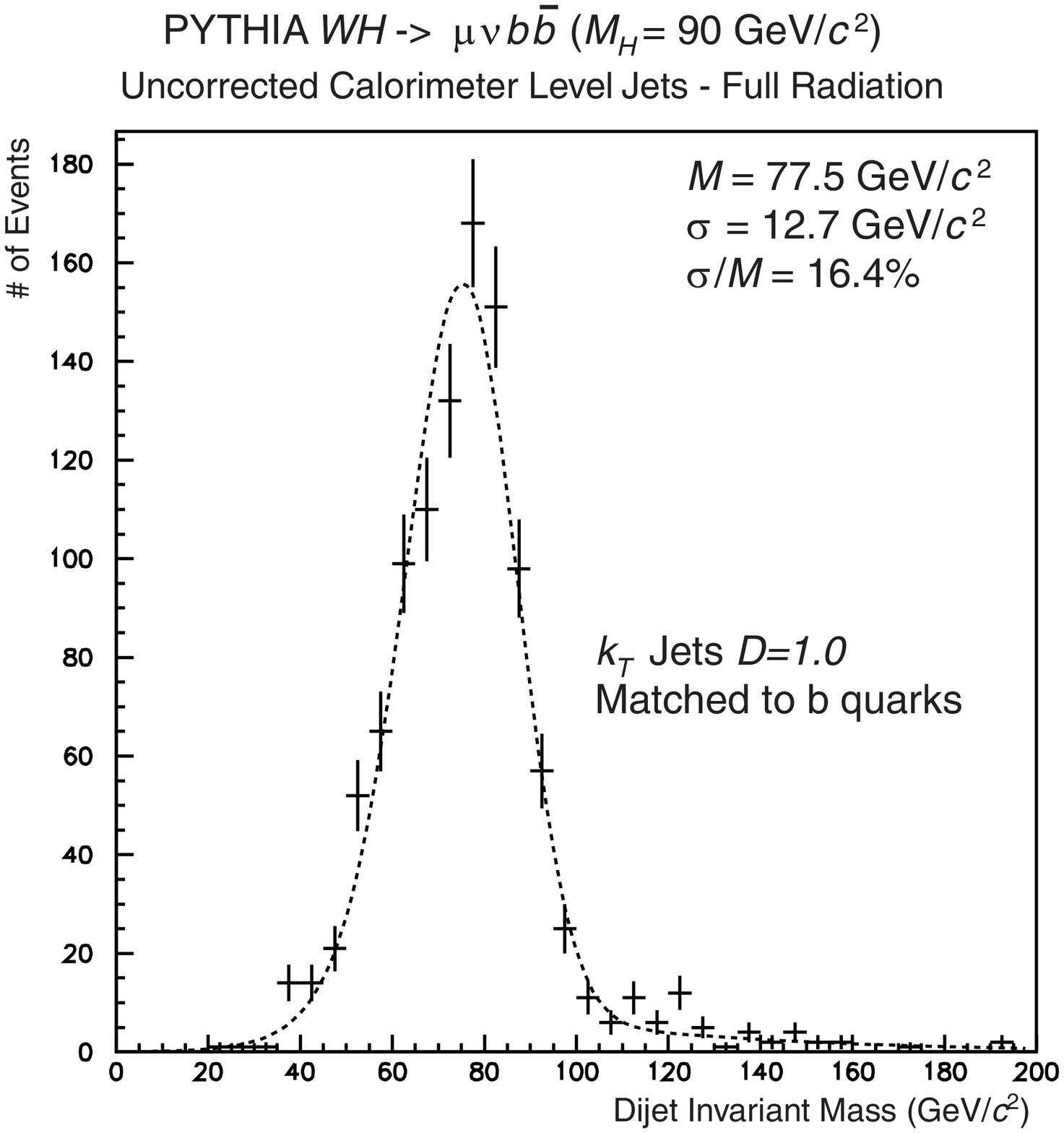}}
  \end{center}
  \caption{The dijet invariant mass calculated using uncorrected calorimeter 
    level $\Delta R=0.7$ (left) and $D=1.0$ (right) $k_T$ jets. The two leading jets 
    are required to be the closest jets to the $b$ and $\bar{b}$ directions.}
  \label{new_t6_15_c_un_11812}
\end{figure}

\vspace{0.2in}
{\bf Luminosity Effects} \\ \nopagebreak

We investigated the effects of multiple interactions per crossing
on the dijet resolution. We overlaid real D\O\ zero-bias events from 
the highest luminosity phase of Run 1 on to our simulated Higgs
events.  A sample with instantaneous luminosity of 
${\cal L} = 2 \times 10^{31}$~cm$^{-2}$s$^{-1}$ was used, which
has approximately 3.3 interactions per crossing on average.
In Run 2b, with 100+ bunches and 132~ns between crossings,
this is equivalent to ${\cal L} = 3 \times 10^{32}$~cm$^{-2}$s$^{-1}$. 
The effect is shown in Figure~\ref{new_zf} and, as can be seen, results in
a significant worsening of the resolution.  We have not attempted to
recover the resolution but it might be possible to do better by imposing 
a threshold cut on calorimeter towers to remove some of the low-energy
pileup.

We also investigated how the mass resolution is affected in overlaid events
as a function of $D$. Figure~\ref{watts-new1} shows the Higgs mass resolutions
for both $D=1.0$ and $D=0.5$.  The smaller size $D$ exhibits less of a 
dependency on the luminosity.  The $k_T$ algorithm's $D$ parameter controls,
approximately, the size of the jet, and smaller jet size will cover less
calorimeter cells.  Multiple interaction background can be approximated as a 
pedestal, distributed almost evenly in all towers of the calorimeter; so the
smaller size $D$ parameter jets are affected less. The effects of multiple
interactions have also been studies in the cone jet algorithms, as shown
in Figure~\ref{watts-new2}.  The loss in resolution is similar for the
$k_T$ $D=1.0$ algorithm.

\begin{figure}
  \begin{center}
    \parbox{3.0in}{\epsfxsize=\hsize\epsffile{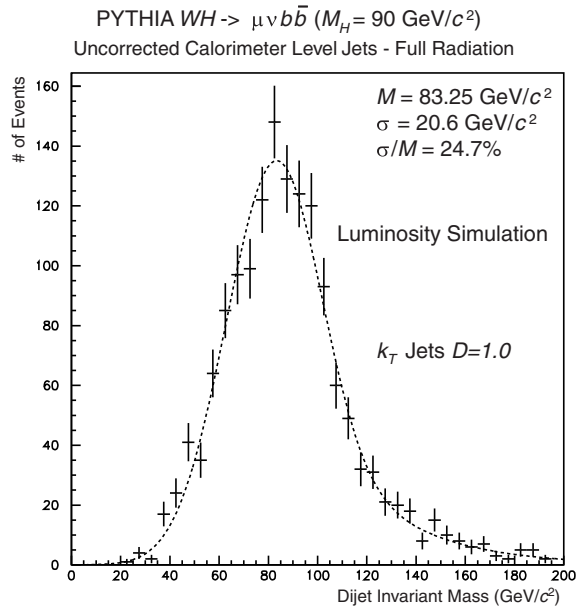}}
  \end{center}
  \caption{The effect of multiple interactions on the calculation of the 
     Higgs mass. This is the dijet mass calculated with calorimeter level 
      $k_T$ jets with $D=1.0$. The simulation contains overlapped data: D\O\ 
      zero-bias data recorded at a luminosity of ${\cal L} = 2 \times
      10^{31}$cm$^{-2}$s$^{-1}$. In Run 2 this will be equivalent to ${\cal L} =
      3 \times 10^{32} $cm$^{-2}$s$^{-1}$ at 132 ns bunch spacing.}
  \label{new_zf}
\end{figure}

\begin{figure}
  \begin{center}
    \parbox{3.0in}{\epsfxsize=3in\epsffile{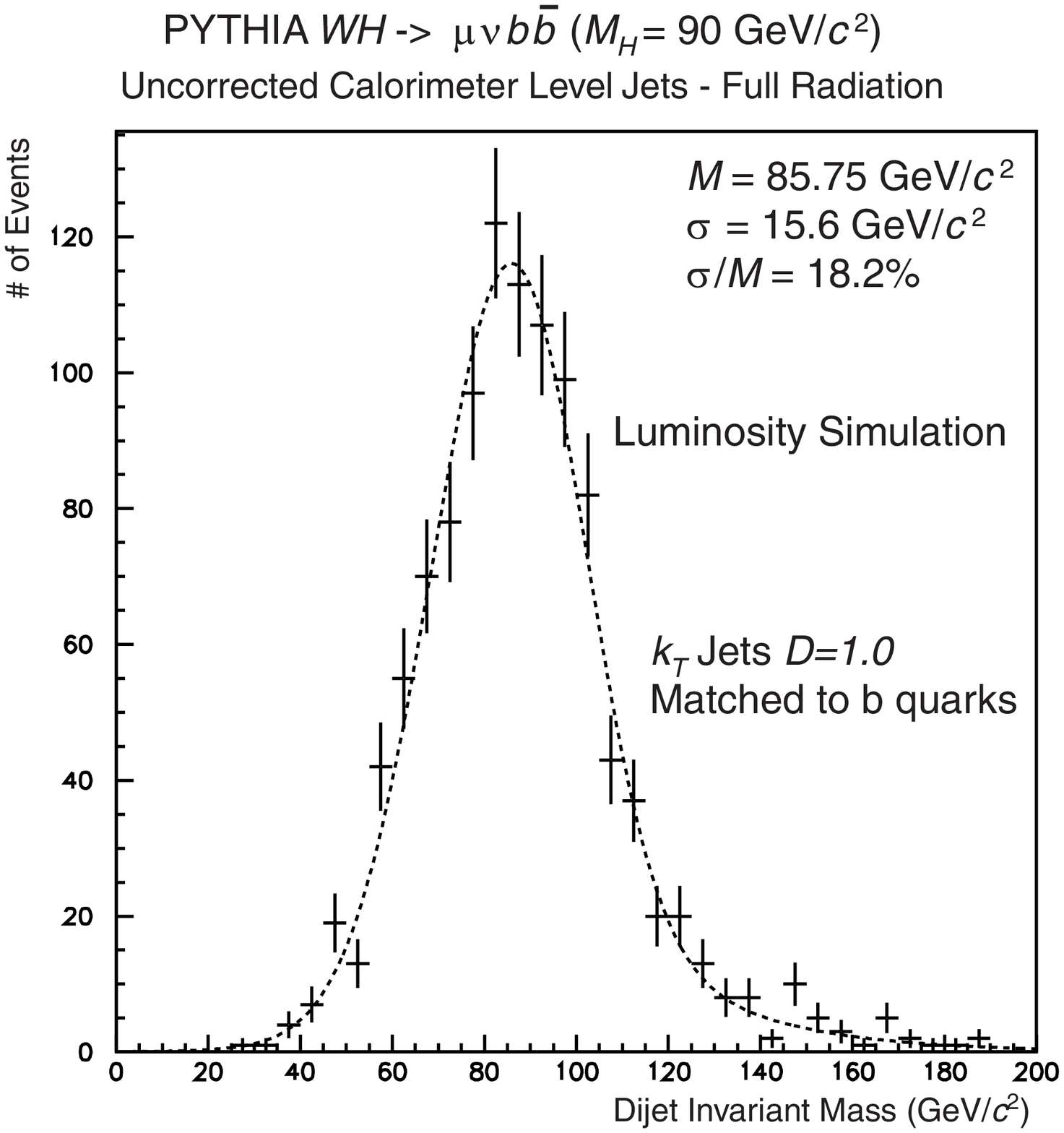}}
    \parbox{3.0in}{\epsfxsize=3in\epsffile{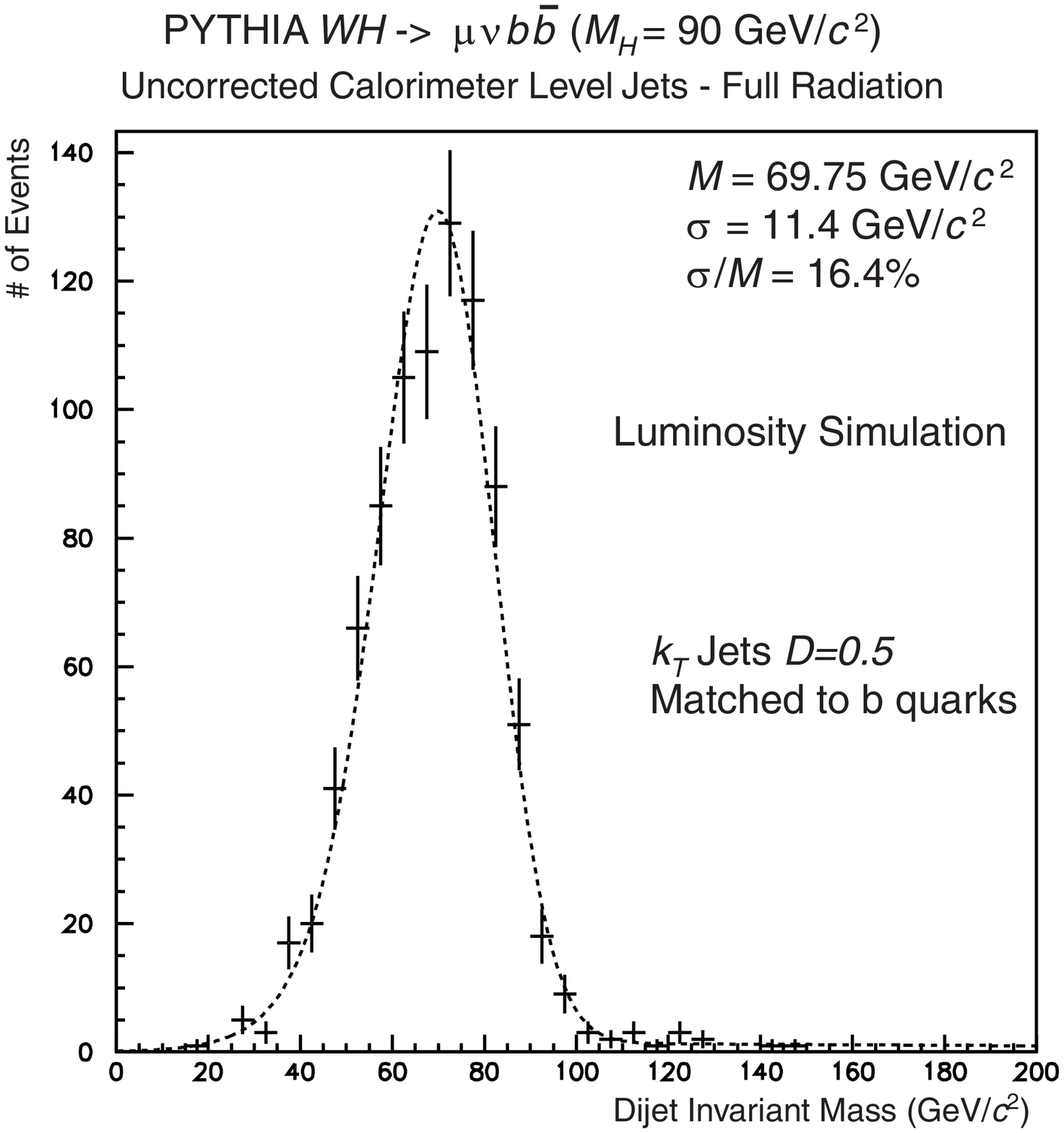}}
  \end{center}
  \caption{The effect of varying the $k_T$ algorithm's $D$ parameter on Higgs 
   mass resolutions in simulated multiple interaction data. The dijet mass of 
   the Higgs is calculated at calorimeter level in both plots, and the leading
   two jets are required to match the b quark direction.}
  \label{watts-new1}
\end{figure}

\begin{figure}
  \begin{center}
    \parbox{3.0in}{\epsfxsize=3in\epsffile{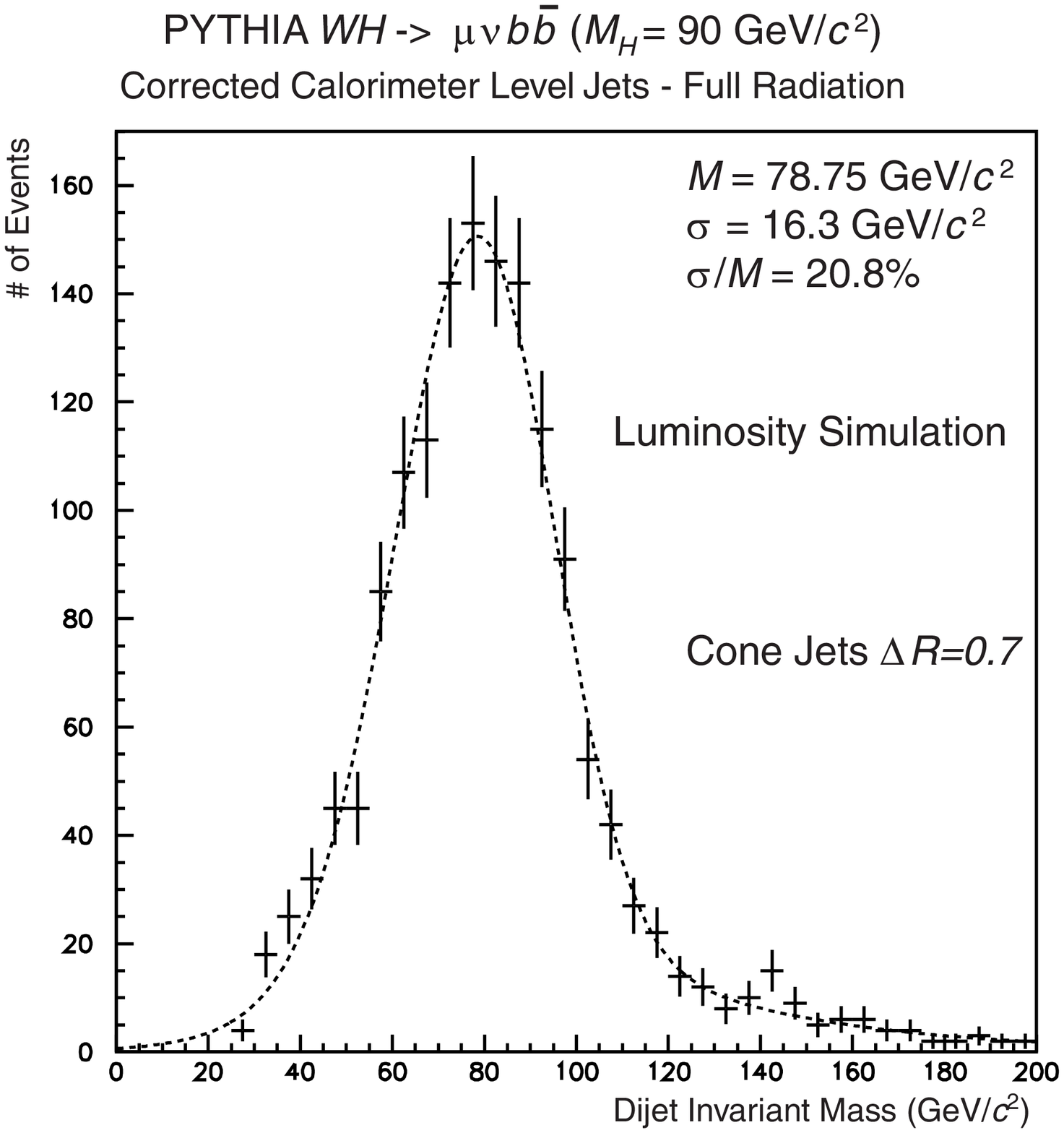}}
    \parbox{3.0in}{\epsfxsize=3in\epsffile{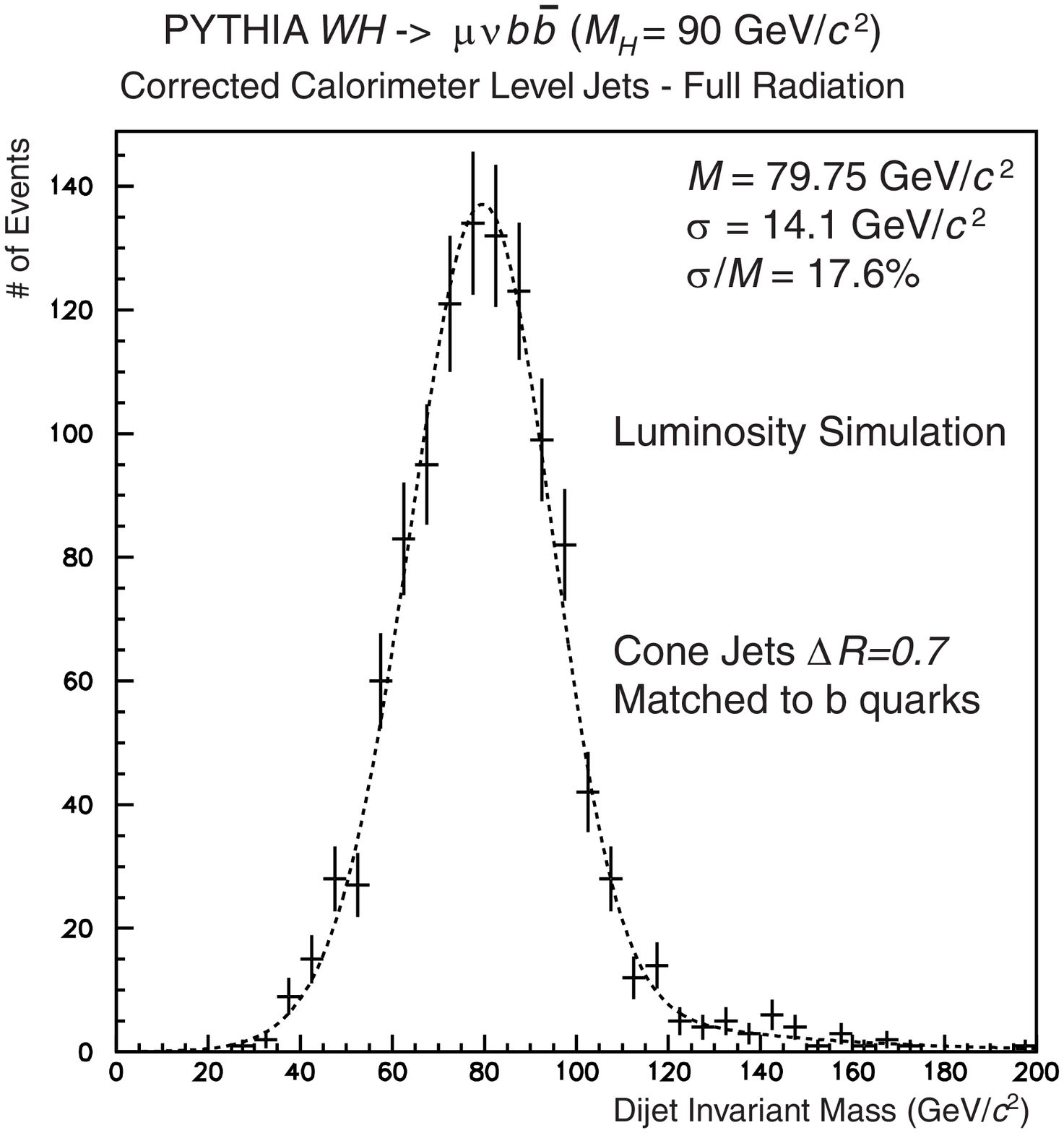}}
  \end{center}
  \caption{The effect of multiple interactions on the calculation of the Higgs
     mass using cone jets. This is the dijet mass calculated with calorimeter 
     level cone jets in overlapped data. The right hand plot requires the 
     leading jets match the direction of a b-quark jet; the left hand plot 
     does not. The jets have had calibration corrections applied.}
  \label{watts-new2}
\end{figure}

\vspace{0.2in}
{\bf Conclusions} \\ \nopagebreak

We have investigated many effects contributing to the invariant mass
resolution.  We find:

\begin{itemize}
\item The $k_T$ algorithm is no silver bullet (at least as implemented
in the D\O\ Run 1 software). In fact it is hard to do much better than a
standard $R=0.7$ cone algorithm, especially if the jets are required
to match the $b$ quark directions.

\item Jet energy corrections will have to be optimized for these 
relatively low-$E_T$ $b$-quark jets. 

\item Maintaining good resolution in the high luminosity environment of 
Run~2 and beyond will be a challenge.  Pileup from minimum bias events
can seriously degrade the energy resolution.  
\end{itemize}

Table~\ref{massconc} summarizes
the resolution and peak position of all the results.
\begin{table}
 \caption{A summary of the mass resolution results.  For each data
   sample, the cone and $k_T$ peak value FWHM are listed for
   a Higgs mass of 90~GeV/c$^2$.}
\label{massconc}
\begin{center}
   \begin{tabular}{|c|c|c|c|c|c|c|} 
\noalign{\vskip-6pt}  
 & \multicolumn{2}{c|}{Cone} &  \multicolumn{2}{c|}{$k_T$ $D=1.0$}
            &  \multicolumn{2}{c|}{$k_T$ $D=0.5$} \\ 
\tableline   
 Data Sample & Peak & Width 
             & Peak & Width 
             & Peak & Width \\
             & (GeV/$c^2$) &       
             & (GeV/$c^2$) &       
             & (GeV/$c^2$) &       \\ 
\tableline   
  Uncorr., no FSR & & & 80.2 & 15.3\% &  & \\
  Uncorr., no ISR & & & 75.7 & 15.0\% &  & \\
  Uncorr., full rad. 
     & 72.5 & 14.6\% & 73.7 & 20.1\% & 67.25 & 22.7\% \\
  Uncorr., full rad., $b$ matching
     & 72.5 & 14.6\% & 77.5 & 16.5\% &       &       \\
  Mult. int., uncorr., full rad.
     & 78.7 & 20.8\% & 83.2 & 24.7\% &       &       \\
  Mult. int., uncorr., full rad., $b$ matching
     & 79.7 & 17.6\% & 85.7 & 18.2\% & 69.75 & 16.4\% \\
  \end{tabular}
\end{center}
\end{table}

    \subsubsection{Study of $b$-jet Tagging} 
\vspace{0.1in}
\small
\begin{center}
{\it M. Roco}
\end{center}
\normalsize \nopagebreak

To identify a jet associated with a $b$ quark, a technique, referred to as
{\em '$b$ tagging'}, is performed by reconstructing the decay vertex of a
long-lived $b$ hadron within the jet.  The algorithm uses
the precise track reconstruction of the silicon tracker to identify
secondary vertices which are significantly displaced from the primary
interaction vertex.  The separation of a secondary vertex from the primary
$p \overline p$ interaction vertex depends on the spatial resolution of the
individual tracks associated to the secondary vertex and on the primary vertex
resolution.

This report presents the results of a study performed to estimate the
$b$-tagging efficiency in $\ttbar$ and Higgs events by finding
secondary vertices from $b$-quark decays.  The study is based on the
MCFAST~\cite{mcfast} package, a useful tool designed to provide a fast
simulation framework for detector design studies.  It has been
interfaced with either PYTHIA or ISAJET event generators for the
parton fragmentation and hadronization.  It performs parameterized
tracking where for each generated track, a covariance matrix is
assembled, which represents all the material and detector planes
traversed by the ideal track.  A reconstructed track is produced by
smearing the generated track parameters according to this covariance
matrix.

\vspace{0.2in}
{\bf Detector Simulation} \\ \nopagebreak

The implementation of the Run 2 D\O \ tracking detector geometry includes a
solenoid which produces a 2 Tesla uniform magnetic field along the $z$-axis.
Within the magnetic field of the solenoid are the tracking detectors for
charged particle identification.
Drift chambers are used to simulate the scintillating central fiber tracker.
A six barrel 4-layer geometry is used for the silicon microstrip tracker (SMT).
For this study layers 1 and 3 of the four innermost barrels have 90$^o$ 
stereo angle.

The outermost barrels have single-sided detectors and therefore only give
$r-\phi$ information.  Layers 2 and 4 of all six barrels have 2$^o$ stereo.
The barrel segments are separated by gaps containing F disks.
The barrels are symmetric around $z=0$ and extend from $-38$ cm to $+38$ cm.

The beam vertex is centered at the origin with a smearing of 30 $\mu m$ in
both $x$ and $y$ and 25 cm in $z$.  In this study only one interaction
per beam crossing is considered.

\vspace{0.2in}
{\bf Tagging Algorithm} \\ \nopagebreak

Given the relatively long lifetime of the $b$ quark, one can select jets
arising from $b$ quark hadronization by 'tagging' long-lived hadrons within
jets.  The decay of a long-lived hadron produces several charged tracks
emanating from a point, or secondary vertex, separated from the primary
$p \overline p$ interaction point.

Track displacement with respect to the primary vertex is measured
using the perpendicular distance at the point of closest approach, or
impact parameter $d_0$ shown in Fig.~\ref{f:ip2d}.  The sign of $d_0$
is given by the location of the beamline in the transverse plane,
relative to the particles' trajectory in this plane.  For positively
charged tracks, the sign of $d_0$ is positive if the location of the
beamline is outside the circle.  The impact parameter resolution
includes contributions coming from the intrinsic detector resolution,
multiple scattering, and uncertainties from the primary vertex
reconstruction.  The impact parameter significance, $\left | d0\right
|/ \sigma (d0)$, is a gaussian of width 1 with non-gaussian tails from
heavy flavor quark content and and tracking errors.  Fig.~\ref{f:ip}
shows the impact parameter and significance distributions for tracks
originating from the primary vertex (top plots) and those tracks
associated with the $b$ jets (bottom plots).  Tracks from the $b$
hadron decays populate the tails of the distributions.  Most particles
in an event originate from the primary vertex.

\begin{figure}
  \begin{center}
    \parbox{5.0in}{\epsfxsize=\hsize\epsffile{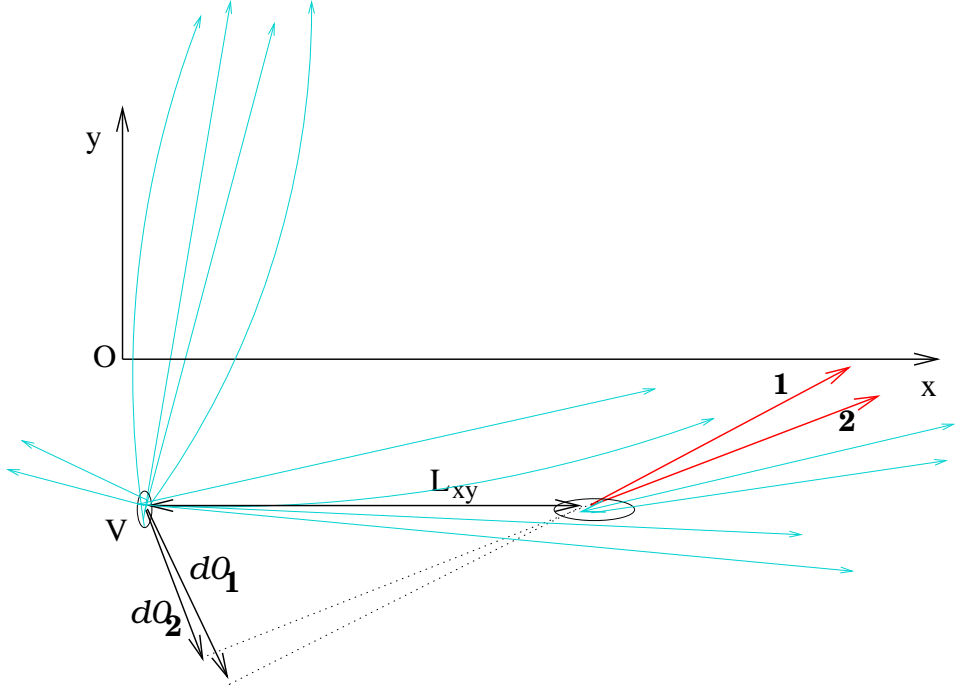}}
  \end{center}
  \caption{The track displacement relative to the primary vertex is measured
	using the perpendicular distance of closest approach, or impact
	parameter $d0$.  The transverse decay length $L_{xy}$ is the
	distance between the primary vertex and the secondary vertex.}
  \label{f:ip2d}
\end{figure}

\begin{figure}
  \begin{center}
    \parbox{5.0in}{\epsfxsize=\hsize\epsffile{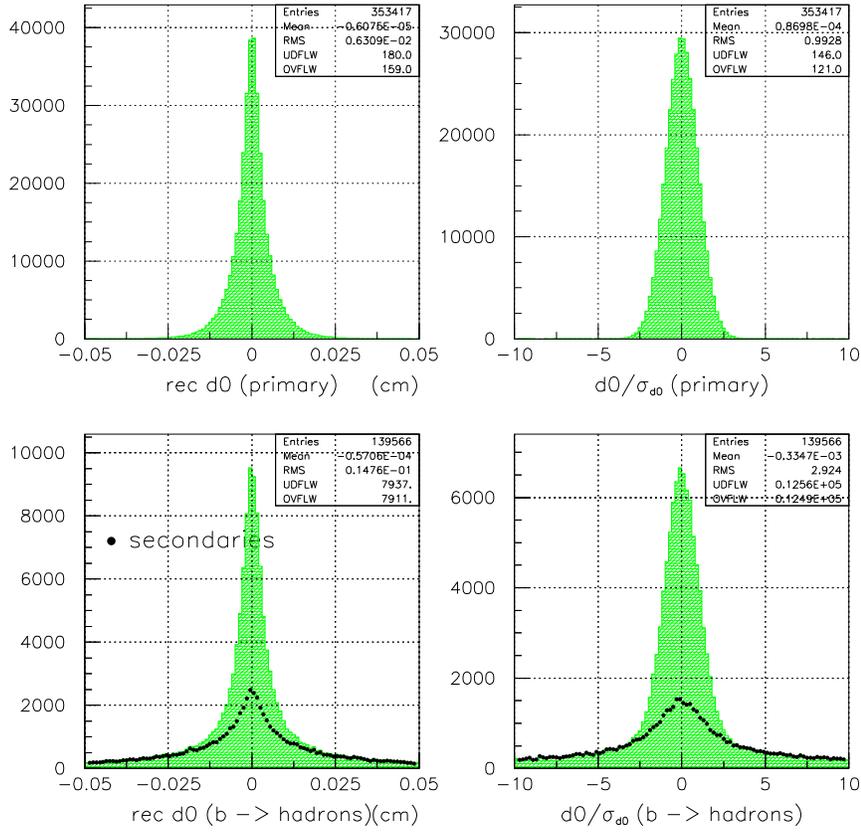}}
  \end{center}
  \caption{The impact parameter and significance distributions of tracks
        originating from the primary vertex are shown in the top plots.
        In comparison, tracks associated with the $b$ jets are shown in
        the bottom plots.  Tracks from the $b$ hadron decays, shown as the
	dots, populate the tails of the distributions.}
  \label{f:ip}
\end{figure}

The algorithm loops over tracks associated to a jet of cone radius $r
> 0.5$ and $p_T >$ 15 GeV/c.   A tight selection criterion is applied
in the first pass.  If there are less than three tracks associated to
the reconstructed decay vertex, a looser selection is applied.  The
selection is based on the track impact parameter significance.
Fig.~\ref{f:rate} shows the efficiency of the track selection using a
sample of ISAJET $\ttbar$ events.  It gives the fraction of jets,
associated with either a $W$ or a $b$ hadron, which have at least
three significantly displaced tracks (solid curves) for increasing
$\left |d0\right |/ \sigma (d0)$ cuts.  The dashed curves show the
corresponding efficiency if the requirement on the minimum number of
tracks is relaxed to two.  These plots show that a tight $+$ loose
tagging scheme requiring at least two significantly displaced tracks
keeps about 55\% of the $b$ jets in the $\ttbar$ sample, while less
than 5\% of the jets associated with $W$ boson production ($W \to
\qq$), mainly from charm hadron decays, are retained.

\begin{figure}
  \begin{center}
    \parbox{5.0in}{\epsfxsize=\hsize\epsffile{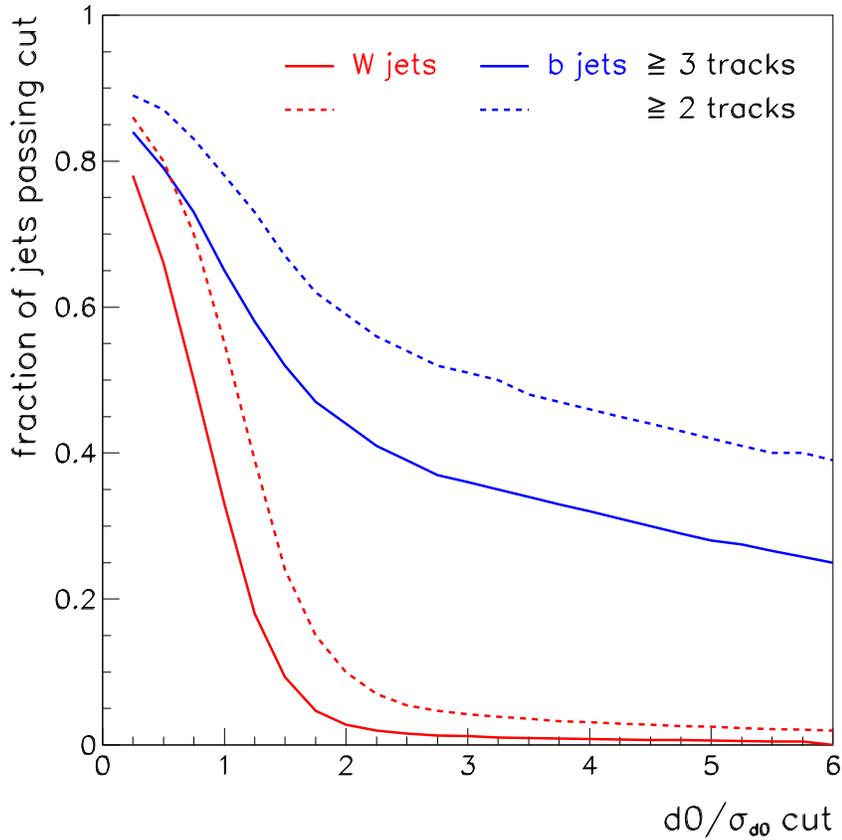}}
  \end{center}
  \caption{Using an ISAJET $\ttbar$ sample,
	the plot shows the fraction of jets, associated with either a
	$W$ or a $b$ hadron, which have two (dashed curves) or three
	(solid curves) significantly displaced tracks for increasing
	$\left |d0\right |/ \sigma (d0)$ cuts.}
  \label{f:rate}
\end{figure}

The selection criteria for the tight and loose algorithms, PASS1 and PASS2,
are enumerated below:

\begin{itemize}
\item[$\bullet$] PASS1 cuts:
\begin{itemize}
\item [-]track $p_T \geq$ 0.5 GeV/c with at least 4 (axial + stereo)
SMT hits
\item [-]require $d0 < 0.15$ cm to remove tracks consistent with $\gamma$
conversions and $K_S$ and $\Lambda$ decays originating from the primary vertex
\item [-]impact parameter significance, $d0/\sigma (d0) = S_{d0} \geq 2.5$
\item [-]give list of at least 3 candidate tracks to vertex finder
\item [-]require at least 3 tracks associated to the vertex
\end{itemize}
\item[$\bullet$] PASS2 cuts:
\begin{itemize}
\item [-]track $p_T \geq$ 1.5 GeV/c
\item [-]impact parameter significance, $d0/\sigma (d0) = S_{d0} > 3.0$
\item [-]give list of at least 2 candidate tracks to vertex finder
\item [-]require at least 2 tracks associated to the vertex
\end{itemize}
\item[$\bullet$] Secondary Vertex:
\begin{itemize}
\item [-]candidate tracks are passed to a 3D constrained vertex fit
\item [-]find a set of pairwise compatible tracks and
\item [-]reject tracks whose contribution to the overall $\chi^2$
of the fit $>$ 16
\item [-]$\chi^2$ per degree of freedom of vertex fit $< 10$
\item [-]calculate the 2D decay length, $L_{xy}$, and its error $\sigma (L_{xy})$
\item [-]sign of $L_{xy}$ = dot product of $L_{xy}$ and the jet direction
\item [-]require $\frac{\left | L_{xy} \right |}{\sigma (L_{xy})} > 3$ and $\left | L_{xy} \right | < 2.5$ cm
\end{itemize}
\end{itemize}

Candidate tracks are constrained to a common vertex to determine the
transverse decay length $L_{xy}$.  A vertex position is calculated and tracks
with a large contribution to the overall fit $\chi^2$ are excluded.  There
should be at least two tracks used in the vertex fit.
The reconstructed vertex is required to be within the innermost
layer of the SMT, with $L_{xy} <$ 2.5 cm, and that the decay length significance
be $\left |L_{xy}\right |/\sigma (L_{xy}) >$ 3.

\begin{figure}
  \begin{center}
    \parbox{5.0in}{\epsfxsize=\hsize\epsffile{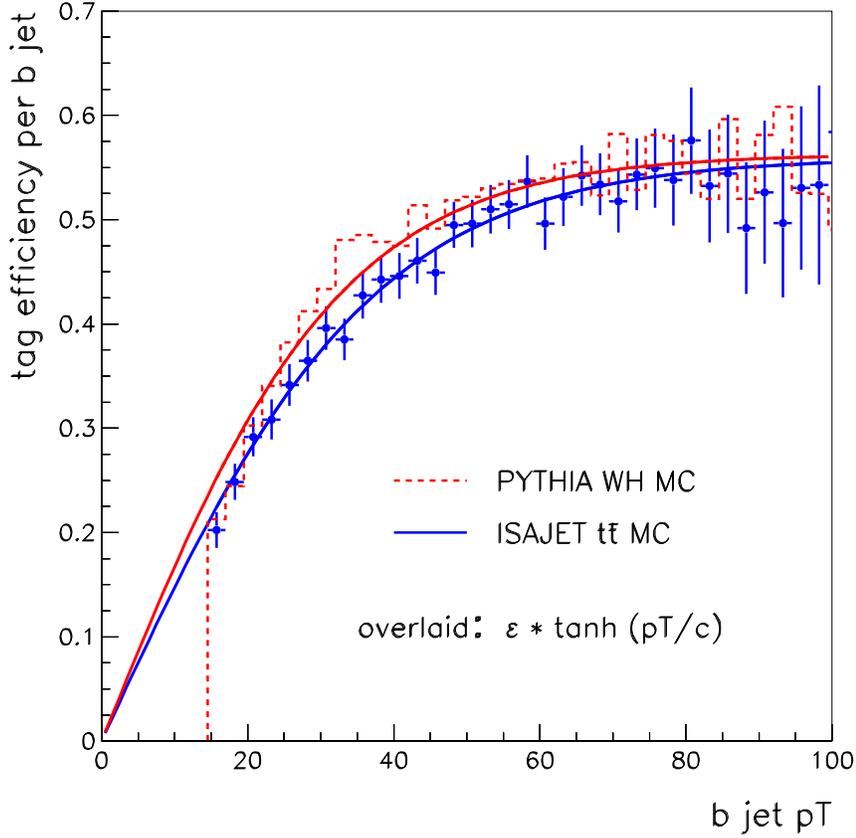}}
  \end{center}
  \caption{Tag efficiency per jet as a function of the $b$-jet $p_T$
	for two different physics samples.  The rates are parameterized
	using the functional form $\epsilon \ \tanh \ (p_T/c)$.}
  \label{f:tagged}
\end{figure}

\begin{figure}
  \begin{center}
    \parbox{5.0in}{\epsfxsize=\hsize\epsffile{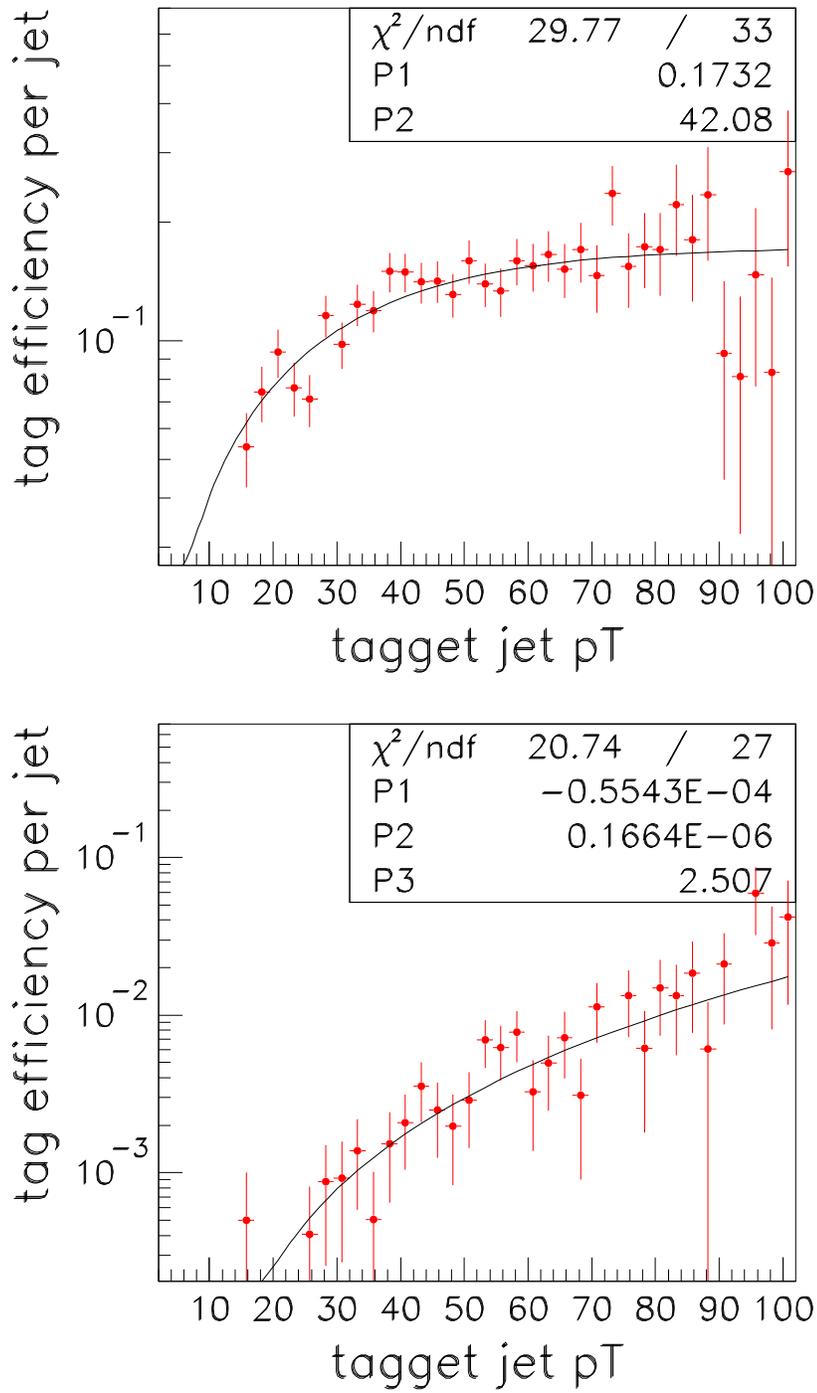}}
  \end{center}
  \caption{Mistag efficiency per jet as a function the jet $p_T$ for
  	$c$-jets (top plot) and $u,d,s,g$ jets (bottom plot).}
  \label{f:mistags}
\end{figure}

\vspace{0.2in}
{\bf Tag Rates} \\ \nopagebreak

The tag rates obtained after applying the selection criteria enumerated above
are shown in Figs.~\ref{f:tagged} and ~\ref{f:mistags}.  The plot in
Fig.~\ref{f:tagged} shows a comparison of the $b$-tag rates for two
different samples, a $\ttbar$ sample generated using ISAJET shown as
the points and a Higgs sample ($\qq \to WH$) generated using PYTHIA
shown as the dashed histogram.  The tag rates are parameterized in terms of
the $b$-jet $p_T$ using the functional form $\epsilon \ \tanh \ (p_T/c)$,
where $\epsilon = 0.57 \pm 0.012$ and  $c=36.05 \pm 1.42$ are the constants
obtained from the average of the two fits.  This plot also shows that
the difference between the rates obtained assuming different fragmentation
and hadronization schemes is small.

Fig. ~\ref{f:mistags} shows the efficiency of mistagging light quark
or gluon jets as $b$-jets.  The tag rate for $c$-jets, shown in the
top plot, is roughly one third of the $b$-jet tag rate. Tag rates for
$u,d,s$ and $g$ jets, shown in the bottom plot, are less than 1\%
for $p_T < 70$ GeV.  However, this still presents a serious problem
when dealing with background QCD multi-jet events with cross sections
which are orders of magnitude larger than the $\ttbar$ or the
expected Higgs production cross sections.

  \subsection{Low-mass Standard Model Higgs Bosons: 90--130 GeV}

    \subsubsection{$\ell\nu\bb$ Channel}		\small
\begin{center}
{\it E. Barberis, 
     W. Bokhari, 
     P. Bhat, 
     R. Gilmartin, 
     H. Prosper, 
     W. Yao}\\
\end{center}
\normalsize \nopagebreak

Previous studies \cite{tev2k,snowmass96,tev33} have indicated that the
$WH$ channel, with $H\rightarrow b\bar{b}$ and $W\rightarrow l\nu$ is
the single most powerful channel to detect a Higgs signal over the
large expected background.  We present here three different analyses
of this channel, each an extension of the previous.  The initial
analysis uses events produced by a well--established Run 1 CDF
detector simulation, QFL$^\prime$ and sequential selection requirements on a
single variable at a time.  The second analysis is similar to the
first, but uses the SHW detector simulation developed for this
workshop.  The final analysis also uses SHW for detector simulation,
but neural network techniques are employed to maximize the
sensitivity.  The good agreement between the first and second analyses
indicates the quality of the SHW simulation, and the third analysis
illustrates potential improvements to the sensitivity of this channel.
      
       \vskip 4mm 
       \centerline{ \large \it 1a. $\ell\nu\bb$, QFL$^\prime$ Simulation
           and Traditional Techniques} \nopagebreak
       \vskip 4mm 
       
This section presents study of $WH$ production, where $W\to\ell\nu$
using a proven Run 1 CDF detector simulation for the signal and background
estimation.

\vspace{0.2in}
{\bf Event Samples} \\ \nopagebreak

Throughout this section, we use Monte Carlo events generated with
Pythia event generator and simulated with current CDF fast detector
simulation, QFL$^\prime$.  This would allow us to have a realistic
calorimeter simulation, and good central lepton identification ($e$,
$\mu$ and isolated tracks from $\tau$ decay).  We do not attempt to
simulate the leptons in the region $1<|\eta|<2$,\footnote{This region
was outside the Run~1 lepton acceptance, but it is being instrumented
for Run~2.}  but rather, simply rescale the number of identified
central leptons by the ratio of leptons between $|\eta|<1.0$ and
$1.0<|\eta|<2.0$ at the generator level.

For the $WH$ signal we generated 30K Pythia $WH$ Monte Carlo events at
each mass, in which the $W$ decays freely and Higgs is forced to decay
into $\bb$.
 
Background events come predominantly from the direct production of $W$
bosons in association with heavy quarks ($W\bb$ ), $\ttbar$, single
top production ($W^*\rightarrow tb$, $gW\rightarrow tb$)~\cite{xsec}
and diboson $WZ$.  The requirement of two $b$ jets, along with the 
$\met$ and isolated lepton requirements, remove essentially all generic 
multijet events.

To estimate the $W$ plus heavy quark backgrounds, we use the Herwig
Monte Carlo program to calculate the fraction of $W$+jet events that
contain heavy quarks and the corresponding tagging efficiencies. The
number of background events in a given jet multiplicity bin is then
obtained as the product of heavy flavor fraction and tagging
efficiency times the observed number of $W$+jet events per 1~fb$^{-1}$.

The top quark contributions ($\ttbar$ and single top) are estimated
using Pythia Monte Carlo and a theoretical calculation of
$\sigma_{\ttbar}=6.5$ pb, $\sigma_{W^*\rightarrow tb}=0.88 $ pb, and
$\sigma_{gW\rightarrow tb} = 2.4 $ pb for a top quark mass of 175
GeV/$c^2$.

The $WZ$ background is calculated using PYTHIA and a leading order
cross section of 3.2 pb.

For the $b$-tagging, we start with the jets observed in the
calorimeter ($E_T> 10$ GeV and $|\eta|<2$) that are matched in a cone
of $\Delta R=0.4$ to the $b$ quark at the four--vector level.  The
$b$-tagging efficiency is parametrized as a function of jet $E_T$
using Run 1 data as described below.
 
\vspace{0.2in}
{\bf $b$-tagging Efficiency}  \\ \nopagebreak

The ability to tag $b$ jets with high efficiency and low mistagging
rate is vital for searching for the decay of $H\rightarrow \bb$. The
technique for $b$-tagging has been well established in CDF using the
Silicon Vertex tagger (SECVTX) and the Soft-Lepton Tagger (SLT), but
will be much improved in the Run~2 Tevatron
detectors~\cite{cdfii,d0upg}.  The full 3-D silicon tracker will
eliminate a large fraction of mistags, allowing greater efficiency.
The stand-alone pattern recognition in the silicon tracker (SVX-II and
ISL) will allow $b$ tagging to extend into the $1<|\eta|<2$ region.

In this section, we use the same $b$ tag efficiencies and mistag rate
per jet inside the SVX fiducial region measured in the CDF Run~1 data
as in Figure~\ref{shw_tl_tag}.

The tight $b$-tagger is the default CDF secondary vertex tagger
(SECVTX).  The loose $b$-tagger includes the SLT tags with
$p_T>2.0$~GeV and the jet-probability tags with a cut at 5\%.

\vspace{0.2in}
{\bf Trigger Requirements} \\
                        
 The events are required to pass the following triggers: 

\begin{itemize} 
  \item High-$p_T$ inclusive lepton trigger ($p_T > 20$ GeV/c$^2$ and $|\eta|<1$).
        (This selects $W\rightarrow l \nu$, $Z\rightarrow l^+l^- $.) 
  \item $\met >20$ GeV plus $b$-tagging: Selects $W\rightarrow l \nu$, 
      $Z\rightarrow  \nu \bar \nu $ and $Z\rightarrow \mu^+ \mu^-$. 
\end{itemize}

\vspace{0.2in}
{\bf Selection Criteria} \\ \nopagebreak

In order to reduce the large standard model backgrounds, we impose the
following conventional selections based on the optimization of
$S/\sqrt{B}$ ratio, where S and B are the number of signal and the
background events passing the selections.
  
\begin{itemize}
  \item Prompt isolated lepton with $p_T > 20$ GeV/c.
  \item Isolated tracks from $\tau$ decay also included ($p_T>15$~GeV, 
        $\Sigma p_T<1.0$~GeV in cone $\Delta R<0.4$). 
  \item $\met> 20$ GeV.
  \item Jet raw $E_T >10$~GeV, $|\eta|<2.0$ in cone $\Delta R<0.4$
  \item Two $b$-tagged jets in $E_T > 10$~GeV and $|\eta| < 2.0$, one tight 
        SECVTX and other loose (SECVTX, JPB, SLT).
  \item One $b$ jet raw $E_T > 25$~GeV
  \item No extra jet with $E_T > 20$~GeV and $|\eta|<2.4$.
  \item Any events containing two isolated tracks ($p_T > 10$ GeV, 
        $|\eta|<2.0$) are removed. 
  \item A requirement that the reconstructed mass of $\bb$ jet system be
        near the target Higgs mass. (Described below.)
\end{itemize}

Figure~\ref{fig-etb} shows the $b$ jet $E_T$ distributions for $WH$
signal and backgrounds.  We also considered requirements on the $H_T$
and the Higgs scattering angle in the $WH$ rest frame, but these did not
significantly improve the result.

\begin{figure}
  \begin{center}
    \parbox{3.0in}{\epsfxsize=\hsize\epsffile{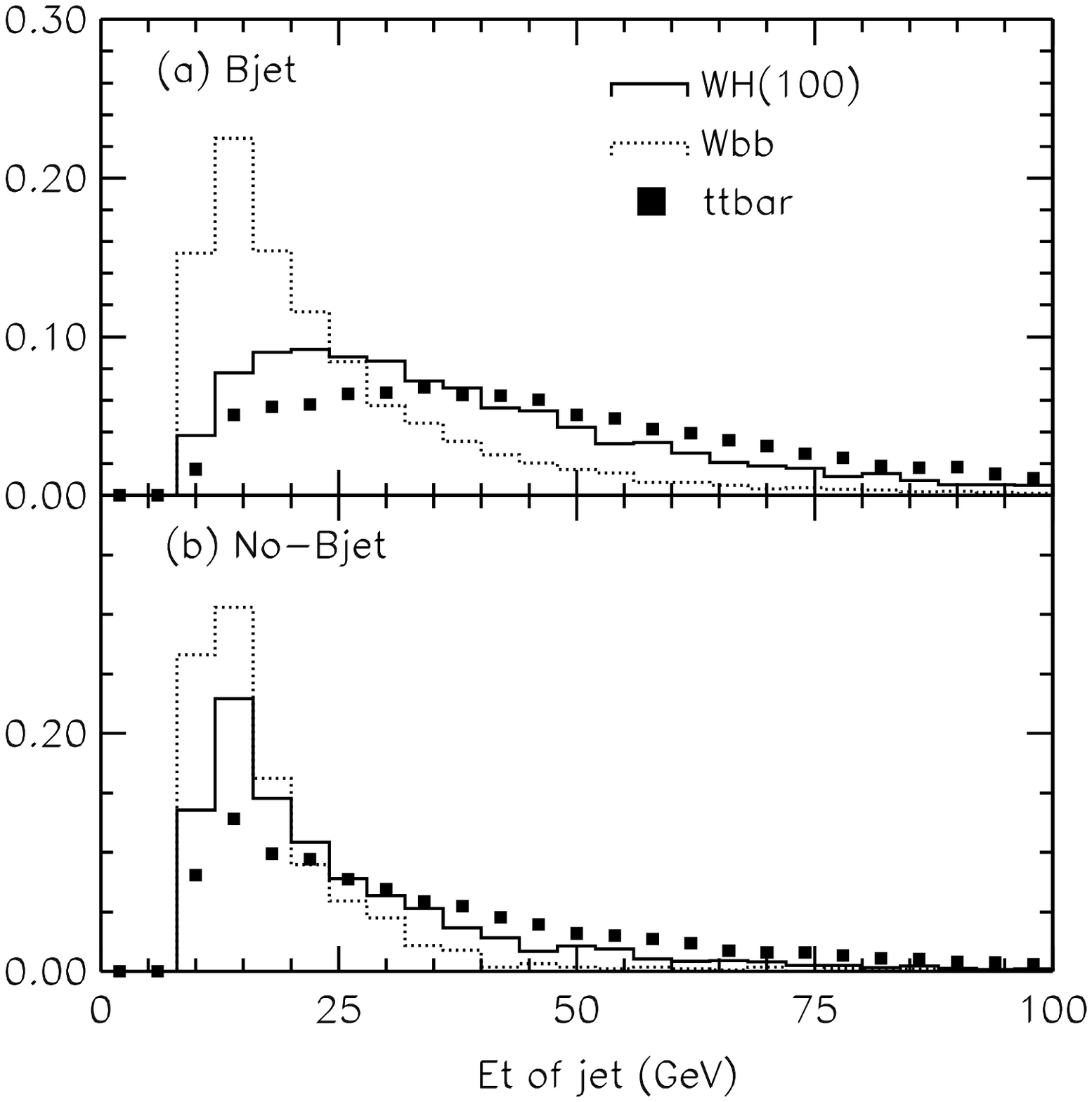}}
    \parbox{3.0in}{\epsfxsize=\hsize\epsffile{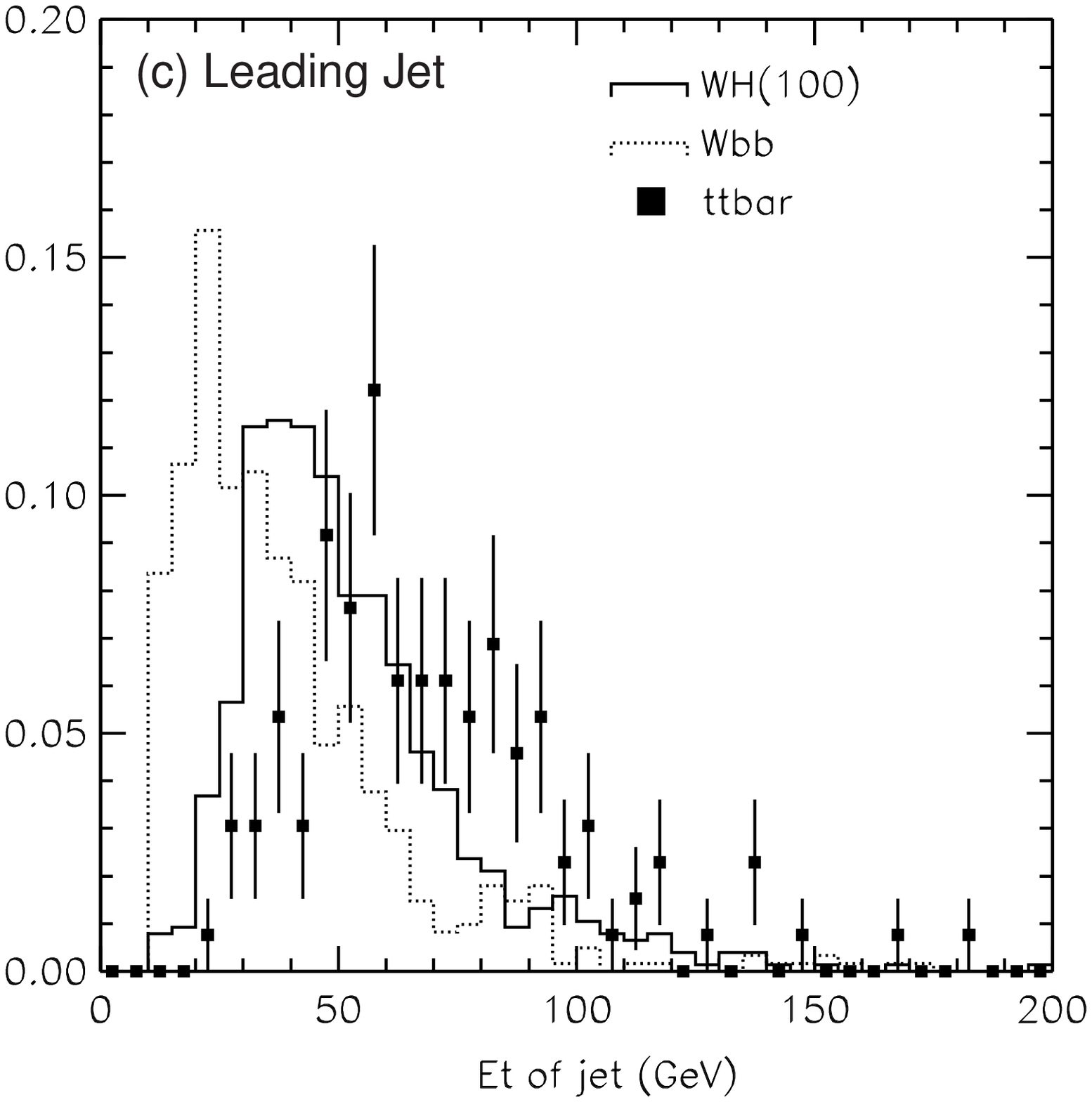}}
  \end{center}
  \caption{$WH\rightarrow \ell\nu\bb$ channel, QFL$^\prime$ analysis.  The comparisons of
           jet $\et$ distribution between signal and backgrounds (a) $b$-jet  
           and (b) non $b$-jet.  The leading jet $\et$ is shown in (c).}
  \label{fig-etb}
\end{figure}

To increase the sensitivity of the search we look for a peak in the
reconstructed two-jet invariant mass distribution using the 4-momenta
of jets as measured by the calorimeter after correction for detector
effects. With the present detector resolution, we found the dijet mass
resolution $\sigma = 16 $ GeV/c$^2$ for the Higgs mass at 110~GeV/c$^2$. 
The expected two-jet invariant mass distributions for $WH$,
$W\bb$ and $\ttbar$ are shown in Figure~\ref{mass}.

\begin{figure}
  \begin{center}
    \parbox{4.0in}{\epsfxsize=\hsize\epsffile{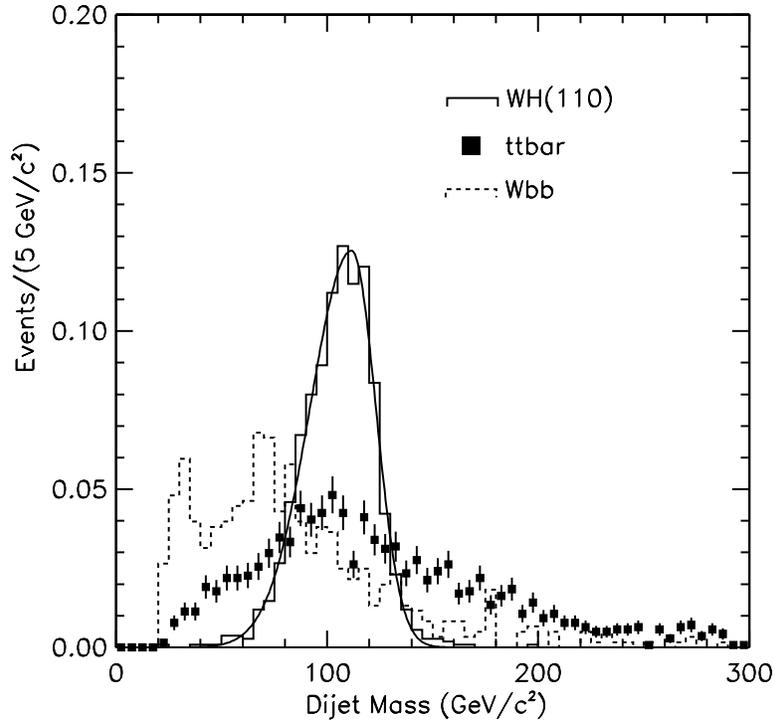}}
  \end{center}
  \caption{$WH\rightarrow \ell\nu\bb$ channel, QFL$^\prime$ analysis.  The dijet mass 
           distributions for Higgs mass $m_H = 110 $ GeV/c$^2$ and backgrounds.} 
  \label{mass}
\end{figure}

The studies described above in this report indicate that significant
improvement in the dijet mass resolution (30\%) is available if the
jet reconstruction uses information from tracking and shower maximum
detectors as well as the calorimeter energies used thus far.  Such an
improvement has been incorporated in this study by rescaling the
default di-jet mass resolution down by 30\%.
 
The mass windows we used to count for signal are listed in
Table~\ref{whlv-yao-table1}.

\vspace{0.2in}
{\bf Results} \\ \nopagebreak

The overall acceptance, including $B(H\rightarrow \bb)$ and dijet mass
requirement, ranges from 2.0\% to 2.3\% and is shown in
Table~\ref{whlv-yao-table1}, for the Higgs mass between 90 to 130
GeV/c$^2$.  The numbers of expected events and the corresponding
background estimates per 1~fb$^{-1}$ are summarized in
Table~\ref{whlv-yao-table1}.  The luminosity required for 95\% CL
exclusion and 5$\sigma$ discovery are shown in Figure~\ref{yao-lum}.
 
  \begin{table}
  \caption{For the $WH\rightarrow \ell\nu\bb$ channel, QFL$^\prime$ analysis, the 
           overall efficiency times branching ratio and expected 
           number of $WH\rightarrow \ell\nu\bb$ signal and background events 
           per 1~fb$^{-1}$. }
    \begin{center}
      \begin{tabular}{lccccc}
$M_{H}$ (GeV/c$^2$)          & 90 & 100 & 110 & 120 & 130 \\ \hline
$\Delta M$  &(76.7,103.3)    & (84.6,116.8) & (91.8,123.3) & (89.3,134.7) & (104.8,144.0) \\ 
$\sigma_{WH}$ (pb)           & 0.42  & 0.30  & 0.22  & 0.16  & 0.12 \\ 
$\epsilon\times $B($H\to\bb)$& 2.0\% & 2.2\% & 2.3\% & 2.3\% & 1.9\% \\ \hline
signal events                & 8.4   & 6.6   & 5.0   & 3.7   & 2.2  \\ 
$W\bb$                       & 21.5  & 20.6  & 18.9  & 19.7  & 15.4 \\ 
$WZ$                         & 7.7   & 7.3   & 4.9   & 2.3   & 0.6 \\ 
$\ttbar$                     & 9.6   & 12.3  & 12.2  & 13.8  & 13.0 \\
$tb$                         & 6.7   & 8.5   & 8.4   & 9.6   & 9.1 \\
$tqb$                        & 2.7   & 3.5   & 3.4   & 3.9   & 3.6 \\ \hline 
background                   & 48.2  & 52.2  & 47.8  & 49.3  & 41.7 \\
$S/\sqrt{B}$                 & 1.2   & 0.91  & 0.72  & 0.53  & 0.34 \\ \hline
    \end{tabular}
  \end{center}
  \label{whlv-yao-table1} 
\end{table}

\begin{figure}
  \begin{center}
    \parbox{4.0in}{\epsfxsize=\hsize\epsffile{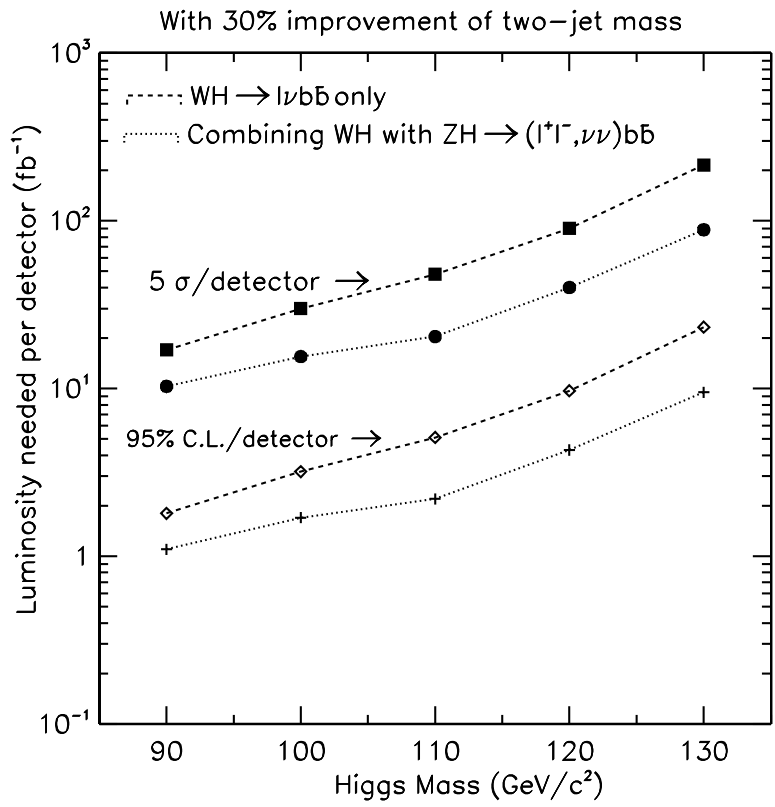}}
  \end{center}
  \caption{$WH\rightarrow \ell\nu\bb$ channel, QFL$^\prime$ analysis.  The luminosity 
           required per detector for 5$\sigma$ discovery or 95\% CL exclusion.}
  \label{yao-lum}
\end{figure}

       \vskip 4mm 
       \centerline{ \large \it 1b. $\ell\nu\bb$, SHW Simulation 
           and Traditional Techniques} \nopagebreak
       \vskip 4mm 
       
The present study differs from the previous studies primarily in that
it uses more detailed Run~2 simulation of detector efficiencies and
resolutions, carried out by the SHW simulation package.  The analysis
uses Monte Carlo generated event samples for the signal and all
backgrounds. Cross section estimates are taken either from data
measurements (such as the $\ttbar$ cross section) or from theoretical
predictions including highest order corrections to the particular
process.  Trigger efficiencies are assumed to be $\sim 100\%$ for this
channel, physics object reconstruction (leptons, jets, $\met$) is done
by SHW.  The efficiency for tagging $b-$jets is parametrized according
to the efficiency measured by the CDF Run 1 detector, and
cross-checked with the result of a simulation of the Run~2 D\O\
Microstrip Silicon Tracker.\footnote{See Section II.B.3.}  The
event shapes for signal and background were carefully studied to
determine the selection which maximizes significance, given in the
gaussian regime as $S/\sqrt{B}$. A significance value is calculated
for fixed Higgs masses and different values of the di-jet (i.e. $\bb$)
mass resolution, as a function of luminosity.
 
\vspace{0.2in}
{\bf Event samples} \\ \nopagebreak

The background processes to the $WH$ channel are $Wjj$, $\ttbar$,
$WZ$, single top production and QCD.  The requirement of double
$b-$tagging reduces the contribution of processes containing mostly
light quark jets, such as QCD and $Wjj$ other than $W\bb$. These
processes will not be considered any further.  Single top consists of
Drell-Yan ($t$-$b$ in the final state) and $W$-gluon fusion ($t$-light
quark-$b$ in the final state) processes. A small contribution comes
also from $t$-light quark final states, which are simulated as well.
The PYTHIA generator~\cite{PYTHIA} is used to simulate $WH$ (for values
of the Higgs mass of 80, 90, 100, 110, 120, and 130 GeV/$c^2$),
$\ttbar$, $WZ$, and single top events; $W\bb$ is generated using the
CompHEP package \cite{comphep}.  
Cross section values for the different processes are
listed in Table~\ref{ela-table1}.

\begin{table*}
\caption{$WH\rightarrow \ell\nu\bb$ channel, SHW analysis.
 Cross section values for $WH$ and background processes.}
\label{ela-table1}
\begin{tabular}{ccccccc}
 & $WZ$ & $Wb\bar{b}$ & $t\bar{t}$ & $tq+tbq$ & $tb$ &  \\ \tableline
$\sigma$ (pb) & 3.2  & 10.6\tablenote{this is the result of multiplying 
the CompHep cross section by approximately a factor 1.4, as 
suggested by a conversation with K. Ellis} & 7.5 & 2.4  & 1.0&\\
\end{tabular}
\end{table*}

Although we are not directly considering reconstructed $\tau$'s,
$W$ decays into $\tau$ are simulated, since the $\tau$ can feed into
the signal and background samples via the $e$ and $\mu$ decay modes.
Detector simulation for all samples was performed using
standard SHW.

\vspace{0.2in} 
{\bf Selection Criteria} \\ \nopagebreak

The following selection criteria are applied:

\begin{itemize}
\item One and only one isolated lepton (either $e$ or $\mu$), veto on 
      additional leptons with $\et>10$ GeV.
\item primary lepton $\et>20$ GeV.
\item $\met > 20$ GeV
\item Jet veto cuts: 
\begin{itemize}
  \item[-] No extra jets with $p_T>30$ and $|\eta|<2.5$. 
  \item[-] $\leq 1$ extra jet with $p_T<30$ and $|\eta|<2.5$. 
 \end{itemize}
\item $\pm2\sigma$ cut on $m_{b\bar{b}}$
\end{itemize}

The double $b-$tagging requirement consists of:
\begin{itemize}
\item One tight $b-$tag, with efficiency parameterized, as a function of $\et$,
  by: 0.57$\times$(tanh(($\et+8$)/26.5)).
\item One loose $b-$tag, with efficiency of a tight tag, enhanced by a 
  factor: 0.35$\times$(tanh($\et$/26.5).
\item $b-$tagged jets $\et>30(15)$ GeV  and $|\eta|<2$.
\end{itemize}

The $b-$tagging efficiencies are obtained from fits to CDF Run 1 data,
and are in good agreement with the results of a MCFAST simulation of
the $D\O$ Run~2 detector, under similar assumptions for the
$b-$tagging algorithms.  Both efficiencies are rescaled by a factor
1.1 to take into account soft lepton tagging. The total double $b-$tag
efficiency is in the range $30-40\%$.  Corrections are applied to
relate the energy of $b-$jets and untagged jets to the initial parton
energies, before QCD radiation.  The sequential effect of the cuts on
the event samples is shown in Table~\ref{ela-table2}.

\begin{table*}[tbp]
\caption{Data reduction in the $WH\rightarrow \ell\nu\bb$ channel, SHW analysis.
         Table shows the effect of the cuts described in the text on the
         signal and background samples.}
\label{ela-table2}
\begin{tabular}{lcccccc}
 & $WH$(110 GeV) & $Wb\bar{b}$ & $WZ$ & $t\bar{t}$ & $tq+tbq$ & $tb$   \\ \tableline
Generated  & 5000           & 80000  & 5000 & 45000  & 7000 & 6227 \\ 
One lepton (veto secondary leptons)&  2663 & 38736 & 2408 & 16149 & 3657 & 1651\\ 
lepton $\et$ & 1981 & 29072 & 1694 & 9761 & 2839 & 847 \\
$\met$ & 1763  & 22880 & 1437 & 8967 & 2558 & 779 \\
double $b-$tag & 666 & 4512 & 497 & 4160 & 236 & 324\\
jet vetoes & 382 & 4432 & 322 & 328 & 55 & 141\\
$\sigma(m_{b\bar{b}})$ cut ($15\%$)& 345 & 1248 & 271 & 130 & 6 & 65\\
\end{tabular}
\end{table*}

The mass resolution of the $\bb$ system, after all corrections are
applied to the SHW jets, is approximately 15\%, and the distribution
itself is shown if Figure~\ref{ela-bb}.  Since it is foreseen in Run
II to achieve better values in dijet mass resolution, the resolution
here is assumed as a free parameter and the analysis is performed for
a set of different hypotheses. The effect of the $m_{\bb}$ resolution
on the significance for a 100 GeV Higgs signal in 10 fb$^{-1}$ is
illustrated in Figure~\ref{figure5}.

\begin{figure}
  \begin{center}
    \parbox{3.5in}{\epsfxsize=\hsize\epsffile{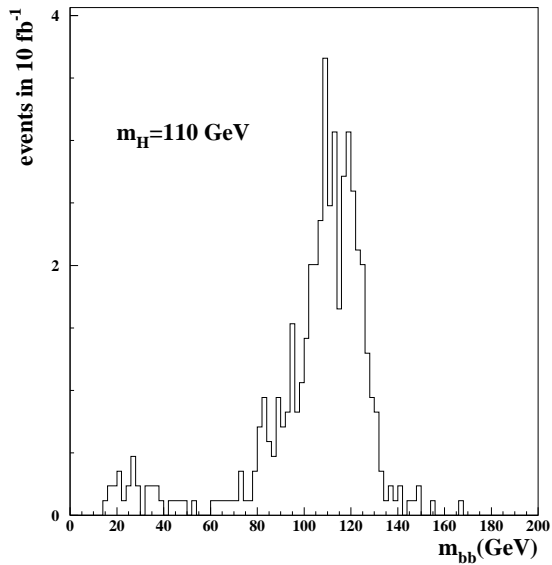}}
  \end{center}
  \caption{$WH\rightarrow \ell\nu\bb$ channel, SHW analysis.  Reconstructed 
           $\bb$ mass distribution for $m_H = 110$~GeV.}
  \label{ela-bb}
\end{figure}

\begin{figure}
  \begin{center}
    \parbox{3.5in}{\epsfxsize=\hsize\epsffile{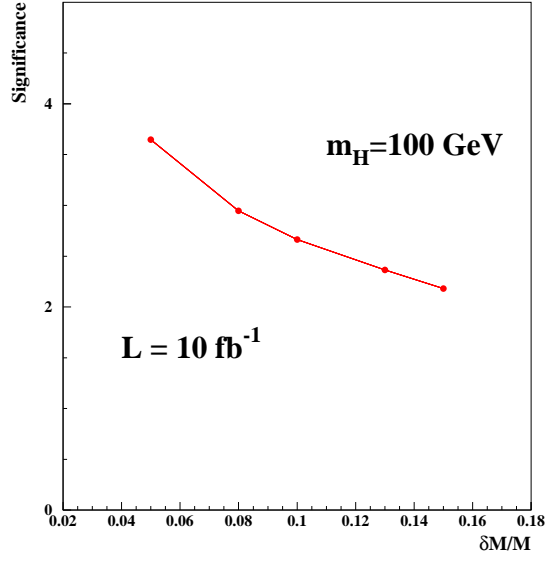}}
  \end{center}
  \caption{$WH\rightarrow \ell\nu\bb$ channel, SHW analysis.  Significance 
           ($S/\sqrt{B}$) as a function of the $b\bar{b}$ mass resolution, for a 
           100~GeV Higgs and 10 fb$^{-1}$.}
  \label{figure5}
\end{figure}

\vspace{0.2in}
{\bf Results and Conclusions} \\ \nopagebreak

The number of signal and background events after all selection cuts are applied
is listed in Tables~\ref{table10}, \ref{table11} and \ref{table12} for three
values of mass resolution and for 1~fb$^{-1}$ of data. The significance values
for each Higgs mass and mass resolution are listed in Table~\ref{table13}.
Figure~\ref{figure10} shows the significance values scaled to 10~fb$^{-1}$ as a
function of Higgs mass for the different mass resolutions.

The luminosity required to observe a Higgs signal in the $WH\rightarrow l\nu
b\bar{b}$ channel at different confidence levels is shown in
Figure~\ref{ela-lum}. The $WH\rightarrow l\nu b\bar{b}$ channel appears to be
the most significant for Higgs masses below 130 GeV/$c^2$.

\begin{figure}
  \begin{center}
    \parbox{3.5in}{\epsfxsize=\hsize\epsffile{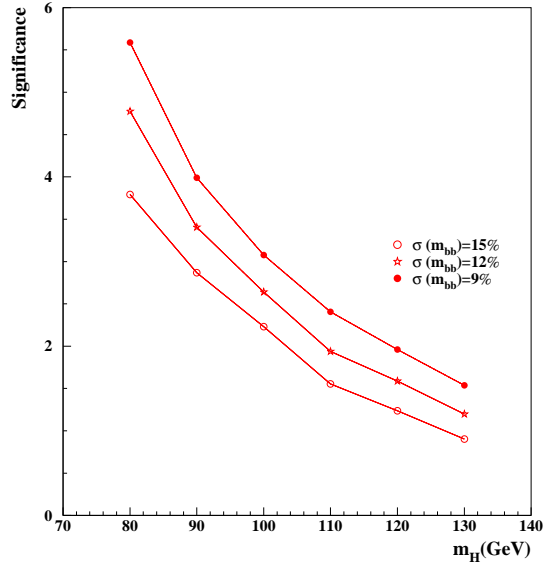}}
  \end{center}
  \caption{$WH\rightarrow \ell\nu\bb$ channel, SHW analysis.  Significance 
           ($S/\sqrt{B}$) as a function of Higgs mass, for 
           10 fb$^{-1}$ and different mass resolutions.}
  \label{figure10}
\end{figure}

\begin{table*}
\caption{$WH\rightarrow \ell\nu\bb$ channel, SHW analysis.  Number of events 
 in 1 fb$^{-1}$ after selection cuts for nominal SHW $m_{b\bar{b}}$ resolution 
 ($\sim 15\%$).}
\label{table10}
\begin{tabular}{lcccccc} 
\multicolumn{1}{c}{$m_H$ (GeV/$c^2$)} & 80 & 90 & 100 & 110 & 120 & 130 \\ \hline
signal events (S) & 14 & 10 &   8 &   5 &   4 &   3 \\ \tableline
$Wb\bar{b}$    & 94& 76& 68& 63 & 59&  59\\
$WZ$    & 10& 11& 11& 11& 11&  10\\
$t\bar{t}$    & 22& 23& 26& 30& 34&  37\\
single top  & 92& 11 & 11& 12& 14 &  15 \\ \tableline
total background (B) & 136 & 121& 117& 117& 119& 122\\ 
\end{tabular}
\end{table*}

\begin{table*}
\caption{$WH\rightarrow \ell\nu\bb$ channel, SHW analysis.  Number of events 
  in 1 fb$^{-1}$ after selection cuts for $12\%$ $m_{b\bar{b}}$ resolution.}
\label{table11}
\begin{tabular}{lcccccc} 
\multicolumn{1}{c}{$m_H$ (GeV/$c^2$)} & 80   & 90  & 100 & 110 & 120 & 130  \\ \hline
signal events (S)    & 14   & 10  &   8 &   5 &   4 &   3  \\ \tableline
$Wb\bar{b}$          & 59   & 54  & 47  & 38  & 32  &  28  \\
$WZ$                 & 8.6  & 10  & 8.8 & 7.2 & 6.0 &  4.4 \\
$t\bar{t}$           & 12   & 16  & 19  & 21  & 24  &  26  \\
single top           & 5.1  & 6.0 & 8.4 & 9.2 & 10  &  11  \\ \tableline
total background (B) &  85  & 86  & 84  & 75  & 72  &  70  \\ \tableline
$S/\sqrt(B)$         & 1.5  & 1.1 & 0.9 & 0.6 & 0.5 & 0.4  \\
\end{tabular}
\end{table*}

\begin{table*}
\caption{$WH\rightarrow \ell\nu\bb$ channel, SHW analysis.  Number of events 
  in 1 fb$^{-1}$ after selection cuts for $10\%$ $m_{b\bar{b}}$ resolution.}
\label{table12}
\begin{tabular}{lcccccc} 
\multicolumn{1}{c}{$m_H$ (GeV/$c^2$)} & 80   & 90   & 100  & 110  & 120  & 130  \\ \hline
signal events(S)    & 14   & 10   &   8  &   5  &   4  &   3  \\  \tableline
$W\bb$                      & 50   & 45   &  36  &  29  &  25  &  19  \\
$WZ$                        &  7   & 10   &   9  &   6  &   4  &   2  \\
$\ttbar$                    &  9   & 14   &  16  &  15  &  20  &  22  \\
single top          &  3   &  6   &   7  &   7  &   9  &   9  \\ \tableline
total background (B)& 69   & 75   &  68  &  57  &  58  &  52  \\ \tableline
$S/\sqrt(B)$        & 1.7  & 1.2  & 0.9  & 0.7  & 0.6  & 0.4  \\
\end{tabular}
\end{table*}

\begin{table*}
\caption{$WH\rightarrow \ell\nu\bb$ channel, SHW analysis.  Significance 
  $S/\sqrt{B}$ at 1 fb$^{-1}$ integrated luminosity as a function of Higgs 
  mass for different $m_{b\bar{b}}$ resolutions.}
\label{table13}
\begin{tabular}{ccccccc} 
$m_H$ (GeV/$c^2$) & 80 & 90 & 100 & 110 & 120 & 130 \\ \tableline
$S/\sqrt{B}$:  & & & & & &  \\
$\sigma$($m_{b\bar{b}})=15\%$ & 1.2 & 0.9 & 0.7 & 0.5 & 0.4 & 0.3\\
$\sigma$($m_{b\bar{b}})=12\%$ & 1.5 & 1.1 & 0.8 & 0.6 & 0.5 & 0.4\\
$\sigma$($m_{b\bar{b}})=10\%$ & 1.7 & 1.2 & 0.9 & 0.7 & 0.6 & 0.4 \\
\end{tabular}
\end{table*}

\begin{figure}
  \begin{center}
    \parbox{3.5in}{\epsfxsize=\hsize\epsffile{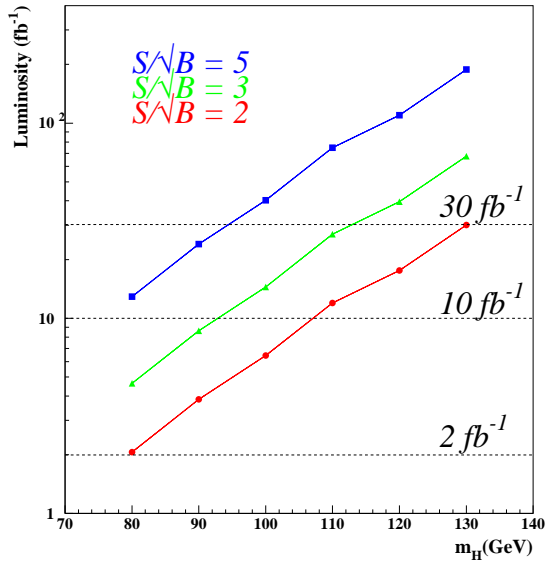}}
  \end{center}
  \caption{Luminosity thresholds as function of Higgs mass for a 95\% confidence
           limit and 3$\sigma$ and 5$\sigma$ signals. The mass resolution assumed in this
           case is $12\%$.}
\label{ela-lum}
\end{figure}

       \newpage
       \vskip 4mm 
       \centerline{ \large \it 1c. $\ell\nu\bb$, SHW Simulation 
           and Neural Network Techniques} \nopagebreak
       \vskip 4mm 
	\def\9{\phantom0}

\vspace{-1pt}
{\bf Optimal Event Selection} \\ \nopagebreak

Given any set of variables, it is useful to construct the discriminant
function $D$, given by $$D =\frac{s(x)}{s(x)+b(x)},$$ where $x$ is the
vector of variables and $s(x)$ and $b(x)$, respectively, are the
$n-$dimensional probability densities describing the signal and
background \cite{prosper_bhat}. 
A cut on the single variable $D$ yields the optimal
demarcation between signal and background. The variable $D = r/(1+r)$
is related to the {\em Bayes discriminant function}, which is
proportional to the likelihood ratio $r \equiv s(x)/b(x)$. The
discrimination is optimal in the sense that it minimizes the
probability to mis-classify events.

Of the several methods available to approximate the discriminant $D$,
neural networks are the most convenient because they approximate $D$
directly~\cite{blum}.  To compute $D$, we have used the JETNET
package~\cite{jetnet} to train three-layer (that is, input, hidden and
output) feed-forward networks. The training was done using the
backpropagation algorithm, with the desired output for the signal set
to one and that for the background set to zero.

\vspace{0.2in}
{\bf Event Samples} \\ \nopagebreak

We have considered final states with a high $p_T$ electron~(e) or
muon~($\mu$) and a neutrino from $W$ decay and two $b$ quarks
($b\bar{b}$) from Higgs decay. The $WH$ events were generated using
the PYTHIA program (version 6.023) for Higgs masses of $M_H$ = 80, 90,
100, 110, 120, 130 and 140 GeV/c$^2$.  The cross section times
branching ratio (BR) for the process $p\bar{p}\rightarrow
WH\rightarrow l\nu b\bar{b}$ where $l=e,\mu$, $\tau$ are listed in
Table~\ref{whlv_nn_table1}.

\begin{table}
\caption{Cross section times branching ratio for the process 
$p\bar{p}\rightarrow WH\rightarrow 
\ell\nu b\bar{b}$ for various  $M_H$, and for the
various backgrounds. Note: For $tb$ and $t\bar{t}$ we give the
total cross section.}
\label{whlv_nn_table1}
\begin{center}
\begin{tabular}{lc}
$M_H$ (GeV/c$^2$) & $\sigma\times BR(\mbox{pb})$ \\ \tableline
\980  & 0.17\9 \\
\990  & 0.119 \\
100 & 0.085 \\
110 & 0.062 \\
120 & 0.045 \\
130 & 0.034 \\
140 & 0.011 \\
\tableline
Backgrounds             &                   \\
$Wb\bar{b}$             &  3.5\9\9\\
$WZ$                              &  0.165  \\
$tbq$                             &  0.8\9\9\\ \tableline
                                  & $\sigma$ (pb)     \\ 
$tb$                              &  1.0\9\9\\
$t\bar{t}$              &  7.5\9\9\\
\end{tabular}
\end{center}
\end{table}

The background processes which are important for this search are
$Wb\bar{b}$, $WZ$, $t\bar{t}$ and single top production ($W^*
\rightarrow tb$, $Wg$ $\rightarrow tqb$).  They have the same
signature, $\ell \nu b \bar{b}$, as the signal.  The $Wb\bar{b}$
sample was generated using CompHEP, a parton level Monte Carlo program
based on exact leading order (LO) matrix elements, with the
fragmentation of partons done using PYTHIA. The single top, $t\bar{t}$
and $WZ$ events were generated using PYTHIA.  The $W^*$ (s-channel)
process was generated using off-shell $W$ production, with
$\sqrt{\hat{s}} > m_t + m_b$, and selecting the $W \rightarrow t b$
final state.  In Table~\ref{whlv_nn_table1} , we list the cross sections we
have used for the background processes.

The SHW program was used for detector simulation.  It provides a
fast (approximate) simulation of the trigger, tracks, calorimeter
clustering, event reconstruction and c and b-tagging. The SHW
simulation predicts a di-jet mass resolution of about 14\% at $M_H$ =
100~GeV/$c^2$, varying only slightly over the mass range of interest.
However, to allow for comparisons with the other $WH$ and $ZH$
studies, we have rescaled the di-jet mass variables for all signal and
background events so that the resolution is 10\% at each Higgs boson
mass.

\vspace{0.2in}
{\bf Selection Criteria} \\ \nopagebreak

We applied the following requirements to produce a base sample to
which the neural network was then applied:
\begin{itemize}
\item
the transverse momentum of the isolated lepton
$p_T^{\ell} > 15$ ~GeV,
\item
the pseudorapidity of the lepton 
$|{\eta}_{\ell}| < 2.0$,
\item
the missing transverse energy in the event 
$\met > 20$~GeV,
\item
two or more jets in the event 
with $\et^{jet} > 10$~GeV and $|{\eta}_{jet}| < 2$.
\end{itemize}
In order to reduce the dominant $Wb\bar{b}$ background, we assumed
that the two jets will be $b$-tagged.  We then searched for variables
that discriminate between the signal and the backgrounds and arrived
at the following set:
\begin{itemize}
\item
$E_T$ of the $b$-tagged jets 
\item
$M_{b\bar{b}}$ -- invariant mass of the $b$-tagged jets 
\item
$H_T$ -- sum of the transverse energy of all selected jets
\item
$E_T$ of the lepton
\item
$\eta$ of the lepton
\item
$\met$ -- missing tranverse energy
\item
$S$ -- sphericity
\item
$\Delta R$ between the $b$-tagged jets
\item
$\Delta R$ between the lepton and first $b$-tagged jet
\item
$\Delta R$ between the lepton and second $b$-tagged jet
\end{itemize}

\begin{table*}[h!]
\caption{Variables used in training the neural networks for
signals against specific backgrounds in the $WH$ study.}
\vspace*{-1pc}
\label{whlv_nn_table2}
\begin{center}
\renewcommand{\arraystretch}{1.2}
\begin{tabular}{ccc}
$Wb\bar{b}$         & $WZ$ & $t\bar{t}$\\ \tableline
$E_T^{b1}$  & $E_T^{b1}$  & $E_T^{b1}$\\  
$E_T^{b2}$  & $E_T^{b2}$  & $E_T^{b2}$\\  
$M_{b\bar{b}}$ & $M_{b\bar{b}}$ & $M_{b\bar{b}}$\\ 
$H_T$       & $H_T$       & $H_T$\\
$E_T^{l}$ & $E_T^{l}$ & $\met$\\ 
$S$  & $S$  & $\Delta R(b_1,l)$\\ 
$\Delta R(b_1,b_2)$ & $\eta_{l}$ & $\Delta R(b_1,b_2)$\\ 
\end{tabular}
\end{center}
\end{table*}
\clearpage

\begin{table*}[h!]
\caption{Results for $WH$ channel.
Note: The network cut was optimized to yield maximum significance 
for each Higgs mass leading to
different background counts at each mass. }
\label{whlv_nn_table3}
\begin{center}
\begin{tabular}{lrrrrrrr}
\multicolumn{1}{c}{$M_H$~GeV/c$^2$}&80&90&100&110&120&130&140\\ \hline
Number of events&&&&&&& \\
 $WH$&     12.71&  8.65&  8.97&  4.81&  4.41&  3.71&  1.22\\ \tableline
 $Wb\bar{b}$&17.51& 12.28& 12.48&  5.84&  9.66& 20.12& 10.26\\
 $WZ$&           6.89&  7.52& 10.32&  1.72&  1.00&  0.97&  0.18\\
 $tqb$&          0.55&  0.51&  0.95&  0.58&  0.71&  0.96&  0.91\\
 $tb$&           2.95&  2.46&  5.40&  3.44&  5.89&  9.33&  8.35\\
 $t\bar{t}$&          3.34&  5.63&  9.89&  7.24&  8.39& 14.49& 14.15\\ \tableline

 Total background &31.23& 28.40& 39.04& 18.81& 25.67& 45.87& 33.84\\ 
\end{tabular}
\end{center}
\end{table*}

The networks were configured with 7 input variables, 9 hidden nodes
and one output node. The subset of variables used to train the
networks, one network for each Higgs mass and for each of the three
dominant backgrounds, are listed in Table~\ref{whlv_nn_table2}. We show the
distributions of some of these variables for $W\bb$ and $\ttbar$ in
Figures~\ref{fig:whvarsbbar} and~\ref{fig:whvarsttbar}.

Figure~\ref{fig:whnetworks} shows the distributions of network output
for each background relative to the signal with $M_H = 100$
GeV/c$^2$. We see that all backgrounds, with the exception of $WZ$,
are well separated from the signal. For Higgs masses close to the $Z$
mass the $WZ$ background is kinematically identical to the signal and
therefore difficult to deal with.  But for Higgs masses well above the
$Z$ mass the discrimination between $WH$ and $WZ$ improves, as does
that between $WH$ and the other backgrounds.

The final selection is a requirement that the neural network output
$f$, which is defined to the range $0\le f\le 1$, satisfy $f>f_0$ in
which $f_0$ is a Higgs--mass dependent value.  The final numbers of
expected signal and background events are given in Table~\ref{whlv_nn_table3}.

\begin{figure}
  \begin{center}
    \parbox{6.0in}{\epsfxsize=\hsize\epsffile{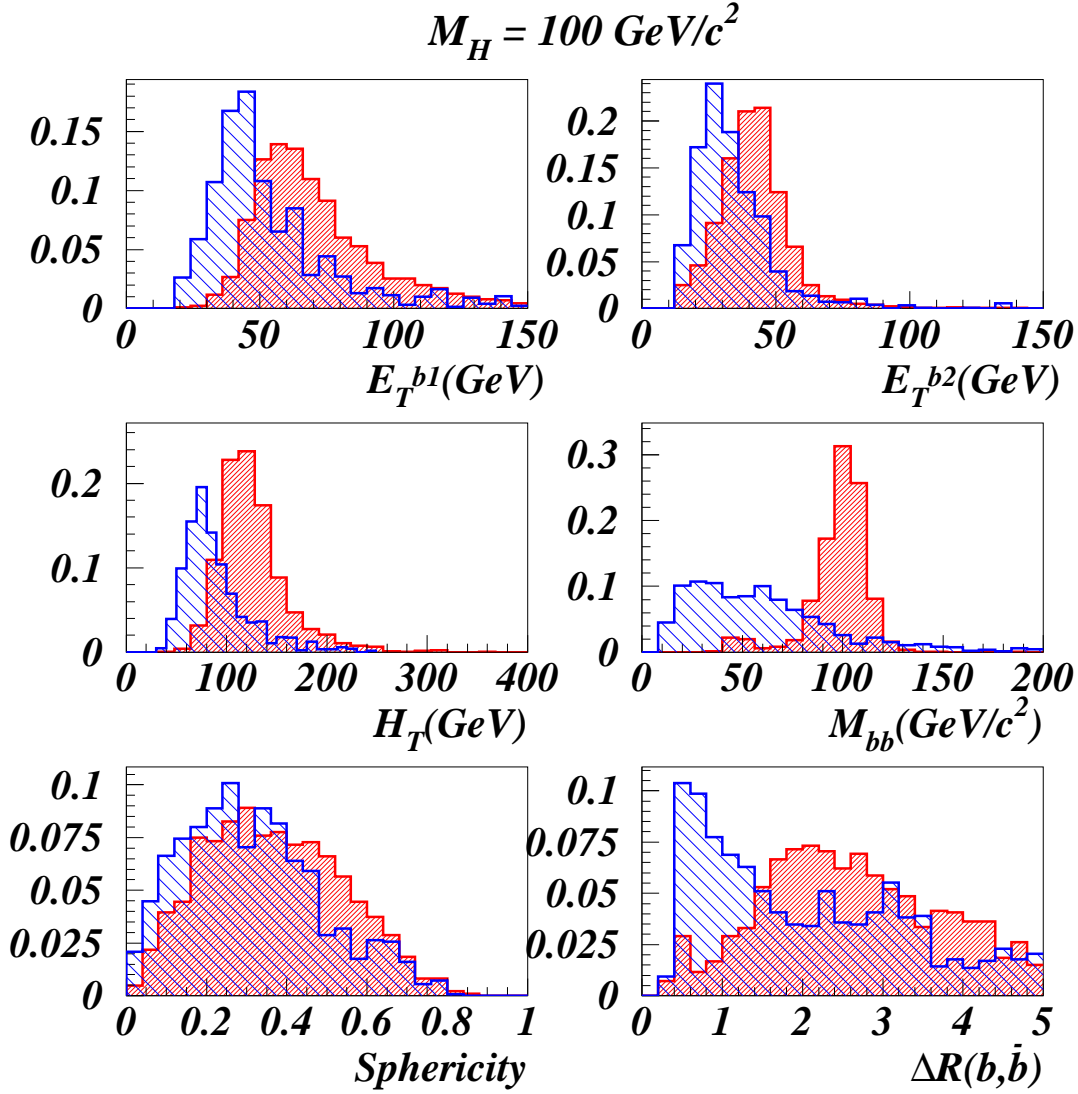}}
  \end{center}
  \caption{$WH\rightarrow \ell\nu\bb$ channel, neural network analysis.
           Distributions of variables used in training the neural networks 
           for signal (heavy shading) and background; $M_H=100$~GeV vs. 
           $Wb\bar{b}$.}
  \label{fig:whvarsbbar}
\end{figure}

\begin{figure}
  \begin{center}
    \parbox{6.0in}{\epsfxsize=\hsize\epsffile{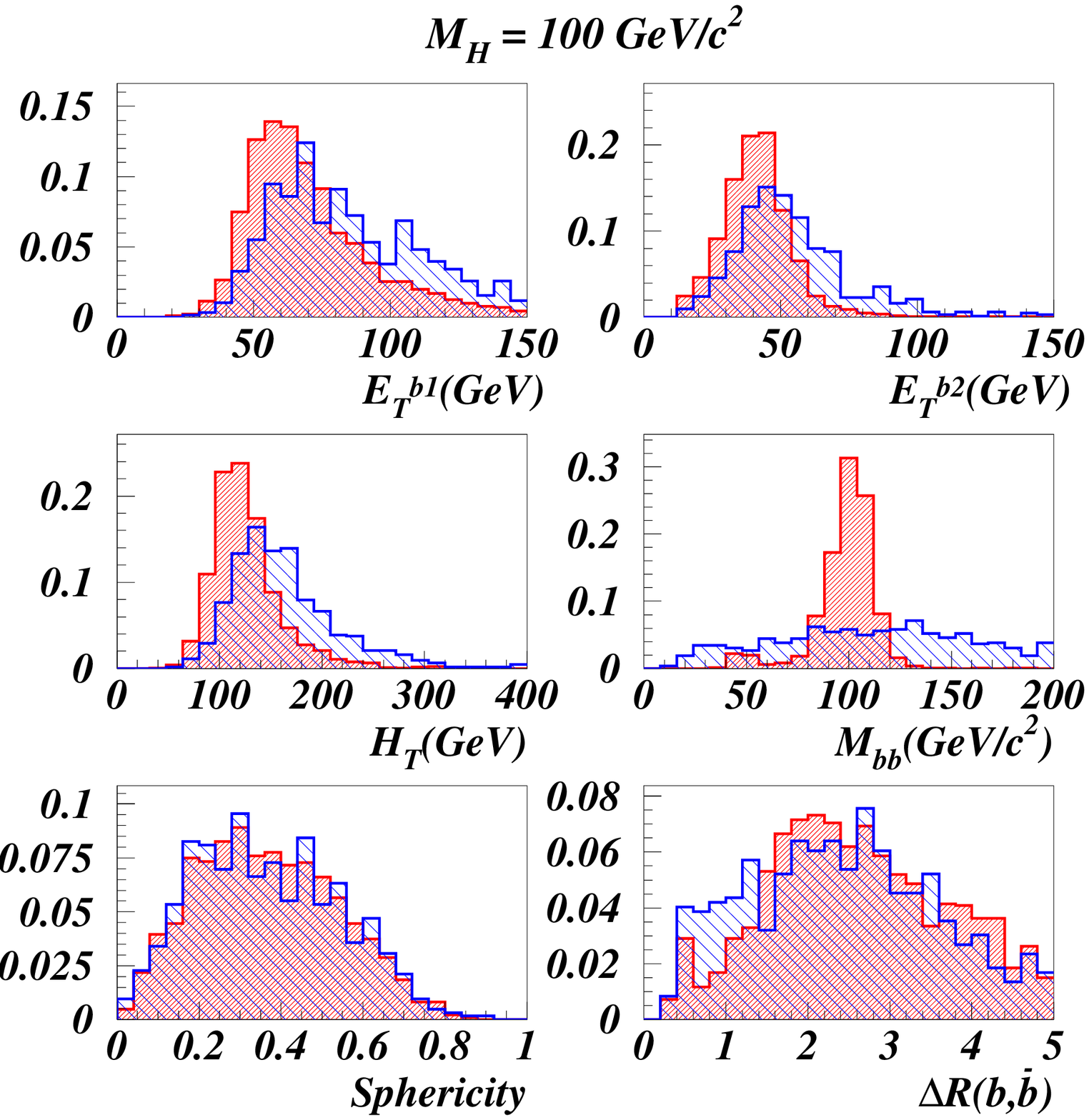}}
  \end{center}
  \caption{$WH\rightarrow \ell\nu\bb$ channel, neural network analysis.
           Distributions of variables used in training the neural networks
           for signal (heavy shading) and background; $M_H=100$~GeV vs. 
           $t\bar{t}$. }
  \label{fig:whvarsttbar}
\end{figure}

\begin{figure}
  \begin{center}
  \epsfxsize=0.9\hsize\epsfbox{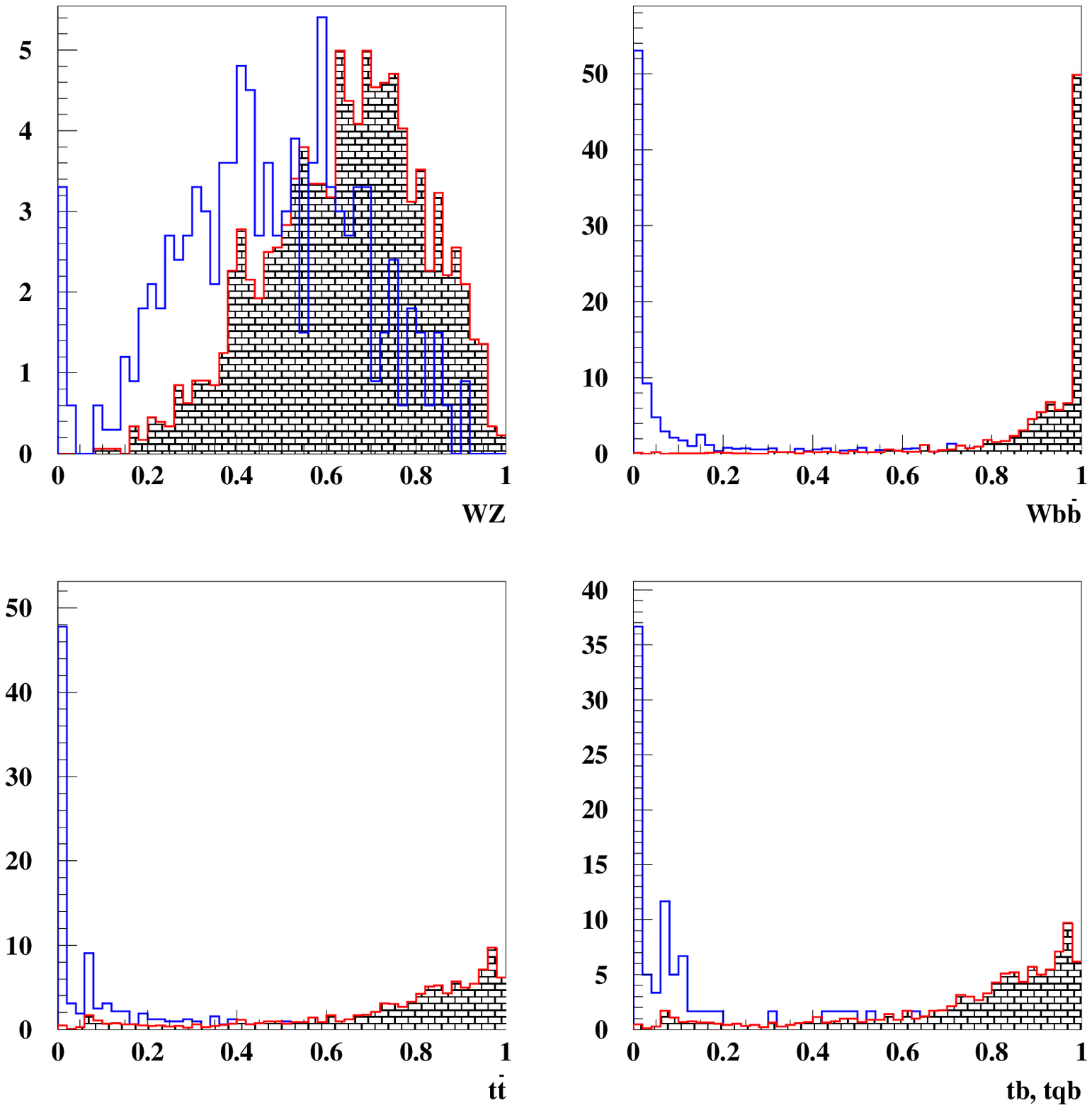}
  \caption{$WH\rightarrow \ell\nu\bb$ channel, neural network analysis.
           Distributions of neural network outputs of networks 
           trained using the backgrounds WZ, $Wb\bar{b}$ and
           $t\bar{t}$.  The heavily shaded histograms are signal 
           and the other histograms are for background.}
  \label{fig:whnetworks}
  \end{center}
\end{figure}

    \subsubsection{$\nn\bb$ Channel}			\small
\begin{center}
{\it P. Bhat,
     W. Bokhari, 
     J. Conway,
     R. Demina, 
     R. Gilmartin, 
     D. Hedin, 
     R. Jesik, 
     M. Kruse,
     H. Prosper,
     V. Sirotenko,
     W. Yao} \\
\end{center}
\normalsize

The cross section times branching ratio for the mode $\pp\rightarrow ZH
\rightarrow \nn\bb$ is comparable to that for the $WH$ mode discussed in the 
previous section.  If the backgrounds at fixed efficiency were similar,
the $\nn\bb$ channel would have as much sensitivity as $\ell\nu\bb$
previous.  Three analyses of this final state are presented here.  The
first is an analysis based on the CDF Run~1 result, using the QFL$^\prime$
simulation.  The second is an analysis selection developed using SHW
simulated events.  The third analysis of this section is an extension
of the SHW analysis applying neural network techniques to the event
selection.

       \vskip 2mm 
       \centerline{ \large \it 2a. $\nn\bb$, QFL$^\prime$ Simulation
           and Traditional Techniques} \nopagebreak
       \vskip 2mm 
       We have repeated the study of $ZH\rightarrow (\nu \bar
\nu,l^+l^-)b\bar b$ reported at Snowmass 96~\cite{snowmass96} using 
QFL$^\prime$ and found consistent results.  This analysis achieves
sensitivity via a tight selection which reduces the background to a
low level. 

\vspace{0.2in}
{\bf Event Samples}\\ \nopagebreak

The event samples used in this analysis were produced as described in
the section for the $WH\rightarrow\ell\nu\bb$ QFL$^\prime$ analysis.

\vspace{0.2in}
{\bf Selection Criteria}\\ \nopagebreak

For the $ZH\rightarrow \nu\bar \nu b\bar b$, We apply the following
kinematic selection criteria:
\begin{itemize}
  \item Raw $\met> 35$ GeV to pass the trigger requirement.
  \item Two $b$ tagged jets with $E_t>15$ GeV in $|\eta|<2.0$. 
  \item Minimum $\delta \phi_{min}>0.5$ between $\met$ 
 and any jets $E_t>8$ GeV and $|\eta|<2.0$. 
  \item No extra jet $Et>8$ GeV in $|\eta|<2.4$.
  \item No isolated track above 10 GeV in the events, where the isolation 
   requirement is $\Sigma P_t <2.0$ GeV over the additional tracks 
   inside the cone $\Delta R=0.4$.
\end{itemize} 

The numbers of observed events and the corresponding background
estimates for 1~fb$^{-1}$ at $\sqrt{s} = 2 $ TeV, are summarized in
Table~\ref{yao-table2}.  As in the $\ell\nu\bb$ channel, the
backgrounds are dominated by vector boson and top quark production.
Generic multijet events are removed very efficiently by the
requirements of two $b$ jets and large missing $\et$.  This background
is extremely difficult and unreliable to simulate due to the large
cross section and very small acceptance, and must be estimated with
real data.  In the CDF Run 1 Higgs search, the $\bb$ dijet background
was about half of the total.  With tighter selection criteria, such as
a back-to-back cut on the jet azimuth, the authors believe that this
background will represent no more than 50\% of the total without loss
of signal sensitivity.

\begin{table}[tbp]
\caption{$ZH\rightarrow \nn\bb$ channel, QFL$^\prime$ analysis.
The expected $ZH\rightarrow (\nu \bar \nu, l^+l^-) b\bar b$ 
 signal and background events per 1 fb$^{-1}$.}
\label{yao-table2}
\begin{center}
\begin{tabular}{lccccc}
$M_{H}$ (GeV/c$^2$)  &  90 & 100 & 110 & 120  & 130 \\ \tableline
$\sigma_{ZH}$ (pb)  & 0.24 & 0.18 & 0.13 & 0.1 & 0.074 \\ 
$\epsilon_{ZH}$ ($\%$) & 1.5 & 1.7 & 2.1 & 1.7 & 1.2 \\ 
signal (S)  & 3.6 & 3.1 & 2.7 & 1.7 & 0.9 \\ \tableline 
backgrounds & & & & &\\ 
$Z\bb$, $Zc\bar{c}$ & 6.53 & 5.20 & 4.9 & 4.83 & 3.77 \\ 
$ZZ$                & 5.43 & 4.90 & 3.4 & 1.33 & 0.0 \\
top                 & 1.63 & 2.2 & 2.2 & 2.2 & 2.2 \\
background$^a$ (B)  & 13.6 & 12.3 & 10.4 & 8.3 & 5.9 \\ 
\end{tabular}
$^a$This does not include $\bb$ dijet production
processes, which must be estimated from data; see text.\\
\end{center}
\end{table}

       \vskip 2mm 
       \centerline{ \large \it 2b. $\nn\bb$, SHW Simulation 
           and Traditional Techniques} \nopagebreak
       \vskip 2mm 
       
In the previous analysis, hard cuts on {\em raw} missing transverse
energy and additional jet activity were used reduce the background to
an acceptable level.  For the analysis presented here, a different
strategy is used in which a looser selection gives both larger signal
and larger background but with approximately the same statistical
power as the previous analysis.

\vspace{0.2in}
{\bf Event samples} \\ \nopagebreak

The signature of $ZH$ decays is two $b$-jets with large invariant mass
and either large missing energy ($\met$), or two opposite sign leptons
with mass near that of the $Z$.  These requirements effectively
eliminate generic multijet events, as shown below, and leave the
following background processes: $Zb\bar{b}$, $ZZ$, $\ttbar$, and
single top production. The $\met$ selection also picks up events with
$W\rightarrow l \nu$ decay, particularly if the leptons escape
detection.  This introduces additional backgrounds from $WZ$ and
$Wb\bar{b}$ processes and adds $WH$ events to the effective signal.
$ZH$ and $WH$ events are generated using PYTHIA (for Higgs mass values
of 90, 100, 110, 120, and 130 GeV$/c^2$). PYTHIA is also used for the
$ZZ$, $WZ$, $\ttbar$, and single $t$ backgrounds, and CompHEP is
used for $Zb\bar{b}$ and $Wb\bar{b}$. The $Zb\bar{b}$ simulation is
checked against ISAJET, and the results agree to within 8\%.  Cross
sections and branching ratios for both the signal and background
processes are taken directly from the generators without any
correction factors applied, except for $\ttbar$ events which are
normalized to the measured cross section.  Detector
simulation for all samples is performed using the SHW package.

As in the previous analysis, no estimate for the level of background 
from $\bb$ dijet production is made here, as it will require actual
data. 
\pagebreak

\vspace{-0.2in}
{\bf Selection Criteria} \\ \nopagebreak

$ZH \rightarrow \nu \bar{\nu} b \bar{b}$ events are selected by the following
criteria:

\begin{itemize}
  \item 2 $b$-jets: 
  \begin{itemize}
    \item[-] one tight $b$-tagged jet with $E_T>20$ GeV and $|\eta|<2$
    \item[-] one loose $b$-tagged jet with $E_T>15$ GeV and $|\eta|<2$ 
  \end{itemize}
  \item missing transverse energy $\met > 35$ GeV
  \item the angle between the $\met$ and its closest jet $\delta\phi$ $>0.5$
  \item reject events with one or more isolated leptons with ($p_T>8$ GeV/c) 
  \item scalar sum of hadronic energy $H_T<175$ GeV
\end{itemize}

The $b$-tagging and lepton identification efficiencies are all taken
from SHW.  The effect of the cuts on both the 100~GeV$/c^2$~Higgs
signal and background samples is shown in Table~\ref{evflow}. As shown
in Figure~\ref{met_ht}, the $\met$ cut removes generic multijet
background, and the additional cut on $\delta\phi$ removes multijet
events with $\met$ due to mismeasured jets.  Combined with the
requirement of two $b$-tagged jets, few generic multijet events
remain; based on Run~1 data, we believe that the background from
multijet events will be no more than 50\% of the total background from
all sources.  The isolated lepton cuts reduce the backgrounds
involving $W$ decay.  However, significant contributions remain due to
muon detection inefficiencies (which also add to the apparent $\met$
of an event). This also means that $WH$ events contribute to the
effective signal.

It should be noted that the $WH$ search mode selection requires the
detection of a high $p_T$ lepton, so the two samples are independent.
The $H_T$ cut helps remove $\ttbar$ background (see
Figure~\ref{met_ht}), and cuts much less of the signal than the third
jet veto used to veto top in previous analyses (increasing the
significance by as much as 30\%). Energy corrections are applied to
the $b$-jets to correct for final state radiation and undetected
energy from semileptonic decays. Only the two highest $E_T$ $b$-jets
in an event are used as the candidate jets from the Higgs decay.

\begin{table}
\caption{$ZH\rightarrow \nn\bb$ channel, SHW analysis.
 Data reduction for 100 GeV$/c^2$~Higgs and backgrounds.}
\label{evflow}
\begin{center}
\begin{tabular}{lccccccccc} 
channel                & $ZH$ & $WH$ & $ZZ$ & $WZ$ & $Zb\bar{b}$ & $Wb\bar{b}$ & $\ttbar$ & $tb$ & $tq+tbq$  \\
\tableline
$\sigma\times BR$ (fb) & 27.0 & 71.5 & 137  & 135  &     668     &     2530    &    6800    & 1000 &    807 \\ \tableline
generated              & 7000 & 7000 & 7000 & 7000 &   20000     &    20000    &  270000    & 6223 &  55000 \\
2 $b$-jets             & 3022 & 2798 & 2437 & 2254 &    3694     &     3802    &  119713    & 2012 &   5036 \\
$\met > 35$ GeV        & 2177 & 1744 &  941 & 1297 &    2573     &     1924    &   55559    &  555 &   3686 \\
$\Delta\phi >0.5$      & 1983 & 1538 &  840 & 1171 &    2330     &     1704    &   48657    &  491 &   3324 \\
$H_T < 175$ GeV        & 1527 &  831 &  668 &  802 &    1592     &     1526    &    2004    &  200 &   1660 \\
no isolated leptons    & 1376 &  364 &  609 &  401 &    1458     &      680    &     767    &   99 &    902 \\ 
$80 < m_{bb} < 125$    & 1014 &  271 &  412 &  268 &     314     &       96    &     286    &   32 &    194 \\
\end{tabular}
\end{center}
\vspace{-1pc}
\end{table}

\begin{figure}
  \begin{center}
    \parbox{6.0in}{\epsfxsize=\hsize\epsfbox{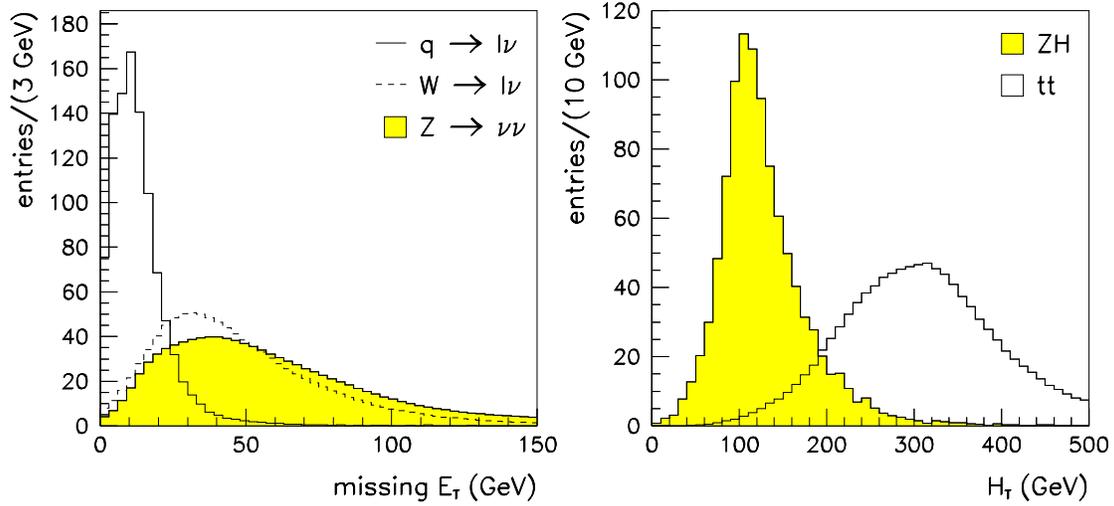}}
  \end{center}
  \caption{$ZH\to\nn\bb$ channel, SHW analysis.  Distributions of
           missing transverse energy and scalar sum of hadronic 
           energy ($H_T$).}
  \label{met_ht}
\end{figure}

\vspace{0.2in}
{\bf Results and Conclusions} \\ \nopagebreak

Signal mass windows for each Higgs mass are determined using the
nominal mass resolution obtained from SHW with $b$-jet corrections
($\approx 15\%$ for 100 GeV$/c^2$~Higgs).  The number of signal and
background events expected in 1~fb$^{-1}$ of data is shown in
Table~\ref{nunu_sig}. Numbers obtained assuming a 10\% 
mass resolution are also shown in the table. These results assume a
100\% trigger efficiency.  Efficiencies close to this should be
obtainable with a multijet-$\met$~trigger with the $\met$ threshold
set as low as the rates will allow. The addition of secondary vertex
tagging at the trigger level would help keep the rates down if
necessary, as would a veto on back--to--back jets.

\begin{table*}[tbp]
\vspace{1cm}
\caption{$ZH\rightarrow \nn\bb$ channel, SHW analysis.  Expected number 
         of $\met b\bar{b}$ signal and background events in 1 fb$^{-1}$ 
         of data (per experiment) for various Higgs masses. Numbers in 
         parentheses are for 10\% $\bb$ mass resolution.}
\label{nunu_sig}
\begin{center}
\begin{tabular}{lccccc}
$m_H$ (GeV$/c^2$) & 90           & 100          & 110          & 120          & 130          \\ \tableline
$\Delta m$        & 70--110      & 80--125      & 85--130      & 90--140      & 95--150      \\ 
                  &(80--105)     &(90--120)     &(95--125)     &(105--135)    &(115--145)    \\ \tableline
$ZH$              & 4.9          & 3.9          & 2.8          & 1.9          & 1.3          \\
$WH$              & 4.0          & 2.9          & 1.8          & 1.3          & 0.8          \\
signal            & 8.9          & 6.7          & 4.6          & 3.2          & 2.1          \\ \tableline
$ZZ$              & 8.5  (8.5)   & 8.0  (6.0)   & 7.1  (2.9)   & 5.8  (0.2)   & 4.4  (0.0)   \\
$WZ$              & 6.0  (5.9)   & 5.9  (5.1)   & 5.4  (2.7)   & 4.7  (0.2)   & 3.6  (0.0)   \\
$Zb\bar{b}$       & 12.3 (7.0)   & 10.3 (6.3)   & 9.2  (5.6)   & 8.7  (4.6)   & 8.1  (3.5)   \\
$Wb\bar{b}$       & 15.1 (8.5)   & 12.1 (7.1)   & 11.1 (6.7)   & 10.5 (5.4)   & 10.1 (4.2)   \\
single $t$        & 7.4  (4.6)   & 8.0  (5.4)   & 7.8  (5.3)   & 8.0  (4.7)   & 7.9  (3.8)   \\
$\ttbar$          & 6.3  (4.0)   & 7.2  (4.7)   & 7.0  (4.7)   & 7.3  (4.3)   & 7.2  (3.5)   \\ 
background$^a$    & 55.6 (38.5)  & 51.5 (34.6)  & 47.6 (27.9)  & 45.0 (19.4)  & 41.3 (15.0)  \\
\end{tabular}
$^a$This does not include generic $\bb$ dijet production
processes, which must be estimated from data; see text.\\
\end{center}
\end{table*}

       \vskip 2mm 
       \centerline{ \large \it 2c. $\nn\bb$, SHW Simulation 
           and Neural Network Techniques} \nopagebreak
       \vskip 2mm 
       
This section describes a neural network analysis applied to the
$ZH\rightarrow\nu\bar{\nu}\bb$ final state.  The technique is
identical to that for the neural network analysis for the
$WH\rightarrow\ell\nu\bb$ section.

\vspace{0.2in}
{\bf Event Samples} \\ \nopagebreak

The $ZH$ events were generated using PYTHIA for Higgs masses of 90, 100, 110,
120 and 130 GeV/c$^2$.  The principal backgrounds to $ZH$ production are $ZZ$,
$Zb\bar{b}$, single top and $t\bar{t}$. The $Zb\bar{b}$ background sample was
generated using CompHEP, with fragmentation done using PYTHIA, while all other
samples were generated using PYTHIA.  We used SHW to simulate the detector
response and, as in the $WH$ study, we assumed that two jets are $b$-tagged.
The cross sections for signal and background are shown in Table~\ref{table6}.
As in the other two analyses, no $\bb$ dijet background is considered.

\begin{table}[h!]
\caption{$ZH\rightarrow \nu\bar{\nu}\bb$ channel, neural network selection. Accepted
         cross section times branching ratio for signal and background processes.}
\label{table6}
\begin{center}
\begin{tabular}{lr}
$M_H$ (GeV/c$^2$) & $\sigma\times BR(\mbox{pb})$ \\ \tableline
90                      &  0.041\\
100                     &  0.030\\
110                     &  0.022\\
120                     &  0.016\\
130                     &  0.013\\ \tableline
Backgrounds             &                   \\
$Zb\bar{b}$             & 0.700\\
$tbq$                             & 0.800\\ \tableline
                                  & $\sigma$ (pb) \\ 
$ZZ$                              & 1.235\\ 
$tb$                              & 1.000 \\
$t\bar{t}$              & 7.500\\ 
\end{tabular}
\end{center}
\end{table}

\vspace{0.2in}
{\bf Selection Criteria} \\ \nopagebreak

We applied the following requirements to produce a base sample to which the
neural network was then applied:
\begin{itemize}
\item
the missing transverse momentum
$\met > 20$ ~GeV 
\item
two or more jets in the event 
with $E_T^{jet} > 10$~GeV and $|{\eta}_{jet}| < 2$
\end{itemize}
In order to reduce the dominant $Wb\bar{b}$ background, we  assumed
that the two jets will be  $b$-tagged.

A network was trained for each Higgs mass
and for each of the backgrounds with the following variables
\begin{itemize}
\item
$E_T$ of the $b$-tagged jets
\item
$M_{b\bar{b}}$ -- invariant mass of the $b$-tagged jets
\item
$H_T$ -- sum of jet transverse energies
\item
$\met$
\item
$S$ -- sphericity
\item
Centrality ($\sum{E_T(jets)}/\sum{E(jets)}$)
\item
$\frac{\met}{\sqrt{E^{b1}_t}}$
\item
Minimum $\Delta\phi$(jet,$\met$)
\end{itemize}

\vspace{0.2in}
{\bf Results and Conclusions} \\ \nopagebreak

Figure~\ref{fig:zhnn-net} shows the network output distributions $Z
\rightarrow \nu\bar{\nu}$, respectively.  From these figures it is
clear that discriminating between $ZH$ and its principal backgrounds
is much more difficult than for $WH$ and that further study is needed
to identify better variables. Nonetheless, the $ZH$ channel
contributes usefully to the final result.  The results for the
$ZH\rightarrow \nu\bar{\nu}\bb$ channel are in Table~\ref{table7}.

\begin{table}[h!]
\caption{$ZH$, $Z\to\nn$ channel, neural network selection.
         Expected number of $\met\bb$ signal and background
         events in 1 fb$^{-1}$ of data (per experiment) for various
         Higgs masses, assuming 10\% $\bb$ mass resolution.}
\label{table7}
\begin{center}
\begin{tabular}{lrrrrr} 
$M_H$ (GeV/c$^2$) &        90 &    100 &   110 &   120 &   130\\ \tableline
number of events &&&&& \\
$ZH$           &  6.66 &    4.37 &   3.53 &   2.76 &   2.16 \\ 
$WH$           &  5.59 &    3.75 &   2.79 &   1.98 &   1.70 \\ \tableline
signal         & 12.25 &    8.12 &   6.32 &   4.74 &   3.86 \\ \tableline 
$Zb\bar{b}$    &  8.12 &    4.97 &   4.83 &   3.85 &   3.92 \\
$Wbb$          & 21.70 &   13.12 &  10.68 &   8.22 &   7.53 \\    
$ZZ$           & 11.24 &    6.14 &   2.59 &   1.05 &   0.59 \\
$WZ$           &  7.95 &    4.49 &   1.99 &   0.90 &   0.54 \\
$tqb$          &  0.63 &    0.27 &   0.37 &   0.24 &   0.29 \\
$tb$           &  6.83 &    2.99 &   4.27 &   5.12 &   6.40 \\ 
$t\bar{t}$     &  5.10 &    2.70 &   3.00 &   3.00 &   4.35 \\ \tableline
background$^a$ & 61.57 &   34.80 &  27.73 &  22.38 &  23.62 \\ 
\end{tabular}
$^a$This does not include generic $\bb$ dijet production
processes, which must be estimated from data; see text.\\ 
\end{center}
\vspace{2pc}
\end{table}

\begin{figure}
\begin{center}
\centerline{\epsfsize=0.9\hsize\epsfbox{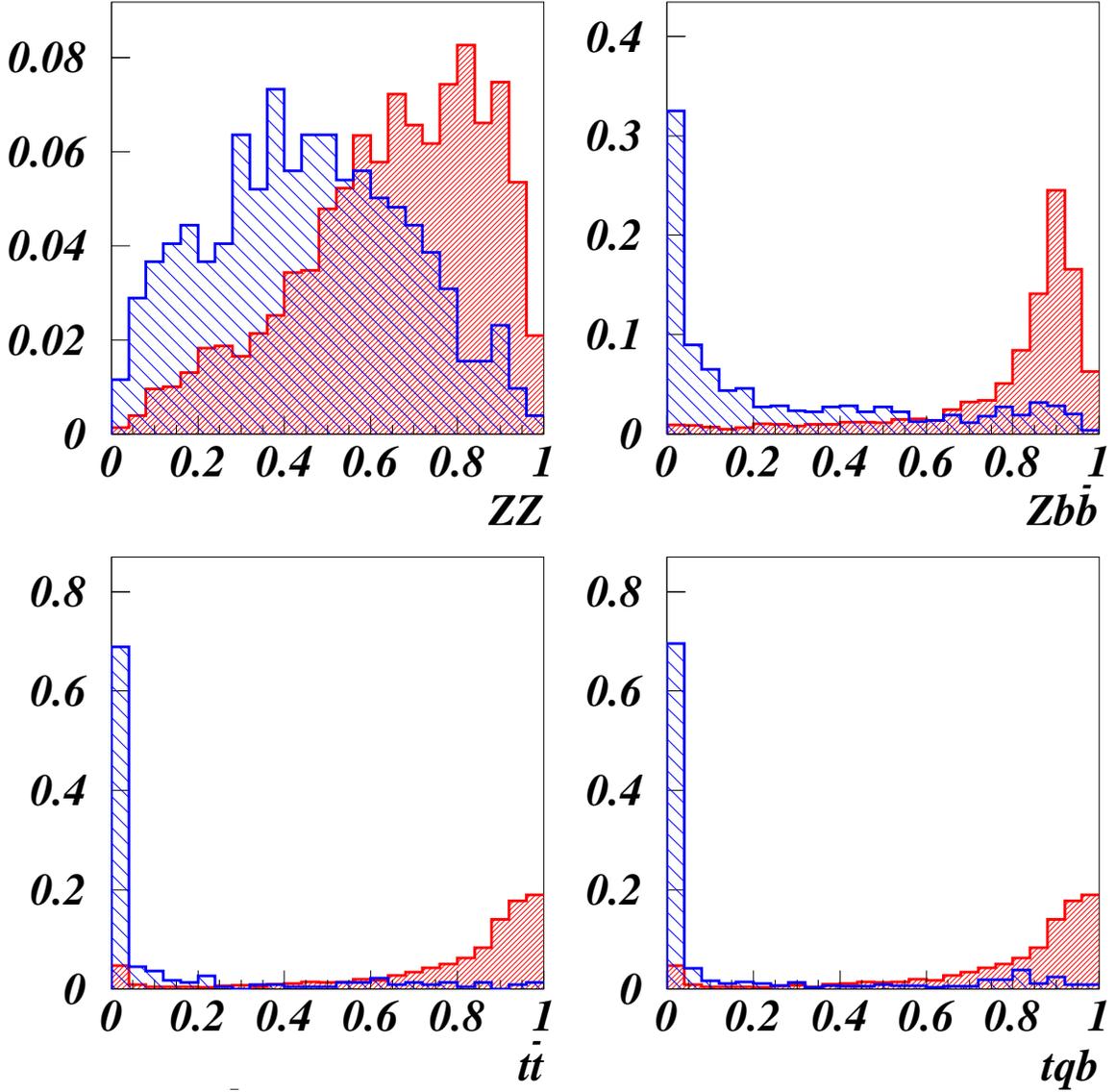} }
\caption{$ZH\rightarrow \nn\bb$ channel, neural network analysis.
Distributions of neural network outputs for
networks trained using signal (heavy shading) vs. 
the backgrounds $ZZ$, $Zb\bar{b}$, $t\bar{t}$, and single top.}
\end{center}
\label{fig:zhnn-net}
\end{figure}

    \subsubsection{$\lpm\bb$ Channel}			\small
\begin{center}
{\it P. Bhat,
     R. Gilmartin, 
     R. Jesik, 
     M. Kruse,
     H. Prosper,
     W. Yao} \\
\end{center}
\normalsize \nopagebreak

Although the cross section times branching ratio for the $ZH\rightarrow\lpm\bb$
final is roughly 1/3 that of the previous two channels, these events are easily
and cleanly identified.  Furthermore, given the high luminosity required for
observation in the higher rate $WH\rightarrow\ell\nu\bb$ and $ZH\rightarrow
\nn\bb$ channels, it is clear that every possible event is valuable to the
Higgs search.

       \vskip 4mm 
       \centerline{ \large \it 3a. $\lpm\bb$, QFL$^\prime$ Simulation
           and Traditional Techniques}
       \vskip 4mm 
	
The selection criteria for this channel are:
 In the $ZH\rightarrow (e^+e^-,\mu^+\mu^-)b\bar b$ channel, we select the 
 dilepton events with the following cuts.
\begin{itemize} 
      \item First lepton  $\et(\pt)>20$ GeV, $|\eta|<1.0$ 
      \item Second lepton $\et(\pt)>10$ GeV, $|\eta|<2.0$ 
      \item $| M_{Z}-M_{l^+l^-}|<15 $ GeV
      \item Two $b$-tagged jets with $\et>15$ GeV and $|\eta|<2.0$ 
\end{itemize} 
The results are included with those for the $ZH\rightarrow\nn\bb$
channel described in section {\it 2a}.

       \vskip 4mm 
       \centerline{ \large \it 1b. $\lpm\bb$, SHW Simulation 
           and Traditional Techniques} \nopagebreak
       \vskip 4mm 
       
\vspace{0.2in}
{\bf Event Samples}\\ \nopagebreak

The event samples used in this SHW--based analysis were generated
simultaneously with the samples used in the $ZH\rightarrow\nn\bb$ SHW--based
analysis described previously.

\vspace{0.2in}
{\bf Event Selection} \\ \nopagebreak

$ZH \rightarrow l^{+} l^{-} b \bar{b}$ events are selected by the following
criteria:

\begin{itemize}
  \item 2 $b$-jets: 
  \begin{itemize} 
     \item[-] one tight $b$-tagged jet with $\et>20$ GeV and $|\eta|<2$ 
     \item[-] one loose $b$-tagged jet with $\et>15$ GeV and $|\eta|<2$ 
  \end{itemize} 
   \item at least two opposite sign leptons ($e$ or $\mu$) of the same flavor,
     \\ (tau's are not explicitly searched for, but they contribute to the 
      $\met$ analysis) 
   \item both leptons with $\pt>10$ GeV/c \item dilepton mass within 10 
       GeV/c$^2$ of the $Z$ mass 
   \item scalar sum of hadronic energy $H_T<175$ GeV
\end{itemize}

\begin{table}[h!]
  \caption{$ZH\rightarrow\lpm\bb$, SHW analysis.
           Expected number of $l^{+}l^{-}b\bar{b}$ signal and background events in 
           1~fb$^{-1}$ (per experiment) for various Higgs masses. Numbers in
           parentheses are for a 10\% $\bb$ mass resolution.}
  \label{ll_sig}
  \begin{tabular}{l|c|c|c|c|c}
\noalign{\vskip-8pt}
$m_H$ (GeV/$c^2$) & 90        & 100          & 110          & 120          & 130 \\
\tableline
$\Delta M$     & 70--110      & 80--125      & 85--130      & 90--140      & 95--150 \\ 
$\Delta M$     & (80--105)    & (90--120)    & (95--125)    & (105--135)   & (115--145) \\
\tableline
$ZH$ signal    & 1.5          & 1.2          & 0.9          & 0.6          & 0.4 \\
\hline
$ZZ$           & 2.2  (2.2)   & 2.1  (1.6)   & 1.9  (0.8)   & 1.5  (0.1)   & 1.2  (0.0)\\
$Zb\bar{b}$    & 2.9  (1.6)   & 2.4  (1.4)   & 2.1  (1.2)   & 2.0  (1.0)   & 1.9  (0.8)\\
$t\bar{t}$     & 1.8  (1.1)   & 1.9  (1.3)   & 1.8  (1.2)   & 1.9  (1.2)   & 1.9  (1.1)\\
background     & 6.9  (4.9)   & 6.4  (4.3)   & 5.8  (3.2)   & 5.4  (2.3)   & 5.0  (1.9) \\
\tableline
$S/\sqrt{B}$   & 0.57  (0.7) & 0.47  (0.6) & 0.37  (0.5) & 0.26  (0.4) & 0.18  (0.3) \\
  \end{tabular}
\end{table}

\vspace{0.2in}
{\bf Results and Conclusions} \\ \nopagebreak

The event flow through these cuts is similar to that in the $\nu
\bar{\nu}$ analysis, except that there are no backgrounds involving
the $W$ in this case. The two leptons in these events provide an easy,
high efficiency trigger. The number of signal and background events
expected in 1~fb$^{-1}$, assuming a 100\% efficient trigger is shown
in Table~\ref{ll_sig}. These events are independent of both the $ZH
\rightarrow \nu \bar{\nu} b\bar{b}$ and $WH \rightarrow l \nu
b\bar{b}$ samples.

Combining both the $ZH\rightarrow\nn\bb$ and $ZH\rightarrow\lpm\bb$
channels, a three sigma observation of a 100 GeV/$c^2$ Higgs would
require 8~fb$^{-1}$ for a single experiment.  If a 30\% improvement in
mass resolution with respect to that of Run~1 can be achieved, only
5~fb$^{-1}$ would be required.  The mass distribution for both signal
and background in 10~fb$^{-1}$ is shown in fig.~\ref{ZH_mass}.


\begin{figure}
  \begin{center}
    \parbox{4.0in}{\epsfxsize=\hsize\epsffile{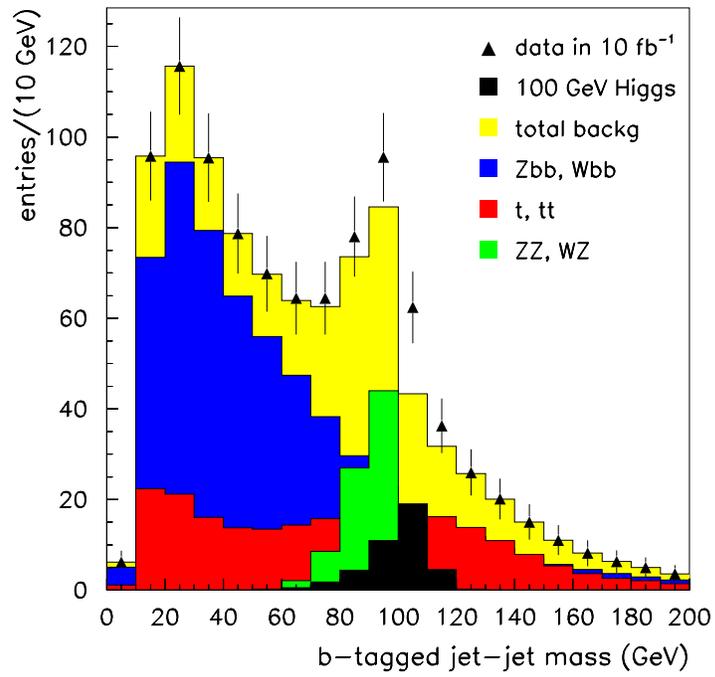}}
  \end{center}
  \caption{Combined $ZH\to\nn\bb,\lpm\bb$ channels, using the SHW analysis.  
      The plot shows the $b\bar{b}$ dijet mass distribution for a 100 GeV/$c^2$ 
      Higgs and background in 10 fb$^{-1}$.}
  \label{ZH_mass}
\end{figure}

       \vskip 4mm 
       \centerline{ \large \it 1c. $\lpm\bb$, SHW Simulation 
           and Neural Network Techniques} \nopagebreak
       \vskip 4mm 
       
\vspace{0.2in}
{\bf Event Samples} \\ \nopagebreak

The $ZH$ events were generated using PYTHIA for Higgs masses of 90,
100, 110, 120 and 130 GeV/$c^2$.  The principal backgrounds to $ZH$
production are $ZZ$, $Zb\bar{b}$, single top and $t\bar{t}$. The
$Zb\bar{b}$ background sample was generated using CompHEP, with
fragmentation done using PYTHIA, while all other samples were
generated using PYTHIA.  We used SHW to simulate the detector response
and, as in the $WH$ study, we assumed that two jets are $b$-tagged.
The cross sections for signal and background are shown in
Tables~\ref{table4} and~\ref{table5}.

\begin{table}[h!]
\caption{$ZH\rightarrow\lpm\bb$ neural network analysis. Accepted cross section
         times branching ratio for signal and background processes.}
\label{table4}
\begin{center}
\begin{tabular}{lr}
$M_H$ (GeV/c$^2$) & $\sigma\times BR(\mbox{pb})$ \\ \tableline
90	&  0.020\\
100 	&  0.015\\ 
110 	&  0.011\\ 
120 	&  0.008\\ 
130 	&  0.006\\ \tableline
backgrounds	&		\\
$Zb\bar{b}$ 	& 0.35\\
$tbq$ 		& 0.80\\ \tableline 
		& $\sigma$ (pb) \\ 
$ZZ$ 		& 1.235\\
$tb$ 		& 1.0\\
$t\bar{t}$ 	& 7.5
\end{tabular}
\end{center}
\end{table}

\begin{table}[h!]
\caption{$ZH, Z\to\lpm$, neural net analysis, signal and
         background expected in 1 fb$^{-1}$.}
\label{table5}
\begin{center}
\begin{tabular}{lrrrrr}
$M_H$ (GeV/c$^2$)&90 &    100 &   110 &     120 &   130\\ \tableline
Number of events &&&&& \\
$ZH$ 	    &   1.26 &   0.87 &   0.79 &   0.80 &   0.58\\ \hline
$Zb\bar{b}$ &   0.61 &   0.45 &   0.61 &   1.50 &   1.42\\
$ZZ$        &   2.04 &   1.44 &   1.42 &   0.83 &   0.31\\
$t\bar{t}$  &   0.28 &   0.05 &   0.23 &   0.44 &   0.18\\ \hline
total background & 2.93 & 1.94 &  2.26 &   2.77 &   1.91\\ \hline
$S/B$        &   0.43 & 0.45 & 0.35 &  0.29 &  0.31\\
$S/\sqrt{B}$ &   0.74 & 0.63 & 0.54 &  0.48 &  0.42
\end{tabular}
\end{center}
\end{table}

A network was trained for each Higgs mass
and for each of the backgrounds with the following variables
\begin{itemize}
\item
$\et$ of the $b$-tagged jets
\item
$\et$ of the two leptons
\item
$M_{b\bar{b}}$ -- invariant mass of the $b$-tagged jets
\item
$M_{l\bar{l}}$ -- invariant mass of the leptons
\item
$H_T$ -- sum of the transverse energy of all the jets
\item
$\Delta R$ between the first lepton and the first $b$-tagged jet
\end{itemize}

\begin{figure}
  \begin{center}
\centerline{\parbox{6.0in}{\epsfxsize=\hsize\epsffile{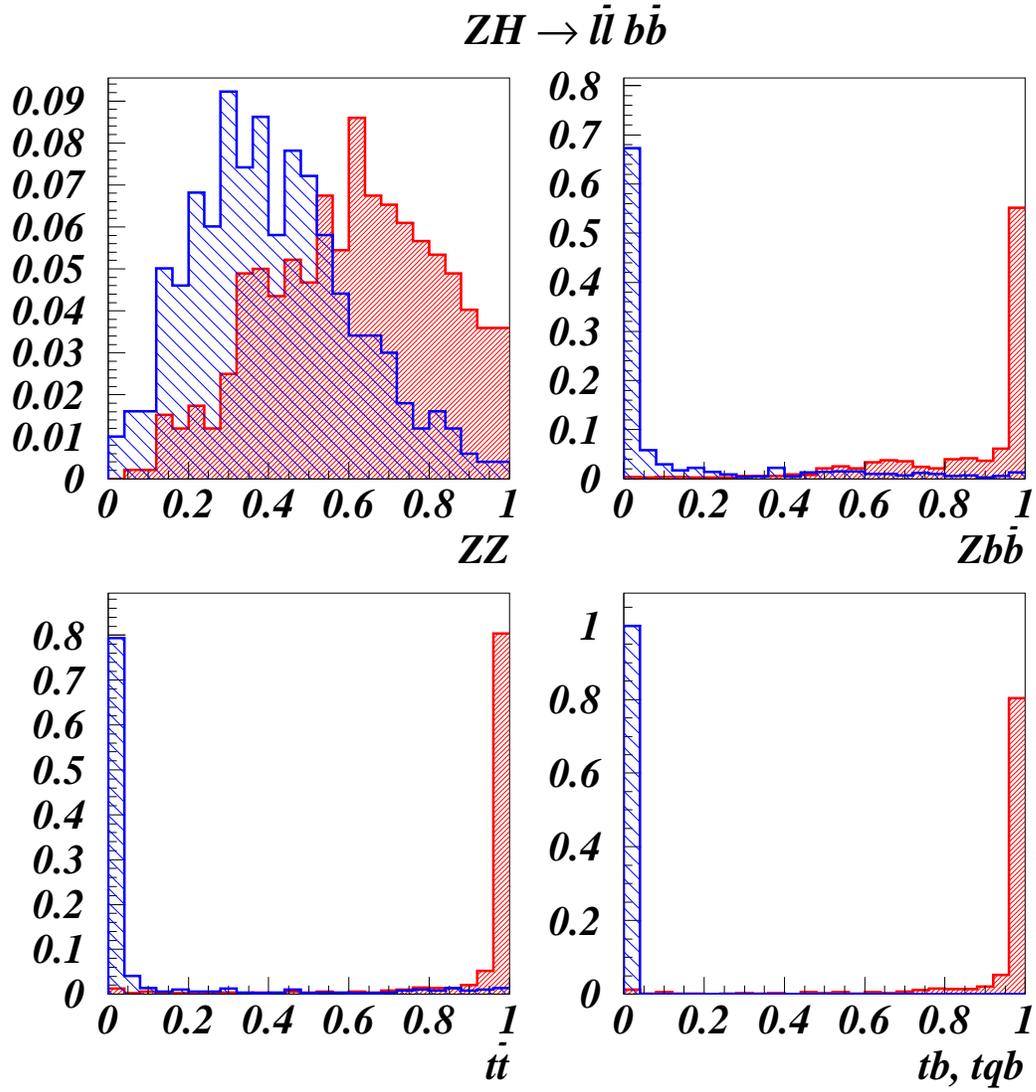}} }
  \end{center}
  \caption{Distributions of neural network outputs for selecting
           $ZH, Z\to\lpm$ events, for networks trained using signal 
           (heavy shading) vs. the backgrounds $ZZ$, $Z\bb$, 
           $t\bar{t}$ and single top.}
  \label{fig:zhll-net}
\end{figure}

\vspace{0.2in}
{\bf Results and Conclusions} \\  \nopagebreak

Figure~\ref{fig:zhll-net} shows the network output distributions.
From these figures it is clear that discriminating between $ZH$ and
the $ZZ$ backgrounds is much more difficult than for $WH$ and its
backgrounds.  Further study is needed to identify better variables if
the neural network technique is to prove as powerful in this channel
as in the others.  We summarize the results in Table~\ref{table5}.

    \subsubsection{$\qq\bb$ Channel}			\small
\begin{center}
{\it A. Goussiou, 
     J. Hobbs,
     J. Valls,
     R. Vilar,
     D. Zeppenfeld} \\
\end{center}
\normalsize \nopagebreak

Earlier workshops~\cite{tev2k,tev33} and this study have
identified the reactions $\pp\rightarrow WH\rightarrow l\nu\bb$ and
$\pp\rightarrow ZH\rightarrow \nn\bb$ as likely modes for discovering the
Higgs boson at the Tevatron.  However, the great majority of events
from the associated production modes $\pp\rightarrow WH$ and
$\pp\rightarrow ZH$ do not give final states in these channels.
Roughly 70\% of the final states arise from events in which the $W$ or
$Z$ has decayed hadronically.  These modes result in a four-jet final
state with two jets from the vector boson decay and two $b$ quark
jets from the Higgs decay.  Moreover, the same $\qq\bb$ final
state arises in the vector boson fusion process: $\qq\to VV +
\qq \to H + \qq$ ($V=W$, $Z$) when $H\to \bb$, the
dominant Higgs decay mode for $m_H\lsim 135$~GeV.  In the case of
$VV$ fusion, because there are two forward jets originating from the
$\qq$ in the final state, one may hope to be able to suppress the
Standard Model backgrounds by appropriate cuts.  Three analyses of
this final state are presented here.  The first is an extrapolation of
a published CDF Run~1 analysis and the second is a prototype analysis
using a combination of SHW and data.  These studies focus on the
$\qq\bb$ final state that arises in $WH$ and $ZH$ production.  The
third is a feasibility study of the vector boson fusion process and is
based on a parton-level Monte Carlo analysis.

 \vskip 4mm
 \centerline{ \large \it 4a. $\qq\bb$, Run 1 Extrapolation}
 \vskip 4mm \nopagebreak
   
CDF has recently published~\cite{juano} a search for Higgs bosons
produced in association with vector bosons in $91 \pm 7$ pb$^{-1}$ of
Run 1 data.  Observations are consistent with background
expectation. 95\% confidence level upper limits are set as a function
of the Higgs mass on $\sigma(p\bar{p} \rightarrow HV) \cdot \beta$ ,
with $\beta$ the branching ratio of the Higgs decays to
$b\bar{b}$. The sensitivity of the search is limited by statistics to
a cross section approximately two orders of magnitude larger than the
predicted cross section for standard model Higgs production. In this
paper we extrapolate the results from~\cite{juano} to the next
Tevatron Run 2 were we hope for an approximately twenty-fold increase
in the total integrated luminosity as well as a significant
improvement in the total acceptances.

\vspace{0.2in}
{\bf Selection Requirements} \\ \nopagebreak

The results from~\cite{juano} 
are based on a data sample recorded with a trigger
which requires four or more clusters of contiguous calorimeter towers,
each with transverse energy $E_T \ge 15$ GeV, and a total
transverse energy $\sum E_T \ge 125$ GeV. Further steps in the data reduction 
are: 1) four or more reconstructed jets with uncorrected
$E_T > 15$ GeV and $|\eta| < 2.1$. Jets are defined as localized 
energy depositions in the calorimeters and are reconstructed 
using an iterative clustering algorithm with a fixed cone of radius
$\Delta R = \sqrt{\Delta\eta^2 + \Delta\phi^2} = 0.4$ in $\eta - \phi$ space.
After this initial selection the sample
contains 207,604 events; 2) at least two among the four highest-$E_T$ 
jets in the event are identified (tagged) as $b$ quark 
candidates~\cite{secvtx}. There are 764 events with four or more jets and two 
or more $b$-tags; 3) a cut on the transverse momentum of the $b\bar{b}$ 
system, $p_T({b\bar{b}})\ge 50$ GeV/c$^2$ strongly discriminate against QCD 
direct production and flavor excitation of heavy quarks. A total of 589 
events remain after this cut.

\vspace{0.2in}
{\bf Backgrounds and efficiencies} \\ \nopagebreak

The main source of background events is QCD heavy flavor production. The 
heavy flavor content of QCD hard processes are conventionally classified in 
three groups: direct production, gluon splitting, and flavor excitation.
The relative contributions have been modelled with the PYTHIA Monte Carlo 
program~\cite{PYTHIA}. 
No attempt to estimate the normalization of this source of background directly 
from Monte Carlo is made due to the large number of uncertainties in the 
predictions. 

Other backgrounds are $t\bar{t}$ production, $Z$ + jets 
events with $Z\rightarrow b\bar{b}/c\bar{c}$
and fake double-tags. The first two are estimated from Monte Carlo and the 
last one from data.
 After trigger, kinematic and $b$-tag requirements 
we expect $26 \pm 7$ and $17 \pm 4$ $t\bar{t}$ and $Z$ + jets background 
events, respectively.

The combined trigger and acceptance efficiency is determined using PYTHIA 
followed by detector and trigger simulations. It depends on the Higgs mass 
and increases from $8\pm 1\%$ for $m_H = 70$ GeV/$c^2$ to $31 \pm 3\%$ 
for $m_H = 140$ GeV/$c^2$.
The SVX double $b$-tagging efficiency is calculated with a combination 
of data and Monte Carlo samples and is $14\pm 3\%$ with 
a small dependence on the Higgs mass.
The total efficiency increases linearly from 
$0.6\pm 0.1\%$ to $2.2\pm 0.6\%$ for Higgs masses ranging from 70 GeV/$c^2$ 
to 140 GeV/$c^2$.

\vspace{0.2in}
{\bf CDF Run 1 Results} \\ \nopagebreak

The shape of the observed $b$-tagged dijet invariant mass distribution is fit,
using a binned maximum-likelihood method, to a combination of  
signal, fake double-tag events, and QCD, $t\bar{t}$ and 
$Z$ + jets backgrounds. 
The QCD and signal normalizations are left free in the fit while
the normalizations of the $t\bar{t}$, $Z$ + jets and fakes are 
constrained by Gaussian functions to their expected values and 
uncertainties. 
The fit yields $\sigma_{VH^0} \cdot \beta = 44 \pm 42$ pb 
for $m_H = 70$ GeV/$c^2$, statistically compatible with zero signal.
For larger masses, zero signal contribution is preferred.
Table~\ref{tab1} shows the result of the fits  
as a function of the Higgs mass and Figure~\ref{fig1} shows the 
$b$-tagged dijet invariant mass distribution for the data 
compared to the results of the fit for $m_H \ge 80$ GeV/$c^2$.

\begin{table*}[t!]
\caption{Summary of the CDF Run 1 hadronic analysis fit results,
standard model predictions for $\beta\sigma$, and 95\% C.L. upper limits.}
\label{tab1}
\begin{tabular}{lcccccccc} 
 & \multicolumn{8}{c}{$m_H$ (GeV/$c^2$)} \\ [0.1cm] \cline{2-9}
\multicolumn{9}{c}{} \\ [-0.2cm]
 & 70 & 80 & 90 & 100 & 110 & 120 & 130 & 140 \\ [0.1cm] \tableline
\multicolumn{9}{c}{} \\ [-0.2cm]
Fit $\beta\sigma$ (pb) & $44 \pm 42$ & $0^{+19}_{-0}$ & $0.0^{+9.7}_{-0.0}$
&$0.0^{+7.6}_{-0.0}$ &$0.0^{+6.3}_{-0.0}$ & $0.0^{+5.9}_{-0.0}$ 
& $0.0^{+5.5}_{-0.0}$ & $0.0^{+5.1}_{-0.0}$ \\ [0.15cm]
SM $\beta\sigma$ (pb) & $1.13$ & $0.76$ & $0.55$ & $0.41$ & $0.30$ & 
$0.20$ & $0.12$ & $0.06$ \\ [0.1cm]
Limit $\beta\sigma$ (pb) & $117.3$& $53.2$& $28.9$& $22.8$& $18.7$
& $17.6$& $16.7$& $15.3$ \\ [0.15cm]
\end{tabular}
\end{table*} 

\begin{figure}[t]
\centerline{\epsfxsize3.2in\epsffile{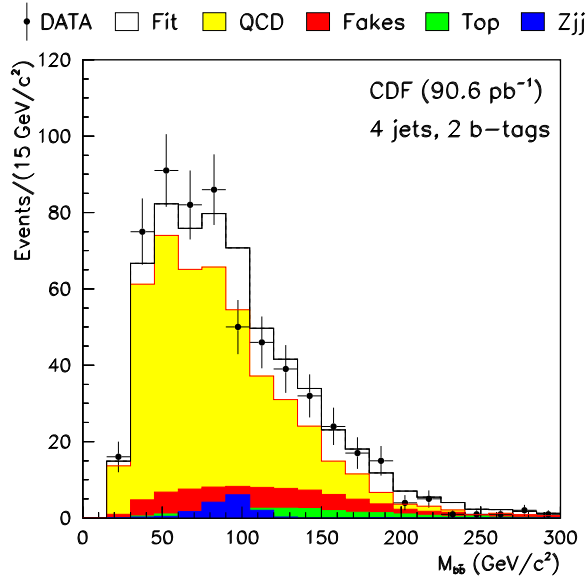}}
\caption[0]{Invariant mass distribution, $M_{b\bar{b}}$, for 90.6 pb$^{-1}$
of CDF data (points) compared to the fit prediction~\cite{juano}. The solid line is the 
sum of the QCD, fakes, $t\bar{t}$, and $Z$ + jets components.}
\label{fig1}
\end{figure}

\begin{figure}[t]
\hspace*{1.26cm}
\begin{minipage}[t]{2.5in}
\hspace*{0.26cm}
\centerline{\epsfxsize3.6in\epsffile{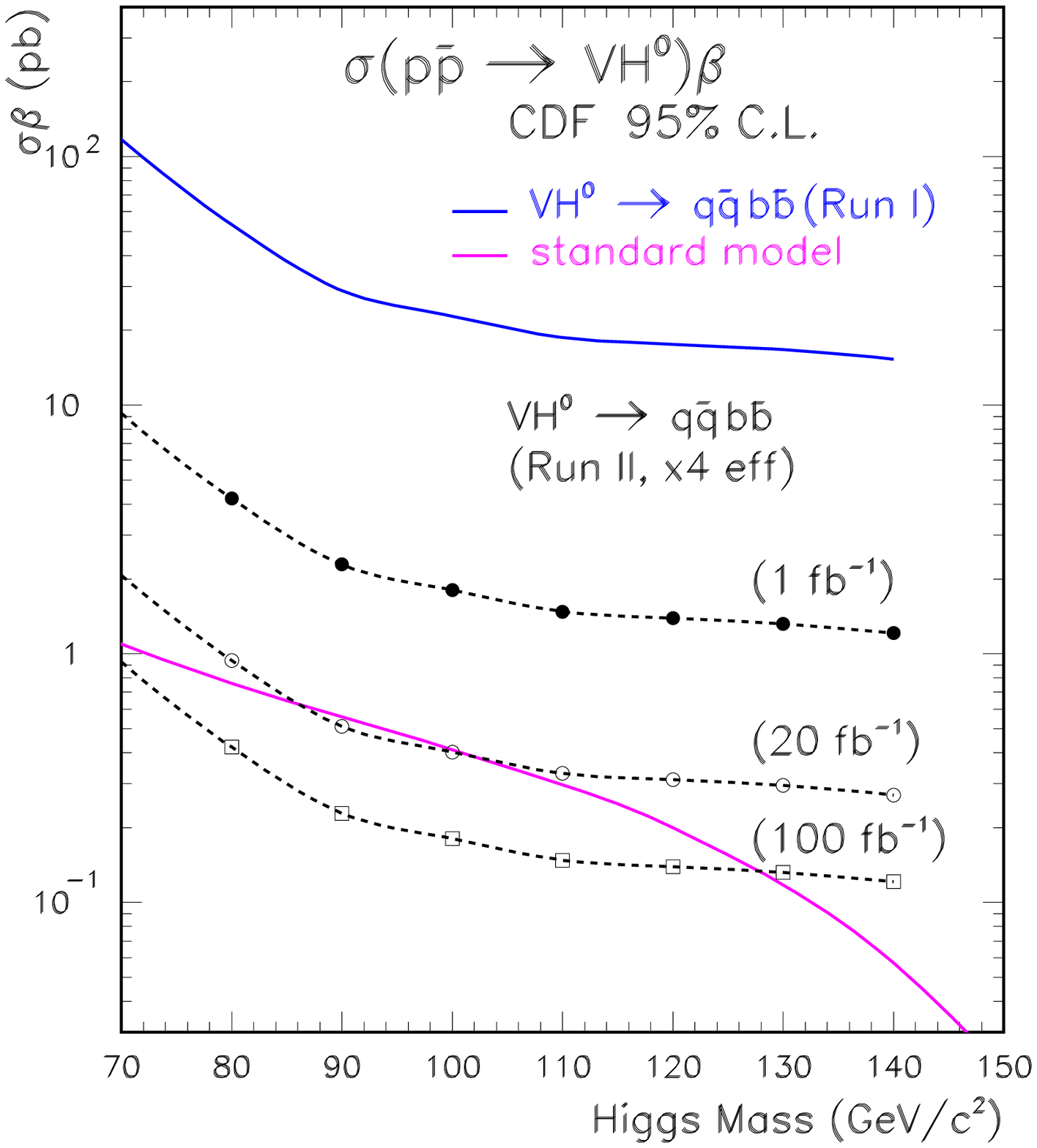}}
\caption[0]{The Run 1 CDF 95\% C.L. upper limits on $\sigma(p\bar{p} 
\rightarrow VH^0) \cdot \beta$ where $\beta = {\cal B}(H^0 \rightarrow 
b\bar{b})$. Also shown is the extrapolation for three different enhanced
luminosity scenarios and Run 2 acceptances.}
\label{fig2}
\end{minipage}
\hspace*{1.5cm}
\begin{minipage}[t]{2.5in}
\centerline{\epsfxsize3.6in\epsffile{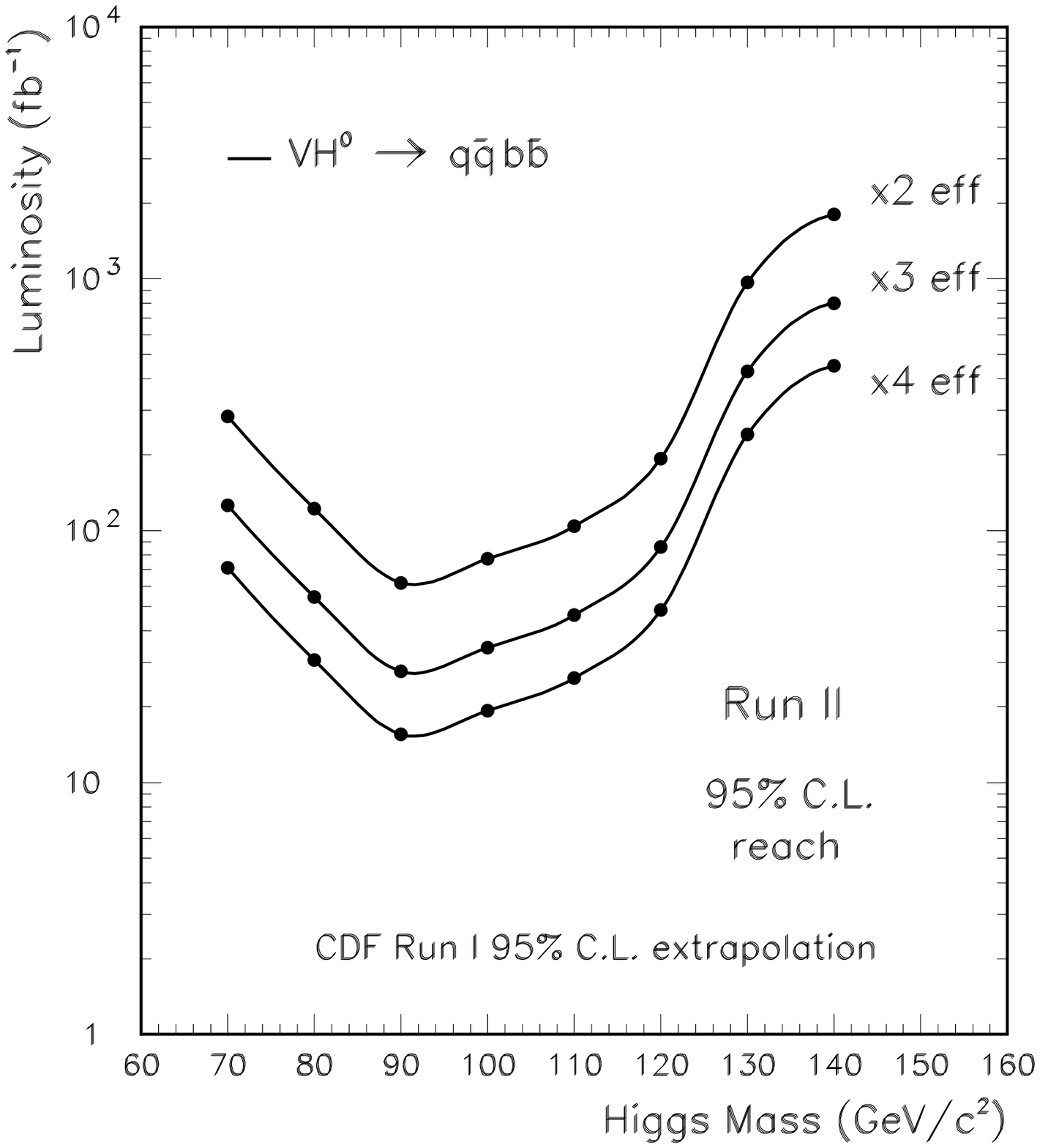}}
\caption{Luminosity as a function of the Higgs mass for a 95\% CL limit reach
in Run 2.}
\label{fig3}
\end{minipage}
\end{figure}

\vspace{0.2in}
{\bf CDF Run 2 Extrapolation} \\ \nopagebreak

Figure~\ref{fig2} shows the CDF Run 1 results for the 95\% CL upper
limits on $\sigma\beta$ as a function of the Higgs mass. In general,
the obtained 95\% CL statistical limits agree very well with the
expected ones, except for the region of very low masses were the
magnitude of the limits is strongly influenced by the fluctuation
structure of the~data.

A simple extrapolation to account for the expected increase in
luminosity and in the total acceptances lead to improved limits by, at
least, a factor $1/\epsilon\sqrt{{\cal L}}$.\footnote{No account is
taken here of the increase in center of mass energy from 1.8 TeV to
2.0 TeV.}  Here $\epsilon$ and ${\cal L}$ represent the increased
acceptances and luminosity factors with respect to Run 1. We consider
three different luminosity scenarios for Run 2 and beyond: 1, 20 and
100 fb$^{-1}$. In addition we consider the increased geometrical
acceptance of the new Run 2 CDF silicon vertex detector SVX II, which
will extend the $b$-tagging capabilities up to the range $|\eta|<2$. A
Monte Carlo study for the process $p\bar{p} \rightarrow V +
H^0\rightarrow jjb\bar{b}$ indicates a total signal improvement of
$\sim 80\%$ in double $b$-tag efficiencies (assuming Run 1 central
$b$-tag intrinsic efficiencies per jet). We finally consider an
optimized multijet trigger for associated Higgs production in Run
2. The requirement of displaced tracks with large impact parameter at
the trigger level makes it possible to relax the Run 1 multijet
criteria to only 3 jets with uneven $E_T$ cuts and a lower $\sum E_T$
threshold. A preliminary study~\cite{petar} shows that these cuts
result on a few-fold increase in signal efficiencies without
compromise the total trigger rates.

Figure~\ref{fig2} shows the CDF Run 1 95\% CL extrapolated limits for
the different luminosity scenarios combined with the roughly four times
better acceptance for Run 2. Figure~\ref{fig3} shows the 95\% CL
upper limit reach as a function of the required luminosity and the
Higgs mass for three different improved acceptances.

 \vskip 4mm 
 \centerline{ \large \it 4b. $\qq\bb$, SHW Analysis}
 \vskip 4mm \nopagebreak
   
\def\ttb{t\overline{t}}
\def\Zbb{Z\rightarrow b\overline{b}}
\def\WHqb{WH\rightarrow q\overline{q'} b\overline{b}}
\def\ZHqb{ZH\rightarrow q\overline{q} b\overline{b}}


\def\insertfig#1{
  \openin 1 #1
  \ifeof 1
    \closein 1
    \typeout{File #1 does not exist} \vskip 1truecm 
    \centerline{\fbox{Missing: #1}} \vskip 1truecm
  \else
    \closein 1
    \typeout{Reading figure in #1}
    \leavevmode\epsfxsize=\textwidth\epsfbox{#1}
  \fi
}
\def\OOO{\hphantom{O}}

The first of these sections describes the data samples and
normalizations used in the analysis.  The second describes the
selection criteria, and the third describes the relationship between
the analysis sensitivity and hypothetical improvements in event
reconstruction.  The last section contains the conclusions.

\newpage
\vspace{0.2in}
{\bf Event Samples} \\ \nopagebreak

For the channels considered here, the dominant background is QCD
production of four jet events, two of which are $b$-quark jets.
Other backgrounds include $\ttb$ and single-$t$ events, $WZ$ events,
$WW$ events, $W+2$~jet events and $Z+2$~jet events.  The relative
cross sections after a basic four-jet selection are given in
Table~\ref{t-bkg-xsec}.  One sees that the QCD multijet events
completely dominate the other backgrounds.  The multijet cross section
in Table~\ref{t-bkg-xsec} was computed by multiplying the total
four-jet cross section observed in a monitor data sample (described
below) by the fraction of simulated four-jet events which have two or
more $b$-quark jets.  The $\ttb$ cross section uses the D\O\
published value. The remaining cross-sections are purely theoretical
and were taken from either CompHep($W+\bb$) or Pythia(all others).
The QCD multijet background is roughly five orders of magnitude larger
than all others.  Because of this, the background calculations are
made using only QCD production.  All others are neglected.

\def\9{\hphantom{0}}
\begin{table}
\caption{$WH\rightarrow jj\bb$, SHW analysis.
 The total cross section times branching ratio to hadronic final
 states and accepted cross section times branching ratio for a basic four jet
 selection for background events.  One sees that the QCD multi-jet background
 is roughly five orders of magnitude larger than all others.   Only the QCD
 background is considered in the remainder of this analysis.\label{t-bkg-xsec} }
\begin{center}
\begin{tabular}{|l||c|c|} 
\noalign{\vskip-6pt}
   & \multicolumn{2}{c|}{Cross Section (fb)} \\
 Sample & Total & 4-jet Accepted \\ \tableline
Multijet   & --- & $\approx10^8$ \\
 $\ttb$    &  2600  & 2600 \\
 $WW$      &  4000  & 2700 \\
 $WZ$      &  1400  & \9900 \\
 $ZZ$      & \9490  & \9350 \\
 $W+\bb$   &   ---  & 2000
\end{tabular}
\end{center}
\end{table}

Run 1 monitor data taken with the D\O\ detector were used to provide the jet
rapidity and momentum distributions for the multijet background.  The sample
represented approximately 1~nb$^{-1}$ of exposure and events were required to
have passed a trigger of at least one L1 trigger tower with $\et\ge 5$~GeV and
at least three L2 trigger jets with $\et\ge 15$~GeV.  In addition standard
quality requirements were made.  The triggered sample contained approximately
$10^5$ events.  The flavor information of each jet in an event was randomly
assigned according to the flavor-by-jet-rank distributions from Pythia+SHW.

Signal samples of $WH$ events were generated using Pythia.  The $W$ boson was
forced to decay to quarks and the Higgs was required to decay to $\bb$.  The
quark flavors in the vector boson decay were distributed according to the
standard model expectation.  The Pythia four-vectors were then input to SHW
for detector simulation.  Samples were generated for Higgs masses of 80, 90,
100, 110, 120, 130 and 140~GeV.

The $ZH$ final states were not directly simulated.  Instead, it was assumed
that the acceptance and distributions were identical to those for the $WH$
modes.  The $W$ mass, $WH$ production cross section, and $W \rightarrow
q\bar{q^{\prime}}$ branching ratio, had of course been replaced by the
corresponding quantities for the $Z$ boson.  Thus, we report results for
inclusive production of $WH$ and $ZH$.

\begin{table}
\caption{$WH\rightarrow jj\bb$, SHW analysis.
         The acceptance as a function of Higgs mass for 
         the basic four-jet selection in the $WH$ channel.}
\label{t-sig-xsec}
\begin{center}
\begin{tabular}{cc} 
 $m_H$ (GeV) & Acceptance\\ \tableline
 \hphantom{1}80 &  0.11 \\
 \hphantom{1}90 &  0.13 \\
            100 &  0.14 \\
            110 &  0.15 \\
            120 &  0.16 \\
            130 &  0.15 \\
            140 &  0.13 
\end{tabular}
\end{center}
\end{table}

\vspace{0.2in}
{\bf Selection Criteria} \\ \nopagebreak

The analysis selection is quite simple.   The requirements are:
\begin{itemize}
  \item $N_{jet}\ge 4$ 
  \item $N_b\ge 2$
  \item $|\Delta\eta_{\bb}|\le 1,\ |\Delta\eta_{jj}|\le 1$
  \item $M_{jj} \ge M_W - \sigma(M_{jj})$\footnote{For the $ZH$ modes, $M_W$ 
    was replaced by $M_Z$.}
  \item $M_{\bb} \ge m_H - \sigma(M_{\bb})$.
\end{itemize}
Here $N_J$ is the number of $\Delta R =0.5$ cone jets in the event
which satisfy $|\eta|\le2.0$ and $p_T\ge15$~GeV.  $N_b$ is the number
of 0.5 cone jets tagged as $b$-quark jets using the SHW $b$-tag
algorithm based on a displaced vertex or a soft lepton.\footnote{This
is a parameterization based on true jet flavor and $p_T$.  It includes
mistag parameterizations for light quarks and $c$-quarks.}  Tag jets
must also satisfy $|\eta|\le2.0$ and $p_T\ge15.0$.  The variables
$\Delta\eta_{\bb}$ and $\Delta\eta_{jj}$ are the difference in
pseudorapidity $\eta$ between the two $b$-tagged jets
($\Delta\eta_{\bb}$) and the two non-tagged jets($\Delta\eta_{jj}$).
The masses $M_{\bb}$ and $M_{jj}$ are respectively the masses of the
$b$-tagged jet pair and the non $b$-tagged jet pair in the event.
For properly assigned flavors, these masses should be close to the
Higgs mass $m_H$ and the $W$ mass $M_W$.  The resolution parameters
$\sigma(M_{\bb})$ and $\sigma(M_{jj})$ are, respectively, the RMS
values of the mass distributions for the Higgs and $W$ decay products
seen in simulated events.


Other simple variables (including the $p_T$ of the $b\bar{b}$ system,
total energy and scalar $H_T$ of the four jets, rapidity boost of the
$b\bar{b}$ and $jj$ systems) have been investigated and found not to
contribute significantly beyond these.  The $\Delta\eta$ requirements
arise because the signal is peaked toward narrower opening angles
than the background.  This arises because the $W$ and Higgs are
boosted, so their decay products will on average be closer together
than any given jet pair from the four jets in multijet backgrounds.

The $WH$($ZH$) events should show peaks in the reconstructed dijet
mass spectra near the $W(Z)$ mass for non $b$-tagged jets and near
the Higgs mass for the $b$-tagged jets.  The multijet background, on
the other hand, should have no such peaks, but instead be a steeply
falling distribution in both masses.  This observation motivates the
last two requirements which simply demand that the reconstructed
masses lie near the $W$($Z$) and Higgs masses.  The distributions are
shown in two-dimensional plots, for $m_H=100$ GeV, in
Figure~\ref{f-mbkg-msig}.  In all figures, the background corresponds
to a luminosity of 1 pb$^{-1}$, while the signal plots correspond to
30.5~fb$^{-1}$.

\begin{figure}
  \begin{center}
    \parbox{6.0in}{\epsfxsize=\hsize\epsffile{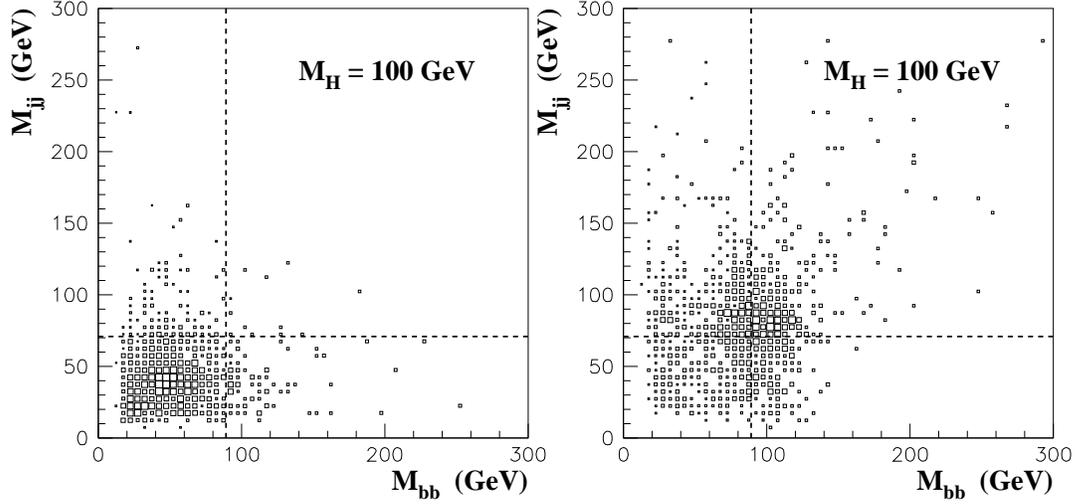}}
  \end{center}
  \caption{$WH\rightarrow jj\bb$, SHW analysis.
           $M_{jj}$ vs. $M_{\bb}$ for the multijet background sample
          (left histogram) and the $WH$ signal (right histogram).
          The background events represent 1 pb$^{-1}$ of integrated
          luminosity. The signal events correspond to 30.5 fb$^{-1}$.
          The dashed lines represent the applied selection criteria 
          on $M_{\bb}$ and $M_{jj}$ for a 100~GeV Higgs.} 
  \label{f-mbkg-msig}
\end{figure}

A number of different ways of imposing the mass condition were
considered.  In addition to the final choice of lower bounds on
$M_{\bb}$ and $M_{jj}$, elliptical regions centered around the masses
and diagonal lines were also considered.  In the absence of
backgrounds other than multijet production, the various methods gave
similar results, so the simple box was chosen.  If additional
backgrounds, particularly top production, would be added, a
requirement which also places an upper bound on the reconstructed
masses might also be needed.

The effects of the selection cuts on the background and signal samples
are shown in Table~\ref{t-effi} as number of events corresponding to
an integrated luminosity of 1 fb$^{-1}$.

\begin{table}
\caption{Number of events corresponding to 1 fb$^{-1}$ integrated
         luminosity, for $WH+ZH$ signal (S) and multijet background (B).
         Basic Jet selection includes the requirements
         on jet multiplicities, jet $E_T$'s and $\eta$'s.\label{t-effi}}
\begin{center}
\begin{tabular}{|c||r|c||r|c||r||c|c|} 
\noalign{\vskip-6pt}
                & \multicolumn{2}{c||}{Basic Jet Selection}
                & \multicolumn{2}{c||}
                 {$\Delta\eta_{\bb}$, $\Delta\eta_{jj}$ Selection}
                & \multicolumn{3}{c|}{Mass Cut} \\ \hline
  $m_H$ (GeV)   & $S\;\;\;$ & $B\;\;\;\;\,$ 
                & $S\;\;\;$ & $B\;\;\;\;\:$ 
                & $S\;\;\;$ & $B\;\;\;\;\:$ & $S/\sqrt{B}$ \\ \hline\hline 
\hphantom{1}80 & 58 & 860000 &27 & 200000 & 11    & 6800 & 0.13 \\
\hphantom{1}90 & 48 &   ''   &23 &   ''   &  8.1  & 5400 & 0.11 \\
           100 & 38 &   ''   &17 &   ''   &  5.6  & 3600 & 0.09 \\
           110 & 28 &   ''   &12 &   ''   &  3.5  & 2800 & 0.07 \\
           120 & 19 &   ''   & 8 &   ''   &  2.5  & 2300 & 0.05 \\
           130 & 10 &   ''   & 5 &   ''   &  1.3  & 2000 & 0.03 \\
           140 &  4 &   ''   & 2 &   ''   &  0.45 & 1700 & 0.01 \\
\hline
\end{tabular}
\end{center}
\end{table}

The data in Table~\ref{t-effi} were used to derive luminosity
requirements to achieve 95\% C.L. exclusion limits, and $3\sigma$ and
$5\sigma$ excesses.  The significance $s$ is defined as $s\equiv
S/\sqrt{B}$.  Here $S$ is the number of signal events for a given
luminosity, and $B$ is the number of background events in the same
luminosity.  This measure of significance corresponds to the
probability that the background fluctuates upward to at least the
number of observed signal events.  In all cases, one-sided, gaussian
confidence intervals were used.  The results are shown in
Figure~\ref{f-answ1} plotted as luminosity required versus Higgs mass.

\begin{figure}
  \begin{center}
    \parbox{4.0in}{\epsfxsize=\hsize\epsffile{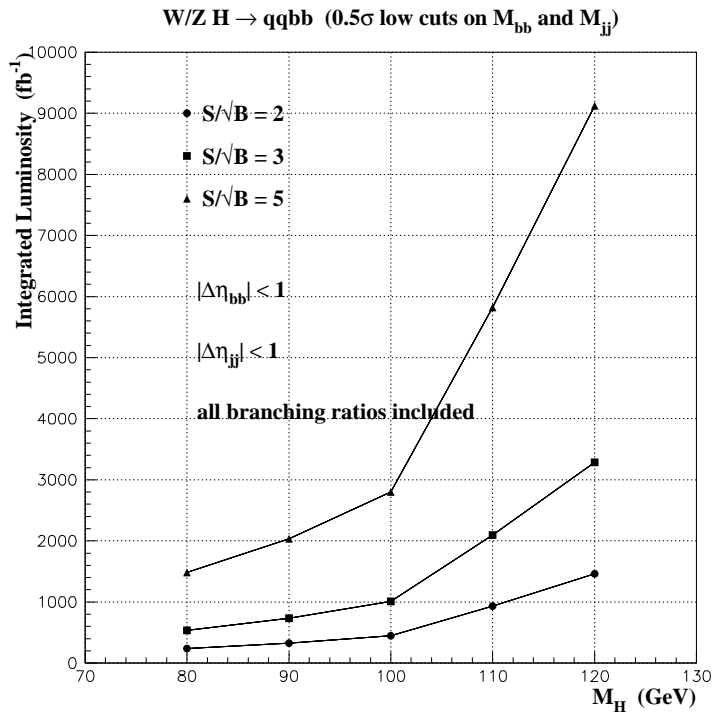}}
  \end{center}
  \caption{Luminosity required to achieve a given signal significance versus
           Higgs mass.}
  \label{f-answ1}
\end{figure}

\vspace{0.2in}
{\bf Improvements} \\ \nopagebreak

There are a number of ways in which this analysis could be improved.  Among
them are:
\begin{enumerate}
  \item the reconstructed mass resolution is improved relative to the SHW
    results,
  \item the $b$-quark jet identification efficiency or purity is improved and
  \item other analysis techniques, particularly multivariate likelihoods or
    neural nets, giving significantly improved background rejection or signal
    acceptance.
\end{enumerate}
We have attempted to assess the improvements these might offer by
computing the luminosity required to reach a given statistical
significance for a hypothetical improvement in discrimination were
made.

The most likely candidate is improved $\bb$ mass resolution.  The
current jet finding algorithm uses only calorimeter information.  It
has already been shown that including additional information will
improve the reconstructed mass.  Also, the jet corrections used in
this analysis were designed only to correctly reproduce the peak
positions of the mass distributions.  No attempt was made to minimize
the resolution.  Figure~\ref{f-mbb-plus} shows the significance versus
mass resolution for $m_H=100$~GeV and 1~fb$^{-1}$.  This result was
derived by improving the signal mass resolution $\sigma$ by a fixed
amount and making the same box selection as the standard analysis.
Because $\sigma$ is smaller, the selection requirement,
$M_{jj}\geq M_W-\sigma$, corresponds to a higher minimum mass.  Thus,
this procedure corresponds to a fixed acceptance (for the signal) and
decreasing background.  One sees that a factor of two improvement in
mass resolution corresponds to a factor of four reduction in
background and thus a factor of two improvement in significance.

\begin{figure}
  \begin{center}
    \parbox{3.0in}{\epsfxsize=\hsize\epsfbox{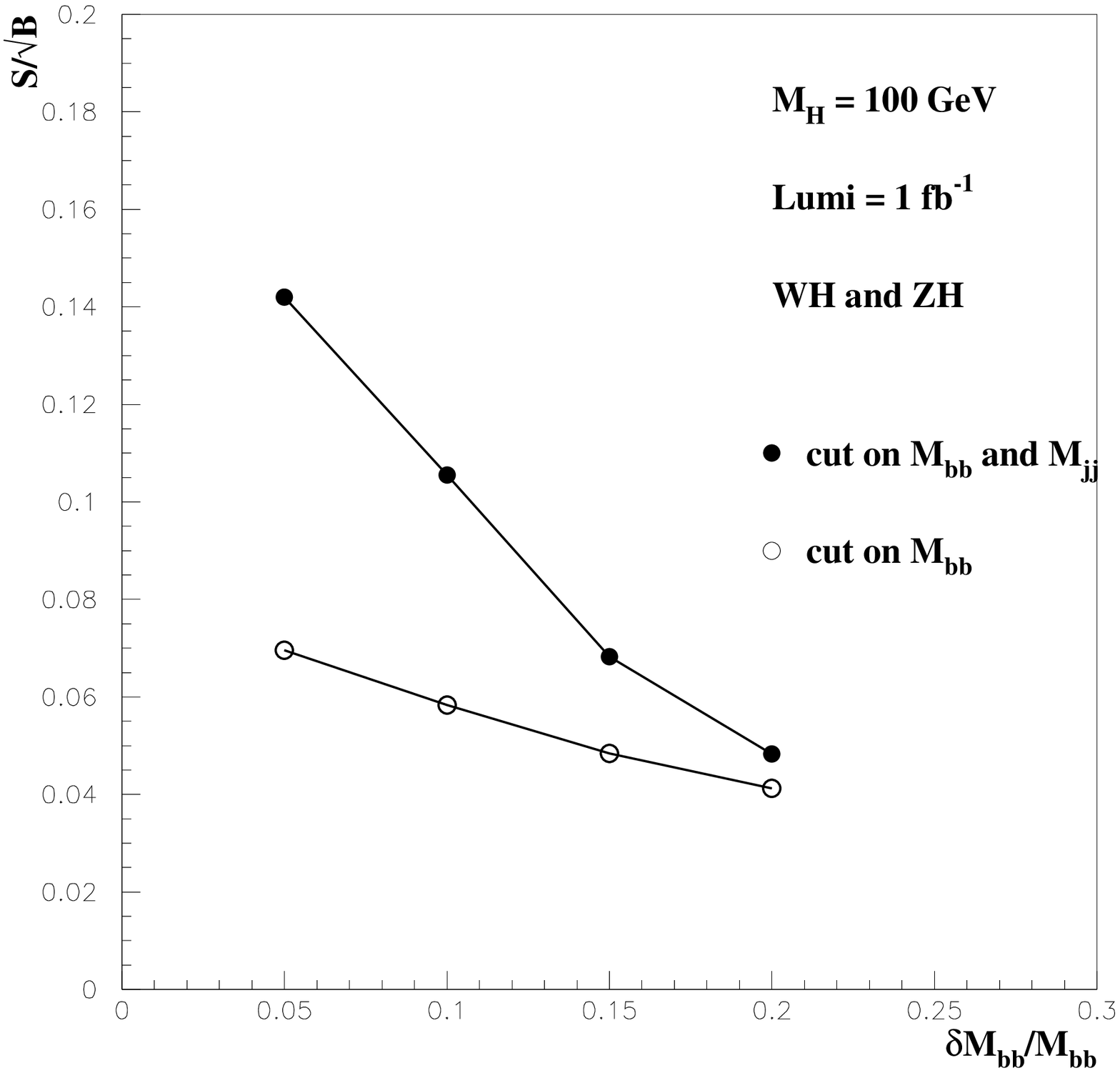}}
    \parbox{3.0in}{\epsfxsize=\hsize\epsfbox{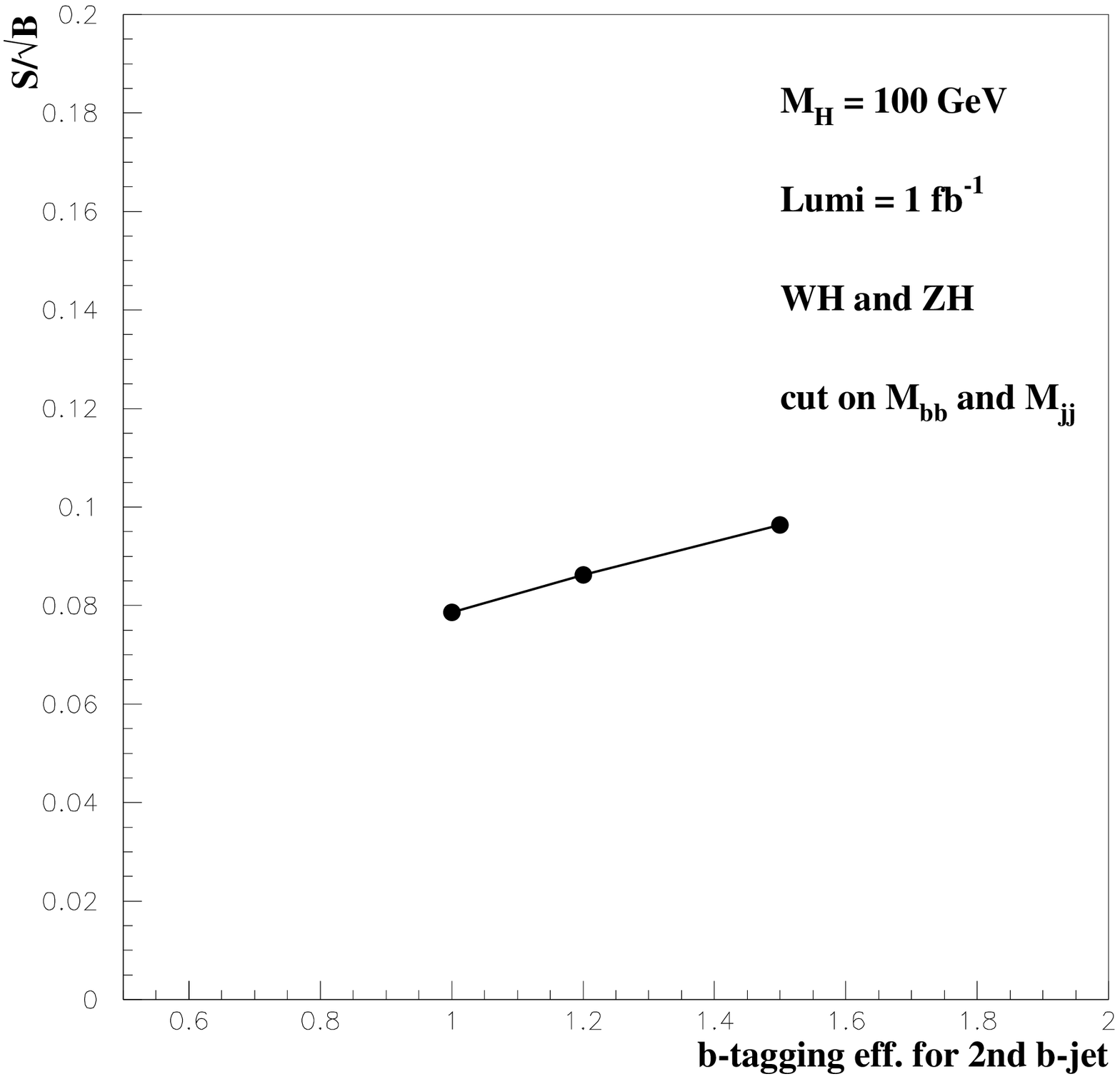}}
  \end{center}
  \caption{Signal significance as a function of mass resolution (left-hand
            panel) and as a function of second-jet tag probability (right-hand panel).
            The standard SHW processing gives $\delta M/M\approx 20$\% for the 
            four-jet final states considered here.}
\label{f-mbb-plus}
\end{figure}

It is also possible that as the Run 2 data are processed, we will be
able to increase the $b$-tagging efficiency or purity.
Figure~\ref{f-mbb-plus} also shows the significance as a function of
tagging efficiency of the second $b$-tagged jet.  The improvement is
realized by multiplying the standard SHW tag probability for the
second jet by the factor shown on the abscissa with the constraint
that the efficiency never exceed unity.  One sees that a 50\%
improvement in efficiency gives only a modest gain in significance.

\vskip 4mm 
 \centerline{ \large \it 4c. $\qq\bb$ with Forward Jet Tags}
 \vskip 4mm \nopagebreak
   
We consider the weak gauge boson fusion process, $\qq\to \qq\hsm$,
where the quark and anti-quark both radiate vector bosons which then
annihilate to produce the Higgs boson.  This includes $WW$ and $ZZ$
fusion originating from all allowed combinations of initial quarks
and/or antiquarks.  The two scattered (anti)quarks are observed in the
detector as forward jets, of $E_T\approx 20$ to 80~GeV and provide a
telltale signature for the process. Tagging these forward jets is a
very powerful tool to suppress backgrounds, and has already been shown
to be a highly promising tool for intermediate mass Higgs boson
studies at the LHC~\cite{wbfLHC}.
 
\vspace{0.2in}
{\bf Selection Criteria} \\ \nopagebreak

Here, we concentrate on the dominant decay mode for the Higgs mass
region of interest, $H\to \bb$. The Higgs boson would appear as a
$\bb$ invariant mass peak in 4-jet events in which two of the jets
are identified as $b$ quarks. Let us consider the case of
$m_H=120$~GeV as an example. The signal rate is given by
\begin{equation}
   \sigma(qq\to qqH)\;B(H\to\bb) = 58\;{\rm fb}\;,
\end{equation}
which is reduced to about 40~fb after minimal acceptance cuts of
\begin{eqnarray}
   E_T(b) &>& 20\;{\rm GeV}\;, \qquad |\eta(b)|<2 \;, \\
   E_T(j) &>& 10\;{\rm GeV}\;, \qquad |\eta(j)|<4 \;, \\
   \Delta R_{jj} &=& \sqrt{(\eta_{j_1}-\eta_{j_2})^2+
  (\phi_{j_1}-\phi_{j_2})^2} > 0.7\,,
\end{eqnarray}
where $\eta$ and $\phi$ is the corresponding pseudorapidity and
azimuthal angle. Here the jet separation cut is applied to all four
jets, including the two $b$-quark jets.

\vspace{0.2in}
{\bf Backgrounds} \\ \nopagebreak

An irreducible background to these signal events arises from real
emission QCD corrections to $\bb$ production. We have calculated
the SM $\bb jj$ cross section with full ${\cal O}(\alpha_s^4)$
matrix elements, i.e. with a tree level parton Monte Carlo program. We
find a cross section of 2.4~nb for this irreducible background within
the minimal cuts given above and for $b$-quark pair invariant masses
above 100~GeV.
 
A significant reduction of the background is possible by exploiting
the characteristic features of the weak boson fusion signal. The two
tagging jets of the signal typically have $E_T(j)>20$~GeV, and they
are widely separated, $\eta_{jj} = | \eta_{j_1}-\eta_{j_2}|>3$, with
the two $b$-quark jets located between these two tagging jets, which
are required to reside in opposite detector hemispheres. In addition
we can increase the transverse momentum requirement for the $b$-quark
jets to 40~GeV without undue loss of signal acceptance. These
additional cuts reduce the $\bb jj$ background to 30~pb for
$m_{bb}>100$~GeV while keeping a signal rate of $\sigma(qq\to
qqH,\;H\to\bb)=9$~fb.

\vspace{0.2in}
{\bf Results and Conclusions} \\ \nopagebreak

The Higgs signal would have to be seen as a $\bb$ invariant mass
peak in this high background environment. Even assuming a $\bb$
invariant mass resolution of $\delta m = \pm 10$~GeV, the signal to
background ratio does not exceed 1/1000 within a 20~GeV invariant mass
bin.  Additional distributions exist which differ somewhat between the
Higgs signal and the $\bb jj$ background. However, we have been
unable to find cuts which significantly improve the statistical
significance of the Higgs signal.  The SM Higgs search via weak gauge
boson fusion does not appear promising at the upgraded Tevatron.

    \subsubsection{Exclusive Higgs Production}		\vspace{0.1in}
\small
\begin{center}
{\it M. Albrow, 
     D. Litvintsev,
     P. Murat,
     A. Rostovtsev
     }
\end{center}
\normalsize \nopagebreak

There is a possibility that a light ($M_H$ less than about 135 GeV)
Higgs can be discovered at the Tevatron by the {\em exclusive}
production process $p + \bar{p} \rightarrow p H \bar{p}$.  This would
lead to events with two $b$ quark jets in the central detector, and in
the far forward and backward directions the scattered proton and
antiproton.  One could select such events by triggering on the central
jets, and by measuring the trajectories of the scattered beam
particles one could reconstruct very precisely the Higgs
four-momentum.

The QCD process $gg \rightarrow b\bar{b}$ is overwhelming if one
simply tries to reconstruct the di-jet invariant mass spectrum.  While
the mass resolution in reconstructing a $\bb$ jet pair at $M_{JJ}$
100-130 GeV is anticipated to be about 10\% as shown in the previous
sections, the actual width of a Higgs in this mass region is less than
10 MeV~\cite{daws}, so if one can improve the mass resolution the
signal to background ratio $S/B$ keeps improving.

In the exclusive channel there is a possibility of a factor about
50-100 improvement, limited mainly by the momentum spread of the
incoming beams.  The visibility of the Higgs by this technique depends
on the size of the cross section for the process where the Higgs is
produced (in the central region) completely exclusively, i.e. the $p$
and $\bar{p}$ go down the beam pipes each having lost about
$\frac{M_H}{2}$ in longitudinal momentum and no other particles are
produced. The mechanism is as usual $gg\rightarrow H$ through
intermediate loops of heavy particles (predominantly a top loop); this
leaves the $p$ and $\bar{p}$ in color-octet states which normally
gives rise to color strings filling rapidity with hadrons. However
some fraction of the time another gluon can be exchanged which
neutralises (in a color sense) the $p$ and $\bar{p}$ and can even
leave them in their ground state.  The difficulty is in calculating
what this fraction is.  It involves non-perturbative QCD: soft gluon
calculations, soft color interactions, or Regge theory etc. In Regge
theory it is the double pomeron exchange ($DPE$) process.

In 1990 Sch\"{a}fer, Nachtmann and Sch\"{o}pf~\cite{scha} made
calculations of diffractive Higgs production at the LHC and SSC, and
concluded that cross sections for the exclusive $DPE$ reaction could
not be determined reliably. They concluded that ``great effort should be
invested experimentally and theoretically to better understand the
properties of the pomeron and to be able to make more precise
predictions.''  In 1991 Bialas and Landshoff~\cite{bial} calculated
from Regge theory that some 1\% of all Higgs events may have the $p$
and $\bar{p}$ in the $DPE$ region of very high Feynman $x_F \approx
0.95$, but did not estimate the fully exclusive process. In 1995
Cudell and Hernandez~\cite{cude} made a lowest order QCD calculation
with the non-perturbative form factors of the proton tuned to
reproduce elastic and soft diffractive cross-section
measurements. They find, for the exclusive (they call this
\emph{elastic}) production of Higgs with $M_H$ = 100(130) GeV at
$\sqrt{s}$ = 1.8 TeV a cross section of about 50(30) fb, with an
uncertainty about a factor of 2.  The total Higgs production cross
section (at 2 TeV) is about 1200(600) fb at these masses~\cite{daws},
so the exclusive fraction is 4\% to 5\%, even larger than that in the
Bialas and Landshoff paper.  However there are issues of ``rapidity
gap survival probability" and whether they are or are not implicitly
included, and renormalization of the ``pomeron flux", initial and
final state interactions, shadowing effects etc. All these phrases are
poorly understood but are probably different ways of saying the same
thing, namely that diffractive cross sections are lower than
na\"{\i}ve Regge expectations by about an order of magnitude in
hadron-hadron collisions (not in $e-p$ collisions). Understanding that
there is a large uncertainty in this estimate, we apply a factor of 10
reduction to the estimate of ref.~\cite{cude} and get about 4 fb (the
increase of $\sqrt{s}$ from 1.8 to 2.0 TeV will also help).

That would mean, in Run~2 (2 fb$^{-1}$) only 8 events (before
applying any efficiencies). It appears that 2 fb$^{-1}$ is very
difficult for \emph{any} Higgs searches, so we consider a Run 3 with
30 fb$^{-1}$ which could give 120 events, or with 50\% acceptance 60
events. This could be a discovery channel if the background is about
140 events or less. Now we come to the key of our idea.

For these exclusive events, with the Higgs produced centrally, the $p$
and $\bar{p}$ each lose about $\frac{M_H}{2}$ in momentum, so they
have $x_F \approx 0.94$.  They could both be detected in Roman pot
detectors and, with high resolution ($\sim$10 $\mu m$) tracking their
momenta can be measured with very high precision ($\frac{\Delta p}{p}
\approx 10^{-4}$). This needs a measurement of the primary vertex (at
the intersection of the two $b$-jet directions, using SVX tracking)
and good knowledge of the quadrupole, electrostatic and dipole fields
from the vertex to the pots. Then the missing mass ($m_{miss}$) to the
measured $p$ and $\bar{p}$ will have a resolution limited by the
spread on the incoming beam particles' momenta. This is about
10$^{-4}$, and gives a mass resolution on the selected $b\bar{b}$
dijet candidates of about 200 MeV\footnote{These numbers are still
largely ``back of the envelope" estimates; a proper simulation has yet
to be done.}. This is a factor about 50 better than the resolution on
the effective mass ($m_{eff}$) of the dijet and hence we get a factor
50 improvement in $S/B$. Furthermore, even if the Higgs were to be
discovered another way, this process has the potential of giving the
best value for $M_H$. We should now try to estimate the $S/B$ ratio to
see whether there is hope that the signal could be visible.

We take CDF's published cross section~\cite{cdfb}
$\frac{d\sigma}{dM_{JJ}}$ for two b-tagged jets, which starts at 150
GeV, and extrapolate down to 100(130) GeV finding 200(40) pb/GeV (in
$|\eta| < 2.0, |cos(\theta^*)| < 2/3$).  From our other $DPE$ studies,
at lower mass, we expect that about $2.10^{-5}$ of these would have a
rapidity gap on each side (assuming this fraction is not
$E_T$-dependent). That gives $\approx $ 4.0(0.8) fb.GeV$^{-1}$, or
0.8(0.16) fb per 200 MeV bin to be compared with a signal of around
5(3) fb in such a bin.  We thus have a $S/B$ of order 6(20) (with the
assumed reduction of the Cudell and Hernandez estimate by a factor
10).  In 10 fb$^{-1}$ this means 30 events on a background of 1.6 for
a 130 GeV Higgs! We have not included a factor for $b$-tagging
efficiency and acceptance, but that affects both signal and background
proportionately, to first order. We have also not put in the
acceptance of the Roman pots for these events, which depends on
Tevatron-dependent factors but should be high if the
$|t|$-distribution of the $p$ and $\bar{p}$ is not much harder than in
generic high mass diffraction.

We think it is very important to study the backgrounds as early as
possible in Run~2. This means covering the region forward of $\eta$ =
3.5 with miniplug calorimeters (supplementing the CLC counters) and
Beam Shower Counters out to the beam rapidity ($\approx 7.5$).  One
trigger is $\ge$ 2 jets with nothing in both forward directions.
(Looser prescaled triggers are needed to measure the non-$DPE$
backgrounds.) We can measure $d\sigma/dM_{JJ}$ vs $M_{JJ}$ for generic
jets and for tagged $b$-jets out to $M_{JJ}$ above 100 GeV but with
poor resolution and relatively wide bins. By this time we should be
better able to judge how promising the Higgs search may be.

If it does look promising then, to do the Higgs search, an additional
arm of Roman pot detectors must be placed on the $p$-side after
dipoles to give acceptance for $x_F \approx 0.94$ down to $p_T$ = 0 as
we have now on the $\bar{p}$-side. At present there is no warm space
(spool-piece) where such pots could go. The simplest way to make such
a space is by removing the Q1 quadrupoles which are redundant in Run~2
(they are being taken out at D\O) and to move the dipoles closer to
the interaction point.  A Beams Division study would be needed, but
probably a lever arm of 1 m for pot detectors could be made, giving
$\frac{\sigma_{\theta}}{\theta} \approx 10^{-5}$ if positioning
accuracy of 10$\mu m$ can be achieved (which is overkill, given the
uncertainty on the incoming momenta). Moving the dipoles closer to B\O
(symmetrically on the $p$ and $\bar{p}$ sides) would mean moving the
CDF detector by about 2 cm; we do not see that as a major problem.

This Higgs search trigger is then a high $x_F$ track in each forward
arm, no hits in any BSC counter or either miniplug (i.e. rapidity gaps
from $3.5 < \eta < 7.5$) and two central jets above about 20 GeV in
$E_T$.  At a later level trigger and/or offline one selects two tagged
b-jets and requires $X_{JJ} = \frac{M_{JJ}}{M_{cen}}$ large, where
$M_{cen}$ is the total effective mass of the central calorimeter
towers. After low missing $E_T$ and other clean-up cuts we would
require $m_{miss} \approx m_{eff}$ and plot $m_{miss}$ in bins
dictated by the resolution. This strategy will only work for single
interactions per bunch crossing, because additional interactions will
spoil the gaps and pile up in $M_{cen}$. The search will be more
difficult in the presence of multiple interactions but should still be
possible.  Precision timing on the pot tracks can (a) check that the
$p$ and $\bar{p}$ came from the same interaction (b) measure the
position in $z$ of that interaction, and (c) correlate with the timing
of the $b$ and/or $\bar{b}$ jets from the time of flight barrel (in
CDF). A fast trigger processor could be used to correlate $m_{miss}$
and $m_{eff}$ and record events when these are comparable. Off-line
the cleanliness of the vertex from which the $b$ and $\bar{b}$ emerged
is used as a cut.

There is another process with a similar diagram to the exclusive Higgs
production but with a larger cross section and this could be used to
test our understanding. This is $p+p\rightarrow p+\chi_b^o+ p $
followed by $\chi_b^o \rightarrow \Upsilon\gamma \rightarrow
\mu^+\mu^-\gamma$.  The $\chi_b^o$ has the same quantum numbers
($I^GJ^{PC} = 0^+0^{++}$) as the vacuum and the Higgs. A measurement
of the fraction of all (low $p_T$) $\chi_b^o$ which are produced in
total isolation would be instructive, and could probably be made with
a 2-$\mu$ 2-gap trigger early in Run~2. One can also study exclusive
$\chi_c^o$ production, using the time of flight barrel in a Level 1
trigger on the $2(\pi^+\pi^-)$ or $\pi^+\pi^-K^+K^-$ final states.

  \subsection{High-mass Standard Model Higgs Bosons: 130--190 GeV}  \vspace{0.1in}
\small
\begin{center}
{\it Tao Han,
     A. Lucotte,
     M. Schmitt,
     A.S Turcot,
     R.-J. Zhang
     }
\end{center}
\normalsize \nopagebreak

\def\mh{m_h^{}}
\def\gev{\rm GeV}
\def\fbi{\rm fb^{-1}}
\def\ww{W^*W^*}
\def\zz{Z^*Z^*}
\def\lsim{\mathrel{\raise.3ex\hbox{$<$\kern-.75em\lower1ex\hbox{$\sim$}}}}
\def\gsim{\mathrel{\raise.3ex\hbox{$>$\kern-.75em\lower1ex\hbox{$\sim$}}}}
\def\ljj{\ell\nu jj}
\def\lljj{\ell\bar\ell jj}
\def\jj{\protect jj}
\def\llnn{\ell\bar\ell\nu\bar\nu}

\newcommand{ \slashchar }[1]{\setbox0=\hbox{$#1$}   
   \dimen0=\wd0                                     
   \setbox1=\hbox{/} \dimen1=\wd1                   
   \ifdim\dimen0>\dimen1                            
      \rlap{\hbox to \dimen0{\hfil/\hfil}}          
      #1                                            
   \else                                            
      \rlap{\hbox to \dimen1{\hfil$#1$\hfil}}       
      /                                             
   \fi}                                             %

\def\ptmiss{\slashchar{p}_{T}}
\def\etmiss{\slashchar{E}_{T}}

\input{epsf}
\ifx\epsffile\undefined
\message{(Uncomment input epsf to include figures)}
\newlength{\epsfysize}
\def\epsffile#1#2#3#4]#5{}
\fi

As previous studies and the present studies have
shown, in the Higgs mass region where the $\bb$ decay mode dominates
the most sensitive channels are those arising from the associated 
production of an electroweak gauge boson and the (SM, or SM-like) 
Higgs boson $h$~\cite{Stange:1994ya,GH,steve}:
\begin{equation}
   p\bar p \to W hX,\ ZhX.
  \label{whzh}
\end{equation}
However, as shown in the first section of this report the leading
production mechanism for a SM-like Higgs boson at the Tevatron is the
gluon-fusion process via heavy quark triangle loops
\begin{equation}
  p \bar p \to gg X\to h X.
  \label{gluon}
\end{equation}

There are also contributions to $h$ production from the vector boson
fusion processes\footnote{Here and henceforth, $W^*(Z^*)$ generically
denotes a $W(Z)$ boson of either on- or off-mass-shell.}
\begin{equation}
  \ww,\ \zz \to h,
  \label{wwzz}
\end{equation}
where $\ww$ and $\zz$ are radiated from the quark partons.  In
Fig.~\ref{fg:4}, we present cross sections for SM Higgs boson
production at the Tevatron for processes (\ref{whzh}), (\ref{gluon})
and (\ref{wwzz}). We see that the gluon fusion process yields the
largest cross section, typically a factor of four above the associated
production (\ref{whzh}). For $\mh>160$ GeV, $WW,ZZ$ fusion processes
become comparable to that of (\ref{whzh}). In calculating the total
cross sections, the QCD corrections have been included for all the
processes~\cite{19,23,32}, and we have used the CTEQ4M
parton distribution functions~\cite{CTEQ}.

\begin{figure}[tbp]
  \begin{center}
    \parbox{5.5in}{\epsfxsize=\hsize\epsffile{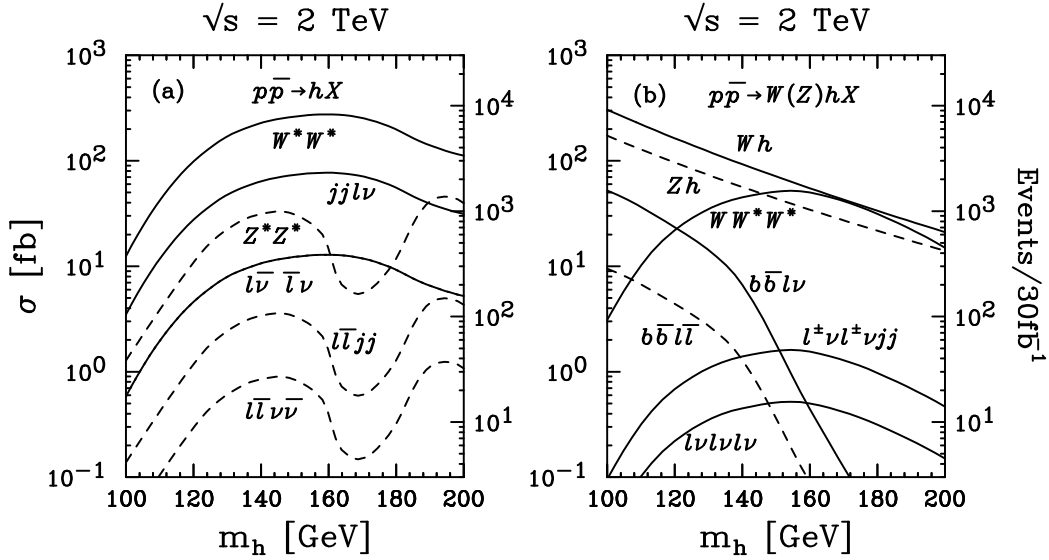}}
  \end{center}
  \caption[]{Standard Model Higgs boson production cross sections
             (in fb) and various subsequent decay modes versus $m_h$ for
             (a) $gg\to h \to W^*W^*$ (solid curves) and $Z^*Z^*$ (dashed
             curves). (b) $q \bar{q}' \to Wh$ with $h \to W^*W^*$ (solid
             curves) and $Zh$ (dashed curves). Also shown at $h \to b \bar{b}$
             with leptonic $W,Z$ decays. The right hand scale indicates the
             number of events per 30 fb$^{-1}$ integrated luminosity (per detector).}
  \label{sigxbr}
\end{figure}

Although the decay mode $h\to b\bar b$ in Eqs.~(\ref{gluon}) 
and (\ref{wwzz}) would be swamped by the QCD background, 
the decay modes to vector boson pairs, 
\begin{equation}
 h\to W^*W^*,\ Z^*Z^* \ \ ,
\label{vv}
\end{equation}
will have increasingly large branching fractions
for $m_h\gsim 130$ GeV and are natural channels to consider
for a heavier Higgs boson. 

In Fig.~\ref{sigxbr}(a), we show the cross sections for $gg\to h$
with $h\to W^*W^*$ and $Z^*Z^*$ versus $\mh$ at $\sqrt s=2$ TeV. 
The leptonic decay channels are also separately shown 
by solid and dashed curves, respectively, for
\begin{eqnarray}
 h \to && \ww \to \ell\nu jj\ \ {\rm and}\ \ 
\ell\bar\nu \bar\ell \nu ,\label{sigww}\\ 
       && \zz \to \ell\bar \ell jj\ \ {\rm and}\ \ 
\ell\bar \ell \nu\bar \nu ,
\label{sigzz}
\end{eqnarray}
where $\ell=e,\mu$ and $j$ is a quark jet.  Although the $\ljj$ mode
has a larger production rate, the $\llnn$ mode is cleaner in terms of
the SM background contamination.  The corresponding modes from $\zz$
are smaller by about an order of magnitude.  At the LHC, the 
$\lpm\nn$ channel is an important discovery mode~\cite{dreiner}.

It is natural to also consider the $h\to \ww$ mode from the $Wh$
associated production in Eq.~(\ref{whzh}).  It was noted \cite{Stange:1994ya}
that associated production with $h\to\ww$ can lead to a unique 
like-sign lepton signature via
\begin{eqnarray}
W^\pm h \to \ell^\pm \nu\  \ww \to 
&& \ell \nu\ \ell\nu\  \ell\nu,
\label{unlikesign}\\ 
&& \ell^\pm \nu\ \ell^\pm \nu\  jj.
\label{likesign}
\end{eqnarray}

The production rates for these modes are denoted by solid curves in
Fig.~\ref{sigxbr}(b).  The trilepton signal is smaller than the
like-sign lepton plus jets signal by about a factor of three due to
the difference of $W$ decay branching fractions to $\ell=e,\mu$ and to
jets.  For comparison, also shown in Fig.~\ref{sigxbr}(b) are $Wh\to
b\bar b\ell\nu $ (solid) and $Zh\to b\bar b\ell\bar \ell$ (dashed) via
$h\to b\bar b$.  We see that the signal rates for these channels drop
dramatically for a higher $\mh$.  Comparing the $h$ decays in
Fig.~\ref{sigxbr}(a) and (b), it makes the gauge boson pair modes of
Eq.~(\ref{vv}) a clear choice for Higgs boson searches beyond 130 GeV.

The purely leptonic channel in Eq.~(\ref{sigww}) has been studied at
SSC~\cite{lnulnu} 
and LHC energies~\cite{dreiner} and at a 4 TeV
Tevatron~\cite{GH}. Despite the difficulty in reconstructing $\mh$ from
this mode due to the two missing neutrinos, the obtained results for
the signal identification over the substantial SM backgrounds were all
encouraging.  More recently for the 2 TeV Tevatron upgrade, there has
been a parton level study~\cite{parton_studies} for the $\ww$ channels
of Eq.~(\ref{sigww}) and a study at the detector level for the purely
leptonic channel~\cite{htz} of Eq.~(\ref{sigww}) and like-sign lepton
plus jets channel~\cite{htz} of Eq.~(\ref{likesign}). The trilepton
signature has been examined for the Tevatron and LHC~\cite{Wells}.

\vspace{0.2in}
{\bf Event Samples} \\

The event samples for this study were generated using the PYTHIA~6.023
event generator and SHW detector simulation.  The signal cross
sections have NLO QCD corrections~\cite{19} included and are based
on the the CTEQ4M parton distribution functions~\cite{CTEQ}. For
pair production of resonances, {\em e.g.}  $WW$, PYTHIA incorporates
the full $2\rightarrow 2 \rightarrow 4$ matrix elements thereby
insuring proper treatment of the final state angular correlations.
Similarly for $h\to WW$, the angular correlations between the four
final state fermions have been taken into account. The full
$Z/\gamma^*$ interference is simulated for $ZZ$ production; however,
the $WZ$ process considers only the pure $Z$ contribution.  For Higgs
boson production in association with a gauge boson in Eq~(\ref{whzh}),
the associated $W$ and $Z$ decay angular distributions are treated
properly.  For some samples, such as $\tt$ where the rejection factor
was high, but the cross section was also large, generator--level
preselections were applied to reduce the sample passed to SHW. The
production cross-sections for the principal background processes were
normalized to $\sigma(WW) = 10.4$ pb, $\sigma(t\bar{t}) = 6.5$ pb,
$\sigma(WZ) = 3.1$ pb, and $\sigma(ZZ) = 1.4$ pb.  The measured D\O\
$W/Z + 3j$ rate~\cite{w3jet} was used to estimate the background from
$W/Z\,jj$ with a third jet faking a electron.

  \subsubsection{$\lpm\nn$ Channel}	       	
The process $gg\to h\to WW^*$ can lead to the purely leptonic channel
in Eq.~(\ref{sigww}), where we identify the final state signal as two
isolated opposite-sign charged leptons and large missing transverse
energy.  Such a signature has been shown to be a sensitive way to
search for the Higgs in the mass range 155-180 GeV/$c^2$ at the LHC \cite{dreiner}.

At the Tevatron the leading SM background processes in this search are
\begin{eqnarray}
p\bar p &\to& W^+W^-\to  \ell \bar \nu \bar \ell \nu,\ \ 
 Z Z(\gamma^*)\to \nu \bar \nu \ell \bar \ell,\ \  
 W Z(\gamma^*)\to \ell \bar \nu \ell \bar \ell,
\nonumber\\
p\bar p &\to& t \bar t\to  \ell \bar \nu \bar \ell \nu b\bar b,
\ \  
p\bar p \to Z(\gamma^*)\to  \tau^+\tau^- \to 
\ell \bar \nu \bar \ell \nu \nu_\tau \bar \nu_\tau .
\label{dy}
\end{eqnarray}

\vspace{0.2in}
{\bf Selection Criteria}\\ 

We first impose basic acceptance cuts for the 
leptons \footnote{The cuts for leptons were chosen to reflect
realistic trigger considerations. It is desirable to
extend the acceptance in $\eta_\ell^{}$.}
\begin{eqnarray}
\nonumber
&&p^{}_T(e)  > 10\ {\gev},\ \ |\eta^{}_e| < 1.5,\nonumber\\
&&p^{}_T(\mu_1)>10\ {\gev},\ \  p^{}_T(\mu_2)>5\ {\gev},\ \ 
|\eta^{}_\mu| < 1.5,\nonumber \\
&&m(\ell\ell)>10\ {\gev},\ \ \Delta R(\ell j)>0.4,\ \ 
\etmiss > 10\ {\gev},
\label{basic}
\end{eqnarray}
where $p^{}_T$ is the transverse momentum and $\eta$ the
pseudo-rapidity.
The cut on the invariant mass $m(\ell\ell)$ is to remove the
photon conversions and leptonic $J/\psi$ and $\Upsilon$ decays.
The isolation cut on $\Delta R(\ell j)$ removes the muon events 
from heavy quark ($c, b$) decays.\footnote{The electron identification
in the SHW simulation imposes strict isolation requirements already.} 

\begin{figure}[t!]
  \begin{center}
    \parbox{3.75in}{\epsfxsize=\hsize\epsffile{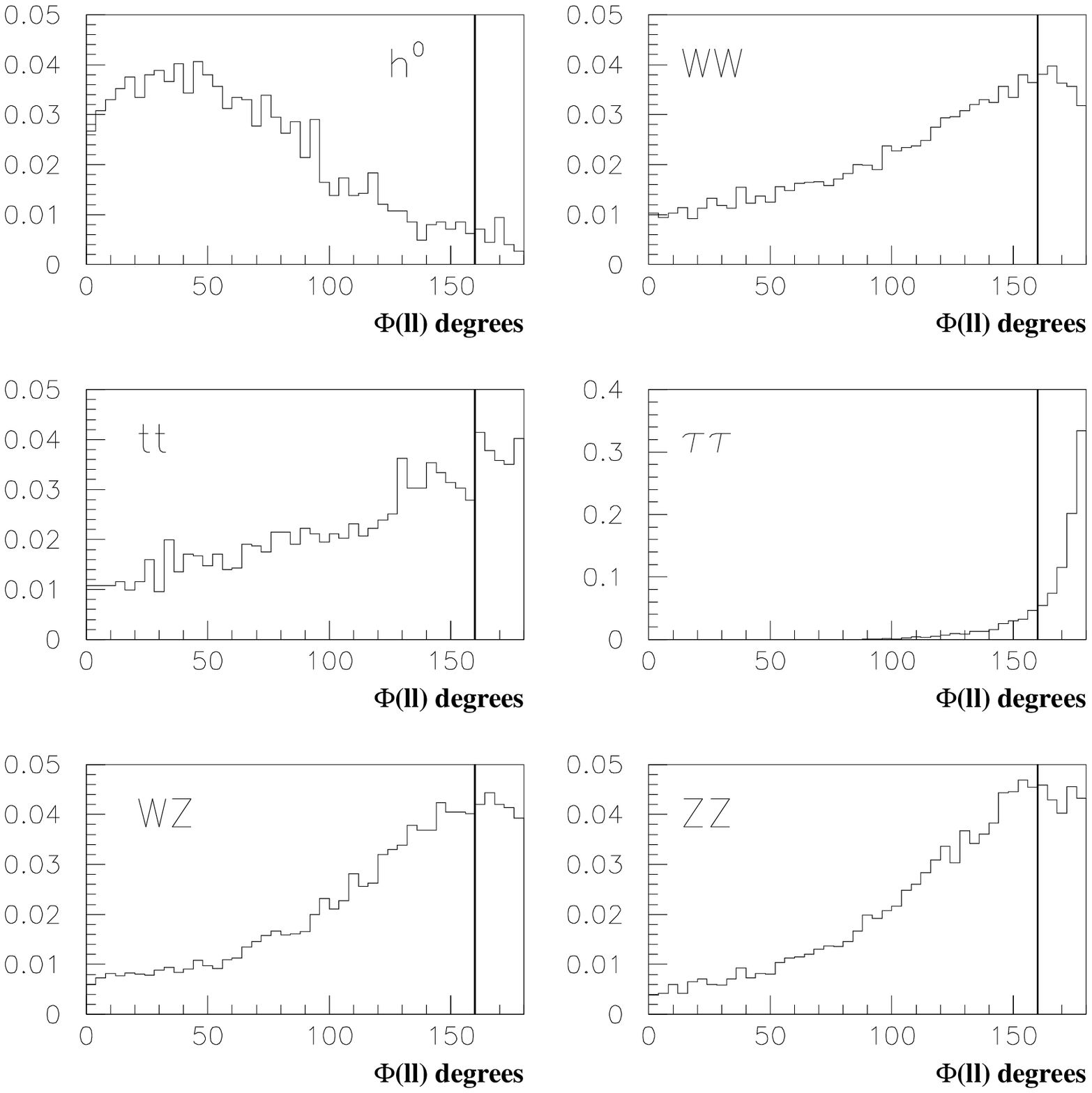}}
  \end{center}
  \caption{Normalized azimuthal angle distributions for the
           signal $gg\to h\to \ww\to\llnn$ with $\mh=170$ GeV
           and backgrounds $WW$, $t \bar t$, $\tau^+\tau^-$, $WZ$ and $ZZ$.}
  \label{phis}
\vskip1pc
  \begin{center}
    \parbox{3.75in}{\epsfxsize=\hsize\epsffile{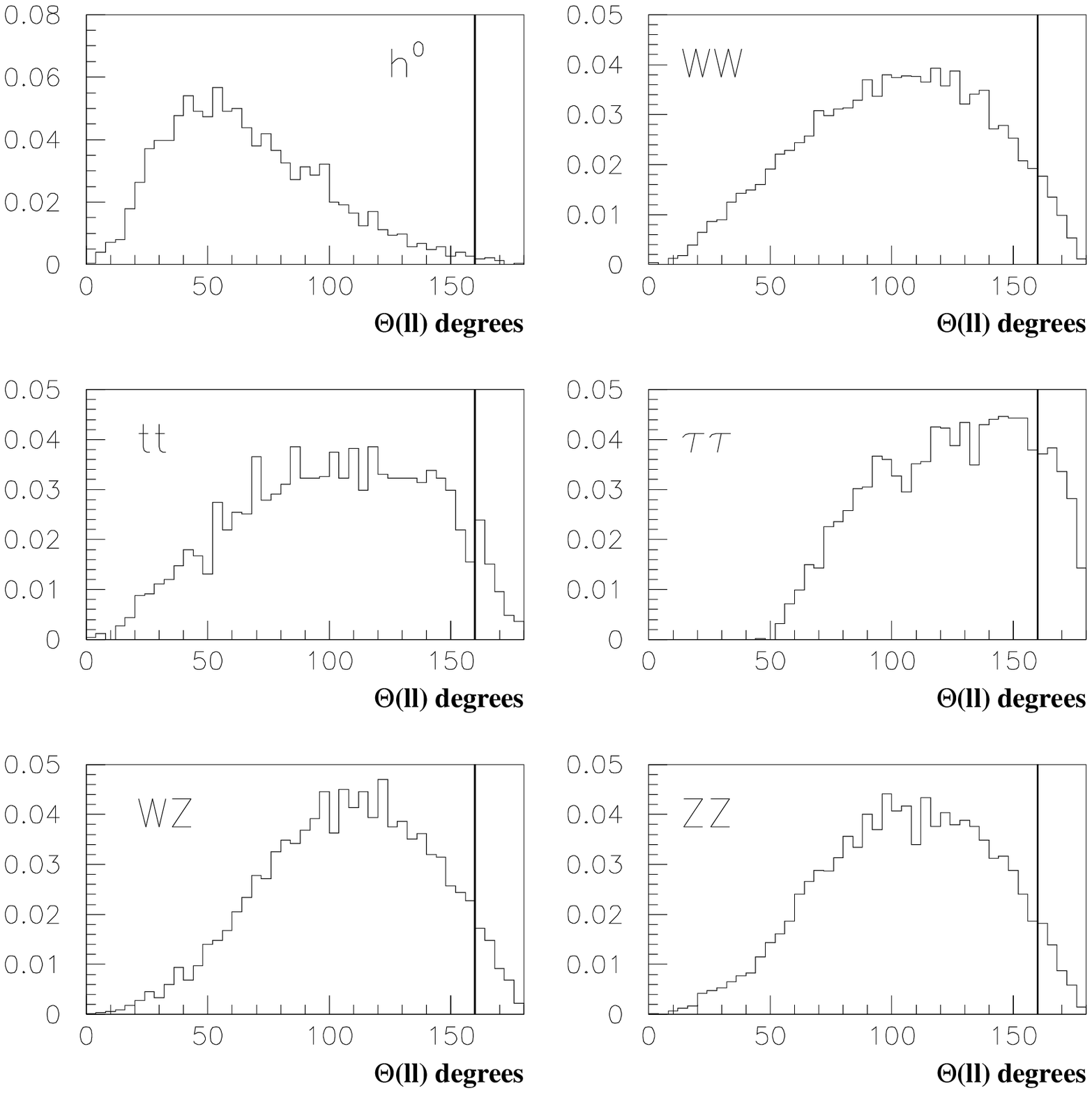}}
  \end{center}
  \caption{Normalized distributions 
           for the opening angle in Eq.~(\ref{phi}) for the signal 
           $gg \to h \to \ww \to \llnn$ with $\mh=170$ GeV and backgrounds $WW$, 
           $t \bar t$, $\tau^+\tau^-$, $WZ$ and $ZZ$. }
  \label{thes}
\end{figure}

At this level, the largest background comes from the Drell-Yan
process for $\tau^+\tau^-$ production. However, the charged
leptons in this background are very much back-to-back and this
feature is also true, although to a lesser extent, for other 
background processes as well.
On the other hand, due to the spin correlation of the Higgs boson
decay products, the two charged leptons tend to move in 
parallel~\cite{dreiner}.
We demonstrate this point in Figs.~\ref{phis} and \ref{thes}
where the distributions of the azimuthal angle in the transverse 
plane [$\phi(\ell\ell)$] and the three-dimensional opening-angle 
between the two leptons [$\theta(\ell\ell)$] for the signal 
and backgrounds are shown.\footnote{Since we are mainly interested
in the shapes of the kinematic distributions, we present them
normalized to unity with respect to the total cross section
with appropriate preceding cuts.}
This comparison motivates us to impose the cuts 
\begin{equation}
\phi(\ell\ell) < 160^\circ,\ \ 
\theta(\ell\ell) < 160^\circ.
\label{phi}
\end{equation}
The $\tau^+\tau^-$ background can be essentially 
eliminated with the help of additional cuts
\begin{equation}
p_T^{}(\ell\ell) > 20\ {\gev},\ \ 
\cos\theta_{\ell\ell-\etmiss} < 0.5,\ \ 
M_T^{}(\ell \etmiss) >20\ {\gev},
\label{ptll}
\end{equation}
where 
$\theta_{\ell\ell-\etmiss}$ is the relative angle between the lepton 
pair transverse momentum and the missing transverse momentum, 
which is close to 180$^\circ$ for the signal and near 0$^\circ$
for the Drell-Yan $\tau^+\tau^-$ background. The
two-body transverse-mass is defined for each
lepton and the missing energy as 
\begin{equation}
M_T^2(\ell \etmiss)=2 p_T(\ell)\etmiss(1-\cos\theta_{\ell-\etmiss}),
\label{tm}
\end{equation}
and the distributions are shown in Fig.~\ref{mtlnu}.
\clearpage


\begin{figure}[t!]
  \begin{center}
    \parbox{3.75in}{\epsfxsize=\hsize\epsffile{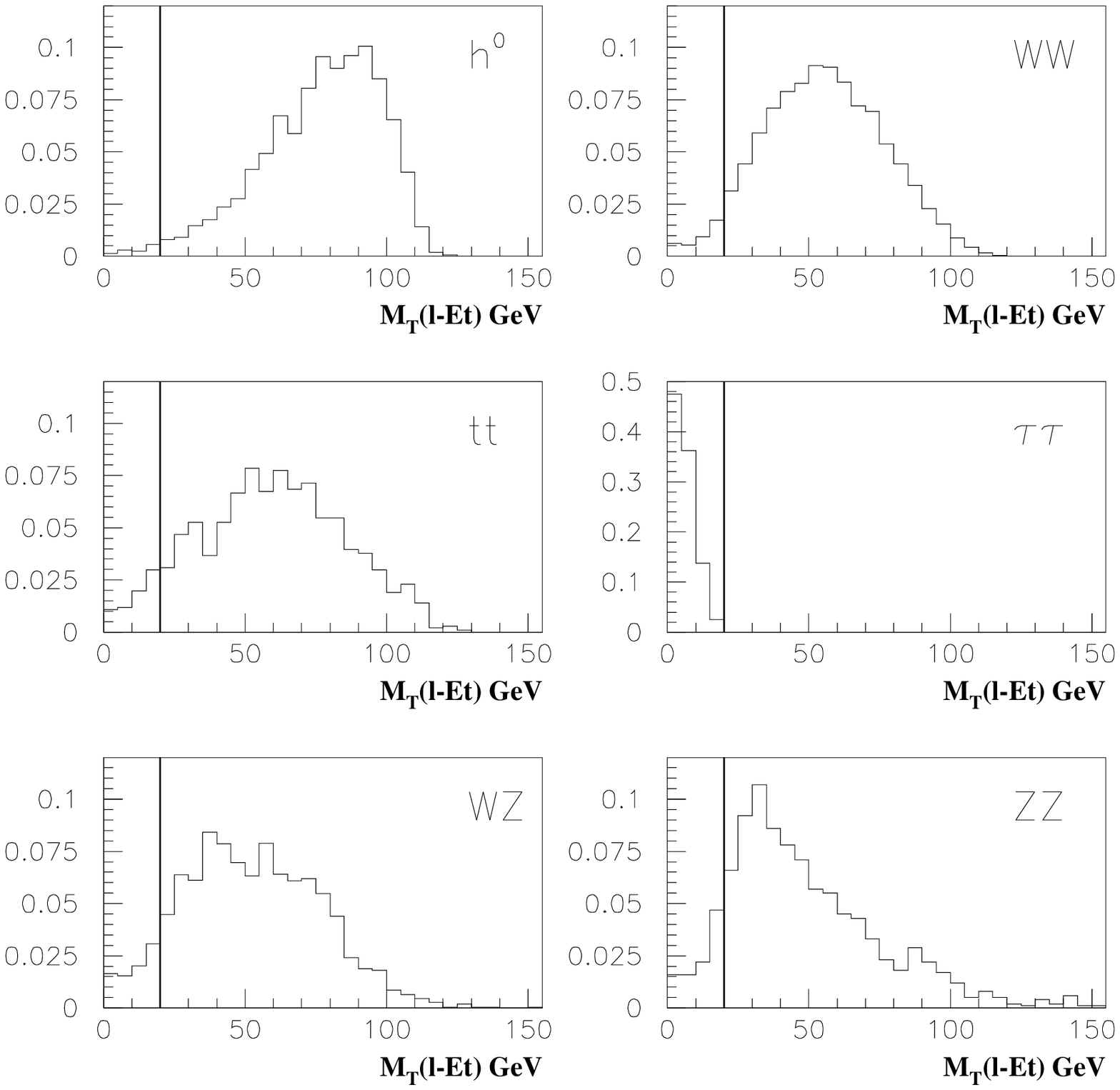}}
  \end{center}
  \caption{Normalized distributions 
           for the two-body transverse-mass defined in Eq.~(\ref{tm}) 
           for the signal $gg \to h \to \ww \to \llnn$ with $\mh=170$ GeV
           and backgrounds $WW$, $t \bar t$, $\tau^+\tau^-$, $WZ$ and $ZZ$. 
           The minimum of $M_T(\ell_1 \etmiss)$ and $M_T(\ell_2 \etmiss)$ is shown.}
  \label{mtlnu}
\end{figure}

We can further purify the signal by removing the high
$m(\ell\ell)$ events from $Z\to \ell\bar \ell$ 
as well as from $t\bar t, W^+W^-$, 
as demonstrated in Fig.~\ref{mlls}. We therefore impose
\begin{eqnarray}
\nonumber
m(\ell\ell)&<& 78\ {\gev}\ \ {\rm for}\ e^+e^-,\mu^+\mu^-,\\ 
m(\ell\ell)&<& 110\ {\gev}\ \ {\rm for}\ e\mu.
\label{zout}
\end{eqnarray}
As suggested in ref.~\cite{5}, the lepton correlation angle
between the momentum vector of the lepton pair and the momentum of
the higher $p_T$ lepton ($\ell_1$) in the lepton-pair rest frame, 
$\theta^*_{\ell_1}$, also has discriminating power between the
signal and backgrounds. This is shown in Fig.~\ref{thetas}. 
We thus select events with
\begin{equation}
-0.3 < \cos\theta^*_{\ell_1} < 0.8.
\label{theta}
\end{equation}
A characteristic feature of the top-quark background is the presence
of hard $b$-jets. We thus devise the following jet-veto 
criteria:
\begin{eqnarray}
{\rm veto\ if}\ \ p_T^{j_1}>95\ {\gev},\ \  |\eta^{}_j| < 3,\nonumber\\ 
{\rm veto\ if}\ \ p_T^{j_2}>50\ {\gev},\ \  |\eta^{}_j| < 3,\nonumber\\ 
{\rm veto\ if}\ \ p_T^{j_3}>15\ {\gev},\ \  |\eta^{}_j| < 3.
\label{jetveto}
\end{eqnarray}

Furthermore, if either of the two hard jets ($j_1,j_2$) is identified
as a $b$ quark, the event will be also vetoed. The $b$-tagging 
efficiency is taken from SHW.

The results up to this stage are summarized in Table~\ref{tabI} for thqqqe signal
$\mh=140$--190~GeV as well as the SM backgrounds.  The acceptance cuts discussed
above are fairly efficient, approximately 35\% of the signal is retained while
backgrounds are substantially suppressed. We see that the dominant background
comes from the electroweak $WW$ production, about a factor of 30 higher than
the signal rate. The sub-leading backgrounds $t\bar t$ and $W$+fake (the
background where a jet mimics an electron with a probability of $P(j\to
e)=10^{-4}$~\cite{fake}) are also bigger than the signal.  We note that
although the $b$-jet veto is effective against the $t\bar t$ background, the
final results are not affected if the veto efficiency is significantly worse.

\begin{figure}[t!]
  \begin{center}
    \parbox{3.75in}{\epsfxsize=\hsize\epsffile{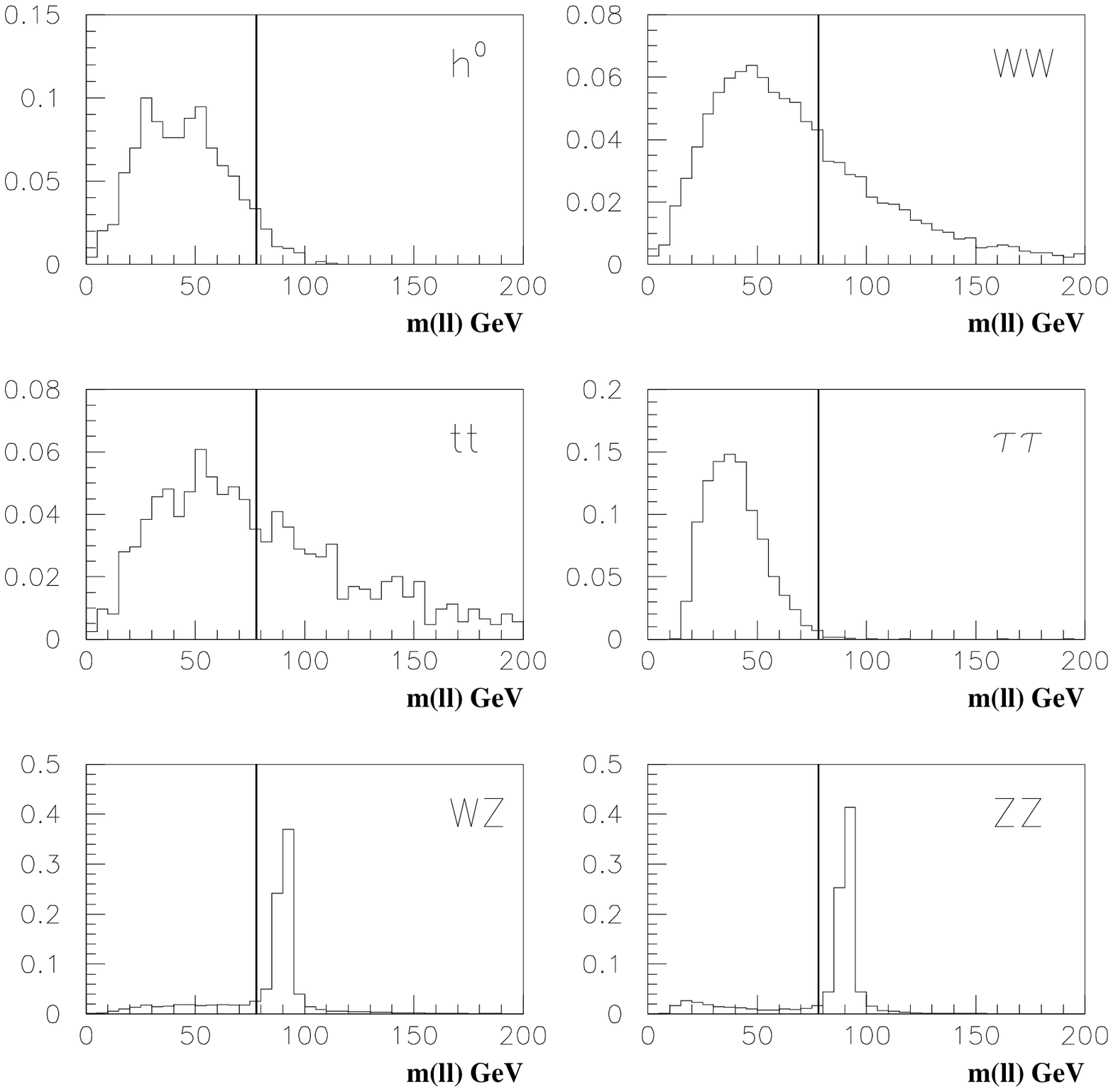}}
  \end{center}
  \caption{Normalized like-flavor lepton-pair invariant mass distributions 
           for the signal $gg \to h \to \ww \to \llnn$ with $\mh=170$ GeV
           and backgrounds $WW$, $t \bar t$, $\tau^+\tau^-$, $WZ$ and $ZZ$.}
  \label{mlls}
\vskip1pc
  \begin{center}
    \parbox{3.75in}{\epsfxsize=\hsize\epsffile{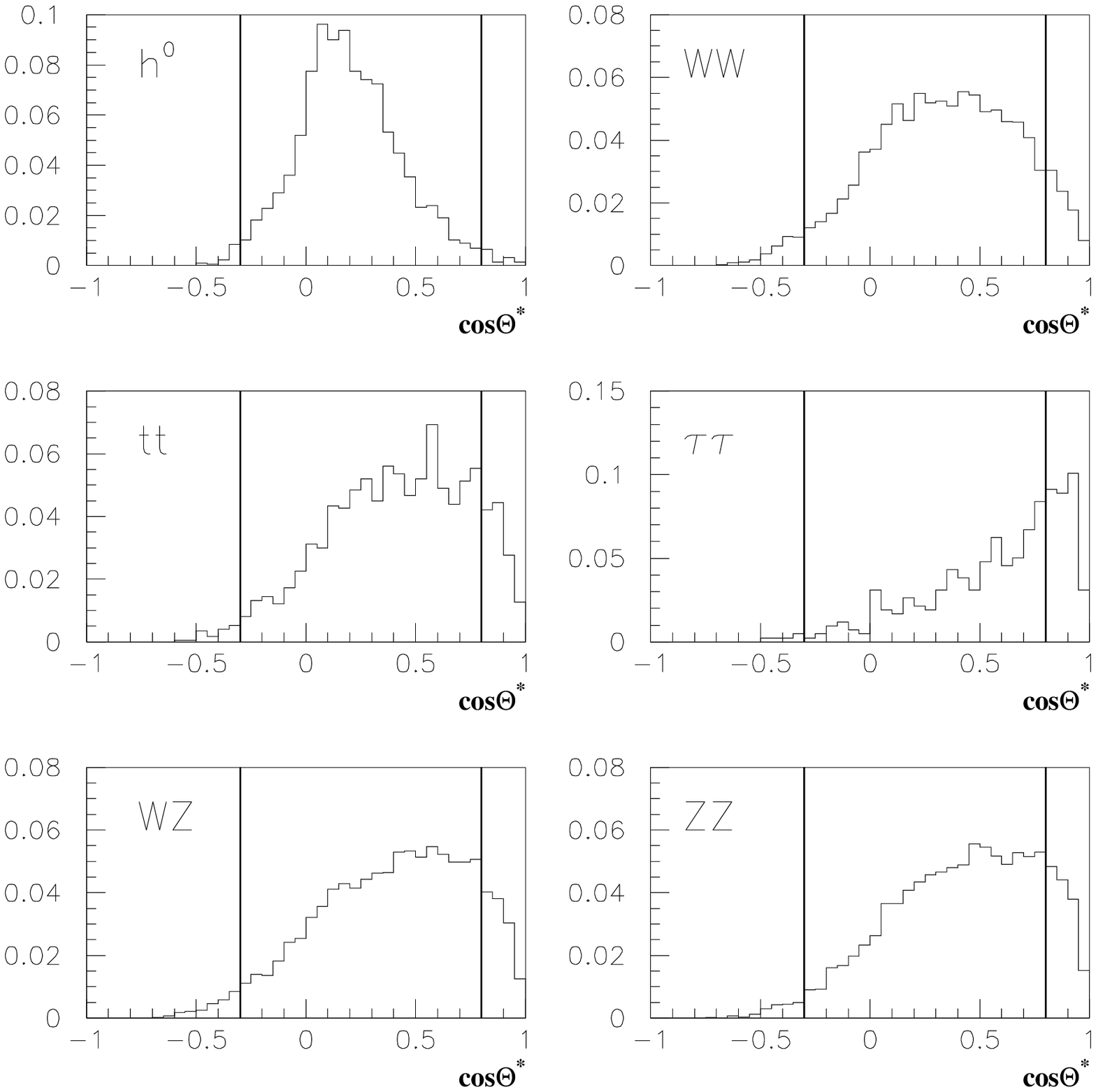}}
  \end{center}
  \caption{Normalized angular distributions for the correlation angle 
           defined above Eq.~(\ref{theta}) for the signal 
           $gg \to h \to \ww \to \llnn$ with $\mh=170$ GeV and backgrounds $WW$, 
           $t \bar t$, $\tau^+\tau^-$, $WZ$ and $ZZ$.}
  \label{thetas}
\end{figure}
\clearpage

\begin{table}[h!]
\caption[]{$h\rightarrow \ww \rightarrow \llnn$ signal cross section
(in fb) for $\mh=140-$190 GeV and various SM backgrounds after 
the kinematic cuts of Eqs.~(\ref{basic})$-$(\ref{jetveto}). 
The signal efficiencies are also shown (in percentage). 
$W+$fake refers to the background where a jet mimics
an electron with a probability of $P(j\to e)=10^{-4}$.
The backgrounds are independent of $\mh$.
}
\label{tabI}
\begin{tabular}{|l|c|c|c|c|c|c|}
\noalign{\vskip-6pt}
 $\mh$ [GeV] & 140 & 150 & 160 & 170 & 180   & 190\\  \tableline
 signal [fb] & 3.9 & 4.4 & 5.2 & 4.8 & 3.6   & 2.5\\  
 efficiency. [\%] & 35  & 34  & 38  & 39  & 36    & 37\\   \hline \hline
{}&$WW$ & $t\bar t$ & $\tau^+\tau^-$& $WZ$ & $ZZ$ & $W+$fake\\ 
\hline
backgrounds [fb]& 130 & 13  & 0   & 4.4 & 2.4   & 18 \\ 
\end{tabular}
\end{table}

\vspace{0.2in}
{\bf Further Selection: Likelihood Analysis}\\ 

One can improve the signal observability by
constructing a likelihood based on some characteristic
kinematic variables. We choose the variables as
\begin{itemize}
\item[1.] $\cos\theta_{\ell\ell}$, the polar angle with respect 
to the beam axis of the dilepton~\cite{dreiner};
\item[2.] $\phi(\ell\ell)$ as in Eq.~(\ref{phi});
\item[3.] $\theta(\ell\ell)$ as in Eq.~(\ref{phi});
\item[4.] $\cos\theta_{\ell\ell-\etmiss}$ as in Eq.~(\ref{ptll});
\item[5.] $p_T^{j1}$ as in Eq.~(\ref{jetveto});
\item[6.] $p_T^{j2}$ as in Eq.~(\ref{jetveto}).
\end{itemize}
We wish to evaluate the likelihood for a candidate event
to be consistent with one of five event classes: a Higgs
boson signal ($140 < \mh <$ 190 GeV), 
$WW$, $t\bar t$, $WZ$ or $ZZ$. For a single variable $x_i$, 
the probability for an event to belong to class $j$ is given by
\begin{equation}
P^j_i(x_i) = \frac{f^j_i(x_i)}{\Sigma_{k=1}^5\ f^k_i(x_i)},
\end{equation}
where $f_i^j$ denotes the probability density for 
class $j$ and variable $i$. 
The likelihood of an event to belong
to class $j$ is given by the normalized products of the individual
$P^j_i(x_i)$ for the $n=6$ kinematic variables:
\begin{equation}
{\cal L}^j = \frac{ \Pi_{i=1}^n\  P^j_i(x_i)}
{\Sigma_{k=1}^5\ \Pi_{i=1}^n\  P^k_i(x_i)},
\label{like}
\end{equation}
The value of ${\cal L}^j$ for a Higgs boson signal hypothesis 
($j=1$) is shown in Fig.~\ref{likeli} 
where it can be seen that a substantial fraction
of the $t\bar t$ and $WW$ background can be removed for a modest loss
of efficiency. The $WZ$ and $ZZ$ backgrounds
have similar distributions to the
$WW$ and have been omitted for clarity. 
We thus impose the requirement
\begin{equation}
{\cal L}^{j=1} > 0.10.
\label{likec}
\end{equation}
The improved results are summarized in Table~\ref{tabII}.

\begin{table}[t!]
\caption[]{$h\rightarrow \ww \rightarrow \llnn$ signal cross section
(in fb) for $\mh=140-$190 GeV and various SM backgrounds after the kinematic 
cuts of Eqs.~(\ref{basic})$-$(\ref{jetveto}) and the likelihood cut
Eq.~(\ref{likec}). $W+$fake refers to the background where a jet mimics
an electron with a probability of $P(j\to e)=10^{-4}$.
The backgrounds are independent of $\mh$.
}
\label{tabII}
\begin{tabular}{|l|c|c|c|c|c|c|}
\noalign{\vskip-6pt} 
$\mh$ [GeV] & 140 & 150 & 160 & 170 & 180   & 190\\  \tableline
 signal [fb] & 3.1 & 3.6 & 4.5 & 4.1 & 2.9   & 2.0\\  \hline
\hline
{}&$WW$ & $t\bar t$ & $\tau^+\tau^-$& $WZ$ & $ZZ$ & $W+$fake\\ 
\hline
backgrounds [fb]& 83 & 4.5  & 0   & 3.1 & 1.8   & 13 \\ [2pt] 
\end{tabular}
\vspace{1pc}
\end{table}

\begin{figure}[h!]
  \begin{center}
    \parbox{3.5in}{\epsfxsize=\hsize\epsffile{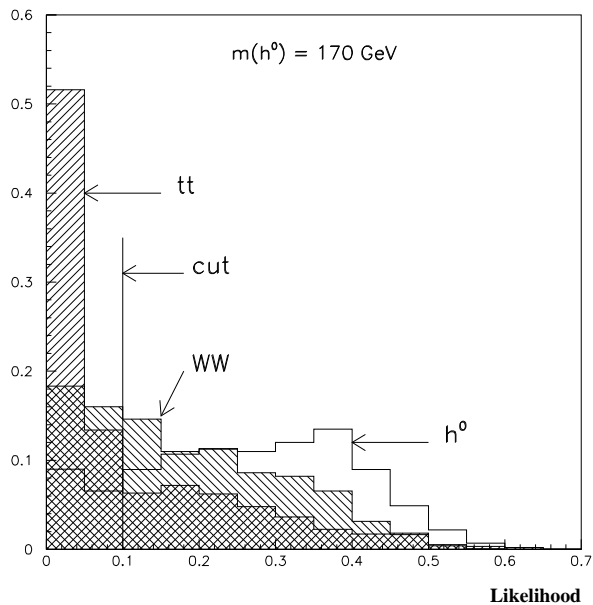}}
  \end{center}
  \caption{Distributions for the likelihood variable
           defined in Eq.~(\ref{like}) for the signal $\mh=170$ GeV
           and the leading SM backgrounds $WW$ and $t\bar t$.}
  \label{likeli}
\end{figure}

In identifying the signal events, it is crucial to
reconstruct the mass peak of $\mh$. Unfortunately,
the $\ww$ mass from the $h$ decay cannot be accurately 
reconstructed due to the two undetectable neutrinos. 
However, both the transverse mass $M_T$ and the 
cluster transverse mass $M_C$ \cite{bho}, defined as
\begin{eqnarray}
M_T &=& 2\sqrt{ p^2_T(\ell\ell)+m^2(\ell\ell)},\label{mt} \\ 
M_C &=&  \sqrt{ p^2_T(\ell\ell)+m^2(\ell\ell)}\ + \etmiss,
\label{mc}
\end{eqnarray}
yield a broad peak near $\mh$ and have a long tail below.
The cluster transverse mass $M_C$ has a Jacobian structure
with a well defined edge at $m_h$.
We show the nature of these two variables for the signal
with $\mh=170$ GeV and the leading $WW$ background
in Fig.~\ref{masses}(a) for $M_T$ and (b) for $M_C$ after application
of the likelihood cut.
For a given $m_h$ to be studied, one can perform additional cut optimization.
In Table~\ref{tabcut}, we list $m_h$-dependent criteria for
the signal region defined as
\begin{equation}
\mh - 60 < M_C < \mh + 5\ \gev.
\end{equation}
\begin{table}
\caption[]{Summary of the optimized cuts additional to those
in Eqs.~(\ref{basic})$-$(\ref{jetveto}) for various Higgs boson mass.
}
\label{tabcut}
\begin{tabular}{|l|c|c|c|c|c|c|}
\noalign{\vskip-6pt}
 $\mh$ [GeV] & 140 & 150 & 160 & 170 & 180   & 190\\  \tableline
 $\cos\theta^*_{\ell_1}$ & - & $<$0.6 & 0.35 & 0.35 & 0.55   & 0.75\\ 
 $\etmiss$ & $>$25 & 25 & 30 & 35 & 40 & 40\\  
min[$M_T^{}(\ell_1 \etmiss),M_T^{}(\ell_2 \etmiss)$]  
& $>$40 & 40 & 75 & 80 & 85   & 75\\  
$M_T^{}(\ell_1 \etmiss)$ & $>$60 & 60 & - & - & - & -\\  
$m(\ell\ell)$ & $<$65 & 65 & 65 & 75 & 85 & -\\  
$p_T^{}(\ell\ell)$ & $>$40 & 50 & 65 & 70 & 70 & 70\\  
$\theta(\ell\ell)$ & $<$100 & 100 & 70 & 70 & 90  & 90\\ 
$M_T^{}$ & - & $>$110 & 120 & 130 & 140 & 140\\
\end{tabular}
\end{table}

We illustrate the effect of the optimized cuts of Table~\ref{tabcut}
in Fig.~\ref{bacuts}, where the cluster transverse mass distribution
for a $\mh=170$ GeV signal and
the summed backgrounds, normalized to 30 $\fbi$, are shown 
before (a) and after the final cuts (b). A clear excess
of events from the Higgs signal can be seen in
Fig.~\ref{bacuts}(b). 
It is important to note that before application of the
final cuts, the dominant backgrounds are $WW$ and the $W$+fake
with other sources accounting for less than 10\% 
of the total. Moreover, for 30 $\fbi$ integrated luminosity, 
the statistical error in the background is less than 2\% 
before application of the final cuts. 
We therefore argue that one should be able 
to normalize the SM background curve ($WW$) with sufficient
precision to unambiguously identify a significant excess 
attributable to Higgs boson signal. It should also be noted that by
selectively loosening the final cuts, it is possible 
to maintain the same $S/\sqrt{B}$ while increasing the accepted 
background by up to factor of 5, and the accepted signal
by a factor of 2.5. This can provide a powerful cross-check
of the predicted background $M_C$ shape and can be used
to demonstrate the stability of any observed excess.

\begin{figure}
  \begin{center}
    \parbox{6.0in}{\epsfxsize=\hsize\epsffile[0 260 530 530]{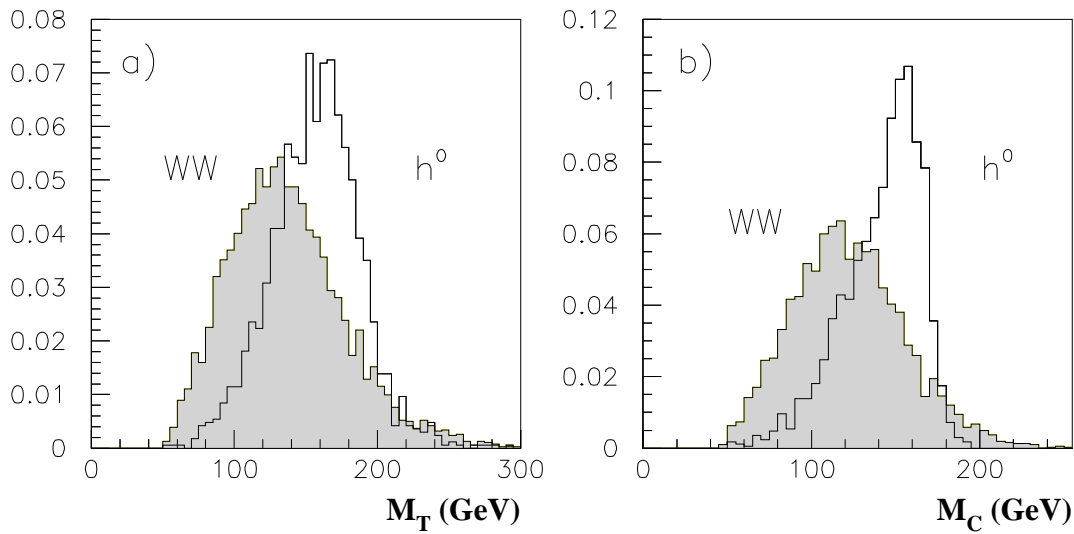}}
  \end{center}
\caption{Normalized mass distributions for the signal 
         $gg \to h \to \ww \to \llnn$ with $\mh=170$ GeV (histogram)
         and the leading $WW$ background (shaded) for (a) the transverse mass
         defined in Eq.~(\ref{mt}), and (b) the cluster transverse mass defined in 
         Eq.~(\ref{mc}).}
  \label{masses}
\end{figure}

Our final results for the channel $h\to \ww\to \ell \bar \nu \bar \ell \nu$ are
summarized in Table~\ref{lnln}. We have included the contributions to $h\to
\ww$ from the signal channels in Eqs.~(\ref{whzh}) and (\ref{wwzz}).  Although
they are small to begin with, they actually increase the accepted signal cross
section by 12--18\%.  We have also included the contribution from $W \to \tau
\nu \to \ell\nu_\ell \nu$.\footnote{From consideration of the
$W$+$(j\rightarrow e)$ background, it should be clear that improving the
sensitivity by incorporating hadronic tau decays will be a difficult task.}  It
can be seen that one may achieve a $S/B$ of at least 6\% for 140 GeV$<\mh< 190$
GeV and reach 45\% for $\mh=170$ GeV. The statistical significance, $S/\sqrt
B$, for 30 fb$^{-1}$ integrated luminosity, is 3$\sigma$ or better for $150<
\mh < 180$ GeV.

\begin{figure}
  \begin{center}
    \parbox{6.0in}{\epsfxsize=\hsize\epsffile[0 260 530 530]{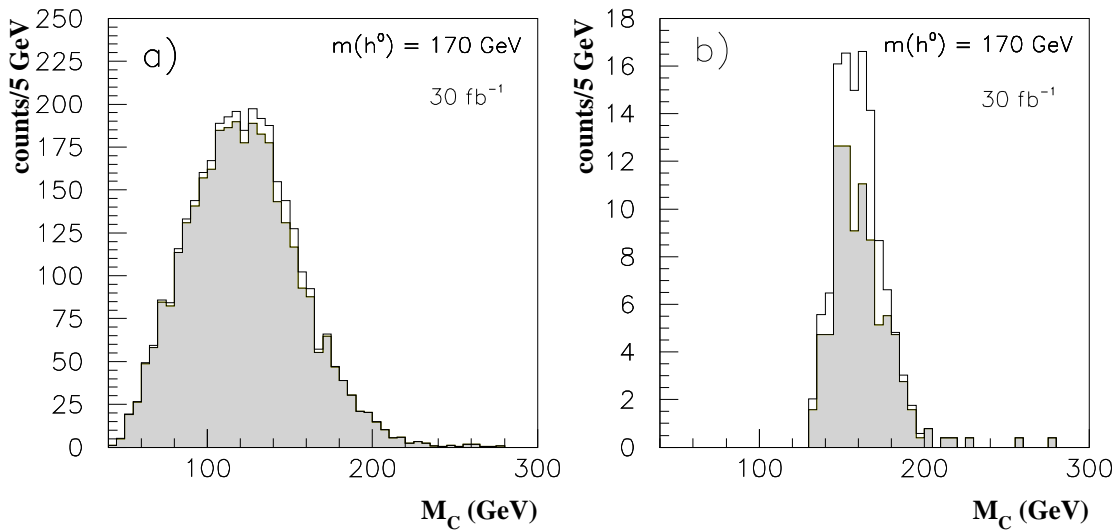}}
  \end{center}
  \caption{Cluster transverse mass distributions for
           the leading $WW$ background (shaded) and the background
           plus the signal $gg \to h \to \ww \to \llnn$ with $\mh=170$ GeV
           (histogram) (a) before the optimized
           cuts in Table~\ref{tabcut} and (b) after the cuts. The
           vertical axis gives the number of events per 5 GeV bin
           for 30 $\fbi$.}
  \label{bacuts}
\end{figure}

\begin{table}[h!]
\caption[]{Summary table for $h\rightarrow \ww \rightarrow \llnn$ signal 
for $\mh=140-$190 GeV and various SM backgrounds after the kinematic 
cuts of Eqs.~(\ref{basic})$-$(\ref{jetveto}) and the likelihood cut
Eq.~(\ref{likec}). The entry ``fake $j\to e$'' refers to the background where a jet mimics
an electron with a probability of $P(j\to e)=10^{-4}$.
The backgrounds are independent of $\mh$.
}
\label{lnln}
\begin{tabular}{|l|c|c|c|c|c|c|}
 $\mh$ [GeV] & 140 & 150 & 160 & 170 & 180   & 190\\  \tableline
 $gg\to h$ [fb] & 2.2 & 2.4 & 1.3 & 0.93 & 0.85 & 0.73\\ 
 associated $VH$ [fb] & 0.26 & 0.31 & 0.13 & 0.09 & 0.06   & 0.06\\ 
 $VV$ fusion [fb] & 0.12 & 0.12 & 0.09 & 0.06 & 0.05 & 0.05\\ 
 signal sum [fb] & 2.6 & 2.8 & 1.5 & 1.1 & 0.96 & 0.83\\ 
\hline
SM backgrounds [fb]& 39 & 27  & 4.1 & 2.3 & 3.8 & 7.0 \\ 
fake $j\to e$ [fb]& 5.1 & 3.4  & 0.34 & 0.15 & 0.08 & 0.45 \\ 
backgrounds sum [fb]& 44 & 30  & 4.4 & 2.4 & 3.8 & 7.5 \\ \hline
$S/B$ & 0.058 & 0.094  & 0.34 & 0.45 & 0.25 & 0.11 \\ \hline
$S/\sqrt B$ for 30 fb$^{-1}$ & 2.1 & 2.8  & 3.9 & 3.8 & 2.7 & 1.7 \\ 
\end{tabular}
\end{table}

\break
In Fig.~\ref{intL}(a), we present the integrated luminosities
needed to reach a 3$\sigma$ significance and 95\% CL exclusion computed
assuming Poisson probabilities as a function of $\mh$.

To assess the effect of inherent systematic uncertainties, we
re-evaluate the corresponding curves in Fig.~\ref{intL}(b) assuming a
10\% systematic error for the signal and SM backgrounds.\footnote{For
the purposes of computing the effects of systematic errors on the
sensitivity to a Higgs signal, we have scaled the expected background
upward by a given percentage and the expected signal downward by the
same percentage simultaneously.}  The results are somewhat degraded,
but they are still encouraging.

\begin{figure}[h!]
  \begin{center}
    \parbox{6.0in}{\epsfxsize=\hsize\epsffile[0 260 530 530]{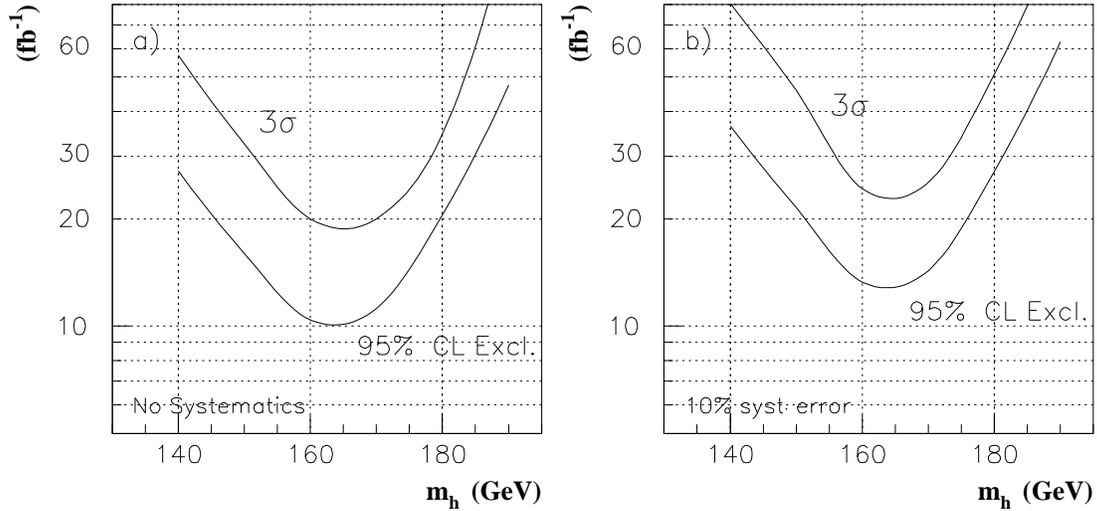}}
  \end{center}
  \caption{The integrated luminosity per experiment required to reach $3\sigma$
           statistical significance and 95\%
           exclusion versus $m_h$ in the $h\to \ww \to \ell\bar \nu
           \bar \ell \nu$ channel for (a) statistical effects only; (b) 10\%
           systematic error for the signal and SM backgrounds included. 
           The contribution from $W\to \tau \to \ell$ decays, associated 
           production and gauge boson fusion have also been included.}
  \label{intL}
\end{figure}

    \subsubsection{$\lsdilep$ Channel}		
\def\mh{m_h^{}}
\def\gev{\rm GeV}
\def\fbi{\rm fb^{-1}}
\def\ww{W^*W^*}
\def\zz{Z^*Z^*}
\def\lsim{\mathrel{\raise.3ex\hbox{$<$\kern-.75em\lower1ex\hbox{$\sim$}}}}
\def\gsim{\mathrel{\raise.3ex\hbox{$>$\kern-.75em\lower1ex\hbox{$\sim$}}}}
\def\ljj{\ell\nu jj}
\def\lljj{\ell\bar\ell jj}
\def\jj{\protect jj}
\def\llnn{\ell\bar\ell\nu\bar\nu}

\def\ptmiss{\slashchar{p}_{T}}
\def\etmiss{\slashchar{E}_{T}}
%
%

For $Wh$ associated production, the $h \to \ww \to \ell\nu jj$ mode
can give like-sign leptons plus two-jets
events~\cite{Stange:1994ya,Wells}.  Compared to the trilepton
signature, the like-sign lepton plus jets process has a three times
larger rate. Moreover, the leading SM source of like-sign and
trileptons, $WZ(\gamma^*)$, will generate little jet activity as both
gauge bosons are required to decay leptonically to give the unique
like sign signature.

The contributing channels to the $\ell^{\pm} \ell^{\pm} j j$
final state include
\begin{eqnarray}
&&Wh \to W\ww \to \ell^\pm\nu \ell^\pm\nu jj,\nonumber\\
&&Wh \to W\zz \to \ell^\pm\nu \ell^\pm\ell^\mp jj,\nonumber\\
&&Wh \to W\zz \to \ell^\pm\ell^\mp \ell^\pm\ell^\mp jj,\nonumber\\
&&Zh \to Z\ww \to \ell^\pm\ell^\mp \ell^\pm\nu jj,\nonumber\\
&&Zh \to Z\zz \to \ell^\pm\ell^\mp \ell^\pm\ell^\mp jj.\nonumber
\end{eqnarray}

The SM backgrounds are
\begin{eqnarray}
\label{vvv}
p\bar p &\to& WWW,\ WWZ,\ WZZ,\ ZZZ,\ t \bar t W,\ t \bar t Z
\to \ell^\pm \ell^\pm jj\  X,\\
\label{wzjj}
p\bar p &\to& W^\pm Z(\gamma^*)+jj\to  \ell^\pm \ell^\pm jj X,\ \ 
Z Z(\gamma^*)+jj\to \ell^\pm \ell^\pm jj X,\ \  
t \bar t\to  \ell \bar \nu jj b\bar b,\\
p\bar p &\to& Wjj,\ Z(\gamma^*)jj +{\rm fake}.\nonumber
\end{eqnarray}
Although the triple gauge boson production \cite{23} in
Eq.~(\ref{vvv}) constitutes an irreducible background, the $WZjj$,
$t\bar t$ through $b$ or $c$ semileptonic decay and the background
from $j\to e$ fakes turn out to be larger.  The background from
$W(Z)jjj$ background with a $j\to e$ fake was estimated using the
measured D\O\ $W/Z + 3j$ rate \cite{w3jet} with $P(j\to e)=10^{-4}$.

\begin{figure}[tbp]
\epsfysize=5.5in
\epsffile[-50 0 560 560]{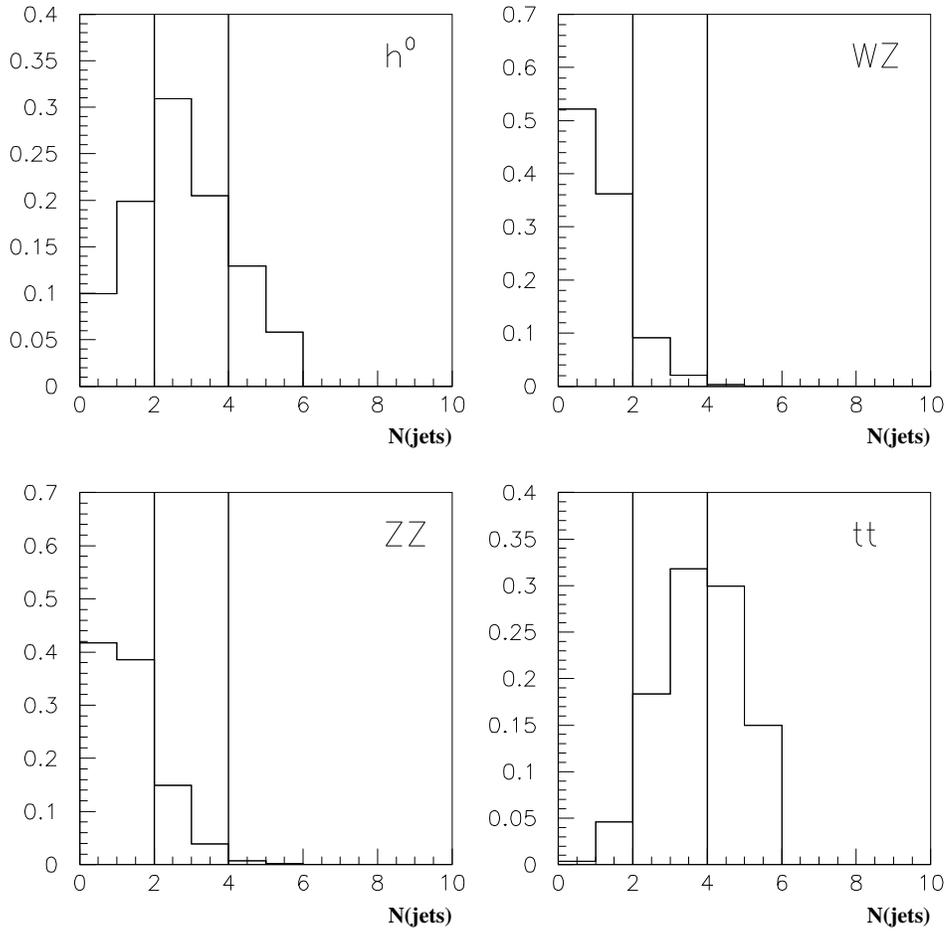}
\caption[]{Normalized jet multiplicity distributions for the signal 
$W^\pm h \to W^\pm\ww\to\ell^\pm\ell^\pm jj$ 
with $\mh=170$ GeV and the backgrounds $WZ$, $ZZ$ and $t\bar t$.
\label{njet}}
\end{figure}

\begin{figure}[tbp]
\epsfysize=5.5in
\epsffile[-50 0 560 560]{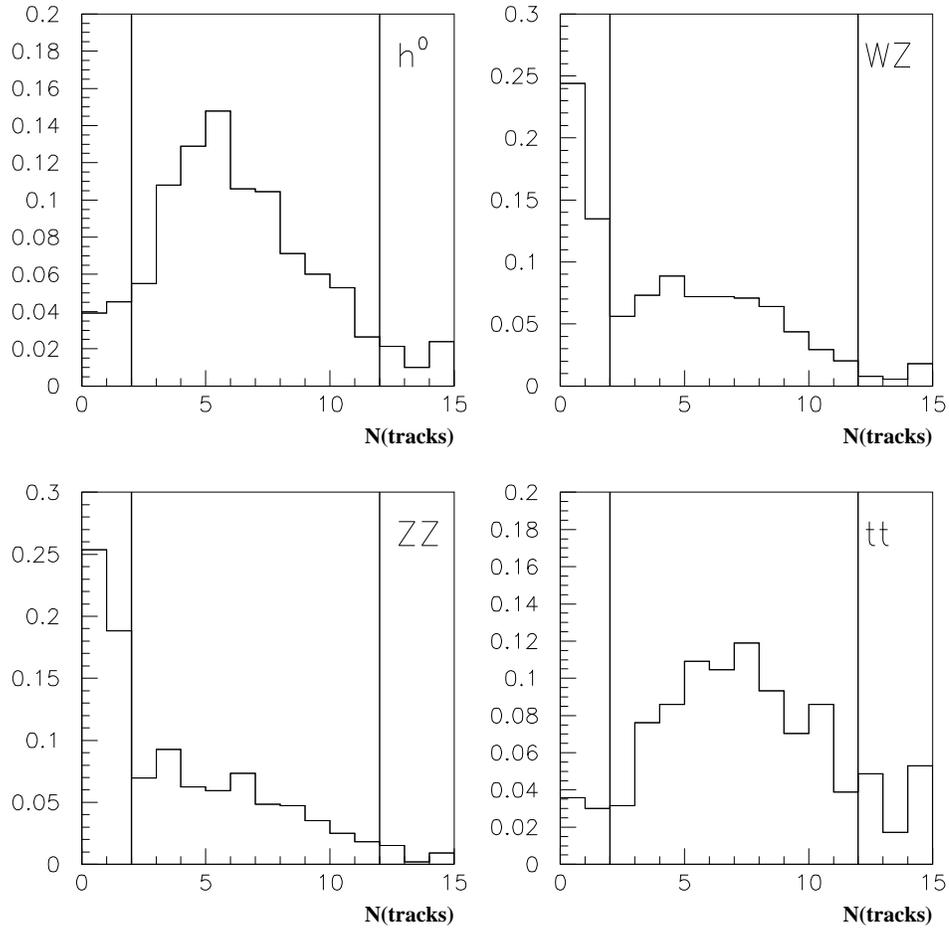}
\caption[0]{Normalized leading jet charged track multiplicity distributions 
for the signal $W^\pm h \to W^\pm \ww \to$ $\ell^\pm\ell^\pm jj$ 
with $\mh=170$ GeV and the backgrounds $WZ$, $ZZ$ and $t\bar t$.
\label{ctmul1}}
\end{figure}

\begin{figure}[tbp]
\epsfysize=5.5in
\epsffile[-50 0 560 560]{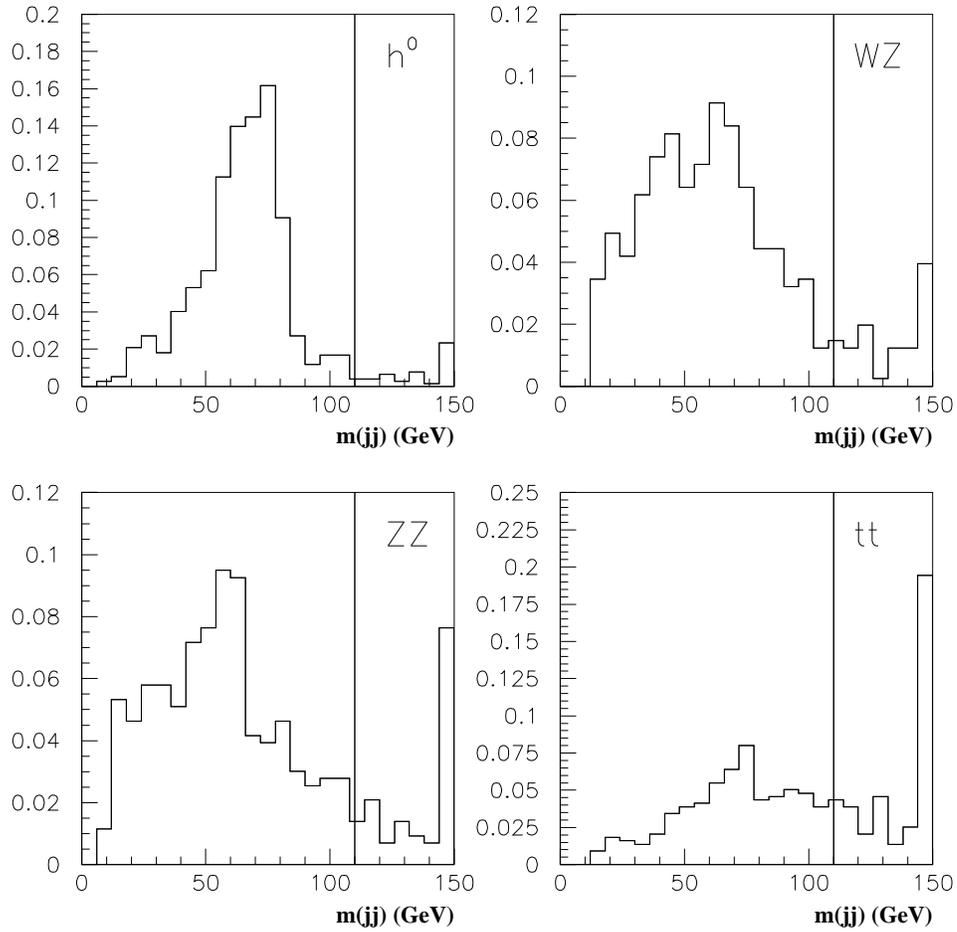}
\caption[]{Normalized di-jet mass distributions for the signal 
$W^\pm h \to W^\pm\ww\to\ell^\pm\ell^\pm jj$ 
with $\mh=170$ GeV and the backgrounds $WZ$, $ZZ$ and $t\bar t$.
\label{mjj}}
\end{figure}

\vspace{0.2in}
{\bf Event Selection} \\ \nopagebreak

We identify the final state signal as two isolated like-sign charged
leptons plus jets. A soft third lepton may be present.  The basic
acceptance cuts required for the leptons are
\begin{eqnarray}
\nonumber
&&p^{}_T(\ell)  > 10\ {\gev},\ \ |\eta^{}_\ell| < 1.5,\ \ 
m(\ell\ell)>10\ {\gev},\nonumber\\
&&0.3 < \Delta R(\ell j) <6,\ \ \etmiss > 10\ {\gev}.
\label{basic2}
\end{eqnarray}
For a muon, we further demand that the scalar sum of additional
track momenta within 30$^\circ$ be less than 60\%
of the muon momentum.
We require that there are at least two jets with
\begin{equation}
p_T^j>15\ {\gev},\ \  |\eta^{}_j| < 3.
\label{jets}
\end{equation}
To suppress the $WZ$ background, we require the leading jet to be
within $|\eta^{}_{j_1}| < 1.5$ and to have a charged track
multiplicity satisfying $2\leq N\leq 12$; while the sub-leading jet to
be within $|\eta^{}_{j_2}| < 2.0$.  The $t\bar t$ background typically
exhibits greater jet activity; we therefore veto events having
\begin{equation}
 p_T^{j_3}>30\ {\gev},
\end{equation}
and events with a fourth jet satisfying Eq.~(\ref{jets}).
The effect of the jet multiplicity and the leading jet
charged track multiplicity requirements on the di-boson backgrounds 
can be seen in Figs.~\ref{njet} and \ref{ctmul1}, respectively.
To suppress backgrounds associated with heavy flavor jets, 
we veto the event if any of the jets have a $b$-tag. 

In Fig.~\ref{mjj}, we present the di-jet mass distributions
for the signal and backgrounds. Since the di-jets in the signal
are mainly from a $W^*$ decay, $m(jj)$ is close to or lower than
$M_W$. This motivates us to further require
\begin{equation}
m(jj)<110\ {\gev},\ \ \Sigma_j |p_T^j|<150\ {\gev}.
\label{mpt}
\end{equation}
Finally, it is interesting to note that the lepton correlation
angle, $\theta^*_{\ell_1}$,
has strong discriminating power to separate the signal from backgrounds 
as shown in Fig.~\ref{thetass}. We then impose a final cut

\begin{equation}
\cos\theta^*_{\ell_1} < 0.95.
\end{equation}

\begin{figure}[tbp]
\epsfysize=5.5in
\epsffile[-50 0 560 560]{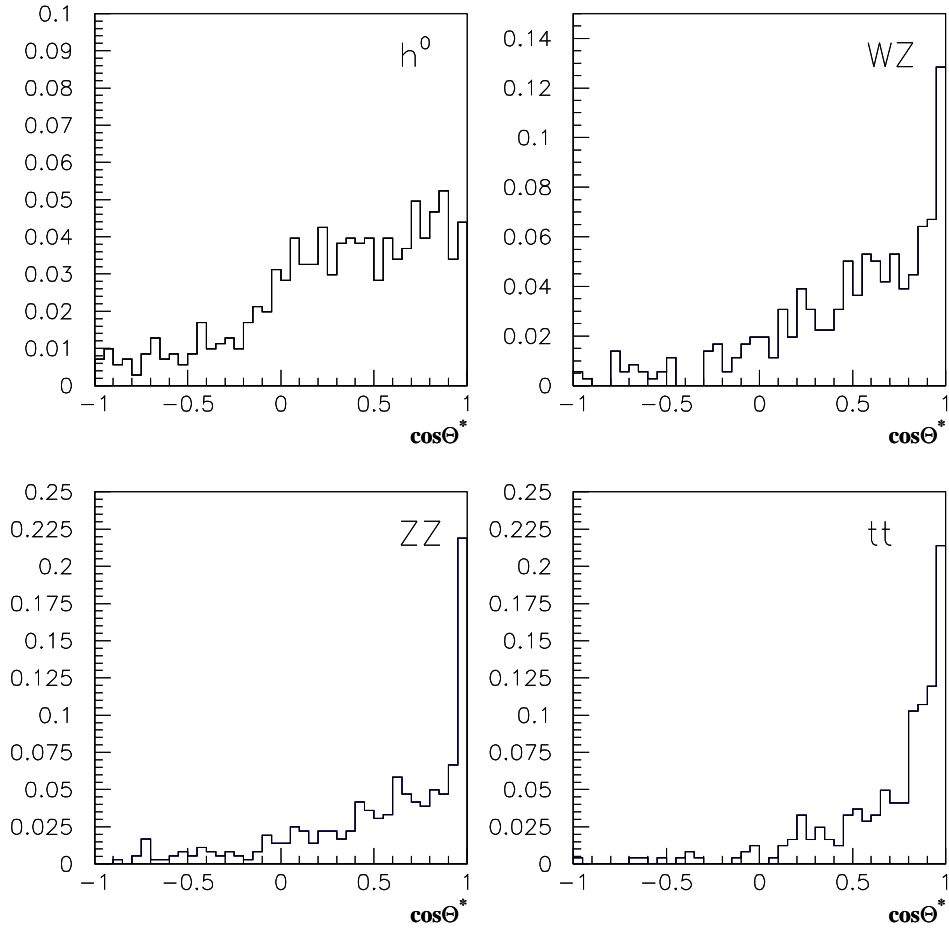}
\caption[]{Normalized distributions 
for the correlation angle 
between the momentum vector of the lepton pair and the momentum of
the higher $p_T$ lepton ($\ell_1$) in the lepton-pair rest frame, 
for the signal 
$W^\pm h \to W^\pm \ww \to \ell^\pm \ell^\pm jj$ 
with $\mh=170$ GeV and backgrounds $WZ$, $ZZ$ and $t \bar t$. 
\label{thetass}}
\end{figure}
\begin{table}[tbp]
\caption[]{$V h \to \ell^\pm\ell^\pm jj$ 
signal for $\mh=120-$200 GeV and the SM backgrounds 
after the kinematic cuts of eqs.~(\ref{basic2})--(\ref{mpt}). 
}
\begin{tabular}{|l|c|c|c|c|c|c|c|c|c|}
\noalign{\vskip-6pt}  
\multicolumn{1}{c}{$\mh$ [GeV]} &120 &130 & 140 & 150 & 160 & 170 & 180 & 190 &200\\ 
\tableline
 signal sum [fb] &0.093 &0.20 &0.34 &0.52 &0.45 & 0.38& 0.29 &0.20 &0.16\\ 
\hline
\hline
background channels&$WZ$&$ZZ$&$WW$&$t\bar t$&$VVV$ &$t\bar t V$ &
$W/Z\ jj+$fake & Sum  &\\
$\sigma$ [fb]& 0.27&0.06 &0.01&0.15&0.07 &0.02 &0.26 &0.83 &\\ 
\hline
$S/B$ [\%] & 11 & 24  & 41 & 63 & 54 & 46 & 35 & 24& 19 \\ \hline
$S/\sqrt B$ [30 fb$^{-1}$] &0.56 &1.2 &2.0 &3.1 &2.7 &2.3&1.7&1.3&0.96\\ 
\end{tabular}
\label{lljj}
\end{table}

With these cuts, we present the results for the signal
and backgrounds in Table~\ref{lljj}. We can see that for a
given $\mh$, the $S/B$ is larger than that for the di-lepton
plus $\etmiss$ signature, reaching as high as 63\%. One can consider
further optimization of cuts with $\mh$ dependence.
However, the rather small signal rate for a 30 fb$^{-1}$
luminosity limits the statistical significance.
Also, the systematic uncertainty in the background 
may be worse than the $\ell\nu\ell'\bar{\nu}$ channel.


%
%
\newcommand{\ppbar}     {{\mathrm{p}\bar{\mathrm{p}}}}
\newcommand{\WpWm}      {{W^+W^-}}
\newcommand{\Zphi}      {{\mathrm{Z}\phi}}
\newcommand{\Et}        {E_T}
\newcommand{\Lumi}  {\cal{L}}
\newcommand{\invfb} {\mathrm{fb}^{-1}}
%
%
\vspace{0.2in}
{\bf Merged Analysis} \\ \nopagebreak
\par
As described in the previous section, a final state with two {\em
same-sign} leptons arises in the associated production process
$\ppbar\rightarrow Wh$ with $h$ decaying to $\WpWm$ (we suppress the
stars for clarity of notation).  One of the $W$'s produced in the $h$
decay will have the same sign as the primary $W$ boson, and if these
two $W$'s both decay leptonically, a particularly clean signature
results.  This signature has been

developed in the search for supersymmetric particles at Run~1
in which very low background rates have been achieved.  We have
attempted to exploit this channel in the search for high-mass
Higgs bosons.  
%
%
\par
This channel was explored in parallel by two members of the group.
One of these efforts has been published~\cite{htz} and is 
described in the previous section.  Later, the two analyses 
were merged and
better performance obtained.  We describe this ``merged analysis''
briefly in this section.
\pagebreak

\par
The definition of leptons and jets is retained from the analysis
described above.  Leptons are identified in the range $|\eta| < 1.5$
and are accepted if $p_t > 10$~GeV.  They must be separated from
all hadronic jets by $dR > 0.3$, and in addition for muons, the
sum of momenta of charged tracks in a $30^\circ$ cone around the
muon must be less than 60\% of the muon's momentum.
Jets must fall in the range $|\eta| < 3$ and have $\Et > 15$~GeV.
\par
The selection requirements are as follows.  There must be exactly
two identified leptons of the same sign, and their invariant mass
must be at least 2~GeV.  Their isolation, computed as the $\Et$ sum
over towers in a cone around the lepton, must be less than 3.4~GeV.
Back-to-back leptons occur in background processes but rarely in
the signal, so we require an acoplanarity $\Delta\phi < 179^\circ$ and 
and acollinearity $\alpha < 174^\circ$. 
\par
In contrast to the trilepton channel, we want the candidate events
to show significant jet activity, expected from the hadronic decay
of the third (opposite-sign) $W$ boson.
There should be at least two jets but no more than six, and
none of them should pass the b-tagging algorithm (impact-parameter
method).  To avoid $\tau$-jets coming from di-boson backgrounds,
we require that the two highest-$\Et$ jets have at least two
charged tracks.  Keeping in mind the mass of the $\mathrm{W}$,
we require that these two jets have an invariant mass of no more
than 200~GeV.  Finally, as further evidence of high-$p_T$ activity,
we require that the scalar $\Et$ sum over all calorimeter towers
and muon tracks be at least 170~GeV.
\par
The numerical values for the cuts listed above were chosen by
a simple optimization program which varied all cuts simultaneously
to obtain the best value for $S/\sqrt{B}$ assuming in integrated
luminosity of $30~\invfb$.  This optimization procedure showed
that some of the cuts of the two original analyses were no longer
effective, and have been dropped.  The signal was computed for
both $Wh$ and $Zh$ production, and included $W \rightarrow
\tau\nu$ decays.
\par
Estimated backgrounds from Drell-Yan, $\WpWm$ and $ZZ$ production
are negligible.  The $WZ$ background is serious however, and
is estimated to be 0.15~fb.  The $t\bar{t}$ background is significantly
reduced with respect to the previous analysis, but still is important
at 0.024~fb.  Background from a $W$ or $Z$ and jets
is 0.15~fb based on D\O\ data.  Contributions from triple-boson and
$t\bar{t}+V$ production have been estimated and are rather small.  The
total background expecation is 0.48~fb.  
\par
For a SM Higgs of mass
$160~\gev$, for which the cuts above were designed, the accepted
cross section is $0.41$~fb, giving $S/B = 0.84$ (Table~\ref{tab:ssdlmerged}).  
For ${\Lumi} = 30~\invfb$, $12$ signal events would be expected,
corresponding to $S/\sqrt{B} = 3.2$, for one experiment.

\begin{table}
\caption[.]{Accepted cross section and significance for the merged
same-sign dilepton selection.  The effective cross section, 
$\sigma_{\mathrm{eff}}$,
includes the leptonic branching ratio factors for $Wh$ 
and $Zh$ production.
\label{tab:ssdlmerged}}
\begin{center}
\begin{tabular}{c|cccc}
\noalign{\vskip-6pt}
$M_h$ & $\sigma_{\mathrm{eff}}$ (fb) & $\sigma_{\mathrm{acc}}$ (fb) &
  $S/B$ & $S/B$ ($30~\invfb$) \cr
  \tableline
 100 &  5.1 & 0.01 & 0.02 & 0.07 \cr
 110 & 15.8 & 0.04 & 0.08 & 0.32 \cr
 120 & 28   & 0.08 & 0.16 & 0.60 \cr
 130 & 59   & 0.15 & 0.30 & 1.15 \cr
 140 & 77   & 0.29 & 0.60 & 2.3  \cr
 150 & 84   & 0.36 & 0.75 & 2.8  \cr
 160 & 84   & 0.41 & 0.84 & 3.2  \cr
 170 & 69   & 0.38 & 0.77 & 3.0  \cr
 180 & 55   & 0.26 & 0.53 & 2.0  \cr
 190 & 44   & 0.20 & 0.41 & 1.55 \cr
 200 & 35   & 0.16 & 0.33 & 1.24 \cr 
\end{tabular}
\end{center}
\end{table}


  \subsection{Higgs Bosons with Enhanced Diphoton Decay Rates}
                           \small
\begin{center}
{\it G. Landsberg,
     K. Matchev}
\end{center}
\normalsize \nopagebreak

The Standard Model (SM) is very economical in the sense that the Higgs
doublet responsible for electroweak symmetry breaking can also be used
to generate fermion masses.  The Higgs boson couplings to the gauge
bosons, quarks, and leptons are therefore predicted in the Standard
Model, where one expects the Higgs boson to decay mostly to b-jets and
tau pairs (for low Higgs masses, $M_h\lsim 140$ GeV), or to $WW$ or
$ZZ$ pairs, (for higher Higgs masses, $M_h \gsim 140$ GeV).  Since the
Higgs boson is neutral and does not couple to photons at tree level,
the branching ratio ${\rm B}(h\rightarrow \gamma\gamma)$ is predicted
to be very small in the SM, on the order of $10^{-3}-10^{-4}$.

In a more general framework, however, where different sectors of the
theory are responsible for the physics of flavor and electroweak
symmetry breaking, one may expect deviations from the SM predictions,
which may lead to drastic changes in the Higgs boson discovery
signatures.  One such example is the so called ``fermiophobic'' (also
known as ``bosophilic'' or ``bosonic'') Higgs, which has suppressed
couplings to all fermions, and may arise in a variety of
models~\cite{fermiophobe}. A variation on this theme is the Higgs in
certain topcolor models, which may couple to heavy quarks
only~\cite{topmodels}. Some even more exotic possibilities have been
suggested in the context of theories with large extra
dimensions~\cite{LED}.

Finally, in the minimal supersymmetric standard model (MSSM), the
width into $b\bar{b}$ pairs can be suppressed due to 1-loop SUSY
corrections, thus enhancing the branching ratios of a light Higgs into
more exotic signatures \cite{CMW1,Mrenna}.  In all these cases, the
Higgs boson decays to photon pairs are mediated through a $W$ or heavy
quark loop and dominate for $M_h\lsim 100$ GeV
\cite{Stange:1994ya,SMW}. In the range $100\lsim M_h\lsim 160$, they
compete with the $WW^\ast$ mode, while for $M_h\gsim 160$ GeV,
$h\rightarrow WW$ completely takes over.  Current bounds from
LEP~\cite{LEP limits} are limited by the kinematic reach of the
machine. The existing Run 1 analyses at the Tevatron have utilized the
diphoton plus 2 jets~\cite{Lauer,D0,Wilson} and inclusive
diphoton~\cite{Wilson} channels and were limited by statistics.  Since
they only looked for a ``bosonic'' Higgs, they did not consider the
Higgs production mechanism through gluon fusion, which can be a major
additional source of signal in certain models~\cite{topmodels}.  Since
$h\rightarrow\gamma\gamma$ is a very clean signature, it will allow
the Tevatron to extend significantly those limits in its next run.

In this study we shall evaluate the Higgs discovery potential of the
upcoming Tevatron runs for several diphoton channels.  We shall
concentrate on the following two questions. First, what is the
absolute reach in Higgs mass as a function of the
$h\rightarrow\gamma\gamma$ branching ratio? Second, which signature
(inclusive diphotons, diphotons plus one jet, or diphotons plus two
jets) provides the best reach.  We believe that neither of those two
questions has been adequately addressed in the literature previously.

\subsubsection{Fermiophobic Higgs Bosons}

Here we consider the case of a fermiophobic Higgs, {\it i.e.} models where
the Higgs couplings to all fermions are suppressed. Then, the main
Higgs production modes at the Tevatron are associated $Wh/Zh$
production, as well as $WW/ZZ$ fusion.  All of these processes have
comparable rates [see fig.~\ref{fg:4}], so it makes sense to consider an
inclusive signature first~\cite{Wilson}.

\vspace{0.2in}
{\bf Inclusive channel: event selection} \\ \nopagebreak

We use the following cuts for our inclusive study: two photons with
$p_T(\gamma)>20$ GeV and rapidity $|\eta(\gamma)| < 2$, motivated by
the acceptance of the CDF or D\O\ detectors in Run~2. Triggering on
such a signature is trivial; both collaborations will have diphoton
triggers that are nearly fully efficient with such offline cuts.

We assume 80\% diphoton identification efficiency, which we apply to
both the signal and background estimates on top of the kinematic and
geometrical acceptance.  Again, this efficiency is motivated by the
CDF/D\O\ EM ID efficiency in Run 1 and is not likely to change in
Run~2.

\vspace{0.2in}
{\bf Inclusive channel: background} \\ \nopagebreak

The main backgrounds to the inclusive diphoton channel come from the
QCD production of dijets, direct photons, and diphotons.  In the
former two cases a jet mimics a photon by fragmenting into a leading
$\pi^0/\eta$ meson that further decays into a pair of photons, not
resolved in the calorimeter.

\begin{figure}[t]
\epsfysize=3.5in
\epsffile[-250 0 100 560]{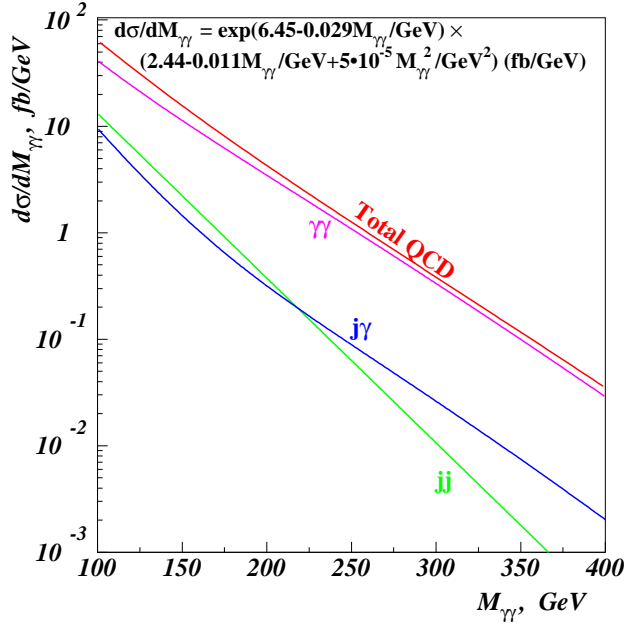}
\begin{center}
\caption[]
{\small The total background in the inclusive diphoton channel,
as well as the individual contributions from $\gamma\gamma$,
$\gamma j$ and $jj$ production. \label{qcd_background}}
\end{center}
\end{figure}

We used the PYTHIA~\cite{PYTHIA} event generator and the
experimentally measured probability of a jet to fake a
photon~\cite{Lauer} to calculate all three components of the QCD
background. The faking probability depends significantly on the
particular photon ID cuts, especially on the photon isolation
requirement~(see, {\it e.g.}, ref.~\cite{Lauer,diboson,monopole}).
For this study we used an $E_T$-dependent jet-faking-photon
probability of $$ P({\rm jet} \rightarrow \gamma) = \exp\left(-0.01\
{E_T\over\mbox{(1 GeV)}}-7.5\right),$$ which is obtained by taking the
$\eta$-averaged faking probabilities used in the D\O\ Run 1
searches~\cite{Lauer}. The fractional error on $P(\mbox{jet}
\rightarrow \gamma)$ is about 25\% and is dominated by the uncertainty
on the direct photon fraction in the $\mbox{jet}+\gamma$ sample used
for its determination.  (For high photon $E_T$, however, the error is
dominated by the available statistics.)  This probability is expected
to remain approximately the same in Run~2 for both the CDF and D\O\
detectors.  We used 80\% ID efficiency for the pair of photons, and
required the photons to be isolated from possible extra jets in the
event.  We accounted for NLO corrections via a constant $k$-factor of
1.34.

\begin{figure}[t]
\epsfysize=3.5in
\epsffile[-250 0 100 560]{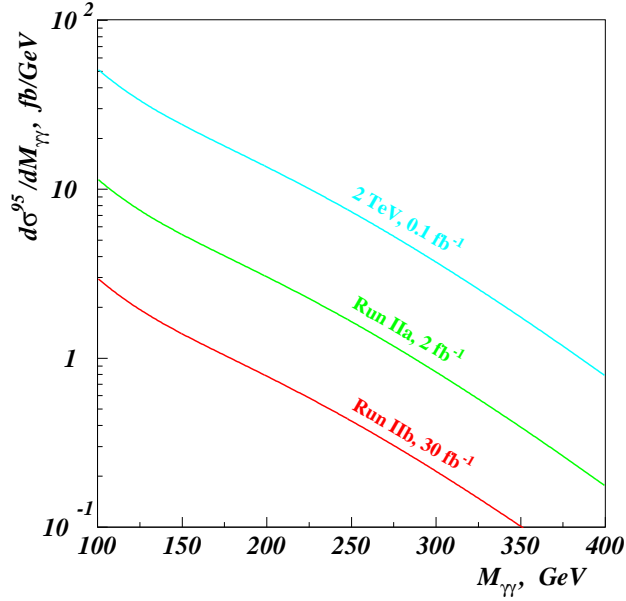}
\begin{center}
\caption[]
{\small The 95\% CL upper limit on the signal cross section after cuts
as a function of the diphoton invariant mass $M_{\gamma\gamma}$,
for several benchmark total integrated luminosities in Run~2. 
\label{ggX}}
\end{center}
\end{figure}

Adding all background contributions, for the total background in the inclusive
diphoton channel we obtain the following parametrization:
$$
{d\sigma\over dM_{\gamma\gamma}}
= \biggl[p_3+p_4 \left({M_{\gamma\gamma}\over 1\ {\rm GeV}}\right)
      +p_5 \left({M_{\gamma\gamma}\over 1\ {\rm GeV}}\right)^2\biggr]\
\exp\,\biggl\{ p_1+
    p_2 \left( {M_{\gamma\gamma}\over 1\ {\rm GeV}}\right)\biggr\},
$$
where $p_1= 6.45$, $p_2=-0.029$, $p_3= 2.44$, $p_4=-0.011$ and $p_5=
0.00005$.  In the region $M_{\gamma\gamma}>100$ GeV it is dominated by
direct diphoton production and hence is irreducible.  The expected
statistical plus systematic error on this background determination is
at the level of 25\%, based on the jet-faking photon probability
uncertainty.  For larger invariant masses, however, the accuracy is
dominated by the uncertainties in the direct diphoton production cross
section, which will be difficult to measure independently in Run~2, so
one will still have to rely on the NLO predictions.  On the other
hand, for narrow resonance searches one could do self-calibration of
the background by calculating the expected background under the signal
peak via interpolation of the measured diphoton mass spectrum between
the regions just below and just above the assumed resonance
mass. Therefore, in our case the background error will be purely
dominated by the background statistics. A combination of the
interpolation technique and the shape information from the theoretical
NLO calculations of the direct diphoton cross section is expected to
result in significantly smaller background error in Run~2.

The total background, as well as the individual contributions from
$\gamma\gamma$, $\gamma j$ and $jj$ production, are shown in
Fig.~\ref{qcd_background}.  Additional SM background sources to the
inclusive diphoton channel include Drell-Yan production with both
electrons misidentified as photons, $W\gamma\gamma$ production,
etc. and are all negligible compared to the QCD background.  The
absolute normalization of the background obtained by the above method
agrees well with the actual background measured by CDF and D\O\ in the
diphoton mode~\cite{Wilson,monopole}.

In Fig.~\ref{ggX} we show the 95\% CL upper limit on the differential
cross section after cuts 
$d(\varepsilon\times\sigma(\gamma\gamma+X))/dM_{\gamma\gamma}$ as a
function of the diphoton invariant mass $M_{\gamma\gamma}$, given the
above background prediction (here $\varepsilon$ is the product of the
acceptance and all efficiencies).  This limit represents $1.96\sigma$
sensitivity to a narrow signal when doing a counting experiment in 1
GeV diphoton mass bins.  This plot can be used to obtain the
sensitivity to any resonance decaying into two photons as follows. One
first fixes the width of the mass window around the signal peak which
is used in the analysis.  Then one takes the average value of the 95\%
C.L. limit in $d\sigma/dM_{\gamma\gamma}$ across the mass window from
Fig.~\ref{ggX} and multiplies it by $\sqrt{w/\mbox{GeV}}$, where $w$
is the width of the mass window\footnote{The square root enters the
calculation since the significance is proportional to the background
to the $-1/2$ power.}, to obtain the corresponding 95\% CL upper limit
on the signal cross-section after cuts.  Similar scaling could be used
if one is interested in the 3$\sigma$ or 5$\sigma$ reach.

\vspace{0.2in}
{\bf Inclusive channel: optimum mass window} \\ \nopagebreak

\begin{figure}[t]
\epsfysize=2.5in
\epsffile[-150 205 200 565]{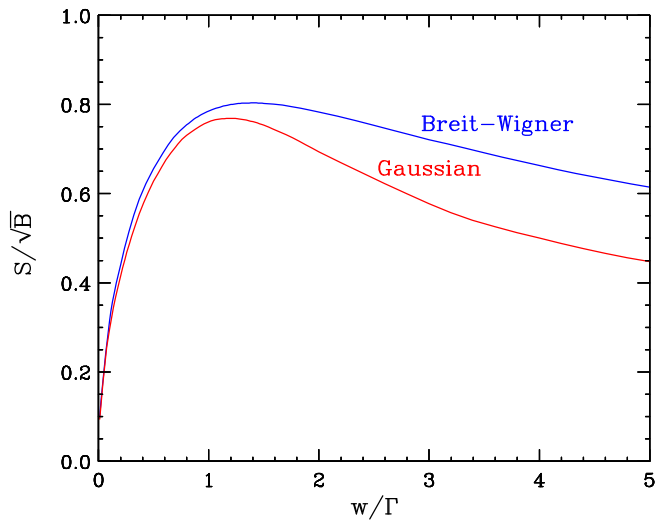}
\begin{center}
\caption[]
{\small Significance $S/\sqrt{B}$ (in arbitrary units),
as a function of the mass window width $w$ (in units of $\Gamma$),
for a Breit-Wigner or Gaussian resonance. \label{fig:significance}}
\end{center}
\end{figure}

When searching for narrow resonances in the presence of large
backgrounds ($B$), the best sensitivity toward signal ($S$) is
achieved by performing an unbinned maximum likelihood fit to the sum
of the expected signal and background shapes.  However, simple
counting experiments give similar sensitivity if the size of the
signal ``window'' is optimized. For narrow resonances the observed
width\footnote{Notice that the width is defined so that the
cross-section at $\pm\Gamma/2$ away from the peak is a factor of 2
smaller than the peak value (FWHM). For a Gaussian resonance the width
is related to the variance $\sigma$ by
$\Gamma=2\sigma\sqrt{\ln4}\simeq 2.35\sigma$.}  $\Gamma$ is dominated
by the instrumental effects, and is often Gaussian.  The background in
a narrow window centered on the assumed position $M_0$ of the peak in
the signal invariant mass distribution could be treated as
linear. Therefore, the Gaussian significance of the signal,
$S/\sqrt{B}$, as a function of the window width, $w$, is given by:
\begin{equation}
{S\over \sqrt{B}} 
\sim {1\over\sqrt{w}}\ 
{1\over \sqrt{2\pi}\sigma}
\int_{M_0-w/2}^{M_0+w/2}d\sqrt{s} 
\exp\left(-{(\sqrt{s}-M_0)^2\over 2\sigma^2}\right)
\sim {1\over\sqrt{w/\Gamma}}\ 
{\rm erf}\left(\sqrt{\ln2}\ {w\over\Gamma}\right),
\label{signgaus}
\end{equation}
where erf$(x)$ is the error function
$$
{\rm erf}(x)={2\over\sqrt{\pi}}\int_0^x e^{-t^2} dt.
$$
The function (\ref{signgaus}) is shown in
Fig.~\ref{fig:significance}, and has a maximum at $w \approx 1.2\Gamma$,
which corresponds to $\pm 1.2(\Gamma/2)$ cut around the resonance maximum.

For resonances significantly wider than the experimental resolution,
the shape is given by the Breit-Wigner function, and in this case
the significance is:
\begin{equation}
{S\over\sqrt{B}}
\sim {1\over \sqrt{w}}
\int_{(M_0-w/2)^2}^{(M_0+w/2)^2}
{ds\over (s-M_0^2)^2+M_0^2\Gamma^2}
\sim {1\over \sqrt{w/\Gamma}}\arctan({w\over\Gamma}).
\end{equation}
This function, also shown in Fig.~\ref{fig:significance},
peaks at a similar value of $w$ ($w \approx 1.4\Gamma$).
We see that for both Gaussian and Breit-Wigner resonances,
the significance does not appreciably change when
using a $w = 1\Gamma-2\Gamma$ cuts. For our analysis
we shall use two representative choices: $w=1.2\Gamma$ 
and $w= 2\Gamma$ for the mass window, which we shall
always center on the actual Higgs mass. 

Clearly, one can do even better in principle, by suitably resizing and
repositioning the mass window around the bump in the combined $S+B$
distribution. Because of the steeply falling parton luminosities,
the signal mass peak is skewed and its maximum will appear somewhat
below the actual physical mass. In our analysis we choose
not to take advantage of these slight improvements,
thus accounting for unknown systematics.

\vspace{0.2in}
{\bf Inclusive channel: results} \\ \nopagebreak

In Tables \ref{1sigma} and \ref{2sigma} we show
the inclusive $\gamma\gamma+X$ background rates in fb
for different Higgs masses, for $w=1.2\Gamma$ and $w=2\Gamma$
mass window cuts, respectively.
\begin{table*}[ht!]
\renewcommand{\arraystretch}{1.5}
\caption{ Background rates in fb for $w=1.2\Gamma$
mass cut, and significance ($S/\sqrt{B}$, for 1 fb$^{-1}$ of data, and
assuming ${\rm B}(h\rightarrow\gamma\gamma)=100\%$) as
a function of the Higgs mass $M_h$. The signal consists of 
associated $Wh/Zh$ production and $WW/ZZ$ fusion. \label{1sigma}}
\begin{tabular}{||c||c||c||c|c|c|c||c|c|c|c||}
\noalign{\vskip-6pt}\hline
      & $\gamma\gamma+X$ 
      & \multicolumn{9}{c||}{Significance $S/\sqrt{B}$}  \\ \cline{3-11}
$M_h$ & bknd &  $\gamma\gamma+X$
      & \multicolumn{4}{c||}{$\gamma\gamma+1$ jet}
      & \multicolumn{4}{c||}{$\gamma\gamma+2$ jets}  \\ \cline{4-11}
(GeV) & (fb) & 
      & $p_T>20$ & $p_T>25$ & $p_T>30$ & $p_T>35$ 
      & $p_T>20$ & $p_T>25$ & $p_T>30$ & $p_T>35$  \\ \hline\hline
100. &271.7 & 16.5 & 31.3 & 34.2 & 36.3 & 35.4 & 31.9 & 36.4 & 35.1 & 31.1 \\ \hline
120. &166.6 & 13.0 & 24.4 & 26.7 & 28.5 & 28.5 & 24.7 & 29.5 & 29.2 & 26.7 \\ \hline
140. &103.0 & 10.7 & 20.1 & 22.4 & 23.9 & 23.9 & 20.6 & 24.0 & 23.7 & 21.0 \\ \hline
160. & 64.3 &  8.9 & 17.0 & 19.1 & 20.1 & 20.4 & 17.2 & 20.2 & 20.8 & 19.2 \\ \hline
180. & 40.5 &  7.3 & 13.5 & 15.0 & 16.2 & 16.5 & 13.6 & 16.3 & 16.7 & 16.2 \\ \hline
200. & 26.1 &  5.8 & 10.6 & 11.9 & 12.5 & 12.7 & 10.4 & 12.3 & 12.7 & 12.0 \\ \hline
250. &  9.4 &  3.7 &  6.7 &  7.5 &  7.9 &  8.2 &  6.6 &  7.9 &  8.7 &  8.4 \\ \hline
300. &  4.8 &  2.2 &  3.8 &  4.3 &  4.7 &  4.7 &  3.6 &  4.2 &  4.9 &  4.7 \\ \hline
350. &  2.3 &  1.5 &  2.8 &  3.1 &  3.3 &  3.3 &  2.4 &  2.8 &  3.0 &  3.4 \\ \hline
400. &  1.2 &  1.0 &  1.8 &  2.0 &  2.0 &  2.1 &  1.7 &  1.7 &  2.1 &  2.4 \\ \hline
\hline
\end{tabular}
\end{table*}

\begin{table*}[h!]
\renewcommand{\arraystretch}{1.5}
\caption{ The same as Table \ref{1sigma}, but for a
$w=2\Gamma$ mass window. \label{2sigma}}
\begin{tabular}{||c||c||c||c|c|c|c||c|c|c|c||}
\noalign{\vskip-6pt}\hline
      & $\gamma\gamma+X$ 
      & \multicolumn{9}{c||}{Significance $S/\sqrt{B}$}  \\ \cline{3-11}
$M_h$ & bknd &  $\gamma\gamma+X$
      & \multicolumn{4}{c||}{$\gamma\gamma+1$ jet}
      & \multicolumn{4}{c||}{$\gamma\gamma+2$ jets}  \\ \cline{4-11}
(GeV) & (fb) & 
      & $p_T>20$ & $p_T>25$ & $p_T>30$ & $p_T>35$ 
      & $p_T>20$ & $p_T>25$ & $p_T>30$ & $p_T>35$  \\ \hline\hline
100. &453.4 & 14.4 & 27.0 & 29.8 & 31.7 & 30.6 & 27.7 & 31.9 & 30.5 & 26.4  \\ \hline
120. &278.1 & 11.3 & 21.3 & 23.5 & 25.1 & 24.9 & 21.9 & 25.5 & 25.1 & 22.5  \\ \hline
140. &171.9 &  9.3 & 17.5 & 19.5 & 21.0 & 21.0 & 17.7 & 20.7 & 21.0 & 18.5  \\ \hline
160. &107.3 &  8.0 & 15.1 & 16.7 & 18.0 & 18.2 & 15.4 & 17.8 & 18.2 & 17.3  \\ \hline
180. & 67.6 &  6.6 & 12.2 & 13.7 & 14.6 & 14.9 & 12.4 & 14.3 & 15.2 & 14.2  \\ \hline
200. & 43.6 &  5.4 & 10.1 & 11.4 & 12.1 & 12.1 &  9.9 & 11.9 & 12.3 & 11.1  \\ \hline
250. & 15.7 &  3.6 &  6.5 &  7.3 &  7.7 &  7.8 &  6.4 &  7.6 &  8.5 &  8.0  \\ \hline
300. &  8.1 &  2.1 &  3.8 &  4.2 &  4.6 &  4.5 &  3.6 &  4.3 &  4.5 &  4.5  \\ \hline
350. &  3.9 &  1.6 &  2.7 &  3.0 &  3.4 &  3.4 &  2.6 &  3.1 &  3.5 &  3.1  \\ \hline
400. &  2.1 &  1.1 &  1.9 &  2.2 &  2.3 &  2.4 &  1.6 &  2.1 &  2.6 &  2.4  \\ \hline
\hline
\end{tabular}
\end{table*}

Here we have added the intrinsic width $\Gamma_h$ and the experimental
resolution $\Gamma_{\rm exp}=2\sqrt{\ln4}\times\sigma_{\rm exp}\simeq
2.35\times0.15\sqrt{2}\sqrt{E(\gamma)} \simeq 0.35\sqrt{M_h}$ in
quadrature: $\Gamma=\left(\Gamma^2_h+\Gamma^2_{\rm exp}\right)^{1/2}$.
The width $\Gamma$ varies between 3.5 GeV for $M_h=100$ GeV and 29.0
GeV for $M_h=400$ GeV. The two tables also show the significance (for
1 fb$^{-1}$ of data, and assuming ${\rm
B}(h\rightarrow\gamma\gamma)=100\%$) in the inclusive diphoton channel
when only associated $Wh/Zh$ production and $WW/ZZ\rightarrow h$
fusion are included in the signal sample. We see that (as can also be
anticipated from Fig.~\ref{fig:significance}) a $w=1.2\Gamma$ cut
around the Higgs mass typically gives a better statistical
significance, especially for lighter (and therefore more narrow) Higgs
bosons.

\begin{figure}[ht!]
\begin{center}
\epsfysize=2.5in
\epsffile[-150 205 200 565]{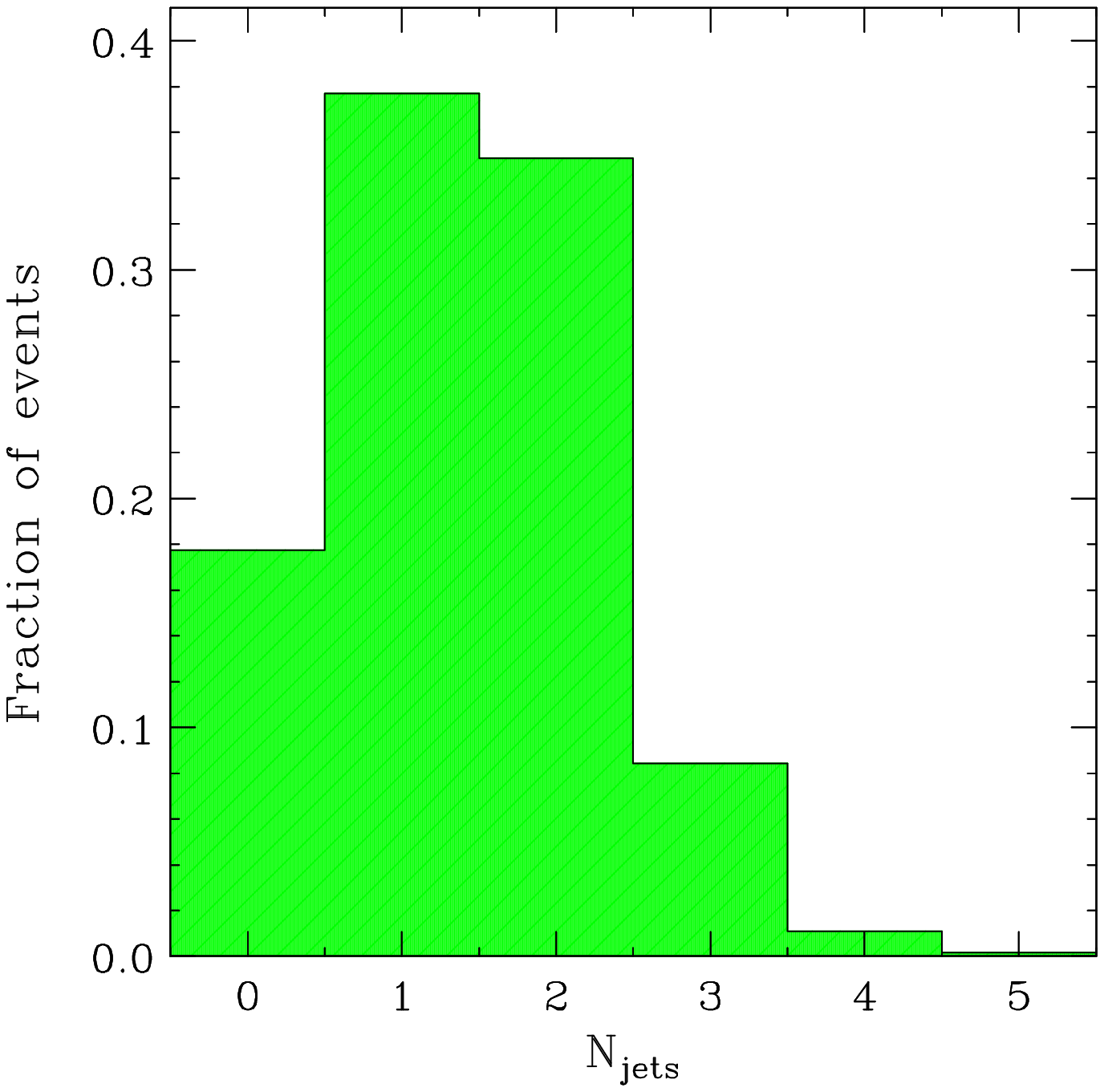}
\end{center}
\caption[]
{\small The number of ``jets'', which stands for
QCD jets, tau jets and electrons, in associated $Wh$
production, once we require the two photons from the
Higgs to pass the photon ID cuts. \label{nobj}}
\end{figure}

\begin{figure}[h!]
\begin{center}			
\epsfysize=3.5in
\epsffile[-20 250 80 550]{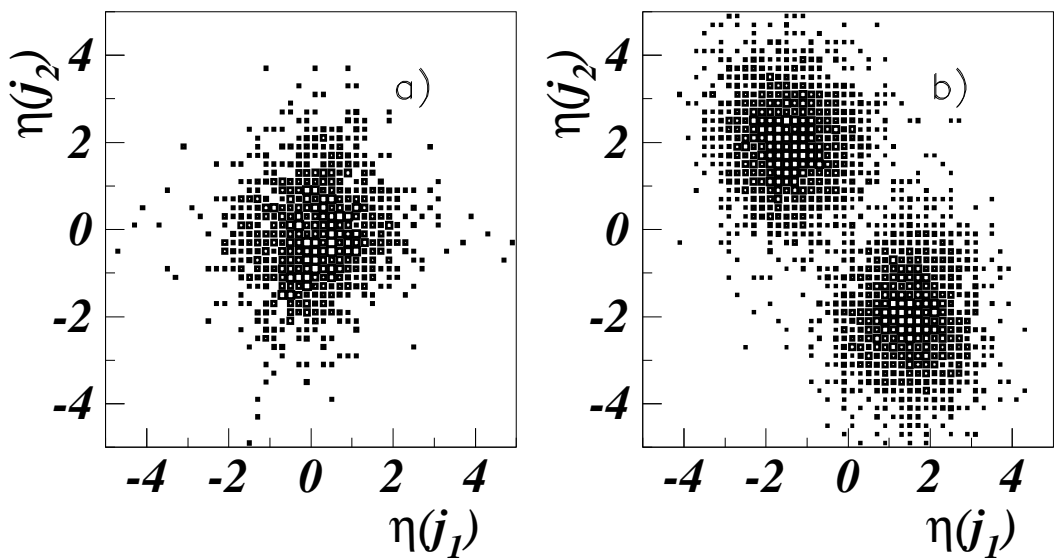}
\end{center}
\caption[]
{\small Pseudorapidity distribution of the two spectator jets
in (a) associated $Wh/Zh$ production and 
(b) $WW/ZZ\rightarrow h$ fusion. Here
$\eta(j_1)$ ($\eta(j_2)$) is the pseudorapidity of
the leading (next-to-leading) jet in the event.
\label{rapidity}}
\end{figure}

\vspace{0.2in}
{\bf Exclusive channels: event selection} \\ \nopagebreak

The next question is whether the sensitivity can be further improved
by requiring additional objects in the event.  The point is that a
significant fraction of the signal events from both associated $Wh/Zh$
production and $WW/ZZ$ fusion will have additional hard objects, most
often QCD jets.  In Fig.~\ref{nobj} we show the ``jet'' multiplicity
in associated $Wh$ production, where for detector simulation we have
used the SHW package with a few modifications as in~\cite{SHWmod}.
Here we treat ``jets'' in a broader context, including electrons and
tau jets as well.

Previous studies \cite{Wilson,D0} have required {\em two} or more
additional QCD jets. Here we shall also consider the signature with at
least {\em one} additional ``jet'', where a ``jet'' is an object with
$|\eta|<2$.  The advantages of not requiring a second ``jet'' are
twofold. First, in this way we can also pick up signal from
$WW/ZZ\rightarrow h$ fusion, whose cross-section does not fall off as
steeply with $M_h$, and in fact for $M_h>200$ GeV is larger than the
cross-section for associated $Wh/Zh$ production\footnote{In the case
of a topcolor Higgs (see the next section) we would also pick up
events with initial state gluon radiation, comprising about 30\% of
the gluon fusion signal, which is the dominant production process for
any Higgs mass.}.  Events from $WW/ZZ\rightarrow h$ fusion typically
contain two very hard forward jets, one of which may easily pass the
jet selection cuts. This is illustrated in Fig.~\ref{rapidity}, where
we show the pseudorapidity distribution of the two spectator jets in
(a) associated $Wh/Zh$ production and (b) $WW/ZZ\rightarrow h$ fusion.
Second, by requiring only one additional jet, we win in signal
acceptance. In order to compensate for the corresponding background
increase, we shall consider several $p_T$ thresholds for the
additional jet, and choose the one giving the largest significance.

For the exclusive channels we need to rescale the background from
Fig.~\ref{qcd_background} as follows.  From Monte Carlo we obtain
reduction factors of $4.6\pm0.5$, $6.2\pm1.0$, $7.6\pm1.4$, and
$8.6\pm1.5$ for the $\gamma\gamma+1$ jet channel, with $p_T(j)>20$,
25, 30 and 35 GeV, respectively. For the $\gamma\gamma+2$ jets channel
the corresponding background reduction is $21\pm5$, $38\pm12$,
$58\pm21$, and $74\pm26$, depending on the jet $p_T$ cuts. These
scaling factors agree well with those from the CDF and D\O\ data from
Run 1.

Notice that we choose not to impose an invariant dijet mass ($M_{jj}$)
cut for the $\gamma\gamma+2$ jets channel.  We do not expect that it
would lead to a gain in significance for several reasons. First, given
the relatively high jet $p_T$ cuts needed for the background
suppression, there will be hardly any background events left with
dijet invariant masses below the (very wide) $W/Z$ mass window.
Second, the signal events from $WW/ZZ$ fusion, which typically
comprise about $25-30\%$ of our signal, will have a dijet invariant
mass distribution very similar to that of the background. Finally, not
imposing the $M_{jj}$ cut allows for a higher signal acceptance
because of the inevitable combinatorial ambiguity for the events with
$>2$ jets.

The significances for the two exclusive channels, with the four
different jet $p_T$ cuts, are also shown in Tables \ref{1sigma} and
\ref{2sigma}.  We see that the exclusive $\gamma\gamma+2$ jets channel
with $p_T(j)>30$ GeV typically gives the largest significance, but our
new exclusive $\gamma\gamma+1$ jet channel is following very close
behind.

\vspace{0.2in}
{\bf Exclusive channels: results} \\ \nopagebreak

We are now ready to present our results for the Run~2 Tevatron reach
for a bosonic Higgs.  In Fig.~\ref{bos} we show the 95\% CL upper
limit on the branching ratio ${\rm B}(h\rightarrow \gamma\gamma)$,
with 0.1 (cyan), 2.0 (green) and 30 ${\rm fb}^{-1}$ (red), as a
function of $M_h$. For each mass point, we compare the significance
for both the inclusive as well as the exclusive channels with all the
different cuts, and for the limit we choose the channel with the set
of cuts providing the best reach.  
\vspace{0.4in}

\begin{figure}[hb!]
\centerline{\psfig{file=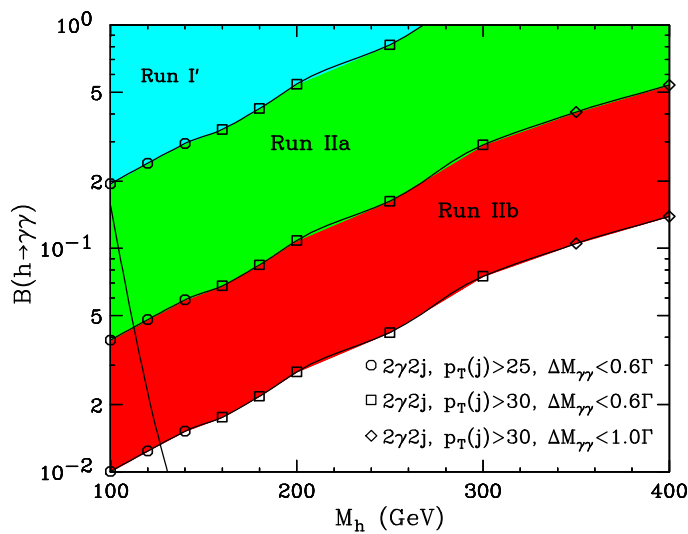,width=9cm}}
\begin{center}
\caption[]
{\small 95\% CL upper limit on the branching ratio
${\rm B}(h\rightarrow \gamma\gamma)$, 
with 0.1 (cyan), 2.0 (green) and 30 ${\rm fb}^{-1}$ (red), 
as a function of $M_h$. For each mass point, we compare the
significance for both the inclusive and the exclusive
channels with different cuts, and for the limit we choose the
set of cuts which provides the best reach, denoted by
o: $2\gamma+2j$, with $p_T(j)>25$ GeV and $w=1.2\,\Gamma$,
$\Box$: $2\gamma+2j$, with $p_T(j)>30$ GeV and $w=1.2\,\Gamma$, and
$\Diamond$: $2\gamma+2j$, with $p_T(j)>30$ GeV and $w=2.0\,\Gamma$.
The solid line is the prediction for 
the branching ratio of a ``bosonic'' Higgs.
\label{bos}}
\end{center}
\end{figure}

It turns out that for this case
the optimal selections are
\begin{itemize}
 \item[o:]          $2\gamma+2j$, with $p_T(j)>25$ GeV and $w=1.2\,\Gamma$, 
 \item[$\Box$:]     $2\gamma+2j$, with $p_T(j)>30$ GeV and $w=1.2\,\Gamma$, and 
 \item[$\Diamond$:] $2\gamma+2j$, with $p_T(j)>30$ GeV and $w=2.0\,\Gamma$. 
\end{itemize} 
In the figure we also show the HDECAY
\cite{hdecay} prediction for ${\rm B}(h\rightarrow \gamma\gamma)$ in
case of a ``bosonic'' Higgs.  The reach shown for 0.1 ${\rm fb}^{-1}$
is intended as a comparison to Run 1, in fact for the 0.1 ${\rm
fb}^{-1}$ curve we scaled down both the signal and background
cross-sections to their values at 1.8 TeV center-of-mass energy,
keeping the efficiencies the same. In other words, the region marked
as Run 1$^\prime$ would have been the hypothetical reach in Run 1, if the
improved Run~2 detectors were available at that time.  As seen from
Fig.~\ref{bos}, the reach for a ``bosonic'' Higgs (at 95\% CL) in
Run~2a and Run~2b is $\sim115$ GeV and $\sim125$ GeV,
correspondingly. This is a significant improvement over the ultimate
reach from LEP \cite{LEP limits} of $\sim 105$ GeV.

\subsubsection{Topcolor Higgs Bosons}

Here we consider the case of a ``topcolor'' bosonic Higgs, where the
Higgs also couples to the top and other heavy quarks~\cite{topmodels}.
We therefore include events from gluon fusion into our signal
sample. We used next-to-leading order cross-sections for gluon
fusion~\cite{hxsec}.

In Tables \ref{1sigmatop} and \ref{2sigmatop} we show the significance
(for 1 fb$^{-1}$ of data, and again assuming ${\rm
B}(h\rightarrow\gamma\gamma)=100\%$) in the inclusive and the two
exclusive channels, for the topcolor Higgs case.  Since gluon fusion,
which rarely has additional hard jets, is the dominant production
process, the inclusive channel typically provides the best reach.
However, the $2\gamma+1j$ channel is again very competitive, since the
additional hard jet requirement manages to suppress the background at
a reasonable signal cost.  We see that our new $2\gamma+1j$ channel
clearly gives a better reach than the $2\gamma+2j$
channel~\cite{Lauer,D0,Wilson}. For Higgs masses above $\sim180$ GeV,
it sometimes becomes marginally better even than the inclusive
diphoton channel. The specific jet $p_T$ cut and mass window size $w$
seem to be less of an issue -- from Tables~\ref{1sigmatop} and
\ref{2sigmatop} we see that $p_T(j)>25$, $p_T(j)>30$ GeV and
$p_T(j)>35$ GeV work almost equally well, and for $M_h\gsim 200$ GeV
both values of $w$ are acceptable.
\vspace{0.4in}

\begin{figure}[h!]
\centerline{\psfig{file=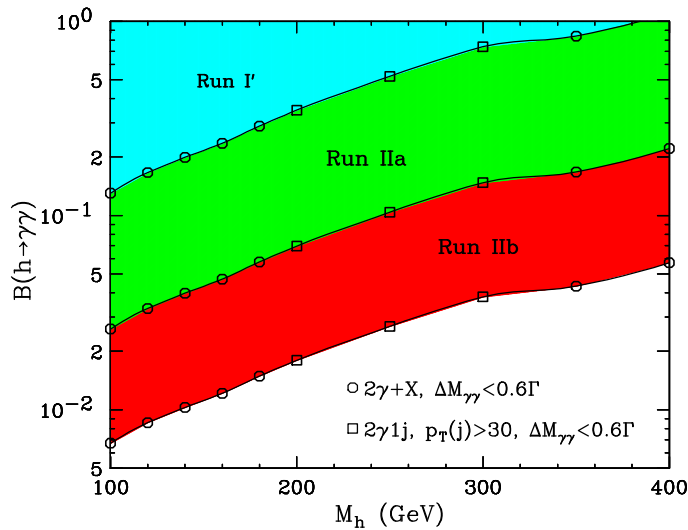,width=9cm}}
\begin{center}
\caption[]
{\small The same as Fig.~\ref{bos}, but for
a topcolor Higgs, {\it i.e.} gluon fusion events are
included in the signal. The channels with the best
$S/\sqrt{B}$ ratio are: 
o: inclusive $2\gamma+X$, and 
$\Box$: $2\gamma+1j$, with $p_T(j)>30$ GeV;
both with $w=1.2\,\Gamma$.
\label{topcolor}}
\end{center}
\end{figure}

\begin{table*}[ht!]
\renewcommand{\arraystretch}{1.5}
\caption{ The same as Table \ref{1sigma}, but for a
topcolor Higgs, {\it i.e.} gluon fusion events are included in the
signal. \label{1sigmatop}}
\begin{tabular}{||c||c||c||c|c|c|c||c|c|c|c||}\noalign{\vskip-6pt}\hline
      & $\gamma\gamma+X$ 
      & \multicolumn{9}{c||}{Significance $S/\sqrt{B}$}  \\ \cline{3-11}
$M_h$ & bknd &  $\gamma\gamma+X$
      & \multicolumn{4}{c||}{$\gamma\gamma+1$ jet}
      & \multicolumn{4}{c||}{$\gamma\gamma+2$ jets}  \\ \cline{4-11}
(GeV) & (fb) & 
      & $p_T>20$ & $p_T>25$ & $p_T>30$ & $p_T>35$ 
      & $p_T>20$ & $p_T>25$ & $p_T>30$ & $p_T>35$  \\ \hline\hline
100. &271.7 & 54.3 & 43.4 & 44.6 & 45.1 & 42.5 & 33.6 & 37.8 & 36.4 & 32.1\\ \hline
120. &166.6 & 42.5 & 35.3 & 36.9 & 37.3 & 35.3 & 26.5 & 30.8 & 30.2 & 27.5\\ \hline
140. &103.0 & 35.5 & 30.4 & 32.1 & 33.0 & 31.6 & 22.3 & 25.6 & 25.1 & 22.0\\ \hline
160. & 64.3 & 30.1 & 27.3 & 28.8 & 29.2 & 28.3 & 19.5 & 22.4 & 22.4 & 20.3\\ \hline
180. & 40.5 & 24.5 & 22.4 & 23.8 & 24.5 & 23.8 & 15.7 & 18.2 & 18.3 & 17.6\\ \hline
200. & 26.1 & 19.9 & 18.6 & 19.9 & 20.3 & 20.0 & 12.7 & 14.4 & 14.3 & 13.3\\ \hline
250. &  9.4 & 12.9 & 12.5 & 13.3 & 13.6 & 13.4 &  8.4 &  9.7 & 10.2 &  9.8\\ \hline
300. &  4.8 &  9.1 &  8.7 &  9.3 &  9.6 &  9.4 &  4.9 &  5.6 &  5.9 &  5.9\\ \hline
350. &  2.3 &  8.4 &  7.9 &  8.4 &  8.4 &  7.9 &  4.0 &  4.1 &  4.5 &  4.5\\ \hline
400. &  1.2 &  6.4 &  6.1 &  6.1 &  6.3 &  6.2 &  3.0 &  3.0 &  3.5 &  3.1\\ \hline
\end{tabular}
\end{table*}

\begin{table*}[h!]
\renewcommand{\arraystretch}{1.5}
\caption{ The same as Table \ref{1sigmatop}, but for a
$w=2\Gamma$ mass window. \label{2sigmatop}}
\begin{tabular}{||c||c||c||c|c|c|c||c|c|c|c||}\noalign{\vskip-6pt}\hline
      & $\gamma\gamma+X$ 
      & \multicolumn{9}{c||}{Significance $S/\sqrt{B}$}  \\ \cline{3-11}
$M_h$ & bknd &  $\gamma\gamma+X$
      & \multicolumn{4}{c||}{$\gamma\gamma+1$ jet}
      & \multicolumn{4}{c||}{$\gamma\gamma+2$ jets}  \\ \cline{4-11}
(GeV) & (fb) & 
      & $p_T>20$ & $p_T>25$ & $p_T>30$ & $p_T>35$
      & $p_T>20$ & $p_T>25$ & $p_T>30$ & $p_T>35$ \\ \hline\hline
100. &453.4 & 47.7 & 37.6 & 38.9 & 39.5 & 36.9 & 29.0 & 33.0 & 31.6 & 27.1\\ \hline
120. &278.1 & 37.7 & 31.2 & 32.6 & 33.1 & 31.2 & 23.5 & 26.6 & 25.9 & 23.1\\ \hline
140. &171.9 & 31.4 & 26.7 & 28.2 & 29.0 & 27.9 & 19.2 & 22.1 & 22.0 & 19.4\\ \hline
160. &107.3 & 26.6 & 24.3 & 25.3 & 26.0 & 25.2 & 17.4 & 19.7 & 19.5 & 18.3\\ \hline
180. & 67.6 & 22.3 & 20.5 & 21.7 & 22.1 & 21.6 & 14.2 & 16.2 & 16.8 & 15.4\\ \hline
200. & 43.6 & 18.8 & 17.7 & 19.1 & 19.4 & 18.9 & 12.0 & 13.7 & 13.8 & 12.5\\ \hline
250. & 15.7 & 12.8 & 12.1 & 13.2 & 13.4 & 13.1 &  8.2 &  9.3 & 10.0 &  9.6\\ \hline
300. &  8.1 &  9.0 &  8.5 &  9.1 &  9.2 &  9.0 &  5.0 &  5.6 &  5.6 &  5.7\\ \hline
350. &  3.9 &  8.3 &  7.6 &  8.1 &  8.1 &  7.9 &  4.0 &  4.7 &  4.6 &  4.4\\ \hline
400. &  2.1 &  6.3 &  6.1 &  6.3 &  6.3 &  6.1 &  2.9 &  3.4 &  3.7 &  3.6\\ \hline
\end{tabular}
\end{table*}

In Fig.~\ref{topcolor} we show the Run~2 reach for the branching ratio
${\rm B}(h\rightarrow \gamma\gamma)$ as a function of the Higgs mass,
for the case of a ``topcolor'' Higgs boson. This time the channels
with the best signal-to-noise ratio are: o: inclusive $2\gamma+X$, and
$\Box$: $2\gamma+1j$, with $p_T(j)>30$ GeV; both with $w=1.2\Gamma$.

\vspace{0.2in}
{\bf Summary} \\ \nopagebreak

We have studied the Tevatron reach for Higgs bosons decaying into
photon pairs (for an alternative study with similar results, see
\cite{MW}). For purely ``bosonic'' Higgses, which only couple to gauge
bosons, the $2\gamma+2j$ channel offers the best reach, but the
$2\gamma+1j$ channel is almost as good.  For topcolor Higgs bosons,
which can also be produced via gluon fusion, the inclusive $2\gamma+X$
channel is the best, but the $2\gamma+1j$ channel is again very
competitive.  We see that in both cases the $2\gamma+1j$ channel is a
no-lose option!

  \subsection{Enhanced MSSM Neutral Higgs Boson Production at 
              Large $\tan\beta$} 			\small
\begin{center}
{\it M. Roco,
     A. Belyaev,
     J. Valls}
\end{center}
\normalsize \nopagebreak

This section of the report presents a study to evaluate the
sensitivity reach of the Fermilab Tevatron for the neutral Higgs
bosons of the Minimal Supersymmetric Standard Model produced in
association with bottom quarks, $\pp\to\bb\phi$ with $\phi\to\bb$
($\phi = h,H,A$).  Due to the strongly enhanced $\bb\phi$ couplings at
large $\tan\beta$, this channel becomes potentially important since it
can be the dominant production and decay modes of the neutral Higgs
bosons over a large region of the parameter space.

The events of interest would have four $b$ jets in the final state,
two of which come from a Higgs resonance.  In this Workshop two
analyses have been performed, emphasizing the different strengths of
the two detectors and different aspects of the signature.  Each
analysis is described in the next two sections, and a comparison made
at the end.

\vspace{0.2in}
\large
{\bf D\O\ Analysis} \\ \nopagebreak
\normalsize

This analysis is based on Monte Carlo simulations which have been
tuned to the existing D\O\ detectors in a well-studied environment.
It is assumed that it will be possible to maintain the excellent
performance of the D\O\ calorimeter during Run 2.  The $b$-tagging
algorithm described in Section II.A.3 is used in this analysis.

\vspace{0.2in}
{\bf Monte Carlo Simulation} \\ \nopagebreak

All signal and background event samples used in this analysis were
generated using the CompHEP package~\cite{comphep}.  CompHEP performs
calculations of complete tree-level matrix elements for all parton
processes with up to five partons in the final state.  It has been
interfaced with Pythia V6.023~\cite{spythia} to simulate the parton
fragmentation and hadronization.  Events were generated at a
center-of-mass energy, $\sqrt s = 2$ TeV.  The generation of the
signal and background Monte Carlo samples and the relevant tree-level
cross section calculations are described in more detail in ref.~\cite{sasha}.

\vspace{0.2in}
{\bf Detector Simulation} \\ \nopagebreak

The signal detection efficiencies and accepted background cross
sections are estimated using a fast detector simulation package,
MCFAST~\cite{mcfast}.  MCFAST performs parameterized tracking where
for each generated track, a covariance matrix is assembled, which
represents all the material and detector planes traversed by the ideal
track.  A reconstructed track is produced by smearing the generated
track parameters according to this covariance matrix.

Particle momenta are smeared as prescribed in SHW using
20$\% / \sqrt E$ and 80$\% / \sqrt E$ for the electromagnetic and
hadronic energy resolutions, respectively.  The reconstruction of
calorimeter clusters (jets) is based on the cone algorithm with a
fixed cone radius $\Delta R=\sqrt {\Delta \eta^2 + \Delta \phi^2} =
0.5$.

The beam vertex is centered at the origin with a smearing of 30 $\mu$m in
both $x$ and $y$ and 25 $cm$ in $z$.  In this study only one interaction
per beam crossing is considered.

\vspace{0.2in}
{\bf Signal Event Samples} \\ \nopagebreak

This analysis uses only the leading order cross sections for the
signal production process.  Next-to-leading order
corrections~\cite{DSSW} can be significant at the Tevatron, ranging
from $-40\%$ for $m_\phi =$ 40 GeV to $+40\%$ for $m_\phi =$ 1000 GeV.
It was also shown in ~\cite{DSSW} that the largest uncertainty in the
$\bb H$ production cross section comes from varying the factorization
scale, $\mu$.  In this analysis, the scale $\mu = m_\phi$ is chosen.
Signal cross sections are determined using a running $b$-quark mass,
with $m_b$ varying from 3.4 to 3.1 GeV in the Higgs mass range between
$m_\phi =$ 90 to 300 GeV.  Table~\ref{t:tab1} lists the the production
cross sections for $\pp\to\bb\phi$ for the pseudoscalar Higgs boson
$\phi = A$, in femtobarns for $\tan\beta=$1.  The cross sections after
applying parton-level kinematic cuts on $p_T$ and $\Delta R$ are also
given in Table ~\ref{t:tab1}.  The decay branching ratios for
$\phi\to\bb$ are obtained from HDECAY~\cite{hdecay}.  It is assumed
that the SUSY particles are heavy enough such that the decays of Higgs
bosons into them are kinematically forbidden.

\begin{table*}
\caption{Signal Cross sections for $\tan \beta =$ 1.}
\label{t:tab1}
\begin{tabular}{ccrr}
$m_A$ & $m_b$ &  \ \ \ \ \ \ \ \ \ \ \ $\sigma$ (fb)  &  \ \ \ \ \ \ \ \ \ \ \ \ \ \ \ \ \ \ \ $\sigma$ (fb) \\ 
 & & & \\
 (GeV/c$^2$) & (GeV/c$^2$) & \ \ no cuts & $p_T >$ 15 GeV/c \\
 & & & and $\Delta R >$ 0.5 \\ \hline
 & & & \\
 90   &  3.40  &  13.820  & 0.7492 \\ 
100   &  3.38  &   8.628  & 0.5224 \\ 
110   &  3.36  &   5.831  & 0.3730 \\ 
120   &  3.34  &   3.730  & 0.2556 \\ 
130   &  3.32  &   2.563  & 0.1924 \\ 
140   &  3.30  &   1.733  & 0.1366 \\ 
200   &  3.22  &   0.255  & 0.0271 \\ 
250   &  3.18  &   0.064  & 0.0076 \\ 
300   &  3.14  &   0.018  & 0.0024 \\
\end{tabular}
\end{table*}

\newpage
\vspace{0.2in}
{\bf Backgrounds} \\ \nopagebreak

The dominant backgrounds come from QCD multi $b$-jet production.
These include the irreducible backgrounds from $\qq,gg\to\bb\bb$ and
$\pp\to Z\bb$, as well as the reducible backgrounds from $\bb jj$ and
$W\bb$ events with the mistagging of a light quark ($u,d,s$), charm or
gluon jet as a $b$-jet.  Table ~\ref{t:tab2} lists the leading order
cross sections in picobarns for the background processes considered in
this study.  The value of $p_T^{min}$ is set to 15 GeV/c for the
$W\bb$ and $Z\bb$ processes.  This value is raised to 25 GeV/c for
$\bb jj$ to obtain statistically significant samples after applying
the selection criteria described below.  From Table ~\ref{t:tab2}, a
serious background arises from QCD $\bb jj$ production which is
several orders of magnitude larger than the signal cross section.
Reducing this overwhelming background requires a $b$-tagging algorithm
with a very high purity.  Other background processes, such as $\ttbar$
production and $Z\cc$, are less important and have negligible
contributions after the analysis cuts are applied.

\begin{table*}
\caption{Background cross sections.}
\label{t:tab2}
\begin{tabular}{cr}
Process & $\sigma (pb)$ \\ & \\ & \ \ \ \ \ \ \ \ \ \ \ \ \ \ \ \ \ \
\ \ \ \ $p_T > p_T^{min}$ \\ & \ \ \ \ \ \ \ \ \ \ \ \ \ \ \ \ \ \ \ \
\ \ \ \ \ \ \ \ \ \ \ \ and $\Delta R >$ 0.5 \\ \hline & \\ $Z\bb$ & 3.3 * 0.15 BR($Z\to\bb$) \\
$W\bb$       &   3.1 * 0.68 BR($W \to q \bar q$)  \\  
$\bb\bb$     &     2.4 \\
$\bb jj$     &  1610.8 \\
\end{tabular}
\end{table*}

\vspace{0.2in}
{\bf Event Selection} \\ \nopagebreak

The $\bb\bb$ final state is characterized by two clear signatures, the
four-jet topology and a high $b$-quark content.  These properties are
the main handles for suppressing the enormous QCD multi-jet
backgrounds.  Events are passed through a filtering algorithm intended
to select hadronic events compatible with the four-jet topology and to
use the $b$-tagging information to reduce the dominant QCD
backgrounds.  The efficient identification of jets arising from $b$
quark production plays an important role in the search for the Higgs
bosons since for sufficiently large $\tan\beta$ values the neutral
Higgs bosons decay predominantly to $\bb$ with a branching ratio of
about 90$\%$.

The significance of the signal can be further improved by
reconstructing the Higgs mass.  The di-jet mass is reconstructed only
from the $b$-tagged jets resulting in a number of possible pair-wise
combinations, $m_{ij}$.  Using these combinations, it may be possible
to define a smooth combinatoric background distribution with an
overall normalization which can be determined from the data.
Otherwise, it becomes difficult to determine the significance of a
mass peak if, for instance, one takes only the $m_{ij}$ combination
closest to the assumed Higgs mass.

\begin{figure}
  \begin{center}
   \parbox{2.6in}{\epsfxsize=\hsize\epsffile{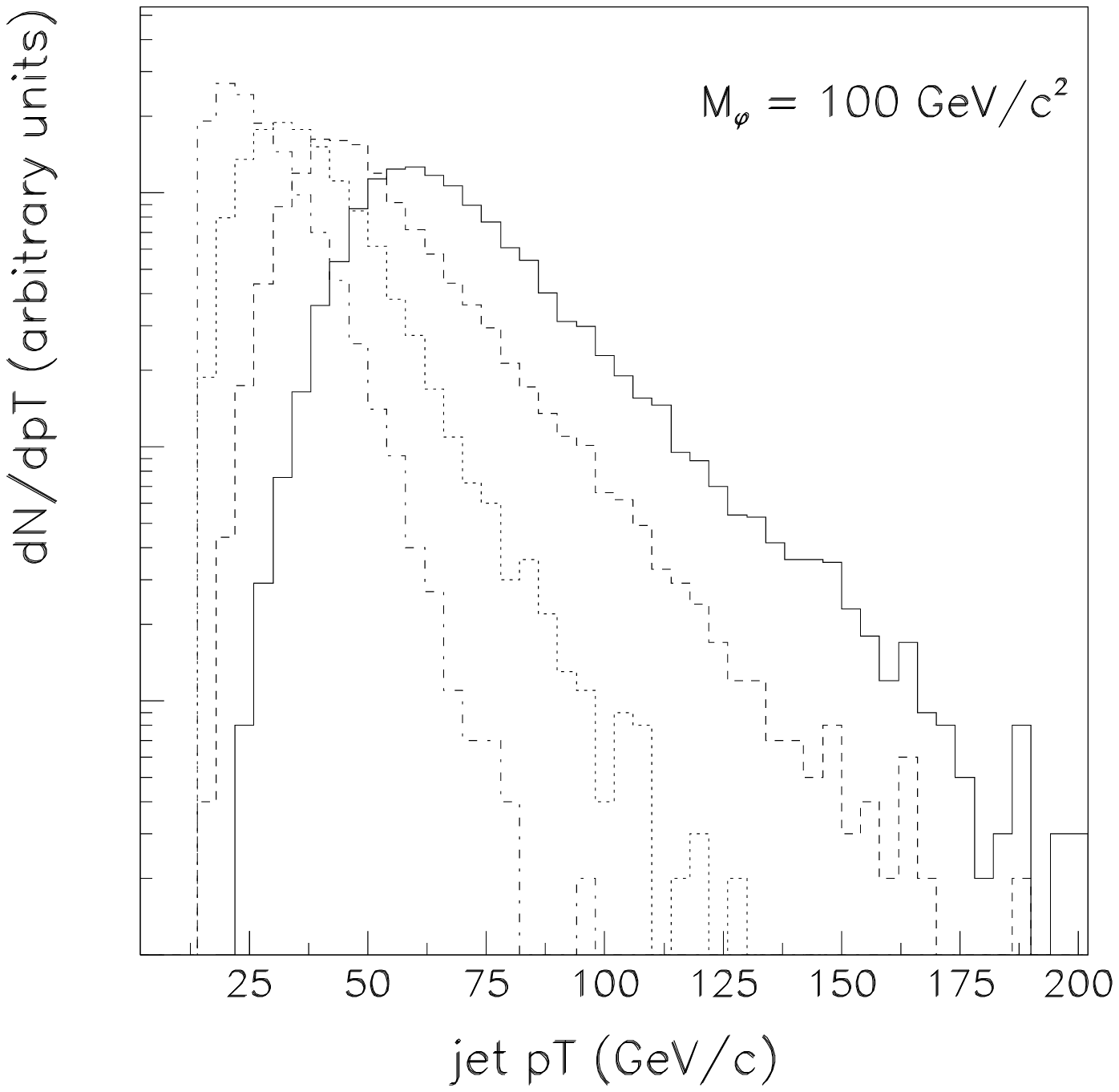}}
   \parbox{2.6in}{\epsfxsize=\hsize\epsffile{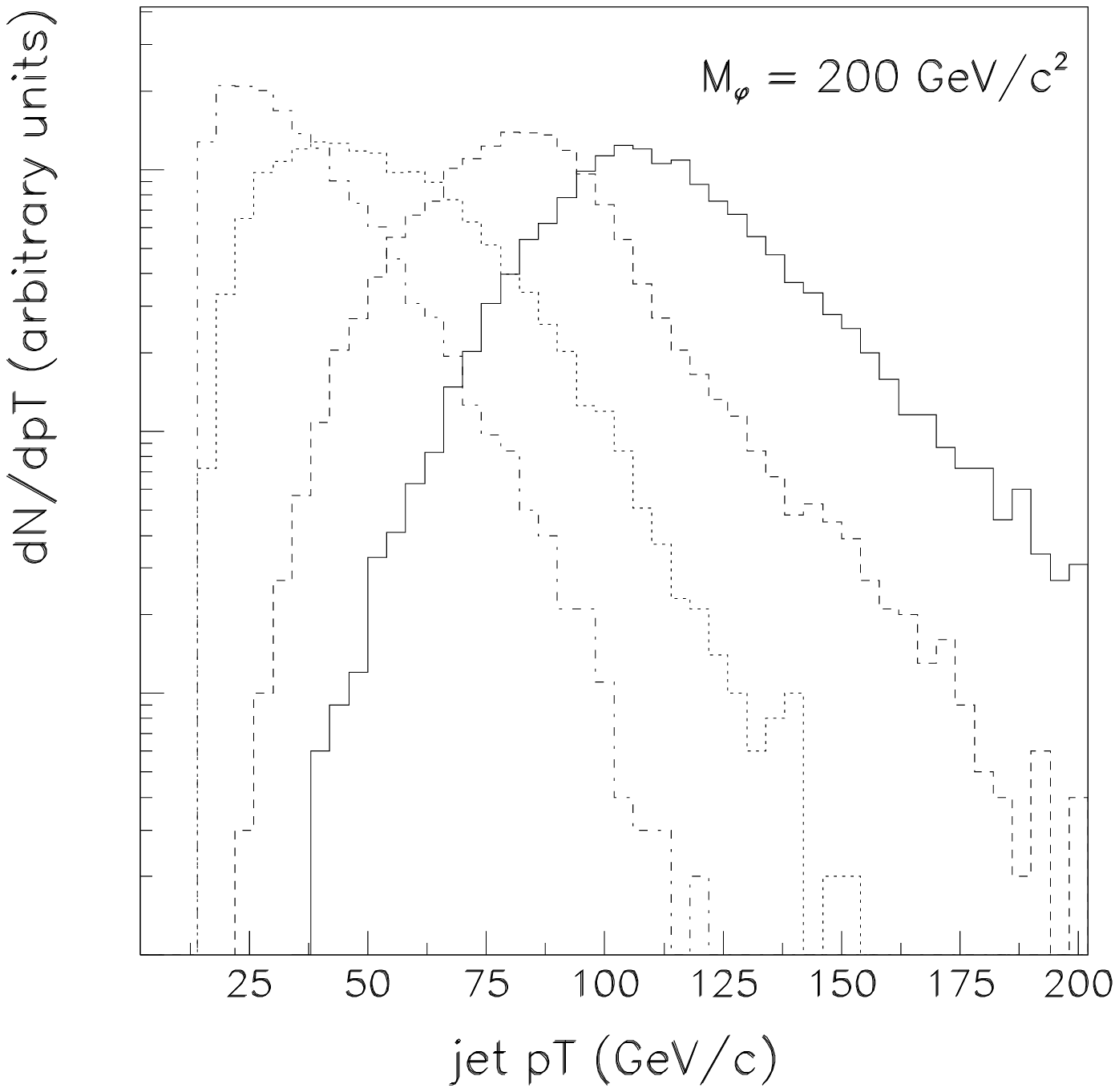}} \\
   \parbox{3.0in}{\epsfxsize=\hsize\epsffile{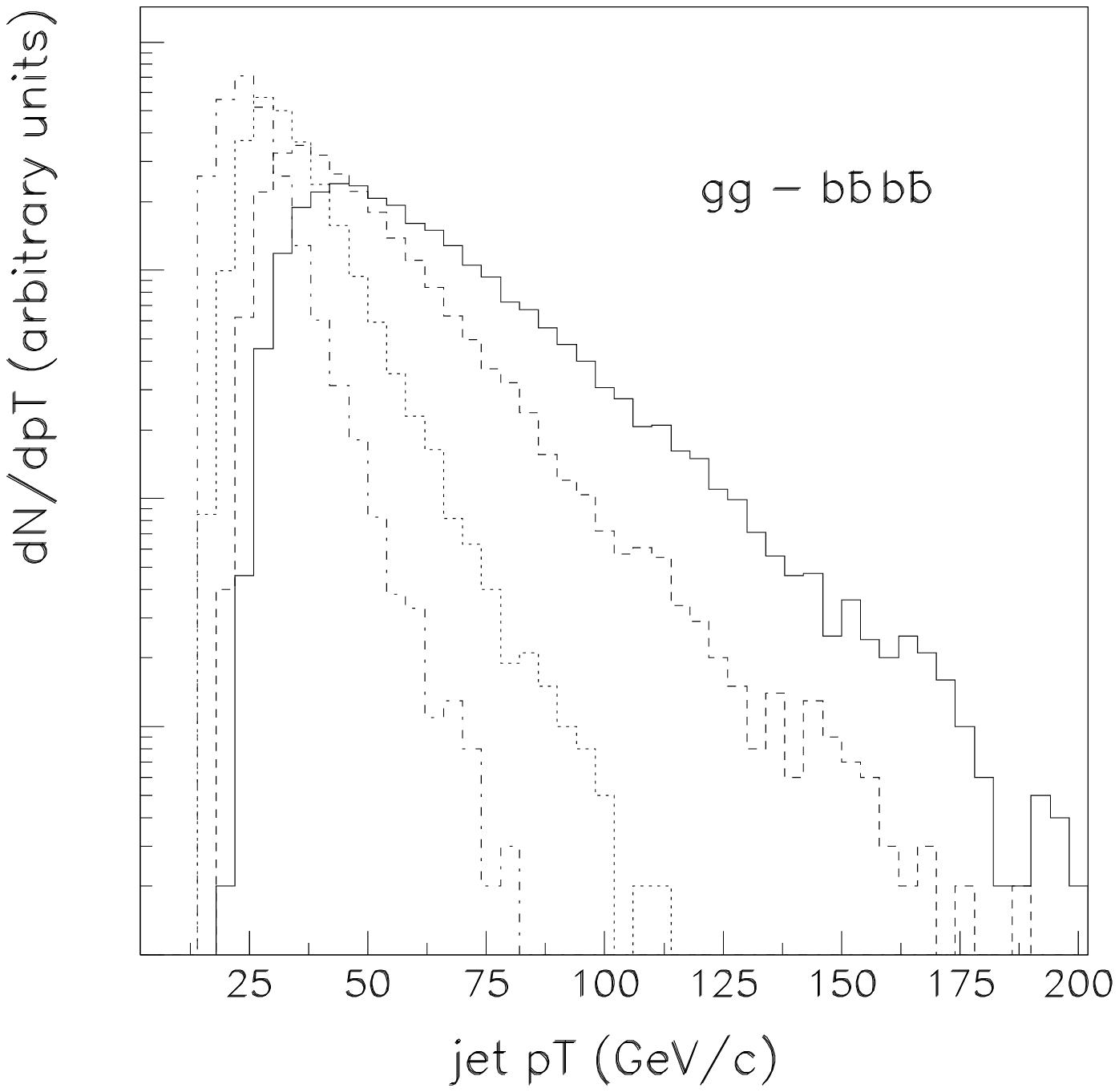}}
   \caption{Distributions of $p_T$ for the four leadings jets in the 
            signal sample from $\bb A$ production for $m_{\phi} = 100$ GeV/c$^2$ 
            (top left), and $m_{\phi} = 200$ GeV/c$^2$ (top right).  
            Also shown are the $p_T$ spectra from the $gg\to\bb\bb$
            background sample.}
  \label{fig-ptspectra}
  \end{center}
\end{figure}

The first level of event selection is performed to filter and select
events which satisfy Run~1 multi-jet trigger requirements of at least
four reconstructed jets in the event with $p_T >$ 15 GeV and
$|\eta|<2$.  The $p_T$ spectra of the leading jets are determined by
the Higgs mass.  This is illustrated in Figure~\ref{fig-ptspectra} where
the $p_T$ distributions of the four leading jets are shown for two
signal samples, with $m_{\phi}$ = 100 GeV/c$^2$ and $m_{\phi}$ = 200
GeV/c$^2$.  For comparison, the bottom plot shows the jet $p_T$
spectra for the irreducible QCD $\bb\bb$ background (this plot was
generated using a sample with a $p_T^{min}$ cut of 15 GeV/c).  The
following selection criteria on the $p_T$ of the leading jets are
chosen to optimize the significance, $S/\sqrt B$, where $S$ and $B$
are the number of signal and background events, respectively,
remaining after cuts: 
\begin{itemize}
\item[$\bullet$] leading jet $p_T > M_{\phi}/2 -5$ 
\item[$\bullet$] second leading jet $p_T >$ max$(30, M_{\phi}/4 +10)$
\item[$\bullet$] third and fourth leading jets $p_T >$ 30 GeV
\end{itemize}
Figure~\ref{fig-ptlcut} shows the optimal values
for the leading jet $p_T$ cut for four different values of the Higgs
mass.

\begin{figure}
  \begin{center}
   \parbox{4in}{\epsfxsize=\hsize\epsffile{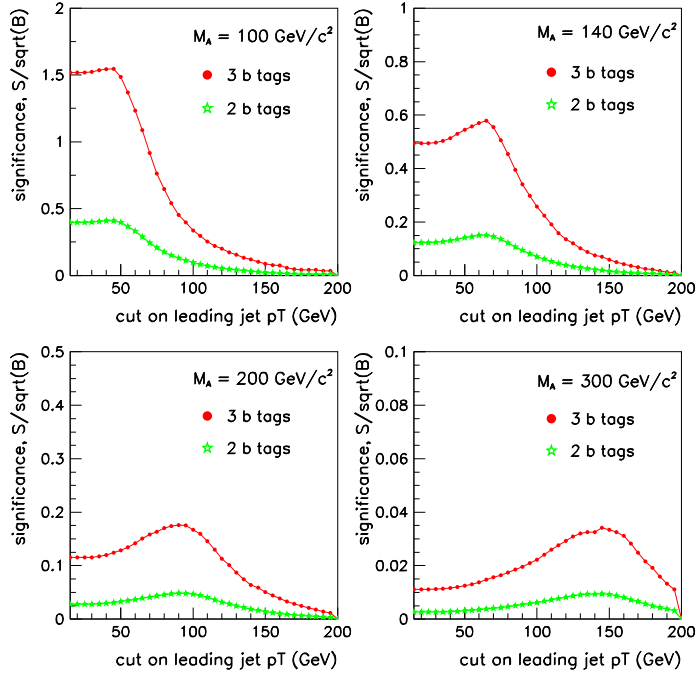}}
   \caption{The selection criteria on the $p_T$ of the leading jets are chosen
            to optimize the statistical significance $S/\sqrt B$.  These plots show
            the significance for different leading jet $p_T$ cuts for four different
            values of the Higgs mass.}
  \label{fig-ptlcut}
  \end{center}
\end{figure}

The $p_T$ requirement on the third and fourth jets is determined by
the value of the $p_T^{min}$ cut used to generate the Monte Carlo
samples.  The selection criteria require that at least three of the
four leading jets are $b$-tagged.  Table~\ref{t:tab3} lists the
signal acceptances for the single channel $\bb A$, where
$\epsilon_{p^j_T}$ is calculated after requiring the multi-jet
triggers.  The $\epsilon_{2b,3b,4b}$ are calculated after the trigger and
jet $p_T$ cuts are applied, and requiring at least two, three or four
$b$-tags, respectively.  The total efficiency $\epsilon_{tot}$ refers
to the remaining events after the nominal selection requiring at least
three $b$-tags.  Table~\ref{t:tab4} lists the number of expected
events passing the selection criteria for the pseudoscalar Higgs
signal and the various background processes, for an integrated
luminosity of ${\cal L}=2~$fb$^{-1}$.  The sensitivities $S/B$ and
$S/\sqrt B$ are given for $\tan\beta$ = 40.

\begin{table*}[ht!]
\caption{Signal acceptances for $p \overline p \to b \overline b A$ in the Higgs mass
range between 90 and 300 GeV/c$^2$.}
\label{t:tab3}
\begin{tabular}{ccccccc}
$m_A$ & $\epsilon_{trig}$ & $\epsilon_{p^j_T}$ & $\epsilon_{2b}$ & $\epsilon_{3b}$ & $\epsilon_{4b}$ & $\epsilon_{tot}$ \\
 (GeV/c$^2$) & ($\%$) & ($\%$) & ($\%$) & ($\%$) & ($\%$) & ($\%$) \\ \tableline
\noalign{\vskip5pt}
 90 & \ 3.46 $\pm$ 0.03 & 27.46 $\pm$ 0.22 & 68.87 $\pm$ 0.24 & 34.36 $\pm$ 0.41 & 5.23 $\pm$ 0.30 & 0.33 $\pm$ 0.01 \\ 
100 & \ 3.97 $\pm$ 0.03 & 30.95 $\pm$ 0.22 & 69.38 $\pm$ 0.26 & 36.79 $\pm$ 0.37 & 5.62 $\pm$ 0.28 & 0.45 $\pm$ 0.01 \\ 
110 & \ 4.39 $\pm$ 0.04 & 33.32 $\pm$ 0.21 & 70.75 $\pm$ 0.28 & 36.90 $\pm$ 0.35 & 6.64 $\pm$ 0.28 & 0.54 $\pm$ 0.01 \\ 
120 & \ 4.81 $\pm$ 0.04 & 35.63 $\pm$ 0.20 & 70.17 $\pm$ 0.30 & 36.27 $\pm$ 0.34 & 6.50 $\pm$ 0.27 & 0.62 $\pm$ 0.01 \\ 
130 & \ 5.39 $\pm$ 0.04 & 36.43 $\pm$ 0.20 & 71.32 $\pm$ 0.30 & 38.14 $\pm$ 0.33 & 6.46 $\pm$ 0.26 & 0.75 $\pm$ 0.02 \\ 
140 & \ 5.78 $\pm$ 0.05 & 37.69 $\pm$ 0.20 & 70.81 $\pm$ 0.31 & 38.37 $\pm$ 0.32 & 6.60 $\pm$ 0.26 & 0.84 $\pm$ 0.02 \\ 
200 & \ 8.20 $\pm$ 0.06 & 38.72 $\pm$ 0.19 & 72.08 $\pm$ 0.33 & 38.26 $\pm$ 0.30 & 6.33 $\pm$ 0.24 & 1.21 $\pm$ 0.02 \\ 
250 & \ 9.47 $\pm$ 0.07 & 38.63 $\pm$ 0.19 & 70.87 $\pm$ 0.33 & 37.12 $\pm$ 0.30 & 7.23 $\pm$ 0.25 & 1.36 $\pm$ 0.03 \\ 
300 &  10.78 $\pm$ 0.08 & 38.15 $\pm$ 0.19 & 71.07 $\pm$ 0.33 & 36.25 $\pm$ 0.30 & 6.80 $\pm$ 0.24 & 1.49 $\pm$ 0.03 \\
\end{tabular}
\vspace*{1pc}
\end{table*}

\begin{table*}[h]
\caption{The expected number of events and sensitivities, with an
integrated luminosity of ${\cal L} = 2$~fb$^{-1}$, for the signal
$p \overline p \to b \overline b A$ and the various background processes.
The sensitivities $S/B$ and $S/\sqrt B$ are given for $\tan \beta =$ 40.}
\label{t:tab4}
\begin{tabular}{ccccccccccc}
Higgs Mass    &    90  &   100  &  110  &  120  &  130  &  140  &  200  &  250 & 300 \\
(GeV/c$^2$)   &        &        &       &       &       &       &       &      & \\ 	\hline
\noalign{\vskip5pt}
$S \ (\tan\beta = 1)$  &  0.090 & 0.078  & 0.063 & 0.046 & 0.038 & 0.029 & 0.006 & 0.002 & 0.000\\
$\times BR$   &  0.83  & 0.81   & 0.77  & 0.68  & 0.53  & 0.34  & 0.003 & 0.001  & 0.001 \\ \hline
\noalign{\vskip5pt}
$S \ (\tan \beta = 40)$&\  144  &\  125  &\  101 &\ \ 74&\ \ 61 &\ \ 46 &\ \ 10 &\ \ \ 3 &\ \ \ 1 \\
$\times BR$   &  0.91  & 0.91   & 0.91  & 0.91  & 0.90  & 0.90  & 0.90  & 0.90 &  0.89 \\ [5pt]
 \noalign{\vskip3pt}
$\bb jj$      &  2384  &  2225  &  2149 &  1999 &  1873 &  1571 &\  598 &\   297 &\  127 \\
$\bb\bb$      & \ 373  &\  349  &\  312 &\  280 &\  243 &\  207 &\ \ 72 &\ \  32 &\ \ 14 \\
$Z\bb$        & \ 111  &\  108  &\  102 &\ \ 95 &\ \ 85 &\ \ 76 &\ \ 36 &\ \  19 &\ \ 10 \\
$W\bb$        &\ \ 27  &\ \ 27  &\ \ 26 &\ \ 24 &\ \ 22 &\ \ 21 &\ \ 13 &\ \ \ 8 &\ \ \ 5\\ [5pt]
Total Background & 2895  & 2709 &  2589 &  2398 &  2223 &  1875 &\  719 &\  356  &\  156 \\ \hline
\noalign{\vskip3pt}  
$S/B$         &  0.050 &  0.046 & 0.039 & 0.031 & 0.028 & 0.025 & 0.014 & 0.008  & 0.005 \\
$S/\sqrt B$   &  2.68  &  2.39  &  1.98 &  1.52 &  1.30 &  1.07 &  0.37 & 0.15   & 0.07 
\end{tabular}
\vspace*{1pc}
\end{table*}

\vspace{0.2in}\break
{\bf Tagging Issues} \\ \nopagebreak

For comparison, the total number of events passing the selection
criteria which require two instead of three $b$-tags are 235 and
1.46$\times 10^5$ events for the signal (assuming $\tan \beta = 40$)
and background, respectively, with $m_A =$ 100 GeV/c$^2$.  Due to the
very small signal-to-background ratio, using only two $b$-tags is not
reliable for this production mode.  Requiring that all four leading
jets to be tagged would have the advantage of eliminating the $W\bb$
and $\bb jj$ backgrounds.  However, the signal event rate becomes
substantially degraded.  This study indicates that the signal to
background ratio $S/B$, at $m_A =$ 100 GeV/c$^2$, increases from 0.046
to 0.113 while the statistical significance of the signal $S/\sqrt B$
decreases from 2.39 to 1.47.  This decrease is due to the fourth jet
being typically considerably softer than the third jet in the signal
events as shown in Figure~\ref{fig-ptspectra}.

Let us examine the case where the $b$-tag requirement is relaxed such
that the three $b$-tags do not necessarily have to be from the four
leading jets.  In this case, for $m_A =$ 100 GeV/c$^2$ and three
$b$-tags, there are 168 signal events with an estimated background of
2898 giving a significance of $S/\sqrt B = 3.12$.  Requiring four
$b$-tags improves the sensitivities with $S/B =$ 0.244 and $S/\sqrt B
=$ 3.52.

To simplify the algorithm for reconstructing the Higgs mass for
candidate events, the selection requires that only $b$-tags from the
four leading jets are considered.  This analysis is performed by
requiring only three $b$-tagged jets on account of the substantial
loss of statistical significance in requiring four $b$-tags.  Tagging
efficiencies are based only on displaced tracks coming from secondary
vertices, not including soft lepton tags.  Soft lepton tagging of $b$
jets will increase the overall $b$-tagging efficiency, however, this
has not been considered in this study.  The analysis performs full
vertex reconstruction and $b$-tagging on an event-by-event basis,
i.e., tagging efficiencies are not based on any rate parameterization.

\vspace{0.2in}
{\bf Higgs Mass Reconstruction} \\ \nopagebreak

One does not know {\em a priori} which $b$-jets resulted from the
Higgs decay.  For each combination of tagged jets $\{i,j\}$, the
di-jet invariant mass $m_{ij}$ is reconstructed.  There are either
four or six possible di-jet combinations for each event.  To reduce
the combinatoric background from incorrect di-jet combinations,
further cuts are applied.  The jets $\{i,j\}$ are required to be well
separated azimuthally: $\phi_1 < \Delta \phi_{ij} < \phi_2$, where the
values of $(\phi_1,\phi_2)$ range from (70$^o$,290$^o$) for $m_{\phi}
=$ 90 GeV to (120$^o$,240$^o$) for $m_{\phi} =$ 300 GeV.
Alternatively, one can also use the variable $\Delta R =
\sqrt{(\phi_i-\phi_j)^2 + (\eta_i-\eta_j)^2}$ to impose jet isolation.

Finally, a cut on the invariant mass $m_{ij}$ of the di-jet candidate
is necessary to further suppress the remaining QCD backgrounds.  The
di-jet mass resolution of the $\bb$ system coming from the Higgs decay
is $\sigma(m_{b \overline b}) = 15\%$.  For the signal event sample,
all di-jet $m_{ij}$ combinations which are reconstructed within $\pm
2.5 \sigma$ of the Higgs mass constitute the {\em signal S}.
Similarly, for the background samples, all di-jet $m_{ij}$
combinations which are reconstructed within $\pm 2.5 \sigma$ of the
assumed Higgs mass constitute the {\em background B}.

\vspace{0.2in}
{\bf Combining $\bb A$, $\bb H$, $\bb h$} \\ \nopagebreak

At sufficiently large $\tan \beta$ values, one of the CP-even Higgs
bosons and the pseudoscalar $A$ have similar couplings and are
degenerate in mass as illustrated in the following plots.  The program
{\tt hmsusytev.f}~\cite{hehprog} is used to derive the couplings and
masses.  In Figure ~\ref{fig-couplings} the ratios of the CP-even
Higgs couplings $g^2$ to $\tan^2 \beta$ as a function of pseudoscalar
mass are shown for the case of maximal and minimal mixing.
Figure~\ref{fig-massdiff} shows the mass differences ($m_A - m_h$) and
($m_H - m_A$) versus the pseudoscalar mass for different $\tan \beta$
values in the maximal (top plots) or minimal (bottom plots) mixing
scenarios.

\begin{figure}
  \begin{center}
   \parbox{4in}{\epsfxsize=\hsize\epsffile{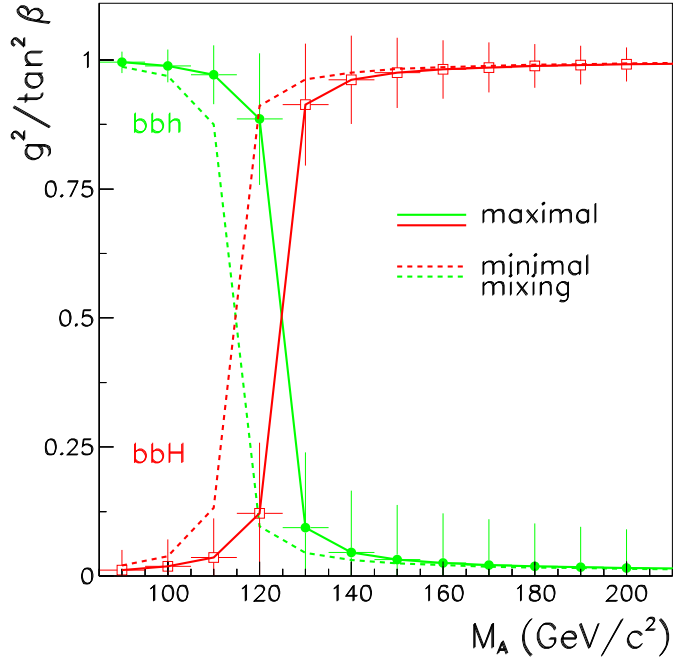}}
   \caption{The ratio of the CP-even Higgs bosons coupling $g^2$ to
            $\tan^2 \beta$ as a function of the pseudoscalar mass $m_A$.}
   \label{fig-couplings}
  \end{center}
\end{figure}

\begin{figure}
  \begin{center}
   \parbox{5in}{\epsfxsize=\hsize\epsffile{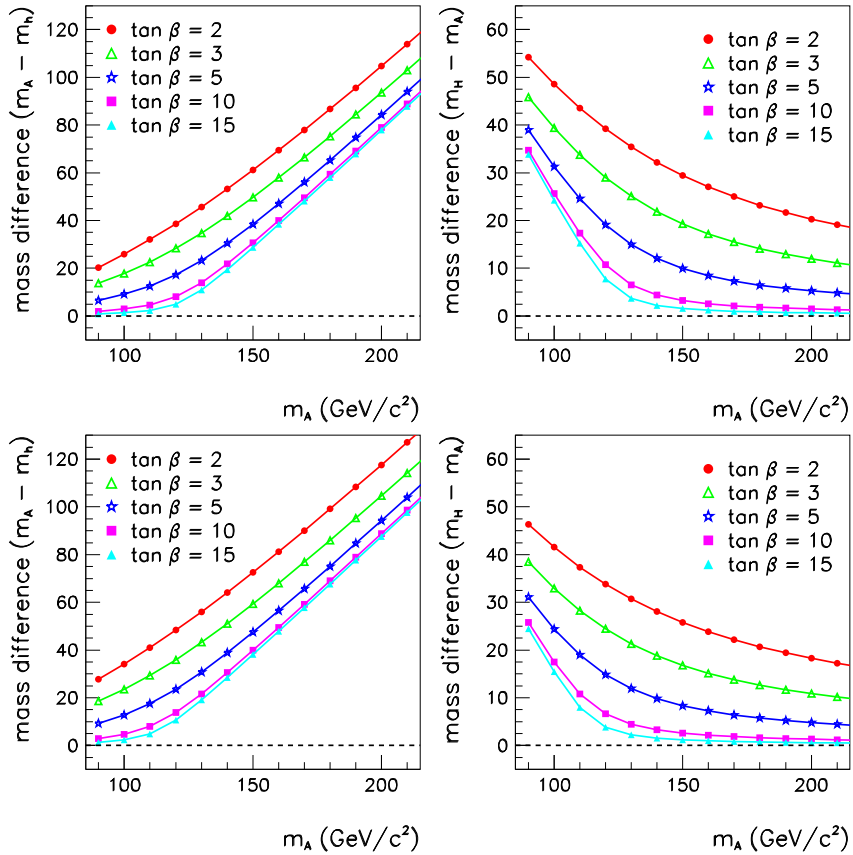}}
   \caption{Contours of $(m_A - m_h$) and $(m_H - m_A$) as a function of the
            pseudoscalar mass $m_A$ for various $\tan \beta$ values in the maximal (top
            plots) or minimal mixing (bottom plots) scenario.}
   \label{fig-massdiff}
  \end{center}
\end{figure}

The $g^2$ couplings and the degree of mass degeneracy determine the
additional contribution from the CP-even Higgs bosons which should be
added to the pseudoscalar Higgs signal.  In the case of maximal
mixing, the light Higgs coupling for $m_A\lsim 115$~GeV is roughly
equal to $\tan^2 \beta$.  This is also the case for the heavy Higgs
coupling for $m_A\gsim 125$~GeV.  However, it is not correct to
simply double the signal from $\bb A$ production since the additional
contribution from $h, H$ may not be within the 2.5$\sigma$ mass
window.  Figure~\ref{fig-massdiff} shows that the mass differences between
the scalar Higgs bosons and pseudoscalar are large for small $\tan
\beta$ values.  Doubling the $\bb A$ production rate would lead to an
overestimate of the signal at low $\tan \beta$.  This analysis assumes
that the shapes of the di-jet mass distributions for the CP-even and
the pseudoscalar Higgs bosons are identical, except for a shift in the
mass peaks due to the mass differences ($m_A - m_h$) and ($m_H -
m_A$).  It is also assumed that the relative normalization of the mass
distributions is given by the ratio $g^2/\tan^2 \beta$.  The
additional contribution from the Higgs scalars is determined by the
fraction of the mass distributions within the 2.5$\sigma$ mass window
around the assumed pseudoscalar mass $m_A$.

\vspace{0.2in}
{\bf Results} \\ \nopagebreak

The signal and background estimates described above are used to
estimate the discovery potential and exclusion reach for the Higgs
bosons of the MSSM.  For a given luminosity and an assumed Higgs mass,
one can determine the minimum value of $\tan \beta$ leading to a
5$\sigma$ discovery ($\sigma = S/\sqrt B$) or exclusion at 95$\%$ CL
(1.96$\sigma$).  The results are shown in Figure~\ref{fig-limits}.  The
sensitivity reach in the $\tan \beta$ and $m_A$ parameter space is
shown for different values of the delivered luminosity per experiment.
These results were obtained assuming maximal mixing, where SUSY
parameters are chosen to give the largest predicted value for the
Higgs mass.  For comparison the LEP excluded region ($m_A < 90$
GeV/c$^2$) is also shown.

In the intermediate Higgs mass range, these results indicate that the
Run 2 sensitivity reach, assuming $\tan \beta = 40$, is about 160
GeV/c$^2$ at 95$\%$ CL exclusion with a delivered luminosity of
2~fb$^{-1}$, extending up to 190 GeV/c$^2$ with 5~fb$^{-1}$.  The
5$\sigma$ discovery reach is up to 115 GeV/c$^2$ at 2~fb$^{-1}$ and 
150~GeV/c$^2$ at 5 fb$^{-1}$.  For larger Higgs masses, the sensitivity
reach degrades significantly due to the decreasing production rates.

\begin{figure}
  \begin{center}
   \parbox{3in}{\epsfxsize=\hsize\epsffile{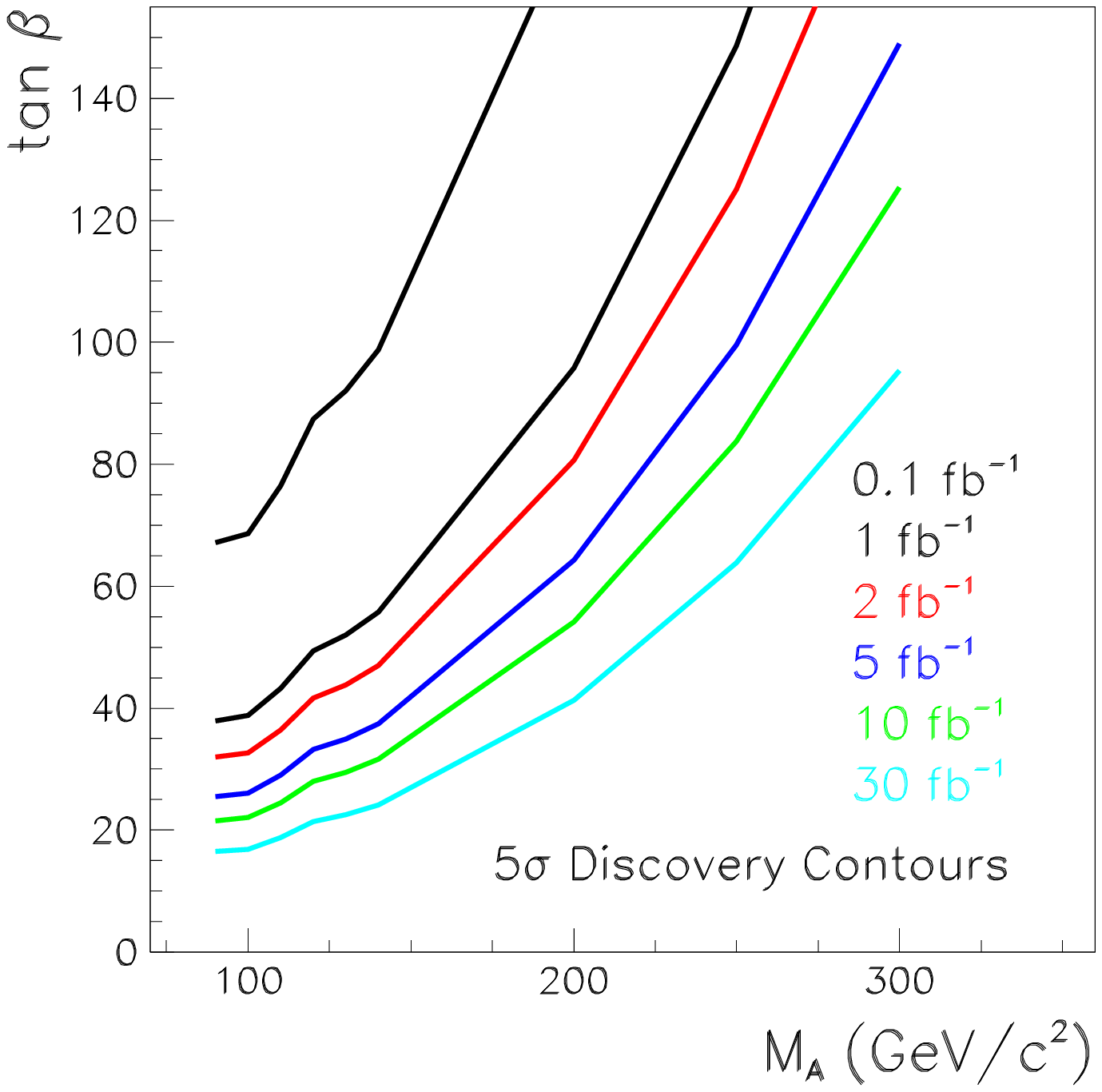}}
   \parbox{3in}{\epsfxsize=\hsize\epsffile{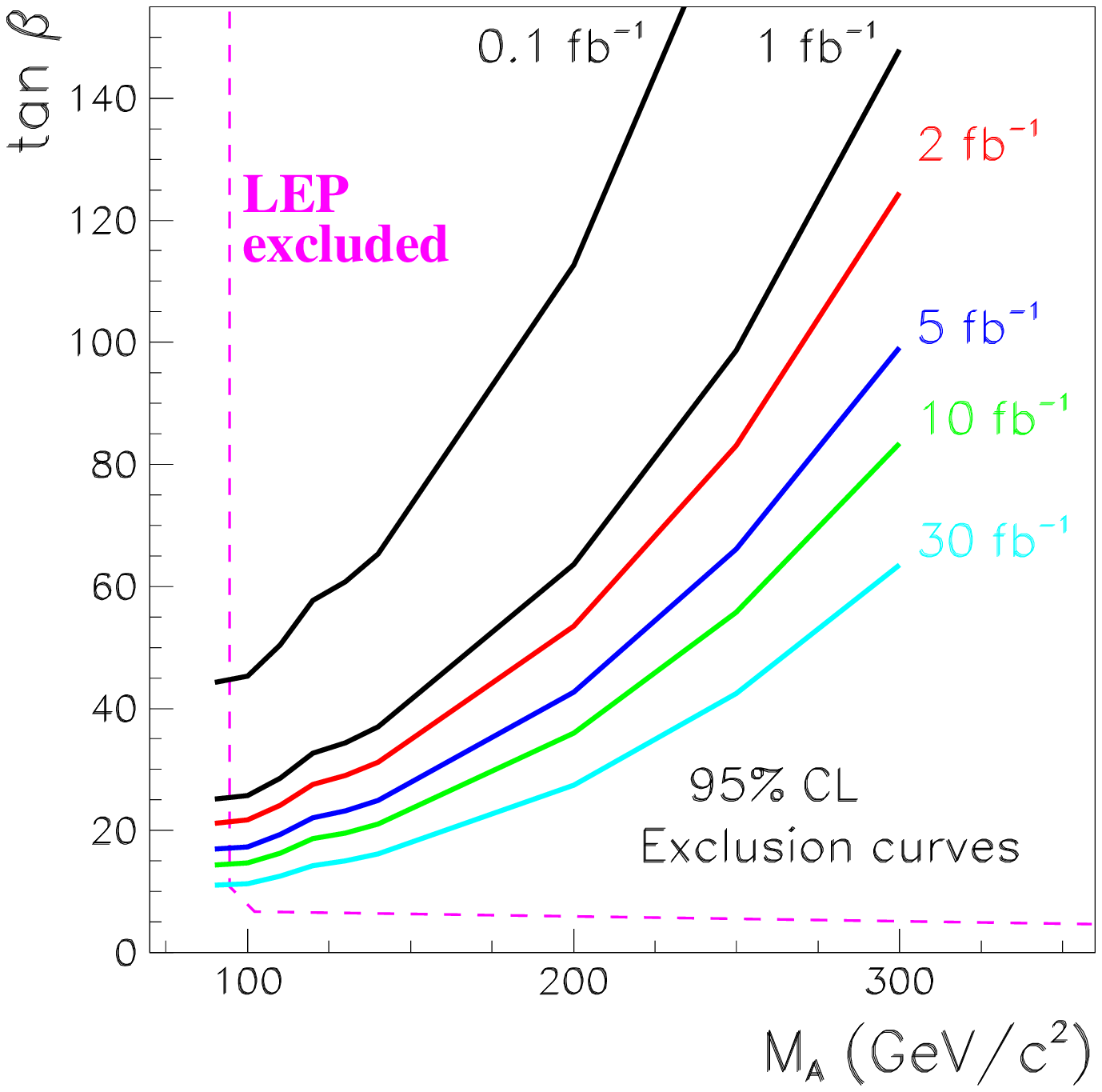}}
   \caption{95$\%$ CL exclusion curves and 5$\sigma$ discovery contours
            for $\pp\to\bb\phi$ with $\phi = h,H,A$.  The curves show the sensitivity 
            reach for the MSSM neutral Higgs bosons in the $\tan \beta$ and 
            $m_A$ parameter space.  The LEP excluded region ($m_A < 90$ GeV/c$^2$) 
            is also shown for comparison.  The results are shown for the case of 
            maximal mixing where SUSY parameters are chosen to give the largest 
            predicted value for the Higgs mass.}
   \label{fig-limits}
  \end{center}
\end{figure}

The estimates for signal efficiencies and backgrounds are affected by
various systematic effects, such as the imperfect modeling of the
physics processes and the detector response.  Uncertainties arising
from the inaccurate description and reconstruction of the energy flow,
as well as the initial and final state radiation have not been
studied.  The main source of uncertainty in this analysis arises from
the evaluation of the QCD multi $b$-jet backgrounds.  Additional
factors which strongly influence the background estimates presented
here include the predictions for QCD multi-jet production, the parton
density distributions, the choice of scale for $\alpha_s$ and
higher-order corrections.  Another critical issue which have not been
evaluated is the effect of high luminosity conditions where the impact
of overlapping interactions may become important.

\vspace{0.2in}
{\bf Summary} \\ \nopagebreak

This report has presented a strategy to search for the neutral Higgs
bosons of the Minimal Supersymmetric Standard Model, where the Higgs
bosons are produced in association with bottom quarks.  Prospects for
5$\sigma$ discovery and 95$\%$ CL exclusion in the process
$\pp\to\bb\phi$, with $\phi = h,H,A$, have been presented.  Results
indicate that the Tevatron Run 2 sensitivity reach, assuming $\tan
\beta = 40$, is about 160 GeV/c$^2$ at 95$\%$ CL exclusion with a
delivered luminosity of 2~fb$^{-1}$, extending up to 190 GeV/c$^2$ with
5~fb$^{-1}$.  The 5$\sigma$ discovery reach is up to 115 GeV/c$^2$ at
2~fb$^{-1}$ and 150 GeV/c$^2$ at 5~fb$^{-1}$.  These results demonstrate
that the Tevatron can significantly constrain a large fraction of the
MSSM parameter space by studying the production of the neutral Higgs
bosons produced in conjunction with $b$ quark pairs.

\vspace{0.2in}
\large
{\bf CDF analysis}\\ \nopagebreak
\normalsize

The CDF-based analysis represents an extrapolation of the technique
used in the analysis of CDF Run~1 data to the new detector geometry
and improved $b$-tagging efficiency anticipated in Run~2.  The
estimates of background are made from Run~1 data, with appropriate
scaling for improved acceptance and efficiency\footnote{No correction
for the increased center-of-mass energy is made to either the Higgs
signal or the background.  This should be conservative in that, all
else being equal, the sensitivity should increase with center-of-mass 
energy.}

Figure~\ref{bbxxs1} show the expected rates for $\bb\phi\to \bb\bb$
production at large $\tan\beta$ as a function of the Higgs mass. In
this regime ($\tan\beta \gtrsim 20$) $\phi$ is either $h$ or $A$ for
masses below the light Higgs upper bound, or either the $H$ or $A$ for
masses above the light Higgs upper limit.  All rates are shown for the
case of vanishing mixing parameters or non-mixing case defined by
$\mu=A_t=A_b=0$ with $\mu$ the Higgs mass parameter and $A_t$ and
$A_b$ the soft SUSY Yukawa breaking parameters.  In all cases, a top
quark mass of $M_t=175$ GeV/c$^2$ and a SUSY mass scale of $M_S=1$ TeV
are assumed. To calculate the Yukawa coupling we use a running bottom
quark mass evaluated at the Higgs mass scale ($M_b\simeq 3$
GeV/c$^2$). All cross sections are evaluated at leading order (LO)
with the CTEQ3 parton distributions functions and the renormalization
scale equal to the Higgs mass.

\begin{figure}[tb]
  \centerline{\parbox{4.0in}{\epsfxsize=\hsize\epsffile{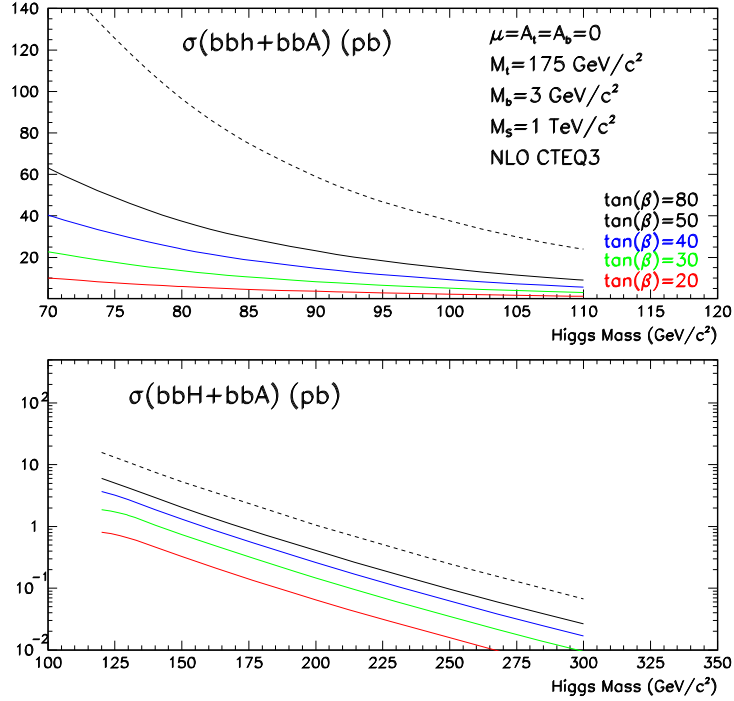}} }
  \caption{Production cross sections for 
           $(b\bar{b}h + b\bar{b}A)\rightarrow b\bar{b} b\bar{b}$ (upper plot) and
           $(b\bar{b}H + b\bar{b}A)\rightarrow b\bar{b} b\bar{b}$ (lower plot) 
           as a function of the Higgs mass.
           The two plots correspond to the different degenerate states,
           $M_h=M_A$ for Higgs masses below the light Higgs upper
           bound, and $M_H=M_A$ for Higgs masses above the light Higgs
           upper bound. Results are shown for $\tan\beta \ge 20$ and the non-mixing
           scenario.}
  \label{bbxxs1}
\end{figure}

\vspace{0.2in}
{\bf Signal Monte Carlo Simulation} \\ \nopagebreak

For signal modelling we used a modified version of the parton level
Monte Carlo program PAPAGENO~\cite{papageno} together with the Lund
PYTHIA V5.6 string fragmentation and hadronization
program~\cite{PYTHIA}.  Fixed weight $\pp\to\bb\phi$ events with
$\phi\to\bb$ are generated at leading order (LO) and fragmented inside
PYTHIA.  The mass of the Higgs has been set to values between
$M_{\phi}$=70 GeV/c$^2$ to 300 GeV/c$^2$. In order to setup parton
showers and fragmentation using PYTHIA, the color-flow information
for colored particles is treated properly.  Both initial and final
state showering are allowed at energy scales corresponding to the
Higgs mass. After fragmentation, events are then passed through the
CDF Run~1 simulation and reconstruction code.

\vspace{0.2in}
{\bf Event Selection} \\ \nopagebreak

The CDF Run~1 multijet trigger requirements are used as the first
step in the data selection. This sample is defined by events that satisfy:
\begin{itemize}
\item[a)] total trigger cluster $\sum E_T^{L2} > 125$ GeV, and
\item[b)] 4 trigger clusters with $E_T^{L2}> 15$ GeV
\end{itemize}
where a level 2 (L2) trigger cluster is defined by a nearest-neighbor
reconstruction algorithm, seeded by a tower of $E_T > 3$ GeV and
including only towers of $E_T > 1$ GeV.  The event selection follows
by requiring events with at least four offline reconstructed jets with
$\et >15$ GeV and $|\eta|\le 2.1$. Jets are defined as localized
energy depositions in the calorimeters and are reconstructed using an
iterative clustering algorithm with a fixed cone of radius $\Delta R =
\sqrt{\Delta\eta^2 + \Delta\phi^2} = 0.4$ in $\eta-\phi$ space. Jet
e\-ner\-gies are then corrected for energy losses in uninstrumented
detector regions, energy falling out\-side the clustering cone,
contributions from underlying event and multiple interactions, and
calorimeter nonlinearities.

\begin{table*}[t!]
\caption{Optimized $\et$ thresholds for the three highest $\et$ jets in
the event as a function of the signal mass.}
\label{eth}
\begin{tabular}{ccccccccccccc} 
 & \multicolumn{12}{c}{$M$ (GeV/c$^2$)} \\ [0.1cm] \hline
 \multicolumn{13}{c}{} \\ [-0.2cm]
 & 70 & 80 & 90 & 100 & 110 & 120 & 130 & 140 &
 150 & 200 & 250 & 300 \\ [0.1cm] \hline
 \multicolumn{13}{c}{} \\ [-0.2cm]
$\et(jet 1)$ (GeV) & $40$ & $40$ & $42$ & $42$ & $46$ & $48$ & $50$ & $54$ & $58$ 
& $75$ & $95$ & $120$ \\ [0.05cm]
$\et(jet 2)$ (GeV) & $32$ & $32$ & $32$ & $34$ & $34$ & $34$ & $38$ & $38$ & $42$ 
& $55$ & $72$ & $85$ \\ [0.05cm]
$\et(jet 3)$ (GeV) & $14$ & $14$ & $14$ & $14$ & $14$ & $14$ & $14$ & $14$ & $14$ 
& $40$ & $55$ & $70$ \\ [0.1cm]
\end{tabular}
\end{table*} 

The typical topology of the signal events consist of two primary $b$
quarks and a Higgs $\phi$, which is radiated from one of the primary
$b$ quarks. There is, thus, a very high energetic primary $b$ quark
with momentum of the order of the mass of the Higgs boson, $M_{\phi}$,
balanced by the Higgs and the other primary $b$ quark, which is
generally much softer.  The Higgs decays into a $\bb$ pair with
typical transverse momentum of the order of $M_{\phi}/2$. The four
highest-$E_T$ (uncorrected) jets in the event are then ordered
in $E_T$:
\begin{equation}
E_T^{1} \ge E_T^{2} \ge E_T^{3} \ge E_T^{4}
\end{equation}
and a mass dependent requirement made for $E_T^{1}$,
$E_T^{2}$, and $E_T^{3}$. The $E_T$ thresholds for these cuts are 
shown in Table~\ref{eth} as a function of the Higgs mass. They have been 
optimized by maximizing the significance of the signal.

Figure~\ref{etj} shows the the $E_T^{j}$, $j=1,2,3,4$ distributions
for two different signal Higgs masses ($M = 120, 250$ GeV/c$^2$)
compared with the QCD background shapes.

\begin{figure}
  \centerline{\parbox{5.8in}{\epsfxsize=\hsize\epsffile[0 0 560 220]{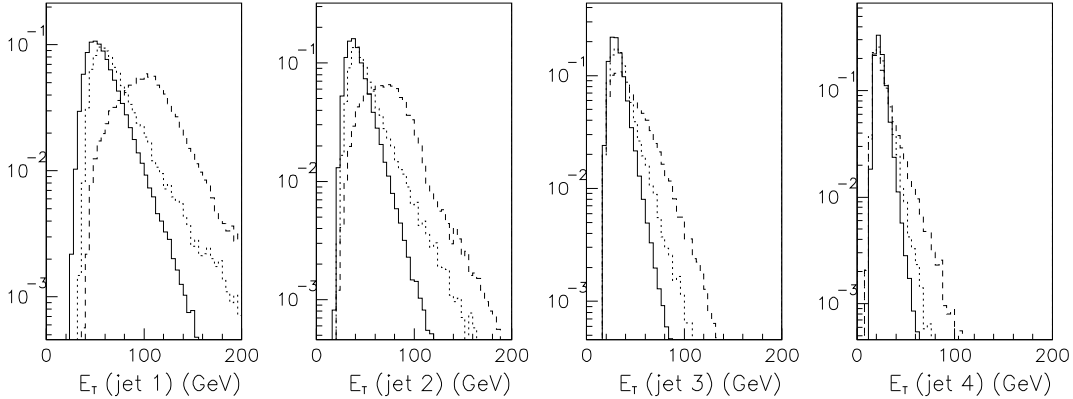}}}
  \caption{Transverse energy distributions for the four leading jets
           in the event. The solid histograms correspond to QCD background. 
           The dashed and dotted histograms correspond to signal masses of 
           $M=250$ and $120$ GeV/c$^2$ respectively.}
  \label{etj}
\end{figure}

We then require that at least three among the four highest-$E_T$ jets
in the event are identified (tagged) as $b$ quark candidates.  We use
the CDF secondary vertex algorithm~\cite{secvtx} with the intrinsic
Run~1 efficiencies and mistag rates per jet inside the SVX fiducial
region. The results are then applied to $b$ jets at generator level
within the increased geometrical acceptance of the Run 2 CDF silicon
vertex detector (SVX-II). The SVX-II extends the $b$-tagging
capabilities up to the range $|\eta|<2$.  The $b$-tagging algorithm
begins by searching for secondary vertices that contain three or more
displaced tracks. If none are found, the algorithm searches for
two-track vertices using more stringent track criteria. A jet is
tagged if the secondary vertex transverse displacement from the
primary one exceeds three times its uncertainty.

Table~\ref{seff} shows the trigger efficiency ($\epsilon_{trig}$),
$E_T^j$ efficiencies ($\epsilon_{E_T^j}$), double $b$-tag
($\epsilon_{\ge 2b}$) and triple $b$-tag ($\epsilon_{\ge 3b}$)
efficiencies, and total efficiencies ($\epsilon_{tot}$) as a function
of the Higgs mass. The double and triple $b$-tag efficiencies are
calculated after the trigger and $E_T^j$ cuts are applied.  The total
efficiency applies to the triple $b$-tag selection.  All errors shown
are statistical only.

Figure~\ref{invm} shows the reconstructed dijet invariant mass 
distribution for different signal masses.
The invariant mass of the dijet system is defined as $M_{ij} = 
\sqrt{2 E_T^i E_T^{j} [\cosh(\Delta\eta)_{ij} - 
\cos(\Delta\phi)_{ij}]}$.
The distributions contain a Gaussian core with a resolution of
$10$--$20$\% depending slightly on the Higgs mass.  To maximize the
signal dijet mass resolutions we use the reconstructed invariant mass
of the highest-$\et$ jets in the event for masses $M\ge 120$
GeV/c$^2$. For masses $M<120$ GeV/c$^2$ we use the invariant mass of
the second and third highest-$\et$ jets in the event. In all cases,
the two highest-$\et$ jets in the event are always required to be
tagged.  The tails of the distributions are dominated by the cases
where the jet assignment in the mass reconstruction is incorrect. In
most of these cases, one of the jets assigned to the Higgs is a
primary $b$ quark jet.

\begin{figure}
  \centerline{\parbox{5.2in}{\epsfxsize=\hsize\epsffile{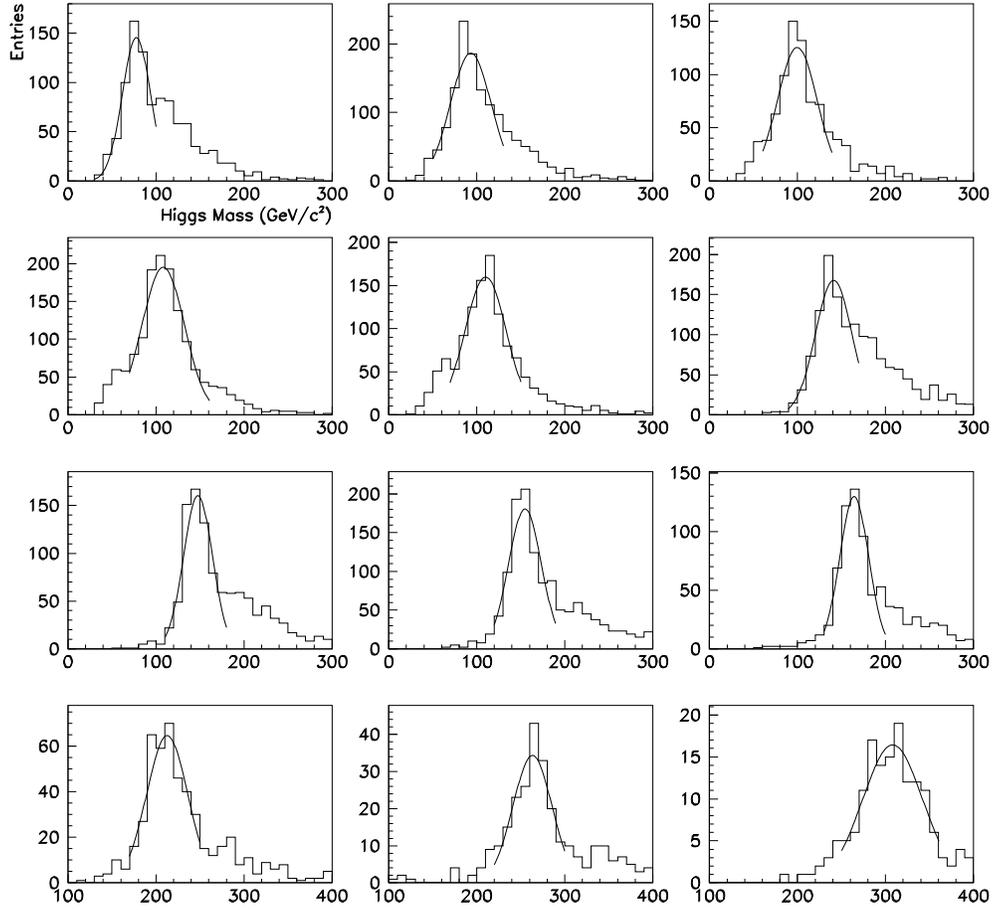}}}
  \caption{Signal dijet invariant mass distributions. From top left the 
           distributions correspond to Higgs masses of 
           $M=70, 80, 90, 100, 110, 120, 130, 140, 150, 200, 250$ 
           and $300$ GeV/c$^2$, respectively.}
  \label{invm}
\end{figure}

\begin{table*}[t!]
\caption{Signal acceptances as a function of the Higgs mass.}
\label{seff}
\begin{tabular}{cccccc}
$M$ & $\epsilon_{trig}$  & $\epsilon_{E_T^j}$ & $\epsilon_{\ge 2b}$ &
$\epsilon_{\ge 3b}$ & $\epsilon_{tot}$   \\ [0.1cm]
(GeV/c$^2$) &  (\%) & (\%) & (\%)& (\%) & (\%) \\ [0.15cm] \hline
& & & & &  \\ [-0.2cm]
70 & $1.25\pm0.02$ &  $92.7\pm0.3$ & $78.9\pm0.7$ &
$28.6\pm0.7$ & $0.33\pm0.01$
 \\ [0.05cm]
80 & $1.48\pm0.02$ &  $93.0\pm0.2$ &  $78.2\pm0.6$ &
$27.7\pm0.6$ & $0.38\pm0.01$ 
 \\ [0.05cm]
90 & $1.70\pm0.02$ &  $89.7\pm0.3$ &  $79.0\pm0.8$ &
$27.6\pm0.8$ & $0.42\pm0.01$ 
 \\ [0.05cm]
100 & $1.90\pm0.03$ & $85.8\pm0.3$ &  $78.8\pm0.6$ &
$27.2\pm0.6$ & $0.44\pm0.01$ 
 \\ [0.05cm]
110 & $2.23\pm0.03$ & $80.7\pm0.3$ &  $76.6\pm0.6$ &
$25.5\pm0.6$ & $0.46\pm0.01$ 
 \\ [0.05cm]
120 & $2.42\pm0.04$ & $79.6\pm0.3$ &  $77.1\pm0.8$ &
$27.0\pm0.7$ & $0.52\pm0.01$ 
 \\ [0.05cm]
130 & $2.65\pm0.04$ & $72.8\pm0.3$ &  $77.5\pm0.7$ &
$26.3\pm0.6$ & $0.51\pm0.02$ 
 \\ [0.05cm]
140 & $2.75\pm0.04$ & $69.5\pm0.3$ &  $77.1\pm0.8$ &
$28.1\pm0.8$ & $0.54\pm0.01$ 
 \\ [0.05cm]
150 & $3.15\pm0.05$ & $61.6\pm0.3$ &  $79.4\pm1.0$ &
$27.3\pm1.1$ & $0.53\pm0.02$ 
 \\ [0.05cm]
200 & $4.35\pm0.07$ & $31.3\pm0.3$ &  $82.6\pm1.3$ &
$31.0\pm1.3$ & $0.42\pm0.02$
 \\ [0.05cm]
250 & $4.95\pm0.07$ & $18.5\pm0.3$ &  $78.7\pm1.5$ &
$25.9\pm1.5$ & $0.24\pm0.01$ 
 \\ [0.05cm]
300 & $6.71\pm0.10$ & $11.5\pm0.2$ &  $79.0\pm1.5$ &
$23.3\pm1.5$ & $0.18\pm0.01$ 
 \\ [0.1cm]
\end{tabular}
\end{table*} 

\vspace{0.2in}
{\bf Backgrounds} \\ \nopagebreak

Backgrounds for the process 
$p\bar{p}\to bb\phi\to \bb\bb$ include all
sources of standard model heavy flavor multijet events. These include
QCD heavy flavor production, fake multitags, $\ttbar$ ($t\to Wb$, 
$W\to q\bar{q}'$), $Z$ + jets ($Z\to \bb/\cc$), 
$W\bb/\cc$ and $Z\bb/\cc$.

\vspace{0.2in}
\underline{QCD} \\ \nopagebreak

The dominant source of background is QCD heavy flavor production.
This is also the most difficult to estimate due to uncertainties in
the predicted total rates as well as heavy flavor jet rates.  Rather
than attempting to estimate the QCD background rates directly from
Monte Carlo, instead we follow the approach used in~\cite{juano} in
the search for standard model Higgs bosons via $VH\to jj\bb$
($V=W,Z$). In~\cite{juano}, the CDF Run~1 multijet data sample was
used to calculate first the double $b$ tagged dijet invariant mass
distribution.  The shape of this distribution is then fit, using a
binned maximum-likelihood method, to a combination of signal, QCD
heavy flavors, fake double tags and other standard model
backgrounds. The QCD heavy flavor and signal normalizations are left
free in the fit while the rest of backgrounds are constrained by
Gaussian functions to their expected values and uncertainties.  Fake
double tags are estimated from data while the other physics
backgrounds are calculated from Monte Carlo.  The double $b$-tag data
selection as well as details of the fit can be found in~\cite{juano}.

The fit yields zero signal contribution for the $\bb\phi\to\bb\bb$
process for all signal masses. The results of the fit are shown in
Table~\ref{fit}. The QCD heavy flavor background expectations are now
corrected for the $E_T^j$ ($j=1,2,3$) and triple $b$-tag acceptances
corresponding to the present analysis.  QCD heavy flavor acceptances
have been calculated with the PYTHIA Monte Carlo program with its QCD
leading order $2\to 2$ hard scattering processes and parton shower
modelling.  A sufficiently large number of events were generated with
a hard scattering $p_T > 40$ GeV/c in order to obtain unbiased and
statistically significant samples. Below this $p_T$ bin no events
satisfy the trigger requirements.  Fake triple $b$ tag rates are
calculated directly from Run~1 data using fake tagging rate
parametrizations.  Table~\ref{extr} shows the expected QCD heavy
flavor and QCD fake rates as obtained from data for the selection
described in this analysis and after extrapolated to a Run 2 total
integrated luminosity of 1~fb$^{-1}$.

As a comparison, Table~\ref{tabqcd} shows the expected initial rates
for the individual and total QCD processes leading to heavy flavors in
the final state as obtained from PYTHIA.  Also shown are the QCD
acceptances and expected events for 1~fb$^{-1}$ integrated luminosity
for the present analysis.  The heavy flavor content in QCD events are
conventionally classified in three groups: direct production
($q\bar{q},gg \to Q\bar{Q}$), flavor excitation ($Qg\to Qg$), and
gluon splitting ($g\to Q\bar{Q}$) in initial or final state shower
evolution. The relative contributions to the total cross sections are
very uncertain and depend on the center of mass energy, the modelling
of initial and final state radiation, shower evolution and
fragmentation.

\begin{table*}[t!]
\caption{Observed and expected double $b$-tag events in 91 pb$^{-1}$ of 
CDF Run~1 data as obtained from the fit.}
\label{fit}
\begin{tabular}{cccccc}
 & data & QCD heavy flavors & Fakes & $\ttbar$ & $Z$ + jets 
\\ [0.1cm] \hline
 & & & & &  \\ [-0.2cm]
events & $589$ & $470 \pm 27$ & $85 \pm 11$ & $23 \pm 6$ & $16 \pm 4$ 
\\ [0.12cm]
\end{tabular}
\end{table*} 

\begin{table*}[t!]
\caption{Expected QCD heavy flavor, QCD fake triple tags and total QCD
events in 1~fb$^{-1}$ total integrated luminosity.}
\label{extr}
\begin{tabular}{c|ccc}
$M$  & QCD heavy flavors & QCD fakes & Total QCD \\ [0.06cm]
(GeV/c$^2$) & \multicolumn{3}{c}{(events per fb$^{-1}$)} \\ [0.1cm]\hline 
     &     &    &    \\ [-0.25cm]
70  & $58.6\pm 12.3$ & $12.1 \pm 10.5$ & $ 70.7 \pm 16.2$ \\ [0.1cm]
80  & $58.6\pm 12.3$ & $12.1 \pm 10.5$ & $ 70.7 \pm 16.2$ \\ [0.1cm]
90  & $56.0\pm 11.9$ & $12.1 \pm 10.5$ & $ 68.1 \pm 15.9$ \\ [0.1cm]
100 & $50.6\pm 11.0$ & $12.1 \pm 10.5$ & $ 62.7 \pm 15.2$ \\ [0.1cm]
110 & $48.0\pm 11.0$ & $ 9.7 \pm  7.8$ & $ 57.7 \pm 13.5$ \\ [0.1cm]
120 & $48.0\pm 11.0$ & $ 7.3 \pm  5.5$ & $ 55.3 \pm 12.3$ \\ [0.1cm]
130 & $45.3\pm 11.1$ & $ 7.3 \pm  5.5$ & $ 52.6 \pm 12.4$ \\ [0.1cm]
140 & $29.3\pm  8.0$ & $ 4.8 \pm  3.4$ & $ 34.1 \pm  8.7$ \\ [0.1cm]
150 & $24.0\pm  7.0$ & $ 2.4 \pm  1.8$ & $ 26.4 \pm  7.2$ \\ [0.1cm]
200 & $ 8.0\pm  3.3$ & $ 1.3 \pm  0.8$ & $  9.3 \pm  3.4$ \\ [0.1cm]
250 & $ 0.0\pm  0.0$ & $ 1.3 \pm  0.8$ & $  1.3 \pm  0.8$ \\ [0.1cm]
300 & $ 0.0\pm  0.0$ & $ 1.3 \pm  0.8$ & $  1.3 \pm  0.8$ \\ [0.13cm]
\end{tabular}
\end{table*} 

\break
\vspace{0.2in}
\underline{$\ttbar\to WbW\bar{b}\to \bb$ + jets} \\ \nopagebreak

Background $\ttbar$ events have been simulated using the
HERWIG~\cite{herwig} v5.6 Monte Carlo generator with $M_t=175$
GeV/c$^2$.  The trigger efficiency has been estimated to be $92.0 \pm
0.8\%$ from simulation.  The measured CDF cross section of $7.6^{+
1.8}_{-1.5}$~\cite{xs} is used to estimate the expected number of
events.  Total $\ttbar$ initial rates and acceptances are summarized
in Table~\ref{tabothers}.  The errors shown include both statistical
and systematic uncertainties.  Systematics are dominated by the effect
of the absolute energy scale, modelling of initial state radiation in
HERWIG, and $b$-tag efficiencies.

\begin{table*}[t!]
\caption{Individual and total cross sections and acceptances for the
different QCD hard scattering dijet subprocesses modelled with
PYTHIA. Events are generated with a hard scattering $p_T > 40$ GeV/c.
The $\et$ thresholds used to calculate $\epsilon_{E_T}$ correspond to the 
selection for $M=100$ GeV/c$^2$. Last column shows the expected
rates for 1~fb$^{-1}$.}
\label{tabqcd}
\setlength{\tabcolsep}{3.6pt}
\begin{tabular}{lccccccc}
 & $\sigma$ (pb) & \multicolumn{5}{c}{} & N\\ [0.1cm]
 & $(p_T\!>\!40$ GeV/c) & $\epsilon_{trig}$ (\%) & $\epsilon_{\et}$ (\%)
& $\epsilon_{\ge 2b}$ (\%) & $\epsilon_{\ge 3b}$ (\%)
& $\epsilon_{tot}$ (\%) & (events/fb$^{-1}$)\\ [0.1cm] \hline
 &  \multicolumn{7}{c}{} \\ [-0.2cm]
$q\bar{q},gg\to q\bar{q}$ & $1.2\times 10^4$ & $1.23 \pm 0.01$ 
& $63.9 \pm 0.3$ & $0.15 \pm 0.04$ & $0.0 \pm 0.0$ & $0.0 \pm 0.0$ 
& $0.0 \pm 0.0$ \\
$qq,gg\to qq,gg$ & $4.9\times 10^5$ & $0.98 \pm 0.01$ 
& $60.6 \pm 0.3$ & $0.45 \pm 0.07$ & $0.04 \pm 0.02$ & $0.0002 \pm 0.0001$ 
& $972.6 \pm 561.5$ \\
$gq\to gq$ & $4.4\times 10^5$ & $1.01 \pm 0.01$ 
& $62.6 \pm 0.3$ & $0.45 \pm 0.07$ & $0.02 \pm 0.02$ & $0.0001 \pm 0.0001$ 
& $593.3 \pm 419.5$ \\
$gb(c)\to gb(c)$ & $3.1\times 10^4$ & $0.76 \pm 0.01$ 
& $60.3 \pm 0.3$ & $3.89 \pm 0.20$ & $0.14 \pm 0.04$ & $0.0006 \pm 0.0002$ 
& $198.3 \pm 57.3$ \\
$q\bar{q},gg\to \bb/\cc$ & $8.2\times 10^3$ & $0.94 \pm 
0.01$ & $62.7 \pm 0.2$ & $7.91 \pm 0.21$ & $0.19 \pm 0.04$ 
& $0.0011 \pm 0.0004$ 
& $91.9 \pm 17.7$ \\ [0.15cm] 
Total & $9.8\times 10^5$ & $0.99 \pm 0.01$ 
& $61.6 \pm 0.6$ & $0.59 \pm 0.05$ & $0.03 \pm 0.01$ 
& $0.0002 \pm 0.0001$ 
& $1856.2 \pm 703.5$ \\ [0.15cm] 
\end{tabular}
\end{table*}

\begin{table*}[t!]
\caption{Cross sections and acceptances for the $\ttbar$ and different 
$W/Z$ + jets backgrounds. The $\et$ thresholds correspond to the 
selection $M=100$ GeV/c$^2$.}
\label{tabothers}
\begin{tabular}{ccccccc}
 & $\sigma\times BR$ (pb) & $\epsilon_{trig}$ (\%) & $\epsilon_{\et}$ (\%)
& $\epsilon_{\ge 2b}$ (\%) & $\epsilon_{\ge 3b}$ (\%)
& $\epsilon_{tot}$ (\%) \\ [0.1cm] \hline
 &  \multicolumn{6}{c}{} \\ [-0.2cm]
$\ttbar$ & $3.49 \pm 0.83$ & $92.0 \pm 0.8$ 
& $91.0 \pm 0.2$ & $12.3 \pm 0.2$ & $0.17 \pm 0.03$ & $0.14 \pm 0.02$ 
\\ [-0.1cm]
($t\to Wb\to jjb$) & & & & & & \\ [0.1cm]
$Z$+jets & $11.8 \pm 1.4$ & $30.7 \pm 1.2$ & $69.8 \pm 0.5$ 
& $7.5 \pm 0.6$ & $0.27 \pm 0.10$ & $0.06 \pm 0.02$ \\ [-0.01cm]
($Z\to \bb/\cc$) & & & & & & \\ [0.1cm]
$W\bb/\cc$ & $237.2 \pm 13.6$ & $28.5 \pm 1.1$ & $69.8 \pm 0.5$ 
& $0.06 \pm 0.03$ & $0.004 \pm 0.003$ & $0.001 \pm 0.001$ \\ [-0.01cm]
($W\to q\bar{q}',c\bar{s}$) & & & & & & \\ [0.1cm]
$Z\bb/\cc$ & $74.6 \pm 6.4$ & $30.5 \pm 1.2$ & $69.8 \pm 0.5$ 
& $0.07 \pm 0.03$ & $0.004 \pm 0.003$ & $0.001 \pm 0.001$ \\ [-0.01cm]
($Z\to q\bar{q}$) & & & & & & \\ [0.12cm]
\end{tabular}
\end{table*} 

\vspace{0.2in}
\underline{$W/Z$ + jets} \\ \nopagebreak

The $W/Z\, + \ge n$ jets background has been studied with a
combination of data and Monte Carlo. The CDF measured cross sections
for $W$ + jets, $W\to e\nu$~\cite{vj1} and $Z$ + jets, $Z\to
ee$~\cite{vj2} are used to normalize Monte Carlo production cross
sections for $W/Z\, + \ge n$ jet events.  Generic $W$ + jets and $Z$ +
jets samples generated with the HERWIG Monte Carlo program~\cite{vj3}
and processed through the CDF simulation package were used to estimate
the expected contributions after our selection cuts.  The predominant
source of heavy flavor production in $W/Z\, + \ge n$ jets events is
$Z\, + \ge n$ jets ($Z\to \bb/\cc$) as well as some contribution of
initial and final state gluon splitting and higher order diagrams
leading to $W\bb/\cc$ and $Z\bb/\cc$ events.

Table~\ref{tabothers} shows the initial cross sections and estimated
acceptances for all the $W/Z$ + jets backgrounds leading to heavy
flavor final states. The errors shown include both statistical and
systematic uncertainties. The systematic uncertainties are dominated
by the normalization of the measured cross sections to the predicted
Monte Carlo cross sections.

Figure~\ref{figeff} shows the individual and total efficiencies as a
function of the Higgs mass for the signal and the different standard
model background contributions.

\begin{figure}[hb!]
  \centerline{\parbox{4.8in}{\epsfxsize=\hsize\epsffile{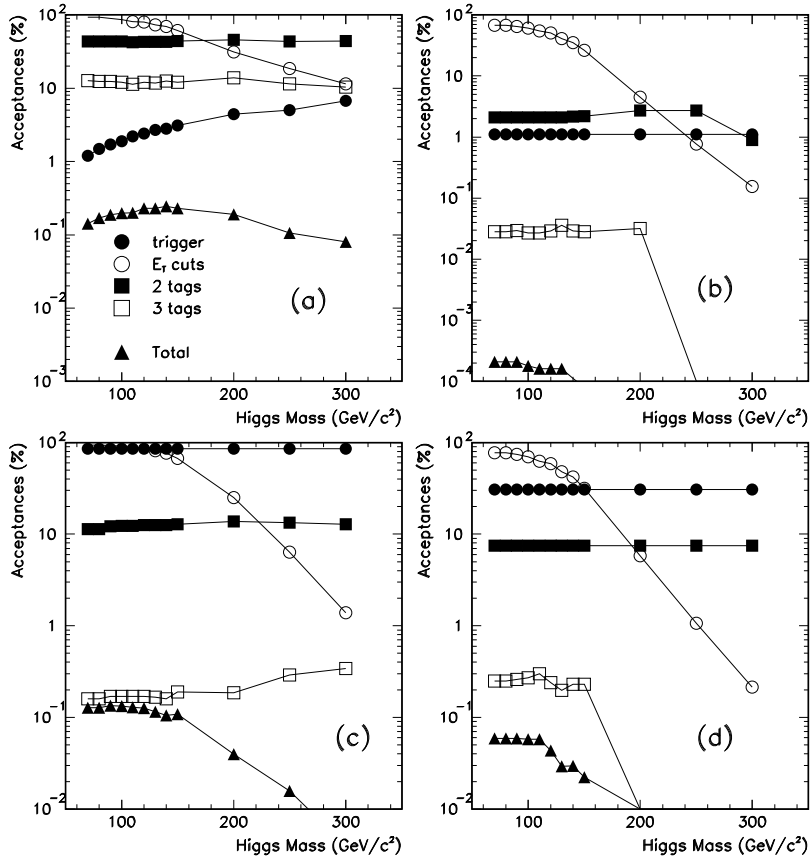}} }
  \caption{Individual and total efficiencies (in \%) as a function
           of the signal mass for (a) signal, (b) QCD, (c) $t\bar{t}$, and 
           (d) $Z$ + jets ($Z\rightarrow b\bar{b}/c\bar{c}$).}
  \label{figeff}
\end{figure}

\begin{table*}[t!]
\caption{Expected number of background events as a function of 
the Higgs mass for 1 fb$^{-1}$.}
\label{backs}
\begin{tabular}{c|cccccc}
$M$  & $N_{QCD}$ & $N_{Zjj}$ & $N_{\ttbar}$
 & $N_{W\bb/\cc}$ & $N_{Z\bb/\cc}$ &
$N_{tot}=B$\\ [0.06cm]
(GeV/c$^2$) & \multicolumn{6}{c}{(events per fb$^{-1}$)} \\ [0.1cm]\hline 
     &     &    &    &  & &\\ [-0.25cm]
70 & $70.7 \pm 16.2$ & $4.4 \pm 1.6$ & $13.2 \pm 3.5$ & $1.5 \pm 0.6$ & 
$0.5 \pm 0.2$ & $90.4 \pm 16.7$ \\ [0.1cm]
80 & $70.7 \pm 16.2$ & $4.4 \pm 1.6$ & $13.5 \pm 3.4$ & $1.5 \pm 0.6$ & 
$0.5 \pm 0.2$ & $90.7 \pm 16.6$  \\ [0.1cm]
90 & $68.1 \pm 15.9$ & $4.6 \pm 1.6$  & $13.4 \pm 3.4$ & $1.5 \pm 0.6$ & 
$0.5 \pm 0.2$ & $88.1 \pm 16.3$  \\ [0.1cm]
100 & $62.7 \pm 15.2$ & $4.5 \pm 1.5$ & $13.2 \pm 3.2$ & $1.4 \pm 0.5$ & 
$0.5 \pm 0.2$ & $82.3 \pm 15.6$  \\ [0.1cm]
110 & $57.7 \pm 13.5$ & $4.1 \pm 1.5$ & $12.8 \pm 3.4$ & $1.3 \pm 0.5$ & 
$0.4 \pm 0.2$ & $76.3 \pm 14.0$  \\ [0.1cm]
120 & $55.3 \pm 12.3$ & $3.4 \pm 1.3$ & $12.6 \pm 3.4$ & $1.2 \pm 0.5$ & 
$0.4 \pm 0.2$ & $72.9 \pm 12.8$  \\ [0.1cm]
130 & $52.6 \pm 12.4$ & $2.1 \pm 1.1$ & $12.2 \pm 3.2$ & $1.0 \pm 0.4$ & 
$0.3 \pm 0.1$ & $68.2 \pm 12.9$  \\ [0.1cm]
140 & $34.1 \pm  8.7$ & $1.7 \pm 0.6$ & $11.3 \pm 3.1$ & $0.8 \pm 0.3$ & 
$0.3 \pm 0.1$ & $48.2 \pm  9.3$  \\ [0.1cm]
150 & $26.4 \pm  7.2$ & $1.3 \pm 0.5$ & $10.8 \pm 2.9$ & $0.6 \pm 0.2$ & 
$0.2 \pm 0.1$ & $39.3 \pm  7.8$  \\ [0.1cm]
200 & $ 9.3 \pm 3.4$ & $0.0 \pm 0.0$ & $4.4 \pm 1.2$ & $0.0 \pm 0.0$ & 
$0.0 \pm 0.0$ & $13.7 \pm  3.6$   \\ [0.1cm]
250 & $ 1.3 \pm 0.8$ & $0.0 \pm 0.0$ & $0.9 \pm 0.3$ & $0.0 \pm 0.0$ & 
$0.0 \pm 0.0$ & $ 2.2 \pm  0.9$   \\ [0.1cm]
300 & $1.3 \pm 0.8$ & $0.0 \pm 0.0$ & $0.3 \pm 0.1$ & $0.0 \pm 0.0$ & 
$0.0 \pm 0.0$ & $ 1.6 \pm  0.8$  \\ [0.13cm]
\end{tabular}
\end{table*}

\begin{table*}[t!]
\caption{Expected number of SM signal events ($\tan\beta = 1$),
MSSM signal events for $\tan\beta = 40$ and significances as a function of 
the Higgs mass for 1 fb$^{-1}$.}
\label{final}
\begin{tabular}{c|cc|cc}
$M$  & $S$ & $S/\sqrt{B}$ & $S$ & $S/\sqrt{B}$ \\ [0.06cm]
(GeV/c$^2$) & \multicolumn{2}{c}{($\tan\beta =1$)} &
\multicolumn{2}{|c}{($\tan\beta = 40$)} \\ [0.1cm]\hline 
     &          &   &          &    \\ [-0.25cm]
70   & $0.039$     & $0.0041$  & $133.2$ &  $14.0$  \\ [0.1cm]
80   & $0.027$     & $0.0028$  & $ 91.9$ &  $ 9.7$  \\ [0.1cm]
90   & $0.018$     & $0.0019$  & $62.4$  &  $6.6$   \\ [0.1cm]
100  & $0.012$     & $0.0013$  & $41.0$  &  $4.5$   \\ [0.1cm]
110  & $0.007$     & $0.0008$  & $26.3$  &  $3.0$   \\ [0.1cm]
120  & $0.005$     & $0.0006$  & $19.1$  &  $2.2$   \\ [0.1cm]
130  & $0.002$     & $0.0002$  & $13.7$  &  $1.7$   \\ [0.1cm]
140  & $0.001$     & $0.0001$  & $10.0$  &  $1.4$   \\ [0.1cm]
150  & $0.0004$    & $0.00006$ & $7.0$   &  $1.1$   \\ [0.1cm]
200  & $0.0   $    & $0.0$     & $1.1$   &  $0.3$   \\ [0.1cm]
250  & $0.0   $    & $0.0$     & $0.15$  &  $0.1$   \\ [0.1cm]
300  & $0.0   $    & $0.0$     & $0.03$  &  $0.02$  \\ [0.13cm]
\end{tabular}
\end{table*} 

\break
\vspace{0.2in}
{\bf Results} \\ \nopagebreak

Table~\ref{backs} shows the expected number of events from each
individual background contribution as well as the total backgrounds
for 1~fb$^{-1}$.  Table~\ref{final} shows the expected number of
signal events for $\tan\beta = 1$ (Standard Model case) and for
$\tan\beta = 40$ as well as the signal significances as a function of
the Higgs mass for 1~fb$^{-1}$.

The reconstructed dijet invariant mass for the highest-$\et$ jets in
the event (with both jets required to be $b$-tagged) and with present
CDF Run~1 dijet resolutions is shown in Figure~\ref{sandb} for both
signal and background.  All rates are normalized to 1~fb$^{-1}$.

\newpage

Figure~\ref{figtb} shows the 95\% CL exclusion limits, $3\sigma$
discovery thresholds and $5\sigma$ discovery thresholds in the 
$\tan\beta$--$M_{\phi}$ plane of the MSSM for four different total integrated
luminosity scenarios, ${\cal L}=91$ pb$^{-1}$, 2 fb$^{-1}$, 
10~fb$^{-1}$ and 30 fb$^{-1}$.
Figure~\ref{figlum} shows the required luminosity for 95\% CL
exclusion limits, $3\sigma$ and $5\sigma$ discovery thresholds as a
function of the Higgs mass for the MSSM scenario with $\tan\beta$ = 10 and 40.

In a theory with a non-minimal Higgs sector (beyond the SM), the $\phi\bb$ 
coupling differs from the corresponding $\hsm\bb$ coupling of the Standard
Model Higgs boson.  As a result, the production cross-section for
$gg$, $q\bar q\to\bb\phi\to\bb\bb$ is given by
\begin{equation}
\sigma_{\bb\bb} = \left( \frac{y_b}{y_b^{\rm SM}} \right)^2
\frac{{\rm BR}(\phi\to \bb)}{{\rm BR}(\hsm\to \bb)} \;
\sigma_{\bb\bb}^{\rm SM}\,,
\end{equation}
where $y_b$ [$y_b^{\rm SM}$] is the $\phi\bb$ [$\hsm\bb$] coupling.  
It is convenient
to define a ratio ${\cal R}$ of the Higgs production amplitude in the
general model divided by the corresponding SM amplitude.  In terms of the
cross-sections, $\sigma_{\bb\bb}\equiv {\cal R}^2 \sigma_{\bb\bb}^{\rm SM}$, 
where
\begin{equation}
{\cal R} = \left( \frac{y_b}{y_b^{\rm SM}} \right)\; 
\left[\frac{{\rm BR}(\phi\to \bb)}{{\rm BR}(\hsm\to \bb)}
\right]^{1/2}\,.
\label{calrdef}
\end{equation}
Figure~\ref{kf} shows the enhancement factor ${\cal R}$
as a function of the Higgs mass, the integrated total luminosity, and
for 95\% CL, $3\sigma$ and $5\sigma$ discovery thresholds.
The ${\cal R}$ factor contains the entire model dependence of the 
cross-section and allows one to present the results in a model-independent way.

\begin{figure}
  \centerline{\parbox{4.0in}{\epsfxsize=\hsize\epsffile{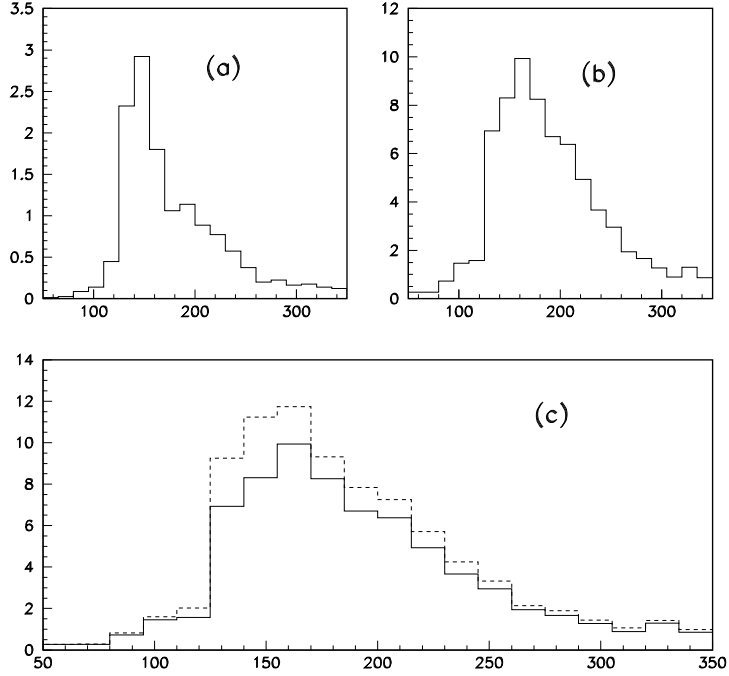}} }
  \caption{Reconstructed dijet invariant mass for the highest-$\et$
           jets in the event for (a) signal ($M=130$ GeV/c$^2$ and $\tan\beta$ = 40), 
           (b) background (QCD, $Z$ + jets with $Z\rightarrow b\bar{b}/c\bar{c}$
           and $t\bar{t}$), and (c) background only (solid histogram)
           and signal + background (dashed histogram). All distributions are 
           normalized to 1 fb$^{-1}$ of total integrated luminosity.}
  \label{sandb}
\end{figure}

\begin{figure}
  \centerline{\parbox{4.8in}{\epsfxsize=\hsize\epsffile{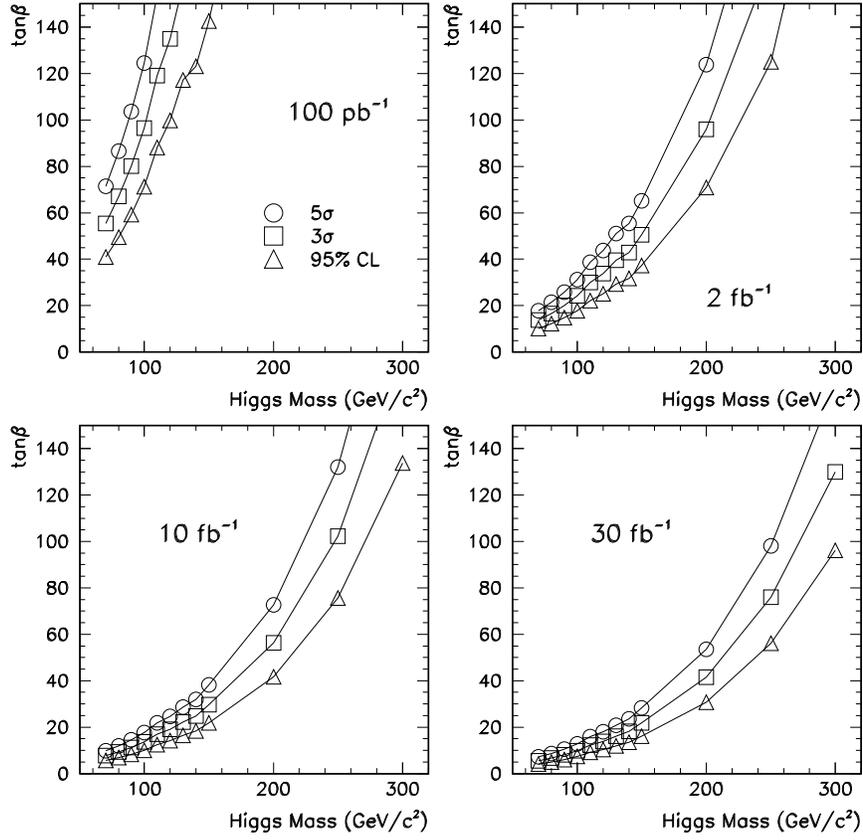}} }
  \caption{Discovery and exclusion contours in the $\tan\beta$--$M_{\phi}$
           plane of the MSSM for total integrated luminosities of 91 pb$^{-1}$,
           2 fb$^{-1}$, 10 fb$^{-1}$ and 30 fb$^{-1}$.}
  \label{figtb}
\end{figure}

\begin{figure}
  \centerline{\parbox{4.8in}{\epsfxsize=\hsize\epsffile[0 0 560 280]{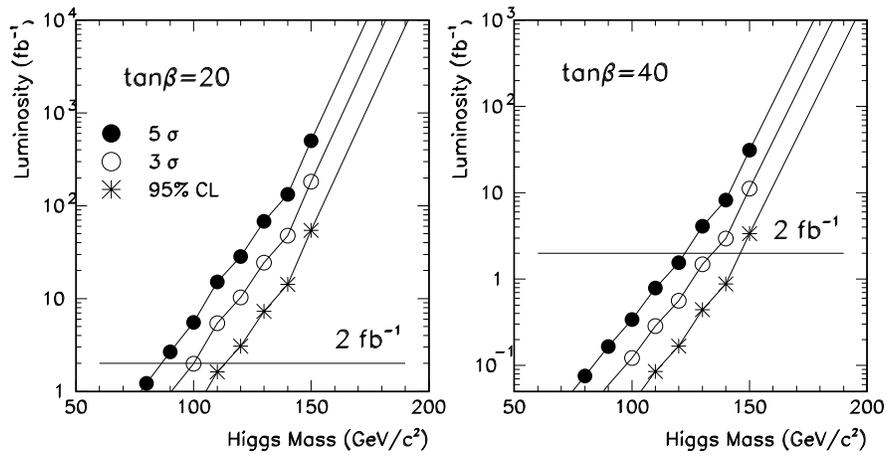}} }
  \caption{Neccesary integrated luminosity for 95\% CL exclusion limits
           and $5\sigma$ and $3\sigma$ discovery thresholds as a 
           function of the Higgs mass for the MSSM scenario with $\tan\beta$ = 10 and 40.}
  \label{figlum}
\end{figure}

\begin{figure}[ht!]
  \centerline{\parbox{4.5in}{\epsfxsize=\hsize\epsffile{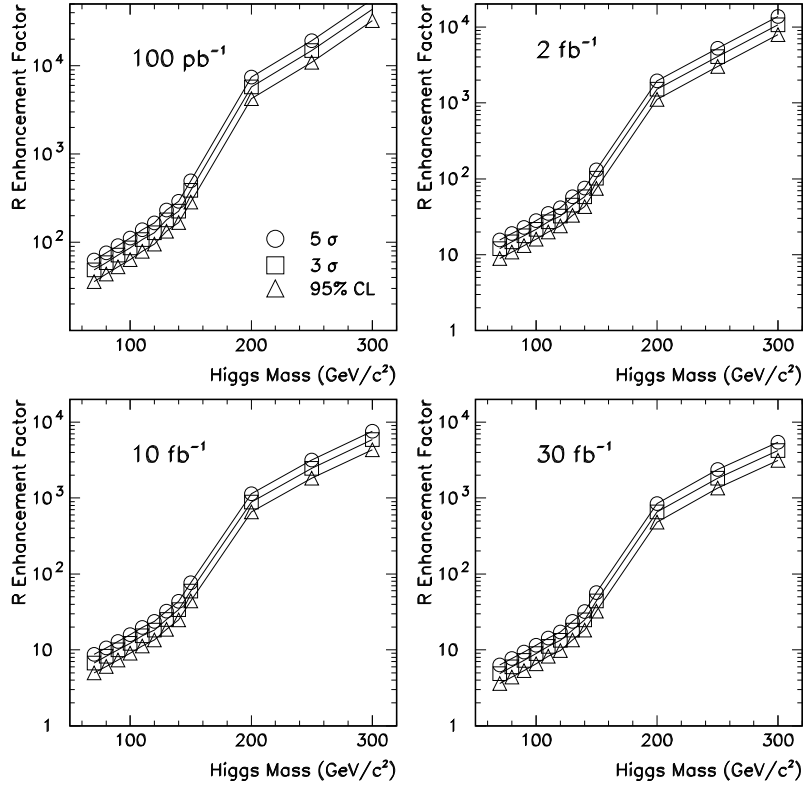}} }
  \caption{Enhancement factor ${\cal R}$ [eq.~(\ref{calrdef})] 
           as a function of the Higgs mass
           for 95\% CL exclusion limits, $5\sigma$ and $3\sigma$ discovery 
           thresholds and different total integrated luminosities.}
  \label{kf}
\end{figure}

\break
\large
{\bf Comparison of analyses} \\
\normalsize \nopagebreak


Starting from quite different input assumptions and methods,
the two $\bb\bb$ analyses discussed above arrive at quite
similar estimates for the exclusion and discovery reach for the 
MSSM scalar and pseudoscalar Higgses at large $\tan\beta$.  

The first main difference is that the analyses use different Monte
Carlo generators; the D\O\ analysis uses {\sc COMPHEP}, whereas the
CDF analysis uses a generator developed during the Workshop from a
modified version of {\sc PAPAGENO} with {\sc PYTHIA} fragmentation.
The cross sections and kinematic distributions from the two generators
agree reasonably well.  


The two kinematic selections proceed along similar lines, both
demanding four jets.  The D\O\ analysis demands a minimum jet
transverse momentum of 30 GeV$/c$ and leading and next-to-leading
jet $p_T$'s which increase with Higgs mass.  The CDF analysis
demands $\sum(E_T) > 125$ GeV and four jets with a minimum of
15 GeV transverse energy.\footnote{This is motivated by the Run 1 trigger
requirements, which may loosen for Run 2 if the experiment uses
the SVT to tag secondary vertices in three-jet events.}


Though there are potentially four taggable $b$ jets in the signal,
both analyses require three tags for optimal sensitivity.  The 
D\O\ analysis assumes a $b$ tagging efficiency based on the Monte
Carlo study of Section II.A.3, which has a maximum efficiency of
55\%.   The CDF analysis uses, conservatively, the Run 1 tagging
efficiencies per taggable jet, but the much larger Run 2 detector 
geometry to determine taggability.  This is perhaps the largest
source of difference between the two analyses, since the cube of the
tagging efficiency determines the signal rate.


By far the largest background comes from QCD $\bb jj$ production, and
the CDF simulations of its total rate did not agree with the observed
rate in analyses of Run~1 data.  The only weapons against it are the
requirement of the third $b$ tag and reconstructing the Higgs mass.
The D\O\ analysis relies on Monte Carlo simulations of the background,
and the CDF analysis uses Monte Carlo scaled by a factor determined
with Run 1 data, and taking into account the increased center of mass
energy in Run 2.  Given the different signal selections, it is
difficult to say how well the two methods agree.


The two analyses also take different approaches to Higgs mass 
reconstruction.  The D\O\ analysis uses all possible $\bb$ combinations
in an event, and a 15\% resolution (for the correct combination) based 
on the MC studies.  The CDF analysis uses jets 1 and 2 at high Higgs 
masses, and jets 2 and 3 at lower Higgs masses.  The assumed resolution 
is that of Run 1, typically 15\% for the correct combination for a 
150 GeV/$c^2$ Higgs mass.  


All these differences nevertheless lead to quite similar estimates for
exclusion and discovery.  Since the reach in $\tan\beta$ goes as the 
square of the cross section limit, the 30-40\% lower reach of the D\O\
analysis implies that the two analyses differ by perhaps 15-20\% in the
end.  Such differences are easily understood in the context of the 
foregoing paragraphs.


Both analyses assume rather conservative $b$ tagging efficiency
and mass resolution; these are quite likely to be better in the 
actual run.  Furthermore, the role of an improved trigger should 
be studied further; the fact that in many events one of the $b$ jets
is very far forward could perhaps mean that it is optimal to require
only three jets in the central region.  

In any case, this search represents the main mode for discovering
or ruling out the MSSM Higgs at large $\tan\beta$, and demonstrates
one of the unique advantages of hadron colliders.

  \subsection{Charged Higgs Bosons}			
\def\lsim{\mathrel{\raise.3ex\hbox{$<$\kern-.75em\lower1ex\hbox{$\sim$}}}}
\def\gsim{\mathrel{\raise.3ex\hbox{$>$\kern-.75em\lower1ex\hbox{$\sim$}}}}

\vspace{0.1in}
\small
\begin{center}
{\it Dhiman Chakraborty}
\end{center}
\normalsize\nopagebreak

In this section we consider the prospects for discovering or excluding
charged Higgs bosons produced in decays of top quarks at the 
Tevatron in Run~2.  If a charged Higgs boson exists, then it can be
directly produced in $p\bar{p}$ collisions.  However, as shown in
Section I.C.5.b, the cross section for inclusive charged Higgs
production is probably too small to be seen at the upgraded Tevatron
if $m_{H^+}\gsim m_t$.  However, if $t\to H^+b$ is kinematically
allowed then it can compete with $t\to W^+b$, which is by far the
dominant decay mode of top quark in the Standard Model (SM).

As previously noted, if $m_{H^+} < m_t - m_b$, and if $\tan\beta$ is
either very small or very large, then a significant fraction of those
events could contain charged Higgs bosons.  We focus on a search for
$p\bar{p}\to\ttbar\to H^+X$ based on the Type II two-Higgs doublet
model (2HDM) where one doublet couples exclusively to the down-type
fermions, and the second doublet couples exclusively to up-type
fermions~\cite{hhg}.\footnote{The MSSM Higgs sector is a Type II 2HDM.
However, the analysis of this section is more general, since the
results depend only on the charged Higgs boson mass and the parameter
$\tan\beta$ which parameterizes the coupling of $H^\pm$ to fermion
pairs [see eq.~\ref{hpmqq}].  No MSSM relations among Higgs sector
parameters are imposed.}

Recent results from searches for charged Higgs bosons can be found in
refs.~\cite{TEV,LEP,CLEO,ALEPH2}.  This study is an extension of a
search performed by D\O\ based on 109 pb$^{-1}$ from Run 1~\cite{TEV}.

\vskip 0.5cm
{\bf Parameter space}
\vskip 0.5cm

In the region $m_{H^+}< m_t - m_b$, the total inclusive cross-section 
for the production of either $H^+$ and/or $H^-$ is given by:
\begin{equation}
     \sigma(p\bar p\to \hpm+X)=\left(1-[{\rm BR}(t\to bW^+)]^2\right)
     \sigma(p\bar p\to t\bar t+X)\,,
\end{equation}
in the approximation that ${\rm BR}(t\to bH^+)+{\rm BR}(t\to bW^+) =
1$.  The parameters $m_{H^+}$ and $\tan\beta$ uniquely fix the top
quark branching ratios, given the top quark mass.  In this analysis,
we use the theoretical prediction that include NLO QCD
corrections~\cite{ttbarxsec}: $\sigma(p\bar{p}\to\ttbar) \simeq 7$ pb
at $\sqrt{s} = 2.0$ TeV.  However, only tree-level formulae for the
top-quark branching ratios have been employed.  Fig.~\ref{fg:params}
shows the region of the [$m_{H^+}$,$\tan\beta$] plane examined in this
study.  The minimum for $m_{H^+}$ is chosen at 50~GeV, somewhat below
the most recent lower limits from LEP.

This search is restricted to $m_{H^+} < 160$~GeV, somewhat less 
than $m_t - m_b$ (assuming $m_t = 175$~GeV); otherwise, the width of 
the charged Higgs $\Gamma(H^+)$ becomes too large ($> 7.5$~GeV) near 
the upper boundary on $\tan\beta$, and leading-order calculations become 
unreliable.  For the same reason, $\Gamma(t)$ is required to be $<15$~GeV.  
Since $\Gamma(t\to W^+b) \approx 1.5$~GeV, irrespective of 
[$m_{H^+}$,$\tan\beta$], this amounts to requiring $B(t\to H^+b) \le 0.9$, 
and thereby excludes from our analysis the dark-shaded regions at the two 
bottom corners of Fig.~\ref{fg:params}.  The cross-hatched regions correspond 
to $B(t\to H^+b) > 0.5$.  Also shown in Fig.~\ref{fg:params} are the decay
modes of $H^+$ that dominate in different parts of the parameter space.  
Analogous charge-conjugate results hold for $H^-$.

\begin{figure}
  \begin{center}
    \parbox{4in}{\epsfxsize=\hsize\epsffile{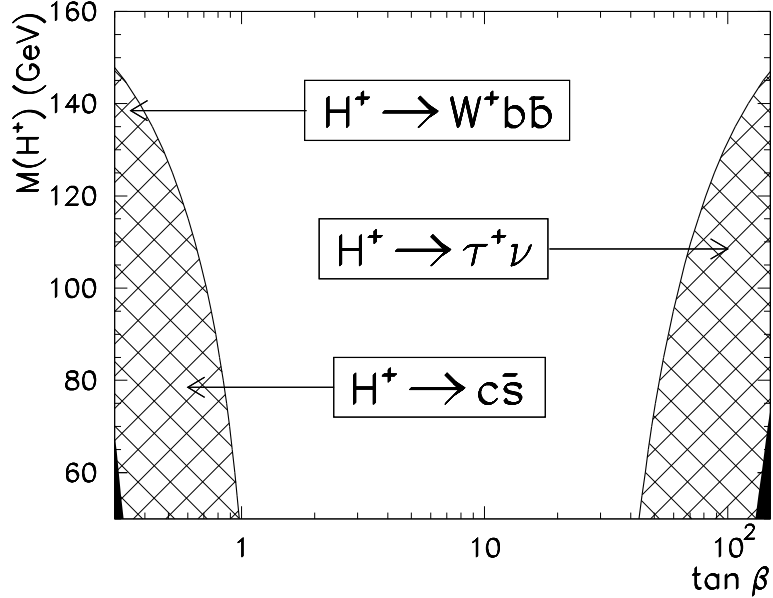}}
  \end{center}
  \caption{The parameter space explored in this analysis.
           Regions where $B(t\to H^+b) > 0.5$ are shown cross-hatched, with 
           the labels for various decay modes of the charged Higgs 
           indicating their regions of dominance.  Regions where 
           $B(t\to H^+b) > 0.9$ (dark shaded areas) are not considered.}
  \label{fg:params}
\end{figure}

\vskip 0.5cm
{\bf Methodology of a disappearance search}
\vskip 0.5cm

A direct search (or ``appearance search'') for charged Higgs bosons
covering the entire parameter space is made difficult by the variety
of final state signatures.  The region dominated by $H^+\to c\bar{s}$
suffers from enormous QCD backgrounds.  While it is possible to look 
directly for $H^+\to\tau^+\nu$, and for $H^+\to W^+\bb$, both D\O\ and 
CDF have found that an indirect ``disappearance'' search was a 
competitive option in Run 1~\cite{TEV}.

With increasing statistics and enhanced detector performance, direct
searches will gain strength faster than a disappearance search, but
the latter, the only one considered in this study, is expected to
dominate until the integrated luminosity exceeds $\sim 2$ fb$^{-1}$.

In a disappearance search, one employs selection criteria optimized to
detect the SM decay of $\ttbar$.  Such criteria are expected to have
relatively low efficiency for decays involving $t\to H^+b$.  The top
quark pair-production cross section is not expected to be
significantly affected by the presence of a charged Higgs.
Consequently, if data agree well with SM-based predictions for
$\sigma_{\ttbar}$, regions of parameter space where $B(t\to H^+b)$ is
large can be excluded because in those regions one would expect fewer
events than seen in data.  We have considered final states containing
one or two charged leptons ($e$ or $\mu$), as was done in Run 1
analyses, although it is possible that all-hadronic final states also
will be utilized in future.

The published values of the $\ttbar$ production cross section
($\sigma_{\ttbar}$) and the mass of the top quark ($m_t$) from direct
observation at D\O\ \cite{d0cs,d0mt} and CDF~\cite{cdfcs,cdfmt} are
based on the assumption of BR($t\to W^+b$) = 1.  The topic of present
interest is based on a violation of this assumption, so those
measurements are not relevant in our analysis.  Hence, in the process
of setting limits in the [$m_{H^+}$,$\tan\beta$] plane, the values of
$\sigma_{\ttbar}$ and $m_t$ serve as input parameters.

\break
\vskip 0.5cm
{\bf Signal, background, and their modeling}
\vskip 0.5cm

Our search for $t \bar t \rightarrow H^\pm X$, are based on selection
criteria optimized for $\ttbar\to W^+ b W^- \bar{b}$ events
which are detailed in refs.~\cite{d0cs,cdfcs}.  $W+$jets and
$Z/\gamma^*+$jets are the main sources of physics background, while
measurement fluctuations resulting in false identification of
electrons, muons, and neutrinos ($\met$) in QCD multijet events
contribute to instrumental background.

All $\ttbar$ events are simulated using ISAJET~\cite{ISAJET} and a
GEANT-based simulation of the Run 1 D\O\ detector.  The efficiencies
are verified using PYTHIA~\cite{PYTHIA}, and improvements for Run 2
are estimated from studies conducted in the course of the technical
design report for D\O\ upgrade, as well as SHW.  $W+$jets and $Z+$jets
events are modeled using the VECBOS tree-level parton generator
interfaced with ISAJET for fragmentation and hadronization.  The
instrumental background is estimated using Run 1 data
\cite{d0cs,d0mt,cdfcs,cdfmt}, on which additional suppression factors
expected from detector improvements are applied.

\vskip 0.5cm
{\bf Event selection}
\vskip 0.5cm

The final states of $\ttbar$ decaying according to the SM are
characterized by the decays of the two $W$ bosons in the event.  We
focus on events where at least one $W$ decays into an $e$, or a $\mu$,
and the associated neutrino.  If the other $W$ also decays into $e
\nu$ or $\mu \nu$, then the event is classified as a ``dilepton''
event.  If it decays into $\qq$, we call it a ``single lepton''
event.  Dilepton events are characterized by two high-$\pt$ isolated
leptons, large $\met$, and two or more jets, both originated by $b$
quarks.  Single lepton events are characterized by one high-$\pt$
isolated lepton, large $\met$, and three or more (nominally, four) jets,
two of which are originated by $b$ quarks.  Tagging of the $b$-quark
jets, either by a decay vertex separated from the primary interaction
vertex, or by an associated non-isolated lepton from semileptonic
decay of $b$ offers strong suppression of backgrounds in single-lepton
channels.  Backgrounds in dilepton channels are so small that the
$b$-tag requirement is not necessary.

\begin{table}
\caption{The event selection criteria.}
\label{tb:cuts}
\begin{tabular}{lcc}
 &
single-lepton &
dilepton \\
\hline
$p_T(\ell)$ & $>$ 20~GeV & $>$ 20~GeV \\
$|\eta(\ell)|$ & $<$ 2.0  & $<$ 2.0  \\
$\met$ & $>$ 20~GeV & $>$ 20~GeV \\
leptons ($n_\ell$)& $\ge 1$ & $\ge 2$ \\
$E_T(j)$ & $>$ 20~GeV & $>$ 20~GeV \\
$|\eta(j)|$ & $<$ 2.0  & $<$ 2.0  \\
jets ($n_j$)& $\ge 3$ & $\ge 2$ \\
$\mu$-tags & 1 or 2 & $\le 2$  \\
Aplanarity & $>$ 0.04 & -- \\[2pt]
$H_T \equiv \sum_{i=1}^{n_j} E_T(j_i)$ & $>$ 110~GeV& $>$ 110~GeV\\[4pt]
\end{tabular}
\end{table}

The selection criteria for dilepton and single-lepton channels are
summarized in Table~\ref{tb:cuts}.  These are similar to those used to
measure $\sigma_{\ttbar}$ in Run 1 \cite{d0cs,cdfcs}.  In addition to
these cuts, any event with an identified $\tau$ lepton (hadronic
decays only) with $E_T > 15$ GeV and $|\eta| < 2.0$ is rejected.
The dilepton and single-lepton selection criteria are combined in a
logical OR, so an event passing both sets of cuts is counted only
once, not twice.

\vskip 0.5cm
{\bf Efficiencies}
\vskip 0.5cm

For each top quark, there are four possible decay modes whose branching 
fractions depend on $m_{H^+}$ and $\tan\beta$: 
(1) $t\to W^+b$; (2) $t\to H^+b$, $H^+\to c\bar{s}$; (3) $t\to H^+b, H^+\to W^+\bb$; 
and (4) $t\to H^+b, H^+\to\tau^+\nu$.  If the decay mode of $t$ ($\bar t$) is
 denoted by $i$ ($j$), then the total acceptance for any set of selection criteria 
is given by
\begin{equation}
    A(m_H^+,\tan\beta) = {\displaystyle \sum_{i,j=1}^{4}} 
                     \epsilon_{i,j}(m_{H^+})  B_i(m_{H^+},\tan\beta) B_j(m_{H^+},\tan\beta),
\label{eq:acc}
\end{equation}
where $\epsilon_{i,j}$ is the efficiency for channel $\{i,j\}$, and
$B_i B_j$ is the branching fraction.  All $B_i$ depend strongly on
both $m_{H^+}$ and $\tan\beta$; $\epsilon_{1,1}$ depends on neither,
and all other $\epsilon_{i,j}$ depend on $m_{H^+}$, but not on
$\tan\beta$.

The efficiencies for various channels, for $m_{H^+}$ = 110 GeV, are given 
in  Table~\ref{tb:effs} in the form of a symmetric $4 \times 4$ matrix.
The 10 independent elements of the matrix are plotted as functions of
$m_{H^+}$ in Figs.~\ref{fg:eff_1} and \ref{fg:eff_2}.  Uncertainties in jet
energy scale, particle identification, and signal modeling are the
main components of systematic uncertainties.  The total systematic
uncertainty is estimated to be 10\% of the efficiency.  Statistical
uncertainties are negligible.

\begin{table}[ht!]
  \caption{The efficiencies $\epsilon_{i,j}$ of our selection criteria (in \%),
           for $m_t$ = 175~GeV and $m_{H^+}$ = 110~GeV, for various decay modes of 
           $\ttbar$.  Here $i$ is the row index and $j$ the column.
           Systematic uncertainties are assumed to be 10\%.
           Statistical uncertainties are negligible.}
  \label{tb:effs}
  \begin{center}
    \begin{tabular}{l|c| c| c| c} 
  &
    $\overline{t} \to W^-\overline{b}$ &
      $\overline{t} \to H^-\overline{b}, H^- \to \bar{c}s$ &
        $\bar{t} \to H^-\bar{b}, H^- \to W^-\bb$ &
          $\bar{t} \to H^-\bar{b}, H^- \to \tau^-\bar{\nu}$ \\ \hline
$t \to W^+b$                   & $4.0$ & $2.7$ & $2.6$ & $1.1$ \\ 
$t \to H^+b, H^+ \to c\bar{s}$ & $2.7$ & $0.1$ & $2.0$ & $1.1$ \\ 
$t \to H^+b, H^+ \to W^+\bb$  & $2.6$ & $2.0$ & $2.0$ & $1.5$ \\ 
$t \to H^+b, H^+ \to \tau^+\nu$ & $1.1$ & $1.1$ & $1.5$ & $0.4$ \\ 
    \end{tabular}
  \end{center}
\end{table}

\begin{figure}[ht!]
  \begin{center}
    \parbox{4in}{\epsfxsize=\hsize\epsffile{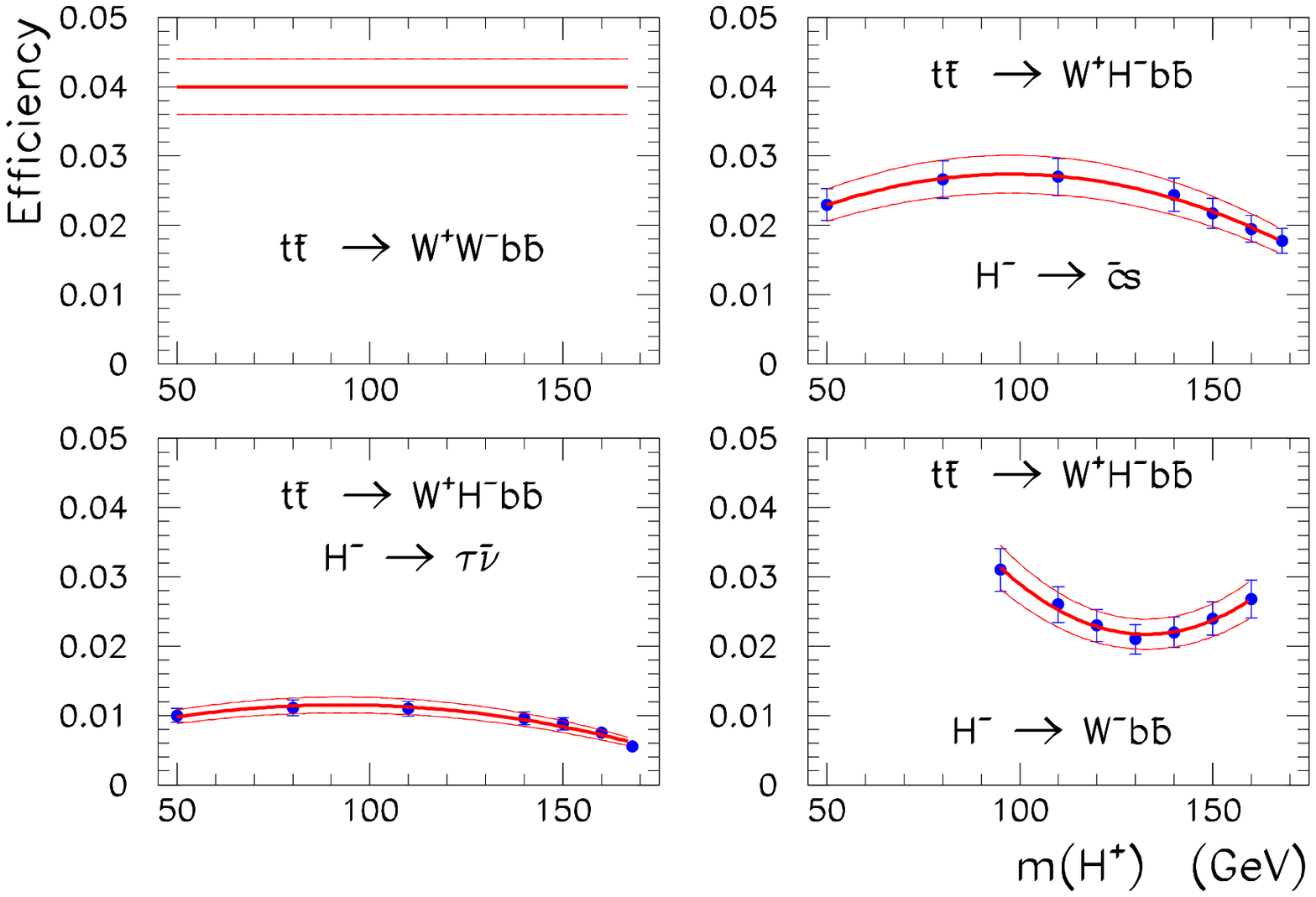}}
  \end{center}
  \caption{The efficiencies of our selection criteria, as functions of 
           $m_{H^+}$, for channels in which no more than one top quark
           decays to charged Higgs.  The thick central curves and the 
           thin curves containing the uncertainty bands are parabolic fits, 
           respectively, to the points and the error bars.}
  \label{fg:eff_1}
\end{figure}

\begin{figure}
  \begin{center}
    \parbox{4in}{\epsfxsize=\hsize\epsffile{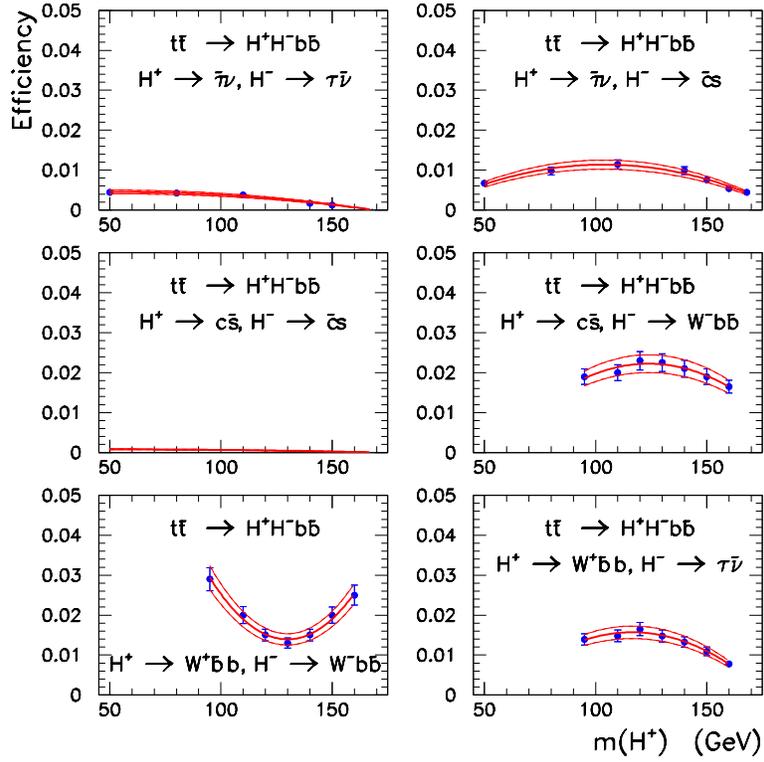}}
  \end{center}
  \caption{The efficiencies of our selection criteria, as functions of 
           $m_{H^+}$, for channels in which both top quarks decay to 
           charged Higgs.  The thick central curves and the thin curves 
           containing the uncertainty bands are parabolic fits, respectively, 
           to the points and the error bars.}
  \label{fg:eff_2}
\end{figure}

It is easy to see why the efficiencies are low in regions of parameter
space dominated by $H\to c\bar{s}$ (small $\tan\beta$, small
$m_{H^+}$).  When both top quarks decay to charged Higgs, and both
charged Higgs decay to $c\bar{s}$, there is no legitimate source for
an isolated $e$ or $\mu$, nor for large $\met$, in the final state.
Compared to the SM, efficiencies are lower in regions dominated by
$H^+\to\tau^+\nu$ (large $\tan\beta$) because $\tau$ leptons decay 
hadronically most of the time, and a veto is applied to events containing 
$\tau$s.  Even when a $\tau$ from an $H^\pm$ decays to an $e$ or a $\mu$, 
the lepton usually has a lower $p_T$ compared to one from a prompt decay of a
$W$.  Regions dominated by $H^+\to W^+\bb$ (small $\tan\beta$, large 
$m_{H^+}$) are more difficult to discriminate against because real $W$ 
bosons are produced.  However, such events are characterized by a large 
number of $b$ jets (up to 6), and can be suppressed somewhat by requiring 
that no more than two of the jets be tagged as $b$.

\vskip 0.5cm
\break

\begin{figure}[ht!]
  \begin{center}
    \parbox{4in}{\epsfxsize=\hsize\epsffile{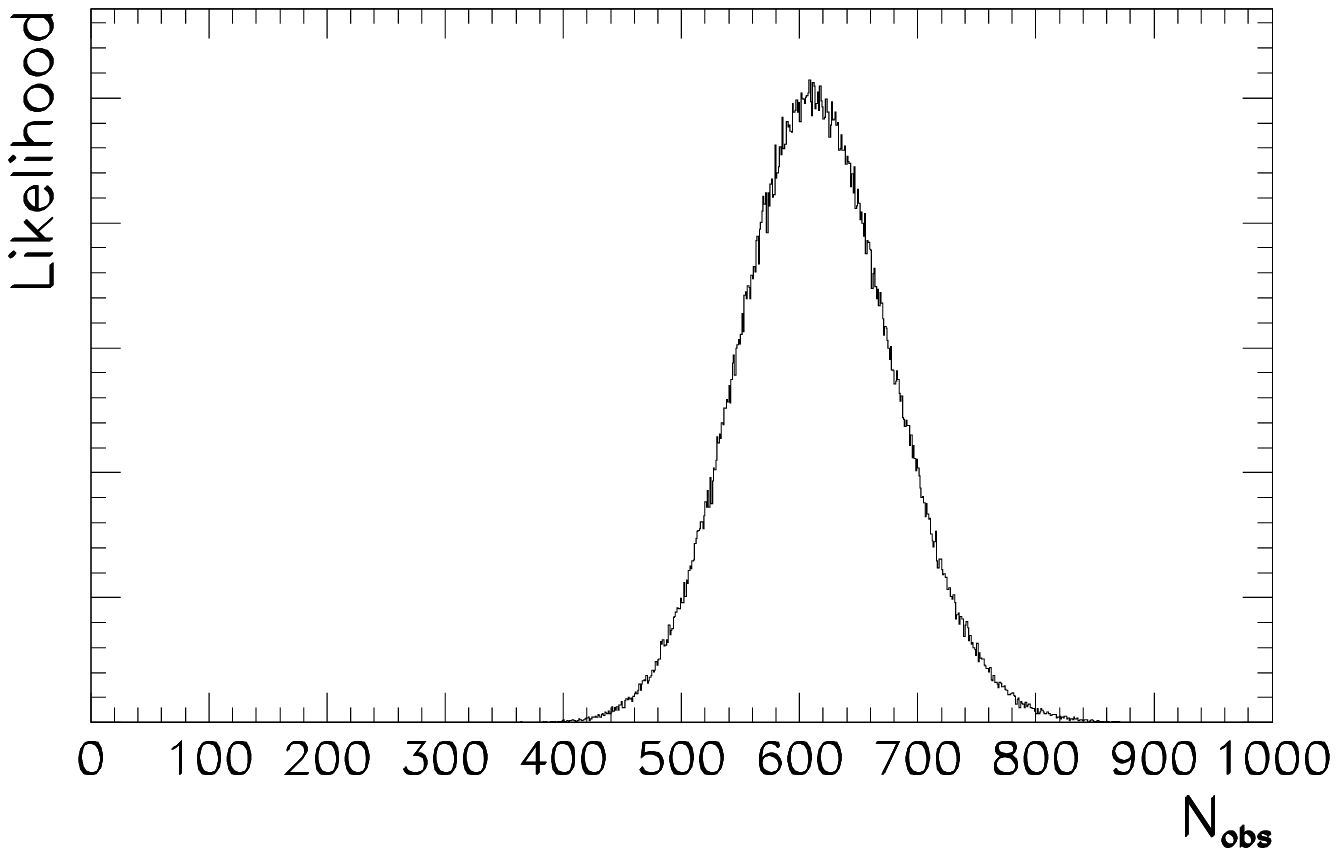}}
  \end{center}
  \caption{Distribution of $n_{\rm obs}$ in simulated experiments with parameters
           given in Table~\ref{tb:exptpars}, assuming $B(t\to W^+b) = 1.$ 
           Lower efficiencies for $t \bar t \rightarrow H^\pm X$ events imply that
           the distribution would peak at smaller values of $n_{\rm obs}$ if 
           $B(t\to H^+b)$ is large.}
  \label{fg:lklhd_nobs}
\end{figure}

\begin{figure}[h!]
  \begin{center}
    \parbox{4in}{\epsfxsize=\hsize\epsffile{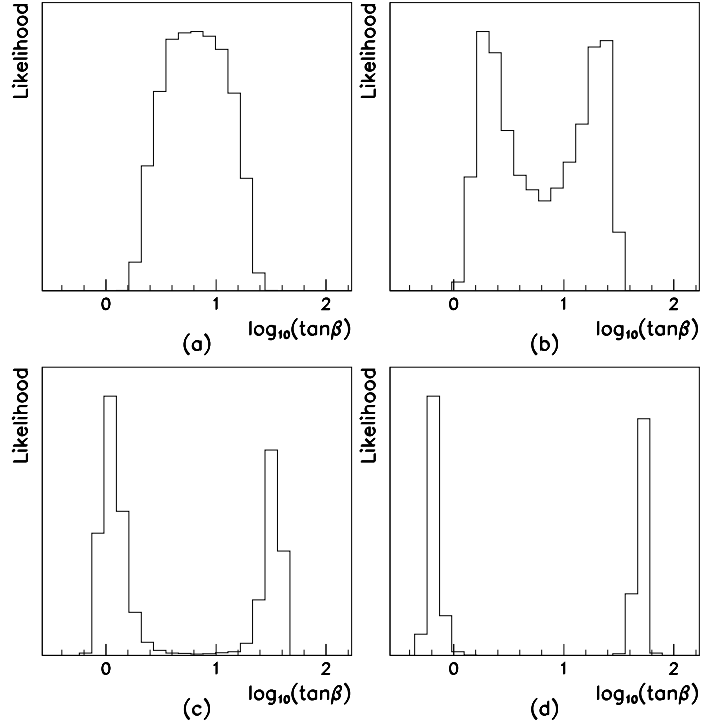}}
  \end{center} 
   \caption{Posterior probability density for $\tan\beta$, for $m_{H^+}$ = 80 GeV, and 
         for four different values of the number of events
         experimentally observed in 2 fb$^{-1}$ integrated luminosity
         (based on simulated experiments with parameters given in
         Table~\ref{tb:exptpars}): (a) $n_{\rm obs}$ = 600, (b)
         $n_{\rm obs}$ = 500, (c) $n_{\rm obs}$ = 400, and (d) $n_{\rm
         obs}$ = 300.  As expected, for $n_{\rm obs}$ = 600, the most
         likely value of $\tan\beta$ is $\sqrt{{m_t}\over{m_b}} \approx 6.5$,
         or, $\log_{10}(\tan\beta) \approx 0.82$, where $B(t\to H^+b)$ is minimum
         (the SM-based prediction peaks at $n_{\rm obs}$ = 610, as
         seen in Fig.~\ref{fg:lklhd_nobs}).  Smaller values of $n_{\rm obs}$ 
        favor larger values of $B(t\to H^+b)$.}
  \label{fg:lklhd_tb}
\end{figure}

{\bf Likelihood analysis}
\vskip 0.5cm

For $n_{\rm obs}$ observed events, the joint posterior probability
density for $m_{H^+}$ and $\tan\beta$ is given by
\begin{equation}
    P(m_{H^+},\tan\beta | n_{\rm obs}) 
    \propto  {\displaystyle \int  G({\cal L}) \int  G(n_B)
    \int}  G(A) P(n_{\rm obs}|\mu) dA dn_B  d {\cal L},
\label{eq:n_obs}
\end{equation}
where $P(n_{\rm obs}| \mu)$, is the Poisson probability of observing
$n_{\rm obs}$ events, given a total (signal + background) expectation of
\begin{equation}
   \mu(m_{H^+},\tan\beta)  =  A(m_{H^+},\tan\beta) \sigma_\ttbar {\cal L}  + n_B,
\end{equation}
and $G$ represents a Gaussian distribution.
The means and widths of the Gaussians for the integrated luminosity $\cal L$,
and the number of background events $n_B$, are given in 
Table~\ref{tb:exptpars},
while those for the acceptance $A(m_{H^+},\tan\beta)$, are calculated using
Eq.~(\ref{eq:acc}), with parametrized functions for 
$\epsilon_{i,j}$, 
and leading order calculations of  $B_i$, $B_j$.

\begin{table}[ht!]
\caption{The integrated luminosity, and the
expectations from background and SM $\ttbar$ signal (assuming $m_t$ = 175~GeV,
$\sigma_{\ttbar}$ = 7.0~pb).}
\label{tb:exptpars}
\begin{tabular}{l|c}
Integrated luminosity, $\cal L $ & $2.0 \pm 0.1$ fb\\
Estimated background, $n_B$ & $50 \pm 5$ \\
Expected signal (SM), $n_S$ & $560 \pm $ 24\\
\end{tabular}
\end{table}

Equation (\ref{eq:n_obs}), which we parametrize as a function of $m_{H^+}$
and $\tan\beta$, gives a Bayesian posterior probability density for those
parameters~\cite{Jaynes}.  The prior distribution is assumed to be
uniform in $m_{H^+}$ and in $\log\tan\beta$.  Assuming instead that the prior is
uniform in $m_{H^+}$ and in $B(H^+\to\tau^+\nu)$ does not significantly 
alter the posterior distribution.  To calculate probabilities, a Monte Carlo
integration is carried out by spanning the parameter space in steps of
5~GeV in $m_{H^+}$ from 50~GeV to 160~GeV, with 25 uniform steps in 
$\log\tan\beta$ covering the range $0.3 < \tan\beta < 150$ at each value 
of $m_{H^+}$, and performing 400,000 trials of Eq.~(\ref{eq:n_obs}) at each step.

\begin{figure}
  \begin{center}
    \parbox{4in}{\epsfxsize=\hsize\epsffile{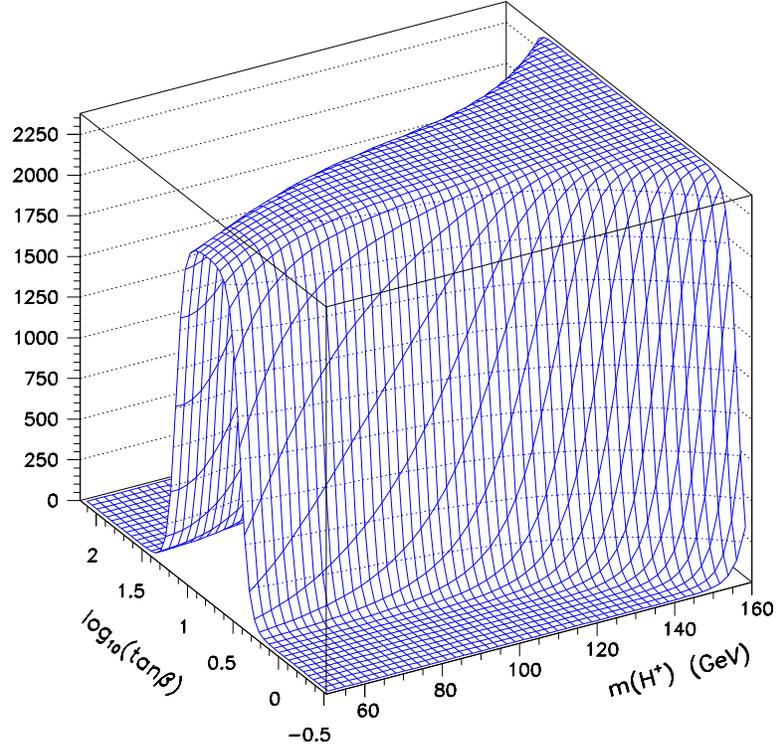}}
  \end{center}
  \caption{The likelihood of $n_{obs}$ = 600 (in arbitrary units), 
           as a function of $m_{H^+}$ and $\tan\beta$ (assuming $m_t$ = 175 GeV 
           and parameters given in Table~\ref{tb:exptpars}).}
  \label{fg:lklhd_surf}
\end{figure}

The likelihood predicted by the SM for observing $n_{\rm obs}$ events
is shown in Fig.~\ref{fg:lklhd_nobs}, while Fig.~\ref{fg:lklhd_tb}
shows the posterior probability density for $\tan\beta$ corresponding
to several values of $n_{\rm obs}$ for $m_{H^+}$ = 80~GeV.  For a
given value of $n_{\rm obs}$, the 95\% CL exclusion boundary in the
[$m_{H^+},\tan\beta$] plane is obtained by integrating the probability
density $P(m_{H^+},\tan\beta|n_{\rm obs})$, given by
Eq.~(\ref{eq:n_obs}), between contours of constant $P$.  The
likelihood surface corresponding to the parameters given in
Table~\ref{tb:exptpars}, and $n_{\rm obs}$ = 600, is shown in
Fig.~\ref{fg:excl_cs_shw} for three different values of $\int {\cal L}
dt$: 109 pb$^{-1}$ (Run 1, $\sqrt{s} = 1.8$ TeV ~\cite{TEV}), 2
fb$^{-1}$, and 10 fb$^{-1}$.

\begin{figure}
  \begin{center}
    \parbox{4in}{\epsfxsize=\hsize\epsffile{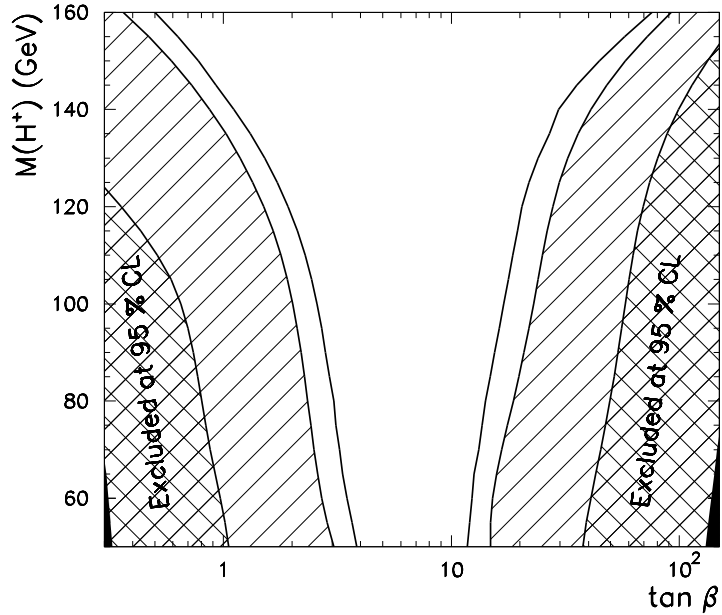}}
  \end{center}
\caption{The 95\% CL exclusion boundaries in the [$m_{H^+},\tan\beta$] plane 
         for  $m_t$ = 175~GeV and several values of the integrated luminosity:
         0.1 fb$^{-1}$ (at $\sqrt{s} = 1.8$ TeV, cross-hatched), 
         2.0 fb$^{-1}$ (at $\sqrt{s} = 2.0$ TeV, single-hatched), and 
         10 fb$^{-1}$ (at $\sqrt{s} = 2.0$ TeV, hollow), 
         if the $n_{\rm obs}$ continues to be where the SM-based prediction peaks.}
  \label{fg:excl_cs_shw}
\end{figure}

\vskip 0.5cm
{\bf Conclusions}
\vskip 0.5cm

Our study shows that, compared to Run 1~\cite{TEV}, the search for
charged Higgs bosons in decays of pair-produced top quarks will have a
much greater reach in Run 2.  A direct appearance search for charged
Higgs bosons in decays of top quark is also possible.  At the
Tevatron, it would primarily involve a search for $H\to\tau\nu$, which
limits its applicability to $\tan\beta > 4$.  CDF has published
results from such searches based on Run 1 data~\cite{CDFch,CDF_H+_2}
which are consistent with its disappearance search results.
Comparable results from an appearance search performed by D\O\ are
expected soon.  An appearance search in the region of $\tan\beta < 1$
will have to be based on $H^+\to c\bar{s}$.  This will be more
challenging since $c$ and $s$ quarks are more difficult to identify
than $\tau$s and $\nu$s.  However, if $m_{H^+}$ is substantially
larger than $M_W$, then it may be possible to distinguish ($t\to H^+b$,
$H\to c\bar{s}$) from ($t\to Wb$, $W\rightarrow q_1 \bar q_2$) based on
event kinematics.  The indirect disappearance search presented here is
expected to yield stronger null results than a direct appearance
search for integrated luminosities up to 2--4 fb$^{-1}$.  If there is a
charged Higgs lighter than the top quark, its presence can be detected
by a disappearance search if $\tan\beta$ is substantially different
from $\sqrt{{m_t}/{m_b}}$, prompting a direct search to confirm
discovery.  Beyond Run 2, appearance searches should lead to superior
results.

\newpage
\section{Interpretation} 

  \subsection{Combined Standard Model Higgs Boson Results}        \vspace{0.1in}
\small
\begin{center}
{\it J. Conway}
\end{center}
\normalsize\nopagebreak

As the results of Section II show, there is no single search channel
for the Higgs which one might call ``golden''; to maximize the
sensitivity of the Higgs search it is necessary to combine the results
of all the channels.  This section presents the results of combining
all Standard Model Higgs search channels, from both experiments, in
terms of the integrated luminosity needed to exclude the Higgs at 95\%
CL (assuming it is not there) or discover it at the $3\sigma$ or
$5\sigma$ level if it is.

\vspace{0.2in}
{\bf Statistical Method} \\ \nopagebreak

The statistical method employed here for combining channels is
discussed in the Appendix.  Briefly, the result of each search channel
is treated as a counting experiment, and for a given outcome there is
some Poisson probability.  For all channels in both experiments, these
probabilities are multiplied together to form a joint likelihood.  This
likelihood can be expressed as a function of the Higgs signal cross
section, and can be used to set 95\% CL limits or discovery 
significances.  To take into account all possible experimental outcomes,
the integrated luminosity threshold quoted below represent those 
values for which the desired statistical threshold is met in 50\%
of all possible outcomes.

\vspace{0.2in}
{\bf Overview of results} \\ \nopagebreak

The results of all the channels studied are summarized in Tables
\ref{t:low} and \ref{t:high}.  The tables show
the expected signal $S$, the expected background $B$, and the
sensitivity $S/\sqrt{B}$ in each channel as a function of the assumed
Higgs mass.  In all the low-mass channels, we have taken the 
numbers from assuming a 10\% resolution in $m_{\bb}$.

The tables indicate that of the low mass channels, the $\ell\nu\bb$
and $\nn\bb$ have the most sensitivity.  Also, while the dilepton mode
adds significantly to the sensitivity, the all-hadronic channel
brings little information to the final combination.

In comparing the different analyses, it is clear that the neural
network technique results in significantly enhanced sensitivity in the
three channels where it has been studied.  Note that the NN- and
SHW-based channel analyses do not take into account trigger
inefficiency for events which otherwise pass the selection; this
should be no problem in the $\ell\nu\bb$ and $\lpm\bb$ cases but may
be a slight overestimate at low masses in the $\nn\bb$ case.

For the high-mass channels, the $\lpm\nn$ channel has the most sensitivity,
whereas the $\lsdilep$ channel has nearly as good sensitivity over a 
broader mass range.  



 \begin{table}
   \begin{center}
   \caption{Summary of low-mass Standard Model Higgs search
                 channel sensitivities used in the combined integrated
                 luminosity threshold calculations.  The values of $S$
                 and $B$ are expressed as the number of
                 events expected per detector in 1 fb$^{-1}$.
                 Here we assume an improved Run 2
                 $m_{\bb}$ resolution of 10\%.  ``SHW'' indicates the
                 analyses based on the SHW simulation, ``NN''
                 indicates the SHW neural-network-based analyses, and
                 ``CDF'' indicates the analyses based on
                 extrapolations from the CDF Run 1 conditions to Run 2
                 detector geometry and efficiencies.  Note that in the
                 $\nn\bb$ analyses the background from $\bb$ dijet
                 production is included by doubling the non-dijet
                 background, as discussed in the text.}

    \label{t:low}
     \begin{tabular}{llccccc}
                      &              & \multicolumn{5}{c}{Higgs Mass (GeV/$c^2$)} \\ \cline{3-7}
\noalign{\vskip3pt}
\multicolumn{1}{c}{Channel} &\multicolumn{1}{c}{Rate}& 
                                        90   &  100  &  110  &  120  &  130 \\ \tableline \tableline
                      & $S$          &  8.4  &  6.6  &  5.0  &  3.7  &  2.2  \\   
 $\ell\nu\bb$ (CDF)   & $B$          &  48   &  52   &  48   &  49   &  42   \\  
                      & $S/\sqrt{B}$ &  1.2  &  0.9  &  0.7  &  0.5  &  0.3  \\  \tableline
                      & $S$          & 10    &  8    &  5    &  4    &  3    \\  
 $\ell\nu\bb$ (SHW)   & $B$          & 75    & 68    & 57    & 58    & 52    \\ 
                      & $S/\sqrt{B}$ &  1.1  &  1.0  &  0.7  &  0.5  &  0.4  \\  \tableline 
                      & $S$          &  8.7  &  9.0  &  4.8  &  4.4  &  3.7  \\  
 $\ell\nu\bb$ (NN)    & $B$          &  28   &  39   &  19   &  26   &  46   \\ 
                      & $S/\sqrt{B}$ &  1.6  &  1.4  &  1.1  &  0.9  &  0.5  \\  \hline  \hline
                      & $S$          &  2.5  &  2.2  &  1.9  &  1.2  &  0.6  \\  
 $\nn\bb$ (CDF)       & $B$          & 20.0  & 18.6  & 16.0  & 13.0  &  9.6  \\ 
                      & $S/\sqrt{B}$ &  0.6  &  0.5  &  0.5  &  0.3  &  0.2  \\  \tableline 
                      & $S$          &  8.9  &  6.7  &  4.6  &  3.2  &  2.1  \\   
 $\nn\bb$ (SHW)       & $B$          &  77   &  69   &  56   &  39   &  30   \\  
                      & $S/\sqrt{B}$ &  1.0  &  0.8  &  0.6  &  0.5  &  0.4  \\  \tableline 
                      & $S$          &  12   &   8   &  6.3  &  4.7  &  3.9  \\   
 $\nn\bb$ (NN)        & $B$          & 123   &  70   &  55   &  45   &  47   \\  
                      & $S/\sqrt{B}$ &  1.1  &  1.0  &  0.8  &  0.7  &  0.6  \\  \hline \hline
                      & $S$          &  1.0  &  0.9  &  0.8  &  0.5  &  0.3  \\    
 $\lpm\bb$ (CDF)      & $B$          &  3.6  &  3.1  &  2.5  &  1.8  &  1.1  \\   
                      & $S/\sqrt{B}$ &  0.5  &  0.5  &  0.5  &  0.4  &  0.3  \\  \tableline 
                      & $S$          &  1.5  &  1.2  &  0.9  &  0.6  &  0.4  \\    
 $\lpm\bb$ (SHW)      & $B$          &  4.9  &  4.3  &  3.2  &  2.3  &  1.9  \\   
                      & $S/\sqrt{B}$ &  0.7  &  0.6  &  0.5  &  0.4  &  0.3  \\  \tableline
                      & $S$          &  1.2  &  0.9  &  0.8  &  0.8  &  0.6  \\    
 $\lpm\bb$ (NN)       & $B$          &  2.9  &  1.9  &  2.3  &  2.8  &  1.9  \\   
                      & $S/\sqrt{B}$ &  0.7  &  0.7  &  0.5  &  0.5  &  0.4  \\  \hline \hline
                      & $S$          &  8.1  &  5.6  &  3.5  &  2.5  &  1.3  \\     
 $\qq\bb$ (SHW)       & $B$          & 6800  & 3600  & 2800  & 2300  & 2000  \\    
                      & $S/\sqrt{B}$ & 0.10  & 0.09  & 0.07  & 0.05  & 0.03  \\  \tableline 
    \end{tabular} 
      \end{center} 
  \end{table}

 \begin{table}
   \begin{center}
   \caption{Summary of high-mass Standard Model Higgs search channel
            sensitivities; all results are based on SHW studies.
            The values of $S$ and $B$ are expressed as the number of 
            events expected per detector in 1 fb$^{-1}$.}
   \label{t:high}
\setlength{\tabcolsep}{9pt}
  \begin{tabular}{llccccccc}
                      &              & \multicolumn{7}{c}{Higgs Mass (GeV/$c^2$)}  \\ \cline{3-9}
\noalign{\vskip3pt}
\multicolumn{1}{c}{Channel} & \multicolumn{1}{c}{Rate}  &
                                         120  &  130  &  140  &  150  &  160  &  170  &  180  \\  \tableline
                      & $S$          &   -   &   -   &  2.6  &  2.8  &  1.5  &  1.1  &  1.0  \\      
 $\lpm\nn$            & $B$          &   -   &   -   &  44   &  30   &  4.4  &  2.4  &  3.8  \\    
                      & $S/\sqrt{B}$ &   -   &   -   &  0.39 &  0.51 &  0.71 &  0.71 &  0.51 \\  \tableline  
                      & $S$          &  0.08 &  0.15 &  0.29 &  0.36 &  0.41 &  0.38 &  0.26 \\      
 $\lsdilep$           & $B$          &  0.58 &  0.58 &  0.58 &  0.58 &  0.58 &  0.58 &  0.58 \\    
                      & $S/\sqrt{B}$ &  0.11 &  0.20 &  0.38 &  0.47 &  0.54 &  0.50 &  0.34 \\
   \end{tabular}
   \end{center}
 \end{table}

\vspace{0.2in}
{\bf Combined channel integrated luminosity thresholds} \\ \nopagebreak

We perform determinations of the integrated luminosity thresholds
combining all low-mass and high-mass channels.  In the combination we
assume that both experiments' data is used by doubling each channel: we
generate separate pseudoexperimental outcomes for each channel in each
experiment, and combine all the results together in the final
likelihood.

To take into account reasonable systematic errors, we incorporate into
the likelihood a relative uncertainty on the background for each
channel which is the smaller of 10\% of the expected background or
$1/\sqrt{LB}$, where $L$ is the integrated luminosity and $B$ is the
expected number of background events in 1 fb$^{-1}$.  Such an
assumption is typical of the level of uncertainty experienced in new
particle searches at the Tevatron.  Note that if one does not let
the systematic error decrease with integrated luminosity, numerical
instability can result.  More importantly, in the real experiments as
the integrated luminosity increases the experimenters will have better
control of the systematic errors, and will in all likelihood harden
the selection criteria to improve the sensitivity while maintaining
tolerable systematic uncertainties.  Without the inclusion of these
systematic errors, the integrated luminosity thresholds are
approximately 30-50\% smaller.

\vspace{0.2in}
{\bf Reliability of the results} \\ \nopagebreak

The nature of a study like this is such that the final results (here
the integrated luminosity thresholds) depend on several unknown
performance factors.  The main factors affecting the result of this
study, in particular the results in the low-mass channels, are
\begin{itemize}
  \item the $b$-tagging efficiency,
  \item the $\bb$ mass resolution, and
  \item the backgrounds in the various channels.
\end{itemize}

All the low mass channels require two $b$-tagged jets, which makes
the result very sensitive to this efficiency.  The signal efficiency 
increases as the square of the $b$-tag efficiency, whereas only the
$W\bb$, $Z\bb$, and QCD $\bb$ dijet backgrounds follow the same 
dependence.  In the gaussian limit, the integrated luminosity
thresholds scale with $B/S^2$, so even if the backgrounds are 
dominated by real $\bb$, increasing the $b$ tag efficiency by 10\%
for example while maintaining the mistag rate would result in a 
20\% reduction in the required integrated luminosity threshold.

Both CDF and D\O\ will have excellent vertex tracking, and new
algorithms and solid understanding of vertex tracking systematics
could result in efficiencies somewhat higher than assumed in this
report.  However, physics effects such as multiple interactions,
and/or unforeseen detector performance issues could adversely affect the
$b$ tag efficiency and purity.  The assumptions made here are
reasonably conservative.

The mass resolution is directly related to the background - indeed the
main effect of improving the resolution is of course to reduce the
number of background events in the window around the Higgs mass.  As
noted above, since the integrated luminosity thresholds scale with
$B/S^2$, increasing the background (or worsening the resolution) by
some relative amount has a the same relative effect on the thresholds.  

As the careful reader will note, in the $\nn\bb$ channel analyses no
estimates were made of the generic $\bb$ dijet background.  This
process has a very large cross section but tiny acceptance, and is
thus not modeled reliably.  In the CDF Run 1 analysis, this
background was about half of the total, and estimated from the data.
With better understanding, this fraction could potentially be reduced.
However, we conservatively assume in the final combination that this
background is equal to the rest of the background, that is, half of
the total.  This is reflected in Table~\ref{t:low} by doubling the 
non-dijet background in the $\nn\bb$ channel analyses.

Perhaps of more concern is the assumption regarding the mass
resolution ultimately attainable.  In the low mass region this has
been assumed to be 10\% (of the Higgs mass).  Even with a perfect
detector the physics of the jet fragmentation and gluon radiation,
along with the missing neutrinos from the $b$ quark decay limit the
$\bb$ mass resolution attainable to a lineshape with a width of about
8\% with a tail on the low side.  If detector-related effects can be
understood and controlled well, and if all the available detector
information is brought to bear, the estimate of 10\% may be reasonable.
Even if the mass resolution is closer to 12\%, though, the resulting
integrated luminosity thresholds will only be 20\% larger, well within
the other uncertainties inherent in this work.  

Furthermore, at this level the relative strengths and weaknesses of
the two detectors will ultimately determine the sensitivity to the
Higgs.  With more elaborate tracking, CDF will most likely excel in
terms of $b$ tagging efficiency, whereas with its more hermetic
calorimeter the D\O\ experiment may attain the best mass resolution.

Given the above discussion, the basic conclusions arrived at below are
not unreasonably aggressive.  Breakthroughs in technique are always
possible, and have indeed been the norm in the past.  For example both
the Higgs search in LEP1 and the top quark search in Run 1 at the
Tevatron exceeded the expectations of studies prior to machine
turn-on.  The studies presented here should be taken as cautiously
optimistic: Using full mass spectrum fits, applying neural network
techniques, improving the trigger efficiencies, adding other search
mode, and improving the mass resolution and tagging efficiency beyond
that projected here may all serve to dramatically improve the
discovery potential for the Higgs at the Tevatron.

\begin{figure}[th]
  \begin{center}
    \parbox{6.0in}{\epsfxsize=\hsize\epsffile[0 200 500 540]
           {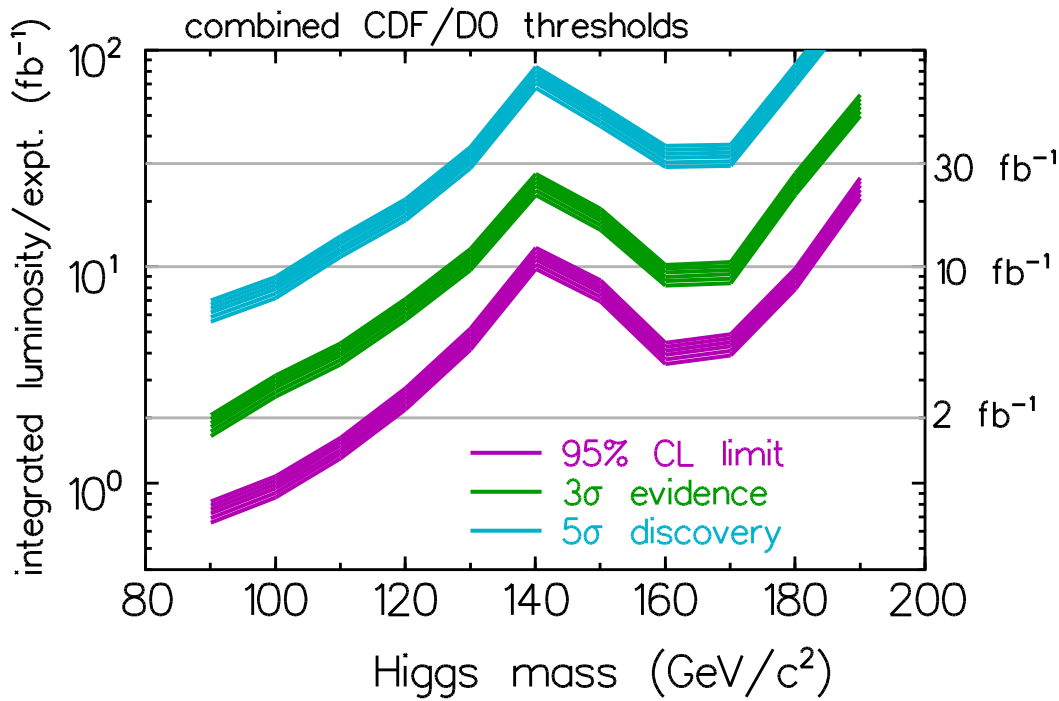}}
  \end{center}
  \caption{The integrated luminosity required per experiment, to
            either exclude a SM Higgs boson at 95\% CL or discover it at the
            $3\sigma$ or $5\sigma$ level, as a function of the Higgs mass.  
            These results are based on the combined
            statistical power of both experiments.  The curves shown  
            are obtained by combining the $\ell\nu b\bar b$,
            $\nu\bar\nu b\bar b$ and $\ell^+\ell^-b\bar b$ channels
            using the neural network selection in the low-mass Higgs region
            ($90~{\rm GeV}~\lsim\mhsm\lsim 130$~GeV), and the
            $\lsdilep$ and $\lpm\nn$ channels in the high-mass Higgs region
            ($130~{\rm GeV}~\lsim\mhsm\lsim 200$~GeV).  The lower edge of 
            the bands is the calculated threshold; the bands extend upward 
            from these nominal thresholds by 30\% as an indication of the
            uncertainties in $b$-tagging efficiency, background rate,
            mass resolution, and other effects.}
  \label{f:combined-final}
\end{figure}

\vspace{0.2in}
{\bf Combined channel integrated luminosity thresholds} \\ \nopagebreak

Fig.~\ref{f:combined-final} shows the integrated luminosity required
to either exclude the SM Higgs boson at 95\% CL or discover it at the
$3\sigma$ or $5\sigma$ level of significance, as a function of Higgs
mass, for the SHW analyses with the neural net selection.  The
integrated luminosity displayed in the plot is the delivered
integrated luminosity {\it per} experiment; whereas the shaded bands shown are
the results obtained by combining the statistical power of {\it both}
experiments.  The required integrated luminosity thresholds for a
single experiment are very close to a factor of two higher than those
for the two combined experiments.  
The bands extend from the calculated threshold
on the low side upward in required integrated luminosity by 30\% to
the high side, as an indication of the range of uncertainty in the
various factors discussed above.
As the plots show, the required integrated luminosity increases
rapidly with Higgs mass to 140 GeV, beyond which the high-mass
channels play the dominant role.  At the upgraded Tevatron with
2~fb$^{-1}$ per detector, the 95\% CL limits will barely extend the
expected LEP2 limits, but with 10 fb$^{-1}$, the SM Higgs boson can be
excluded up to 180 GeV in the absence of a Higgs signal.

If a SM Higgs boson exists with mass less than 180 GeV, the combined
sensitivity of CDF and D\O\ will yield evidence for the Higgs boson at
the $3\sigma$ level with 20 fb$^{-1}$.  (The Higgs mass region around
140 GeV may require slightly more luminosity depending on the
magnitude of systematic errors due to uncertainties in $b$-tagging
efficiency, background rate, the $b\bar b$ mass resolution, {\it
etc.})  With 30~fb$^{-1}$ integrated luminosity delivered per
detector, a $5\sigma$ Higgs boson discovery may be possible for Higgs
masses up to about 130 GeV, a significant extension of the LEP2
Standard Model Higgs search.  The latter figure of merit is
particularly significant when applied to the search for the lightest
Higgs bosons of the minimal supersymmetric model.  We address this
case in detail in the following section.

  \subsection{Higgs Mass Reach in Low-Energy Supersymmetry}
\makeatletter
\def\@cite#1#2{{[{#1}]\if@tempswa\typeout
{IJCGA warning: optional citation argument
ignored: `#2'} \fi}}


\newcount\@tempcntc
\def\@citex[#1]#2{\if@filesw\immediate\write\@auxout{\string\citation{#2}}\fi
  \@tempcnta\z@\@tempcntb\m@ne\def\@citea{}\@cite{\@for\@citeb:=#2\do
    {\@ifundefined
       {b@\@citeb}{\@citeo\@tempcntb\m@ne\@citea\def\@citea{,}{\bf ?}\@warning
       {Citation `\@citeb' on page \thepage \space undefined}}%
    {\setbox\z@\hbox{\global\@tempcntc0\csname b@\@citeb\endcsname\relax}%
     \ifnum\@tempcntc=\z@ \@citeo\@tempcntb\m@ne
       \@citea\def\@citea{,}\hbox{\csname b@\@citeb\endcsname}%
     \else
      \advance\@tempcntb\@ne
      \ifnum\@tempcntb=\@tempcntc
      \else\advance\@tempcntb\m@ne\@citeo
      \@tempcnta\@tempcntc\@tempcntb\@tempcntc\fi\fi}}\@citeo}{#1}}
\def\@citeo{\ifnum\@tempcnta>\@tempcntb\else\@citea\def\@citea{,}%
  \ifnum\@tempcnta=\@tempcntb\the\@tempcnta\else
   {\advance\@tempcnta\@ne\ifnum\@tempcnta=\@tempcntb \else \def\@citea{--}\fi
    \advance\@tempcnta\m@ne\the\@tempcnta\@citea\the\@tempcntb}\fi\fi}
\makeatother

\def\mpl{M_{\rm PL}}
\def\beq{\begin{equation}}
\def\eeq{\end{equation}}
\def\beqa{\begin{eqalignno}}

\def\eeqa{\end{eqalignno}}
\def\beqno{\begin{eqalignno}}
\def\eeqno{\end{eqalignno}}
\def\ifmath#1{\relax\ifmmode #1\else $#1$\fi}
\def\half{\ifmath{{\textstyle{1 \over 2}}}}
\def\lsim{\mathrel{\raise.3ex\hbox{$<$\kern-.75em\lower1ex\hbox{$\sim$}}}}
\def\gsim{\mathrel{\raise.3ex\hbox{$>$\kern-.75em\lower1ex\hbox{$\sim$}}}}
\def\eq#1{eq.~(\ref{#1})}
\def\fig#1{fig.~\ref{#1}}
\def\Fig#1{Fig.~\ref{#1}}
\def\figs#1#2{figs.~\ref{#1} and \ref{#2}}
\def\Figs#1#2{Figs.~\ref{#1} and \ref{#2}}
\def\Ref#1{ref.~\cite{#1}}
\def\Refs#1#2{refs.~\cite{#1} and \cite{#2}}
\def\Rref#1{Ref.~\cite{#1}}
\def\Rrefs#1#2{Refs.~\cite{#1} and \cite{#2}}
\def\eqs#1#2{eqs.~(\ref{#1})--(\ref{#2})}
\def\Eq#1{Eq.~(\ref{#1})}
\def\Eqs#1#2{Eqs.~(\ref{#1})--(\ref{#2})}
\def\eqns#1#2{eqs.~(\ref{#1}) and (\ref{#2})}
\def\tanb{\tan\beta}
\def\sinb{\sin\beta}
\def\cosb{\cos\beta}
\def\sina{\sin\alpha}
\def\cosa{\cos\alpha}
\def\sinbma{\sin(\beta-\alpha)}
\def\cosbma{\cos(\beta-\alpha)}
\def\hsm{h_{\rm SM}}
\def\mhsm{m_{h_{\rm SM}}}
\def\hl{h}
\def\ha{A}
\def\hh{H}
\def\hpm{H^\pm}
\def\mha{m_{\ha}}
\def\mhl{m_{\hl}}
\def\mhh{m_{\hh}}
\def\mhpm{m_{\hpm}}
\def\mphi{m_\phi}
\def\mhmax{m_h^{\rm max}}
\def\mz{m_Z}
\def\mw{m_W}
\def\mww{m_W^2}
\def\mzz{m_Z^2}
\def\mt{m_t}
\def\mb{m_b}
\def\mstopa{M_{\widetilde t_1}}
\def\mstopb{M_{\widetilde t_2}}
\def\msusy{M_{\rm S}}
\def\msusyy{M_{\rm S}^2}
\def\mgut{M_{\rm X}}
\def\ie{{\it i.e.}}
\def\eg{{\it e.g.}}
\def\etc{{\it etc.}}
\def\vs{{\it vs.}}
\def\opcit{{\it op.~cit.}}
\def\SM{Standard Model}
\def\phm{\phantom{-}}
\def\rexp{$R_{\rm exp}$}
\def\rth{$R_{\rm th}$}
\def\ls#1{\ifmath{_{\lower1.5pt\hbox{$\scriptstyle #1$}}}}
\vspace{0.1in}
\small
\begin{center}
{\it M. Carena, H.E. Haber, S. Mrenna and C.E.M. Wagner}
\end{center}
\normalsize\nopagebreak

\def\tightenlines{\def\baselinestretch{2}\small\normalsize}

\subsubsection{Reinterpretation of the Standard Model Analysis in
Extended Higgs Sector Theories} 

The Standard Model Higgs boson search simulations presented in this
report can be interpreted more generally in the framework of an
extended Higgs sector.  Generically, extended Higgs sectors contain
additional scalar states beyond $\hsm$ with different couplings from
those found in the Standard Model.  The Higgs boson search strategies
of Sections~II.C and II.D can be applied to a CP-even neutral Higgs
boson, and those of Section~II.F to both CP-even and CP-odd neutral Higgs
bosons.  If CP is not conserved, then the neutral scalar states can be
considered to be mixtures of CP-even and CP-odd scalars, and the
discussion that follows must be modified accordingly
\cite{cpcarlos,cpcarlos2}.  Charged Higgs
boson states may also be present; the corresponding search strategies
were discussed in Section~II.G.  Finally, signatures not considered in
this report are also possible, such as the associated production of
CP-even and CP-odd neutral scalars and Higgs boson decays into $\tau$ leptons.
These additional signatures deserve a closer examination and might
improve the Higgs discovery reach beyond the results given in this section.

Before considering specific extensions of the Higgs sector, it is
useful to express the Standard Model results in a more  
model-independent fashion.  We focus on the analyses based on
Higgs processes in
which the production and decay depend primarily on a single Higgs
boson coupling.  To simplify the presentation, several
fixed values of integrated luminosity ($1,2,5,10,20,30$~fb$^{-1}$)
were chosen as well as two statistical measures (95\% Confidence Level
exclusion, 5$\sigma$ discovery) for the significance of a Higgs boson
signal.

As a function of the Higgs boson mass, the experimental
sensitivity can be described by the ratio of the chosen level of
statistical significance and the actual statistical significance for a
fixed value of integrated luminosity.  This ratio is denoted by \rexp\
and depends on the assumed integrated luminosity,\footnote{Following
the conventions of this report, the integrated luminosity is taken to
be {\it per experiment}, corresponding to the total
luminosity delivered to the collider.  All plots shown in this
section assume that equal data sets from the CDF and D\O\ detectors are
combined.} 
the chosen statistical measure and the Higgs boson mass.
For example, if the discovery criterion is $5\sigma$, and the expected
significance of a Standard Model Higgs boson with mass 125 GeV is
$3.3\sigma$ with 10~fb$^{-1}$ of data, then \rexp$=1.5$ at 125 GeV.
This means that a 125 GeV Higgs boson from some extended Higgs sector
could be discovered with 10~fb$^{-1}$ of data if the production cross
section times branching ratio were 1.5 times the Standard Model value.
Alternatively, $(1.5)^2=2.25$ times more data is required for a
$5\sigma$ discovery of a Standard Model Higgs boson at this mass 
(in the limit of Gaussian statistics, for a background  
limited measurement, when the significance scales
as the square-root of integrated luminosity).

Once the \rexp-values are obtained, they can be applied to a particular
model of an extended Higgs sector.  For a given extended Higgs model,
a {\it theoretical} value \rth~at a fixed $\mphi$ 
can be computed for the set of
parameters that define the model.  {The Higgs mass, 
$\mphi$, may be one of the defining input parameters of the extended
Higgs model.}  
To be more precise, at a fixed Higgs boson mass,
for a given Higgs boson production process
$X$ ({\it e.g.}, $q\bar q' \to W\phi$ or $gg\to\phi$)
with the subsequent decay $\phi\to Y$, we define
\beq \label{ratiodef}
R_{\rm th}(X\,;\,Y)\equiv {\sigma_X(\phi) {\rm BR}(\phi\to Y)\over
\sigma_X(\hsm){\rm BR}(\hsm\to Y)}\,,
\eeq
where the numerator corresponds to the Higgs cross section times
branching ratio predicted in the extended Higgs sector model, and the
denominator corresponds to the same quantity in the Standard Model.
If the theoretical value \rth~lies below \rexp~for a particular $\mphi$,
then there is no experimental
sensitivity to this point of the model parameter space for
the assumed luminosity and statistical significance.
Clearly, if \rth=1, then the Higgs mass reach is precisely that of the
Standard Model, whereas if \rth$>1$ [\rth$<1$] the corresponding Higgs boson mass
reach is increased [decreased] with respect to that of the Standard Model.


Consider first the
production of a neutral scalar, $\phi$,
in association with a vector boson ($V=W$ or $Z$), where $\phi\to
b\bar b$.  For the Tevatron Standard Model Higgs search, 
this mechanism is relevant
in the mass range $90~{\rm GeV}\lsim\mphi\lsim 130$~GeV.  In this case, the
Higgs search simulations of Sections~II.C.1--4 apply.  
The resulting \rexp-curves corresponding to a 95$\%$ CL exclusion
and discovery at the 5$\sigma$ level are shown 
in \fig{rplot95}.
\begin{figure}[ht]
\centering
\centerline{\psfig{file=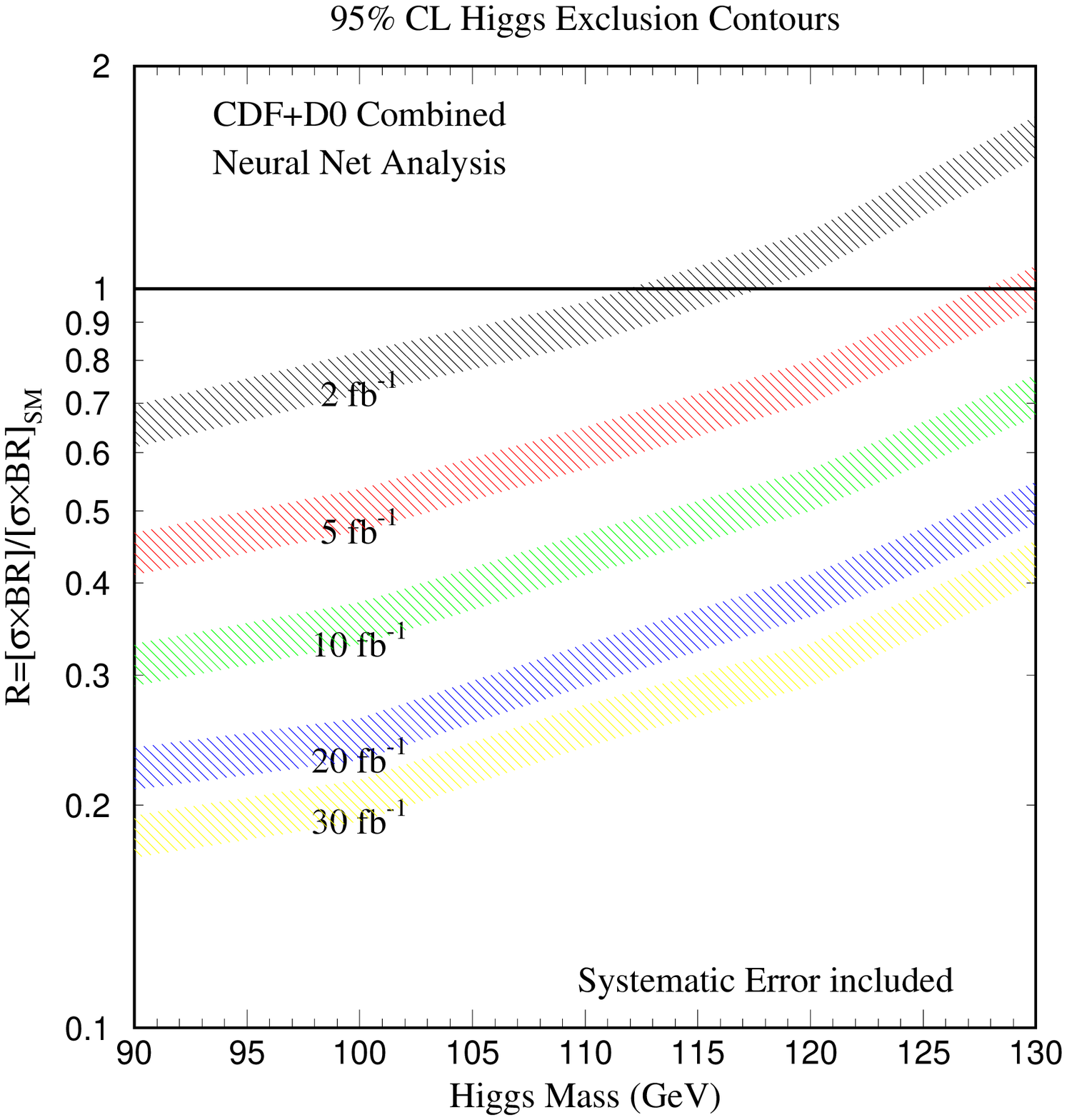,width=9cm,angle=0}
\hfill 
\psfig{file=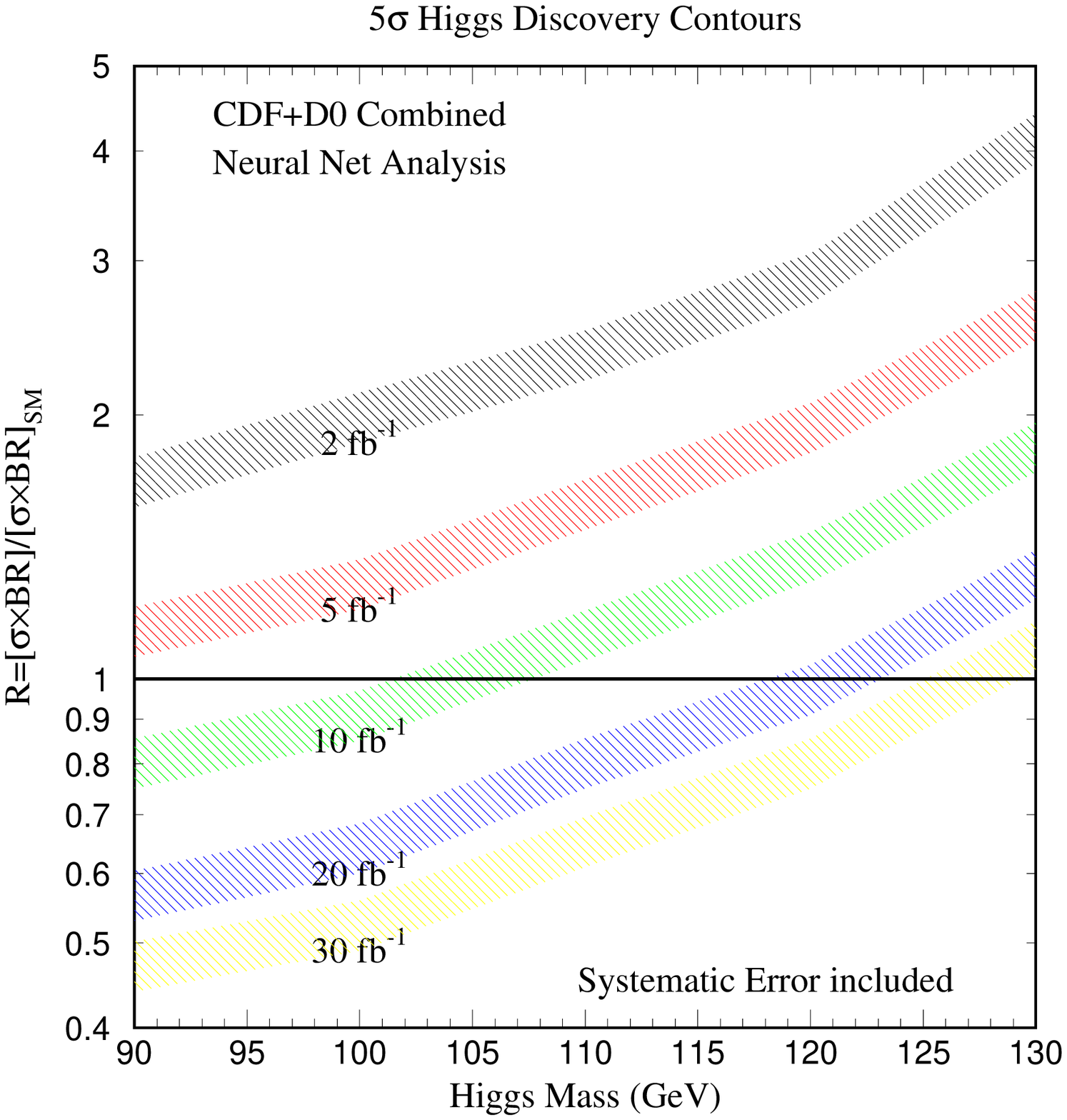,width=9cm,angle=0}}
\vskip2pc
\caption[0]{\label{rplot95} The ratio \rexp
$\equiv \sigma(q\bar q'\to V\phi){\rm BR}(\phi\to b\bar b)/\sigma(q\bar q'\to V\hsm){\rm BR}(\hsm\to b\bar b)$ 
as a function of Higgs boson mass in GeV.
Each shaded region represents the minimal value of \rexp~necessary at
a given Higgs boson
mass to (a)~exclude a Higgs boson signal at $95\%$ CL or (b)~discover
a Higgs boson at the $5\sigma$ level, 
for the integrated luminosity per detector as
indicated (combining the statistical power of both experiments). 
As in \fig{f:combined-final}, the thickness of the bands,
extending upwards from
the estimated thresholds, indicate the uncertainties
in $b$-tagging efficiency, background, mass resolution, and other
effects.}

\end{figure}
The cross section is
governed by the $VV\phi$ coupling $g_{VV\phi}$,\footnote{It is assumed 
that the ratio of $\phi W^+ W^-$ and $\phi ZZ$ couplings is
the same as in the Standard Model.  This assumption is correct in
multi-doublet Higgs models, but can be violated in models with higher
Higgs representations \cite{wudka}.}  
while the rate for the observed final state  
($b\bar b V$, where $V$ decays leptonically) depends on
the branching ratio of $\phi\to b\bar b$.  The relevant 
\rth-curves are obtained by taking $X=q\bar q'\to V\phi$ and $Y=b\bar b$ in
\eq{ratiodef}:
\beq 
R_{\rm th}(q\bar q'\to V\phi\,;\,b\bar b)\equiv 
\left({g^2_{VV\phi} \over g^2_{VV\hsm}}\right)
\left[{{\rm BR}(\phi\to b\bar b)\over {\rm BR}(\hsm\to b\bar b)}\right]\,.
\eeq
These curves will be used in the next section to examine the Higgs
mass reach of the upgraded Tevatron in the framework of the MSSM.

In the Higgs boson mass range of $130~{\rm GeV}\lsim\mphi\lsim 200$~GeV,
the main Tevatron discovery signature for a Standard Model Higgs boson 
comes from two channels: $(i)$ $gg\to\phi\to WW^*$ (where both $W$'s decay
leptonically) and $(ii)$ $q\bar q'\to V\phi\to VWW^*$ [$V=W^\pm$ or
$Z$] (where two of the three vector bosons decay leptonically and the
third decays hadronically; the same-sign dilepton final states
provide the largest statistical weight).   The two production mechanisms
depend upon different Higgs boson couplings and are treated separately.
The simulations of Section~II.D are used to obtain the separate
\rexp-curves shown in \figs{ratwwi}{ratwwii}.
Separate theoretical \rth-curves are defined for cases $(i)$ and $(ii)$: 
\beqa
&\phantom{i}(i)\qquad \phantom{V}R_{\rm th}(gg\to\phi\,;\,WW^*)\, \equiv \,  
\left({\Gamma(\phi\to gg) \over \Gamma(\hsm\to gg)}\right)
\left[
{{\rm BR}(\phi\to WW^*)\over{\rm BR}(\hsm\to WW^*)}\right]\,, \nonumber \\
&(ii) \qquad R_{\rm th}(q\bar q'\to V\phi\,;\,WW^*) \,\equiv \, 
\left({g^2_{VV\phi} \over g^2_{VV\hsm}}\right)
\left[{{\rm BR}(\phi\to WW^*)\over{\rm BR}(\hsm\to WW^*)}\right]\,.
\label{ratiowwdef}
\eeqa

\begin{figure}
\centering
\centerline{\psfig{file=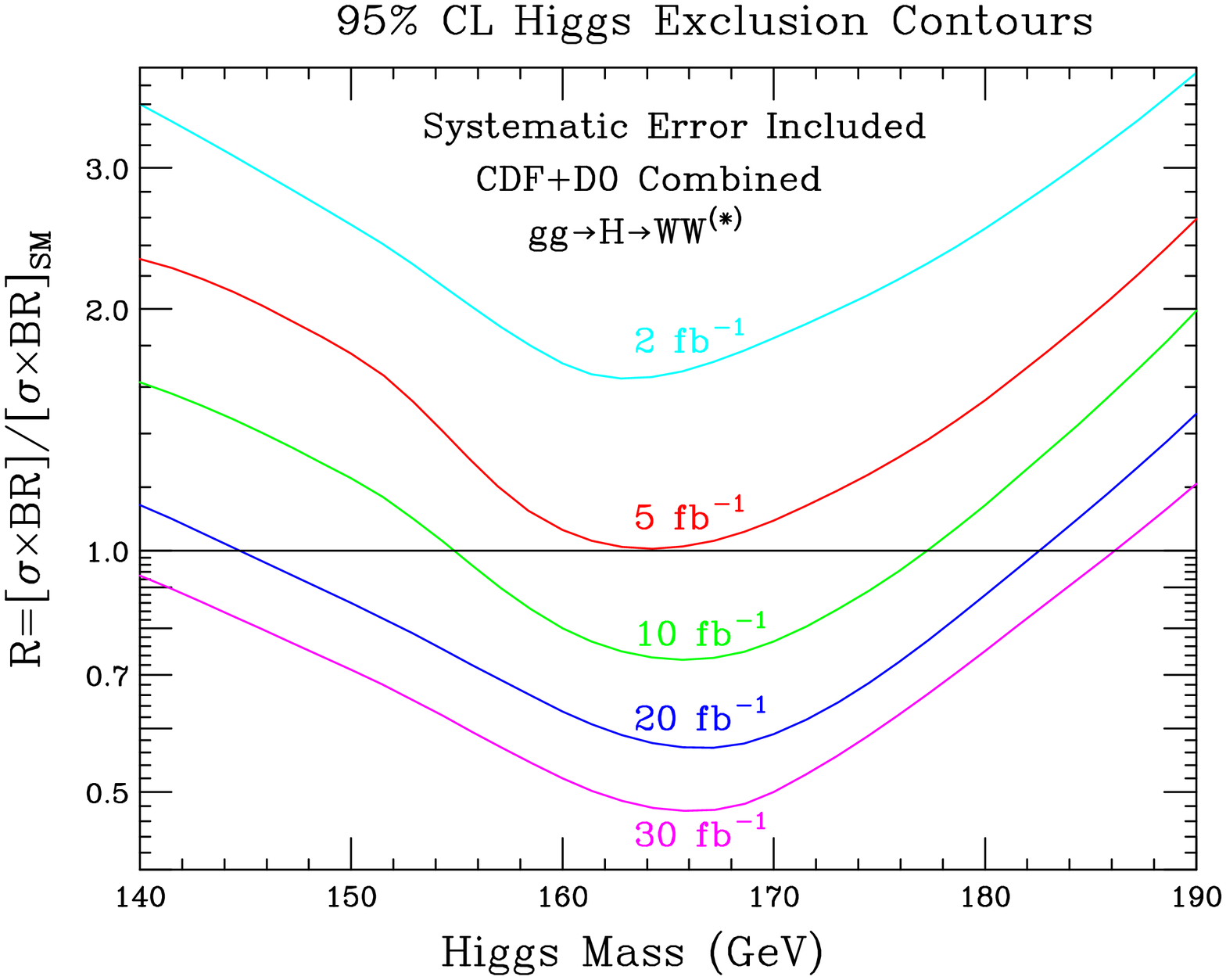,width=8cm,angle=90}
\hfill
\psfig{file=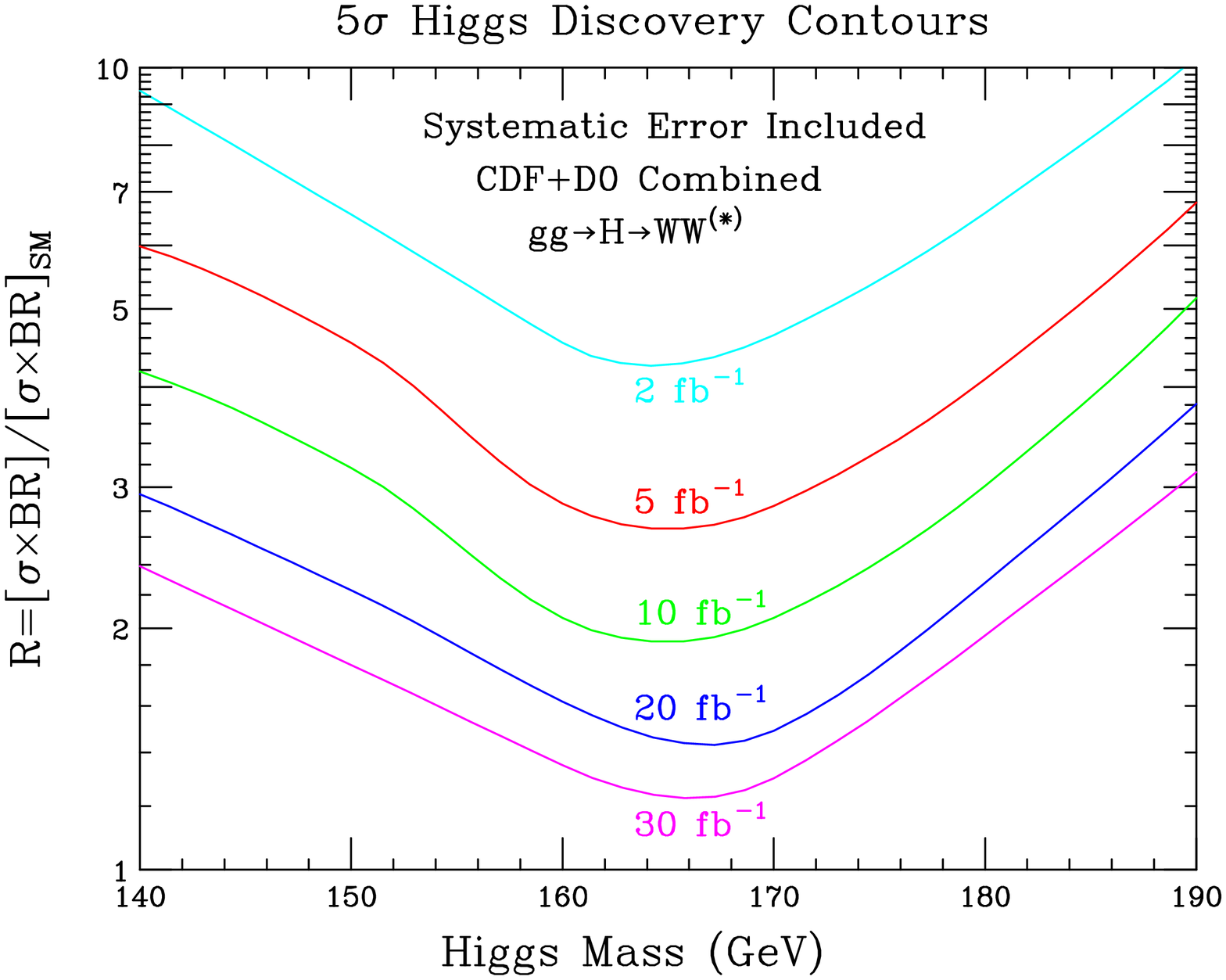,width=8cm,angle=90}}
\vskip2pc
\caption[0]{\label{ratwwi} The ratio \rexp$\equiv 
\sigma(gg\to \phi){\rm BR}(\phi\to WW^*)/\sigma(gg\to \hsm){\rm
BR}(\hsm\to WW^*)$ as a function of Higgs boson mass in GeV
necessary for
(a) excluding a Higgs boson signal at $95\%$ CL or (b) discovering
a Higgs boson at the $5\sigma$ level, 
for the integrated luminosity per detector as
indicated (combining the statistical power of both experiments). }
\end{figure}

\begin{figure}
\centering
\centerline{\psfig{file=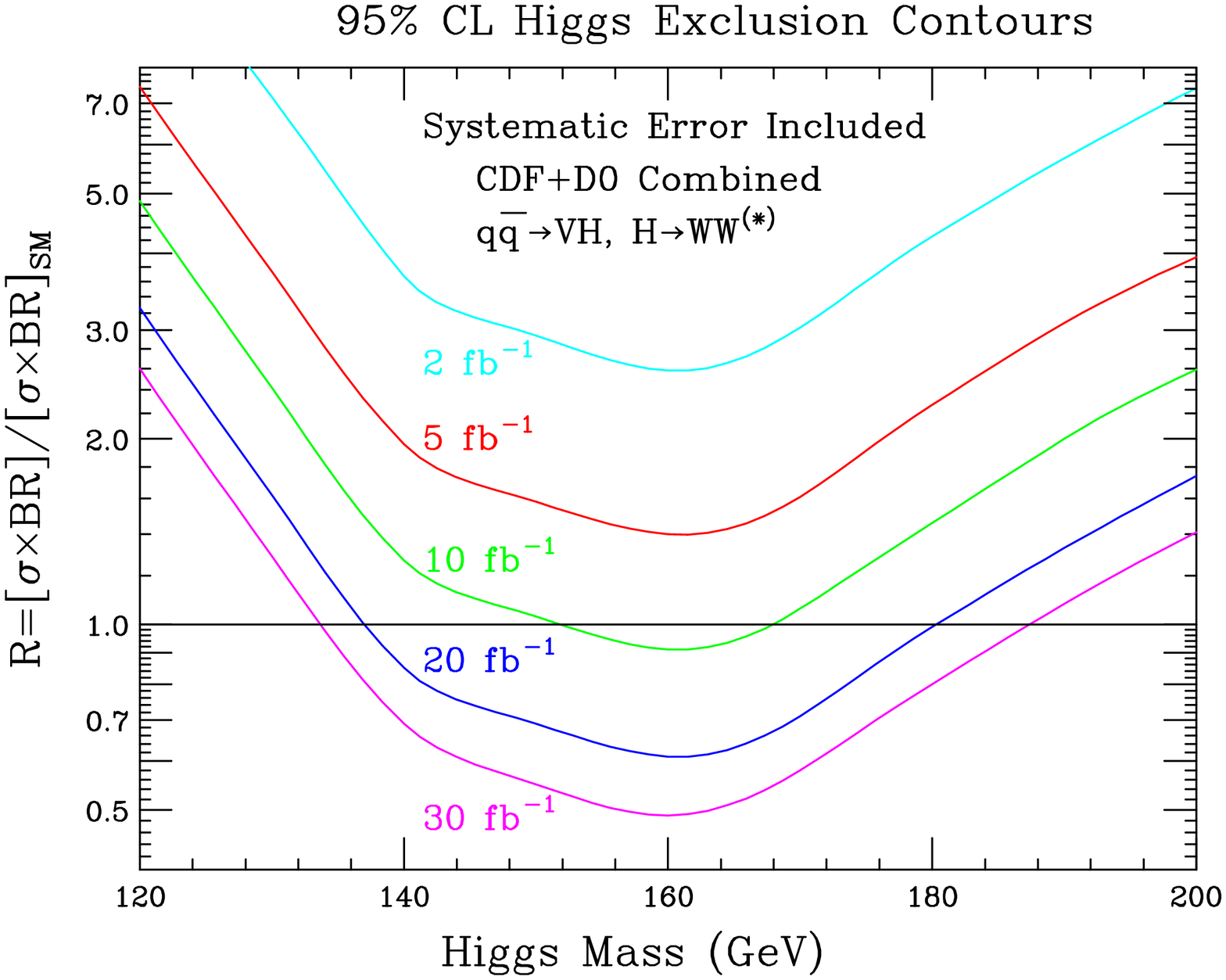,width=8cm,angle=90}
\hfill
\psfig{file=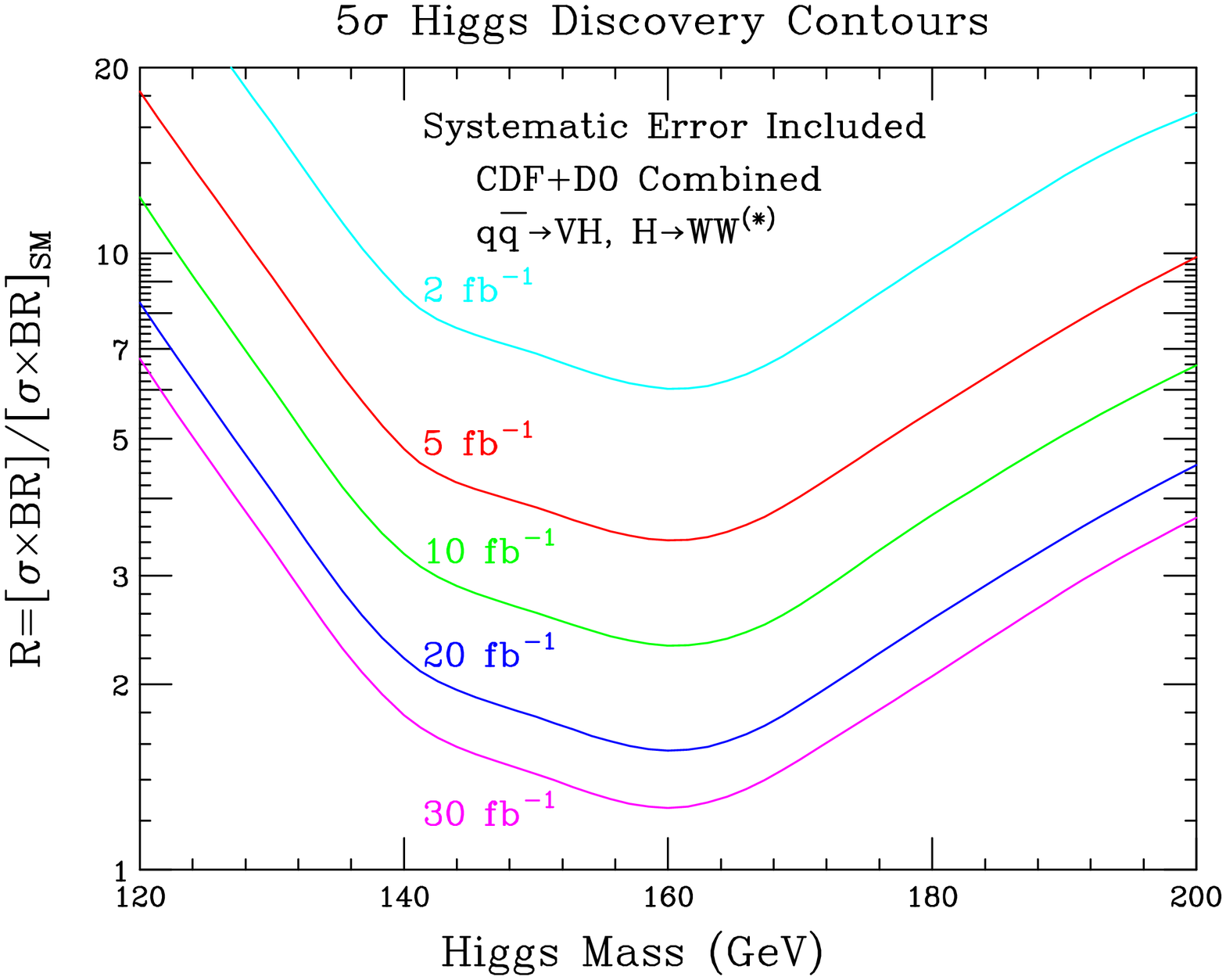,width=8cm,angle=90}}
\vskip2pc
\caption[0]{\label{ratwwii} Same as \fig{ratwwi}, but for \rexp$\equiv 
\sigma(q\bar q' \to V\phi){\rm BR}(\phi\to WW^*)/\sigma(q\bar q'\to V\hsm){\rm
BR}(\hsm\to WW^*)$.}
\end{figure}

Clearly, a combination of both channels $(i)$ and $(ii)$ yields
the maximal sensitivity for Higgs discovery reach (or exclusion).
However, this must be done on a model-by-model basis,
since different couplings may be involved in the calculation of
the two \rth~values in any given extension of the Standard Model.
Note that in the MSSM, the $VV\phi$ coupling is
significantly suppressed for $\mphi>130$~GeV (in this mass range, 
$\phi$ is the heavier CP-even Higgs boson,
$\hh$).  Thus, in the MSSM, the predicted values of \rth$(i)$ and
\rth$(ii)$ are substantially smaller than 1.\footnote{$\sigma(gg\to\phi)$ 
is suppressed [enhanced] relative to the Standard
Model for moderate [large] values of $\tan\beta$.  But, at large
$\tan\beta$,  the suppression of ${\rm BR}(\phi\to WW^*)$ 
is more significant than the enhancement of
the $gg\to\phi$ rate, resulting in \rth$(i)\ll 1$.} 
As a result, the high mass Tevatron Standard Model
Higgs search has no relevance for
the MSSM analysis, and it will not be employed in the next section.  
In models of extended Higgs sectors with exotic
Higgs representations, parameter ranges in which \rth$(i)$ and/or \rth$(ii)$
are larger than one can occur.  As an extreme example, 
models of non-minimal Higgs sectors have been constructed in which the
tree-level coupling of the Higgs bosons to fermions is absent (or
highly suppressed).  In such ``fermiophobic'' Higgs 
models \cite{fermiophobe}, the
Higgs coupling to gluons (which normally occurs at one-loop via
the fermion loop graph) is approximately zero and \rth$(i)$ vanishes.

The last search channel considered here is Higgs production 
via $gg$, $q\bar q\to b\bar
b\phi$, followed by the decay $\phi\to b\bar b$.  
The ratio \rth~is
defined using \eq{ratiodef} with $X=q\bar q,gg\to b\bar b\phi$
and $Y=b\bar b$.
Since $\sigma(gg,\,q\bar q\to b \bar b\phi)$ is proportional to the 
square of the $\phi b\bar b$ Yukawa coupling, $g_{\phi b\bar b}^2$, it
follows that:
\beq \label{rbbbar}
R_{\rm th}(q\bar q,gg\to b\bar b\phi\,;\, b\bar b)= 
\left( g_{\phi b\bar b}^2\over g_{\hsm b\bar b}^2 \right)
\left[{{\rm BR}(\phi\to b\bar b)\over {\rm BR}
(\hsm\to b\bar b)}\right]  \,.
\eeq
The resulting curves corresponding to a 95$\%$ CL Higgs mass exclusion
and Higgs discovery at the 5$\sigma$ level, based on the CDF and
D\O\ analyses
of  Section~II.F, are shown in \fig{rplotmaria}.
Note that the D\O\ analysis has greater sensitivity at large values of $\mphi$
than the CDF analysis, whereas the CDF analysis has greater sensitivity at
low values of $\mphi$.  The case of Standard Model Higgs boson
production, $b\bar b\hsm\to b\bar b b \bar b$, would correspond to 
\rexp$=1$ in \fig{rplotmaria}.  Clearly, neither analysis shows
any sensitivity to the production of a Standard
Model Higgs boson in the $b\bar b\phi\to b\bar b b\bar b$ channel for
any reasonable luminosity.
\begin{figure}[ht]
\centering
\centerline{\psfig{file=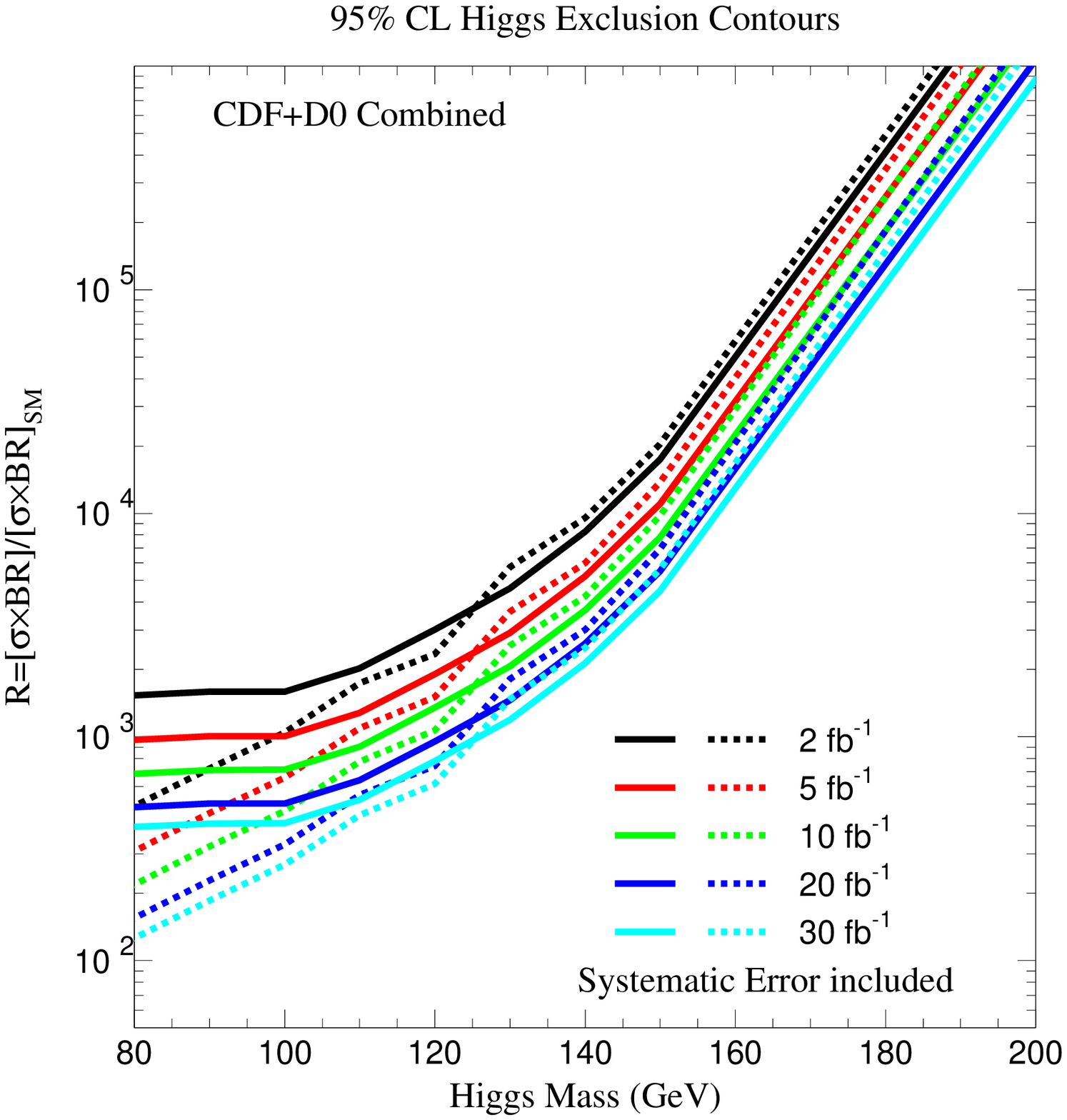,width=9cm,angle=0}
\psfig{file=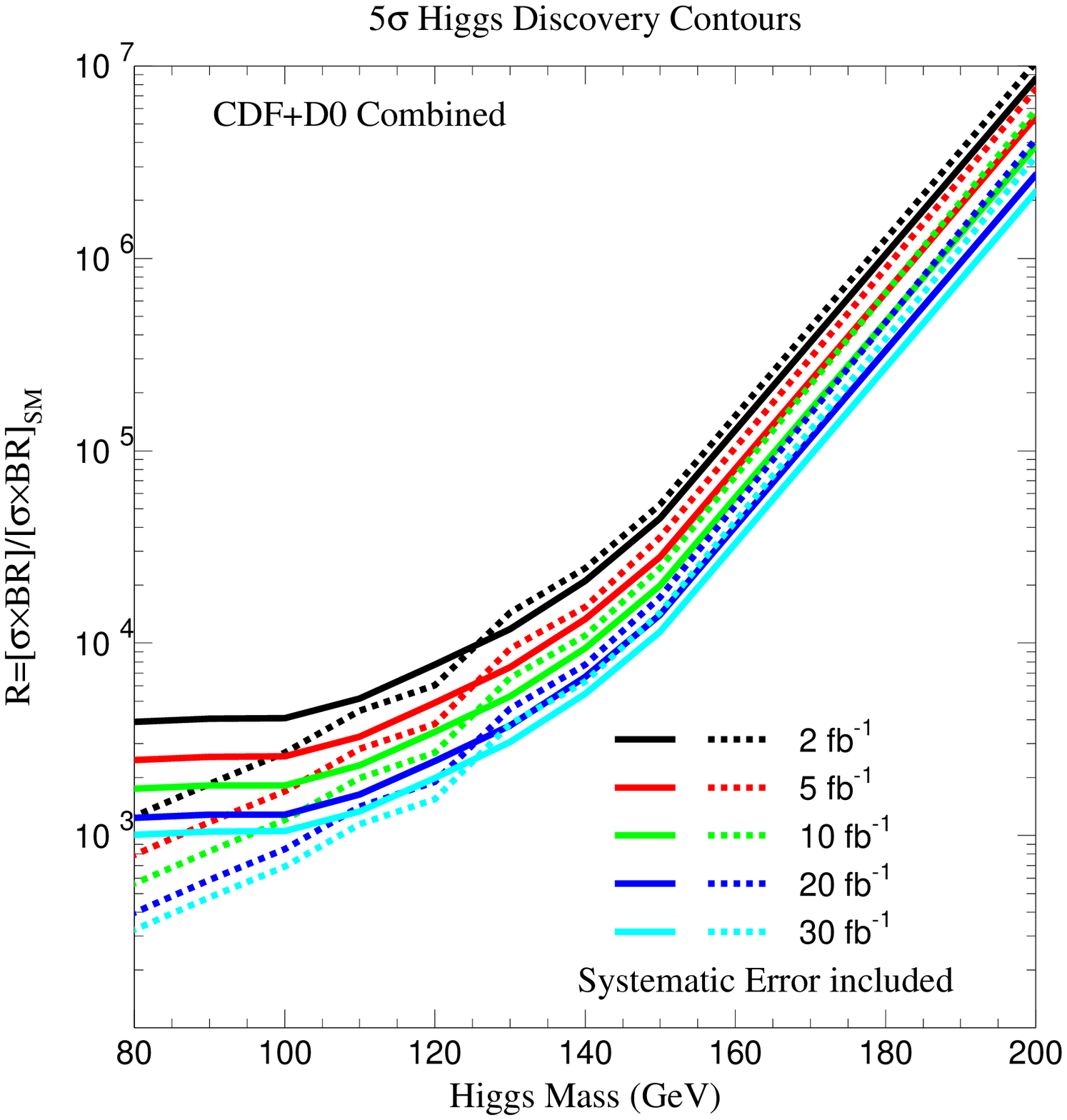,width=9cm,angle=0}}
\vskip2pc
\caption[0]{\label{rplotmaria} The same as \fig{ratwwi}, but for
\rexp$\equiv 
\sigma(gg,\,q\bar q\to b\bar b\phi){\rm BR}(\phi\to b\bar b)/
\sigma(gg,\,q\bar q\to \hsm){\rm BR}(\hsm\to b\bar b)$ and
for the CDF (dashed lines) and D\O\ (solid lines) analyses
described in Section~II.F.}

\end{figure}

Although the values of \rth~must be quite large in order to have any
sensitivity to the $b\bar b\phi\to b\bar b b\bar b$ channel, such
enhancements can readily occur in extensions of the
Standard Model.  The main limitation of the SM searches in this
channel, is that ${\rm BR}(\hsm\to b\bar b)$ is greatly
suppressed
once $\mhsm \gsim$ 135 GeV and decays to W or Z boson pairs become
important.
In the MSSM at large $\tan\beta$, \rth~can be significantly enhanced.
In Section I.C.1 [below
\eq{hpmqq}], we showed that the tree-level
$\phi b\bar b$ coupling is enhanced 
(relative to the Standard Model value) by a factor of $\tan\beta$
for either $\phi=\hl$, $\ha$ or $\phi=\hh$, $\ha$ (depending on the
value of $\mha$).  
Evaluating \eq{rbbbar} for the case where $\phi$ is one of the neutral Higgs
bosons with an enhanced coupling to $b\bar b$, and taking
${\rm BR}(\phi\to b\bar b)\simeq 1$, one finds~\footnote{The value of
\rth~can be modified at large $\tan\beta$ in some regions of
the MSSM parameter space.}
\beq
R_{\rm th}(q\bar q,gg\to b\bar b\phi\,;\, b\bar b)\simeq 
\left[{\rm BR}(\hsm\to b\bar b)\right]^{-1}
\tan^2\beta\,.
\eeq
Since $[{\rm BR}(\hsm\to b\bar b)]\ll 1$ for $\mphi\gsim 150$~GeV,
and sensible parameter regimes exist where $\tan\beta\gg 1$ is possible,
there is sensitivity to the $b\bar b\phi\to b\bar b b\bar b$ channel 
for an interesting range of Higgs masses and $\tan\beta$ for
integrated luminosities as low as 2${\rm~fb}^{-1}$.  This will be
discussed further in Section III.B.3.
Whereas the \rexp-curves in \fig{rplotmaria} are based on the production 
of a CP-even
neutral scalar, they also apply to the CP-odd Higgs boson,
because the corresponding cross-section and decay branching ratios
are very similar [see \figs{fg:2}{fg:5}].
Thus, one can reliably employ
the \rexp-curves obtained above to $b \bar b \ha$ production in the MSSM.

\subsubsection{MSSM Higgs Sector Results:
$q\bar q\to V\phi$ [$\phi=\hl$, $\hh$], $\phi\to b\bar b$}

The general MSSM parameter space involves many {\it a priori} unknown
parameters.  
In practice, only a small subset of these parameters govern the 
properties of the Higgs sector.
A full scan of this reduced subset is a formidable task.
However, a detailed study of a few appropriately chosen points
of the parameter space can help determine the ultimate MSSM Higgs
discovery reach of the upgraded Tevatron.

In order to select a representative set of parameters, it is useful to
study approximate analytic expressions for
the elements of the CP-even Higgs boson squared-mass matrix.
At tree-level, all MSSM Higgs masses and couplings are functions of
$\mha$ and $\tan\beta$.  As shown in Section~I.C.2 and I.C.3, the
radiative corrections to Higgs masses and couplings depend primarily 
on the third generation squark mass and mixing parameters 
(with a somewhat weaker dependence on the gluino mass).
In the ``generic'' MSSM analysis, we derive exclusion and discovery
regions in the $\mha$--$\tan \beta$ plane for various
benchmark values of these parameters.

In the generic MSSM analysis, all the relevant MSSM parameters that
govern the Higgs sector masses and couplings are regarded as
independent.   These parameters arise primarily from the 
supersymmetry-breaking sector of the theory.
However, there is no underlying theoretical model of
supersymmetry-breaking that yields precisely the set of benchmark
parameters that we have employed in the generic MSSM analysis.  An
alternative approach is to choose a specific model of
supersymmetry-breaking, and then derive the values of the
relevant supersymmetric parameters (as a function of more fundamental
model parameters).  We shall illustrate this procedure in the 
minimal supergravity (mSUGRA) model and in simple
models of gauge-mediated supersymmetry-breaking.  

\vskip6pt\noindent
\underline{a. Generic MSSM Analysis} \\[0.2cm]

The results of Sections~II.C.1--4, based on the SHW simulation improved by
neural network techniques, 
show that a Standard Model Higgs boson with $\mhsm\lsim
130$~GeV can be excluded at the 95\% CL [discovered
at the 5$\sigma$ level] 
for an integrated luminosity of 
5~fb$^{-1}$ [30~fb$^{-1}$].
In Section I.C.2, it was noted that there is an upper bound
to the value of the lightest CP-even neutral Higgs boson, $\hl$, of
the MSSM.  Assuming an average squark mass parameter of $M_S\lsim
2$~TeV,  then $\mhl\leq \mhmax\simeq 130$~GeV.
This suggests that a
significant region of the MSSM parameter space can be probed by the
Tevatron Higgs search with enough integrated luminosity.
This can be understood by
considering limiting cases of the MSSM in which some or all of the
properties of one of the neutral CP-even Higgs bosons approach those
of $\hsm$.  For example, the limit $\mha\gg m_Z$ corresponds to the
case in which the heavy CP-even, the CP-odd and the charged Higgs
 decouple from the low energy effective theory and $\hl\simeq\hsm$.  
In this
``decoupling'' region of MSSM parameter space, the Standard Model
simulations apply directly.  
Other
regions of MSSM parameter space exist in which $\hh$ behaves in part
like a Standard Model Higgs boson.  As discussed in Section I.C.2, in
the region of $\tan\beta\gg 1$ and $\mha<\mhmax$, one finds that
$\mhh\simeq \mhmax\lsim 130$~GeV and the coupling of $\hh$ to $VV$
pairs is nearly equal to the corresponding $\hsm$ coupling.  As
long as ${\rm BR}(\hh\to b\bar b)$ is not significantly suppressed,
the Standard Model Higgs simulation can be applied to $\hh$ in this
MSSM parameter regime.

In the rest of the MSSM parameter space, both $\hl$ and 
$\hh$ couplings to vector bosons and fermions differ from those of $\hsm$.
For the generic MSSM analysis, 
\rth$(q\bar q'\to V\phi\,;\,\phi\to b\bar b)$ 
[$\phi=\hl$, $\hh$] is computed as
a function of $\mha$ and $\tan\beta$.  Applying \eq{ratiodef} to the MSSM
yields
\beqa \label{rvhsusy}
R_{\rm th}(q\bar q'\to V\hl\,;\,\hl\to b\bar b) 
&= \left[\frac{{\rm BR}(\hl\to b\bar b)}{{\rm
BR}(\hsm\to b\bar b)}\right] \sin^2(\beta-\alpha)
\, , \nonumber \\[6pt]
R_{\rm th}(q\bar q'\to V\hh\,;\,\hh\to b\bar b) 
&= \left[\frac{{\rm BR}(\hh\to b\bar b)}{{\rm
BR}(\hsm\to b\bar b)}\right] \cos^2(\beta-\alpha)\,.
\eeqa
The above equations demonstrate the complementarity of the 
$V\hl \to V b\bar b$ and  $V\hh \to V b\bar b$
 channels in
the coverage of the $\mha$--$\tan \beta$ plane coming from 
the dependence on $\sin^2(\beta-\alpha)$ and $\cos^2(\beta-\alpha)$
 of the ratio of MSSM to SM
cross sections in each case.  The ratio of branching ratios has a more
complicated behavior as a function of the supersymmetric parameters.   
In a large region of the MSSM parameter space,\footnote{This is
especially true in the region of large $\tan\beta$  
where $g_{\phi b\bar b}$  [$\phi=\hl$ or $\hh$ depending on $\mha$]
may be enhanced relative to the Standard Model.} 
${\rm BR}(\phi\to b\bar b)/{\rm BR}(\hsm\to b\bar b)$ with  
[$\phi=\hl$, $\hh$] is 
slightly above 1 (over the Higgs mass range of interest).
Thus, \rth$(q\bar q'\to V\phi\,;\,\phi\to b\bar b)$ may be slightly above 1 if
$\sin^2(\beta-\alpha)\simeq 1$ [for $\phi=\hl$] or
$\sin^2(\beta-\alpha)\simeq 0$ [for $\phi=\hh$].
In contrast, in the region where $\sin^2(\beta-\alpha)\simeq
\cos^2(\beta-\alpha)\simeq 0.5$, 
the~\rth$(q\bar q'\to V\phi\,;\,\phi\to b\bar b)$ values for
both $\phi=\hl$ and $\hh$ are significantly below 1, which severely limits
the sensitivity of the Tevatron Higgs search.  Finally, 
small regions of MSSM parameter space exist (as discussed in
Section I.C.3), in which $g_{\phi b\bar b}$  
[$\phi=\hl$ or $\hh$] is strongly suppressed relative to the Standard Model,
even though $g_{VV\phi}$ is near the Standard Model value.
In these regions, \rth$(q\bar q'\to V\phi\,;\,\phi\to b\bar b)\ll 1$ 
and the $V\phi\to Vb\bar b$ channel is not a promising search mode.

Once the values of the squark
mass and mixing parameters are fixed, the two ratios
\rth$(q\bar q'\to V\phi\,;\,\phi\to b\bar b)$ [$\phi=\hl$ and $\hh$] can
be computed as a function of $\mha$ and $\tan\beta$.  If the
\rth~value is larger than \rexp~at the same $\mphi$ in
\fig{rplot95}(a) (for a given integrated luminosity), then this
point of parameter space can be excluded at the 95\% CL.  If
\rth$(q\bar q'\to V\phi\,;\,\phi\to b\bar b)$ is larger than \rexp~at
the same $\mphi$ in \fig{rplot95}(b), then $\phi$ can be discovered
at the 5$\sigma$ level.  Whereas in most of the $\mha$--$\tan \beta$
plane, either $h$ or $H$ is the relevant Higgs boson,
there are some regions of MSSM parameter space where 
where both Higgs bosons have comparable rates.
If $\mhh-\mhl\lsim 10$~GeV, then the individual \rth~values of the $\hl
V$ and $\hh V$ signals are added in quadrature before comparing to
\rexp.  This procedure is meant to crudely represent the fact that the
two signals will add constructively as long as the mass difference is
within the experimental resolution.\footnote{One would expect some
loss of significance, depending on the actual mass splitting, since some of
the signal will leak out of the {\it a priori} chosen mass window.}
Note that the near degeneracy of $\hl$ and $\hh$ occurs only in the
narrow region of MSSM parameter space when $\mha\simeq m_{h}^{max}$.

\begin{table}
\caption{\label{benchmarks}
Description of the values of the MSSM parameters for the benchmark scenarios 
considered in this report \protect\cite{benchmark2}.  
In all cases, the $\widetilde W$-gaugino (``wino'') mass parameter
has been taken to be $M_2=200$~GeV.}
\renewcommand{\arraystretch}{1.2}
\begin{tabular}{lcccccc} 
& \multicolumn{5}{c}{Mass parameters [TeV]} & (GeV) \\ \cline{2-6}
\noalign{\vskip3pt}
Benchmark   & $\mu$ &$X_t\equiv A_t-\mu\cot\beta$ & $A_b$ & $M_{SUSY}$ 
   & $M_{\tilde g}$\quad &  $m_h^{max}$ \\ [2pt]
\tableline
maximal mixing &  $-0.2$ & $\sqrt{6}$ & $A_t$ & 1.0 & 1.0 & 129
\\
no-mixing   &  $-0.2$ & $0$ & $A_t$ & 1.0 & 1.0 & 115 \\
suppressed $V \phi \to V b\bar b$ production
   &    1.5 & $-1.5 (1+\cot\beta)$  &0& 1.0 & 1.0 &  120 \\[3pt]
\end{tabular}
\end{table}

Table~\ref{benchmarks} defines the set of benchmark parameters for the
relevant supersymmetric parameters that govern the Higgs radiative
corrections.  These include the supersymmetric Higgs mass parameter
$\mu$, the third generation squark mixing parameters, $A_t$ and $A_b$,
the gluino mass $M_{\tilde g}$, and the diagonal
soft-supersymmetry-breaking third generation squark squared-masses
(which we take for simplicity to be degenerate and equal to $\MSUSY$).
The maximal mixing benchmark scenario, as the name suggests, is
defined as the one in which the squark mixing parameters are such that
they maximize the value of the lightest CP-even Higgs boson mass for
fixed values of $\mha$, $\tan \beta$ and $\MSUSY$.
In contrast, in the no-mixing scenario, the parameters $\mu$ and $A_t$
are chosen such that the combination $X_t\equiv A_t-\mu\cot\beta$
vanishes.  In this limit, the value of the lightest CP-even Higgs boson
is at its minimum for fixed values of 
$\mha$, $\tan \beta$ and $\MSUSY$.  
This corresponds to the minimal mixing case studied in Section~I.C.2.
[See \fig{mhxt} for the dependence of
$\mhl$ on the squark mixing parameter $X_t$.]

One might conclude that the maximal mixing scenario, in
which $\mhl$ can attain its maximal value for a given set of 
MSSM parameters, is the most conservative case for ascertaining the
Higgs mass discovery reach of the upgraded Tevatron.
However, a scan through the MSSM
parameters away from the maximal squark mixing point reveals
regions of parameter space in which the CP-even neutral Higgs boson with
Standard-Model-like couplings to $VV$ has suppressed couplings to
$b\bar b$. The benchmark scenario denoted by
``suppressed $V \phi \to V b\bar b$ production''  
in table~\ref{benchmarks} is an example of this behavior.  
The regions of strongly
suppressed ${\rm BR}(\phi\to b \bar b)$ correspond to a suppressed
$\hh b\bar b$ coupling at lower $\mha$ and a suppressed
$\hl b\bar b$ coupling at larger $\mha$.  In particular,
the suppression for large $\tan\beta$ extends to
relatively large values of $\mha$, indicating a delay in the onset of
the decoupling limit (where the $\hl b\bar b$ coupling approaches the
Standard Model value). 
This is the case when one of the Higgs bosons
becomes mostly $\Phi^0_u$, with no coupling to down-type quarks except those
induced by supersymmetry-breaking effects [see \eq{couplings}].  
Such a Higgs boson has SM--like couplings
to the $W,Z$ and $t$, but vanishing couplings to $b\bar b$ and
$\tau^+\tau^-$.  The necessary condition for this limit to
arise is explained in Section~I.C.3.  
In fact, in the suppressed $V\phi\to Vb\bar b$ benchmark 
scenario, {\it all} the Higgs couplings to $b\bar b$ are generally
suppressed, since $0<\Delta_b\ll 1$ and $\sin 2\alpha\simeq 0$ 
[see eqs.~(\ref{bhCP}), (\ref{barhb}) and (\ref{tildehb})].
From the analytic formulae, it can be deduced that $\mu A_t<0$ and 
large values of $|A_t|$, $|\mu|$ and $\tan\beta$ are needed.  


\begin{figure}[ht!]
\begin{center}
\centerline{\psfig{file=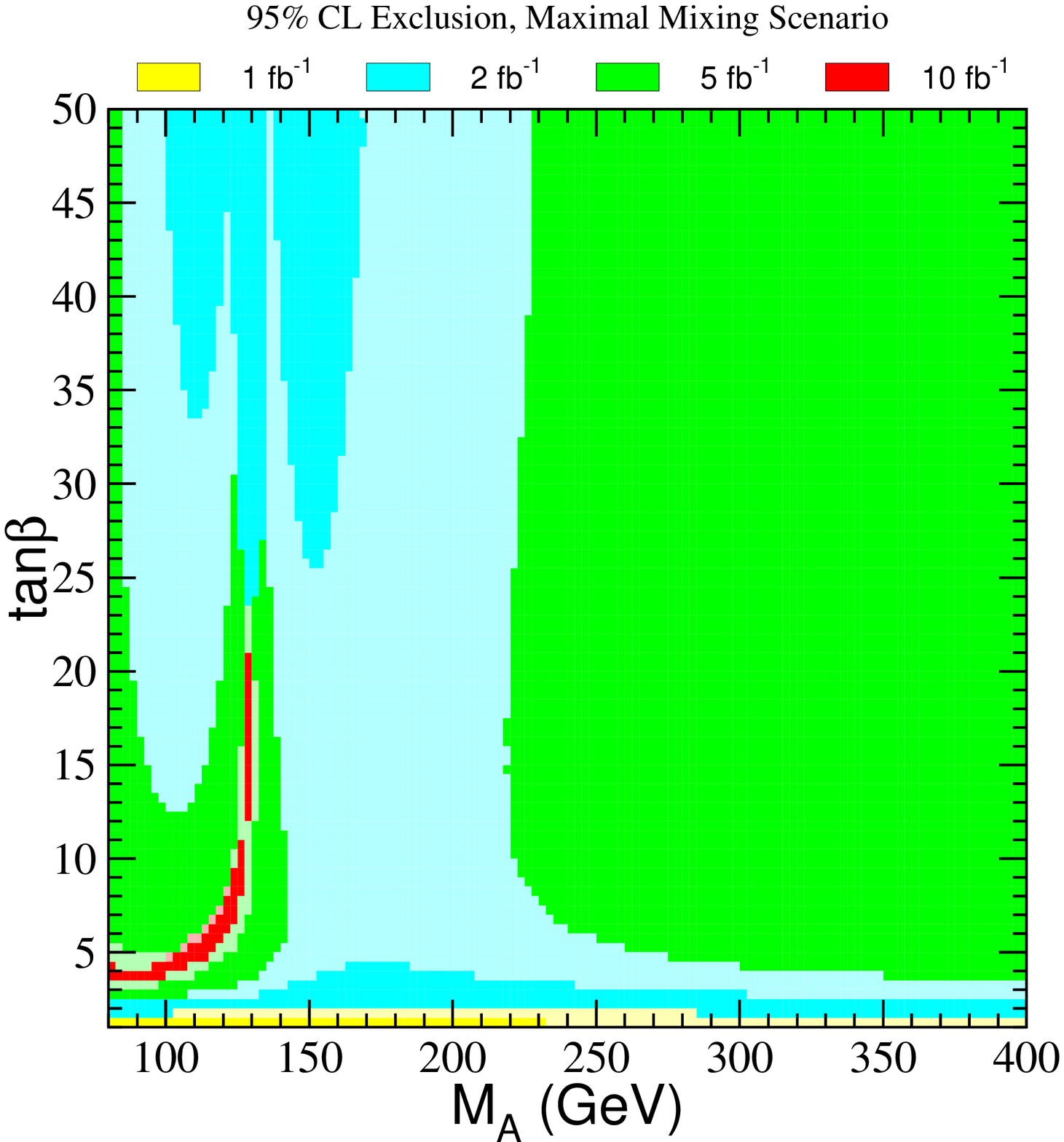,width=8cm,angle=0}
\hfill
\psfig{file=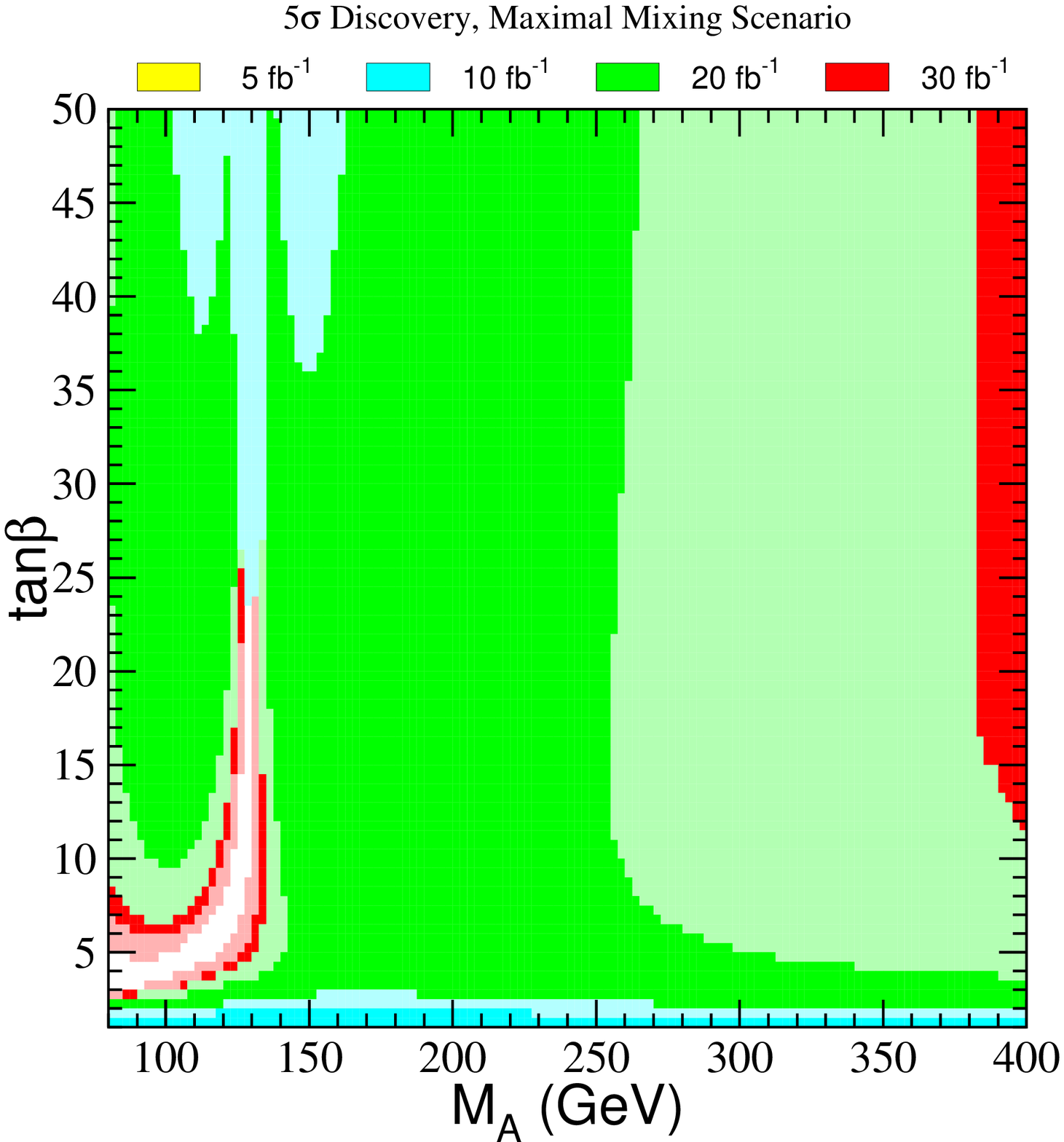,width=8cm,angle=0}}
\end{center}
\caption[0]{\label{amaxnn} Regions of MSSM parameter space, 
for the maximal mixing benchmark scenario of table~\ref{benchmarks},
that correspond to 
(a) the exclusion of a Higgs boson signal at $95\%$ CL or (b) the discovery
a Higgs boson at the $5\sigma$ level. 
As one increases the integrated luminosity,
the corresponding shaded areas successively cover the plane, as
indicated (combining the statistical power of both experiments),
based on the improvement of the SHW simulation employing neural
network techniques.  The darker shading of a given color corresponds
to  a degradation in the coverage of the MSSM parameter space due to 
the uncertainties in  $b$-tagging efficiency, background, mass
resolution and other effects.}
\end{figure}

\Fig{amaxnn} shows
contour plots demonstrating regions of 95$\%$ exclusion and 
$5\sigma$ discovery for
the maximal mixing benchmark scenario  
as defined in table~\ref{benchmarks}.
This illustrates more
generally the fact that the MSSM parameter space coverage 
in the $V\phi\to Vb\bar b$ channel depends
sensitively on the maximal value of the lightest Higgs boson 
mass for a particular set of MSSM parameters.
The darker shading of a given color illustrates how a net $30 \% $
reduction in the detection efficiency, due to   
the uncertainties in $b$-tagging efficiency, background, mass
resolution and other effects,  
 translates in a the coverage of the MSSM 
$\tan \beta$--$m_A$ plane.

A closer examination of \fig{amaxnn} shows three regions of the MSSM
parameter space that raise difficulties for the Tevatron 
$V\phi\to Vb\bar b$ search.  In the region of large $\tan\beta$ and
$\mha\lsim \mhmax$ (where $\mhh\simeq\mhmax$), 
the $\hl VV$ coupling is highly suppressed, 
so that $V\hh \to V b\bar b $ is the relevant process. 
%
%
For large $\mha$ and large $\tan\beta$ the incomplete parameter space
coverage is
simply due to the fact that the lightest Higgs has SM like couplings 
and its mass is at the maximal value.
Another problematical region of MSSM parameter space corresponds to
the case of $\mha\simeq\mhmax$ 
and moderate $\tan\beta$, where $\sin^2(\beta - \alpha) \simeq
\cos^2(\beta - \alpha) \simeq 0.5$. 
If $\tan\beta$ is sufficiently large,
the two CP--even Higgs bosons tend to have similar masses, and both
states contribute to the Higgs signal in the same mass bin.
However, a hole in the coverage of
the $\mha$ {\vs} $\tan\beta$ plane persists in a small region  
for $\tan\beta$ between 5 and 20, since in this case the
mass difference is large and $\sin^2(\beta-\alpha)\simeq 0.5$. 
In contrast, 
a signal can be observed with less integrated luminosity
in the region of $\mha \simeq 150$ GeV,
since $\sin^2(\beta - \alpha)\simeq 1$ 
and the $\hl b\bar b$ coupling
is strongly enhanced with respect to the Standard Model
case, implying an increase in the branching ratio of this Higgs boson
into bottom quarks.

\begin{figure}[hp!]
\begin{center}
\centerline{\psfig{file=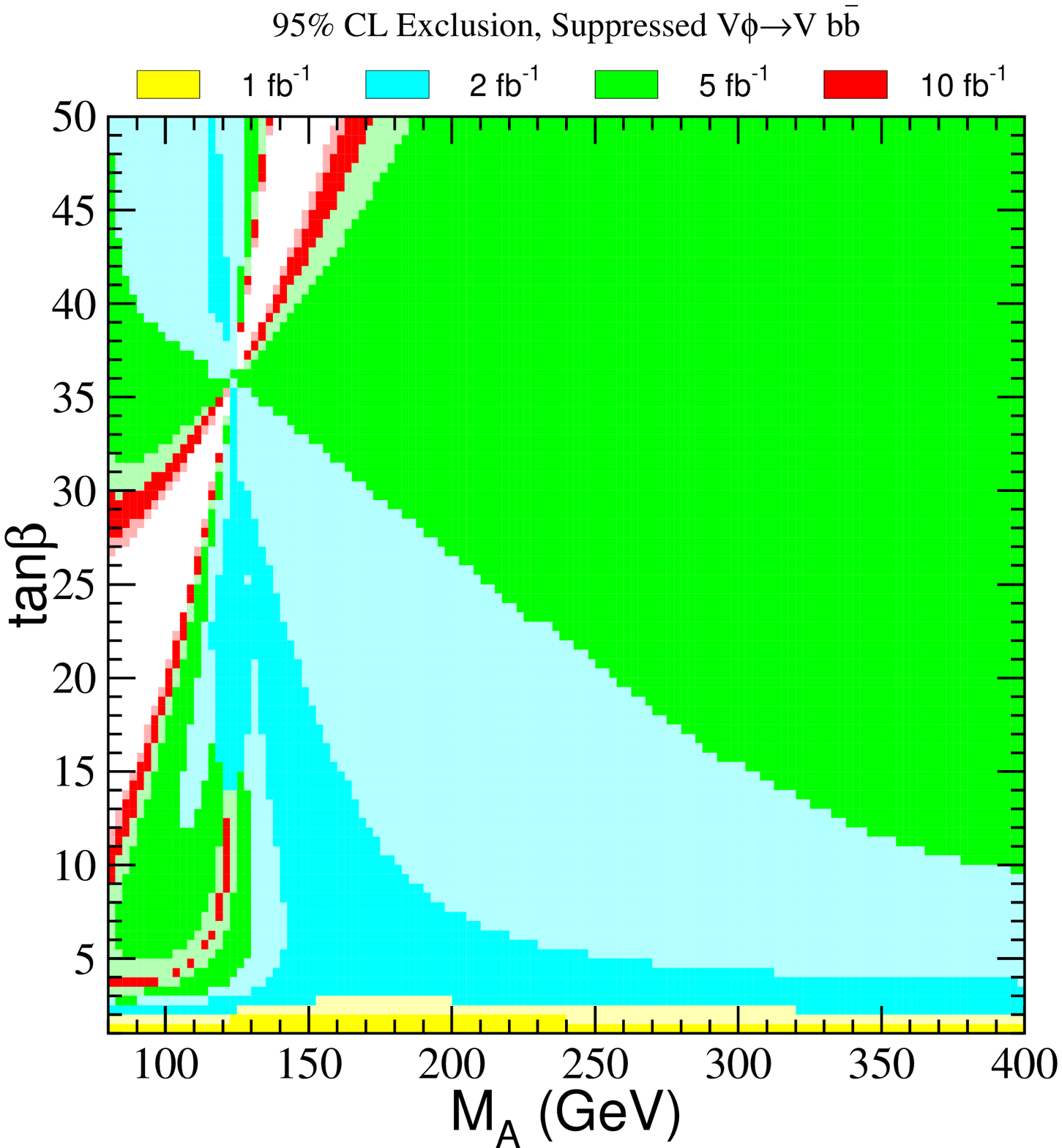,width=8cm,angle=0}
\hfill
\psfig{file=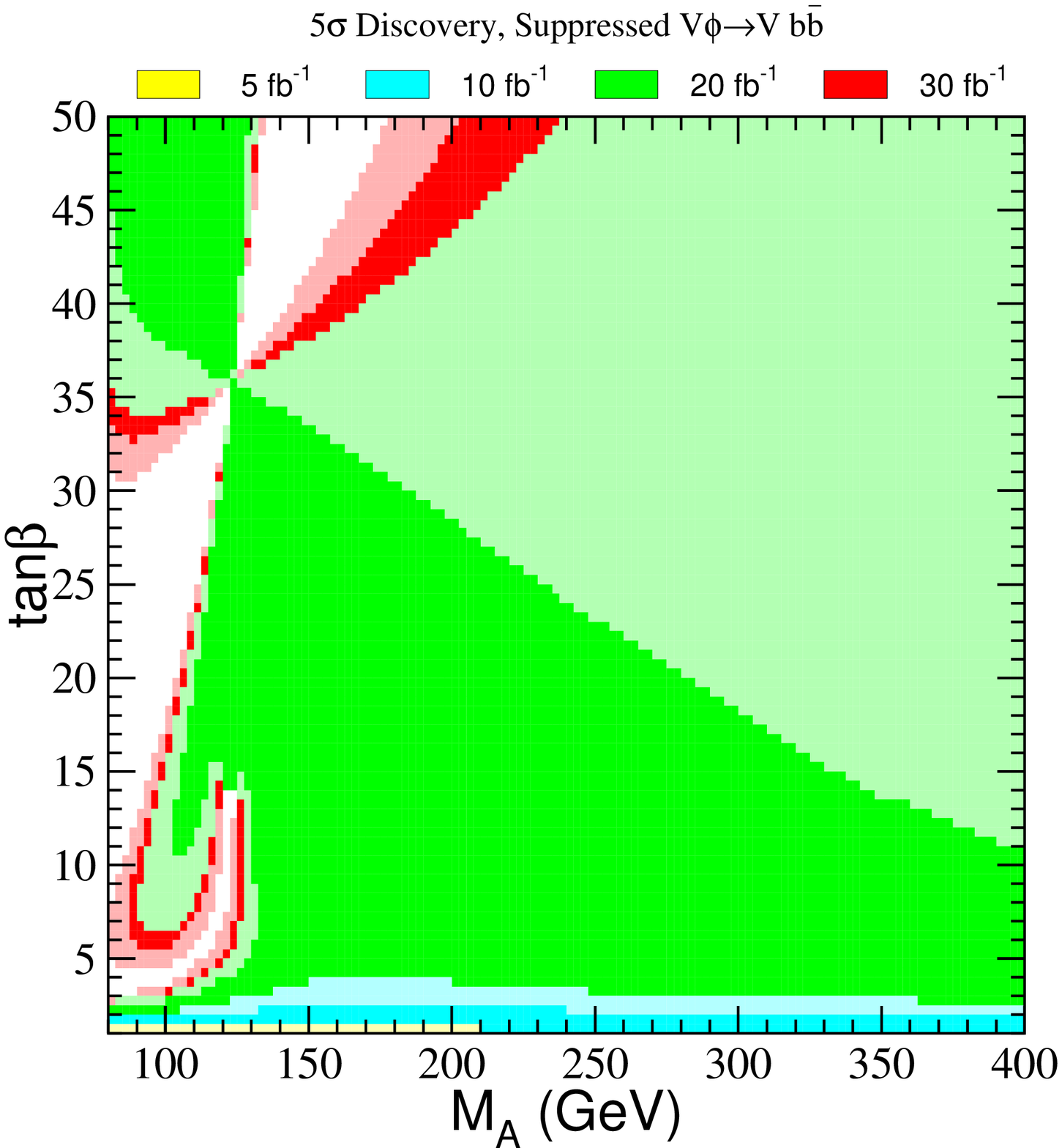,width=8cm,angle=0}}
\end{center}
\caption[0]{\label{holenn}  The same as \fig{amaxnn} but for the
``suppressed $V \phi \to V b\bar b$ production'' benchmark scenario of
table~\ref{benchmarks}.}
\end{figure}

\begin{figure}[htb]
\begin{center}
\centerline{\psfig{file=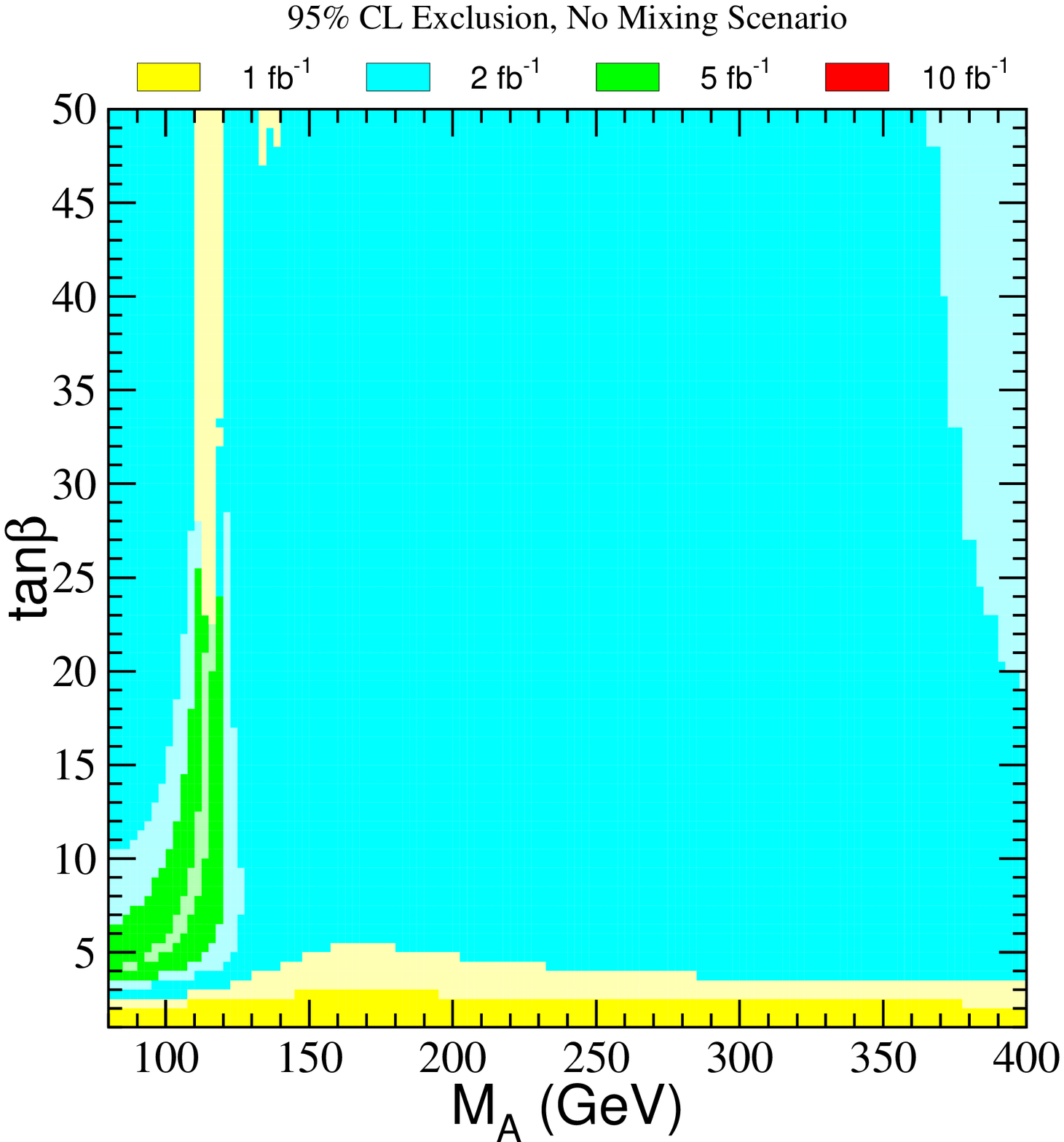,width=8cm,angle=0}
\hfill
\psfig{file=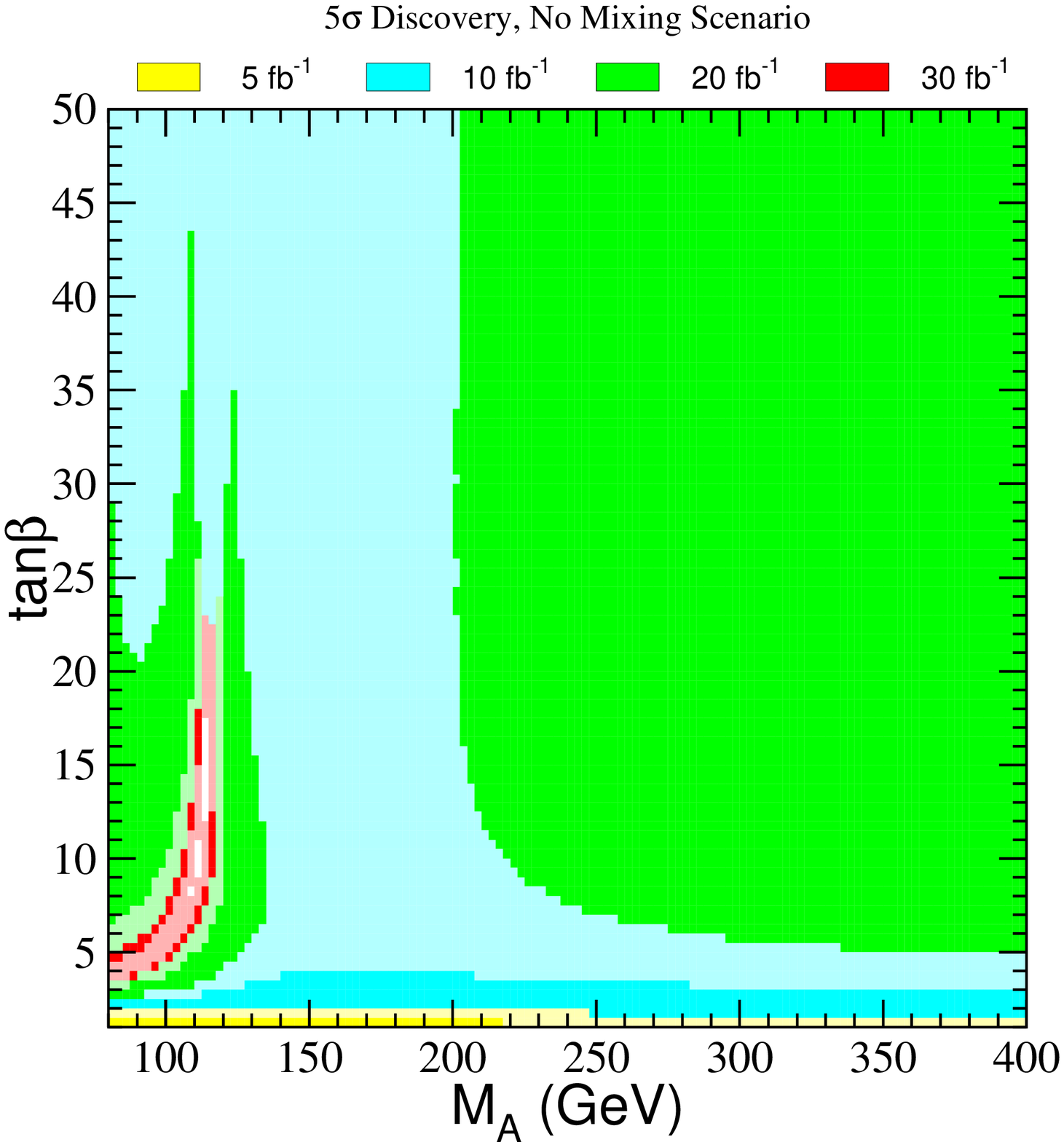,width=8cm,angle=0}}
\end{center}
\caption[0]{\label{aminnn}   The same as \fig{amaxnn} but for 
the no-mixing benchmark scenario of table~\ref{benchmarks}.}
\end{figure}

\Fig{holenn} illustrates the effect of the suppressed $\phi b\bar b$
couplings for the Tevatron MSSM Higgs search for the ``suppressed
$V\phi\to Vb\bar b$''
benchmark scenario of table~\ref{benchmarks}.  
The small
hole in the coverage of the $\mha$ {\vs} $\tan\beta$ plane for
$\tan\beta$ between 5 and 15 corresponds to the same difficult region of MSSM
parameter space noted in the maximal mixing case above.  However, in the
present case, two new larger holes open up corresponding to the
regions of strongly suppressed $\phi b\bar b$ coupling noted above.
Since in this case one has $A_t  \mu < $0, there is a suppression
of the  $\phi b\bar b$ coupling via a suppression of the radiatively
corrected value of $\sin \alpha$. On the other hand, although the
gluino mass and $\mu$ are large, the  vertex
corrections to the bottom Yukawa coupling have the effect of
suppressing the  $\phi b\bar b$ coupling, 
since $\mu > $0 and, hence, $\Delta_b$ is also positive. 
Again, as explained above, the darker shading of a given color 
shows the uncertainty in the discovery and exclusion potential 
due to uncertainties in the experimental simulations.

\Fig{aminnn} shows contours of $5\sigma$ discovery and 95\% CL exclusion
for the case of the ``no-mixing'' benchmark scenario defined in 
table~\ref{benchmarks}, which yields the smallest value for the
mass of the Higgs boson that has SM--like couplings to
the $W$ and $Z$ bosons as a function of the squark mixing effects.
This is an optimistic
scenario for the Tevatron, in which only modest values of integrated 
luminosity are necessary 
to cover a significant portion of the $\mha$--$\tan \beta$
plane.  Note, however, that 
the maximal mass for the lightest Higgs boson is about 115 GeV.
Much of this parameter regime will have been probed at LEP by the end of
its final run.

\break
\vskip6pt\noindent
\underline{b. Model-Dependent MSSM Analysis}

\vspace{0.1in}
\small
\begin{center}
{\it  H. Baer, B.W. Harris and X. Tata}
\end{center}
\normalsize\nopagebreak


The Higgs sector masses and couplings 
are determined from a set of
soft-supersymmetry-breaking parameters that include 
three gaugino mass parameters, scalar
squared-mass parameters for squarks, sleptons and Higgs bosons, 
the $A$ parameters that govern cubic scalar interactions, and 
the $B$ parameter that mixes the two Higgs scalars.
In addition, there is also dependence on the 
supersymmetric Higgsino mass parameter $\mu$.  The latter two
parameters can be exchanged for the two Higgs vacuum expectation
values, under the assumption that the electroweak symmetry is broken
in the usual manner.
The origin of the soft-supersymmetry-breaking parameters is not
presently known.  Nevertheless, one may consider specific models of
supersymmetry breaking, where the soft-supersymmetry-breaking
parameters are specified.  In such approaches, one often makes 
simplifying assumptions in order to reduce the total number of free
parameters.  Although such assumptions are ad-hoc, the resulting
models do provide examples of how the soft-supersymmetry-breaking
parameters could be correlated.  The results of such correlations
would constrain the Higgs sector of a generic MSSM.  In particular, the
discovery of the Higgs bosons of the MSSM would place interesting
constraints on the underlying supersymmetric parameters.

In the {\it minimal} supergravity (mSUGRA)
framework \cite{Nilles85,sugra,sugrawork}, the
supersymmetry breaking takes place in a hidden sector, and 
supersymmetry-breaking effects are communicated to the observable sector via
Planck-scale interactions. In the minimal rendition of this model
the soft-supersymmetry-breaking parameters take a particularly
simple form when evaluated at a very high energy scale (usually taken
to be either the Planck scale or the grand unification scale).
At this high-energy scale, all scalars have a common mass $m_0$,  
all gauginos have a common mass $m_{1/2}$, and all trilinear scalar couplings
unify to $A_0$.  The low-energy Higgs and supersymmetric parameters that
govern the observed Higgs and supersymmetric phenomenology
are obtained via renormalization group running of soft-supersymmetry-breaking
masses and couplings from the high-energy scale down to the
electroweak scale.  In a successful mSUGRA model, electroweak symmetry 
breaking is radiatively generated, since at least one of the Higgs
squared-masses is driven negative in the renormalization-group evolution.

One can count the number of independent parameters in the
mSUGRA framework.  In addition to usual Standard Model parameters
(excluding the Higgs mass), one must specify
$m_0$, $m_{1/2}$, $A_0$, and Planck-scale values for the $\mu$ and
$B$-parameters (denoted by $\mu_0$ and $B_0$).  In principle,
$A_0$, $B_0$ and $\mu_0$ can be complex, although in
the mSUGRA approach, these parameters are taken (arbitrarily)
to be real.  As previously noted,
renormalization group evolution is used to compute the low-energy
values of the mSUGRA parameters, which then fixes all the
parameters of the low-energy MSSM.  In particular, the two Higgs
vacuum expectation values (or equivalently, $m_Z$ and $\tan\beta$)
can be expressed as a function of the Planck-scale supergravity
parameters.  The simplest procedure is to express $m_Z$ and $\tan\beta$ 
in terms of the low-energy $\mu$ and $B$ parameters and the other
free mSUGRA parameters (the sign of $\mu$ is not fixed in this
process).  The resulting MSSM spectrum
and its interaction strengths are then determined by five parameters:
$m_0$, $A_0$, $m_{1/2}$, $\tan\beta$, and the
sign of $\mu$, in addition to the usual parameters of the Standard Model.

Recently, attention has been given to a class of supergravity models in
which the common tree-level gaugino mass is absent.  In such models,
one finds a model-independent contribution to the
gaugino mass \cite{anomalymed},
whose origin can be traced to the super-conformal
(super-Weyl) anomaly which is common to all supergravity models.
This approach has been called {\it anomaly-mediated} supersymmetry
breaking.  In the simplest scenario, all supersymmetric particle
masses are determined by one overall mass parameter 
(the gravitino mass, $m_{3/2}$).
However, anomaly-mediation cannot be the sole
source of supersymmetry-breaking in the MSSM (since in the simplest of such
models, slepton squared-masses are negative).  Moreover,
one still needs a model for the origin of the $\mu$ and $B$
parameters in order to generate electroweak symmetry breaking and a
realistic Higgs sector.  One simple approach is to add to the theory a
universal scalar mass parameter $m_0$ and $\mu$ and $B$ by hand.
The resulting model depends on four parameters: $m_{3/2}$, $m_0$,
$\tan\beta$ and ${\rm sign}(\mu)$.  A preliminary study of the Higgs boson
discovery reach of the upgraded Tevatron in this scenario has been
presented in \Ref{shufang}.  

In models of {\it gauge-mediated} super\-symmetry breaking \cite{dine},
super\-symmetry breaking is transmitted to the MSSM via gauge forces.
A typical structure of such models involves a hidden sector
where supersymmetry is broken, a ``messenger sector'' consisting of
particles (messengers) with SU(3)$\times$SU(2)$\times$U(1) quantum
numbers, and the visible sector consisting of the fields of the
MSSM \cite{gmsbreview}.
The direct coupling of the messengers to the hidden sector generates a
super\-symmetry breaking spectrum in the messenger sector.
Finally, super\-symmetry
breaking is transmitted to the MSSM via the virtual exchange of the
messengers.

In the gauge-mediated supersymmetry-breaking (GMSB) approach,
the low-energy scalar and gaugino mass parameters are generated
through loop-effects (while the resulting $A$-parameters are
suppressed).  In minimal GMSB models,
the supersymmetric particle masses are proportional to the parameter
$\Lambda = F/M$, where $\sqrt{F}$ is the supersymmetry-breaking
scale and $M$ the mass scale for the messenger particles.  The origin
of the $\mu$ and $B$-parameters is model-dependent and lies somewhat
outside the ansatz of gauge-mediated supersymmetry-breaking and must
be taken to be additional free parameters.  The minimal gauge-mediated
(mGMSB) models of this type are parameterized in terms
of $\Lambda$, $M$, $n_5$, $C_{\rm grav}$, $\mu$ and
$B$, where $n_5$ is the number of complete $SU(5)$ messenger
multiplets ($n_5
\leq 4$ if $M\lsim 1000$~TeV to avoid the presence of Landau poles
below $M_{GUT}$) and 
$C_{\rm grav}$ is the ratio of hidden sector to messenger sector vacuum 
expectation values of auxiliary fields \cite{pedro}.   
As in the case of mSUGRA,
one can re-express $\mu$ and $B$ in terms of $m_Z$ and $\tan\beta$
after the electroweak symmetry breaking conditions are imposed.  

The mSUGRA and mGMSB models are incorporated in the event generator
ISAJET \cite{ISAJET}.  The Higgs boson mass spectrum is obtained in
ISAJET by minimizing the renormalization-group-improved one-loop
effective potential. 
The minimization is performed at an optimized
scale choice $Q\simeq (M_{\widetilde t_L}M_{\widetilde t_R})^{1/2}$,
which effectively includes the dominant two-loop
contributions  to the Higgs mass computations.  
The latest improvements including finite two loop
effects through vertex corrections to the top and bottom
Yukawa couplings are not taken into account in these analyses.
Supersymmetric parameter
values are used by ISAJET to calculate the various  
supersymmetric particle masses and mixing angles, and the Higgs boson 
masses and branching fractions \cite{bisset}. The calculations
presented here are an update of work previously presented in
\Ref{bht}.

\begin{figure}[hp!]
\begin{center}
\centerline{\psfig{file=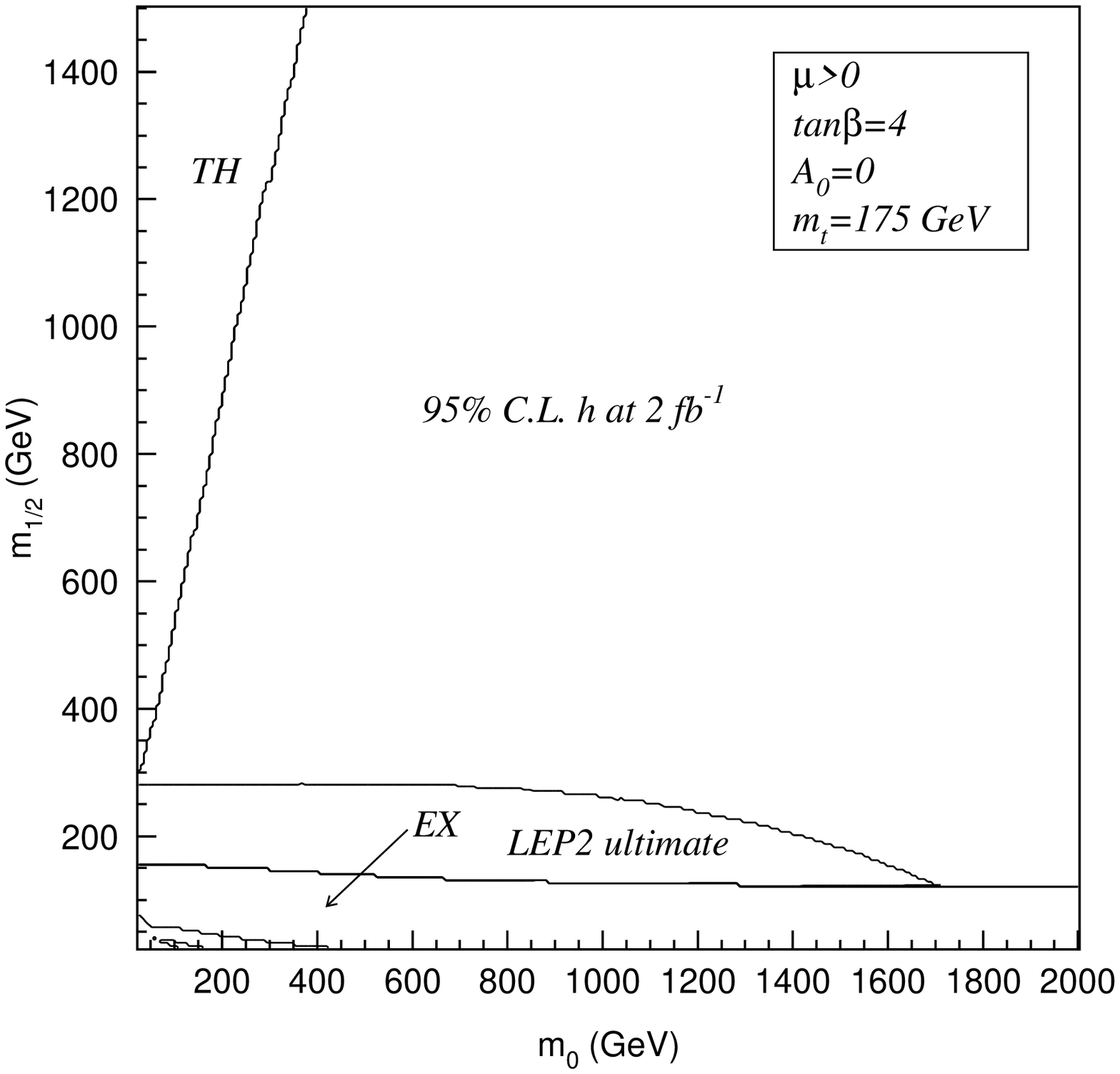,width=9cm}
\psfig{file=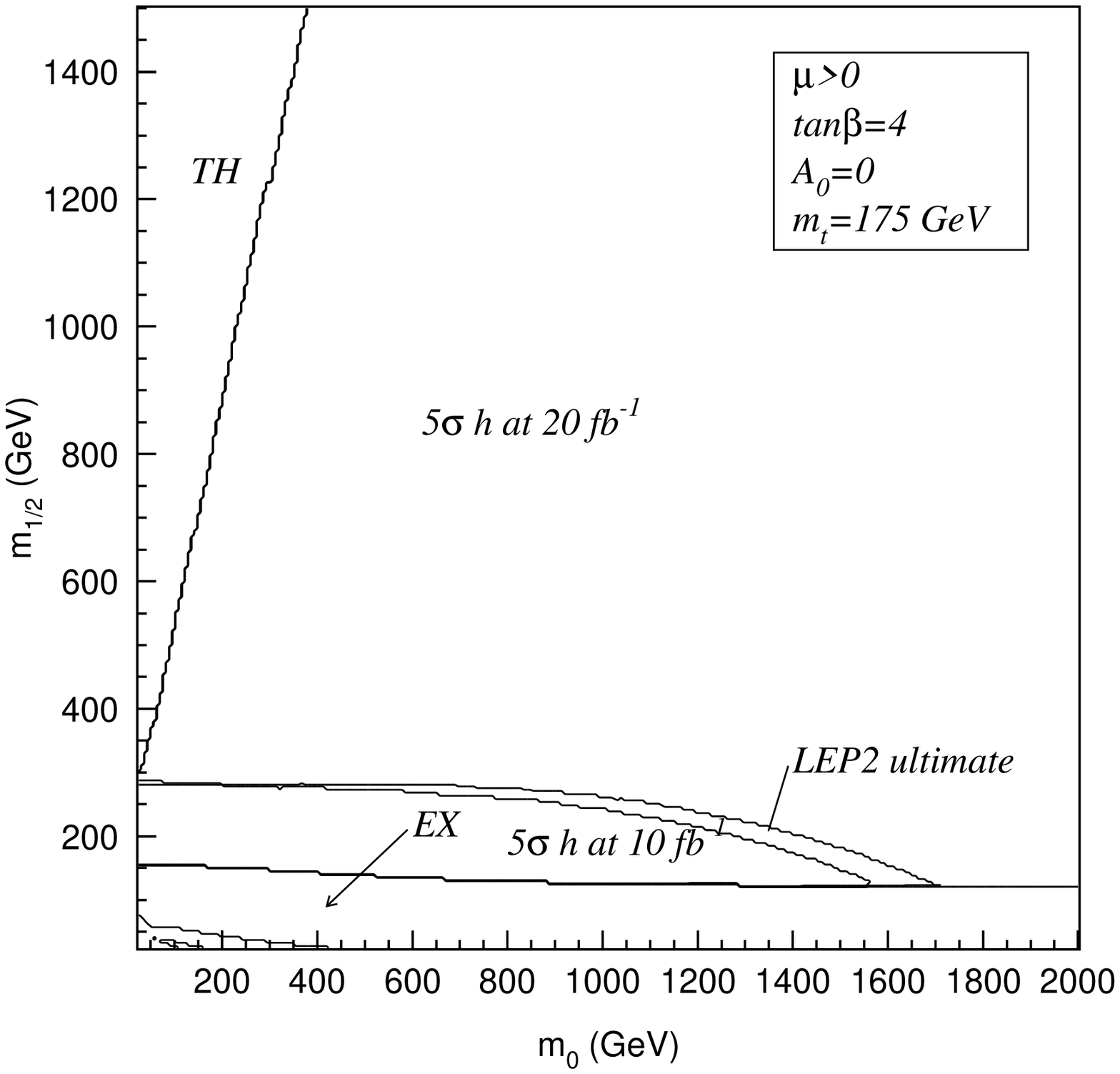,width=9cm}}
\end{center}
\vskip-2pc
\caption[0]{\label{sugra:1} 
Regions of mSUGRA model parameter space corresponding to 
(a) 95\% CL exclusion and (b)
5$\sigma$ discovery that are accessible to 
the upgraded Tevatron with different integrated luminosities.
In this case $A_0=0$, $\tan\beta =4$ and $\mu > 0$.  The TH region is 
excluded either due to lack of proper 
electroweak symmetry breaking or because the lightest neutralino 
is not the lightest supersymmetric particle.  
The EX region is excluded from the experimental bound on the 
lightest chargino mass, and the marked LEP2 region is explorable by
direct Higgs boson searches at LEP.}
\vskip 2pc
\vskip-6pc
\begin{center}
\centerline{\psfig{file=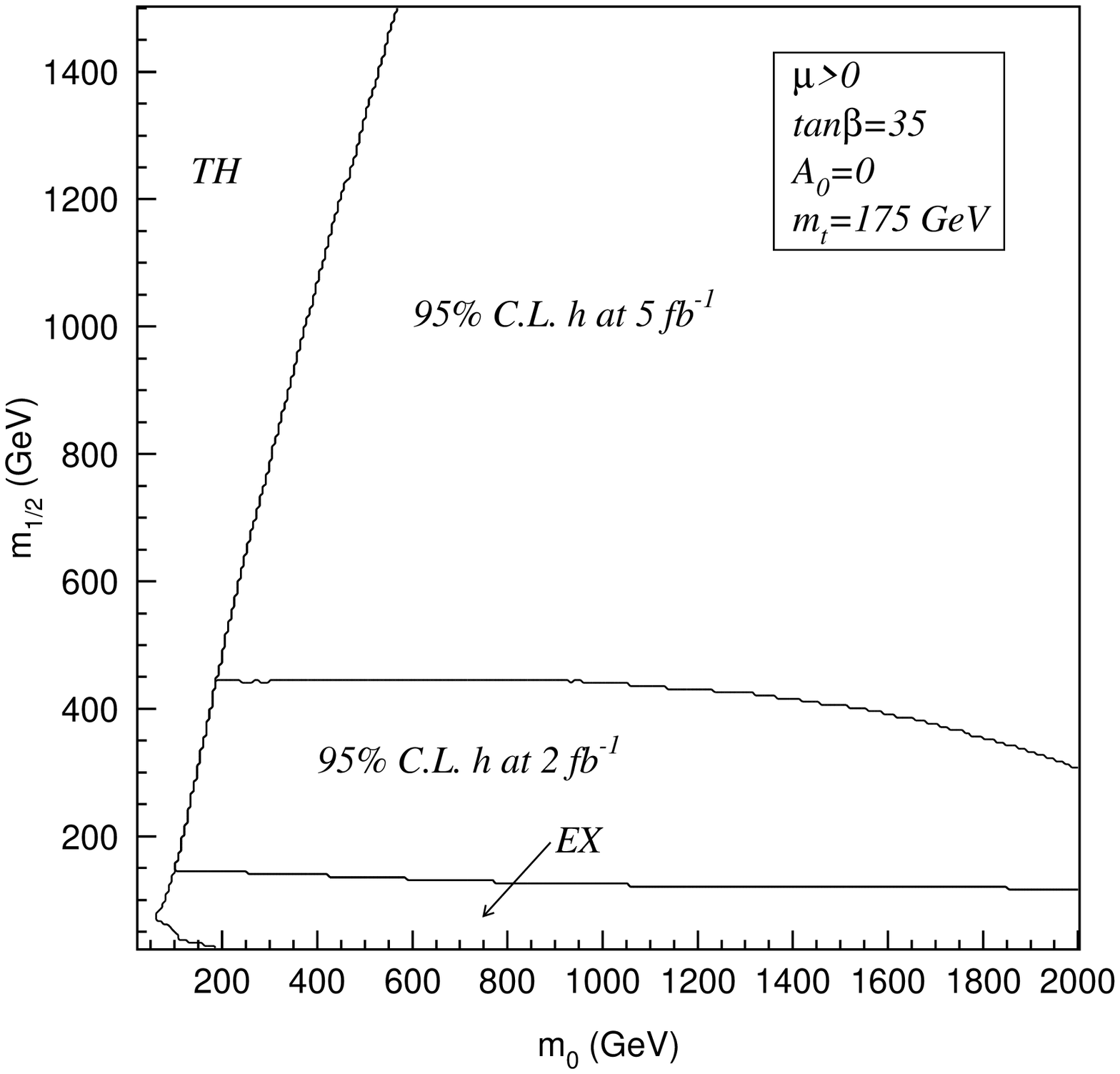,width=9cm}
\psfig{file=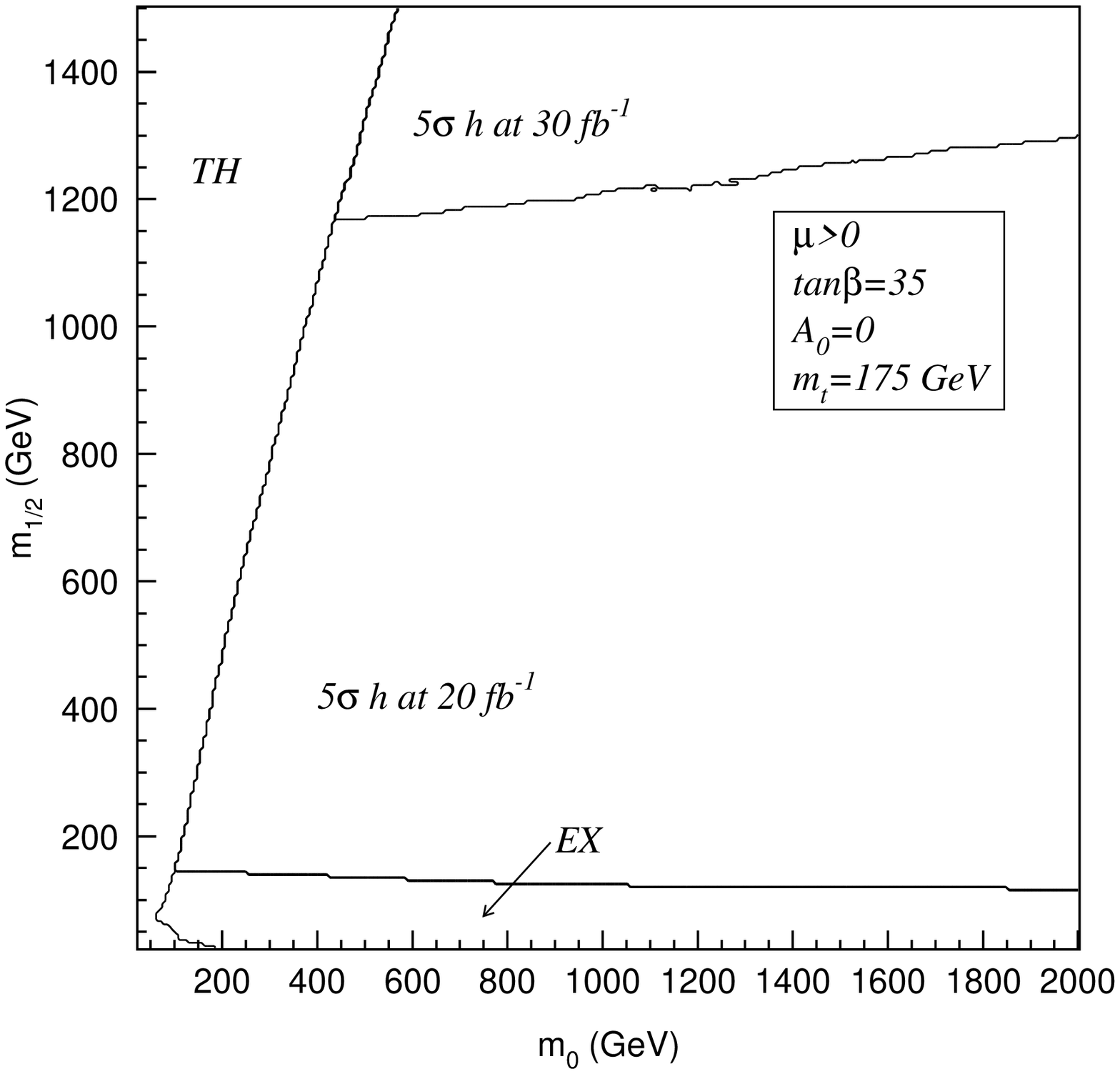,width=9cm}}
\end{center}
\vskip-2pc
\caption[0]{\label{sugra:3} 
As in \fig{sugra:1}, but for
the case $A_0=0$, $\tan\beta =35$ and $\mu > 0$.}
\end{figure}

The 95\% exclusion and 5$\sigma$ discovery regions for the mSUGRA model are 
shown in \fig{sugra:1} for $\tan\beta =4$,  $A_0=0$ and  
$\mu>0$. The TH region is excluded either due to a lack of 
appropriate electroweak symmetry breaking, or because the lightest
neutralino is not the lightest supersymmetric particle.
The EX region is excluded by the negative results from LEP2 searches
for chargino pair production, which require 
$m_{\widetilde\chi^\pm_1} > 100$ GeV. 
The LEP reach is shown here as indicative, assuming a maximal Higgs
mass discovery potential of 108 GeV.
For the region of mSUGRA parameters considered in \fig{sugra:1}, the 
 Fermilab Tevatron experiments should be 
able to explore the entire plane shown at 95\% CL with 2~fb$^{-1}$ or
at $5\sigma$ with 20~fb$^{-1}$.

The corresponding mSUGRA plots for $\tan\beta =35$ are shown in
\fig{sugra:3}.  In contrast to the small $\tan\beta$ case just discussed,
at large $\tan\beta$ the mass of $\hl$ lies above
the LEP2 reach.  The Tevatron experiments should
be able to probe the entire plane shown at 95\% CL with 5~fb$^{-1}$ of
data.  The discovery potential is more limited and 
requires an integrated luminosity as 
high as 30~fb$^{-1}$ (for very large $m_{1/2}$)
to ensure Higgs boson discovery.

\begin{figure}[htp]
\vskip-5pc
\begin{center}
\centerline{\psfig{file=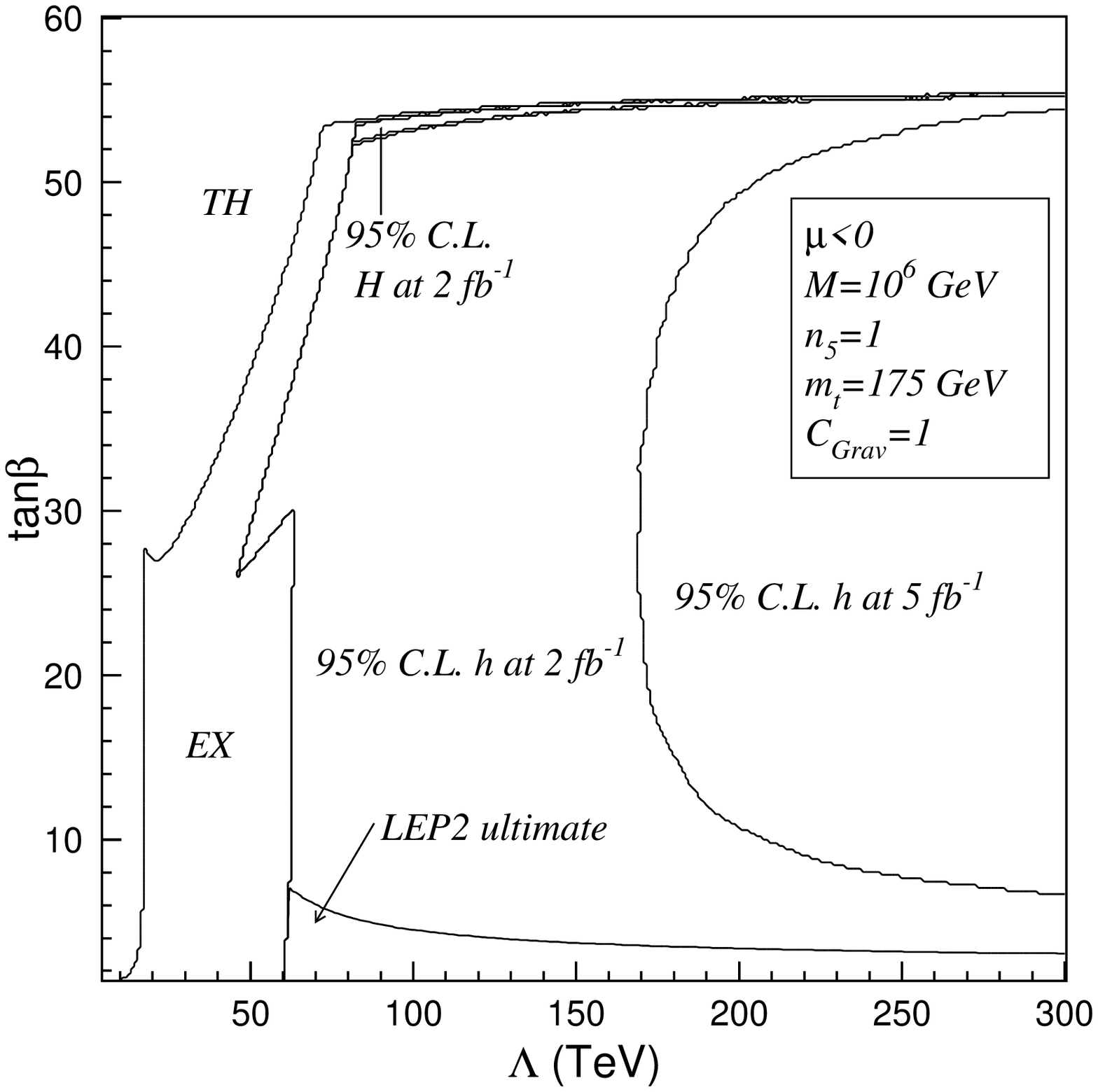,width=9cm}
\psfig{file=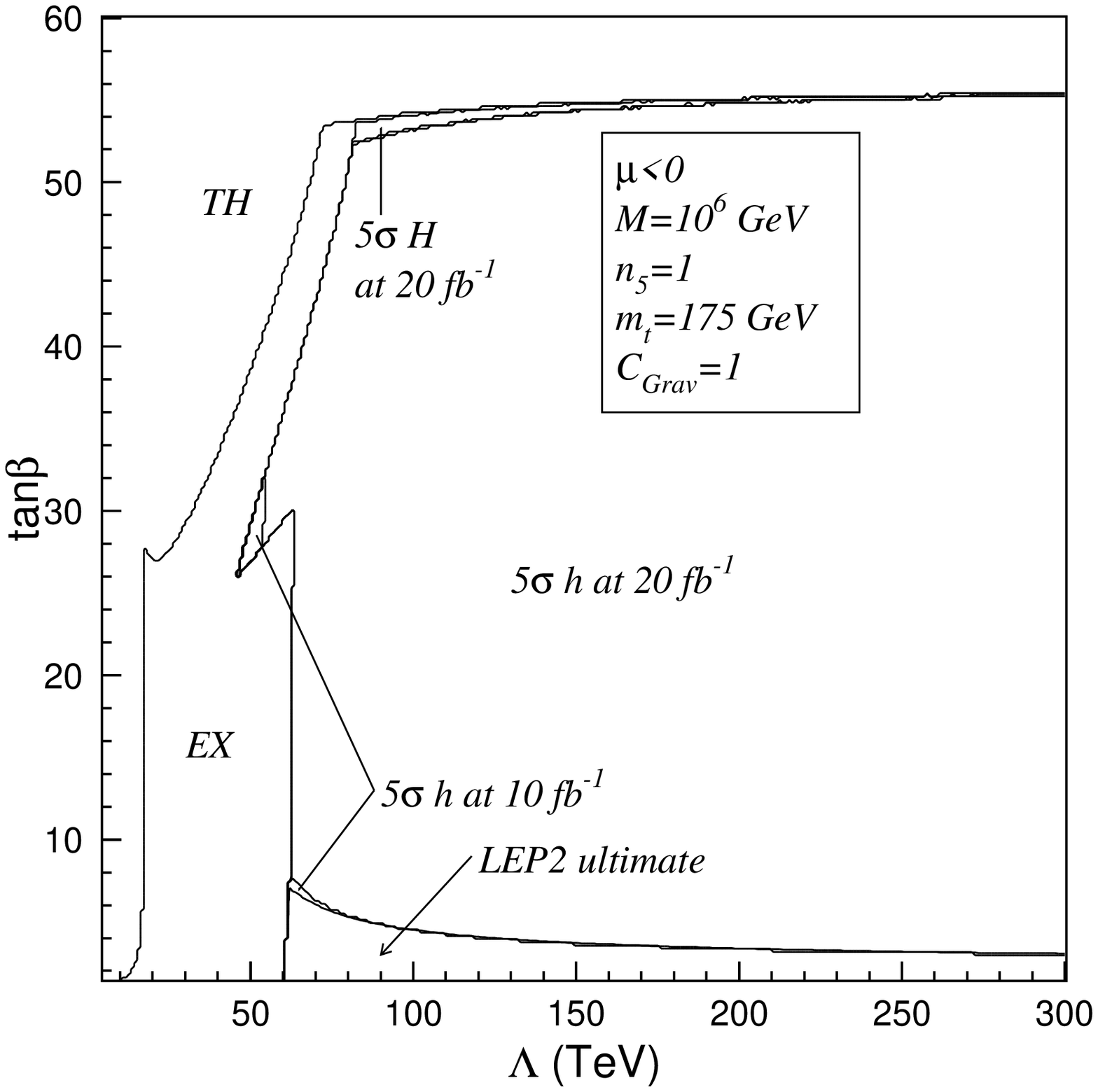,width=9cm}}
\end{center}
\vskip-2pc
\caption[0]{\label{gmsb:5} 
As in \fig{sugra:1}, but 
for a mGMSB model  with the parameter choice
$M=10^6$ GeV, $n_5 =1$ and $\mu < 0$.  The EX 
regions are excluded by experimental bounds on the mass of the 
lightest neutralino or the stau.}
\end{figure}

The 95\% exclusion and 5$\sigma$ discovery regions for the minimal
gauge mediated supersymmetry breaking model, in the
$\Lambda$--$\tan\beta$ plane, are shown in \fig{gmsb:5}, for $n_5=1$
and messenger scale $M=10^6$ GeV.  The TH region is excluded due to a
lack of appropriate electroweak symmetry breaking.  The EX regions are
excluded by experimental bounds on the mass of the lightest neutralino
($m_{\widetilde\chi^0_1} > 84$ GeV).  In these cases, the whole plane
can be probed by Tevatron experiments at the 95\% CL with 5~fb$^{-1}$
of data.  At the $5\sigma$ discovery level, the Tevatron experiments
would need 20~fb$^{-1}$ per experiment to be able to probe this model.

\goodbreak

\subsubsection{MSSM Higgs Sector Results:
$gg$, $q\bar q\to b\bar b\phi$ [$\phi=\hl$, $\hh$, $\ha$]}

In many extensions of the Standard Model (including the MSSM), there exists
at least one non--Standard Model--like Higgs boson: its couplings to gauge
bosons may vanish at tree--level or be suppressed through mixing, 
and their relative couplings to different fermions are not strictly 
set by the corresponding fermion masses.
Such Higgs bosons will exist in models with many Higgs doublets when 
one of the doublets is mostly responsible for electroweak symmetry breaking.
In the MSSM,  
the CP-odd scalar $\ha$ differs from the Standard Model Higgs boson,
with enhanced couplings at large $\tan\beta$
to down--type fermions and suppressed couplings to
up--type ones.  Additionally, when one of the CP-even Higgs bosons 
(either $\hl$ or $\hh$) has Standard Model--like couplings to the gauge
bosons, the other one has a pattern of enhanced and suppressed
couplings to fermions similar to that of the $\ha$.

We consider the prospects for discovering such a non--Standard Model--like
Higgs boson at the Tevatron, based on the
$b\bar b\phi\to b\bar b b\bar b$ signature studied in Section~II.F.
In the case of the MSSM, $\phi$ can in principle be one of three
neutral Higgs bosons: $\hl$, $\hh$ and $\ha$.
If $\tan\beta$ is large, two of the Higgs bosons are
produced with enhanced rates.\footnote{As emphasized
below \eq{hpmqq}, at large $\tan\beta$ the production rate of 
$b\bar b\phi$ is enhanced for two of the neutral MSSM Higgs bosons: 
$\phi=\ha$ and $\phi=\hl$ [$\phi=\hh$] if $\mha\lsim\mz$
[$\mha\gg\mz$].}  
If these two Higgs bosons are very
close in mass, then the corresponding signals should be added as they
will not be distinguishable within the experimental resolution.  
However, over some range of $\mha$ and $\tan\beta$ values,
the two Higgs bosons with enhanced production rates are
significantly split in mass.  For example, for
moderate values of $\tan\beta$, the corresponding mass splitting 
can be as much as $10$--$20$ GeV, and the
procedure for combining the signals must reflect this fact.

The previous considerations
illustrate one of the limitations of encoding all the analysis
information in a \rexp-curve.  In some cases, knowledge of the signal
size is also necessary.  The expected signal in a given model is
simply obtained by multiplying the Standard Model signal by the
\rth-value for the mass $m_\phi$.  Of course,
the expected background in a mass
window centered on the mass $m_\phi$ is unchanged by
choosing a different model.  The experimental sensitivity in the
$b\bar b\phi$ channel to a given set of MSSM parameters is estimated
using the following algorithm.  First, the CP even Higgs boson $\phi$
with the smallest value of $g^2_{VV\phi}$ is chosen as the
non--Standard Model--like one.  The expected signal for the CP even
Higgs boson $S_{CP}$ is calculated by multiplying the Standard Model
signal by the \rth-value for the CP-even Higgs boson, whereas the
expected background $B_{CP}$ in the mass window centered on the CP
even Higgs boson mass $M_{CP}$ is the same as in the Standard Model.
The signal $S_A$ and background $B_A$ for $\ha$ is determined in a
similar fashion.

The invariant mass distributions of the individual signals $S_{CP}$
and $S_A$ are assumed to be well described by Gaussian distributions.
To determine if the signals overlap, the mass splitting $\Delta
M=|M_A-M_{CP}|$ is compared to the quantity
$\sigma_4=(M_A+M_{CP})\times 0.15\times 2$.  The latter is the sum of
two standard deviations in the invariant mass distribution of each
signal assuming a $b\bar b$ mass resolution of 15\%.  If $\Delta
M<\sigma_4$, then the two signals overlap within the experimental
resolution.  In this case, the mass window is expanded to contain both
signals ($2\sigma$ below the lightest Higgs boson mass and $2\sigma$
above the heavier one), and the number of background events within the
window is calculated: $B=(B_{CP}+B_A)[1-\frac{1}{2}(1-\Delta
M/\sigma_4)]$.  The systematic error on the background is assumed to
be ${\rm min}(0.1 B,\sqrt{B})$.  This choice represents the Run 1
experience that systematic errors are roughly 10\% of the number of
background events, but also that the systematic errors do not increase
without bound and tend to saturate at a level comparable to the
statistical errors.  The full error on the background, $\Delta B$, is
the quadrature of statistical and systematic errors, and the
significance is defined as $(S_A+S_{CP})/\Delta B$.  If the two
signals are well separated within the experimental resolution, then
the significance of {\it each} signal is calculated separately:
$S_A/\Delta B_A$ and $S_{CP}/\Delta B_{CP}$.  These two quantities are
added in quadrature to determine the full significance.  This
algorithm is somewhat biased.  For the case when both signals overlap
(which coincides with the range of $\tan\beta$ where the experiments
might observe a signal), one must ``know'' that the best mass window
is a larger one.  It may be more realistic to choose the average Higgs
boson mass as the center of the mass window of size $\pm
2\sigma$.\footnote{In the case when the signals do not overlap (which
is not of practical interest since $\tan\beta$ must be so small that
the production cross section is highly suppressed), one should
consider background fluctuations over all reasonable mass windows that
do not overlap.  Given the experimental resolution, and the maximal
mass splitting possible, this is not a serious concern.}

\begin{figure}[ht!]
\vskip-2pc
\begin{center}
\centerline{\psfig{file=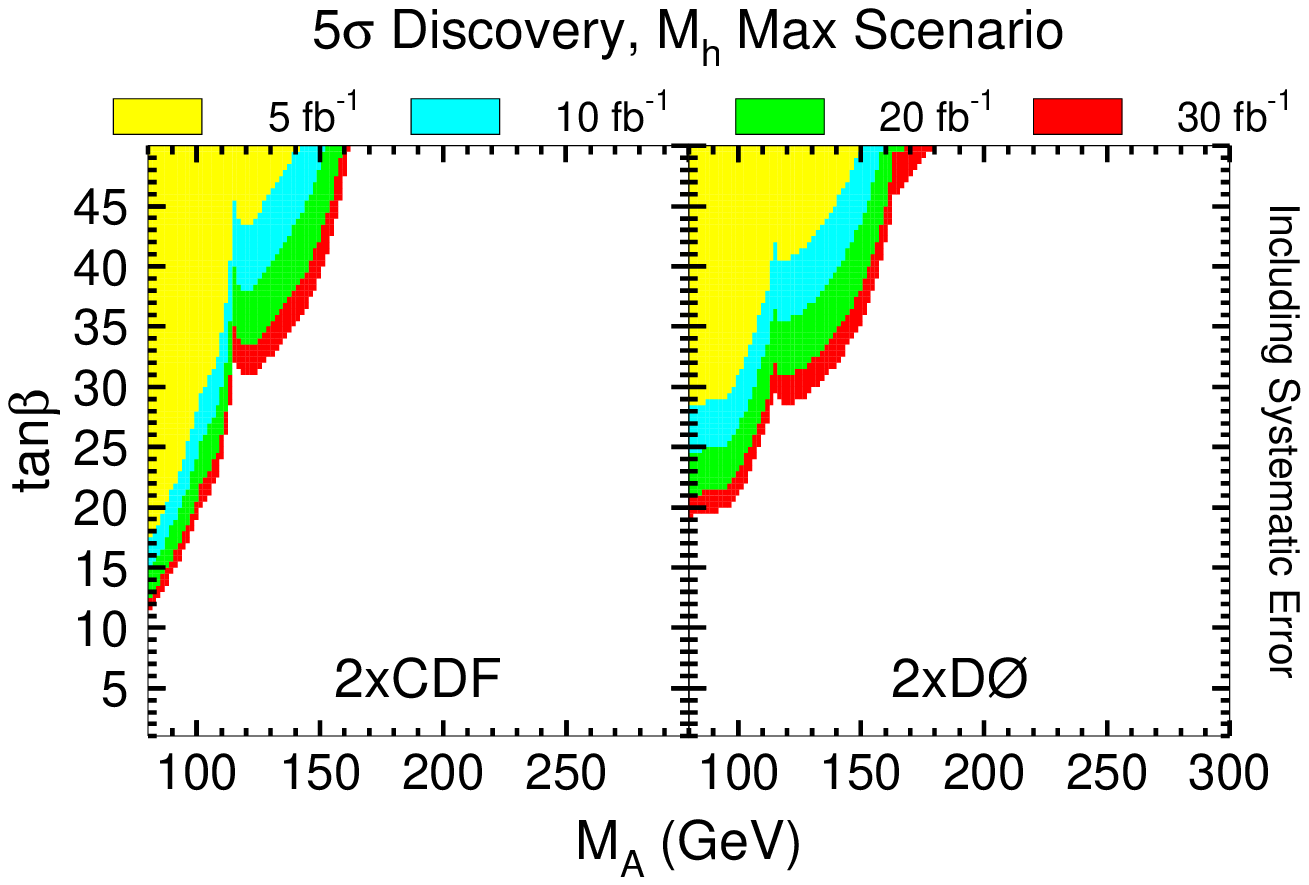,width=14cm}}
\end{center}
\vskip-2pc
\caption[0]{\label{b4max5} The $5\sigma$
discovery reach for the maximal-mixing benchmark scenario 
of table~\ref{benchmarks} for the
signature $q\bar q,gg\to b\bar b\phi\to b\bar b b\bar b$ for different total
integrated luminosities: (a) CDF analysis and  (b) D\O\ analysis.}
%
\begin{center}
\centerline{\psfig{file=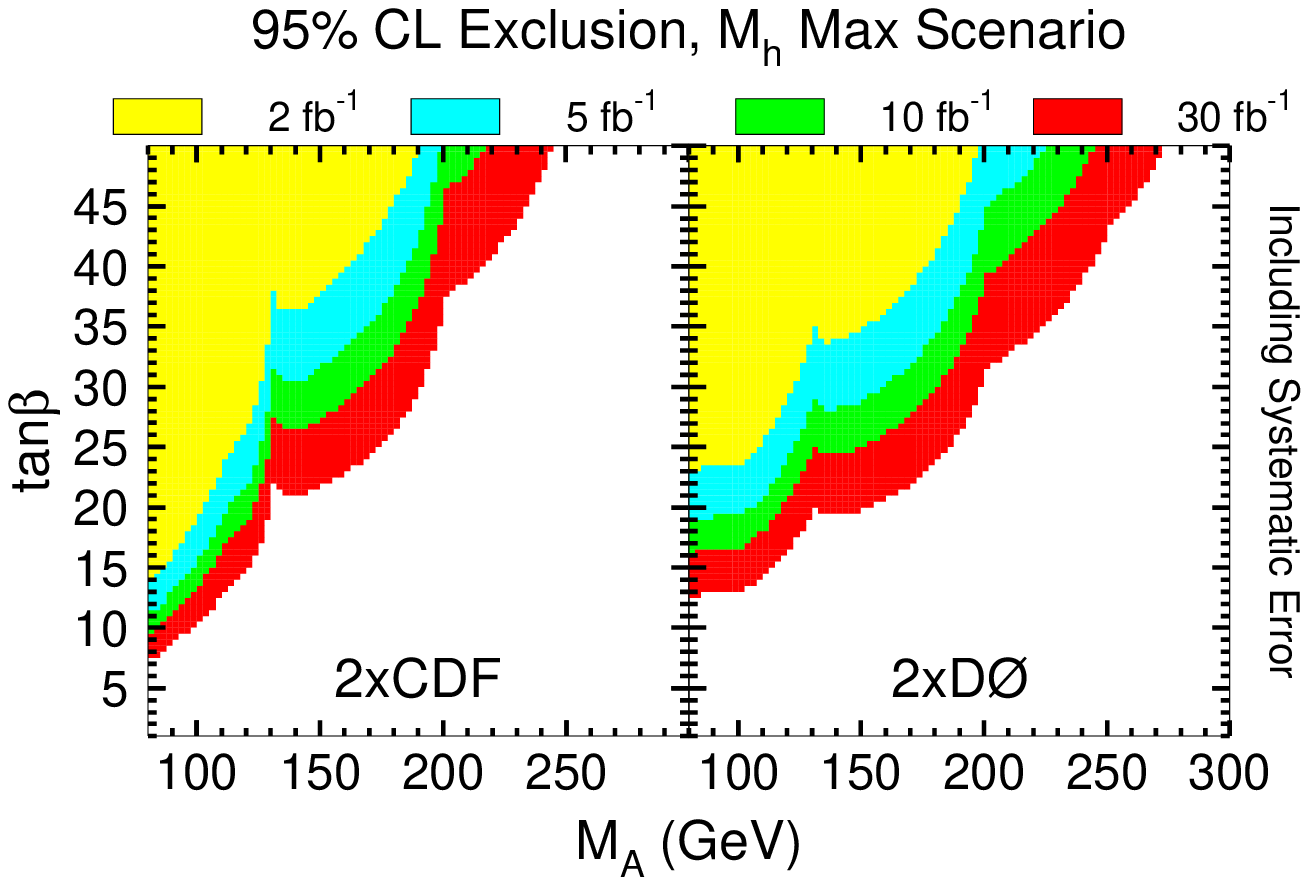,width=14cm}}
\end{center}
\vskip-2pc
\caption[0]{\label{b4max95} Same as \fig{b4max5}, but for 95\%
CL exclusion contours.}
\vskip-2pc
\end{figure}

\begin{figure}[htbp]
\vskip-2pc
\begin{center}
\centerline{\psfig{file=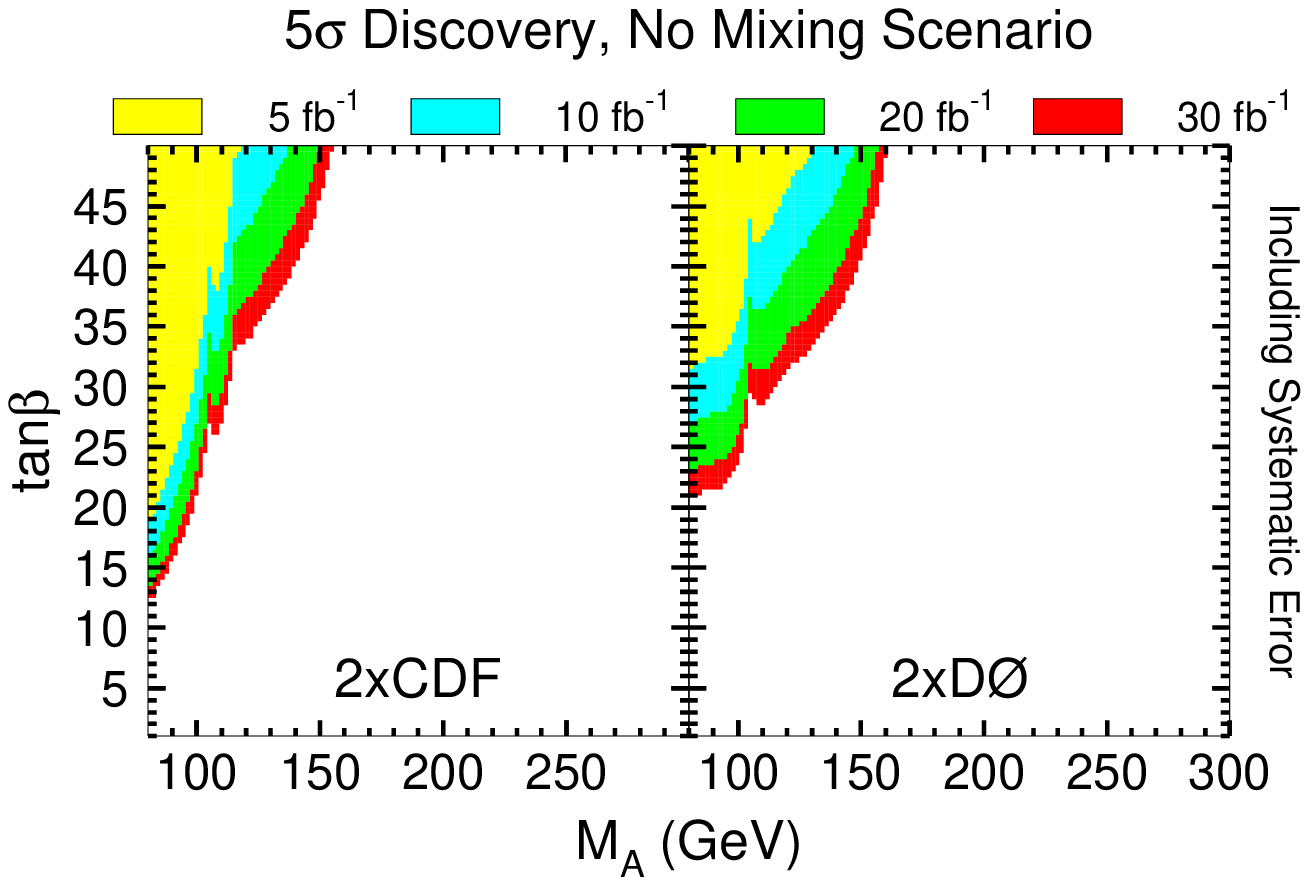,width=14cm}}
\end{center}
\caption[0]{\label{b4nomix5} Same as \fig{b4max5}, but for 
the no-mixing benchmark scenario of table~\ref{benchmarks}.}
%
\begin{center}
\centerline{\psfig{file=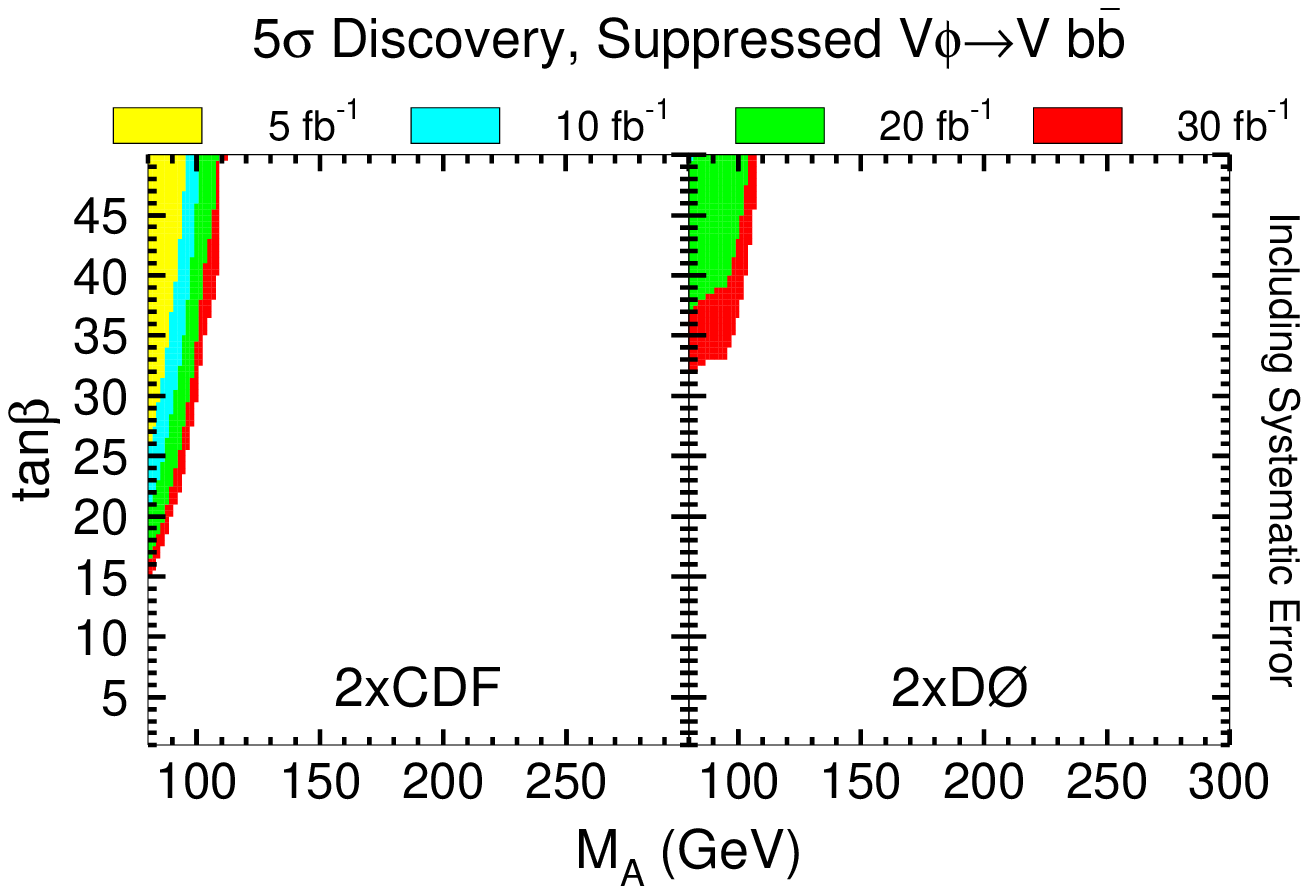,width=14cm}}
\end{center}
\caption[0]{\label{b4mix5} Same as \fig{b4max5}, but for the  
``suppressed $V \phi \to V b\bar b$ production'' benchmark scenario
of table~\ref{benchmarks}.}
\end{figure}

\Figs{b4max5}{b4max95} are
contour plots demonstrating regions of $5\sigma$ discovery and 95\%
C.L. exclusion for
the maximal mixing benchmark scenario. 
Clearly, large values of $\tan\beta$, for which the CP-odd Higgs
boson and one of the CP-even Higgs bosons has enhanced couplings to
the bottom quarks, are necessary to render  
the  $b\bar b\phi$ channel experimentally viable.  In \fig{b4nomix5},
we show the parameter space coverage for $5\sigma$ discovery
in the no-mixing benchmark scenario.  Comparison with \fig{b4max5}
shows that the distinction between the maximal squark mixing and
no squark mixing cases is small.  This can be attributed to the fact
that in both cases, the
parameter $\mu$ was taken to be rather small; hence, the effects of 
the vertex corrections to the bottom Yukawa coupling are small. 
However,  this does not mean
that the $b\bar b\phi$ channel is, in general, 
robust under variation of the MSSM parameters.
Precisely at large values of $\tan\beta$, for which 
the  $b\bar b\phi$ channel is 
promising, the vertex corrections to the bottom Yukawa coupling may
become quite relevant. 
As explained in Section~I.C., depending 
on the sign and magnitude of $\Delta_b$ [\eq{deltamb}], 
large vertex corrections to the bottom Yukawa 
coupling may be induced  varying drastically the coverage  of the 
$\mha$--$\tan \beta$ plane through the 
$q\bar q,gg\to b\bar b\phi \to b\bar b b\bar b$ channel.
%
For example, in the suppressed $V\phi\to Vb\bar b$ benchmark 
scenario, since all the Higgs couplings to $b\bar b$ are generally
suppressed, the MSSM parameter coverage is reduced as shown in 
\fig{b4mix5}.

\begin{figure}[htbp]
\vskip-2pc
\begin{center}
\centerline{\psfig{file=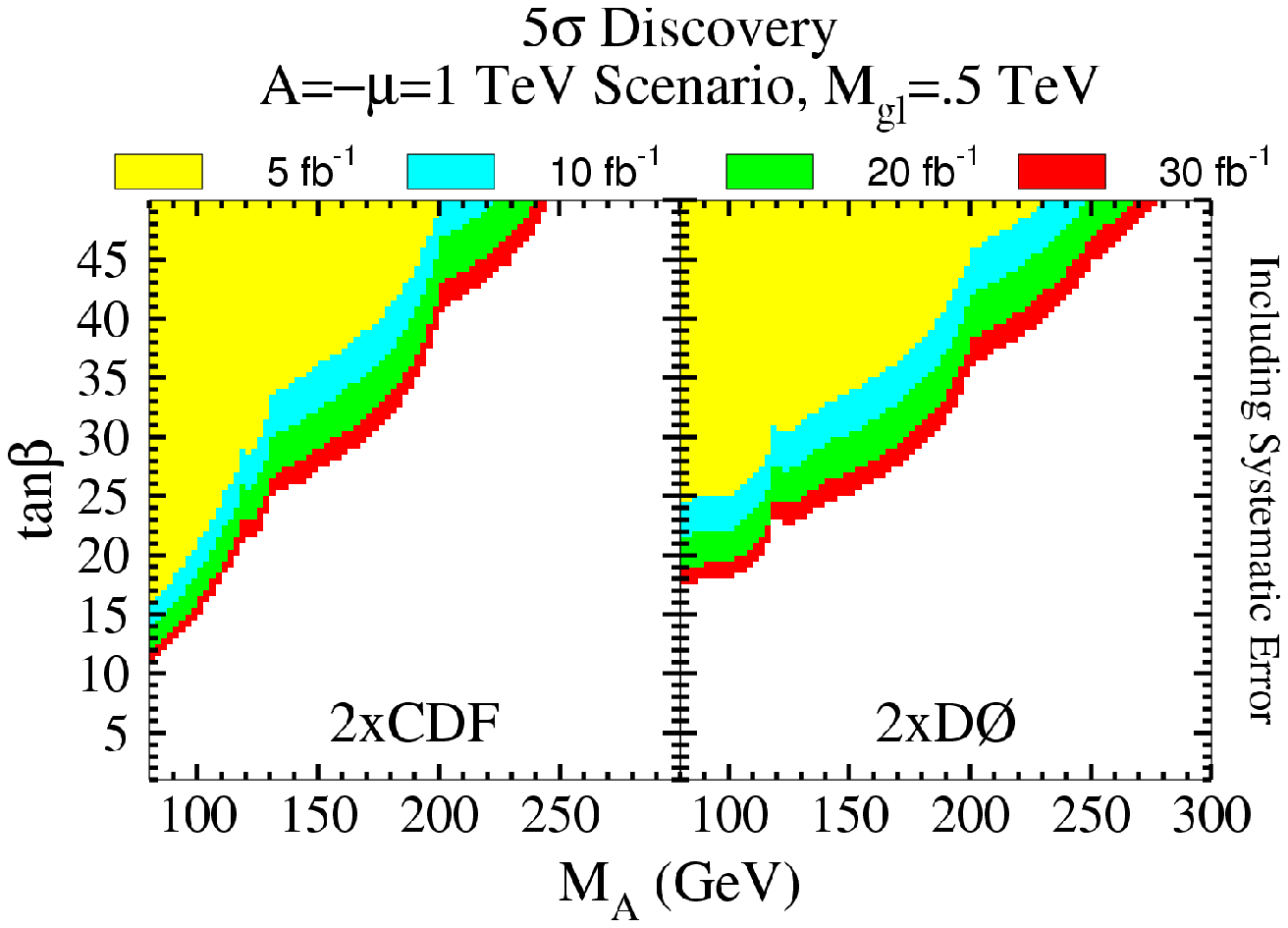,width=14cm}}
\end{center}
\caption[0]{\label{b4susy1}
The $5\sigma$ discovery reach for $b\bar b\phi$ production, for MSSM
parameters which induce radiative corrections that enhance the coupling
of the CP-odd Higgs and one of the CP-even Higgs to the bottom quarks:
(a) CDF analysis, (b) D\O\ analysis}
%
\begin{center}
\centerline{\psfig{file=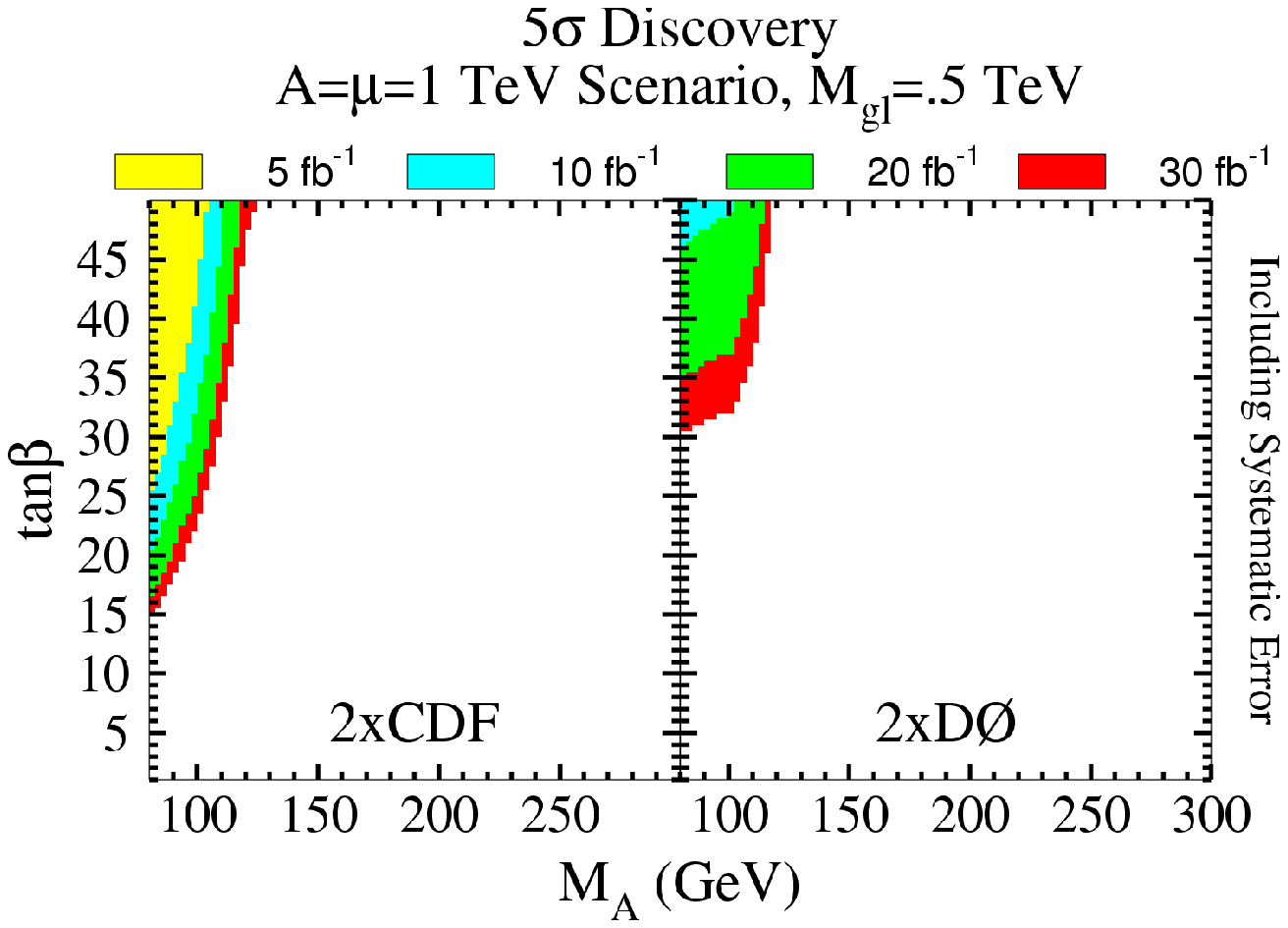,width=14cm}}
\end{center}
\caption[0]{\label{b4susy2}
The $5\sigma$ discovery reach for $b\bar b\phi$ production, for MSSM
parameters which induce radiative corrections that suppress the coupling
of the CP-odd Higgs and one of the CP-even Higgs to the bottom quarks:
(a) CDF analysis, (b) D\O\ analysis}
\end{figure}

As previously noted, 
the leading one-loop corrections to the bottom Yukawa coupling
suppress [enhance] the $b\bar b\phi$ coupling 
for $\Delta_b>0$
[$\Delta_b <0$], thus degrading [improving] 
the Higgs discovery potential in the 
$q\bar q,gg\to b\bar b\phi\to b\bar b b\bar b$ channel.
In Section~I.C.3, the dependence of $\Delta_b$ on the supersymmetric
parameters is exhibited [\eq{deltamb}] and its phenomenological
implications are discussed.  Here, we remind the reader that the sign
of $\Delta_b$ is governed by the sign of $\mu$ (in the convention
where $M_{\tilde{g}}$ is positive), as the dominant contribution to
$\Delta_b$ from bottom-squark loops is proportional to $M_{\tilde{g}}\mu$.
The top-squark loop contribution to $\Delta_b$ adds a sub-dominant
contribution proportional to $A_t\mu$, which increases [decreases] the
overall magnitude of $\Delta_b$ for $A_t>0$ [$A_t<0$].
The dependence of the parameter space coverage on the sign of $\mu$ is
illustrated in
\figs{b4susy1}{b4susy2}, which exhibit discovery contours for the
case  $A_t = |\mu|$ = 1 TeV, $M_S$= 1 TeV
and  $M_{\tilde{g}}=0.5$ TeV.
%
These plots illustrate the two possible choices 
for the sign of $\mu$, which
correspond to different signs for the loop corrections to the 
Higgs--bottom-quark
Yukawa coupling.  Moreover, since $A_t>0$, 
the leading one-loop corrections are further enhanced for
$\mu <0$, thus improving the Higgs discovery potential in the 
$q\bar q,gg\to b\bar b\phi\to b\bar b b\bar b$ channel.
Thus, we see that there can be strong dependence of the potential for
$q\bar q,gg\to b\bar b\phi\to b\bar b b\bar b$ discovery on
the specific choice of MSSM parameters.
Observation of a signal in this channel will clearly suggest that
$\tan \beta$ is large, and can place some useful constraints on the
allowed regions of MSSM parameter space.

\subsubsection{MSSM Higgs Sector Summary}

We may combine the results for the various channels
obtained in the generic MSSM analysis  to provide summary plots
of the MSSM Higgs discovery reach of the upgraded Tevatron collider.
For comparison, we show in each case the expected LEP final
coverage of the $\mha$--$\tan\beta$
plane obtained from the search mode
$e^+ e ^- \to Z \phi$ with the subsequent decay of $\phi = \hl$ or
$\hh$ into $b\bar b$ or $\tau^+\tau^-$. 
Moreover, although not 
explicitly shown in the final combined plots presented here, 
the absence of a Higgs signal at LEP2 in the associated production process
$e^+ e ^- \to \ha \phi$ would rule out the parameter space region
defined by $\mha\lsim 95$~GeV and $\tan\beta\gsim 5$ as shown in
\fig{mh200} in the maximal mixing scenario.  
The reach of the latter channel has not yet been presented in a model
independent way by the LEP collaborations.  Nevertheless, for the
three benchmark scenarios of table~\ref{benchmarks}, the excluded
region at low $\mha$ and large $\tan\beta$ will not be appreciably
different from the one quoted above.

\begin{figure}[hb!]
\centering
\centerline{\psfig{file=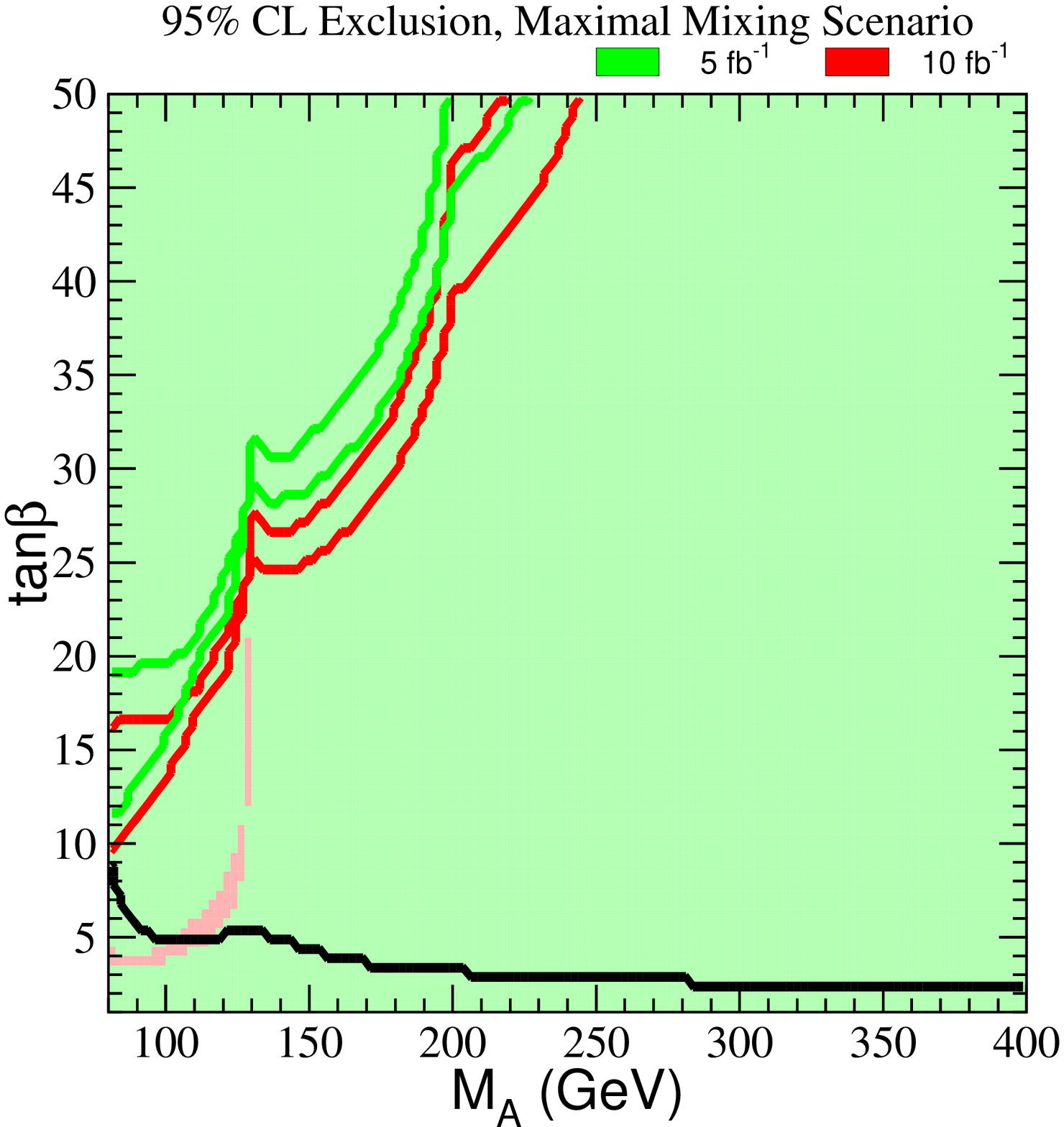,width=8cm}
\hfill
\psfig{file=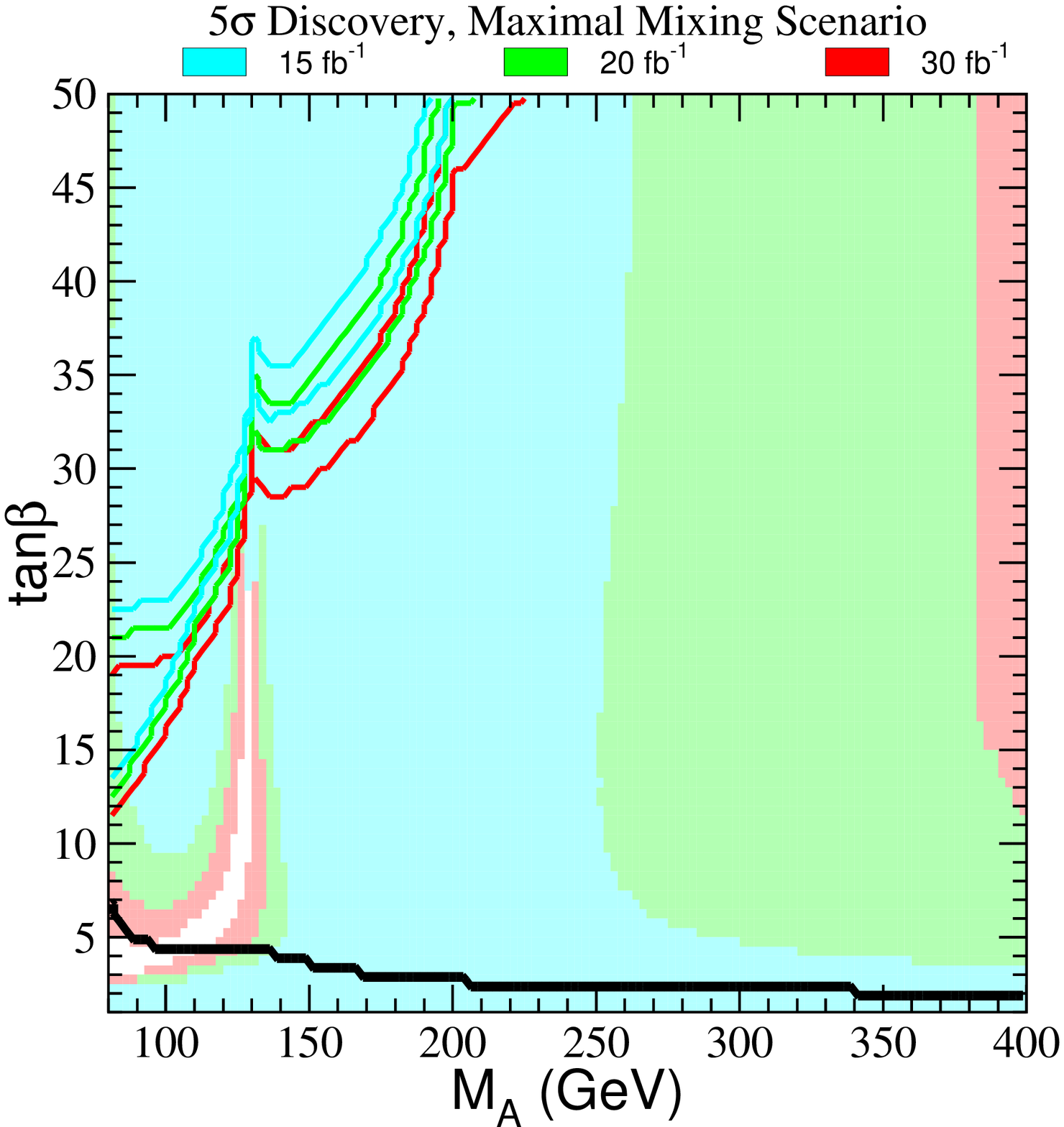,width=8cm}}
\vskip2pc
\caption[0]{\label{fullmhmax95} (a) 95$\%$ CL exclusion region
and (b) $5\sigma$ discovery region on the $\mha$--$\tan \beta$
plane, for the maximal mixing benchmark scenario of table~\ref{benchmarks}.  
and two different search channels:
$q\bar q\to V\phi$ [$\phi=\hl$, $\hh$], $\phi\to b\bar b$
(shaded regions) and 
$gg$, $q\bar q\to b\bar b\phi$ [$\phi=\hl$, $\hh$, $\ha$],
$\phi\to b\bar b$ (region in the upper left-hand corner bounded by the
solid lines).  Different integrated 
luminosities are explicitly shown by the color coding.
The two sets of lines (for a given color) 
correspond to the CDF and D\O\ simulations, 
respectively.  The region below the solid black line near the bottom
of the plot is excluded by the absence of observed $e^+e^-\to Z\phi$
events at LEP2.
}
\end{figure}

\begin{figure}[hp!]
\begin{center}
\centerline{\psfig{file=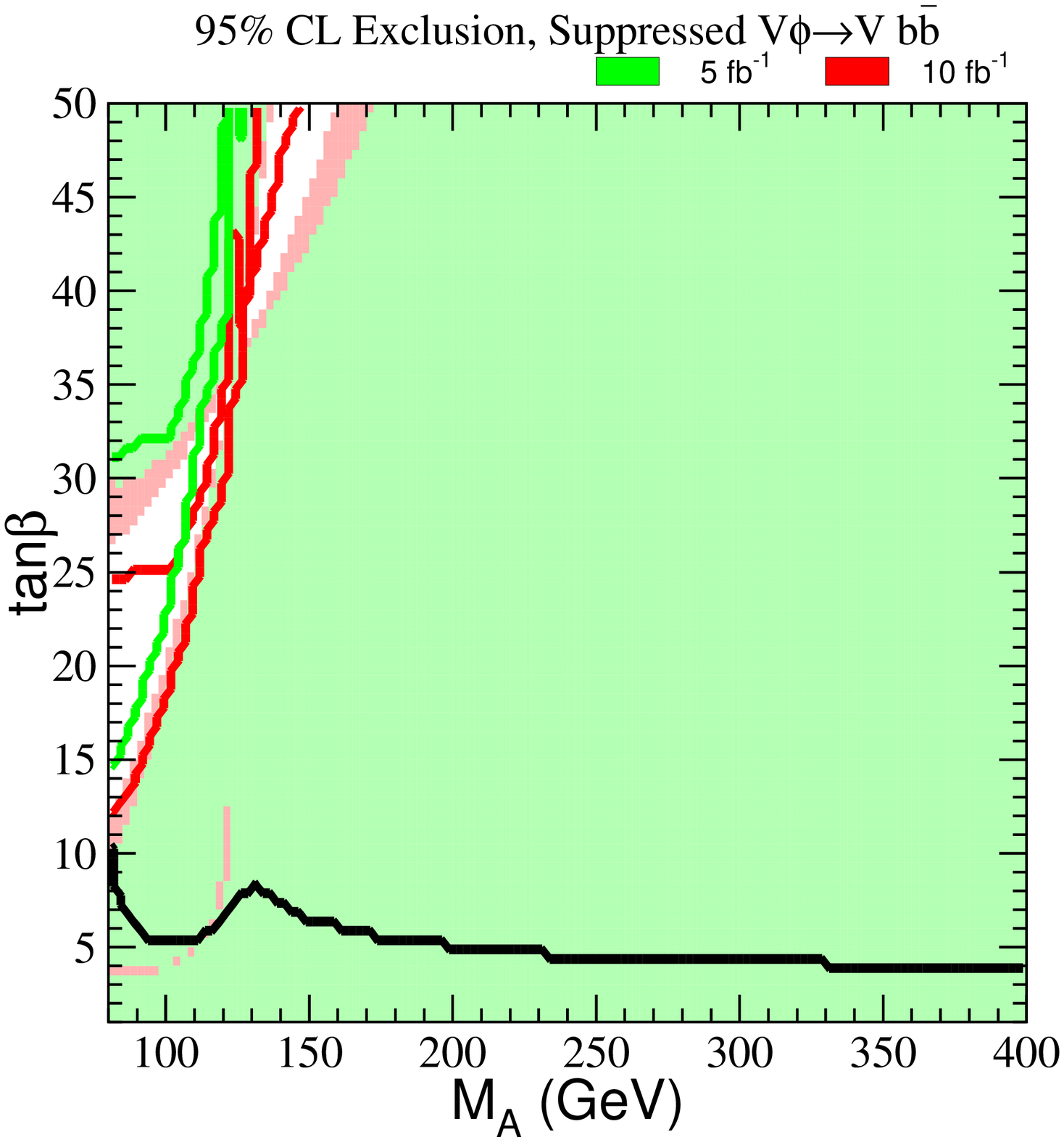,width=8cm}
\hfill
\psfig{file=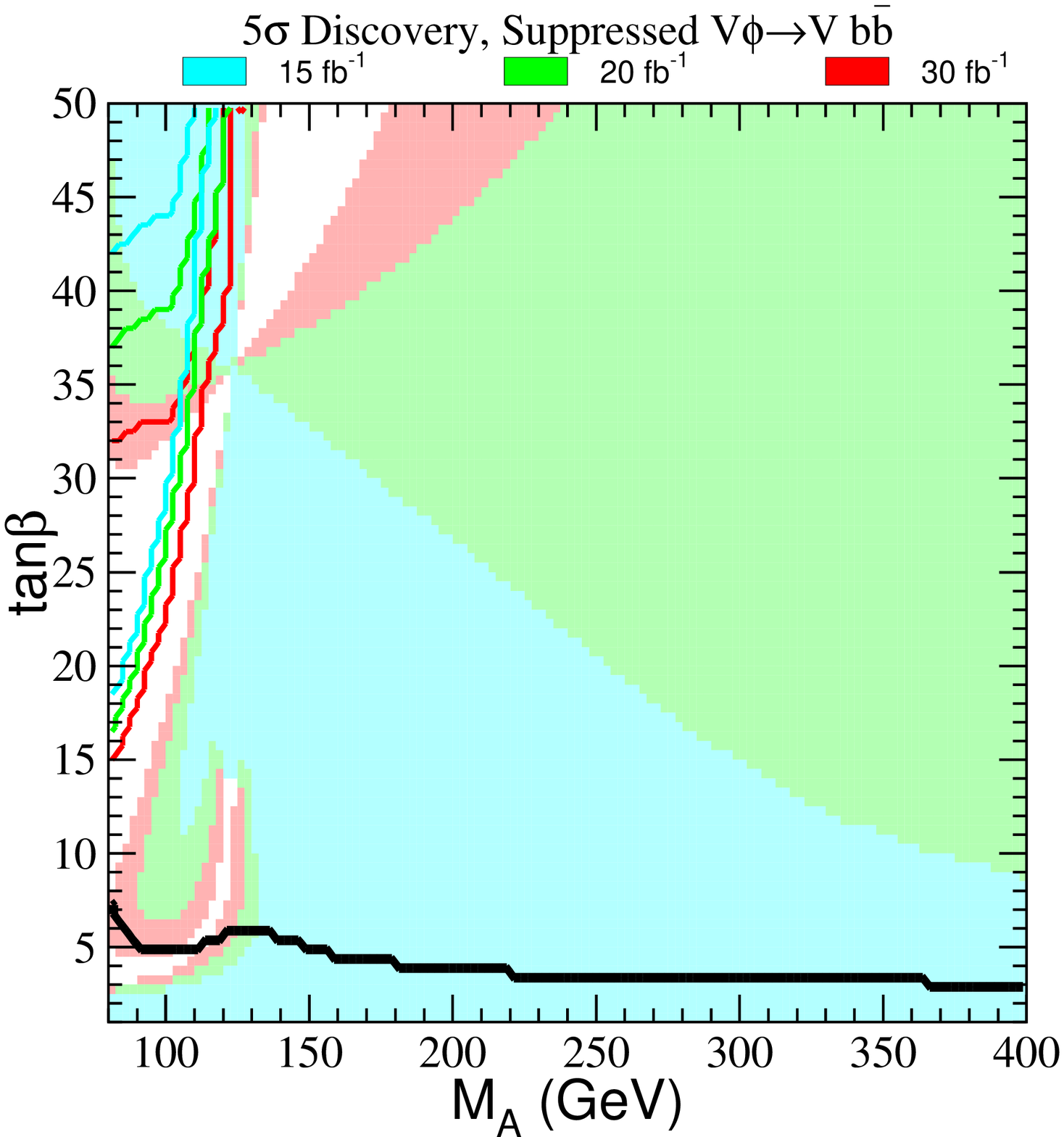,width=8cm}}
\end{center}
\caption[0]{\label{fullmix95}
The same as \fig{fullmhmax95} but for the ``suppressed 
$V \phi \to V b\bar b$ production'' benchmark scenario
of table~\ref{benchmarks}. 
}
%
           \vskip2pc
%
\begin{center}
\centerline{\psfig{file=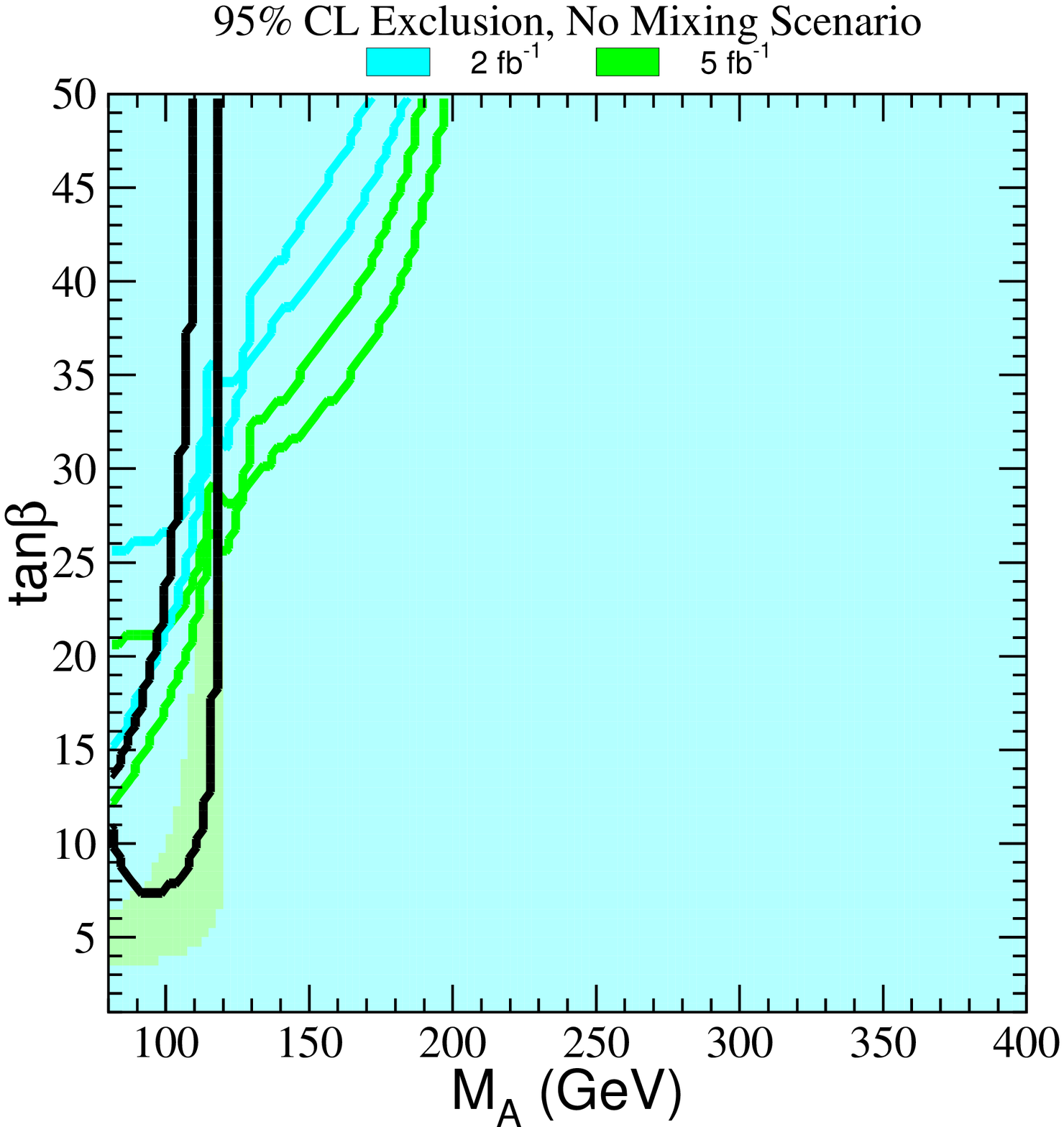,width=8cm}
\hfill
\psfig{file=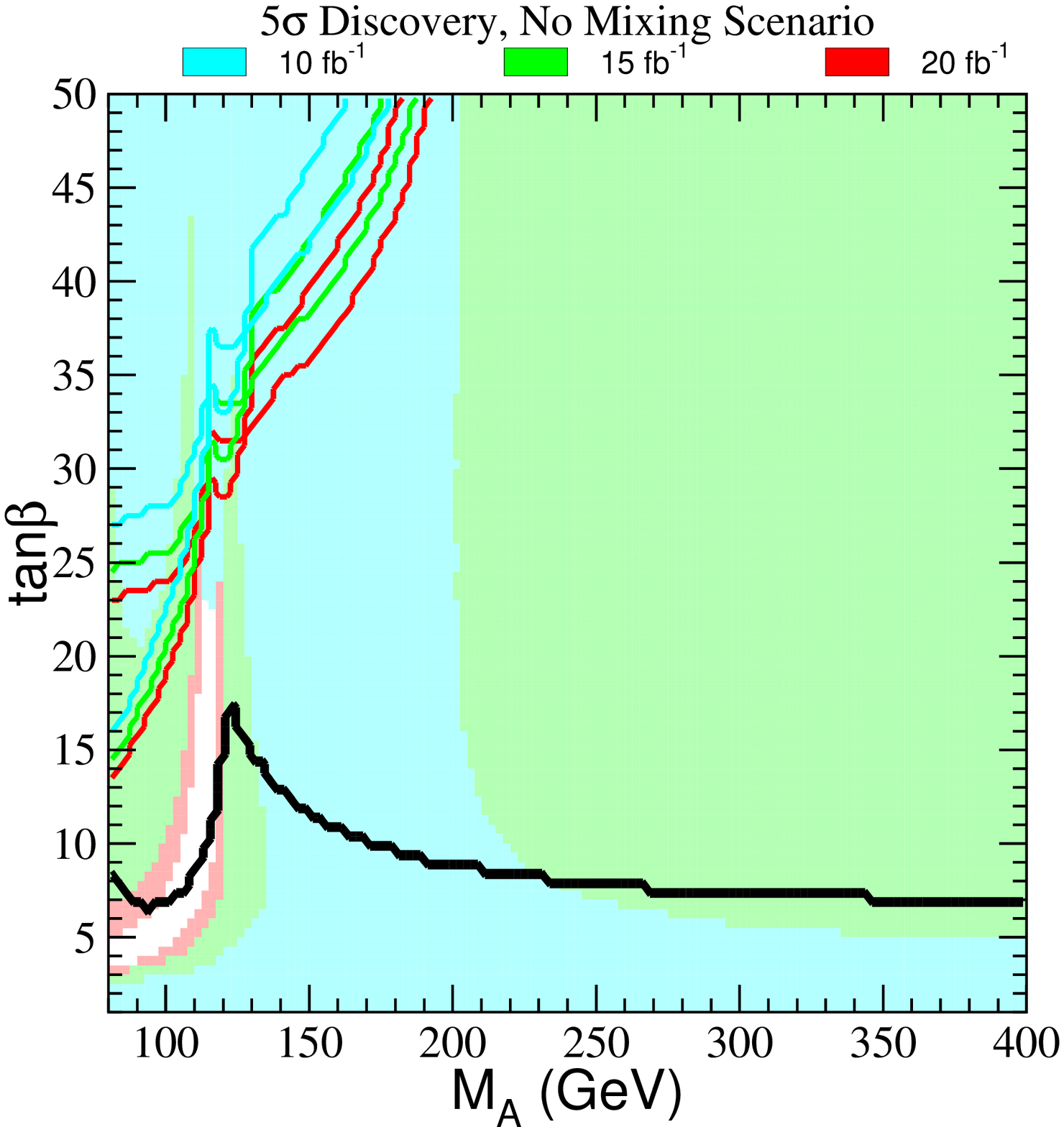,width=8cm}}
\end{center}
\caption[0]{\label{fullnomix95}
The same as \fig{fullmhmax95} but for the the no-mixing benchmark 
scenario of table~\ref{benchmarks}.  In (a), 
the region between the two solid black lines
is {\it not} excluded by the absence of observed $e^+e^-\to Z\phi$
events at LEP2, while in (b) the region unexcluded by LEP2 lies above
the solid black line.  
}
\end{figure}

In figs.~\ref{fullmhmax95}--\ref{fullnomix95}, 
we show the regions of 95$\%$ CL Higgs exclusion  
and $5\sigma$ Higgs discovery on the $\mha$--$\tan \beta$
plane, for representative MSSM parameter choices,
via the search of neutral Higgs bosons in the 
channels: $q\bar q\to V\phi$ [$\phi=\hl$, $\hh$], $\phi\to b\bar b$
(shaded regions) and 
$gg$, $q\bar q\to b\bar b\phi$ [$\phi=\hl$, $\hh$, $\ha$],
$\phi\to b\bar b$ (region in the upper left-hand corner bounded by
the solid lines), for different integrated 
luminosities as indicated by the color coding.
The two sets of lines (for a given color) bounding the regions accessible
by the $b\bar b\phi$ search correspond to the CDF and D\O\ simulations, 
respectively.  Finally, the solid black line (usually near the bottom
of each plot) reflects the upper limit of $\tan\beta$ (as a function
of $\mha$) deduced from the absence of observed $e^+e^-\to Z\phi$
events at LEP2.
The shaded regions presented in these figures reflects the results
of the SHW simulation of $q\bar q\to V\phi$
improved by neutral network techniques, as described in the generic
MSSM analysis presented in Section~III.B.2.  We do not display here
the effects of systematic error due to uncertainties in $b$-tagging 
efficiency, background, mass resolution and other effects.  The
implications of a 30\% reduction in detection efficiency has already
been illustrated by comparing lighter and darker color shading in
figs.~\ref{amaxnn}--\ref{aminnn}.

From our results, it follows that in the most demanding 
scenarios, (shown in \figs{fullmhmax95}{fullmix95}),
5~fb$^{-1}$ of integrated luminosity per experiment will allow one to
test nearly all of the MSSM Higgs parameter space at 95\% CL. 
Note the importance of the complementarity between the
$q\bar q\to V\phi$ and $q\bar q\to b\bar b\phi$ channels 
for improving the coverage of the
MSSM parameter space in the low $\mha$ region in \fig{fullmix95},
where there is a strong
reduction of ${\rm BR}({\phi \to b \bar b})$ for the  
CP-even Higgs boson $\phi$ with strongest coupling to the vector 
gauge bosons. 

To assure discovery of a CP-even Higgs boson at the 5$\sigma$ level,
the luminosity requirement becomes very important.
Figs.~\ref{fullmhmax95}(b), \ref{fullmix95}(b) and \ref{fullnomix95}(b) 
show that 
a total integrated luminosity of about 20~fb$^{-1}$ per experiment is 
necessary in order to assure a significant, although not exhaustive,
coverage of the MSSM parameter space. In general, we observe that
the complementarity between the two channels,  
$q\bar q\to V\phi$ and  $q\bar q\to b\bar b\phi$,   
is less effective in assuring discovery of a Higgs boson as compared
with a 95\% CL Higgs exclusion. 
This is due to 
the much higher requirement of total integrated  
luminosity combined with the existence of MSSM parameter regimes
which can independently suppress both Higgs production channels.
\Fig{fullmix95} exhibits one of the most difficult regions of
MSSM parameter space for Higgs searches at the Tevatron collider.
Nevertheless, even in this case,
a very high luminosity experiment can cover a significant fraction
of the available MSSM parameter space.

In conclusion, Run 2 of the Tevatron (with an integrated luminosity of
2~fb$^{-1}$) cannot significantly improve the MSSM Higgs mass limits
obtained at LEP.  However, if the integrated luminosity can be
increased to 10~fb$^{-1}$ and beyond, then substantial improvements
beyond the LEP reach are achievable.  Although 
the upgraded Tevatron cannot probe the entire MSSM Higgs parameter
space, integrated luminosities of 20--30~fb$^{-1}$ would provide close
to complete coverage for a 5$\sigma$ discovery of the Higgs boson.




\subsection{New Higgs Physics Beyond the Standard Model/MSSM}

\vspace{0.1in}
\small
\begin{center}
{\it M.C. Gonzalez-Garcia, S.M. Lietti and S.F. Novaes}
\end{center}
\normalsize\nopagebreak

Effective Lagrangians provide a formalism for a model-independent
description of new TeV-scale physics beyond the Standard Model.
An effective Lagrangian for gauge boson--Higgs boson
interactions was presented in \eqs{lagrangian}{H}.  
A set of new effective Higgs couplings are thus introduced
that in principle can be measured in future
experiments.  In the Standard Model, these effective couplings are
generated at the loop level.  Thus, to discover physics beyond the Standard
Model, one would have to measure these couplings with sufficient
accuracy to detect deviations from Standard Model predictions.

In this section, we discuss the present limits on anomalous Higgs
couplings and evaluate the improved limits attainable from searches at
the upgraded Tevatron.  Since the Higgs boson in the model-independent
approach is not strictly the Standard Model Higgs boson (although the
deviation from Standard Model behavior may be small), we shall
henceforth denote the Higgs boson by $H$ for the remainder of this
section. 

\vskip6pt\noindent
{\bf {Present Bounds from Searches at Tevatron Run 1 and LEP2}}

\vskip3pt
Anomalous Higgs boson couplings have been studied in Higgs and
$Z$ boson decays in $e^+ e^-$
\cite{ee,our:lep2aaa,our:NLCWWA,our:NLCZZA} and in $p \bar{p}$
collisions
\cite{our:tevatronjj,our:tevatronmis,our:tevatron3a,our:combined}.
The combined bounds on anomalous Higgs
boson interactions are derived from the following data:
\begin{equation}
\begin{array}{lll}
~~~ \mbox{Process  } &
~~~~~ \mbox{Anomalous Higgs Contribution}  &  ~~~ \mbox{Exp.\ Search}\\
&&\\
p\, \bar p  \rightarrow j \, j \, \gamma\, \gamma &  ~~~~~
p \bar{p} \to  W (Z) (\to j\, j) + H (\to \gamma \gamma) & ~~~~~
\mbox{D\O \protect\cite{d0jj}}
\\
p\, \bar p \rightarrow  \gamma \,\gamma \,+ \,\not \!\! E_T \nonumber &~~~~~
p \bar{p} \to  Z^0 (\to \nu \bar{\nu}) + H (\to \gamma \gamma) & ~~~~~
\mbox{D\O\protect\cite{d0miss}}  \\
& ~~~~~ p \bar{p} \to  W (\to [\ell] \nu) + H (\to \gamma \gamma) & ~~~~~
\nonumber \\
p\, \bar p  \rightarrow  \gamma\, \gamma\, \gamma & ~~~~~
p \bar p \rightarrow \gamma +H(\rightarrow \gamma \gamma) & ~~~~~
\mbox{CDF\protect\cite{cdf}} \\
e^+\, e^-  \rightarrow \gamma\, \gamma\, \gamma & ~~~~~
e^+ e^-  \rightarrow \gamma +H(\rightarrow \gamma \gamma) & ~~~~~
\mbox{OPAL  \protect\cite{opal:ggg} }
\\
\label{proc}
\end{array}
\end{equation}
Events containing two photons plus missing transverse energy 
($\not \!\! E_T$), additional photons or charged fermions represent 
a signature for several theories involving physics beyond the SM.
In the framework of anomalous Higgs couplings presented 
in \eqs{lagrangian}{H}, they
can arise from the production of a Higgs boson which
subsequently decays in two photons [second column in
Eq.(\ref{proc})].  
Recent analyses of these
signatures showed a good agreement with  the expectations from
the SM. Thus, these negative experimental  results can be used
to constrain new anomalous couplings in the bosonic sector of the
SM.
\begin{figure}
\centering
\centerline{\psfig{file=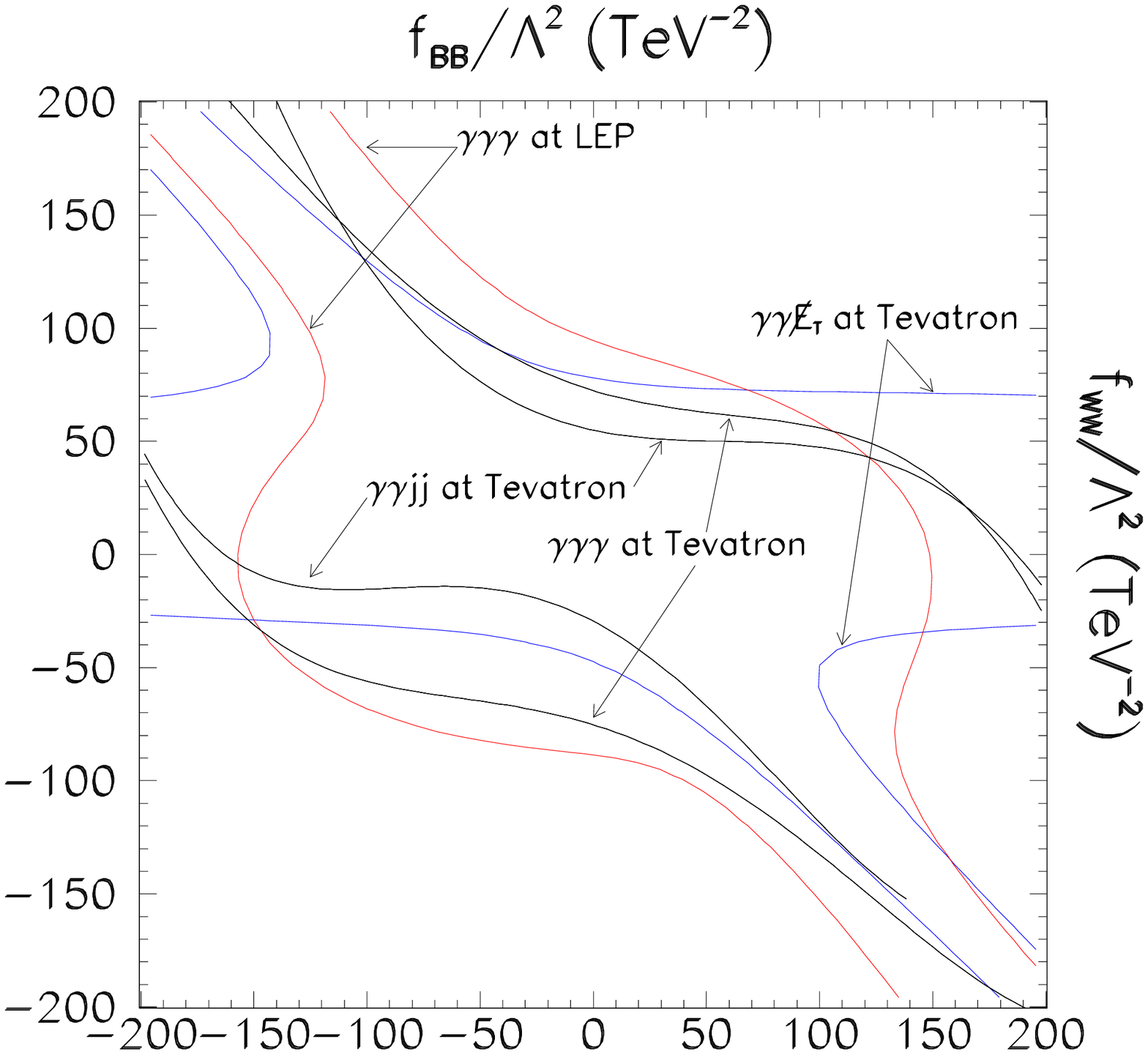,width=8cm}
\hfill
\psfig{file=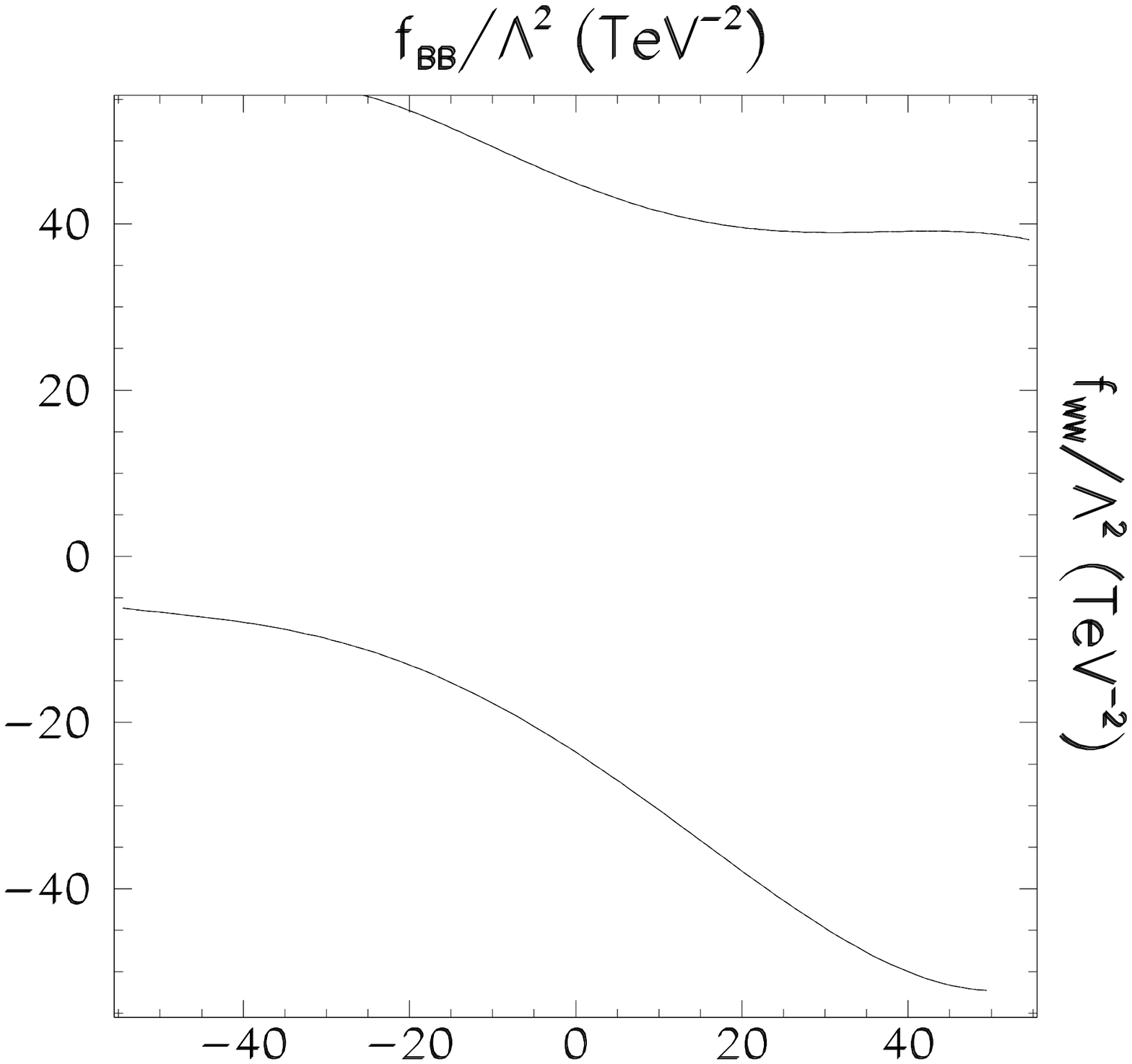,width=8cm}}
\caption[0]{\label{conchafig1} (a) Exclusion region outside the curves in the
$f_{BB} \times f_{WW}$ plane, in TeV$^{-2}$, based  on the D\O
~analysis \protect\cite{d0jj} of $\gamma\gamma j j$ production,
on  the D\O ~analysis \protect\cite{d0miss} of $\gamma\gamma \not
\!\! E_T$, on the CDF analysis \protect\cite{cdf} of
$\gamma\gamma\gamma$ production, and on the OPAL analysis
\protect\cite{opal:ggg} of $\gamma\gamma\gamma$ production,
always assuming $ M_H = 100$ GeV.  The curves show the 95\% CL
deviations from the SM total cross section. (b) Same as
(a) for the combined analysis.}
\end{figure}


D\O\ has searched for high invariant--mass 
photon pairs in $p \bar{p} \to \gamma
\gamma j j$ events \cite{d0jj} at $\sqrt{s}=1.8$ TeV and $100$
pb$^{-1}$ of integrated luminosity.
No events were observed with two--photon invariant mass  
in the range $100 < M_{\gamma\gamma}\lesssim 220$,
which constrains the anomalous Higgs boson couplings defined in
\eq{H} that lead to the process $p\bar{p} \to
W (Z) (\to j\, j)  + H (\to \gamma \gamma)$ \cite{our:tevatronjj}.
In the analysis of \Ref{concha},
the cuts of \Ref{d0jj} were applied and included the particle
identification and  trigger efficiencies. The searches
discussed here are limited to a
Higgs boson with mass in the range $100 < M_H \lesssim 220$,
since after the $WW(ZZ)$ threshold is reached the diphoton
branching ratio of Higgs is quite reduced.

D\O\ has also searched for events with two photons and large missing
transverse energy \cite{d0miss}.  No events were observed in the invariant
mass range $100 < M_{\gamma\gamma} \lesssim 2 M_W$.  This can
be used to constrain operators leading to the processes
$p\bar p\to Z(\to\nu\bar\nu) H(\to\gamma\gamma)$ {\it and}
$p\bar p\to W(\to\ell\nu) H(\to\gamma\gamma)$, where the
lepton is not detected \cite{our:tevatronmis}.
To compare with the D\O\ results 
\cite{d0miss}, the same cuts as in the last article in
\Ref{d0miss} were applied,
including the particle identification and trigger efficiencies
which vary from 40\% to 70\% per photon.
After these cuts, 80\% to 90\%
of the signal comes from associated $ZH$ production while
10\% to 20\% arises from $WH$ production. 

CDF has searched for events with three photons, requiring two photons in the
central region of the detector, with a minimum transverse energy
of 12 GeV, plus an additional photon with $E_T > 25$ GeV, and
a photon separation of at least $15^\circ$ \cite{cdf}.  These
results have been used to constrain operators leading to
$p\bar p\to \gamma H(\to\gamma\gamma)$.
These same operators can be constrained by results 
for electron--positron collisions \cite{our:lep2aaa}, in particular
the recent OPAL results \cite{opal:ggg}, where data
taken at several energy points in the range $\sqrt{s}=130$--$172$ GeV 
were combined.

The calculations have included all SM (QCD plus
electroweak), and anomalous contributions that lead to these
final states. The SM one-loop contributions to the
$H\gamma\gamma$ and $H Z\gamma$ vertices were introduced through
the use of the  effective operators with the corresponding form
factors in the coupling.  Neither the narrow--width approximation
for the Higgs boson  contributions, nor the effective $W$ boson
approximation were employed. The interference effect
between the anomalous signature and the SM
background have been included consistently. 
For all $p\,\bar p$ analyses, the  MRS (G)
\cite{mrs} set of proton structure functions were employed with the scale
$Q^2=\hat{s}$.

\begin{figure}
\centerline{\psfig{file=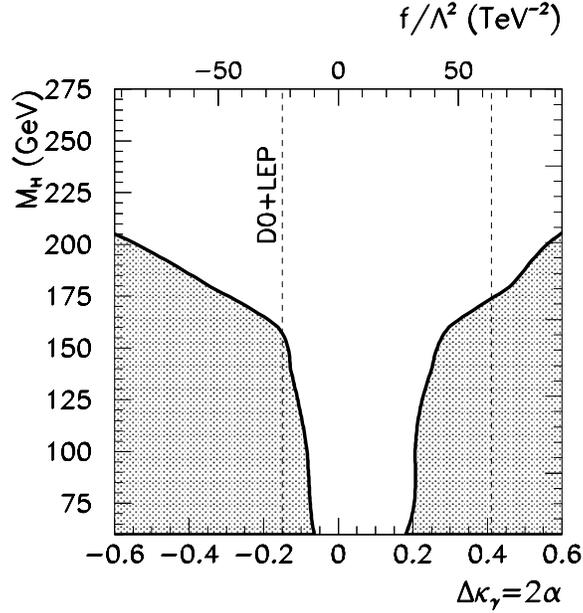,width=8cm}}
\caption{Excluded region in the $f \times M_H$
plane from the combined analysis of the LEP2 and Tevatron
searches.}
\label{kappa}
\end{figure}

The coupling $H\gamma\gamma$ [see \eq{g}] involves $f_{WW}$ and
$f_{BB}$ \cite{hagiwara2}.  As a consequence, the anomalous
signature $f\bar f \gamma\gamma$ is only possible when those
couplings are not vanishing. The couplings $f_B$ and $f_W$, on
the other hand, affect the production mechanisms for the Higgs
boson.  \Fig{conchafig1} (a) shows the
excluded  region in the $f_{WW}$, $f_{BB}$ plane from the
different channels  studied for $M_H = 100$ GeV, assuming that
these are the only non--vanishing couplings. Since the anomalous
contribution to  $H\gamma\gamma$ is zero for $f_{BB} = - f_{WW}$
(see \eq{g}), the bounds become very weak close to this
line, as is clearly shown in \fig{conchafig1}. 
Note that these bounds are based
on Poisson statistics. 

The results obtained from the analysis of the four reactions
(\eq{proc}) can be statistically  combined in order to constrain
the value of the coefficients $f_i$, $i=WW, BB,W, B$ of
(\eq{lagrangian}) \cite{our:combined}. 
\Fig{conchafig1} (b) exhibits the 95\% CL exclusion region in the plane $f_{BB}
\times f_{WW}$ obtained from combined results.

In order to reduce the number of free parameters, one can make the
assumption that all blind operators affecting the Higgs
interactions have a common coupling $f$.
\Fig{kappa} shows the combined limits for the coupling
constant $f = f_{BB}= f_{WW}= f_{B} = f_{W}$ for Higgs boson masses
in the range of $100 \leq M_H \leq 220$ GeV.

\vskip6pt\noindent
{\bf {Attainable Bounds at the Upgraded Tevatron}}

\vskip3pt
The effect of the anomalous operators becomes more evident with
the increase of energy, and therefore, higher sensitive to
smaller values of the anomalous coefficients can be achieved by
studying their contribution to different processes at the
upgraded Tevatron collider. 
Estimates for the upgraded Tevatron, 
with 1~fb$^{-1}$ and 10 fb$^{-1}$ respectively are considered here. 
For the reactions $p \bar{p} \to\gamma \gamma \not
\!\! E_T$ and $p \bar{p} \to \gamma \gamma j j$, Run 1 cuts
and efficiencies are used.
For the $\gamma\gamma\gamma$ final state, the effects of
additional kinematic cuts to
improve the sensitivity have been considered \cite{our:tevatron3a}.
The best results in this case are obtained for the following set of cuts:
$E_{T_{1}} > 40$ GeV, with $E_{T_{2,3}} > 12$ GeV, where 
the three photons are ordered according to their transverse energy,
{\it i.e.\/} $E_{T_{1}} > E_{T_{2}} > E_{T_{3}}$, and
$|\eta_{i}| < 1$.
For these estimates, the CDF Run 1 detection \cite{cdf}
efficiency for photons has been assumed.

\begin{table*} [h!]
\caption{95\% CL allowed range for $f/\Lambda^2$,  from
$\gamma\gamma\gamma$, $\gamma\gamma + E_T $,  $\gamma\gamma
j j $ production at the upgraded Tevatron 
with an integrated luminosity of 1~fb$^{-1}$ [10~fb$^{-1}$]
assuming all $f_i$
to be equal.  Limits worse than $|f|=100$
TeV$^-2$ are denoted by ---.}
\label{tab:run2}
\vskip6pt
\renewcommand{\arraystretch}{1.1}
\begin{tabular}{ccccc}
  &\multicolumn{4}{c}{$f/\Lambda^2$(TeV$^{-2}$)} \\ \cline{2-5}
\noalign{\vskip3pt}
$\;\;M_H$(GeV)\qquad & $p \bar{p} \to \gamma \gamma \gamma$  & $p \bar{p}
  \to \gamma \gamma +  E_T $  &
  $p \bar{p} \to \gamma \gamma j j$  & Combined\\[3pt]
\tableline
\noalign{\vskip3pt}
100     & ($-$24, 24)  [$-$13, 15]
        & ($-$16, 36)  [$-$9.4, 26]
        & ($-$9.2, 22) [$-$3.3, 5.6]
        & ($-$7.6, 19) [$-$3.0, 5.6]
\\
120     & ($-$26, 26)  [$-$14, 14]
        & ($-$20, 39)  [$-$15, 27]
        & ($-$8.6, 21) [$-$3.4, 5.9]
        & ($-$7.4, 18) [$-$3.3, 5.9]
\\
140     & ($-$30, 31)  [$-$15, 16]
        & ($-$25, 44)  [$-$14, 30]
        & ($-$10, 23)  [$-$4.5, 8.9]
        & ($-$9.1, 20) [$-$4.0, 8.7]
\\
160     & ($-$36, 38)  [$-$17, 19]
        & ($-$29, 50)  [$-$14, 33]
        & ($-$11, 24)  [$-$6.0, 14]
        & ($-$9.9,22)  [$-$5.1, 13]
\\
180     & (---, ---)  [---, ---]
        & ($-$63, 72) [$-$46, 53]
        & ($-$26, 34) [$-$16, 24]
        & ($-$24, 33) [$-$16, 24]  \\
200     & (---, ---)  [---, ---]
        & ($-$87, 90) [$-$50, 53]
        & ($-$33, 40) [$-$17, 23]
        & ($-$32, 39) [$-$17, 23] \\
220     & (---, ---)  [---, ---]
        & (---, ---)  [---, ---]
        & ($-$42, 45) [$-$19, 26]
        & ($-$42, 45) [$-$19, 26] \\[3pt]
\end{tabular}
\end{table*}

Table \ref{tab:run2} contains the 95\% CL  limit on the
anomalous couplings for the upgraded Tevatron 
for each individual process. All couplings are assumed equal  ($f =
f_{BB}= f_{WW}= f_{B} = f_{W}$) and the Higgs boson mass is
varied in the range $100 \leq M_H \leq 220$ GeV. Combination
of the results obtained from the analysis of the first three
reactions in \eq{proc} leads to the improved bounds given
in the last column of   Table \ref{tab:run2}. Comparing these
results with those in \fig{kappa}, we observe an improvement
of about a factor $\sim$ 2--3 [$\sim$ 4--6] for the combined
limits at the upgraded Tevatron with 1~fb$^{-1}$ [10~fb$^{-1}$]  
integrated luminosity.

As mentioned previously for linearly realized effective Lagrangians,
the modifications introduced in the Higgs boson and in the vector
boson sectors are related to each other.  Consequently, the
bounds on the new Higgs couplings should also restrict the
anomalous gauge--boson self interactions.  Under the assumption
of equal coefficients for all anomalous Higgs operators, 
the common Higgs boson anomalous coupling $f$ can
be related to the
conventional parameterization of the vertex $WWV$ ($V = Z^0$,
$\gamma$) \cite{hhpz},
\begin{equation}
\Delta \kappa_\gamma =\frac{M_W^2}{\Lambda^2}~ f =
\frac{2 \cos^2 \theta_W }{1 - 2 \sin^2 \theta_W}~
\Delta \kappa_Z= 2 \cos^2 \theta_W ~\Delta g^Z_1
\label{trad}
\end{equation}
A different set of parameters has been also used by the LEP
Collaborations in terms of  three independent couplings,
$\alpha_{B \Phi}$, $\alpha_{W \Phi}$, and $\alpha_{W}$, These
parameters are related to the parameterization of Ref.\
\cite{hhpz} through $\alpha_{B \Phi} \equiv \Delta \kappa_\gamma
- \Delta g^Z_1 \cos^2 \theta_W$, $\alpha_{W \Phi} \equiv \Delta
g^Z_1 \cos^2 \theta_W$, $\alpha_{W} \equiv \lambda_\gamma$.
Under  the relations given in \eq{trad} they verify
$\alpha_{B \Phi}=\alpha_{W \Phi}\equiv \alpha= \Delta \kappa_\gamma/2$.

The current experimental limit on these couplings
from combined results on double gauge
boson production at Tevatron and LEP2 \cite{vancouver}
is $-0.15\;<\;\Delta\kappa_\gamma\;=\;2\alpha \;<\;0.41$
at 95 \% CL. This limit is derived under the relations given in
\eq{trad} \cite{hisz} and it is plotted in \fig{kappa}
for comparison with the bounds obtained from this analysis.

Table \ref{tab:f5} contains the 95\% CL limit on the
anomalous coupling $\Delta\kappa_\gamma$ using the limits on
$f/\Lambda^2$ obtained  through the analysis of the processes
in \eq{proc}.  The expected bounds that are attainable
at the upgraded Tevatron are also shown. These 
results show that the
present combined limit from the Higgs production analysis
obtained here is comparable with the existing bound
from gauge boson production for $M_H \leq 170$ GeV.
\begin{table*}[t!]
\caption{95\% CL allowed range for the anomalous triple gauge boson
couplings derived from the limits obtained for the anomalous Higgs boson
coupling $f$ for $M_H= 100$ GeV.}
\label{tab:f5}
\vskip6pt
\setlength{\tabcolsep}{1.85pc}
\begin{tabular}{lc}
\multicolumn{1}{c}{Process}  & $ \Delta\kappa_\gamma = 2 \,\alpha= 2 \,\alpha_{B \Phi} =
2 \,\alpha_{W \Phi}$ \\ [3pt]
\tableline
$\;$Combined Tevatron Run 1 + LEP2         & ($-$0.084, 0.204) \\[0.1cm]
$\;$Combined Tevatron Run 2 (1 fb$^{-1}$)  & ($-$0.048, 0.122) \\ [0.1cm]
$\;$Combined Tevatron Run 2 (10 fb$^{-1}$) & ($-$0.020, 0.036) \\ [3pt]
\end{tabular}
\end{table*}


\section{Summary and Conclusions}                 The next run of the Tevatron proton-antiproton collider at Fermilab
will see enormous increases in luminosity, initially more than an
order of magnitude larger than those in Run~1.  With the new Main
Injector and Antiproton Recycler rings a total integrated luminosity
of 2 fb$^{-1}$ should be delivered to the CDF and D\O\ collider
experiments by the end of 2002.  Continuous running over multiple years 
and additional improvements to the
collider could result in a total data sample exceeding 10 fb$^{-1}$
prior to the start of the LHC.
Both CDF and D\O\ will have greatly improved
detectors,  with all new charged particle tracking and vertexing, 
improved calorimetry and triggering, and new offline analysis
software.   These improvements
to the accelerator and the detectors enhance the
sensitivity of the experiments for new particle searches.
In particular, given sufficient integrated luminosity, the discovery
of the Higgs boson at the upgraded Tevatron collider may be within reach.

In Run~1, the CDF and D\O\ collaborations began developing the
techniques for searching for the Higgs bosons of the Standard Model
(SM) and Minimal Supersymmetric Standard Model (MSSM).  The Tevatron
Run~1 integrated luminosity of about 100 pb$^{-1}$ was too small to
yield sufficient sensitivity to exclude or discover the SM Higgs.  In
the MSSM, certain Higgs fermion-couplings are enhanced (relative to
their Standard Model values) in some regions of MSSM parameter space.
As a result, one can already deduce some weak restrictions on the MSSM
Higgs parameters based on the Run~1 MSSM Higgs searches for neutral
Higgs production (in association with $b\bar b$ pairs) and charged
Higgs bosons (resulting from top decays).  This Report capitalizes on
the experience gained in Run 1 in estimating the discovery and
exclusion reach for the SM and MSSM Higgs bosons in Run~2, assuming
that CDF and D\O\ combine their results to maximize the sensitivity.

Estimating the discovery and exclusion reach for Higgs bosons requires
accurate knowledge of the search acceptance and expected backgrounds
in the various channels.  No detailed
simulation existed prior to this Workshop, 
and so the participants developed SHW---a simple Monte
Carlo simulation---to represent an ``average'' of the CDF and D\O\
experiments, taking into account the improvements to the detectors.
SHW simulates the detector response to
all the individual particles in the event, including tracking and
calorimeter efficiency, resolution, and geometry, but does not
simulate the effects of multiple interactions, the magnetic field,
lateral shower development, secondary interactions, or details of the
microvertex detectors.  Comparisons of the simulation with the Run 1
full simulation and data are encouraging, and show agreement at the
15-20\% level in acceptance.

Without the real Run~2 data, it is also difficult to get an
accurate estimate of the $b$ quark jet tagging efficiency and mistag
rate, the $\bb$ jet-jet mass resolution, and certain backgrounds that
must be estimated from actual data.  The philosophy in this study is
to make an optimistic yet realistic estimate of the Higgs boson 
discovery and exclusion reach, assuming improvements to 
these performance parameters
of the detectors and analyses.  These estimates are based on Run~1
experience, albeit with less capable detectors, and
detailed simulations of $b$-tagging at the planned Run~2 detectors.
Further work will be necessary to understand some of the relevant 
backgrounds with higher level of accuracy. The completion of the
background analyses will also require direct handling of the actual 
Run 2 data.  A more accurate estimate of the Run 2 Higgs discovery
and exclusion reach can only be achieved after the Tevatron
Run 2 experiments start collecting data.


\vspace{0.2in}
{\bf Search for the Standard Model Higgs Boson} \\

In the mass region of interest to the Tevatron 
(100~GeV$\lsim\mhsm\lsim 200$~GeV),
the SM Higgs boson is produced most copiously via $gg$ fusion, with a cross
section from about 1--0.1~pb.  For $\mhsm\lsim 135$~GeV, the Higgs boson decays
dominantly to $\bb$.  Since the cross section for the QCD
production of $\bb$ dijet events is orders of magnitude larger than
the Higgs production cross section, the  $gg \to h_{SM} \to b \bar b$  
channel is not a promising channel and thus
has not been investigated in this Report.
For $\mhsm\gsim 135$~ GeV, the Higgs boson decays
dominantly to $WW^{(*)}$ (where $W^{*}$ is a virtual $W$),
and the channel $gg\to \hsm\to WW^{(*)}$ is accessible to the
Tevatron Higgs search.

The next largest Higgs cross section is for the production of the SM
Higgs boson in
association with a vector boson.  The combined $W\hsm$ and $Z\hsm$ cross
sections is about 0.2---0.5~pb in the mass region of interest
($100\lsim\mhsm\lsim 135$~GeV), while the dominant Higgs decay is
$\hsm\to b\bar b$.  These processes lead to four main final
states: $\ell\nu\bb$,
$\nn\bb$, $\lpm\bb$, and $\qq\bb$.  Of these channels the first three
have distinct signatures on which the experiments can trigger
(high $p_T$ leptons and/or missing $E_T$) and the backgrounds are
controllable, typically dominated by vector-boson pair production,
$t\bar t$ production and QCD dijet production.  The backgrounds can in
all cases be estimated and/or cross-checked with actual data, allowing
the systematic error to remain roughly a constant fraction of
the total background. 
The signal efficiencies and backgrounds have all been estimated with
both the CDF Run~1 detector simulation and with the simple SHW
simulation.  In addition, the selection was optimized using neural
network techniques, resulting in a demonstrable gain 
in the significance of the Higgs signal for the
$\ell\nu\bb$ and $\nn\bb$ channels.

As noted above, the $b$-tagging efficiencies and the $\bb$ mass
resolution play a key role in determining the ultimate efficiency and
background rejection.  Much work remains, using real data studies, to
optimize the performance in both these areas.  In this regard, we have
been optimistic in this Report, assuming a ``loose'' $b$-tag
efficiency of 75\% at large jet $E_T$, and a ``tight'' $b$-tag
efficiency of 60\% at large jet $E_T$.  We believe that these are
realistic numbers, based on simulation studies.  For the mass
resolution, we have assumed in the final results a 10\% resolution on
the ``core'' of the distribution; inefficiencies due to the low side
tail remain after this rescaling of the simulation output.  This is
aggressive, but simulations suggest that a 10\% $b\bar b$ mass
resolution may be achievable.  Moreover, given CDF's demonstrated 12\%
$\bb$ mass resolution in $Z\to\bb$ events at Run~1, we believe that an
ultimate goal of a 10\% mass resolution is not overly optimistic.

For backgrounds we have used Monte Carlo estimates everywhere except
in the $\nn\bb$ channel, where there is a significant contribution
from QCD $\bb$ dijet production.  This background comes from the
extreme tail of a very large cross section, and is thus very difficult
to model.  In the CDF Run~1 analysis of this channel, QCD $\bb$ dijets
constituted half of the total background.  
D\O\ searches involving the same topologies have considerably lower   
background levels, despite the absence of a $b$ tag requirement,              
due to their better missing $E_T$ resolution.
To be conservative, we have taken the unknown
QCD $b\bar b$ dijet background to the $\nn\bb$ channel to be equal in
size to the sum of all other contributing background processes.

Larger Higgs masses ($\mhsm\gsim 135$~GeV) had not been studied in previous
Tevatron studies.  This Workshop produced several analyses aimed at
exploiting the distinct signatures present when the Higgs 
boson decay branching ratio to $WW^{(*)}$ becomes appreciable.
In this case, there are final states with $WW$ (from the
gluon-fusion production of a single Higgs boson), and
$WWW$ and $ZWW$ arising from associated vector boson--Higgs boson
production. Three
search channels emerged in the Workshop as potentially sensitive at
these high Higgs masses: like-sign dilepton plus jets ($\lsdilep$)
events, high-$p_T$ lepton pairs plus missing $E_T$ ($\lpm\nn$), and
trilepton ($\trilep$) events.  Of these, the first two were found to
be most sensitive.

Note that in this study, the same analysis selection has been assumed
for integrated luminosities ranging over an order of magnitude.  This
is not the typical pattern in new particle searches in high energy
physics: as the integrated luminosity increases, better understanding of
signal and background separation typically leads to the tightening of
selection cuts in order to keep background low while maintaining the signal
efficiency.  Of course, some backgrounds will remain irreducible, but it
is possible that the integrated luminosity thresholds could be
lowered with respect to those reported here, as experience is gained with
real data.

The final combined result of the integrated luminosity required to
exclude or discover the SM Higgs boson, as a function of the Higgs
mass, depicted in \fig{f:combined-final}, is obtained by combining the
sensitivity of all the channels described above, and assuming that the
data from the CDF and D\O\ experiments can be combined.  This result
is achieved by forming a joint likelihood from the product of the
Poisson probabilities of single-channel counting of signal and
background in each channel.  A nominal systematic error on the
expected background was taken into account by Bayesian integration.

The final result shows that for an integrated luminosity of
10~fb$^{-1}$, if the SM Higgs boson does not exist in the energy range
explorable at the Tevatron, then one can attain a 95\% CL exclusion
for masses up to about 180~GeV.  Moreover, if the SM Higgs happens to
be sufficiently light ($\mhsm\lsim 125$~GeV), then a tantalizing
$3\sigma$ effect will be visible with the same integrated luminosity.
With about 25~fb$^{-1}$ of data, $3\sigma$ evidence for the Higgs
boson can be obtained for the entire Higgs mass range up to 180~GeV.
However, the discovery reach is considerably more limited for a
5$\sigma$ Higgs boson signal.  Even with 30~fb$^{-1}$ of integrated
luminosity, only Higgs bosons with masses up to about 130~GeV can be
detected with 5$\sigma$ significance.  

Suppose that the Higgs boson mass is 115~GeV, which lies just above
the 95\% CL exclusion limit achieved by LEP \cite{tomjunk}.  At the
Tevatron, with 5~fb$^{-1}$ of integrated luminosity per experiment,
there would be sufficient data to see a 3$\sigma$ excess above
background, providing evidence for a Higgs boson.  With 15~fb$^{-1}$
of integrated luminosity per experiment, a $5\sigma$ discovery of the
Higgs boson would be possible.

The above integrated luminosity requirement crucially depend on the
success of a greatly improved $\bb$ mass resolution and $b$-tagging
rates (as compared to Run~1).  Nevertheless, these results clearly
provide strong motivation to the accelerator division to deliver the
largest possible integrated luminosity, and to the experimental
collaborations to optimize the data analyses and reconstruction
techniques, and continue to study new methods and ideas for enhancing
the sensitivity in this search.

Other Higgs signatures could help improve the sensitivity of the Higgs
search at the Tevatron.  In this Report, 
none of the channels corresponding to the $\tau^+\tau^-$ decay mode of
the Higgs have been investigated, as the small branching ratio 
(less than $8\%$) makes the corresponding signal rates small.  
Still, a significant improvement of $\tau$-lepton identification 
could lead to a viable Higgs signal in the low Higgs-mass 
($\mhsm\lsim 130$~GeV) regime.  This possibility deserves further
study.  Perhaps more promising is the detection of the Higgs boson via
$t\bar t\hsm$ production (the Higgs boson is radiated off the
top-quark), followed by $\hsm\to b\bar b$.  Initial studies
\cite{rainwater} suggest that this channel could be observable at the
upgraded Tevatron, with a statistical significance  
comparable to the Higgs signals in the $W\hsm$ and
$Z\hsm$ channels.  Clearly, if this is confirmed by a more detailed
analysis, it would improve the prospects for a Higgs signal at the
Tevatron, perhaps even lowering significantly the total integrated
luminosity required for evidence of a Higgs signal or a discovery.

\vspace{0.2in}
{\bf Search for Higgs Bosons of Low-Energy Supersymmetry} \\

The possibility that nature is supersymmetric 
at the TeV scale opens up several new
experimental alternatives in the search for Higgs bosons.  The Minimal
Supersymmetric Standard Model (MSSM) requires two Higgs doublets, 
which yields five physical scalars: two CP-even neutral scalars
($\hl$ and $\hh$), a CP-odd scalar ($\ha$) and a
charged Higgs pair ($ \hpm$).  At tree level, 
the masses and couplings of these particles are
governed by two parameters, $\tan\beta\equiv v_2/v_1$ (the ratio
of the two neutral Higgs field vacuum expectation values)
and one of the Higgs masses (usually chosen to be $\mha$).
When loop effects are taken into account, one finds that 
Higgs masses and couplings depend additionally on various
supersymmetric mass parameters.  Such effects can be significant in
certain regions of the MSSM parameter space.   These include
an increase in the upper bound of the mass of the lightest
CP-even Higgs boson ($\mhl\lsim 130$~GeV, assuming an average top
squark mass no larger than 1~TeV), a possible suppression or
enhancement of the dominant decay
rate $\phi\to b\bar b$ (where $\phi=\hl$, $\hh$, and/or $\ha$),
and CP-violating effects that can mix the CP-even and CP-odd eigenstates.
 
In various regimes of the $\mha$---$\tan\beta$ parameter space, the
phenomenology of the Higgs bosons can be very different.  For example,
in the regime where $\mha\gg m_Z$, known as the decoupling limit, the
light scalar $\hl$ possesses the same production cross-sections and
decay rates as $\hsm$, and is thus indistinguishable from the SM Higgs
boson.  In this case, the SM Higgs search analyses apply, while all
other Higgs bosons are too heavy and not observable at the Tevatron.
In contrast, if $\mha\sim {\cal O}(m_Z)$, then the $\hl$ properties
differ significantly from those of $\hsm$.  In this parameter range,
it may be possible to detect more than one Higgs state at the
Tevatron, or demonstrate that an observed Higgs signal does not
correspond to the one expected from the SM Higgs boson.

In the latter case, we can still make use of the Standard Model Higgs
analyses as follows.  For each Higgs production/decay channel,
we first define theoretical ratios (denoted
collectively by $R$) of $\sigma\times {\rm BR}$ for Higgs production
and decay in the MSSM relative to its Standard Model value.   
Then, we can derive a lower limit for the required luminosity $L$ in order
to obtain a Higgs signal of fixed significance ({\it e.g.} 95\% CL
limit or a $5\sigma$ discovery).  If $R=1$, the result for $L$ is
simply the one obtained in the Standard Model Higgs search analysis.
If $R\neq 1$, then we expect $R\sqrt{L}$ to be constant in the
Gaussian limit, which yields a lower limit for $L$ as a function of
$R$ (say, at fixed Higgs mass).  In the regime where Poisson
statistics is required, this scaling is somewhat modified; the
corresponding $R$-plots were given in Section III.B.1.

The MSSM analysis proceeds as follows.  The Higgs signatures arising from 
associated vector boson--Higgs boson production followed by Higgs decay
to $b\bar b$, treated in the SM analysis,
also apply to the case of the CP-even Higgs bosons of the MSSM.  For a
given point in supersymmetric parameter space (which depends on
$\tan\beta$, $\mha$, and other supersymmetric mass parameters, such as
the third generation squark masses, the gaugino masses, the 
trilinear couplings $A_t$, $A_b$ and higgsino mass parameter $\mu$,
which govern the size of the
radiative corrections), one can compute the corresponding $R$ values
for the associated vector boson--Higgs boson production and subsequence
Higgs decay to $b\bar b$.  Comparing with the $R$-plots previously
obtained, one can then either exclude or discover a CP-even Higgs
boson of a given mass for a fixed integrated luminosity.  In this way, one can
map out the MSSM Higgs parameter space in the $\mha$---$\tan\beta$
plane (under various assumptions for the values of the other supersymmetric
parameters that govern the impact of the loop-effects) and determine
which parameter regimes are accessible to the Tevatron Higgs search at
a given luminosity.

We have studied three examples: 
(i)~the top-squark (stop) mixing parameters is set to zero, which
minimizes the maximal value of the radiatively-corrected light CP-even
Higgs mass (the so-called ``minimal-mixing'' or ``no-mixing'' scenario);
(ii)~the stop mixing parameters are chosen to maximize 
the value of the lightest CP-even Higgs mass
(the ``maximal-mixing'' scenario); and (iii)~the stop mixing 
parameters and the gluino mass are chosen in order to suppress 
the $b\bar b$ coupling of the Higgs state that is primarily
produced in association with the gauge bosons
(the ``suppressed $V\phi\to Vb\bar b$ production'' scenario).
Case~(iii) exemplifies a possible worst case scenario for the MSSM Higgs
search at the upgraded Tevatron.
In the case of minimal mixing, a Higgs boson produced in association
with the gauge bosons may be excluded at the 95$\%$ CL with as little as
5~fb$^{-1}$ of data per experiment, 
and can be discovered at the $5\sigma$ level
over nearly the entire region of the $m_A$---$\tan\beta$
plane with less than 20~fb$^{-1}$ per experiment.  However, in the other 
two cases mentioned above, our results
show that an integrated luminosity
of order 20---30~fb$^{-1}$ is required to obtain 
a 5$\sigma$ discovery over a significant portion of the parameter
space.   Yet, even in these cases, 
10~fb$^{-1}$ is sufficient to exclude the Higgs sector of the MSSM at 
the 95$\%$ CL, for most values of $m_A$ and $\tan\beta$
(independent of the values of the stop mixing parameters, assuming
that the average top squark mass is not much larger than 1~TeV), if no
Higgs signal is detected.  However, in the suppressed 
$V\phi\to Vb\bar b$ production scenario, there is a non-negligible
region of the MSSM parameter space in which even a 95$\%$ CL Higgs 
exclusion limit is not achievable at 30~fb$^{-1}$.
Note that if the Higgs signature due to $t\bar t\hsm$ (with $\hsm\to
b\bar b$), examined in \Ref{rainwater}, proves to be a viable
discovery mode, then the analogous MSSM Higgs signatures could improve the
upgraded Tevatron coverage of the MSSM Higgs parameter space described above.

In the large $\tan\beta$ regime of the MSSM, some of the
neutral Higgs boson couplings to down-type fermions (such as
$b\bar b$) are enhanced, and lead to production cross
sections for Higgs bosons that are significantly larger than those of the
Standard Model Higgs boson.
As a result, in general two of the neutral Higgs
bosons of the MSSM can also be discovered in an additional channel:
$b\bar b\phi$ production, followed by $\phi\to b\bar b$ (where
$\phi$ is either the CP-odd Higgs boson or the neutral CP-even
Higgs boson with enhanced coupling to $b\bar b$,
at large $\tan\beta$).  The corresponding channel cannot be observed
in the case of the SM Higgs boson, since the $\hsm b\bar b$ coupling
is suppressed by a factor of $m_b/m_W$ and has no additional
enhancement.  The production of a neutral MSSM Higgs boson in association
with a $\bb$ pair followed by Higgs decay to $b\bar b$ leads to a
distinctive final state with four high-$E_T$ $b$ jets.  These events
can be observed by positively tagging at least three of these
$b$-jets, which dramatically reduces the QCD multijet background.  The
cross section for this process is proportional to $\tan^2\beta$, and
is thus greatly enhanced at large $\tan\beta$.  The studies presented
in this Workshop show that an interesting region of the MSSM parameter
space at high $\tan\beta$ is accessible in Run~2.  In particular, some
limits already emerge with 2~fb$^{-1}$ of data in the region
of high $\tan\beta$ and low $\mha$.  The region of sensitivity grows
to slightly lower values of $\tan\beta$ and larger values of $\mha$ as
the luminosity is increased further.  However, one finds that
there are large supersymmetric and QCD corrections that must be taken
into account, which will require detailed theoretical analysis in the
coming years.  Nevertheless, the enhanced $b\bar b\phi$ production
process increases the overall sensitivity of the Tevatron MSSM Higgs
search, and provides some complementary coverage in the
$\mha$---$\tan\beta$ plane with respect to the associated vector
boson--Higgs boson production discussed above.

Finally, if the charged Higgs boson is light enough so that the decay
$t\to H^+b$ is allowed, then the Tevatron Higgs boson search at Run~2 
can discover the charged Higgs boson if $\tan\beta$ is sufficiently
larger than 1 (or smaller than 1, although the latter is disfavored by
theoretical considerations).  Again, it is possible to see a signal at
Run~2 at large values of $\tan\beta$ and small values of $\mha$ (the
latter is required in order that $\mhpm<m_t-m_b$).  With additional
luminosity, the region of sensitivity in $\tan\beta$ also grows,
although the ultimate region of MSSM parameter space
at $\tan\beta$ accessible by the charged Higgs search is not as great
as the corresponding region accessible by the search for enhanced
$b\bar b\phi$ production.

\vspace{0.2in}
{\bf Conclusions} \\

The search for the Higgs boson and the dynamics responsible for
electroweak symmetry breaking is the central goal of high
energy physics today.  The Tevatron experiments, if given sufficient
data, are poised to make major advances in meeting this goal.  A
great deal of effort remains in order to 
raise the performance of the accelerator to the level 
demanded by the Higgs search.  The nominal yearly integrated
luminosity for Run~2 is 2~fb$^{-1}$ per year.  However,
10---30~fb$^{-1}$ of data (per experiment) is needed to significantly
improve the Higgs search beyond the current LEP2 limits, with the
higher luminosity crucial for maximizing the coverage both in the
Standard Model and its supersymmetric extension.

Further challenges must be met in bringing the detectors on line and
fully operational, and in developing the techniques and understanding,
particularly in $\bb$ jet-jet mass reconstruction and $b$ jet tagging,
necessary to extract the faint signal of the Higgs boson from the
larger Standard Model background.  In some cases, we were perhaps
optimistic with regard to the expected capabilities of the detector.
However, this Report clearly points to some important goals that the
experimental detectors and analysis methods must achieve if the Higgs
search at the Tevatron is to be successful.  In some cases, the
magnitude of the Standard Model backgrounds are not known at the
required level of accuracy.  Additional theoretical work along with
background studies once higher luminosity data become available will
be crucial for improving the Higgs search strategies and maximizing
the chances for uncovering the Higgs signal.  We believe
that the results obtained in this Report provide a useful attempt to
devise a realistic search strategy for Higgs bosons, 
and provide a benchmark for future improvements.
 
The LHC is expected to begin its physics run in 2006.  The search for
the Higgs boson is one of the primary missions of the ATLAS and CMS
detector collaborations.  With much larger annual luminosities and
energies, they will discover the Higgs boson (in most
theoretical scenarios) if it has not yet been observed.  There is a window of
opportunity for the Tevatron that will be open for the next six to
seven years.  With a little luck and a great deal of determination,
the Higgs boson can be discovered at a sufficiently upgraded Tevatron.
The potential physics payoff is great, and provides a strong motivation to
do the work necessary to meet this challenge in the pre-LHC era.

\newpage\appendix
\large
\centerline{\bf APPENDIX}\normalsize
\section{Statistical Method for Combining Channels} \small
\begin{center}
{\it John Conway, 
     Howard Haber,
     John Hobbs,
     Harrison Prosper}
\end{center}
\normalsize\nopagebreak

The statistical method employed here for combining channels follows an
essentially Bayesian approach by forming a joint likelihood for a
given set of experimental outcomes, as a function of a dummy
multiplier $f$ on the theoretical cross section.

For each search channel $i$ assume that we expect $s_i$ events from
the Higgs signal and $b_i$ events from all backgrounds, in 1 fb$^{-1}$
integrated luminosity.  These values, then, can equivalently be thought
of as the accepted signal and background cross sections in units of
femtobarns.

In the actual experiments, after some integrated luminosity L has been
recorded, one expects L$s_i$ signal and L$b_i$ background in a given
channel $i$, and observes some number $n_i$.  The Poisson probability
${\cal P}(n_i|\mu_i)$ for observing $n_i$ events, where
$\mu_i\equiv{\rm L}(s_i+b_i)$ events are expected in this channel is
well known:
\begin{eqnarray}
     {\cal P}(n_i|\mu_i) = \frac{\mu_i^{n_i} e^{-\mu_i}}{n_i!} \ \ .
\end{eqnarray}
The joint likelihood ${\cal L}(n_1,n_2,...|s_1,s_2,...,b_1,b_2,...)$ 
for all channels is simply the product of these probabilities:
\begin{eqnarray}
     {\cal L}(n_1,n_2,...|s_1,s_2,...,b_1,b_2,...) = 
              \prod_i \frac{\mu_i^{n_i} e^{-\mu_i}}{n_i!} \ \ .
\end{eqnarray}
This likelihood tells us the absolute probability of our particular
experimental outcome, but gives no information about whether or not
a signal is actually present.  To accomplish this one can introduce 
a factor $f$ which multiplies the expected signal, and determine 
confidence intervals on its true value from a probability density
function (p.d.f.) based on the joint likelihood above.  Thus, we
now have $\mu_i(f)\equiv{\rm L}(fs_i+b_i)$, and from Bayes'
Theorem we find that this ``posterior'' p.d.f. can be written as
\begin{eqnarray}
     {\cal P}(f|\bar{n},\bar{s},\bar{b}) = 
          \frac{\displaystyle \prod_i \mu_i(f)^{n_i} e^{-\mu_i(f)}/n_i!}
               {\displaystyle \prod_i \int_0^\infty \mu_i(f^\prime)^{n_i} e^{-\mu_i(f^\prime)}/n_i! df^\prime}
\end{eqnarray}
where we have expressed the sets $\{n_1,n_2,...\}$, $\{s_1,s_2,...\}$,
and $\{b_1,b_2,...\}$ as vectors for clarity.  If there is no Higgs
then the function ${\cal P}(f|\bar{n},\bar{s},\bar{b})$ will be
maximum at $f=0$, and fall more and more sharply as the integrated
luminosity increases.  If, on the other hand, there is a Higgs (at 
the mass were are considering), then the posterior p.d.f. will be
more and more sharply peaked at $f=1$ as L increases.  

For a given integrated luminosity L and a given experimental outcome,
then, how can one use the posterior p.d.f. in $f$ to either claim a
discovery at some level of significance or exclude the predicted Higgs
cross section with come confidence?  

For discoveries, one usually quotes significance in units of $\sigma$,
(say, $3\sigma$ or $5\sigma$) where it is understood that this refers
to equivalent Gaussian standard deviations.  In the method used here
one finds that value of $f$ at which ${\cal P}(f|\bar{n},\bar{s},\bar{b})$ 
is a maximum, ${\cal P}_{max}$, and
compares with the value ${\cal P}_0$ evaluated for $f=0$.  That is,
one compares the maximum likelihood to that of the null
hypothesis. One can then find the number of ``equivalent Gaussian
$\sigma$'' from the ratio of these values:
\begin{eqnarray}
  \frac{{\cal P}_{max}}{{\cal P}_0} = \frac{e^0}{e^{-x^2/2\sigma^2}}
\end{eqnarray}
and so the significance $x$ can be written as
\begin{eqnarray}
  x = \left( 2 \ln \frac{{\cal P}_{max}}{{\cal P}_0} \right)^{\frac{1}{2}}\sigma \ \ .
\end{eqnarray}

For exclusion limits, or confidence intervals on $f$ in general, one 
simply performs integrals of the posterior p.d.f., which by definition
is normalized to unity.  Thus if we desire a 95\% confidence level
exclusion, we need to find that value of $f$ for which
\begin{eqnarray}
   \epsilon = \int_f^\infty {\cal P}(f^\prime |\bar{n},\bar{s},\bar{b}) df^\prime
\end{eqnarray}
where $\epsilon=0.05$.  If, for example, there is no Higgs and we find
that we can exclude at 95\% CL any value of $f>1$, then we conclude that
the Higgs is excluded at that confidence level, since $f=1$ corresponds
to the Standard Model prediction for the Higgs cross section.

One can also take into account systematic errors using this method, by
convoluting Gaussians representing the relative uncertainties in the
$s_i$ and $b_i$ with the expression for the joint likelihood.  In the
method used here this is approximated by a coarse numerical integration
in the likelihood calculation.

The main goal of the Higgs studies presented here, however, is to
estimate the integrated luminosity needed to reach certain thresholds
of significance.  But clearly for a given set of expected number of
signal and background events, at a certain integrate luminosity, there
are many possible future experimental outcomes, some of which will 
and some of which will not meet the desired level of significance.
Following the precedent of the LEP-II Working Group, we simply seek
that integrated luminosity at which 50\% of the possible future 
outcomes would meet the desired level.

To actually calculate these integrated luminosity thresholds requires
simulating a large ensemble of pseudoexperiments, each representing a
certain outcome, distributed according to their actual probability of
occurrence.  Then, one varies the integrated luminosity until the 50\%
criterion is met.   Note that in the case of determining 95\% CL
limits, the program generates no signal events in each
pseudoexperiment, whereas in the case of determining discovery
thresholds, the program does generate signal events in each
pseudoexperiment.

A few notes about the statistical method are in order.  Firstly, in
the case of a single channel with no systematic errors, for setting
95\% CL limits, the Bayesian method outlined above is in exact
agreement with the commonly used frequentist method described in the
1996 PDG Review of Particle Properties.   In that approach one
excludes at some confidence level $1-\epsilon$ any value of $f$
greater than that for which
\begin{eqnarray}
   \epsilon = \frac {\displaystyle \sum_{j=0}^n (fs+b)^n e^{-(fs+b)}/n! }
                    {\displaystyle\sum_{j=0}^n b^n e^{-b}/n! } \ \ .
\end{eqnarray}
This formula simply asks for that level of expected signal such that
the probability of seeing $n$ or fewer events, and having $n$ or fewer
background, is less than some value $\epsilon$.

Also, in the limit of high statistics, the results of the Bayesian
method outlined above agree closely with those from what one might
call the ``naive Gaussian approximation.''  This approximation uses
the values of the $s_i$ and $b_i$ to form ``significances''
$S_i(L)\equiv \sqrt{L}s_i/\sqrt{b_i}$, and then finds those values of L
for which $S(L)=\sqrt{\sum_i S_i^2(L)}$ exceeds some value, namely 3 or
5 for $3\sigma$ and $5\sigma$ discoveries, respectively, and 1.96 for
a 95\% CL limit.  This commonly used approximation breaks down, of
course, in the low-statistics regime where the distributions are
Poisson and not well approximated by Gaussians.




\begin{references}
\bibitem{benchmark2}
The maximal mixing and no-mixing
benchmark scenarios introduced in \Ref{benchmark} are
employed here with one minor change---the gluino mass is chosen 
in this section to be 1~TeV rather than 800~GeV.  
This has a very small impact on the final Higgs mass results.
\end{references}
\end{document}